%% file: intro-pub.tex
\documentclass[reqno,11pt]{amsbook}
\usepackage[utf8]{inputenc}
\usepackage{amsmidx}
\usepackage[linktocpage]{hyperref}
\usepackage{graphicx}
\usepackage{amscd}
\usepackage{slashed}
\usepackage{amssymb}
\usepackage{esint}
\usepackage{enumitem}
\usepackage{dsfont}
\usepackage[off]{auto-pst-pdf} % schnelle Version, Bilder müssen dazu schon vorliegen

\usepackage{tikz}
\usetikzlibrary{cd} % for commutative diagrams

\usepackage{tikz}

\usepackage[mathscr]{eucal}

\numberwithin{equation}{chapter}
\allowdisplaybreaks[1]
\definecolor{labelkey}{gray}{.65}

\title[Introduction to the Fermionic Projector and Causal Fermion Systems]{An Introduction to
the Fermionic Projector and Causal Fermion Systems}

\newtheorem{Exercise}{Exercise}[chapter]

\newtheorem{Def}{Definition}[section]
\newtheorem{Thm}[Def]{Theorem}
\newtheorem{Prp}[Def]{Proposition}
\newtheorem{Lemma}[Def]{Lemma}
\newtheorem{Remark}[Def]{Remark}
\newtheorem{Corollary}[Def]{Corollary}
\newtheorem{Example}[Def]{Example}

\definecolor{brown}{rgb}{.6,.3,0}
\definecolor{green}{rgb}{0,0.5,0}
\definecolor{yellow}{rgb}{1,1,0}
\definecolor{darkyellow}{rgb}{0.7,0.9,0}
\definecolor{red}{rgb}{0.7,0,0}

\newcommand{\Cb}{\textcolor{blue!80!black}}

\newcommand{\beq}{\begin{equation}}
\newcommand{\eeq}{\end{equation}}
\newcommand{\beqNo}{\begin{equation*}}
\newcommand{\eeqNo}{\end{equation*}}
\newcommand{\Proof}{\begin{proof}}
\newcommand{\QED}{\end{proof} \noindent}
\newcommand{\QEDrem}{\ \hfill $\Diamond$}
\newcommand{\la}{\langle}
\newcommand{\ra}{\rangle}
\newcommand{\bra}{\mathopen{<}}
\newcommand{\ket}{\mathclose{>}}
\newcommand{\Sl}{\mbox{$\prec \!\!$ \nolinebreak}}
\newcommand{\Sr}{\mbox{\nolinebreak $\succ$}}

\newcommand{\C}{\mathbb{C}}
\newcommand{\R}{\mathbb{R}}
\newcommand{\1}{\mathds{1}}
\newcommand{\Z}{\mathbb{Z}}
\newcommand{\N}{\mathbb{N}}
\newcommand{\Pdd}{\mbox{$\partial$ \hspace{-1.2 em} $/$}}
\newcommand{\Aslsh}{\slashed{A}}

\newcommand{\V}{\mathscr{V}}

\DeclareMathOperator{\Tr}{Tr}
\DeclareMathOperator{\tr}{tr}
\renewcommand{\O}{{\mathscr{O}}}
\renewcommand{\L}{{\mathcal{L}}}
\newcommand{\Sact}{{\mathcal{S}}}
\newcommand{\T}{{\mathcal{T}}}
\newcommand\B{{\mathscr{B}}}
\newcommand{\U}{\text{\rm{U}}}
\newcommand{\SU}{\text{\rm{SU}}}
\newcommand{\SO}{\text{\rm{SO}}}

\newcommand{\Cisc}{C^\infty_{\text{\rm{sc}}}}
\newcommand{\Cisco}{C^\infty_{\text{\rm{sc}},0}}
\newcommand{\Dir}{{\mathcal{D}}}
\DeclareMathOperator{\supp}{supp}
\renewcommand{\H}{\mathscr{H}}
\newcommand{\nuslsh}{\slashed{\nu}}

\newcommand{\Lin}{\text{\rm{L}}}
\newcommand{\F}{{\mathscr{F}}}

\DeclareMathOperator{\Symm}{\mbox{\rm{Symm}}}
\newcommand{\nablaLC}{\nabla^\text{\tiny{\tt{LC}}}}
\newcommand{\DLC}{D^\text{\tiny{\tt{LC}}}}

\DeclareMathOperator{\norm}{| \hspace*{-0.1em}| \hspace*{-0.1em}|}
\newcommand{\K}{{\mathscr{K}}}
\newcommand{\D}{\mathscr{D}}
\DeclareMathOperator{\re}{Re}
\DeclareMathOperator{\im}{Im}
\newcommand{\SD}{S^\text{\tiny{\rm{D}}}}
\newcommand{\I}{{\mathcal{I}}}
\DeclareMathOperator{\Pexp}{Pexp}

\DeclareMathOperator{\sdot}{\cdot}

\newcommand{\spc}{\qquad}

\newcommand{\reg}{\text{\rm{reg}}}
\newcommand{\as}{{\mathfrak{a}}}
\newcommand{\bs}{{\mathfrak{b}}}

\newcommand{\Sig}{\mathscr{S}}
\newcommand{\scrM}{\myscr M}
\newcommand{\scrN}{\myscr N}
\newcommand{\scrU}{{\mathscr{U}}}
\newcommand{\scrA}{{\mathscr{A}}}
\newcommand{\m}{{\mathfrak{m}}}
\newcommand{\n}{{\mathfrak{n}}}
\newcommand{\M}{{\mathbb{M}}}
\newcommand{\itemD}{\item[{\raisebox{0.125em}{\tiny $\blacktriangleright$}}]}
\newcommand{\PP}{\text{\rm{PP}}}
\newcommand{\J}{\mathfrak{J}}
\renewcommand{\u}{\mathfrak{u}}
\renewcommand{\v}{\mathfrak{v}}
\newcommand{\w}{\mathfrak{w}}
\newcommand\calB{{\mathcal{B}}}
\newcommand{\Jdiff}{\mathfrak{J}^\text{\rm{\tiny{diff}}}}
\newcommand{\Jtest}{\mathfrak{J}^\text{\rm{\tiny{test}}}}
\renewcommand{\div}{{\rm{div}}\,}
\newcommand{\Jin}{\mathfrak{J}^\text{\rm{\tiny{in}}}}

\newcommand{\Jlin}{\mathfrak{J}^\text{\rm{\tiny{lin}}}}
\newcommand{\Jvary}{\mathfrak{J}^\text{\rm{\tiny{vary}}}}
\newcommand{\Gdiff}{\Gamma^\text{\rm{\tiny{diff}}}}
\newcommand{\Gtest}{\Gamma^\text{\rm{\tiny{test}}}}

\newcommand{\Ctest}{C^\text{\rm{\tiny{test}}}}

\newcommand{\pseudo}{\Gamma}

\newcommand{\G}{\mathscr{G}}

\newcommand{\Spin}{\text{\rm{Spin}}}
\newcommand{\p}{\mathfrak{p}}
\newcommand{\q}{\mathfrak{q}}
\newcommand{\lla}{\langle}
\newcommand{\rra}{\rangle}
\newcommand{\llb}{\big\langle\!\langle}
\newcommand{\rrb}{\rangle\!\big\rangle}
\renewcommand{\j}{\mathfrak{j}}
\newcommand{\s}{{\mathfrak{s}}}
\newcommand{\GL}{{\rm{GL}}}
\newcommand{\Comm}{{\mathscr{C}}}

\newcommand{\tmax}{{t_{\max}}}
\newcommand{\A}{\myscr A}

\newcommand{\E}{\mathrm{e}}
\newcommand{\Diff}{\mathrm{d}}

\newcommand{\loc}{\mathrm{loc}}
\newcommand{\cI}{\mathrm{i}}

\DeclareMathSymbol\dAlembert  {\mathop}{AMSa}{"03}
\newcommand{\dyn}{{\text{\rm{dyn}}}}
\newcommand{\fermi}{{\mathrm{{f}}}}
\newcommand{\circa}{\overset{\,\circ\!}{A}}
\newcommand{\NORM}{\|}

\newcommand{\nindex}{\index{notation}}
\newcommand{\tindex}{\index{theme}}
\newcommand{\sindex}{\index{subject}}
\makeindex{notation}
\makeindex{theme}
\makeindex{subject}

\newcommand{\bitem}{\begin{itemize}[leftmargin=2em]}
\newcommand{\eitem}{\end{itemize}}

\DeclareFontFamily{OT1}{rsfso}{}
\DeclareFontShape{OT1}{rsfso}{m}{n}{ <-7> rsfso5 <7-10> rsfso7 <10-> rsfso10}{}
\DeclareMathAlphabet{\myscr}{OT1}{rsfso}{m}{n}

\newcommand\Felix[1]{}
\newcommand\Sebastian[1]{}
\newcommand\Joh[1]{}
\newcommand\JH[1]{}
\newcommand\Evtl[1]{}

\newcommand{\todo}[1]{}

\newcommand{\ci}{\mathrm{i}\;\!}

\newcommand{\dd}{\,\mathrm{d}}

\begin{document}

%\maketitle
\frontmatter
\title{{{\vspace*{2.0cm}
\Huge{Causal Fermion Systems} \\[0.3em]
\Large{An Introduction to Fundamental Structures, \\ Methods and Applications} \\[1.5em]
\large{Felix Finster, Sebastian Kindermann and Jan-Hendrik Treude} \\[1.5em]
{\large{\href{https://www.cambridge.org/core/books/causal-fermion-systems/CCA6DE1E1F4DA3AC0EF6729664A5D5B9}{Cambridge Monographs on Mathematical Physics} \\
\rm{Cambridge University Press, October 2025}}} \\ \large{\rm{ISBN: 978-1009632621}} \\[12cm]
\large{\rm{Former working title: \\[-0.8em] An Introductory Course on Causal Fermion Systems}}
}}}
\maketitle

\vspace*{-4cm}

\noindent
Felix Finster \\
Fakult\"at f\"ur Mathematik \\
Universit\"at Regensburg \\
Regensburg, Germany \\[0.2em]
{\tt{finster@ur.de}} \\ \\
Sebastian Kindermann \\
Comenius Gymnasium Deggendorf \\
Deggendorf, Germany \\[0.2em]
{\tt{sebastian.kindermann@schule.bayern.de}} \\ \\
Jan-Hendrik Treude \\
Fachbereich Mathematik und Statistik \\
Universit\"at Konstanz \\
Konstanz, Germany \\[0.2em]
{\tt{jan-hendrik.treude@uni-konstanz.de}} \\[2cm]

\begin{tabular}{l}
\\[6cm]
\end{tabular}

\setcounter{page}{4}

\mainmatter

%!TEX root = cfs.tex
\frontmatter
\setcounter{page}{5}

\input{preface}

\tableofcontents

\mainmatter
\setcounter{chapter}{0}

\include{part1-p}

\include{part1-m}
\include{part2}
\include{part3}
\include{part4}
\include{appendix}
\backmatter

\bibliographystyle{amsplain}
%\bibliography{../felix}
\bibliography{intro}

\Printindex{notation}{Notation Index -- in order of appearance}
\Printindex{theme}{Notation Index -- thematic order}
\Printindex{subject}{Subject Index}

\include{backmatter}

\end{document}

%% file: preface.tex
%!TEX root = intro.tex
\chapter*{Preface}
The theory of causal fermion systems is an approach to fundamental physics.
In different limiting cases, causal fermion systems give rise to the standard model of particle physics
and gravity on the level of classical field theory~\cite{cfs} as well as to quantum field theory~\cite{fockfermionic,
fockdynamics}. In view of these results, causal fermion systems are a promising
candidate for a unified physical theory. The dynamics of a causal fermion system is described by a novel variational principle: the causal action principle.
From the mathematical perspective, causal fermion systems provide a general framework for describing
non-smooth geometries and for formulating and analyzing dynamical equations in this non-smooth setting.

This book is intended as an easily accessible introduction to the theory of
causal fermion systems. After giving the physical and mathematical
background (Part 1), the theory of causal fermion systems is introduced (Part 2).
We proceed by providing mathematical methods that can be regarded as
a toolbox for analyzing causal fermion systems (Part 3).
We conclude with an outlook on the applications (Part 4).

In order to address as large an audience as possible, the book contains extensive preliminaries that cover both
physical and mathematical aspects. We have two typical audiences in mind when
writing these preliminaries: physicists with only basic knowledge of mathematics and mathematicians 
without physical background.

The book is based on three main resources: First, the lecture notes of the spring school ``Relativistic Fermion Systems'' held in
Regensburg in April 2013, adapted for the spring school ``Causal Fermion Systems'' held in Regensburg in March 2016.
Second, the lecture ``Causal Variational Principles''
given at the University of Regensburg in the summer semester 2017.
Finally, the online course ``An Introduction to Causal Fermion Systems'' held in the summer semester 2021.

We would like to thank the participants of the spring schools and the
students in the above lectures for valuable feedback.
In particular, we are grateful to Jonas Bierler,
David Cherney, Franz Gmeineder, Stefan Lippoldt, Marcin Napi{\'o}rkowski, Simon
Reinhardt, Julien Sabin and Andrea Sch\"atzl for valuable feedback.
Moreover, we are grateful to Sami Abdallah, Marvin Becker, Shane Farnsworth, Patrick Fischer,
Christoph Krpoun, Magdalena Lottner, Valter Moretti,
Heiko von der Mosel, Claudio Paganini, Marco van den Beld Serrano and Johannes Wurm
for helpful comments on the manuscript.
A special thanks goes to Johannes Kleiner and Marco Oppio for helping with
the lecture notes and providing many exercises.
We are grateful to the Deutsche Forschungsgemeinschaft (DFG) for financial support.
Finally, we would like to thank Nicholas Gibbons and the publishing team of Cambridge University
Press for the excellent collaboration.

\hspace*{1cm}

\hfill November 2024

\hfill Felix Finster, Sebastian Kindermann and Jan-Hendrik Treude

\chapter*{How to use this book}
This book is addressed to both mathematicians and physicists interested in the subject.
We want to address young students and researchers on the master or graduate level.
But the book should be helpful to the senior researcher as well.

The book is divided into four parts, each consisting of several chapters.
Part~\ref{partone} provides the necessary physical and mathematical preliminaries. Here the presentation
is quite brief, and we refer to the standard textbooks for more details. We selected the material
with the focus on what is most essential for causal fermion systems.
We also introduce the conventions and notations which will be used later in the book.
The content of Chapter~\ref{Geometry} can be omitted by a reader who wants to concentrate
on systems without gravity in Minkowski space.

Part~\ref{parttwo} introduces the main concepts and structures. In Chapter~\ref{secbrief}
we motivate and define causal fermion systems and explain the fundamental structures.
This chapter is essential for all the later parts of the book and should be read first.
In the following chapters of Part~\ref{parttwo} the structures of a causal fermion system
are explained in more detail, also setting the state for the later analysis.

In Part~\ref{partthree} we introduce the mathematical methods for the analysis of causal fermion systems.
The different methods can be understood as a toolbox, from which the reader may choose
depending on her interests and needs. The chapters in this part are self-contained,
except for obvious dependencies (for example, the energy methods for the linearized
field equations in Chapter~\ref{seclinhyp} build on similar methods for symmetric hyperbolic
systems in Chapter~\ref{secshs}). We note that the methods presented in this book are by no means
exhaustive; we concentrate on the main methods that have been fruitful so far.

Part~\ref{partfour} provides additional examples and gives an outlook on the physical applications.
Here the presentation is a bit more sketchy than in Parts~\ref{parttwo} and~\ref{partthree}.
The reason is that, after being familiar with Parts~\ref{parttwo} and~\ref{partthree} of
the present book, the reader should be well-prepared for delving into the research articles.
Moreover, the content of Chapter~\ref{seccl} is covered
in detail in the textbook~\cite{cfs}. Therefore, the purpose of this chapter is merely to give a non-technical overview.
The content of Chapter~\ref{secQFT}, on the other hand, is a field of active research.
Therefore, it seems preferable to present this material systematically and in more detail
at a later stage in a separate textbook.

Every chapter is supplemented by a section with exercises.
Studying these exercises is important for getting familiar and deepening the understanding of
the material. Hints on how to solve the problems should simplify the self-study.

We finally note that part of the material of this book is complemented by videos of an online
course, which are available on the website \\
\centerline{\href{https://www.causal-fermion-system.com/online\_course}{\rm{www.causal-fermion-system.com/online\_course}}}
We hope that the reader will enjoy reading and learning from this book.
Feedback is always welcome.

%% file: part1-p.tex
%!TEX root = intro.tex
\tindex{ba@{\bf{Bilinear and Sesquilinear Forms:}}}%
\tindex{d0@{\bf{Norms:}}}%
\tindex{f0@{\bf{Function Spaces:}}}%

\part{Physical and Mathematical Background} \label{partone}

\chapter{Physical Preliminaries}
\label{PhysicalPreliminaries}
In this chapter, we summarize some basics on quantum
mechanics and relativity theory as needed in order to understand the
physical content and context of the theory of causal fermion systems.
We also fix our conventions and introduce the notation that will be used consistently
throughout this book. Clearly, reading this summary cannot
replace studying quantum mechanics and relativity theory in detail. To
this end, we will cite various standard physics textbooks along the way.

\section{The Schr\"odinger Equation}
We begin by recalling a few basics of non-relativistic quantum
mechanics. For more details, we refer to standard textbooks
like~\cite{sakurai, schwabl1, landau3}.

The state of a quantum mechanical particle without spin is described
by its {\em{wave function}}
\sindex{wave function}%
$\psi: \R \times \R^3 \to \C$,
$(t,\vec{x}) \mapsto \psi(t,\vec{x})$, where~$t \in \R$ describes time
and~$\vec{x} \in \R^3$ position. Its absolute square~$|\psi(t,\vec{x})|^2$
has the interpretation as the {\em{probability density}}
\sindex{probability density}%
of the particle to be located at the position~$\vec{x}$ at time~$t$. For this
interpretation to be sensible, the integral over the probability
density must be equal to one,
\begin{equation}
    \label{normalize}
    \int_{\R^3} |\psi(t, \vec{x})|^2 \dd^3x = 1\:.
\end{equation}
This equation must hold at any time~$t \in \R$. This entails that the
dynamical equations must preserve the integral in~\eqref{normalize}, as
will be discussed in more detail shortly.

A basic tenet of quantum mechanics is the {\em{superposition principle}}.
\sindex{superposition principle}%
It states that for any wave functions~$\psi$ and~$\phi$, also
their complex linear combination
\beq \label{clin}
\tilde{\psi} = \alpha \psi + \beta \phi \qquad \text{with~$\alpha, \beta \in \C$}
\eeq
(defined pointwise by~$\tilde{\psi}(t,\vec{x}) = \alpha \psi(t,\vec{x}) + \beta \phi(t,\vec{x})$)
is a physically admissible wave function. Thus, in more mathematical terms,
the physical wave functions form a complex vector space.
Evaluating~\eqref{normalize} for the wave function~$\tilde{\psi}$ and
using that the probability integral must be preserved in time for
all~$\alpha$ and~$\beta$, one concludes that the integral
\begin{equation}
    \label{print0}
    \int_{\R^3} \overline{\phi(t, \vec{x})} \,\psi(t,
    \vec{x}) \dd^3x
\end{equation}
must be time independent for any wave functions~$\psi$ and~$\phi$. The
procedure to deduce~\eqref{print0} from~\eqref{normalize} is sometimes
referred to as {\em{polarization}}.
\sindex{polarization}%
The integral~\eqref{print0}
defines a scalar product on the wave functions, which we denote by
\sindex{scalar product!on wave functions}%
\begin{equation}
    \label{sprod}
    \la \phi | \psi \ra_\H
    := \int_{\R^3} \overline{\phi(t, \vec{x})} \,\psi(t, \vec{x})
    \dd^3x \:.
\end{equation}
From the mathematical point of view, the most natural complex vector
space for this scalar product is the Hilbert space~$L^2(\R^3, \C)$ of
square-integrable functions, which we also denote
\sindex{Hilbert space!of wave functions}%
by~$(\H, \la .|.\ra_\H)$ (for basics on Hilbert spaces see Section~\ref{sechilbertintro} below).
This Hilbert space contains for instance all smooth
functions with compact support (i.e.\ which vanish outside a compact set;
here {\em{smooth}} means that the function is differentiable to every order).

The dynamics of the wave function is described by a linear evolution
equation on~$\H$, the \textit{Schr\"odinger equation},
\sindex{Schr\"odinger equation}%
which is of first order in time and whose general form is
\begin{equation}
    \label{scham}
   \ci \partial_t \psi = H \psi \:.
\end{equation}
Here~$H$, the so-called {\em{Hamiltonian}}, is a linear operator
\sindex{Hamiltonian}%
acting on the Hilbert space~$\H$. The linearity of the Schr\"odinger
equation is essential in order to ensure that the time
evolution is compatible with the superposition principle. The
requirement that the scalar product~\eqref{sprod} must be time independent
implies that
\begin{equation}
    \label{symmc}
    0 = \partial_t \la \phi | \psi \ra_\H
    = -\ci \big( \la H \phi | \psi \ra_\H - \la \phi | H \psi \ra_\H \big)
\end{equation}
for all wave functions~$\psi, \phi$. In other words, the Hamiltonian
must be a {\em{symmetric}} operator on the Hilbert space~$\H$
(for mathematical basics see Definition~\ref{defsymmetric} below; all
mathematical issues like domains and the distinction between
symmetric and selfadjoint operators are postponed to
Section~\ref{secspectral}).

In the simplest setting (more precisely, for a particle without spin), the
Hamiltonian has the form
\beq
    H = - \frac{1}{2m}\: \Delta + V \:,
\eeq
where we chose units such that Planck's constant and the speed or light are equal to one,
\beq \hbar=c=1 \eeq
(we will do so throughout this book).
Here, $\Delta = \partial_1^2+\partial_2^2+\partial_3^2$ is the
Laplacian on~$\R^3$, and~$V(t,\vec{x})$ is a real-valued potential
which acts on wave functions by multiplication. The parameter~$m>0$
is the rest mass of the particle.

The Schr\"odinger equation can be analyzed using various methods. If
the potential is time independent, it can be
solved by exponentiating,
\beq
    \psi(t) = \E^{-\ci H t}\: \psi(0) \:,
\eeq
where the exponential may be defined using the spectral theorem (for
details see Section~\ref{secspectral} below). In this case, the
dynamics of~$\psi$ can be related to spectral properties of the
Hamiltonian. Another method, which has the advantage that it also
applies if the potential depends on time, is to make use of the fact
that the time evolution forms a strongly continuous semigroup of
operators (see for example~\cite[Section~34]{lax}). Alternatively, one
can analyze the Schr\"odinger equation as a parabolic partial
differential equation. Since our focus is the relativistic equations,
these methods are not covered in this book. But we refer the interested
reader to the textbooks~\cite[Chapter~6]{taylor1}
or~\cite[Section~II.7.1]{evans}.

\section{Special Relativity and Minkowski Space}
\label{SRT}
We now give a brief introduction to special relativity. For more
details, in particular on the physical background, we recommend the
textbooks~\cite{rindlerMink, naber, landau2}.

In special relativity, space and time are combined into a
four-dimensional {\em{spacetime}}. Mathematically, this
four-dimensional spacetime is described by Minkowski space
\sindex{Minkowski space}%
\sindex{spacetime!Minkowski}%
\nindex{aaa@$(\scrM, \la.,. \ra)$ -- Minkowski space}%
$(\scrM, \la.,. \ra)$, a real four-dimensional vector space endowed
with an inner product~$\la . , . \ra$ of signature
$(+ \ \!\! - \ \!\! - \ \! - )$. In~$\scrM$ one may always choose a
basis~$(e_i)_{i=0,\ldots,3}$ satisfying~$\la e_0,e_0 \ra = 1$ and
$\la e_i,e_i \ra = -1$ for~$i=1,2,3$. Such a basis is called a
pseudo-orthonormal basis or \emph{reference frame}, 
\sindex{reference frame}%
since the corresponding coordinate system~$(x^i)$ describes time and space 
for an observer in a system of inertia. We also refer to
$t := x^0$ as time and denote spatial coordinates by
$\vec{x}=(x^1, x^2, x^3)$. Representing two vectors
$\xi,\eta \in \scrM$ in such a basis as
$\xi = \sum_{i=0}^3 \xi^i e_i$ and~$\eta = \sum_{i=0}^3 \eta^i e_i$,
the inner product takes the form
\sindex{Minkowski metric}%
\nindex{aab@$\la .,. \ra$ -- Minkowski metric}%
\tindex{bb@$\la .,. \ra$ -- Minkowski metric}%
\nindex{aac@$g_{jk}$ -- Minkowski metric in components}%
\begin{equation}
    \label{minsp}
    \la \xi, \eta \ra
    = \sum_{j,k=0}^3 g_{jk}\: \xi^j\: \eta^k\:,
\end{equation}
where~$g_{jk}$, the {\em{Minkowski metric}}, is the diagonal matrix
\beq g={\mbox{diag}}\,(1,-1,-1,-1) \:. \eeq
We note that the origin of Minkowski space is not distinguished (apart from the fact that it
can be regarded as the origin of the observer). This could be formalized mathematically by regarding~$\scrM$
as an {\em{affine vector space}}. Here, we prefer to regard~$\scrM$ simply as a vector space, noting
that the translation~$\scrM \rightarrow \scrM + u$ by a vector~$u \in \scrM$ corresponds to a symmetry of
spacetime.

In what follows we usually use \emph{Einstein's summation convention},
\sindex{Einstein summation convention}%
according to which one omits the sign for sums and always sums over any
pair of indices appearing twice, one being an upper and one a lower
index. For instance, with this convention, the relation~\eqref{minsp} is written
simply as~$\la \xi,\eta \ra = g_{jk}\xi^j \eta^k$. By~$g^{ij}$ we
denote the inverse of the Minkowski metric, which in a
pseudo-orthonormal basis is again the diagonal matrix~$\text{diag}(1,-1,-1,-1)$.
We \emph{raise and lower indices} using the Minkowski metric and its
inverse, meaning that for a vector~$\xi = \xi^i e_i$, we set
$\xi_i := g_{ij} \xi^j$ for any~$i=0,\ldots,3$, and we also write
$\partial^j = g^{jk} \partial_k$. Finally, we sometimes abbreviate the
Minkowski inner product by writing~$\xi \eta := \la \xi, \eta \ra$ and
$\xi^2 := \la \xi, \xi \ra$.

\begin{figure}[tb]
    \psscalebox{1.0 1.0} % Change this value to rescale the drawing.
    {
      \begin{pspicture}(0,27.804865)(5.2980175,31.474598)
          \definecolor{colour3}{rgb}{0.8,0.8,0.8}
          \pspolygon[linecolor=colour3, linewidth=0.04, fillstyle=solid,fillcolor=colour3](0.058286134,27.834866)(1.8582861,29.634865)(3.658286,27.834866)
          \pspolygon[linecolor=colour3, linewidth=0.04, fillstyle=solid,fillcolor=colour3](0.058286134,31.434866)(3.658286,31.434866)(1.8582861,29.634865)
          \psline[linecolor=black, linewidth=0.04](0.058286134,31.434866)(3.658286,27.834866)
          \psline[linecolor=black, linewidth=0.04](0.058286134,27.834866)(3.658286,31.434866)
          \psline[linecolor=black, linewidth=0.03, arrowsize=0.05291667cm 2.0,arrowlength=1.4,arrowinset=0.0]{<->}(4.058286,31.434866)(4.058286,30.234865)(5.258286,30.234865)
          \psbezier[linecolor=black, linewidth=0.02, arrowsize=0.05291667cm 2.0,arrowlength=1.4,arrowinset=0.0]{->}(3.5582862,29.094866)(3.3282862,29.214867)(3.0882862,29.164865)(2.7582862,28.89486572265625)
          \rput[bl](2.15,29.5){$0$}
          \rput[bl](1.75,30.5){$I^\vee$}
          \rput[bl](1.75,28.5){$I^\wedge$}
          \rput[bl](4.15,31.1){$\xi^0$}
          \rput[bl](4.95,30.35){$|\vec{\xi}|$}
          \rput[bl](3.65,28.95){$L$}
      \end{pspicture}
    }
    \caption{The causal structure of Minkowski space.}
    \label{figlightcone}
\end{figure}
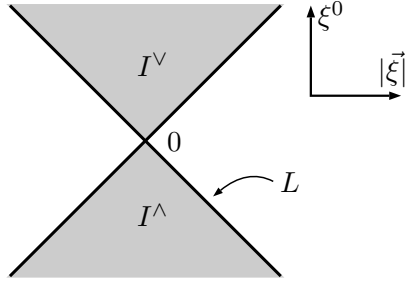

The sign of the Minkowski metric encodes the causal structure
of spacetime. Namely, a vector~$\xi \in \scrM$ is said to be
\sindex{causality!in Minkowski space}%
\sindex{causal structure!of Minkowski space}%
\sindex{spacelike vector}%
\sindex{timelike vector}%
\sindex{lightlike vector}%
\begin{equation}
    \label{causalMink}
    \left\{
        \begin{array}{cll}
          {\mbox{\em{timelike}}} &\quad& {\mbox{if~$\la \xi, \xi \ra >0$}} \\
          {\mbox{\em{spacelike}}} && {\mbox{if~$\la \xi, \xi \ra <0$}} \\
          {\mbox{\em{lightlike}}} && {\mbox{if~$\la \xi, \xi \ra =0$}}\:.
        \end{array}
    \right.
\end{equation}
Lightlike vectors are also referred to as {\em{null}} vectors.
\sindex{null vector}%
Moreover, the term {\em non-spacelike} refers to timelike or lightlike
vectors. The timelike and null vectors form a double cone. Its
boundary
\beq
    L := \{ \xi \in \scrM \,|\, \la \xi, \xi \ra =0\}
\eeq
\nindex{aad@$L$ -- light cone}%
is referred to as the {\em{light cone}}. \sindex{light cone}%
Physically speaking, the
light cone is formed of all light rays through the origin of~$\scrM$. Similarly, the timelike vectors correspond to velocities
slower than the speed of light; they form the
\beq
    \text{\em{interior light cone}} \qquad
    I := \{ \xi \in \scrM \,|\, \la \xi, \xi \ra > 0 \} \:.
\eeq
\sindex{light cone!interior}%
\nindex{aae@$I$ -- interior light cone}%
Likewise, the non-spacelike vectors form the 
\beq
    \text{\em{closed light cone}} \qquad
    J := \{ \xi \in \scrM \,|\, \la \xi, \xi \ra \geq 0 \} = I \cup L\:.
\eeq
\sindex{light cone!closed}%
\nindex{aaf@$J$ -- closed light cone}%
We denote the future and past light cones by superscripts~$\vee$
and~$\wedge$, i.e.\
\begin{align}
  J^\vee &:= \{ \xi \in \scrM \,|\, \la \xi, \xi \ra \geq 0,\;
           \xi^0 \geq 0 \} \\
  J^\wedge &:= \{ \xi \in \scrM \,|\, \la \xi, \xi \ra \geq 0,\;
             \xi^0 \leq 0 \} \:,
\end{align}
\nindex{aag@$J^\vee, J^\wedge$ -- closed future and past light cone}%
\nindex{aah@$I^\vee, I^\wedge$ -- open future and past light cone}%
\sindex{future!in Minkowski space}%
\sindex{past!in Minkowski space}%
\sindex{light cone!future}%
\sindex{light cone!past}%
and similarly for~$I$. These notions are illustrated in
Figure~\ref{figlightcone}.

The spacetime trajectory of a moving object is described by a curve
$q(\tau)$ in Minkowski space (with~$\tau$ an arbitrary parameter). 
We always assume that the parametrization is {\em{regular}}, meaning
that the tangent vector~$\dd q/\dd\tau$ to the spacetime curve is non-zero for all~$\tau$.
\sindex{regular curve}%
We say that the curve~$q(\tau)$ is timelike if its tangent vector
is everywhere timelike. Spacelike, null, and non-spacelike curves are defined analogously. 
The usual statement of causality, which says that no information can travel faster than the speed of
light, can then be expressed as follows:
\sindex{causality!in Minkowski space}%
\medskip
\begin{quote}
{\em{Causality}}: \quad \parbox[t]{9.5cm}{
Information can be transmitted only along
non-spacelike curves.}
\end{quote}\medskip
In view of this notion, a non-spacelike curve is also referred to as a {\em{causal curve}}.
\sindex{causal curve}%
The set of all points that can be joined to a given spacetime point
$x$ by a non-spacelike curve is precisely the closed light cone
centered at~$x$, denoted by~$J_x := J - x$. It is the union of the two
single cones
\begin{align}
J_x^\vee &= \{ y \in \scrM \,|\, (y-x)^2 \geq 0,\;
(y^0-x^0) \geq 0 \} \\
J_x^\wedge &= \{ y \in \scrM \,|\, (y-x)^2 \geq 0, \;
(y^0-x^0) \leq 0 \} \:,
\end{align}
interpreted as the points in the causal future and
past of~$x$, respectively. Therefore, we refer to~$J^\lor_x$ and
$J^\land_x$ as the closed {\em{future}}
\sindex{light cone!future}%
and {\em{past light cones}}
\sindex{light cone!past}%
centered at~$x$, respectively. The
sets~$I^\lor_x$, $I^\land_x$ and~$L^\lor_x$, $L^\land_x$ are introduced similarly.
We remark that the resulting relations like ``lies in the timelike future of''
or ``lies in the causal future of'' are transitive (see Exercise~\ref{extransitive}).

Special relativity demands that physical equations be {\em{Lorentz invariant}}.
\sindex{Lorentz invariance}%
Qualitatively speaking, this means that they must be
formulated in a manner independent of the choice of reference frame.
More concretely, this independence can be formulated in terms of transformation laws:
Recall that a reference
frame is an orthonormal basis of Minkowski space. Any two reference
frames~$(e_i)_{i=0,\ldots,3}$ and~$(\widetilde{e}_i)_{i=0,\ldots,3}$
are related to each other by a linear transformation~$\Lambda \in \Lin(\scrM)$
(here~$\Lin(\scrM)$ are the linear transformations on~$\scrM$; clearly, $\Lambda$
can be written as a~$4 \times 4$-matrix with real-valued entries)
which preserves the Minkowski metric, i.e.\ one has (again using the Einstein
summation convention)
\begin{equation}
    \label{eq:lorentz-transformation-of-reference-frame}
    \widetilde{e}_i = \Lambda_i^j \,e_j
    \qquad\text{and}\qquad
    \Lambda^\ell_j \,\Lambda^m_k \,g_{\ell m} = g_{jk} \,.
\end{equation}
The coordinates in the old and new reference frames and the
corresponding partial derivatives are related to each other by
\begin{equation}
    \label{eq:lorentz-transformation-of-coordinates-and-derivatives}
    \widetilde{x}^i = \Lambda^i_j\, x^j
    \qquad\text{and}\qquad
    \frac{\partial}{\partial \widetilde{x}^i}
    = \Lambda_i^j\, \frac{\partial}{\partial x^j} \,.
\end{equation}
The Lorentz transformations form a group
(with the group operation being the composition of the linear transformations or, equivalently,
matrix multiplication of the corresponding matrices), the so-called {\em{Lorentz group}}. 
\sindex{Lorentz group}%
The Lorentz transformations that
preserve both the time direction and the spatial orientation form a
subgroup of the Lorentz group, the {\em{orthochronous proper Lorentz group}}.
\sindex{Lorentz group!orthochronous proper}%

If one wants to formulate a physical equation, the {\em{principle of Lorentz invariance}} demands that its
\sindex{Lorentz invariance!principle of}%
explicit form must be invariant under the joint transformations~\eqref{eq:lorentz-transformation-of-reference-frame}
and~\eqref{eq:lorentz-transformation-of-coordinates-and-derivatives}. The simplest
example is the {\em{Klein-Gordon equation}}%
\sindex{Klein-Gordon equation!in the vacuum}
\begin{equation}
    \label{KG1}
    (-\Box - m^2)\: \psi = 0 \:,
\end{equation}
where~$\Box := \partial_j \partial^j = g^{jk}\partial_k\partial_j$ is the scalar wave operator
\sindex{wave operator!scalar}%
\nindex{aai@$\Box$ -- scalar wave operator}%
and~$\psi: \scrM \to \C$.

Using~\eqref{eq:lorentz-transformation-of-coordinates-and-derivatives},
one verifies by a short computation that this equation takes the same form in
any reference frame. The Klein-Gordon equation describes a scalar
particle (i.e.\ a particle without spin) of mass~$m$. If the particle
has an electric charge~$e$, one also has to take into account the
interaction with the electromagnet field. One finds empirically that
the correct equation to describe the influence of the field on the
particle is
\begin{equation}
    \label{KG2}
    -(\partial_k - \ci e A_k) (\partial^k - \ci e
    A^k)\:\psi = m^2 \psi \,,
\end{equation}
where~$A$ is the electromagnetic potential.
\sindex{Klein-Gordon equation!in the electromagnetic field}%

\section{The Dirac Equation}
\label{secdirac}
The Schr\"odinger equation~\ref{scham} is not Lorentz invariant. Therefore,
it is not suitable to describe a relativistic quantum particle. Although
being Lorentz-invariant, the Klein-Gordon equation is also not suitable
for this purpose because the interpretation of the absolute value of its
solutions as probability density is not sensible (for example because 
for general solutions of~\eqref{KG2}, the spatial integral of~$|\psi(t,\vec{x})|^2$ is
not conserved in time). The correct
relativistic generalization of the Schr\"odinger equation is the so-called
Dirac equation, which will now be introduced. More on its physical
background can be found, for example, in the classic
textbooks~\cite{bjorken, landau2, peskin+schroeder}.

In order to describe a relativistic particle with spin, Dirac had the
idea to work with a first order differential operator
$\gamma^j\partial_j$ whose square is the wave operator. The
coefficients of this operator are the {\em{Dirac matrices}}
\sindex{Dirac matrix}%
\nindex{aaj@$\gamma^j$ -- Dirac matrix}%
$\gamma^j$, which are $4 \times 4$-matrices characterized by the {\em{anti-commutation
relations}}
\sindex{anti-commutation relations!for Dirac matrices|textbf}%
\begin{equation}
    \label{f:0b}
    2\: g^{jk}\:\1
    \stackrel{!}{=} \{ \gamma^j,\: \gamma^k\} := \gamma^j \gamma^k +
    \gamma^k \gamma^j \:. 
\end{equation}
Using these relations, one finds that the square of the operator
$\gamma^j \partial_j$ indeed gives the wave operator,
\begin{equation}
    (\gamma^j
    \partial_j)^2 = \gamma^j \gamma^k\: \partial_j \partial_k =
    \frac{1}{2}\: \{\gamma^j, \gamma^k \}\: \partial_{jk} = \Box \label{ac}
\end{equation}
(of course, here the operator~$\gamma^j\partial_j$ acts on wave
functions with four components, also called \emph{spinorial wave functions}, 
\sindex{wave function!spinorial}%
and the wave operator has to be
understood as acting on each component separately). There are different
possible choices for $4 \times 4$-matrices~$\gamma^j$
satisfying~\eqref{f:0b} which are all related by a change of spinor basis. For
convenience, we shall always work in the \emph{Dirac representation};
that is, we choose (see Exercise~\ref{exdiranti})
\sindex{Dirac matrix!in the Dirac representation}%
\begin{equation}
    \label{Dirrep}
    \gamma^0 =
    \begin{pmatrix}
        \1 & 0 \\
        0 & -\1
    \end{pmatrix}
    ,\qquad
    \gamma^i =
    \begin{pmatrix}
        0 & \sigma^i \\
        -\sigma^i & 0
    \end{pmatrix}
    \:,
\end{equation}
where~$\sigma^i$ are the three Pauli matrices
\begin{equation}
    \label{pauli}
    \sigma^1 =
    \begin{pmatrix}
        0 & 1
        \\ 1 & 0
    \end{pmatrix}
    , \qquad
    \sigma^2 =
    \begin{pmatrix}
        0 & -\ci \\
        \ci & 0
    \end{pmatrix}
    , \qquad
    \sigma^3 =
    \begin{pmatrix}
        1 & 0
        \\ 0 & -1
    \end{pmatrix}.
\end{equation}
Including the mass~$m \geq 0$, the Dirac equation in the vacuum
\sindex{Dirac equation!in the vacuum}%
\sindex{Dirac spinor!in Minkowski space}%
(i.e.\ without any interaction) reads
\begin{equation}
     \label{Dirac1}
    \left( \ci \gamma^k \frac{\partial}{\partial
          x^k} - m \right) \psi(x) = 0 \;,
\end{equation}
where the {\em{Dirac spinor}}
\sindex{Dirac spinor}%
$\psi: \scrM \to \C^4$ has four complex components. If we
multiply~\eqref{Dirac1} by the operator
$(\ci \gamma^j \partial_j + m)$ and use~\eqref{ac}, we find that each
component of~$\psi$ satisfies the Klein-Gordon equation~\eqref{KG1}.

Following the standard conventions in physics, we denote contractions
with Dirac matrices by a slash, i.e.\ $\slashed{u} = \gamma^j u_j$ for
$u$ a vector of Minkowski space and~$\Pdd = \gamma^j \partial_j$.
\nindex{aak@$\slashed{u}$, $\Pdd$ -- Feynman dagger}%
The action of the matrix~$\slashed{u}$ on a spinor~$\psi$ as
described by the mapping~$u \mapsto \slashed{u}\psi$ is
often referred to as {\em{Clifford multiplication}} by the vector~$u$.
\sindex{Clifford multiplication}%

A quantum particle described by a solution of the Dirac equation is
called Dirac particle. 
\sindex{Dirac particle}%
The leptons and quarks in the standard model
are Dirac particles. Thus, on the fundamental level, all matter is
described by the Dirac equation.
\sindex{Dirac equation}%

In the presence of an electromagnetic field with electromagnetic potential~$A$, the Dirac equation is
modified to
\begin{equation}
    \label{Dirac2} 
    \ci \gamma^k (\partial_k -\ci A_k) \psi = m \psi
\end{equation}
(here for convenience we absorbed the electromagnetic coupling constant into the potential).
\sindex{Dirac equation!in the electromagnetic field}%
Similar as mentioned for the Klein-Gordon equation after~\eqref{KG2}, the coupling
to the electromagnetic field can again be understood from the compatibility with local gauge transformations of electrodynamics (see Exercise~\ref{exgauge2}).
Multiplying by the operator~$(\ci \gamma^j (\partial_j -\ci A_j) + m)$
and using again the anti-commutation relations, we obtain the equation
\beq
    \Big( -(\partial_k - \ci A_k) (\partial^k - \ci A^k) +
    \frac{\ci}{2}\: F_{jk} \gamma^j \gamma^k - m^2 \Big) \psi = 0\:,
\eeq
where~$F_{jk} = \partial_j A_k - \partial_k A_k$ (see
Exercise~\ref{exKGDirac}). This differs from the Klein-Gordon
equation~\eqref{KG2} by the extra term
$\frac{\ci}{2} F_{jk} \gamma^j \gamma^k$, which describes the coupling
of the spin to the electromagnetic field.

A Dirac spinor takes value in~$\C^4$. This four-dimensional complex
vector space is also referred to as the \emph{spinor space}, and its elements
are referred to as \emph{spinors}. An important structure on the spinor
space is an indefinite inner product of signature~$(2,2)$, which we
call {\em{spin inner product}} and denote by
\sindex{spinor space!in Minkowski space}%
\sindex{spinor!in Minkowski space}%
\sindex{spin inner product!in Minkowski space}%
\nindex{aal@$\Sl . \vert . \Sr, \Sl . \vert . \Sr_x$ -- spin inner product}%
\tindex{bb@$\Sl . \vert . \Sr, \Sl . \vert . \Sr_x$ -- spin inner product}%
\begin{equation}
    \label{f:0c}
    \Sl \psi | \phi \Sr := \sum_{\alpha=1}^4 s_\alpha\:
    (\psi^\alpha)^\dagger\: \phi^\alpha \:,\qquad s_1=s_2=1,\;
    s_3=s_4=-1\:,
\end{equation}
where for a spinor~$\psi \in \C^4$, we denote by~$\psi^\dagger$ the
\Felix{Note that this is different and not to be confused with the signature of the
Minkowski metric.}%
componentwise complex conjugate. In physics textbooks, the spin inner product is often
written as~$\overline{\psi} \phi$ with the so-called adjoint spinor
$\overline{\psi} := \psi^\dagger \gamma^0$. By the
{\em{adjoint}}~$A^*$ of an operator~$A$ acting on spinors we always
mean the adjoint with respect to the spin inner product. Thus, it is
defined by the relation
\beq
    \Sl A^* \psi \,|\, \phi \Sr = \Sl \psi \,|\, A \phi \Sr \qquad
    {\mbox{for all~$\psi, \phi \in \C^4$}}.
\eeq
\sindex{adjoint!with respect to spin inner product}%
In an obvious way, this definition of the adjoint gives rise to the
the notions of a {\em{symmetric}},
\sindex{operator!symmetric}%
{\em{anti-symmetric}} and {\em{unitary operator}}.
\sindex{operator!unitary}% 
With these notions, the Dirac matrices are symmetric, meaning that
\begin{equation}
    \label{gammasymmMin}
    \Sl \gamma^l \psi \,|\, \phi \Sr = \Sl \psi
    \,|\, \gamma^l \phi \Sr \qquad {\mbox{for all~$\psi, \phi \in \C^4$}} \:.
\end{equation}
From this it follows that also Clifford multiplication~$\slashed{u}$
by a vector~$u \in \scrM$ is symmetric.
We note for clarity that, in the setting of finite-dimensional matrices considered here,
symmetric operators can be referred to equivalently as {\em{selfadjoint}} operators.
\sindex{operator!selfadjoint}%
We usually prefer the notion of a symmetric operator,
leaving the notion of a selfadjoint operator to the setting of densely defined operators
on infinite-dimensional Hilbert spaces.

To every solution~$\psi$ of the Dirac equation, we can associate a
vector field~$J^k$ by
\begin{equation}
    \label{dc}
    J^k = \Sl \psi \,|\, \gamma^k\:
    \psi \Sr \;,
\end{equation}
referred to as the {\em{Dirac current}}.
\sindex{Dirac current|textbf}%
It is either timelike or lightlike (see Exercise~\ref{excurtime}). Moreover,
it is divergence-free, as the following computation shows,
\begin{equation}
    \label{divfree}
    \begin{split}
        \partial_k J^k
        &= \partial_k \:\Sl \psi \,|\, \gamma^k\: \psi \Sr
        = \Sl \partial_k \psi \,|\, \gamma^k\: \psi \Sr +
        \Sl \psi \,|\, \gamma^k \partial_k \: \psi \Sr \\
        &= \ci \left( \Sl \ci \Pdd \psi \,|\, \psi \Sr - \Sl \psi \,|\, \ci \Pdd \psi \Sr \right) \\
        &= \ci \left( \Sl (\ci \Pdd+\Aslsh-m) \psi \,|\, \psi \Sr
            - \Sl \psi \,|\, (\ci \Pdd+\Aslsh-m) \psi \Sr \right) = 0\:.
    \end{split}
\end{equation}
This property is referred to as {\em{current conservation}}.
\sindex{current conservation!in Dirac equation}%

Current conservation is closely related to the probabilistic
interpretation of the Dirac wave function, as we now explain. Suppose
that~$\psi$ is a smooth solution of the Dirac equation with suitable
decay at spatial infinity (for example of spatially compact support;
see Section~\ref{secspinorbundlemink}). Then current conservation
allows us to apply the Gau{\ss} divergence theorem in a
spacetime-region~$[t_1,t_2] \times \R^3$ to obtain
\begin{equation}
    \label{cc}
    \begin{split}
        0 &= \int_{t_1}^{t_2} \dd t \int_{\R^3} \dd^3x\;
        \partial_k \Sl \psi \,|\, \gamma^k \psi \Sr(t, \vec{x}) \\
        &= \int_{\R^3} \Sl \psi \,|\, \gamma^0 \psi \Sr(t_2, \vec{x})
        \dd^3x - \int_{\R^3} \Sl \psi \,|\, \gamma^0 \psi \Sr(t_1,
        \vec{x}) \dd^3x \:.
    \end{split}
\end{equation}
We remark that this argument works similarly on a
region~$\Omega \subset M$ whose boundary consists of two space-like
hypersurfaces. Polarizing similar as explained after~\eqref{clin},
we conclude that for any two solutions~$\phi, \psi$ of the Dirac
equation, the spatial integral
\begin{equation}
    \label{printMink}
    ( \phi | \psi ) := \int_{\R^3} \Sl
    \phi \,|\, \gamma^0 \psi \Sr(t, \vec{x}) \dd^3x
\end{equation}
\nindex{aam@$(.\vert.), (.\vert.)_m$ -- scalar product on Dirac solutions in Minkowski space}%
\tindex{bb@$(.\vert.), (.\vert.)_m$ -- scalar product on Dirac solutions in Minkowski space}%
\sindex{scalar product!on Dirac solutions}%
is time independent. Since the inner product~$\Sl . | \gamma^0 . \Sr$
is positive definite, the integral~\eqref{printMink} defines a scalar
product. We denote the Hilbert space corresponding to this scalar
product by~$\H = L^2(\R^3)^4$. In analogy to the integrand
in~\eqref{sprod} in non-relativistic quantum mechanics, the
quantity~$\Sl \psi | \gamma^0 \psi \Sr(t,\vec{x})$ can be interpreted
as the probability density of the particle being located at the
spacetime point~$(t,\vec{x})$. Current conservation~\eqref{cc}
ensures that the probability integral is time independent.

The previous considerations generalize immediately to the situation in
the presence of an {\em{external potential}}. 
\sindex{external potential}%
\nindex{aan@$\B$ -- external potential}%
\sindex{Dirac equation!in an external potential}%
To this end, we replace
the operator~$\Aslsh$ in the Dirac equation~\eqref{Dirac2} by a
multiplication operator~$\B: \scrM \to \C^{4 \times 4}$, which may
even depend (smoothly) on the spacetime coordinates and which we
assume to be symmetric with respect to the spin inner product, i.e.
\begin{equation}
    \label{Bsymm}
    \Sl \B(x) \psi | \phi \Sr = \Sl \psi | \B(x) \phi \Sr
    \qquad {\mbox{for all~$\psi, \phi \in \C^4, x \in \scrM$}}.
\end{equation}
We then write the Dirac equation with a Dirac operator~$\Dir$ as
\begin{equation}
    \label{Dout}
    (\Dir -m)\, \psi = 0 \qquad
    \text{where} \qquad \Dir
    := \ci \Pdd + \B \:.
\end{equation}
The symmetry assumption~\eqref{Bsymm} is needed for current
conservation to hold (as one sees immediately if in~\eqref{divfree}
one replaces~$\Aslsh$ by~$\B$).

Similar to the Schr\"odinger equation~\eqref{scham}, also the Dirac equation can be rewritten
with a symmetric operator~$H$ acting on the Hilbert space~$\H$.  To this end, we
multiply~\eqref{Dout} by~$\gamma^0$ and isolate the $t$-derivative on
one side of the equation,
\begin{equation}
    \label{Hamilton}
    \ci \partial_t  \psi = H \psi
    \qquad \text{where} \qquad
    H := -\gamma^0 (\ci
    \vec{\gamma} \vec{\nabla} + \B - m)
\end{equation}
Note here
that~$\gamma^j \partial_j = \gamma^0 \partial_0 + \vec{\gamma}
\vec{\nabla}$. We refer to~\eqref{Hamilton} as the Dirac equation in
the {\em{Hamiltonian form}}.
\sindex{Dirac equation!in the Hamiltonian form}%
\nindex{aao@$H$ -- Dirac Hamiltonian}%
Now we can again apply~\eqref{symmc} to
conclude that the Hamiltonian is a symmetric operator on~$\H$.

We remark that in the Hamiltonian formulation, one often absorbs the
prefactor~$\gamma^0$ in~\eqref{Hamilton} into the other Dirac matrices
and instead works with the new matrices
\beq
    \beta := \gamma^0 \qquad\text{and}\qquad
    \vec{\alpha} := \gamma^0 \vec{\gamma} \:.
\eeq
This is convenient because these new matrices are Hermitian with
respect to the standard scalar product on~$\C^4$. In this book
however, we shall not work with~$\alpha$ and~$\vec{\beta}$. We prefer
the notation~\eqref{Hamilton}, because it is more visible which parts
of the operators are Lorentz invariant. For calculations using~$\beta$
and~$\vec{\alpha}$ we refer
for example to the monograph~\cite{thaller}. \\

In addition to integrating over space~\eqref{printMink}, one can also
introduce an inner product on spinorial wave functions by integrating
the spin inner product over all of spacetime,
\begin{gather}
    \label{stipMin}
    \bra \psi|\phi \ket = \int_\scrM \Sl \psi | \phi \Sr_x \dd\mu_\scrM\:.   
\end{gather}
This inner product will, in general, not be well-defined on solutions of
the Dirac equation, because (even for ``normalized'' solutions for which the
spatial integrals are finite) the time integral may diverge. But the inner product can
be considered, for example, on spinorial wave functions that are
compactly supported in spacetime (but are no solutions of the Dirac
equation). This {\em{spacetime inner product}} will be important for
the constructions in Chapter~\ref{secFSO}.
\sindex{spacetime inner product}%
\nindex{aap@$\bra . \vert . \ket$ -- spacetime inner product}%
\tindex{bb@$\bra . \vert . \ket$ -- spacetime inner product}%
In this context, it is very useful that the Dirac operator is symmetric with respect to the
spacetime inner product, meaning that
\begin{equation}
    \label{Dirsymm} \bra
    \Dir \psi|\phi \ket = \bra \psi| \Dir \phi \ket
\end{equation}
for all spinorial wave functions that decay sufficiently fast at
spatial infinity and for large times. Indeed, the symmetry
property~\eqref{Dirsymm} holds in curved spacetime as well (see the
explanation after~\eqref{stipintro} below).

So far, Dirac spinors were introduced in a given reference frame. Let
us verify that our definitions are in fact independent of the choice
of reference frame. 
\sindex{Lorentz invariance!of Dirac equation}%
To this end we consider two reference frames~$(x^j)$ and
$(\tilde{x}^l)$ with the same orientation of time and space. Then the
reference frames are related to each other by an orthochronous proper
Lorentz transformation~$\Lambda$; that is, in components,
\sindex{Lorentz transformation}%
\beq \tilde{x}^l = \Lambda^l_j\: x^j \;,\qquad \frac{\partial} {\partial
      x^j} = \frac{\partial \tilde{x}^l}{\partial x^j}
    \frac{\partial}{\partial \tilde{x}^l}= \Lambda^l_j\:
    \frac{\partial}{\partial \tilde{x}^l} \:,
\eeq
and~$\Lambda$ leaves the Minkowski metric invariant, \beq \label{Lt}
\Lambda^l_j \: \Lambda^m_k\: g_{lm} = g_{jk}\:. \eeq Under this change
of spacetime coordinates, the Dirac operator
$\cI \gamma^j (\frac{\partial}{\partial{\tilde{x}^j}} - \cI A_j)$
transforms to
\begin{equation}
    \label{transdir}
    \cI \tilde{\gamma}^l \left( \frac{\partial}{\partial \tilde{x}^l}
        - \cI \tilde{A}_l \right) \qquad {\mbox{with}} \qquad
    \tilde{\gamma}^l = \Lambda^l_j \gamma^j \quad \text{and} \quad
    \tilde{A}_l = \Lambda^k_l A_k
    \:.
\end{equation}
This transformed Dirac operator does not coincide with the Dirac operator
$\cI \gamma^l (\frac{\partial}{\partial{\tilde{x}^l}}	-\cI \tilde{A}_l)$
as defined in the reference frame~$(\tilde{x}^l)$ because the new Dirac matrices have
a different form. However, the next lemma shows that the two Dirac operators
do coincide after a suitable unitary transformation of the spinors.

\begin{Lemma}
    \label{Dirtrans}
    For any orthochronous proper Lorentz transformation~$\Lambda$
    there is a unitary matrix~$U(\Lambda)$ (unitary with respect to
    the spin inner product~\eqref{f:0c}) such that
    \beq
        U(\Lambda)\: \Lambda^l_j \gamma^j\: U(\Lambda)^{-1} = \gamma^l\:.
    \eeq
\end{Lemma}
\Proof Since~$\Lambda$ is orthochronous and proper, we can write it in
the form~$\Lambda=\exp(\lambda)$, where~$\lambda$ is a suitable
generator of a rotation and/or a Lorentz boost.
Then~$\Lambda(s):=\exp(s \lambda)$ with~$s \in \R$, is a family of Lorentz
transformations, and differentiating~\eqref{Lt} with respect to~$s$
as~$s=0$, we find that
\beq \lambda^l_j\:g_{lk} = -g_{jm}\: \lambda^m_k \: \eeq
(note that~$\Lambda(s)^l_j = \delta^l_j + s\, \lambda^l_j + \cdots$).
Using this identity together with the fact that the Dirac matrices are
symmetric, it is straightforward to verify that the matrix
\beq u := \frac{1}{4}\: \lambda^l_k\: \gamma_l \,\gamma^k \eeq
is anti-symmetric. As a consequence, the family of matrices
\beq U(s) := \exp \left( s u \right) \eeq
is unitary. We now consider for a fixed index~$l$ the family of matrices
\beq A(s) := U(s)\: \Lambda(s)^l_j \,\gamma^j\: U(s)^{-1}\:. \eeq
Clearly, $A(0)=\gamma^l$. Furthermore, differentiating with respect to~$s$
gives
\beq \frac{\dd}{\dd s}\:A(s) = U\: \Lambda^l_j \left\{ u \:\gamma^j
- \gamma^j \:u + \lambda^j_k \gamma^k \right\} U^{-1}\:, \eeq
and a short calculation using the commutation relations
(see Exercise~\ref{exdiraccompute})
\beq \label{dirrel1}
\big[ \gamma_l \,\gamma^k, \gamma^j \big]
= 2 \:\big( \gamma_l\: g^{kj} - \delta^j_l\: \gamma^k \big)
\eeq
shows that the curly brackets vanish.
We conclude that~$A(1)=A(0)$, proving the lemma.
\hspace*{2cm} \QED
Applying this lemma to the Dirac operator in~\eqref{transdir}, one
sees that the Dirac operator is invariant under the joint transformation
of the spacetime coordinates and the spinors
\beq x^j \;\longrightarrow\; \Lambda^j_k x^k\:,\qquad
\psi \;\longrightarrow\; U(\Lambda)\: \psi \:. \eeq
Moreover, since the matrix~$U(\Lambda)$ is unitary,
the representation of the spin inner product~\eqref{f:0c} is valid in any
reference frame. We conclude that our definition of spinors is indeed
Lorentz invariant.
%\sindex{Lorentz invariance!of Dirac equation}%

For what follows, it is important to keep in mind that, in contrast to
the spin inner product, the
combination~$\psi^\dagger \phi = \Sl \psi | \gamma^0 \phi \Sr$ is
{\em{not}} Lorentz invariant. Instead, it is the zero component of a
Minkowski vector. Consequently, the integrand in~\eqref{printMink} is
not a scalar. Its spatial integral, on the other hand, is again Lorentz invariant
due to current conservation.

As a combination of all the Dirac matrices one can form the so-called
{\em{pseudo-scalar matrix}}~$\pseudo$ by
\begin{equation}
    \label{rhodef}
    \pseudo = \frac{\cI}{4!}\: \epsilon_{jklm} \gamma^j \gamma^k \gamma^l
    \gamma^m = \cI \gamma^0 \gamma^1 \gamma^2 \gamma^3 \,.
\end{equation}
\sindex{pseudo-scalar matrix}%
\nindex{aaq@$\pseudo$ -- pseudo-scalar matrix}%
(In the physics literature, this matrix is usually denoted by
$\gamma^5$). Here~$\epsilon_{jklm}$ is the totally anti-symmetric
symbol (i.e.\ $\epsilon_{jklm}$ is equal to~$\pm 1$ if~$(j,k,l,m)$ is
an even and odd permutation of~$(0,1,2,3)$, respectively, and vanishes
otherwise).
\sindex{totally anti-symmetric symbol}%
\nindex{aar@$\epsilon_{jklm}$ -- totally anti-symmetric symbol}%
A short calculation shows that the pseudo-scalar matrix is anti-symmetric
and that~$\pseudo^2=\1$ (see Exercise~\ref{exdiraccompute}). As a consequence, the
matrices
\begin{equation}
    \label{chidef}
    \chi_L = \frac{1}{2} \left(\1 - \pseudo
    \right) \;,\qquad \chi_R = \frac{1}{2} \left(\1 + \pseudo \right)
\end{equation}
satisfy the relations (see again Exercise~\ref{exdiraccompute})
\beq \label{chiLRrel}
\begin{split}
   \chi_{L\!/\!R}^2 = \chi_{L\!/\!R} \;,&\qquad
    \pseudo \:\chi_{L} = -\chi_{L} \;,\qquad
    \pseudo \:\chi_{R} = \chi_{R} \;,\\
    &\chi_L^* = \chi_R \;,\qquad \chi_L+\chi_R=\1\:.
\end{split}
\eeq
They can be regarded as the spectral projectors of the
matrix~$\pseudo$ and are called the {\em{chiral projectors}}.
\sindex{chiral projector}%
\nindex{aas@$\chi_L, \chi_R$ -- chiral projectors}%
The projections~$\chi_L \psi$ and~$\chi_R \psi$ are referred to as the
{\em{left-}} and {\em{right-handed}}
\sindex{spinor component!left-handed}%
\sindex{spinor component!right-handed}%
components of
the spinor, respectively. A matrix is said to be
{\em{even}} and {\em{odd}}
\sindex{even matrix}% 
\sindex{odd matrix}%
if it commutes or anti-commutes with~$\pseudo$, respectively. It is
straightforward to verify that the Dirac matrices are odd, and
therefore
\beq \label{chiralflip}
\gamma^j \: \chi_{L\!/\!R} = \chi_{R\!/\!L}\: \gamma^j \:.
\eeq
By multiplying the Dirac equation~\eqref{Dirac2} from the left
by~$\chi_{L\!/\!R}$, one can rewrite it as a system of equations for
the left- and right-handed components of~$\psi$,
\beq \cI \gamma^k (\partial_k - \cI A_k) \:\chi_L \psi = m \:\chi_R \psi \;,\qquad
\cI \gamma^k (\partial_k - \cI A_k) \:\chi_R \psi = m \:\chi_L \psi \:. \eeq
If~$m=0$, these two equations decouple, and we get separate equations
for the left- and right-handed components of~$\psi$.
This observation is the starting point of the 2-component Weyl spinor
formalism.
\sindex{Weyl spinor}%
Here we shall not use this formalism. Instead, we will
describe chiral massless particles (like massless neutrinos)
by the left- or right-handed components of a Dirac spinor.

\section{The Hilbert Space of Dirac Solutions}
\label{secspinorbundlemink}
We now express the structures of Dirac theory in a convenient notation,
which harmonizes with the structures of causal fermion systems to be introduced
later on (in Chapter~\ref{secbrief}) and also generalizes to curved spacetime (in Chapter~\ref{Geometry}).
In Minkowski space, the
Dirac wave functions are four-component complex wave functions. More
generally, one can consider them as sections of a vector bundle. In
view of these general concepts (which will be introduced in
Section~\ref{secbundle} below), we denote the Cartesian product
\beq
    S\scrM := \scrM \times \C^4 \:,
\eeq
as the {\em{spinor bundle of~$\scrM$}}. 
\sindex{spinor bundle!in Minkowski space}%
\nindex{aat@$S \scrM$ -- spinor bundle in Minkowski space}%
For every~$x \in \scrM$, the
subset~$S_x \scrM := \{x\} \times \C^4$ is referred to as the
{\em{spinor space at the spacetime point~$x$}}. Clearly, the spinor
bundle is the disjoint union of all the spinor spaces,
\beq
    S\scrM = \bigcup_{x \in \scrM} S_x \scrM \,,
\eeq
and in the more general language of vector bundles the spinor spaces
at the individual spacetime points are referred to as \emph{fibers} of
the spinor bundles. The spin inner product~\eqref{f:0c} can now be
regarded as an inner product on each fiber~$S_x\scrM$, as we often
clarify by an additional subscript~$x$ (although here the inner
product does not actually depend on~$x$),
\beq
    \Sl .|. \Sr_x \::\: S_x \scrM \times S_x \scrM \rightarrow \C \:.
\eeq
A Dirac wave function can be considered as a mapping
$\psi: \scrM \to S\scrM$ with the property that for every
$x \in \scrM$ one has
\beq \psi(x) \in S_x\scrM \simeq \C^4 \:. \eeq
Such a mapping is referred to as a {\em{section}} of the spinor bundle.
\sindex{section!of the spinor bundle}%

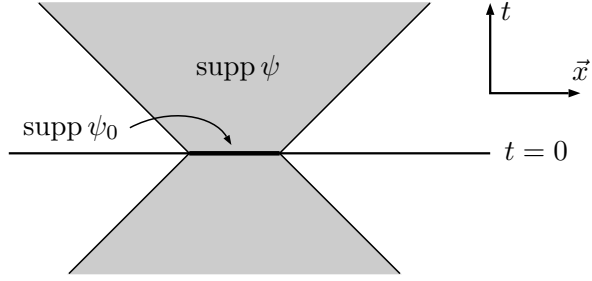
\begin{figure}[tb]
\psscalebox{1.0 1.0} % Change this value to rescale the drawing.
{
\begin{pspicture}(0,27.809866)(7.6497316,31.469597)
\definecolor{colour3}{rgb}{0.8,0.8,0.8}
\pspolygon[linecolor=colour3, linewidth=0.02, fillstyle=solid,fillcolor=colour3](0.41,31.429865)(5.61,31.429865)(3.61,29.429865)(5.21,27.829866)(0.81,27.829866)(2.41,29.429865)
\psline[linecolor=black, linewidth=0.03, arrowsize=0.05291667cm 2.0,arrowlength=1.4,arrowinset=0.0]{<->}(6.41,31.429865)(6.41,30.229866)(7.61,30.229866)
\psline[linecolor=black, linewidth=0.03](0.01,29.429865)(6.41,29.429865)
\psline[linecolor=black, linewidth=0.02](2.41,29.429865)(0.41,31.429865)
\psline[linecolor=black, linewidth=0.02](3.61,29.429865)(5.61,31.429865)
\psline[linecolor=black, linewidth=0.02](3.61,29.429865)(5.21,27.829866)
\psline[linecolor=black, linewidth=0.02](2.41,29.429865)(0.81,27.829866)
\psline[linecolor=black, linewidth=0.06](2.41,29.429865)(3.61,29.429865)
\psbezier[linecolor=black, linewidth=0.02, arrowsize=0.05291667cm 2.0,arrowlength=1.4,arrowinset=0.0]{->}(1.62,29.769865)(2.15,30.049866)(2.74,29.929865)(2.97,29.53986572265625)
\rput[bl](6.55,31.2){$t$}
\rput[bl](7.5,30.4){$\vec{x}$}
\rput[bl](2.5,30.4){$\supp \psi$}
\rput[bl](0.2,29.6){$\supp \psi_0$}
\rput[bl](6.6,29.335){$t=0$}
\end{pspicture}
}
\caption{A spatially compact Dirac solution.}
\label{figprop}
\end{figure}

We next highlight the obtained analytic structures, again anticipating
concepts and results to be introduced later in this book. The scalar
product~\eqref{printMink} on the Dirac solutions gives rise to a
Hilbert space structure on an appropriate class of solutions (for a
mathematical introduction to Hilbert spaces see
Section~\ref{sechilbertintro} below). In order to construct this
appropriate class of solutions, one can begin by solving the Cauchy
problem for smooth initial data~$\psi_0$ of compact support given for
example on the hypersurface~$\{t=0\}$. Rewriting the Dirac equation as
a linear symmetric hyperbolic system (see Chapter~\ref{secshs} below),
one sees that this Cauchy problem has a unique global solution in
Minkowski space. Moreover, this solution is smooth and, due to finite
propagation speed, has compact support on any other
hypersurface~$\{t=\text{const}\}$ (see Figure~\ref{figprop}).
One says that the solution has {\em{spatially compact support}}. 
\sindex{Dirac wave function!of spatially compact support}%
\sindex{wave function!spatially compact}%
\nindex{aau@$\Cisc(\scrM, S\scrM)$ -- spatially compact spinorial wave functions}%
\tindex{ff@$\Cisc(\scrM, S\scrM)$ -- spatially compact spinorial wave functions}%
More
generally, the set of all smooth and spatially compact sections of the
spinor bundle (not necessarily being solutions of the Dirac equation)
is denoted by~$\Cisc(\scrM, S\scrM)$. Clearly, for spatially compact
solutions, the scalar product~\eqref{printMink} is well-defined and
finite. Taking the completion of the set of all spatially compact
solutions with respect to the scalar product~\eqref{printMink}, one
obtains a Hilbert space denoted by~$(\H_m, (.|.))$, where~$m$ denotes
the mass parameter of the Dirac equation (for details on the
completion and Hilbert spaces see Section~\ref{sechilbertintro} and
Exercise~\ref{ex:complete}). By construction, we know that
\beq
    \Cisc(\scrM, S\scrM) \cap \H_m \qquad \text{is dense in~$\H_m$} \:.
\eeq
We note for clarity that the wave functions in the completion~$\H_m$ are not necessarily
differentiable. Therefore, they do not satisfy the Dirac equations. But they are
{\em{weak solutions}} in the sense that the Dirac equation holds after 
taking the spacetime inner product~\eqref{stipMin} with a test wave function~$\phi$ and
formally integrating by parts; that is,
\beq \int_\scrM \Sl (\Dir - m) \phi \,|\, \psi \Sr \: \dd\mu_{\scrM} = 0 \qquad \text{for all~$\phi \in C^\infty_0(\scrM, S\scrM)$}\:, \eeq
where~$C^\infty_0(\scrM, S\scrM)$ denotes the space of smooth wave functions with compact support.
\sindex{Dirac wave function!of compact support}%
\sindex{wave function!of compact support}%
\nindex{aav@$C^\infty_0(\scrM, S\scrM)$ -- spinorial wave functions of compact support}%
\tindex{ff@$C^\infty_0(\scrM, S\scrM)$ -- spinorial wave functions of compact support}%
For the reader familiar with the theory of partial differential
equations, we finally remark that the solutions in~$\H_m$ can also be
characterized in terms of Sobolev spaces (see for
example~\cite[Section~II.5]{evans}). More precisely, the vectors in~$\H_m$ are weak solutions in~$H^{1,2}_\text{loc}(\scrM, S\scrM)$. By the trace theorem (see for
example~\cite[Section~II.5.2]{evans}), their restriction to a
hypersurface~$\{t=\text{const}\}$ is in~$L^2_\text{loc}(\R^3, \C^4)$.
As a consequence, the integrand of the spatial integral
in~\eqref{printMink} is locally integrable. The solutions in~$\H_m$
have the additional property that their restriction to the
hypersurfaces are even in~$L^2(\R^3, \C^4)$, so that the integral
in~\eqref{printMink} exists and is finite.

\section{Dirac's Hole Theory and the Dirac Sea}
\label{secsea}
The Dirac theory gives rise to anti-matter and pair creation, as we
now briefly explain. For the present purposes, it suffices to consider
the Dirac equation in the vacuum~\eqref{Dirac1}. It can be solved by
the plane wave ansatz (for more details see for
example~\cite[Section~3.1]{bjorken},
\cite[Section~3.3]{peskin+schroeder} or~\cite[Section~1.4.1]{thaller})
\sindex{plane wave ansatz}%
\beq
    \psi(x) = \chi(k)\: \E^{-\ci k x} \:,
\eeq
where~$k \in \scrM$ is the four-momentum, and~$k x = \la k,x\ra$ is
the Minkowski inner product (for a mathematically precise treatment in
terms of the Fourier transform see Section~\ref{secfourier} below). Using
this ansatz in~\eqref{Dirac1} yields the (zeroth order) linear system
\begin{equation}
    \label{Diracmom}
    (\slashed{k}-m) \: \chi(k) = 0
\end{equation}
for the vector~$\chi(k) \in \C^4$. Multiplying by the
matrix~$\slashed{k}+m$ and using the anti-commutation
relations~\eqref{f:0b} gives the necessary condition
\begin{equation}
    \label{dispersion}
    k^2 = m^2 \:.
\end{equation}
If this condition is satisfied, the matrix~$\slashed{k}-m$ has a
two-dimensional kernel, which coincides with the image of the
matrix~$\slashed{k}+m$. Thus the general solution of~\eqref{Diracmom}
can be written as
\beq
    \chi(k) = (\slashed{k}+m)\: \phi \qquad \text{with} \qquad \phi \in \C^4\:.
\eeq

\begin{figure}[tb]
\psscalebox{1.0 1.0} % Change this value to rescale the drawing.
{
\begin{pspicture}(0,27.799109)(5.2568016,31.480354)
\psline[linecolor=black, linewidth=0.02](0.017070312,31.440624)(3.6170702,27.840624)
\psline[linecolor=black, linewidth=0.02](0.017070312,27.840624)(3.6170702,31.440624)
\psline[linecolor=black, linewidth=0.03, arrowsize=0.05291667cm 2.0,arrowlength=1.4,arrowinset=0.0]{<->}(4.0170703,31.440624)(4.0170703,30.240623)(5.21707,30.240623)
\psbezier[linecolor=black, linewidth=0.02, arrowsize=0.05291667cm 2.0,arrowlength=1.4,arrowinset=0.0]{->}(2.9370704,29.260624)(2.7070704,29.380623)(2.4670703,29.330624)(2.1370704,29.060623779296876)
\psbezier[linecolor=black, linewidth=0.02](0.12707031,31.430624)(0.92474705,30.655556)(1.2926164,30.26745)(1.8226335,30.260623779296875)(2.3526506,30.253798)(3.0807612,31.002176)(3.5070703,31.410624)
\psbezier[linecolor=black, linewidth=0.06](3.5070703,27.830624)(2.7093935,28.605692)(2.3415244,28.993797)(1.8115072,29.000623779296873)(1.2814901,29.00745)(0.55337936,28.259071)(0.12707031,27.850624)
\pscircle[linecolor=black, linewidth=0.04, fillstyle=solid,fillcolor=black, dimen=outer](1.4070703,30.360624){0.06}
\pscircle[linecolor=black, linewidth=0.01, fillstyle=solid, dimen=outer](2.4370704,28.800623){0.06}
\pscircle[linecolor=black, linewidth=0.01, fillstyle=solid, dimen=outer](2.0470703,28.970623){0.06}
\pscircle[linecolor=black, linewidth=0.04, fillstyle=solid,fillcolor=black, dimen=outer](2.1870704,30.360624){0.06}
\pscircle[linecolor=black, linewidth=0.04, fillstyle=solid,fillcolor=black, dimen=outer](2.8670702,30.830624){0.06}
\psbezier[linecolor=black, linewidth=0.02, arrowsize=0.05291667cm 2.0,arrowlength=1.4,arrowinset=0.0]{->}(2.2070704,31.110624)(2.4370704,31.170624)(2.6570704,30.990623)(2.7470703,30.920623779296875)
\psline[linecolor=black, linewidth=0.02, linestyle=dashed, dash=0.17638889cm 0.10583334cm](0.017070312,30.250624)(1.7970703,30.240623)
\psline[linecolor=black, linewidth=0.02, linestyle=dashed, dash=0.17638889cm 0.10583334cm](0.027070312,29.040625)(1.8170704,29.030624)%
\rput[bl](0.8,31.03){electron}
\rput[bl](3.1,29){positron}
\rput[bl](4.15,31.3){$\omega$}
\rput[bl](5,30.4){$\vec{k}$}
\rput[bl](-1.3,30.2){$\omega=m$}
\rput[bl](-1.5,28.9){$\omega=-m$}
\end{pspicture}
}
\caption{The Dirac sea with particles and anti-particles.}
\label{figshell}
\end{figure}
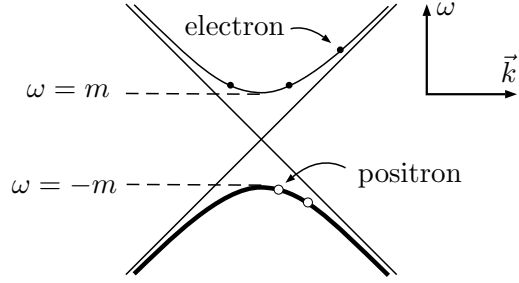

The zero component~$\omega := k^0$ of the four-momentum is physically
interpreted as~$2 \pi$ times the frequency of the wave.
Equation~\eqref{dispersion}, also referred to as the {\em{dispersion relation}}, can then be written as
\sindex{dispersion relation}%
\beq
    \omega^2 = \big| \vec{k} \big|^2 + m^2 \qquad \text{or} \qquad
    \omega = \pm \sqrt{ \big| \vec{k} \big|^2 + m^2 } \:.
\eeq
Here the plus and the minus sign correspond to positive and negative
frequency, respectively. The corresponding solutions are said to be
{\em{on the upper and lower mass shell}} (see Figure~\ref{figshell}).
\sindex{mass shell, upper and lower}%
Using Planck's relation~$E = \hbar \omega$, the frequency can be
related to the {\em{energy}} of the solution. Thus the solutions on
the upper and lower mass shell have positive and negative
energy, respectively.

At first sight, the occurrence of solutions of negative energy seems
problematic, because particles of negative energy have never been
observed. Moreover, at least in a naive consideration, solutions of
arbitrarily large negative energy should make the physical system
unstable, because by bringing a particle into a state of larger and
larger negative energy, one could extract more and more positive
energy from the system by the principle of conservation of energy.
This problem was resolved by Dirac in 1930~\cite{dirac2} and led to
the prediction of {\em{particle creation}} and {\em{anti-matter}}, as
\sindex{anti-matter}%
we now outline (an excellent and more detailed explanation can be
found in~\cite[Section~5.1]{bjorken}). We work in the setting of
non-interacting many-particle quantum mechanics, where the
many-particle wave function is described by a product of one-particle
wave functions. In other words, the quantum state is described by
occupying many one-particle states. Dirac's concept is that in the
vacuum, all the states of negative energy should be occupied, forming
the so-called {\em{Dirac sea}}.
\sindex{Dirac sea|textbf}%
According to the {\em{Pauli exclusion principle}},
\sindex{Pauli exclusion principle|textbf}%
each state may be occupied by at most one electron.
Therefore, adding particles to the system, the additional particles
must occupy states of positive energy, giving rise to
{\em{electrons}}. By convention, the electrons have {\em{negative}}
electric charge. Moreover, one can create ``holes'' in the Dirac sea.
The resulting ``hole in a sea of negative energy'' appears as a
particle of again positive energy, but with the opposite and thus
{\em{positive}} electric charge. These ``holes'' can be observed as
{\em{positrons}}. 
\sindex{positron}%
Furthermore, starting from the completely filled
Dirac sea, one can ``excite'' a particle of the sea by a transition
from a state of negative energy to a state of positive energy. As a
result, one obtains a particle (=electron) plus a hole (=positron).
This explains why matter and anti-matter can be created in pairs in a
process called {\em{pair creation}}.
\sindex{pair creation}%

The above intuitive picture of the Dirac sea has important observable
consequences, because it explains fundamental physical phenomena like
anti-matter and pair creation. Nevertheless, the naive picture suffers
from the problems that the Dirac sea has an {\em{infinite negative
    charge density}} and an {\em{infinite energy density}}. In modern
quantum field theory, these problems are bypassed by introducing a
suitable vacuum state and working ``relative'' to this vacuum state.
Here we shall not enter these constructions. Instead, we shall take
Dirac's concept of a ``sea of interacting particles'' seriously, as
will be explained in more detail in Section~\ref{secphysicalconcepts}.

\section{Exercises}%

\begin{Exercise} \label{extransitive} {\em{
Show that the relations ``lies in the timelike future of'' and ``lies in the causal future of''
are transitive in the following sense, \\
\begin{align}
y \in I^\vee_x \text{ and } z \in I^\vee_y \quad &\Longrightarrow \quad z \in I^\vee_x \\
y \in J^\vee_x \text{ and } z \in J^\vee_y \quad &\Longrightarrow \quad z \in J^\vee_x \:.
\end{align}
}} \end{Exercise}

\begin{Exercise} \label{exgauge1}  (Local gauge transformations I) {\em{
\sindex{local gauge transformation|textbf}%
Show that the Klein-Gordon equation~\eqref{KG2} is invariant under joint transformations
of the electromagnetic potential and the wave function according to
\beq \label{gauge1}
A_j(x) \rightarrow A_j(x) + \partial_j \Lambda(x) \:,\qquad
\psi(x) \rightarrow \E^{-\ci\Lambda(x)}\: \psi(x) \:.
\eeq
Moreover, show that the electromagnetic field tensor~$F_{jk} := \partial_j A_k - \partial_k A_j$
remains unchanged under these transformations.

We remark that the transformations~\eqref{gauge1} are the classical gauge transformations of electrodynamics.
They give rise to local phase transformations of the quantum mechanical wave functions,
which have no physical significance because all measurable quantities involve the
product of the wave function with its complex conjugate.
}} \end{Exercise}

\begin{Exercise} \label{exgauge2} (Local gauge transformations II) {\em{
%\sindex{local gauge transformation}%
Show that the Dirac equation~\ref{Dirac2} 
is invariant under joint transformations~\eqref{gauge1}
of the electromagnetic potential and the Dirac wave function.
}} \end{Exercise}

\begin{Exercise} \label{exdiranti} (Anti-commutation relations) {\em{
\sindex{anti-commutation relations!for Dirac matrices}%
\bitem
\item[(a)] Verify by direct computation that the Dirac matrices~\eqref{Dirrep}
satisfy the anti-commu\-ta\-tion relations~\eqref{f:0b}.
\item[(b)] Why is it not possible to satisfy these anti-commutation relations with $2 \times 2$- or~$3\times 3$-matrices? \emph{Hints:} The case of odd-dimensional matrices can be ruled out by
computing the square and the trace of the matrix~$\gamma^0 \gamma^1$.
For~$2 \times 2$-matrices, a similar argument shows that the matrix~$\gamma^0 \gamma^1$
is diagonalizable, making it possible to proceed in an eigenvector basis.
\eitem
}} \end{Exercise}

\begin{Exercise} \label{exdirsymm} {\em{
Show that the Dirac matrices in the Dirac representation~\eqref{Dirrep}
are symmetric with respect to the spin inner product~\eqref{f:0c}.
Show that this symmetry property is equivalent to the statement that the matrices~$\gamma^0 \gamma^j$
are Hermitian.
}}
\end{Exercise}

\begin{Exercise} \label{exKGDirac} {\em{
Show that, multiplying the Dirac equation~\eqref{Dirac2} by the operator
$(\ci \gamma^j (\partial_j -\ci A_j) + m)$ and using the anti-commutation
relations, we obtain the equation
\beq \Big( -(\partial_k - \ci A_k) (\partial^k - \ci A^k)
+ \frac{\ci}{2}\: F_{jk} \gamma^j \gamma^k - m^2 \Big) \psi = 0\:. \eeq
This differs from the Klein-Gordon equation~\eqref{KG2} by the extra term
$\frac{\ci}{2} F_{jk} \gamma^j \gamma^k$, which describes
the coupling of the spin to the electromagnetic field. }}
\end{Exercise}

\begin{Exercise} \label{excurtime} {\em{
In this exercise, we shall verify that for any non-zero spinor~$\psi$, the
corresponding Dirac current vector~$J^k = \Sl \psi | \gamma^k \psi \Sr$ is non-spacelike.
\sindex{Dirac current}%
\bitem
\item[(a)] Show that the matrix~$\gamma^0 \gamma^1$ is Hermitian
and has eigenvalues~$\pm 1$. Deduce that
\beq \la \psi, \gamma^0 \gamma^1 \psi \ra_{\C^4} \leq \|\psi\|^2_{\C^4} \:. \eeq
\item[(b)] Show that the last inequality implies that~$|J^1| \leq J^0$.
\item[(c)] Use the rotational symmetry of the Dirac equation to conclude
that~$J^0 \geq |\vec{J}|$ (where~$\vec{J}=(J^1, J^2, J^3) \in \R^3$).
\eitem
}} \end{Exercise}

\begin{Exercise} \label{exdiraccompute} {\em{
This exercise has the purpose of getting more familiar with the computation rules for Dirac matrices.
\bitem
\item[(a)] Derive~\eqref{dirrel1} from the anti-commutation relations.
\item[(b)] Derive from the anti-commutation relations and the symmetry of the Dirac matrices
that the pseudo-scalar matrix~$\pseudo$ in~\eqref{rhodef} is anti-symmetric
and that~$\pseudo^2=\1$.
\item[(c)] Show that the chiral projectors~$\chi_L$ and~$\chi_R$ defined by~\eqref{chidef}
satisfy the relations~\eqref{chiLRrel} and~\eqref{chiralflip}.
Show that the Dirac equation in the presence of an external field~\eqref{Dirac2}
can be rewritten as a system of equations for the left- and right-handed components of~$\psi$,
\beq \ci \gamma^k (\partial_k - \ci A_k) \,\chi_L \psi = m \,\chi_R \psi \:, \qquad
\ci \gamma^k (\partial_k - \ci A_k) \,\chi_R \psi = m \,\chi_L \psi \:. \eeq
What happens in the limiting case~$m=0$?
\eitem
}} \end{Exercise}

\begin{Exercise} {\em{
This exercise explains how the {\em{causal structure}} of Minkowski space is encoded in the
Dirac matrices. This method generalizes to partial differential equations of so-called {\em{symmetric
hyperbolic type}} as will be introduced later in this book (see Chapter~\ref{secshs}).
For this exercise no knowledge on symmetric hyperbolic systems is needed.
\sindex{symmetric hyperbolic system!connection to causality}%
But we use the same notions to be introduced in Chapter~\ref{secshs}.
Show that the Dirac equation~$( \cI \slashed{\partial}-m)\psi=0$  can be rewritten as a so-called
\emph{symmetric hyperbolic system}; that is, in the form
\beq \big( A^0(x)\,\partial_0 + A^\alpha(x)\,\partial_\alpha + B(x) \big)\psi = 0,\quad\mbox{with }\ (A^i)^\dagger = A^i\ \mbox{ and }\ A^0(x)\ge c\,\1 \eeq
with a constant~$c>0$ (where the dagger means transposition and complex conjugation).
For such systems a notion of \emph{causality} can be introduced as follows. A vector~$\xi\in \R^4$ is said to be \emph{time-like} or \emph{light-like} at~$x\in\R^4$, if the matrix~$A(x,\xi):=A^i(x)\,\xi_i$ is definite (either positive or negative) or singular, respectively. 

Find the matrices~$A^i$ and~$B$ for the Dirac equation and show that the above notions of time-like and light-like vectors coincide with the corresponding notions in Minkowski space.
{\em{Hint:}} Do not be surprised if the naive choice~$A^j=\gamma^j$ does not work. }}
\end{Exercise}

\begin{Exercise} (Decomposition of Dirac solutions into positive and negative energy) {\em{
\sindex{negative-energy solutions of Dirac equation}%
For a momentum~$\vec{k}\in\R^3$  we define the 
energy~$\omega(\vec{k}):=\sqrt{\vec{k}^2+m^2}$ and the matrices
\beq
p_\pm(\vec{k}):=\frac{\slashed{k}+m}{2\,k^0}\,\gamma^0\bigg|_{k^0=\pm \omega(\vec{k})}\in \mathrm{Mat}(4,\C) \quad (\mbox{with }\slashed{k}:=k^0\gamma^0-\vec{k}\cdot\vec{\gamma}).
\eeq 
\bitem
\item[{\rm{(i)}}] Referring to the standard scalar product of~$\C^4$, show that the matrices~$p_\pm(\vec{k})$ are symmetric, idempotent, add up to the identity and have orthogonal images. Conclude that the spinor space~$\C^4$ can be decomposed into the orthogonal direct sum
\beq
\C^4=W_{\vec{k}}^+\oplus W_{\vec{k}}^-,\quad \mbox{with}\quad W_{\vec{k}}^\pm:= \im p_\pm(\vec{k}).
\eeq
\item[{\rm{(ii)}}] Let~$\varphi\in C_{\mathrm{sc}}^\infty(\R^4,\C^4)$ be a smooth solution of the Dirac equation with spatially compact support; that is,
\beq
(\ci\gamma^0\partial_0+\ci\gamma^\alpha\partial_\alpha-m)\varphi=0,\quad\mbox{with }\ \varphi(t,\,\cdot\,)\in C_0^\infty(\R^3,\C^4) \mbox{ for all~$t\in\R$}.
\eeq
Let~$\hat{\varphi}$ be the smooth function on~$\R^4$ defined by taking the Fourier transform of~$\varphi$ in the spatial variables only. Find~$h\in C^\infty(\R^3,\mathrm{Mat}(4,\C))$ such that 
\beq
\ci\partial_t\, \hat{\varphi}(t,\vec{k})= h(\vec{k})\cdot \hat{\varphi}(t,\vec{k})\quad\mbox{for all~$t\in\R,\ \vec{k}\in\R^3$}. 
\eeq
Show that~$h(\vec{k})$ is also symmetric with respect to the standard scalar product of~$\C^4$ and satisfies
\beq
h(\vec{k})p_\pm({\vec{k}})=\pm \omega(\vec{k})p_\pm({\vec{k}}).
\eeq
In particular, $\pm \omega(\vec{k})$ form  the spectrum of~$h(\vec{k})$.
\item[{\rm{(iii)}}] Referring to point (ii), conclude that
\beq
\varphi(t,\vec{x})=\int_{\R^3}\frac{\dd^3\vec{k}}{(2\pi)^{3/2}}
\: \Big(p_-(\vec{k})\hat{\varphi}(0,\vec{k})\,\E^{\ci\omega(\vec{k})t}+p_+(\vec{k})\hat{\varphi}(0,\vec{k})\,\E^{-\ci\omega(\vec{k})t}\Big)\:\E^{\ci\vec{k}\cdot\vec{x}}
\eeq
\emph{Hint:} You can use the fact that the Cauchy problem admits unique smooth solutions.
\eitem
From a mathematical point of view, the \emph{Dirac sea}
\sindex{Dirac sea}%
 is described by the Hilbert space generated by all the smooth solutions with spatially compact support and the property that~$p_-(\vec{k})\,\hat{\varphi}(0,\vec{k})=\hat{\varphi}(0,\vec{k})$. }}
\end{Exercise}

%% file: part1-m.tex
%!TEX root = intro.tex
\chapter{Mathematical Preliminaries}
\label{cha:mathematical-preliminaries}

In this chapter, we summarize some mathematical preliminaries that we believe are important for studying causal 
fermion systems. In particular, the basic definitions of measure theory and functional analysis
are needed already for the very definition of a causal fermion system. 
For mathematicians, the material covered in this chapter will probably be rather familiar. For 
physicists, depending on one's background, this might be different, and we hope that this chapter helps to make the 
core contents of this book accessible. The presentation is mostly a summary without detailed proofs, but we do 
provide explanations, examples and exercises to get acquainted with the material. For a more thorough coverage,
we  again provide references to various standard textbooks throughout the chapter.

\section{Basics on Topology}
\label{sectopology}
We here recall a few basic concepts from topology. A more systematic
treatment can be found in many good elementary textbooks like, for
example,  \cite{schubert} or~\cite{jaenich}. In a topological space the
fundamental concept is that of an open set. 
Since topological spaces are rather abstract concepts, we prefer to begin with
metric spaces.
\begin{Def}
    \label{defmetricspace}
    Let~$E$ be a set. A mapping
    \beq d \::\: E \times E \rightarrow \R^+_0 \eeq is called
    {\bf{metric on~$E$}} if it has the following properties:
    \bitem
\item[{\rm{(i)}}] Positivity: For all~$x,y \in E$,
    \beq d(x,y) \geq 0 \qquad \text{and} \qquad d(x,y) = 0 \;\;\; \Longleftrightarrow \;\;\; x=y \:. \eeq
\item[{\rm{(ii)}}] Symmetry: For all~$x,y \in E$,
    \beq d(x,y) = d(y,x) \:. \eeq
\item[{\rm{(iii)}}] Triangle inequality: For all~$x,y,z \in E$,
    \beq d(x,y) \leq d(x,z) + d(z,y) \:. \eeq \eitem If~$d$ is a metric on~$E$, then the pair~$(E, d)$ is called a {\bf{metric space}}.
\sindex{metric space}%
\nindex{aba@$d(x,y)$ -- metric}%
\end{Def}

\noindent A simple example of a metric space is~$E=\R^3$ with the
Euclidean distance function
\beq
    d(x,y) = \|x-y\| := \bigg( \sum_{\alpha=1}^3 \big|
    x^\alpha-y^\alpha \big|^2 \bigg)^\frac{1}{2} \:,
\eeq
or also~$\R^n$ with the analogously defined distance function. In view
of this example, $d(x,y)$ is sometimes also referred to as
the {\em{distance}} between~$x$ and~$y$. More examples of metric spaces will be given
in the following section.

\begin{figure}[tb]
{
\begin{pspicture}(0,-4.2325)(5.21,-1.3675)
\definecolor{colour1}{rgb}{0.9019608,0.9019608,0.9019608}
\definecolor{colour0}{rgb}{0.7019608,0.7019608,0.7019608}
\psframe[linecolor=black, linewidth=0.02, dimen=outer](5.21,-1.3675)(0.0,-4.2325)
\psbezier[linecolor=black, linewidth=0.02, fillstyle=solid,fillcolor=colour1](1.2454545,-1.8275)(0.23023655,-1.5793246)(0.26668856,-2.308141)(0.76,-2.78)(1.2533114,-3.251859)(2.0992365,-2.9723558)(2.0383637,-3.7325)(1.9774908,-4.4926443)(3.9880388,-2.7637026)(4.32,-2.065625)(4.6519613,-1.3675473)(2.2606726,-2.0756755)(1.2454545,-1.8275)
\pscircle[linecolor=black, linewidth=0.02, fillstyle=solid,fillcolor=colour0, dimen=outer](1.38,-2.4325){0.49}
\pscircle[linecolor=black, linewidth=0.04, fillstyle=solid,fillcolor=black, dimen=outer](1.405,-2.4225){0.055}
\psline[linecolor=black, linewidth=0.02](1.43,-2.4125)(1.82,-2.2725)
\rput[bl](4.8,-1.75){$E$}
\rput[bl](3,-2.8){$\Omega$}
\rput[bl](1.2,-2.3){$x$}
\rput[bl](1.57,-2.6){$r$}
\end{pspicture}
}
\caption{An open set~$\Omega \subset E$.}
\label{figopen}
\end{figure}
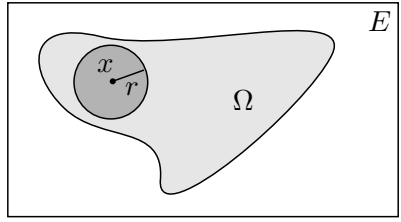

A metric gives rise to a corresponding topology:
\begin{Def}
    Let~$(E,d)$ be a metric space. For any~$x \in E$ and~$r > 0$, the
    set
    \beq
        B_r(x) := \big\{ y \in E \:\big|\: d(x,y)<r \big\}
    \eeq
    is referred to as the {\bf{open ball of radius~$r$ centered
        at~$x$}}. A subset~$\Omega \subset E$ is {\bf{open}} if for
    every~$x \in \Omega$, there is some radius~$r>0$ such
    that~$B_r(x) \subset \Omega$ (see Figure~\ref{figopen}). The
{\bf{metric topology}}~$\O$ of~$(E,d)$ is defined as the family of
\sindex{topology!metric}%
\nindex{abb@$\O$ -- topology}%
    all open subsets,
    \beq
        \O := \big\{ \Omega \subset E \mid \Omega \textrm{ is open} \big\} \subset
        {\mathfrak{P}}(E)
    \eeq
    (here~${\mathfrak{P}}(E)$ denotes the power set of~$E$; that is, the
    set of all subsets of~$E$).
\end{Def}
  
\noindent The open sets in a metric space satisfy certain properties
(we omit the proof, which is an easy exercise and can also be found in
many textbooks):
\begin{Lemma} \label{lemmatopology} Given a metric space~$(E,d)$, the corresponding metric topology~$\O$
has the following properties:
\bitem
\item[{\em{(i)}}] $\varnothing, E \in \O$.
\item[{\em{(ii)}}] Closedness under finite intersections: For any~$n \in \N$ and~$\Omega_1,\ldots,\Omega_n \subset E$,
    \beq
        \Omega_1, \ldots, \Omega_n \in \O
        \quad \Longrightarrow \quad
        \Omega_1 \cap \cdots \cap \Omega_n \in \O \:.
    \eeq
\item[{\em{(iii)}}] Closedness under arbitrary unions: For any
    (possibly infinite)
    family~$(\Omega_\lambda)_{\lambda \in \Lambda}$ of subsets of~$E$,
    \beq
        \Omega_\lambda \in \O \;\; \textrm{for all } \lambda \in \Lambda
        \quad \Longrightarrow \quad
        \bigcup_{\lambda \in \Lambda} \Omega_\lambda \in \O \:.
    \eeq
\eitem
\end{Lemma}

\begin{figure}[tb]
{
\begin{pspicture}(0,28.043255)(4.5714207,29.899439)
\definecolor{colour3}{rgb}{0.8,0.8,0.8}
\psbezier[linecolor=black, linewidth=0.02, fillstyle=solid,fillcolor=colour3](1.0485302,29.448092)(0.3994741,29.791529)(-0.30084538,29.37465)(0.17853028,28.638092041015625)(0.65790594,27.901535)(1.3004439,28.104355)(1.5585303,28.578093)(1.8166167,29.05183)(1.6975864,29.104654)(1.0485302,29.448092)
\psbezier[linecolor=black, linewidth=0.02, fillstyle=solid,fillcolor=colour3](3.0085304,29.568092)(2.407358,28.768972)(2.543345,28.226103)(3.5985303,28.078092041015626)(4.6537156,27.93008)(4.6677537,28.932787)(4.4185305,29.168093)(4.1693068,29.403397)(3.6097026,30.367212)(3.0085304,29.568092)
\pscircle[linecolor=black, linewidth=0.02, fillstyle=solid,fillcolor=black, dimen=outer](3.3235302,29.003092){0.045}
\pscircle[linecolor=black, linewidth=0.02, fillstyle=solid,fillcolor=black, dimen=outer](0.69353026,28.923092){0.045}%
\rput[bl](0.85,28.8){$x$}
\rput[bl](3.5,29){$y$}
\rput[bl](1.45,29.4){$U$}
\rput[bl](4.3,29.5){$V$}
\end{pspicture}
}
\caption{The Hausdorff property.}
\label{fighausdorff}
\end{figure}
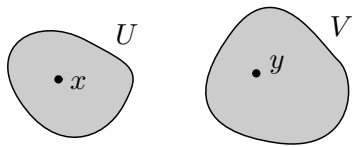

A topological space is now defined by turning exactly these properties
into a definition.
\begin{Def}
    A set~$E$ together with a distinguished family of
    subset~$\O \subset {\mathfrak{P}}(E)$ satisfying the
    properties~{\rm{(i)}}--{\rm{(iii)}} in Lemma~\ref{lemmatopology}
    is referred to as a {\bf{topological space}}. The family of
    subsets~$\O$ is called the {\bf{topology}} of~$E$. The sets in~$\O$ are
    called {\bf{open subsets of~$E$}} (with respect to~$\O$). A
    topology~$\O$ on~$E$ with the additional property that for any
    distinct point~$x,y \in E$, there are disjoint open sets~$U, V \in \O$ with~$x \in U$ and~$y \in V$ is called
    {\bf{Hausdorff}} (see Figure~\ref{fighausdorff}).
\sindex{topology}%
\sindex{topological space}%
\sindex{topology!Hausdorff}%
\nindex{abb@$\O$ -- topology}%
\sindex{set!open}%
\end{Def} \noindent

Clearly, topological spaces are a general and abstract concept. In
particular, the topology of a topological space does not necessarily
need to come from an underlying metric. Note that the topology coming
from a metric is always Hausdorff, as one sees immediately by
choosing~$U=B_r(x)$ and~$V=B_r(y)$ with~$r=d(x,y)/3$.

The significance of the definition of a topological space lies in the
fact that many notions from analysis can be formulated purely in
topological terms and thus be generalized to arbitrary topological
space. We conclude by recalling a few of such topological definitions:

A set~$A \subset E$ is called {\em{closed}} if its
\sindex{set!closed|textbf}%
complement~$E \setminus A$ is open. The properties in
Lemma~\ref{lemmatopology} can be restated for closed sets by saying
that the empty set and~$E$ are closed, and that finite unions as well
as arbitrary intersections of closed sets are again closed (see
Exercise~\ref{exclosed}). The {\em{closure}}~$\overline{A}$ of a
subset~$A \subset E$ is defined by
\sindex{closure of a set}%
\nindex{abd@$\overline{A}$ -- closure of~$A$}%
\beq
    \overline{A}
    := \bigcap \big\{ B \subset E \mid
    A \subset B \text{ and } B \text{ is closed} \} \:.
\eeq
It is by definition the smallest closed set containing~$A$. Similarly,
the {\em{interior}}~$\overset{\,\circ\!}{A}$ of a set~$A \subset E$ is
defined as the largest open set contained in~$A$; that is,
\sindex{interior of a set}%
\nindex{abe@$\circa$ -- interior of~$A$}%
\beq
    \overset{\circ\!\!}{A}
    := \bigcup \big\{ B \subset E \mid
    B \subset A \text{ and } B \text{ is open} \} \:.
\eeq

A subset~$K \subset E$ is called {\em{compact}} if for every
\sindex{compactness}%
collection~$\{U_i\}_{i \in I}$ of open sets of~$E$ with
$K \subset \bigcup_{i \in I} U_i$ there exist finitely many
$i_1,\ldots,i_n \in I$ such that still
$K \subset \bigcup_{k=1}^n U_{i_k}$ (every open cover of~$K$ has a
finite subcover).

A sequence~$(x_n)_{n \in \N}$ in~$E$ \emph{converges to a point~$x \in E$} if for any open set~$U \subset E$ with~$x \in E$ there is
\sindex{convergence!in topological space|textbf}%
some~$N \in \N$ with~$x_n \in U$ for all~$n \geq N$. In this notion of
convergence the Hausdorff property is important because it
guarantees uniqueness of the limit point~$x$ (if it exists).

The {\em{support of a function}}~$f : E \rightarrow \R$ (or, more generally, mapping to a vector space)
is defined as the closure of the set where it is non-zero,
\beq \label{suppfdef}
\supp f := \overline{ \{ x \in E \:|\: f(x) \neq 0 \} } \:.
\eeq
\sindex{support!of a function}%
In the applications one often encounters functions with {\em{compact support}}.
\sindex{function!with compact support}%
Finally, a mapping~$f : E \rightarrow F$ between two topological
spaces~$(E, \O_E)$ and~$(F, \O_F)$ is {\em{continuous}} if the
\sindex{continuity!in topological space}%
\sindex{function!continuous}%
pre-image of any open set is open; that is, if for any~$\Omega \subset F$
\beq
\label{defcont}
\Omega \in \O_F \quad \Longrightarrow \quad f^{-1}(\Omega) \in \O_E
\:.
\eeq
A continuous mapping that is invertible and whose inverse is also continuous
is referred to as a {\em{homeomorphism}}.
\sindex{homeomorphism|textbf}%

In metric spaces, this definition of continuity is equivalent to the
usual~$\varepsilon$-$\delta$-criterion (see Exercise~\ref{extop1}).
The topological definition has the advantage that in proofs it fits together nicely with other topological notions.
Many important theorems from real analysis have topological
generalizations. For example, every real-valued continuous function on
a compact topological space attains its maximum (see
Exercise~\ref{extop2}).

\section{Banach Spaces, Hilbert Spaces and Linear Operators} \label{sechilbertintro}
In this section, we consider complex vector spaces equipped with additional structures
like a norm and a scalar product (most definitions can be adapted in a straightforward way
to real vector spaces). We also recall the notion of completeness and
introduce linear operators. For more details and further reading we recommend the
textbooks~\cite{rudin, reed+simon, lax}.

\begin{Def} \label{defnormedspace} Let~$V$ be a complex vector space.
A {\bf{norm on~$V$}} is a mapping
\beq \| \cdot \| \::\: V \rightarrow \R^+_0 \eeq
with the following properties:
\bitem
\item[\rm{(i)}] Homogeneity: For all~$x \in V$ and~$\lambda \in \C$,
    \beq
        \big\| \lambda x \big\|= |\lambda|\: \|x\| \:.
    \eeq
\item[\rm{(ii)}] Definiteness: For all~$x \in V$,
\beq \|x\|=0 \; \Longleftrightarrow \; x=0 \:. \eeq
\item[\rm{(iii)}] Triangle inequality: For all~$x,y \in V$,
    \beq \|x+y\| \leq \|x\| + \|y\| \:. \eeq \eitem If~$||\cdot\|$ is a
    norm on~$V$, then the pair~$(V, \|\cdot\|)$ is called a
    {\bf{normed space}}.
\end{Def}
\sindex{normed space|textbf}%
\nindex{abf@$\NORM . \NORM$ -- norm}%
\tindex{dd@$\NORM . \NORM$ -- norm}%

\noindent Every normed space is naturally a metric space (see
Definition~\ref{defmetricspace}) with the metric defined
by~$d(x,y) := \|x-y\|$ for all~$x,y \in V$.

To give a concrete example, on~$\C^n$ for
any~$p \in [1,\infty) \cup \{\infty\}$, one obtains a norm~$\|\cdot\|_p$
as
\beq \label{lpnorm}
    \| x \|_p := \bigg( \sum_{i=1}^n |x_i|^p \bigg)^{\frac{1}{p}}
    \quad \text{ for } p < \infty \qquad \text{and} \qquad
    \| x \|_\infty := \max\{|x_1|, \ldots |x_p|\} \:.
\eeq
The same norms can be considered on~$\R^n$ instead of~$\C^n$, and in the case~$p=2$, this gives the Euclidean length of a vector~$x \in \R^n$.
As a related, but infinite-dimensional example, on the vector space of compactly supported
continuous functions~$C^0_0(\R^n)$, one defines the integral norms
\beq \label{Lpnorm}
    \|f\|_p := \left( \int_{\R^n} |f(x)|^p \Diff x\right)^{\frac{1}{p}}
    \quad \text{ for } p < \infty \qquad \text{and} \qquad
    \|f\|_\infty := \sup_{x \in \R^n} |f(x)| \:.
\eeq
For more details on these examples see Exercise~\ref{ex:norm}.
The norm~$\| .|\|_p$ will be introduced on an abstract measure space
in Section~\ref{secbasicmeasure}.

We next recall the notion of completeness, which is a property about
convergence of sequences. First recall that a sequence
$(x_n)_{n \in \N}$ in a metric space~$E$ {\em{converges}} to a point
\sindex{convergence}%
$x \in E$ if for any~$\varepsilon > 0$ there is~$N \in \N$ such
that~$d(x_n,x) < \varepsilon$ for all~$n \geq N$. Similarly, a
sequence~$(x_n)_{n \in \N}$ in~$E$ is a {\em{Cauchy sequence}} if for
any~$\varepsilon>0$ there is some~$N \in \N$ such
that~$d(x_n, x_m) < \varepsilon$ for all~$m,n \geq N$. Any sequence
converging to a point of~$E$ is a Cauchy sequence. If conversely also
\sindex{Cauchy sequence}%
any Cauchy sequence converges to a point of~$E$, one calls~$E$ a
\emph{complete metric space}. 
\sindex{completeness}%
A normed space which as metric space is
complete is referred to as a {\em{Banach space}}.
\sindex{Banach space|textbf}%
A few examples of Banach spaces are given in Exercise~\ref{ex:banach}.
Completeness is an important and very useful property.
Therefore, we would like to restrict attention to complete metric
spaces. This is no major restriction because any non-complete metric
space can be regarded as subset of a corresponding complete metric space, its
so-called {\em{completion}} (for details see
Exercise~\ref{ex:complete}).
\sindex{completion!of metric space}%

We next specialize the setting by turning from a norm to a scalar
product.
\begin{Def}
    \label{defspspace}
    Let~$V$ be a complex vector space. A {\bf{scalar product on~$V$}}
    is a mapping
    \beq
        \la .|. \ra \::\: V \times V \rightarrow \C
    \eeq
    with the following properties: \bitem
\item[\rm{(i)}] Linearity in the second argument: For all~$u,v,w \in V$
    and~$\alpha, \beta \in \C$,
    \beq
        \la u \,|\, \alpha v + \beta w \ra
        = \alpha\: \la u | v \ra + \beta\: \la u | w \ra\:.
    \eeq
\item[\rm{(ii)}] Hermitian symmetry: For all~$u,v \in V$,
    \beq
        \overline{ \la u | v \ra } = \la v | u \ra \:.
    \eeq
\item[\rm{(iii)}] Positive definiteness: For all~$u \in V$,
    \beq
        \la u | u \ra \geq 0
        \qquad \text{and} \qquad
        \la u | u \ra = 0
        \; \Longleftrightarrow \; u=0 \:.
    \eeq
    \eitem If~$\la .|. \ra$ is a scalar product on~$V$, the
    pair~$(V, \la .|. \ra)$ is a {\bf{scalar product space}}.
\end{Def}
\sindex{scalar product}%
\sindex{scalar product space}%
\nindex{abg@$\la . \vert . \ra$ -- complex scalar product}%
\tindex{bb@$\la . \vert . \ra$ -- complex scalar product}%

\noindent Every scalar product space~$(V,\la .|. \ra)$ is also a
normed space with the norm being defined
by~$\|u\|:= \sqrt{\la u | u \ra}$ for all~$u \in V$ (see
Exercise~\ref{ex:hilbertnorm}). The {\em{Cauchy-Schwarz inequality}} 
\beq \big| \la u | u \ra \big| \leq \|u\|\:\|v\| \eeq
\sindex{Cauchy-Schwarz inequality}%
bounds the scalar product in terms of the corresponding norms; it is a direct
consequence of the above properties of a scalar product. A scalar product space that is a
Banach space; that is, which is complete, is called a {\em{Hilbert
    space}}. We usually denote a Hilbert space by~$\H$ or~$(\H, \la .|. \ra)$.
\sindex{Hilbert space}%

A simple example of a scalar product space is~$\C^n$ with the scalar product defined by
\beq
    \la u | v \ra := \sum_{i=1}^n \overline{u_i}\; v_i \,.
\eeq
An infinite-dimensional example is the space~$C_0(\R^n, \C)$ of complex-valued test functions
with the scalar product
\beq
    \la f | g \ra_{L^2}
    := \int_{\R^n} \overline{f(x)} \,g(x) \Diff x \:.
\eeq
\sindex{basis!of Hilbert space}%
\sindex{basis!orthonormal}%
The corresponding norms give us back~\eqref{lpnorm} or~\eqref{Lpnorm}, respectively, in the case~$p=2$.

Throughout this book, all Hilbert spaces will be {\em{separable}}, meaning that there is a countable
subset~$D \subset \H$ which is {\em{dense}} in the sense that its closure is the whole Hilbert space,
$\overline{D}=\H$. In a separable Hilbert space, one can choose an
{\em{orthonormal Hilbert space basis}}~$(e_i)_{i \in I}$ characterized by the following properties:
\bitem
\item[{\rm{(i)}}] The index set~$I$ is at most countable.
\item[{\rm{(ii)}}] The system~$(e_i)_{i \in I}$ is {\em{orthonormal}}, that is
\beq \la e_i | e_j \ra = \delta_{ij} \:, \eeq
where~$\delta_{ij}$ is the {\em{Kronecker delta}} defined by
\sindex{Kronecker delta}
\nindex{abg2@$\delta_{ij}$ -- Kronecker delta}%
\beq \delta_{ij} = \bigg\{ \begin{array}{cl} 1 & \text{if~$i=j$} \\ 0 & \text{otherwise}\:. \end{array} \eeq
\item[{\rm{(iii)}}] The system~$(e_i)_{i \in I}$ is {\em{complete}}, meaning that every vector~$u \in \H$
has the representation as a (possibly infinite) linear combination
\beq u = \sum_{i \in I} c_i\: e_i \qquad \text{with} \qquad c_i \in \C \:, \eeq
where the series converges in~$\H$.
\eitem
Using property~(ii), the coefficients~$c_i$ in the representation in~(iii) can be computed by~$c_i = \la e_i | u \ra$.
Thus, every vector~$u \in \H$ can be written as
\beq %\label{ucomplete}
u = \sum_{i \in I} \la e_i | u \ra \: e_i \:. \eeq
The convergence of this series is guaranteed by {\em{Bessel's inequality}}
\sindex{Bessel's inequality}%
\beq \sum_{i \in I} \big| \la e_i | u \ra \big|^2 \leq \|u\| \:, \eeq
which holds for any orthonormal system~$(e_i)_{i \in I}$ (even if not complete)
as a direct consequence of the properties of the scalar product.
Moreover, the cardinality of the index set~$I$ does not depend on the choice of the basis, making it
possible to define the {\em{dimension}} of the Hilbert space
\beq \dim \H := \# I \;\in\; \N_0 \cup \{ \infty \} \:. \eeq
\sindex{dimension!of Hilbert space}%

Now we turn our attention to linear operators.
\begin{Def} \label{defbounded}
Let~$(V, \| \cdot \|_V)$ and~$(W, \| \cdot \|_W)$ be normed spaces. A mapping
\beq A : V \rightarrow W \eeq
is a {\bf{bounded linear operator}} from~$V$ to~$W$ if it has the following properties:
\sindex{operator!bounded linear}%
\bitem
\item[\rm{(i)}] Linearity: For all~$u,v \in V$ and~$\alpha, \beta \in \C$,
\beq A \big( \alpha u + \beta v \big) = \alpha\, A(u) + \beta\, A(v) \:. \eeq
\item[\rm{(ii)}] Boundedness: There is a constant~$c>0$ such that for
    all~$u \in V$,
\beq \big\| A(u) \big\|_W \leq c\: \|u\|_V\:. \eeq
\eitem
\end{Def} \noindent 
Usually, one also writes~$A(u)$ simply as~$Au$.
We remark that for linear operators, boundedness is equivalent to continuity
(see Exercise~\ref{ex:linbound}).

The set of all bounded linear operators between two complex vector spaces~$V$ and~$W$ 
forms again a complex vector space with the vector operations defined pointwise. Thus for any bounded linear
operators~$A,B: V \to W$ and~$\alpha,\beta \in C$, the operator~$\alpha A + \beta B: V \to W$ is defined by
\beq
    \big( \alpha A + \beta B \big)(u) := \alpha \,Au + \beta\, Bu
    \qquad\text{for any } u \in V \:.
\eeq
The resulting vector space of bounded linear operators from~$V$ to~$W$ is denoted
by~$\Lin(V, W)$. One obtains a norm on this vector space by setting
\nindex{abh@$( \Lin(V, W), \NORM . \NORM)$ -- Banach space of linear bounded operators}%
\tindex{ff@$( \Lin(V, W), \NORM . \NORM)$ -- Banach space of linear bounded operators}%
\beq \|A\| := \sup_{u \in V, \|u\|_V = 1} \big\| Au \big\|_W \:, \eeq
referred to as the $\sup$-norm or the {\em{operator norm}}.
\sindex{$\sup$-norm}%
\sindex{operator norm}%
\nindex{abi@$\NORM . \NORM$, $\NORM . \NORM_\H$ -- $\sup$-norm, operator norm}%
\tindex{dd@$\NORM . \NORM$, $\NORM . \NORM_\H$ -- $\sup$-norm, operator norm}%
With this
norm, the space~$\Lin(V,W)$ is complete if and only if~$W$ is. For
details we refer to Exercise~\ref{ex:LVW}.

We will be concerned mainly with bounded linear operators acting on a Hilbert
space~$(\H, \la .|. \ra)$. Two cases are of specific interest:
mappings from~$\H$ to the complex numbers, and mappings from~$\H$ back
to itself. In the first case, the resulting operator is also referred
to as a  {\em{bounded linear form}}. These operators
form the so-called {\em{dual space}}~$\H^*$,
\beq \H^* := \Lin(\H, \C) \:. \eeq
\sindex{dual space}%
\sindex{linear form!bounded}%
An example of a bounded linear form is the mapping~$\H \ni u \mapsto \la v | u \ra \in \C$
obtained by taking the scalar product with a fixed vector~$v \in \H$.
In fact, every bounded linear form can be written in this way,
making it possible to canonically identify the dual space of a Hilbert space 
with the Hilbert space itself:
\sindex{Fr{\'e}chet-Riesz theorem}%
\begin{Thm} {\bf{(Fr{\'e}chet-Riesz)}} \label{thm-FR}
Let~$\H$ be a Hilbert space. Then for any bounded linear functional~$\phi \in \H^*$ there is a unique vector~$v \in \H$
such that
\beq \label{eq:FR}
\phi(u) = \la v | u \ra \qquad \text{for all~$u \in \H$} \:.
\eeq
\end{Thm} \noindent
In the case of a separable Hilbert space of interest here, this theorem can be understood
in simple terms by expanding vectors in an orthonormal Hilbert space basis~$(e_i)_{i \in I}$. Writing a vector~$u \in \H$ as~$u = \sum_{i \in I} \la e_i | u \ra\: e_i$, we have
\beq
    \phi(u) = \sum_{i \in I} \la e_i | u \ra\: \phi(e_i) \,,
\eeq
where in the infinite-dimensional case, we have to use continuity (boundedness) of~$\phi$ to
make sure that we can pull out the (infinite) sum. Now the idea is to rewrite the last term as
\beq
    \sum_{i \in I} \la e_i | u \ra\: \phi(e_i)
    = \Big\la \sum_{i \in I} \overline{\phi(e_i)}\, e_i \,\Big|\, u \Big\ra
\eeq
(where we used linearity and continuity of the scalar product),
and then simply define the vector~$v \in \H$ we are looking for by
\beq \label{vsum}
    v = \sum_{i \in I} \overline{\phi\big( e_i \big)}\: e_i \:.
\eeq
In finite dimensions, this computation is fine. In  infinite dimensions, however, one must show that the series 
defining~$v$ converges in~$\H$ \emph{before} one can compute as above. Fortunately, with a little trick, this 
convergence is not hard to see: Let~$I = \N$ and note first that for~$v_n := \sum_{i=1}^n \phi(e_i)\, e_i$, we get
\beq
    \sum_{i=1}^n \big| \phi(e_i) \big|^2
    = |\phi(v_n)|
    \leq \| \phi \| \cdot \|v_n\|
    = \| \phi \| \cdot \Big( \sum_{i=1}^n |\phi(e_i)|^2 \Big)^{\frac{1}{2}} \,,
\eeq
and therefore~$\sum_{i=1}^n |\phi(e_i)|^2 < \|\phi\|^2$. Using the completeness of~$\H$, one readily finds that 
the series in~\eqref{vsum} also converges as desired. It is also straightforward to verify that~$\|v\| = \|\phi\|$.

The Fr\'echet-Riesz theorem gives the mathematical justification for the \emph{bra/ket notation} commonly
used in quantum mechanics. 
\sindex{bra/ket notation}%
In this notation elements of~$\H$ are denoted by~$|v\rangle$ and referred to as \emph{kets}, and elements of the 
dual space~$\H^*$ are denoted by~$\langle v|$ and referred to as~\emph{bra}s. As a consequence of the
Fr\'echet-Riesz theorem, for any ket~$|v\rangle$ one may form the corresponding bra~$\langle v|$ and vice-versa. 
This notation resembles the role of the inner product~$\la \cdot | \cdot \ra$, as for any bra~$\la v|$ and
any ket~$|w\ra$ one may form the bra-(c)ket~$\la v|w \ra$.

Another application of the Fr{\'e}chet-Riesz theorem is to define the {\em{adjoint}} of a bounded
linear operator: Given two Hilbert spaces~$(\H_1, \la .|. \ra_{\H_1})$
and~$(\H_2, \la .|. \ra_{\H_2})$ as well as an operator~$A \in \Lin(\H_1, \H_2)$, for every~$u \in \H_2$ we
can define the linear functional on~$\H_1$
\beq \phi \::\: \H_1 \rightarrow \C \:,\qquad v \mapsto \la u \,|\, Av \ra_2 \:. \eeq
The estimate~$|\phi(v)| \leq \|u\|_{\H_2}\,\|A\|\,\|v\|_{\H_1}$ shows that this functional is bounded,
making it possible to represent it uniquely by a vector~$w \in \H_1$; that is,
\beq \la u | A v \ra_{\H_2} = \la w | v \ra_{\H_1} \qquad \text{for all~$v \in \H_1$}\:. \eeq
By direct computation one verifies that the resulting mapping~$v \mapsto w$ is linear and bounded.
The resulting operator~$A^* \in \Lin(\H_2, \H_1)$, referred to as the {\bf{adjoint}} of~$A$,
\sindex{operator!adjoint of}%
\sindex{adjoint!of operator}%
\nindex{abi1@$A^*$ -- adjoint of operator}%
\sindex{operator!of finite rank|textbf}%
satisfies and is uniquely determined by the relation
\beq \label{Addef}
\la u |A v \ra_{\H_2} = \la A^*u | v \ra_{\H_1} \qquad \text{for all~$u \in \H_2$ and~$v \in \H_1$}\:.
\eeq

We finally consider the space~$\Lin(\H, \H)$ of bounded linear endomorphisms.
For brevity, this space is also denoted by~$\Lin(\H)$.
In the context of such linear endomorphisms, the following additional notions are important.
\begin{Def} \label{defsymmetric}
\sindex{operator!symmetric|textbf}%
\sindex{operator!unitary|textbf}%
\sindex{operator!of finite rank|textbf}%
A bounded linear operator~$A \in \Lin(\H)$ is {\bf{symmetric}} if
\beq \la A u \,|\, v \ra = \la u \,|\, A v \ra \qquad \text{for all~$u,v \in \H$}\:. \eeq
An operator~$U \in \Lin(\H)$ is {\bf{unitary}} if it has a bounded inverse and if
\beq \la U u \,|\, U v \ra = \la u \,|\, v \ra \qquad \text{for all~$u,v \in \H$}\:. \eeq
It has {\bf{finite rank}} if its image~$A(\H)$ is a finite-dimensional subspace of~$\H$.
\end{Def} \noindent
For clarity, we mention that there is also the notion of an operator being \textbf{selfadjoint}.
For bounded linear operators, this is equivalent to being symmetric.
For unbounded operators, however, being selfadjoint is a stronger property than being symmetric.
Selfadjointness is required, for instance, in the proof of the spectral theorem (see Section~\ref{secspectral}).
With this in mind, when talking about bounded operators, in this book we usually prefer the notion of
a {\em{symmetric}} operator.

If the Hilbert space~$\H$ is finite-dimensional, symmetric and unitary operators
can be diagonalized by choosing an orthonormal basis of eigenvectors
(for details see standard textbooks on linear algebra like~\cite{halmosla, strang}).
The eigenvalues of a symmetric operator are all real, whereas the eigenvalues of a unitary
operator have modulus one. The generalization of this result to the infinite-dimensional setting
is provided by the {\em{spectral theorem}}. The general spectral theorem for bounded symmetric
operators, which will be needed mainly in Chapter~\ref{secFSO},
will be treated in Section~\ref{secspectral} below.
In large parts of this book, however, we only deal with symmetric operators having {\em{finite rank}} (on possibly infinite-dimensional Hilbert spaces).
For such operators the results of linear algebra carry over in a straight-forward manner, as we now explain.

Thus, let~$A \in \Lin(\H)$ be a symmetric bounded operator of finite rank.
Then, by definition, its image~$I:=A(\H)$ is finite-dimensional. Its orthogonal
complement
\sindex{orthogonal complement}%
\nindex{abj@$I^\perp$ -- orthogonal complement of~$I \subset \H$}%
\beq \label{Iperp}
I^\perp := \big\{v \in \H \:|\: \la v, u\ra=0 \text{ for all~$u \in I$} \big\}
\eeq
is a closed subspace of~$\H$, and every vector~$u \in \H$ can be decomposed uniquely as
\beq \label{udecomp}
u = u^{||} + u^\perp \qquad \text{with} \qquad u^{||} \in I, u^\perp \in I^\perp
\eeq
(see Exercise~\ref{ex:ortho}). Moreover, for any~$u \in I^\perp$, the computation
\beq 0 = \la A^2 u \:|\: u \ra = \la A u \:|\: A u \ra \eeq
shows that~$Au=0$. Therefore, $A$ vanishes identically on~$I^\perp$, and so it suffices to consider the restriction~$A|_I : I \rightarrow I$. Being an operator on
a finite-dimensional vector space, it can be diagonalized as in linear algebra.

When taking products of symmetric operators~$A_1, \ldots, A_n$ of finite rank, one chooses~$I$ as the
finite-dimensional vector space spanned by the images of all the operators.
Then the restrictions~$A_k|_I$ with~$k=1,\ldots,n$ all map~$I$ to itself, making it possible to work
again in a finite-dimensional subspace of~$\H$.

\section{Basics on Abstract Measure Theory} \label{secbasicmeasure}
The basic object of a causal fermion system is a measure.
For this reason, we now provide the necessary basics on
abstract measure theory. For more details we refer to
good standard textbooks like~\cite{rudin, halmosmt, bogachev, folland, bauer}.

Let~$\F$ be a set. A measure is a mapping which associates a non-negative number to a subset of~$\F$ ,
which can be thought of as the ``volume'' of the set. In order to get into a mathematically sensible setting,
one cannot define the measure on any subset, but only on a distinguished family of subsets of~$\F$.
This family must form a $\sigma$-algebra, which we now define.
\begin{Def} A system~${\mathfrak{M}}$ of subsets of~$\F$ is a {\bf{$\sigma$-algebra}} if
it has the following properties:
\bitem
\item[\rm{(i)}] $\varnothing \in {\mathfrak{M}}$.
\item[\rm{(ii)}] ${\mathfrak{M}}$ is closed under taking complements: For any subset~$A \subset \F$,
\beq A \in {\mathfrak{M}} \quad \Longrightarrow \quad \F \setminus A \in {\mathfrak{M}} \:. \eeq
\item[\rm{(iii)}] ${\mathfrak{M}}$ is closed under at most countable unions: For any sequence~$(A_n)_{n \in \N}$ of subsets of~$\F$,
\beq A_n \in {\mathfrak{M}} \text{ for all } n \in \N \quad \Longrightarrow \quad
\bigcup_{n \in \N} A_n \in {\mathfrak{M}} \:. \eeq
\eitem
The sets in~${\mathfrak{M}}$ are also referred to as the {\bf{measurable}} sets.
\end{Def} \noindent
\sindex{$\sigma$-algebra}%
\sindex{set!measurable}%
\nindex{abk@${\mathfrak{M}}$ -- $\sigma$-algebra of measurable sets}%
Using De Morgan's laws, it follows that a $\sigma$-algebra is closed even under at most countable
intersections and, more generally, under at most countable set operations.

We next introduce a measure as a mapping that to every measurable set
associates its ``volume.'' This mapping is compatible with at most countable set operations,
as is made precise by the notion of $\sigma$-additivity.
\begin{Def} \label{defmeasure}
A {\bf{measure}}~$\rho$ is a mapping from a $\sigma$-algebra to the non-negative numbers
or infinity,
\beq \rho \::\: {\mathfrak{M}} \rightarrow \R^+_0 \cup \{\infty\} \:, \eeq
which has the following properties:
\bitem
\item[\rm{(i)}] $\rho(\varnothing)=0$.
\item[\rm{(ii)}] $\rho$ is {\bf{$\sigma$-additive}}: For any sequence~$(A_n)_{n \in \N}$ 
of pairwise disjoint measurable sets,
\sindex{$\sigma$-additivity}%
\beq \rho \Big( \bigcup_{n \in \N} A_n \Big) = \sum_{n=1}^\infty \rho \big( A_n \big) \:. \eeq
\eitem
The structure~$(\F, {\mathfrak{M}}, \rho)$ is a {\bf{measure space}}.
\sindex{measure space}%
\end{Def} \noindent
By choosing almost all of the sets~$A_n$ in (ii) as empty sets, one sees
that $\sigma$-additivity implies {\em{finite additivity}}. In particular~$\rho(A \cup B) = \rho(A) + \rho(B)$
for any disjoint measurable sets~$A$ and~$B$. As a consequence, a measure is {\em{monotone}}
in the sense that~$\rho(A) \leq \rho(B)$ for any measurable sets~$A \subset B$ (see Exercise~\ref{ex:measure-basic-properties}).

A set of measure zero is also referred to as a {\em{null set}}. By monotonicity, every measurable
subset of a null set is again a null set. If every subset of a null set is measurable, then the measure
is called {\em{complete}}.
\sindex{completeness!of measure}%

On a measure space~$(\F, {\mathfrak{M}}, \rho)$, a notion of {\em{integration}} is introduced as follows.
\sindex{integration}%
We begin with complex-valued functions that take only a finite number of values,
also referred to as {\em{step functions}}. A step function~$f : \F \rightarrow \C$ can be written as
\beq f = \sum_{n=1}^N c_n \chi_{A_n} \eeq
with~$N \in \N$, coefficients~$c_n \in \C$ and measurable sets~$A_n \in {\mathfrak{M}}$
(here~$\chi_A$ is the characteristic function defined by~$\chi_A(x)=1$ if~$x \in A$ and~$\chi_A(x)=0$ otherwise).
Its integral is defined in a natural way by
\beq \int_\F f\: \Diff\rho := \sum_{n=1}^N c_n \: \rho \big(A_n \big) \:. \eeq
This integral can be extended to more general functions as follows.
A function~$f : \F \rightarrow \C$ is {\em{measurable}}
\sindex{function!measurable}%
if the pre-image of any open set is measurable.
For a measurable function~$f: \F \to [0,\infty]$ taking real and nonnegative values or the value plus infinity,
one defines
\begin{align}
&\int_\F f\: \Diff\rho \notag \\
&:= \sup \bigg\{ \int_\F s\: \Diff \rho \,\bigg|\,
    s: \F \to [0,\infty) \text{ is a step function with } s \leq f \bigg\}
    \in \R^+_0 \cup \{\infty\}\,.
\end{align}
This expression is allowed to be infinite. If it is finite, one calls
$f$ \emph{integrable}. From here on, the further generalization is straightforward:
For a measurable function~$f: \F \to \R$, one defines its positive and negative~$f^+,f^-: \F \to [0,\infty)$ by~$f^+(x) := 
\max\{f(x),0\}$ and~$f^-(x) := \max\{-f(x),0\}$. Then, clearly, $f = f^+ - f^-$, and one can show that~$f^+,f^-$ are again 
measurable. If at least one of them is integrable, one defines
\beq
    \int_\F f \Diff \rho
    := \int_\F f^+ \Diff \rho - \int_\F f^- \Diff \rho
    \in \R \cup \{-\infty,\infty\} \,.
\eeq
Again, $f$ is defined to be integrable if this integral is finite (note that no
cancellation between positive and negative terms can happen, as
positive and negative part of~$f$ are considered separately). Finally,
for a complex-valued measurable function~$f: \F \to \C$ whose real and
imaginary part are both integrable (they are always measurable), one defines
\beq
    \int_\F f \Diff \rho
    := \int_F \re(f) \:\Diff \rho
    + \cI \int_\F \im(f)\: \Diff \rho \in \C \,.
\eeq
A complex-valued measurable function on~$\F$ whose real and imaginary part are both integrable is again called 
integrable. One can combine all these notions of integrability by demanding that the absolute value
is integrable,
\beq \label{intfinite}
\int_\F |f(x)|\: \Diff\rho(x) \;\in\; \R^+_0 \cup \{ \infty \} \:.
\eeq
\nindex{abl@$L^p(\F, \Diff\rho)$ -- $L^p$-spaces with~$1 \leq p \leq \infty$}%
\tindex{ff@$L^p(\F, \Diff\rho)$ -- $L^p$-spaces with~$1 \leq p \leq \infty$}%
This condition can be understood immediately from the requirement that
in integrals one must always avoid expressions of the form~``$\infty - \infty$.''

The integrable functions form a vector space denoted by~$L^1(\F, \Diff\rho)$.
Similarly, the measurable functions~$f$ whose power~$|f|^p$ with~$p \in (1, \infty)$
is integrable, form a vector space~$L^p(\F, \Diff\rho)$. Finally, the space~$L^\infty(\F, \Diff\rho)$
is defined as the functions that are essentially bounded in the sense that there is
a number~$c>0$ such that the pre-image~$|f|^{-1}((c, \infty])$ has $\rho$-measure zero.
The spaces~$L^p(\F, \Diff\rho)$ with~$p \in [1, \infty]$ are almost normed spaces if endowed with the corresponding norms
\begin{align}
\|f\|_p &:= \bigg( \int_\F |f|^p\: \Diff\rho \bigg)^{\frac{1}{p}} \qquad \text{if~$p \in [0, \infty)$} \\
\|f\|_\infty &:= \inf \bigg\{ c \geq 0 \:\Big|\: \rho \Big( |f|^{-1} \big( (c, \infty] \big) \Big)=0 \bigg\} \:.
\end{align}
The only issue here is that functions that vanish almost everywhere
(i.e., which are non-zero only on a set of $\rho$-measure zero) have norm zero.
In order to resolve this issue, in the $L^p$-spaces one quotients out these functions.
We thus obtain a normed space denoted by $L^p(\F,\Diff\rho)$, which even turns out to be a Banach space.
Although the vectors in these~$L^p$-spaces are equivalence classes of functions that differ on sets of measure zero, for simplicity, one usually refers to vectors in~$L^p(\F, \Diff\rho)$ as {\em{functions}} and understands
implicitly that they may be changed arbitrarily on sets of measure zero.
The space~$L^2(\F, \Diff\rho)$ is even a Hilbert space, endowed with the scalar product
\beq \label{scalprod}
\la f | g \ra_{L^2(\F, \Diff\rho)} := \int_\F \overline{f(x)} \:g(x)\: \Diff\rho(x) \:.
\eeq
\nindex{abm@$\la . \vert . \ra_{L^2(\F, \Diff\rho)}$ -- $L^2$-scalar product}%
\tindex{bb@$\la . \vert . \ra_{L^2(\F, \Diff\rho)}$ -- $L^2$-scalar product}%
We remark that these constructions generalize immediately to functions taking values in a Banach
or Hilbert spaces, if one simply replaces the absolute value in~\eqref{intfinite} by the norm on the Banach space
and the inner product~$\overline{f} g$ in~\eqref{scalprod} by the Hilbert space scalar product.
Finally, we remark that integration over a subset~$A \subset \F$ is defined as
\beq
    \int_A f \,\Diff \rho
    := \int_\F f \chi_A \,\Diff \rho \,,
\eeq
where~$\chi_A$ is the characteristic function of~$A$
(defined by~$\chi_A(x)=1$ if~$x \in A$ and~$\chi_A(x)=0$ otherwise).
\sindex{characteristic function}%
\nindex{abn@$\chi_A$ -- characteristic function of~$A$}%

We now specialize the setting by considering a class of measures that will be
of major importance in this book, namely Borel measures.
To this end, we assume that~$(\F, \O)$ is a topological space. 
Then, the {\em{Borel algebra}} is defined as the smallest
$\sigma$-algebra that contains all the open sets (see Exercise~\ref{ex:borel}).
An element of the Borel algebra is a {\em{Borel set}}.
\sindex{Borel set}%
\sindex{set!Borel}%
\sindex{Borel algebra|textbf}%
A measure on the Borel algebra is referred to as a {\em{Borel measure}}.
\sindex{Borel measure}%
\sindex{measure!Borel}%
\sindex{Borel measure!regular|textbf}%
\sindex{Borel measure!locally finite}%
The Borel measures of relevance to us will typically harmonize with the topology in the following sense.
\begin{Def} A Borel measure~$\rho$ on~$\F$ is called {\bf{regular}} if for any measurable set~$A$,
\beq \rho(A) = \sup_{K \subset A \text{\rm{ compact}}} \rho(K) = \inf_{\Omega \supset A \text{\rm{ open}}} \rho(\Omega)\:. \eeq
It is {\bf{locally finite}} if every point of~$\F$ has an open neighborhood~$\Omega$ with~$\rho(\Omega)< \infty$.
Regular, locally finite Borel measures are also referred to as {\bf{Radon measures}}.
\sindex{Radon measure}%
\end{Def} \noindent
The Lebesgue measure on~$\R^n$ is a Radon measure. We remark that a Borel measure in general is {\em{not complete}}, because a Borel null set may have subsets that are not Borel sets.
One could improve the situation by forming the {\em{completion}} of the measure
(see Exercise~\ref{exmeasurecomplete}). 
\sindex{completion!of measure|textbf}%
However, completeness of the measure is not
important for most applications, and it will often be more convenient to work with Borel measures.

One of the advantages of the notion of integration introduced above (compared to, for example,
the Riemann integral)
is that various (easy to use and prove) results regarding convergence of integrals hold. For instance, for a
sequence~$(f_n)_{n \in \N}$ of measurable functions~$f_n: \F \to \C$ that converge pointwise to a function~$f: \F \to \C$, the function~$f$ is again measurable and one has
\beq \int_\F f_n \,\Diff \rho \to \int_\F f \,\Diff \rho \eeq
if all functions only take nonnegative values and~$f_n \leq f_{n+1}$ for all~$n \in \N$ (\emph{Lebesgue's monotone convergence theorem}), 
\sindex{Lebesgue's monotone convergence theorem}%
\sindex{Lebesgue's dominated convergence theorem}%
\sindex{monotone convergence theorem}%
\sindex{dominated convergence theorem}%
or if there exists an integrable function~$g: \F \to [0,\infty)$ with~$|f_n| \leq g$ for all~$n \in \N$ (\emph{Lebesgue's dominated convergence theorem}). Another important result in integration theory is \emph{Fubini's theorem},
\sindex{Fubini's theorem}%
which is about iterated integrals. If~$(\F,\mathfrak{M},\rho)$ and~$(\mathcal{G},\mathfrak{N},\nu)$ are two measure 
spaces, then on the product space~$\F \times \mathcal{G}$, there is a natural~$\sigma$-algebra containing all product 
sets~$M \times N$ with~$M \in \mathfrak{M}$, $N \in \mathfrak{N}$, and a measure~$\rho \times \nu$ on this 
$\sigma$-algebra 
such that~$(\rho \times \nu)(M \times N) = \rho(M) \nu(N)$ for all~$M \in \mathfrak{M}$, $N \in 
\mathfrak{N}$. Now Fubini's theorem now states that if~$f: \F \times \mathcal{G} \to \C$ is integrable with respect to 
this measure, then
\begin{equation}
    \label{eq:fubini-theorem}
    \int_{\F \times \mathcal{G}} f \:\Diff (\rho \times \nu)
    = \int_{\mathcal{G}} \left( \int_{\F} f(x,y)\: \Diff \rho(x) \right) \Diff \nu(y)
    = \int_\F \left( \int_{\mathcal{G}} f(x,y)\: \Diff \nu(y) \right) \Diff \rho(x) \,.
\end{equation}
A variant of Fubini's theorem, referred to as {\em{Tonelli's theorem}},
\sindex{Tonelli's theorem}%
states that if~$f$ is measurable and non-negative (but not necessarily integrable),
then again~\eqref{eq:fubini-theorem} holds, but now the integrals could take the value plus infinity.

We conclude this section by introducing a few other notions that will be needed later on.
The first notion is the support of a measure.
\begin{Def} \label{defsupp}
Let~$(\rho, {\mathfrak{M}})$ be a measure on the topological space~$(\F, \O)$.
\sindex{support!of a measure}%
The {\bf{support}} of~$\rho$ is defined as the complement of the largest open set of measure zero; that is,
\beq \label{suppdef}
\supp \rho := \F \setminus \bigcup \big\{ \text{$\Omega \subset \F$ \,\big|\,
$\Omega$ is open and~$\rho(\Omega)=0$} \big\} \:.
\eeq
\end{Def} \noindent
Note that the support is by definition a closed subset of~$\F$. In integrals, one can always restrict to the support of a measure in the sense that the identity
\beq
    \int_\F f \:\Diff \rho
    = \int_{\supp \rho} f \:\Diff \rho
\eeq
holds for any integrable function~$f$ on~$\F$.

Suppose we want to compare two Radon measures~$\rho$ and~$\tilde{\rho}$ on~$\F$.
A natural idea is to consider the difference of the measures~$\rho - \tilde{\rho}$.
The difficulty is that for a measurable set~$A \subset \F$, its measures~$\rho(A)$
and~$\tilde{\rho}(A)$ could both take the value~$+\infty$, in which case their difference would be ill-defined.
In order to avoid this problem, we use the regularity of a Radon measure and exhaust by compact sets.
Assuming that~$\F$ is locally compact, the fact that Radon measures are locally finite
implies that Radon measures of compact sets are always finite. This leads us to the following definition:

\begin{Def} \label{deftotvar}
Given two Radon measures~$\rho$ and~$\tilde{\rho}$ on a locally compact topological space~$\F$, we
define the Borel measures~$\mu^\pm$ by
\begin{align}
\mu^+(A) &= \sup_{K \subset A \text{\rm{ compact}}} \big( \tilde{\rho}(K) - \rho(K) \big) \\
\mu^-(A) &= \sup_{K \subset A \text{\rm{ compact}}} \big( \rho(K) - \tilde{\rho}(K) \big)
\end{align}
for any Borel subset~$A \subset \F$.
The difference of measures~$\rho-\tilde{\rho}$ is said to have {\bf{bounded total variation}}
if the measures~$\mu^\pm$ are finite, i.e.\ if
\beq \mu_+(\F), \mu_-(\F) < \infty \:. \eeq
If this is the case, the {\bf{total variation measure}}~$|\rho - \tilde{\rho}|$ is defined by
\beq |\rho - \tilde{\rho}| = \mu^+ + \mu^- \:. \eeq
\end{Def} \noindent
We remark that~$\rho-\tilde{\rho}$ can also be defined in the context of signed measures;
we refer the interested reader to~\cite[\S28]{halmosmt} or~\cite[Section~6.1]{rudin}.
\Evtl{Mache signierte Ma{\ss}e vielleicht als Aufgabe.}%
\sindex{total variation!bounded}%
\sindex{total variation measure}%

Another notion of measure theory that we will use frequently is the push-forward measure,
which we now define (for more details see for example~\cite[Section~3.6]{bogachev}
or Exercise~\ref{expushforward}).
To this end, let~$(\F, {\mathfrak{M}}, \rho)$ be a measure space, and suppose we are given a mapping~$f \::\: \F \rightarrow X$,
where~$X$ is any set. Then~$f$ induces a measure on~$X$ as follows: Let~${\mathfrak{M}}_X$
be the set of all subsets~$\Omega \subset X$ whose pre-image~$f^{-1}(\Omega)$ is $\rho$-measurable.
Using the elementary identities for inverse images of unions and complements,
% \beq f^{-1} \bigg( \bigcup_{n \in \N} \Omega_n \bigg) = \bigcup_{n \in \N} f^{-1} \big(\Omega_n \big) \qquad \text{and} \qquad
% f^{-1} \big( X \setminus \Omega \big) = \F \setminus f^{-1} (\Omega) \:, \eeq
one verifies that~${\mathfrak{M}}_X$ is indeed a~$\sigma$-algebra on~$X$. On this $\sigma$-algebra, the {\em{push-forward measure}}~$f_* \rho$ is defined by
\sindex{measure!push-forward|textbf}%
\beq (f_* \rho)(\Omega) := \rho\big( f^{-1}(\Omega) \big) \:. \eeq
Using again the above-mentioned identities for inverse images, one verifies that~$f_*\rho$ is indeed a measure.

\section{Distributions and Fourier Transform} \label{secfourier}
We now recall a few basics on distribution theory and the Fourier transform.
For more details, we recommend the textbook~\cite{friedlander2},
\cite[Sections~V.3 and~IX]{reed+simon}
or~\cite[\S2.1, \S2.2 and Appendix~A]{rauch}.

The theory of distributions describes a generalization of the concept of a function on~$\R^n$ (or, similarly, on a 
bounded domain or smooth manifold). Moreover, the differential calculus for smooth functions 
is extended to objects that are more singular than functions. The desire for such objects can be motivated for 
instance by the classical problem in electrostatics to determine the electric field generated by a distribution of 
charges. In the continuum formulation, a distribution of charges is described by a charge density~$\rho: \R^3 \to \R$ 
(typically compactly supported) having the interpretation that for any domain~$\Omega \subset \R^3$ the integral 
$\int_\Omega 
\rho(x) \Diff^3 x$ describes the total charge contained inside of~$\Omega$. The electromagnetic field 
$E: \R^3 \to \R^3$ generated by~$\rho$ can then be computed as~$E = - \nabla \phi$, where~$\phi$ is a (suitable) 
solution of Poisson's equation~$\Delta \phi = - \rho$. Now suppose that in this formulation one wants to deal with a 
\emph{point particle} whose complete charge~$Q$ is concentrated at
a single point, say the origin. Then the corresponding density~$\rho$ would need to satisfy
\beq \label{chicharge}
\int_\Omega \rho(x) \:\Diff^3 x
    =
    \begin{cases}
        Q & 0 \in \Omega \,, \\
        0 & 0 \notin \Omega
    \end{cases}
    \qquad \text{ for any } \Omega \subset \R^3 \,.
\eeq
It is not difficult to see that such a function~$\rho$ cannot exist (see Exercise~\ref{exnofunc}).
Intuitively speaking, this function would need to vanish outside of the origin.
At the origin, however, its value would have to be ``so (infinitely) large'' that an integral over a region
containing the origin still gives a nonzero contribution. The most common way to rigorously deal with such singular 
objects is to understand them as linear functionals on certain spaces of smooth functions on~$\R^3$, referred to as 
{\em{test functions}}.
\sindex{test function}%
In order to motivate this functional, we write~\eqref{chicharge} more generally as
\beq \int_{\R^3} \rho(x)\, f(x) \:\Diff^3 x = Q\, f(0) \:. \eeq
Indeed, choosing~$f$ as the characteristic function~$f=\chi_\Omega$
(defined by~$\chi_\Omega(x)=1$ if~$x \in \Omega$ and~$\chi_\Omega(x)=0$ otherwise),
we recover~\eqref{chicharge}. But now~$f$ can be a more general function.
Restricting to smooth function gives rise to Dirac's $\delta$ distribution (or simply $\delta$ distribution),
as is explained in more detail in the next example (for simplicity in one dimension).

 \begin{Example} {\bf{(The $\delta$ distribution)}} \label{exdelta}
 {\em{ The prime example of a distribution is Dirac's {\em{$\delta$ distribution}},
 which in physics textbook is introduced as a ``function''~$\delta(x)$ that is zero everywhere except at the origin,
 where it takes the value~$\infty$. The infinite contribution at the origin is ``normalized'' by demanding that
 its integral is equal to one. These properties can be summarized by saying that
 \beq \label{deltadef}
 \int_{-\infty}^\infty f(x)\: \delta(x)\: \Diff x = f(0) \:.
 \eeq
 There are various ways to make mathematical sense of this equation. One method is to
 regard the combination~$\delta(x)\: \Diff x$ as a measure~$\delta_0$ supported at the origin of total volume one,
 that is, $\delta_0(\R)=1$. In this way, \eqref{deltadef} makes sense if~$f$ is any continuous function.
 An alternative method is to take~\eqref{deltadef} as the definition of a linear functional on a space
 of suitable test functions~$f$. The latter method has the advantage that it makes it possible to even define
 the derivative of the~$\delta$ distribution by
 \beq \label{deltaderiv}
 \int_{-\infty}^\infty f(x)\: \delta'(x)\: \Diff x = -f'(0) \:.
 \eeq
 In order to allow for distributions to include an arbitrary number of derivatives,
 we choose a space of {\em{smooth}} test functions. Distributions will be defined as linear
 functionals on this space of test functions. Derivatives of distributions can be defined similar
 as in~\eqref{deltaderiv} by ``formally integrating by parts.''
 Since a function~$g$ defines a linear functional~$T_g$ by integrating,
 \beq \label{Tint}
 T_g(f) := \int_{-\infty}^\infty g(x)\: f(x)\: \Diff x \:,
 \eeq
 every function gives rise to a corresponding distribution. In this sense, distributions are generalized functions.
 In order to make sure that the integral in~\eqref{Tint}
 exists and is finite, it is a good idea to assume that the test function~$f$ has suitable decay properties at infinity.

 In order to make these ideas mathematically precise, we need to
 specify the space of test functions. Moreover, we need to endow this space of test function with
 a topology. Then, we can introduce distributions as the space of linear functionals on the test functions.
 As we shall see, working with the right space of test functions, one can make mathematical sense of
 the Fourier transform for distributions.
 }} \QEDrem
 \end{Example}

After this motivation, we now turn to the mathematical definition of distributions.
In preparation of our constructions, we recall the {\em{multi-index notation}} in~$\R^n$. A {\em{multi-index}}
is an $n$-tuple $\alpha=(\alpha_1, \ldots, \alpha_n) \in \N_0^n$ of non-negative integers.
For such a multi-index~$\alpha$, we define the corresponding monomial~$x^\alpha$ and  combination of partial derivatives~$\partial^\alpha$ by
\beq x^\alpha := \big( x_1 \big)^{\alpha_1} \cdots \big( x^n \big)^{\alpha_n} \qquad \text{and} \qquad
 \partial^\alpha := \partial_x^\alpha := \Big( \frac{\partial}{\partial x^1} \Big)^{\alpha_1} \cdots \Big( \frac{\partial}{\partial x^n} \Big)^{\alpha_n} \:. \eeq
The {\em{order}}~$|\alpha|$ of the multi-index~$\alpha$ is defined by
\beq |\alpha| := \alpha_1 + \cdots + \alpha_n \:. \eeq
\sindex{multi-index}%

We say that a function~$f: \R^n \to \C$ is {\em{smooth}} if all its partial derivatives
\sindex{function!smooth}%
exist to every order. The space of smooth, complex-valued functions is denoted by~$C^\infty(\R^n, \C)$.
For a smooth function~$f \in C^\infty(\R^n, \C)$ we define its {\em{Schwartz norms}}~$\| f \|_{p,q}$
\sindex{Schwartz norm}%
\nindex{abo@$\NORM f \NORM_{p,q}$ -- Schwartz norm of~$f$}%
\tindex{dd@$\NORM f \NORM_{p,q}$ -- Schwartz norm of~$f$}%
with~$p,q \in \N_0$ by
\beq \|f\|_{p,q} := \max_{\substack{\alpha \in \N_0^n\\ |\alpha| \leq p}}\;\;
\max_{\substack{\beta \in \N_0^n\\ |\beta| \leq q}}  \;\; \sup_{x \in \R^n} \big| x^\alpha\, \partial^\beta f(x) \big| \:. \eeq
The {\em{Schwartz space}}~${\mathcal{S}}(\R^n)$ is formed of all smooth functions for which all the
Schwartz norms are finite; that is,
\nindex{abp@${\mathcal{S}}(\R^n)$ -- Schwartz space}%
\tindex{ff@${\mathcal{S}}(\R^n)$ -- Schwartz space}%
\sindex{Schwartz space|textbf}%
\beq {\mathcal{S}}(\R^n) := \Big\{ f \in C^\infty \big(\R^n, \C \big) \:\Big|\: \|f\|_{p,q} < \infty \text{ for all~$p,q \in \N_0$}
\Big\} \:. \eeq
We always consider complex-valued functions, but the constructions work similarly for real-valued functions.
Defining the vector operators pointwise, ${\mathcal{S}}(\R^n)$ is a complex vector space.
The functions in~${\mathcal{S}}(\R^n)$ are referred to as {\em{Schwartz functions}}.
\sindex{Schwartz function}%
These functions have the property that they as well as all their partial derivatives have {\em{rapid decay}} in the sense that
multiplying them by a polynomial of arbitrary order still gives a bounded function. In particular, one has~$C^\infty_0(\R^n) \subset \mathcal{S}(\R^n) \subset C^\infty(\R^n)$, where both inclusions are strict. An example of a Schwartz function without compact support is a Gaussian~$f: \R^n \to \R, x \mapsto \E^{-x^2}$.

The Schwartz norms induce a {\em{topology}} on~${\mathcal{S}}(\R^n)$ as follows.
\sindex{topology!on Schwartz functions}%
We say a set~$\Omega \subset {\mathcal{S}}(\R^n)$
is {\em{open}} if for every~$f \in \Omega$ there exists~$p,q \in \N_0$ and~$r>0$ such that
the open $r$-ball corresponding to the norm~$\|.\|_{p,q}$ is contained in~$\Omega$; that is,
\beq \label{Stop}
\big\{ g \in {\mathcal{S}}(\R^n) \:\big|\: \|f-g\|_{p,q}< r \big\} \subset \Omega \:.
\eeq
For many purposes, it is sufficient to have in mind how convergence is expressed concretely by the Schwartz norms. For a sequence~$(f_n)_{n \in \N}$ in~$\mathcal{S}(\R^n)$ and~$f \in \mathcal{S}(\R^n)$, one has (see Exercise~\ref{ex:Stop})
\beq \label{locconvex}
f_n \rightarrow f \text{ in~${\mathcal{S}}(\R^n)$} \quad \Longleftrightarrow \quad \|f_n - f\|_{p,q} \rightarrow 0
\text{ for all~$p,q \in \N_0$} \:.
\eeq

Here we point out that for a set~$\Omega \subset \mathcal{S}(\R^n)$ to be an open neighborhood of~$f \in \mathcal{S}(\R^n)$, it suffices that condition~\eqref{Stop} is satisfied for
\emph{some}~$p,q \in \N_0$. 
If instead one uses the stronger condition that~\eqref{Stop} must hold for \emph{all}~$p,q \in \N_0$, one obtains a \emph{coarser} topology on~$\mathcal{S}(\R^n)$, meaning that there are fewer
open sets. In contrast, demanding~\eqref{Stop} merely for some~$p,q \in \N_0$ gives a
finer topology. Working with a {\em{finer topology}} has the following purpose:
For a finer topology, fewer sequences converge (as is obvious in~\eqref{locconvex}, where
the sequence must converge for {\em{all}} Schwartz norms).
As a consequence, there are more continuous linear functionals, simply because sequential
continuity must be verified for fewer sequences.
In other words, choosing a finer topology on a vector space has the effect that
its dual space becomes larger (where the dual space is defined as the
space of all continuous linear functionals). Since distributions will now be defined as such a dual space, this is desirable because it will ensure a sufficiently rich and general class of objects.

\begin{Def} The space of {\bf{tempered distributions}} denoted by~${\mathcal{S}}'(\R^n)$
is defined as the dual space of the Schwartz space,
\sindex{distribution!tempered}%
\nindex{abq@${\mathcal{S}}'(\R^n)$ -- space of tempered distributions}%
\tindex{ff@${\mathcal{S}}'(\R^n)$ -- space of tempered distributions}%
\beq {\mathcal{S}}'(\R^n) := {\mathcal{S}}^*(\R^n) = \Lin \big( {\mathcal{S}}(\R^n), \C \big)\:. \eeq
\end{Def} \noindent
For a linear functional~$T : {\mathcal{S}}(\R^n) \rightarrow \R$, continuity
means that there are~$p,q \in \N_0$ and a constant~$c>0$ such that
(see again Exercise~\ref{ex:Stop})
\beq \label{Tcontinuous}
\big| T(f) \big| \leq c\: \|f\|_{p,q} \qquad \text{for all~$f \in {\mathcal{S}}(\R^n)$} \:.
\eeq

As the first example, we return to the \emph{$\delta$ distribution} from our motivating Example~\ref{exdelta}.
It is the tempered distribution~$\delta \in \mathcal{S}'(\R^n)$
\sindex{$\delta$ distribution}%
\sindex{Dirac's $\delta$ distribution}%
\nindex{abq2@$\delta$ -- (Dirac's) $\delta$ distribution}%
given by~$\delta(f) := f(0)$. The linearity of~$\delta$ is obvious and the estimate
\beq
    |\delta(f)| = |f(0)| \leq \|f\|_{0,0}
\eeq
shows that~$\delta$ is continuous.

Next, we explain how any (bounded, measurable) function can be naturally viewed as a distribution, thus explaining why tempered distributions can be regarded as generalized functions. To this end, let~$g \in L^\infty(\R^n)$ be a bounded, measurable function. We define a linear functional~$T_g$
on~$\mathcal{S}(\R^n)$ by
\beq T_g(f) := \int_{\R^n} g(x) f(x) \Diff^n x \:. \eeq
This functional is continuous because
\beq \label{intes}
\begin{split}
\big| T_g(f) \big| &\leq \int_{\R^n} |g(x) f(x)| \Diff^n x
= \int_{\R^n} \frac{|g(x)|}{\big(1 + |x|^2 \big)^\frac{n+1}{2}} \: |f(x)|\: \big(1 + |x|^2 \big)^\frac{n+1}{2}\: \Diff^n x \\
&\leq C(n)\: \|g\|_{L^\infty(\R^n)}\: \|f\|_{n+1, 0}\: \int_{\R^n} \frac{\Diff^nx}{\big(1 + |x|^2 \big)^\frac{n+1}{2}} \\
&\leq C'(n)\: \|g\|_{L^\infty(\R^n)}\: \|f\|_{n+1, 0}\: \,.
\end{split}
\eeq
Since the remaining integral is finite, it follows that the inequality~\eqref{Tcontinuous}
holds for a suitable constant~$c>0$, $p=n+1$ and~$q=0$.
Therefore, $T_g \in {\mathcal{S}}'(\R^n)$ is a tempered distribution.

In this way, every function~$g \in L^\infty(\R^n)$ gives rise to a corresponding tempered distribution~$T_g$. Let us verify that the corresponding linear mapping
\beq \label{Tembed}
T \::\: L^\infty(\R^n) \rightarrow {\mathcal{S}}'(\R^n) \:,\quad g \mapsto T_g \qquad \text{is injective} \:.
\eeq
(Note that, strictly speaking, we are dealing with equivalence classes, identifying two functions that differ only on a set of measure zero.)
To this end, let~$g \in L^\infty(\R^n)$
be non-zero. Then~$A := \{x \in \R^n \mid |g(x)| \geq \|g\|_\infty/2\}$ has nonzero measure. By inner regularity of the 
Lebesgue measure, the same is true for~$A \cap B_R(0)$ for~$R > 0$ sufficiently large. Choose~$\eta \in 
C^\infty_0(\R^n)$ with~$\eta|_{B_R(0)} \equiv 1$. Then, $\eta f$ is bounded and compactly supported, hence square 
integrable, and one easily checks that~$\|\eta g\|_{L^2} \neq 0$. Using that~$C^\infty_0(\R^n)$ is dense 
in~$L^2(\R^n)$, we conclude that there is a function~$f \in C^\infty_0(\R^n)$
with
\beq 0 \neq \la f,\eta g\ra_{L^2(\R^n)} = \int_{\R^n} f(x)\: \big(\eta g)(x) \Diff^n x = T_g(\eta f) \:. \eeq
Hence~$T_g \neq 0$, and we conclude that the mapping~$T$ in~\eqref{Tembed}
is indeed injective.

The fact that the mapping~\eqref{Tembed} is an embedding means that distributions
can be regarded as ``generalized functions.'' Distributions that can be represented in
the form~$T_g$ with~$g \in L^\infty(\R^n)$ are referred to as {\em{regular distributions}}.
\sindex{distribution!regular}%
We finally remark that~$T_g$ can be defined more generally for functions~$g$ that
increase at most polynomially at infinity. But we do not need this generalization here.

In order to speak about convergence of distributions and related things, one needs to endow the 
space~$\mathcal{S}'(\R^n)$ 
with a topology. Being defined as a dual space of functionals on~$\mathcal{S}(\R^n)$, the usual 
choice is the so-called \emph{weak*-topology},
\sindex{topology!weak*}%
which is the coarsest topology such that for every~$f \in \mathcal{S}(\R^n)$ the evaluation map~$\mathcal{S}'(\R^n) \ni T \mapsto T(f) \in \C$ is continuous. This means that for a sequence~$(T_n)_{n \in \N}$
in~$\mathcal{S}'(\R^n)$ and~$T \in \mathcal{S}'(\R^n)$, one has
\beq
    T_n \to T \text{ in } \mathcal{S}'(\R^n)
    \quad\Longleftrightarrow\quad
    T_n(f) \to T(f) \text{ for all } f \in \mathcal{S}(\R^n) \,.
\eeq
With respect to this topology, the map~\eqref{Tembed} is continuous.

In the applications, it is important to differentiate distributions.
For example, one wants to construct distributional solutions of partial differential equations
(like the Poisson equation with a $\delta$ distribution or the Green's kernels of the Dirac equation
to be considered in Chapters~\ref{seclinhyp} and~\ref{secgreen}).
It turns out that distributions can always be differentiated, as we now explain. The idea behind the definition of the derivative of a distribution is to generalize the integration-by-parts formula, which for two Schwartz functions~$f,g \in \mathcal{S}(\R^n)$ states that
\begin{align}
T_{\partial^\alpha f}(g) &= \int_{\R^n} (\partial^\alpha f)(x) \:g(x)\: \Diff^nx \notag \\
&= (-1)^{|\alpha|} \int_{\R^n} f(x) \:(\partial^\alpha g(x)) \:\Diff^nx
= (-1)^{|\alpha|} \:T_f(\partial^\alpha g)
\end{align}
(note that there are no boundary terms due to the rapid decay of~$f$ and~$g$ at infinity).

\begin{Def} For a tempered distribution~$T \in \mathcal{S}'(\R^n)$ and multi-index~$\alpha \in \N_0^n$, we define
the {\bf{distributional derivative}}~$\partial^\alpha T \in \mathcal{S}'(\R^n)$ by
\sindex{distribution!derivative of}%
\sindex{distributional derivative}%
\beq \big( \partial^\alpha T \big)(f) := (-1)^{|\alpha|}\: T \big( \partial^\alpha f \big) \qquad \text{for all~$f \in {\mathcal{S}}(\R^n)$} \:. \eeq
\end{Def} \noindent
Using the continuity estimate~\eqref{Tcontinuous} for~$T$, we have
\beq \big| \big( \partial^\alpha T \big)(f) \big| =
\big| T \big( \partial^\alpha f \big) \big| \leq c\: \big\| \partial^\alpha f \big\|_{p,q}
\leq c\: \big\| f \big\|_{p,q+|\alpha|} \quad \text{ for all } f \in \mathcal{S}(\R^n)\,, \eeq
which shows that~$\partial^\alpha T$ is indeed a {\em{continuous}} linear functional again.

We now come to the {\em{Fourier transformation}}. We first introduce it for Schwartz functions.
\begin{Def} \label{deffourier} For~$f \in {\mathcal{S}}(\R^n)$, we define the Fourier transform~$({\mathcal{F}} f): \R^n \to \C$
and the adjoint Fourier transform~$({\mathcal{F}}^* f): \R^n \to \C$ by
\sindex{Fourier transform|textbf}%
\nindex{abr@${\mathcal{F}}, {\mathcal{F}}^*$ -- Fourier transform}%
\begin{align}
({\mathcal{F}} f)(p) &= \int_{\R^n} f(x)\: \E^{\cI px}\: \Diff^nx \label{fourier} \\
({\mathcal{F}}^* f)(x) &= \int_{\R^n} f(p)\: \E^{-\cI px}\: \frac{\Diff^np}{(2 \pi)^n} \:, \label{fourierback}
\end{align}
where~$x, p \in \scrM$, and~$px = \la p, x \ra$ denotes the Minkowski inner product.
\end{Def}

\begin{Lemma} The Fourier transform and its adjoint map Schwartz functions to Schwartz functions and yield continuous linear maps
\beq \label{Fmapsto}
{\mathcal{F}}, {\mathcal{F}}^* \::\: {\mathcal{S}}(\R^n) \rightarrow {\mathcal{S}}(\R^n) \:.
\eeq
\end{Lemma}
\Proof In order to prove~\eqref{Fmapsto}, we differentiate~\eqref{fourier} to obtain
\begin{align}
p^\alpha\, \partial_p^\beta ({\mathcal{F}} f)(p) 
&= p^\alpha \int_{\R^n} f(x)\: x^\beta \:\E^{\cI px}\: \Diff^nx
= (-\cI)^{|\alpha|} \int_{\R^n} f(x)\: x^\beta \:\big( \partial_x^\alpha \E^{\cI px} \big)\: \Diff^nx \\
&= \cI^{|\alpha|} \int_{\R^n} \big( \partial_x^\alpha f(x) \big) \: x^\beta \: \E^{\cI px}\: \Diff^nx \:,
\end{align}
where in the last step, we integrated by parts (to justify differentiation under the integral one can use the dominated convergence theorem). Taking the absolute value, we obtain the estimate
\beq \big| p^\alpha\, \partial_p^\beta ({\mathcal{F}} f)(p) \big| 
\leq \int_{\R^n} \big| \partial^\alpha f(x) \big| \: \big| x^\beta \big| \: \Diff^nx 
\overset{(\star)}{\leq} C(n) \cdot \| \partial^\alpha f \|_{|\beta| + n+1,0} \leq \|f\|_{|\beta| + n+1, |\alpha|} \:, \eeq
where in~$(\star)$ the integral can be estimated similar to that in~\eqref{intes}.
This estimate shows that the Fourier transform of a Schwartz function is again a Schwartz function. It also follows 
directly from this estimate that~$\mathcal{F}:\mathcal{S}(\R^n) \to \mathcal{S}(\R^n)$ is continuous (as linearity 
clearly holds).
The estimate for the inverse Fourier transform is similar.
\QED

\begin{Thm} {\bf{(Fourier inversion formula)}} \label{lemmafourierinverse}
The Fourier transform and its adjoint on Schwartz functions are inverses of each other,
\beq %\label{fourierinverse}
{\mathcal{F}} \circ {\mathcal{F}}^* = {\mathcal{F}}^* \circ {\mathcal{F}} = \1_{{\mathcal{S}}(\R^n)} \:. \eeq
\sindex{Fourier inversion formula}%
\end{Thm}
A detailed proof of this lemma can be found in~\cite[Theorem~2.2.4]{rauch} or~\cite[Theorem~8.2.2]{friedlander2}.
With this in mind, we only give a sketch of the proof in one dimension. Writing everything out explicitly, for~$f \in \mathcal{S}(\R^n)$ we get
\beq
    \mathcal{F}^* \big( {\mathcal{F}} f \big)(x)
    = \int_{-\infty}^\infty \left( \int_{-\infty}^\infty f(y) \E^{\cI p y} \Diff y\right)
    \E^{-\cI px} \:\frac{\Diff p}{2\pi} \,.
\eeq
The basic idea is to exchange the order of integration. However, the problem is that the resulting function is not 
integrable in the $p$-variable. One way to make this mathematically sound is by inserting a convergence-generating 
factor~$\E^{-\varepsilon p^2}$. More precisely, using dominated convergence and Fubini, one computes
\begin{align}
{\mathcal{F}}^* \big( {\mathcal{F}} f \big)(x) &= \lim_{\varepsilon \searrow 0}
\int_{-\infty}^\infty \big( {\mathcal{F}} f \big)(p)\: \E^{-\varepsilon p^2} \:\E^{-\cI px}\: \frac{\Diff p}{2 \pi} \notag \\
&= \lim_{\varepsilon \searrow 0} \int_{-\infty}^\infty \bigg( \int_{-\infty}^\infty f(y)\: \E^{\cI py}\: \Diff y \bigg) \:
\E^{-\varepsilon p^2} \:\E^{-\cI px}\: \frac{\Diff p}{2 \pi} \notag \\
&= \lim_{\varepsilon \searrow 0} \int_{-\infty}^\infty f(y) \bigg\{ \int_{-\infty}^\infty \E^{\cI p (y-x) -\varepsilon p^2}\: \frac{\Diff p}{2 \pi}
\bigg\} \: \Diff y \:. \label{generate-convergence}
\end{align}
The integral inside the curly brackets is Gaussian and can be computed explicitly.
The resulting family of Gaussians tends to the $\delta$ distribution~$\delta(x-y)$ (see Exercise~\ref{ex:deltasequence}), and thus altogether one obtains~$f(x)$ in the limit~$\varepsilon \searrow 0$ as desired.

Having given the proof of the Fourier inversion formula, we return once more to the formal computation from the beginning,
\begin{align}
\mathcal{F}^* \big( {\mathcal{F}} f \big)(x)
&= \int_{-\infty}^\infty \left( \int_{-\infty}^\infty f(y) \E^{\cI p y} \Diff y\right) \E^{-\cI px} \frac{\Diff p}{2\pi} \notag \\
&= \int_{-\infty}^\infty f(y) \left( \int_{-\infty}^\infty \E^{\cI p (y-x)} \frac{\Diff p}{2 \pi} \right) \Diff x\:.
\end{align}
Although the right-hand side is ill-defined as an integral, knowing that~$\mathcal{F}^* \big( {\mathcal{F}} f \big)(x) = f(x)$ holds, one may interpret it as the distributional identity
\beq \label{fourier-delta}
\int_{-\infty}^\infty \E^{\cI p(x-y)}\: \Diff p = 2\pi\: \delta(x-y) \:.
\eeq

Having the Fourier transform for Schwartz functions at our disposal, we can now introduce the
Fourier transform of tempered distributions. Similar to the definition of the
distributional derivative, the idea is to let the Fourier transform act on the test function. To see how to do this concretely, we again first consider a regular distribution~$T_{{\mathcal{F}} g}$
corresponding to the Fourier transform of a Schwartz function~$g$. Then, for any~$f \in \mathcal{S}(\R^n)$, using Fubini, we have 
\begin{align}
T_{{\mathcal{F}} g} (f) &= \int_{\R^n} ({\mathcal{F}} g)(p)\: f(p)\: \Diff^np =
\int_{\R^n} \bigg( \int_{\R^n} g(x)\: \E^{\cI px}\: \Diff^nx \bigg) \:f(p)\: \Diff^np \notag \\
&= \int_{\R^n} g(x) \bigg( \int_{\R^n} \: \E^{\cI px}\: f(p) \:\Diff^np \bigg) \: \Diff^nx 
= T_g\big({\mathcal{F}} f \big) \:.
\end{align}
The right side can now be used to {\em{define}} the Fourier transform of a tempered distribution.
\begin{Def} The Fourier transform and the adjoint Fourier transform of a tempered
distribution~$T \in {\mathcal{S}}'(\R^n)$ are defined by
\beq {\mathcal{F}}, {\mathcal{F}}^* \::\: {\mathcal{S}}'(\R^n) \rightarrow {\mathcal{S}}'(\R^n) \:,\qquad
\big( {\mathcal{F}} T \big)(f) = T \big( {\mathcal{F}} f)
\:,\quad \big( {\mathcal{F}}^* T \big)(f) = T \big( {\mathcal{F}}^* f) \:. \eeq
\end{Def}
Note that for a tempered distribution~$T$, the maps~$\mathcal{F}T = T \circ \mathcal{F}$ and~$\mathcal{F}^*T = T 
\circ \mathcal{F}^*$ are again linear and continuous as the composition of two linear and continuous maps. Hence,
they are tempered distributions again. A direct computation shows that~${\mathcal{F}}^*$ is the inverse 
of~${\mathcal{F}}$
(see 
Exercise~\ref{ex:fourierinverse}).
Examples for how to compute Fourier transforms of distributions can be found in the exercises
(see Exercise~\ref{ex:fourierdelta}).
\Evtl{Mehr Beispiele für Distributionen in \"Ubungsaufgaben?}%

We now come to an operation on functions and distributions which we will use a few
times in this book: the {\em{convolution}}.
\sindex{convolution}
\nindex{abr1@$f * g$ -- convolution of Schwartz functions}%
For two Schwartz functions~$f, g \in \mathcal{S}(\R^n)$, the convolution~$f * g \in \mathcal{S}(\R^n)$ is defined
by (see Exercise~\ref{exconv1})
\begin{equation}
    \label{eq:convolution}
    (f*g)(x)
    := \int_{\mathbb{R}^n} f(x-y)\: g(y) \dd y \:.
\end{equation}
One immediately notes that convolution is commutative; that is, $f*g = g*f$ (it is also associative).
Interestingly, taking the Fourier transform, the convolution goes over to multiplication; that is,
\beq \label{convF}
{\mathcal{F}}\big( f * g \big)(p) = ({\mathcal{F}} f)(p)\: ({\mathcal{F}} g)(p)
\eeq
(for the derivation see Exercise~\ref{exconvF}). Conversely, the Fourier transform of a product
is the convolution of the individual Fourier transforms.
The convolution can be extended to an operation involving tempered distributions by interpreting the right hand side of~\eqref{eq:convolution} as the action of a regular distribution~$T_g$.
Namely, given~$T \in \mathcal{S}'(\R^n)$ and~$f \in \mathcal{S}(\R^n)$, one defines
\beq \label{fTconv}
\big(f * T \big)(x) := T(f_x) \qquad \text{with} \qquad f_x(y) := f(y-x) \:.
\eeq
\nindex{abr2@$f * T$ -- convolution of Schwartz function with tempered distribution}%
This is even a smooth function (see Exercise~\ref{exconv2}). With this in mind, a
convolution can be used to ``smoothen'' or ``mollify'' functions and distributions.
More details and examples can be found for example in~\cite[Chapter~5]{friedlander2}.

We close with two remarks. First, it is often very useful to consider the Fourier transform
on~$L^2$-functions, where the Fourier transform is unitary:
\begin{Thm} {\bf{(Plancherel)}} \label{thmplancherel} For any~$f \in \mathcal{S}(\R^n)$,
\sindex{Plancherel's theorem}%
\beq \| {\mathcal{F}} f \|_{L^2(\R^n)} = (2 \pi)^{\frac{n}{2}} \:\| f \|_{L^2(\R^n)} \:. \eeq
Furthermore, the Fourier transform and the adjoint Fourier transform extend to isomorphisms~$\mathcal{F},\mathcal{F}^*: L^2(\R^n) \to L^2(\R^n)$, which are inverse to each other.
\end{Thm}
We again omit the proof, which can be found in~\cite[Theorem~2.3.4]{rauch} or~\cite[Theorem~9.2.2]{friedlander2}.
On a formal level, Plancherel's formula is obtained by a direct computation using again~\eqref{fourier-delta}.
Similar to~\eqref{generate-convergence}, this computation can be made mathematically sound by
introducing a convergence-generating factor. The extension to~$L^2(\mathbb{R}^n)$ then simply follows using that~$\mathcal{S}(\R^n) \subset L^2(\mathbb{R}^n)$ is dense.

Finally, we remark that in some applications (for example when working in local charts on a manifold or in a bounded domain),
it is not feasible to work with functions defined in all of~$\R^n$ having suitable decay properties.
In this case, instead of the Schwartz functions, one considers the space of {\em{test functions}}; that is,
smooth functions with compact support, denoted by
\nindex{abs@${\mathcal{D}}(\R^n)$ -- space of test functions}%
\tindex{ff@${\mathcal{D}}(\R^n)$ -- space of test functions}%
\beq {\mathcal{D}}(\R^n) = C^\infty_0(\R^n) \:, \eeq
endowed with the topology induced by the family of norms~$\| . \|_{q}$
with~$q \in \N_0$ given by
\beq \|f\|_{q} := \max_{\substack{\beta \in \N_0^n\\ |\beta| \leq q}}  \;\; \sup_{x \in \R^n} \big| \partial^\beta f(x) \big\| \:. \eeq
Its dual space~${\mathcal{D}}'(\R^n)$ is referred to as the space of {\em{distributions}}.
\sindex{distribution}%
\nindex{abt@${\mathcal{D}}'(\R^n)$ -- space of distributions}%
\tindex{ff@${\mathcal{D}}'(\R^n)$ -- space of distributions}%
Here the dual space is again defined as the space of all continuous linear functionals,
where continuity of a linear functional~$T: \mathcal{D}(\R^n) \to \C$ means in analogy to~\eqref{Tcontinuous} that
there is~$q \in \N_0$ and a constant~$c>0$ such that
\beq %\label{Tcont2}
\big| T(f) \big| \leq c\: \|f\|_{q} \qquad \text{for all~$f \in {\mathcal{D}}(\R^n)$} \:. \eeq
Differentiation of distributions can be defined similarly to tempered distributions. However, the Fourier transform 
cannot be defined on all of~$\mathcal{D}'(\R^n)$ since the Fourier transform of a test function need not be 
compactly supported again. Test functions and distributions can also be defined on any open subset~$U \subset 
\R^n$ instead of all of~$\R^n$ (and with a little bit of extra work also on manifolds).

\section{Manifolds and Vector Bundles} \label{secbundle}
We now recall the basic definitions of a manifold and a vector bundle. Since the machinery of differential geometry is 
loaded with many definitions and quite subtle (notational) conventions, we must restrict to the very basics. For more 
details we refer to good textbooks like~\cite{langDG, lee-smooth} or to the basic definitions in~\cite[\S1 and 
\S2]{milnor+stasheff}.
\begin{Def} 
A {\bf{topological manifold}} of dimension~$n \in \N$ is a Hausdorff topological space~$\scrM$
which is $\sigma$-compact (i.e.\ which can be written as an at most countable union of compact subsets)
and has the property that every point in~$\scrM$ has an open neighborhood which is homeomorphic
\sindex{manifold!topological}%
to an open subset of~$\R^n$.
\end{Def} \noindent
More specifically, for every~$p \in \scrM$, there is an open neighborhood~$U \subset \scrM$
of~$p$ and a map
\beq \phi \::\: U \rightarrow \R^n \:, \eeq
such that the image~$\phi(U)$ is an open subset of~$\R^n$ and the mapping~$\phi : U \rightarrow \phi(U)$ is
a homeomorphism (that is, it is continuous, invertible, and its inverse is also continuous).
\sindex{homeomorphism}%
We refer to~$(\phi, U)$ as a {\em{chart}} around~$p$ (see Figure~\ref{figchart}).
\sindex{chart!of a topological manifold}%
\begin{figure}[tb]
{
\begin{pspicture}(0,27.843513)(8.882229,31.04277)
\definecolor{colour1}{rgb}{0.9019608,0.9019608,0.9019608}
\definecolor{colour0}{rgb}{0.7019608,0.7019608,0.7019608}
\pspolygon[linecolor=colour1, linewidth=0.01, fillstyle=solid,fillcolor=colour1](0.042229004,31.004627)(0.212229,30.924627)(0.447229,30.849628)(0.737229,30.824627)(1.277229,30.829628)(1.657229,30.869627)(2.062229,30.929628)(2.442229,30.979628)(2.787229,30.994627)(3.017229,30.964628)(3.117229,30.914627)(3.052229,30.639627)(2.972229,30.264627)(2.912229,29.889627)(2.852229,29.439629)(2.827229,29.099628)(2.822229,28.714628)(2.842229,28.479628)(2.882229,28.249628)(2.967229,27.989628)(2.717229,27.919628)(2.427229,27.874628)(2.102229,27.879627)(1.727229,27.904627)(1.372229,27.964628)(1.002229,28.024628)(0.602229,28.089628)(0.337229,28.104628)(0.15222901,28.109627)(0.052229002,28.099628)(0.172229,28.449627)(0.237229,28.829628)(0.267229,29.314629)(0.257229,29.644629)(0.222229,30.059628)(0.13722901,30.559628)
\psbezier[linecolor=black, linewidth=0.01, fillstyle=solid,fillcolor=colour0](0.922229,30.289627)(0.72385263,30.08004)(0.7996588,29.25552)(1.302229,29.089627685546876)(1.8047992,28.923737)(2.3709583,29.3787)(2.202229,29.829628)(2.0334997,30.280554)(1.1206053,30.499216)(0.922229,30.289627)
\pscircle[linecolor=black, linewidth=0.02, fillstyle=solid,fillcolor=black, dimen=outer](1.482229,29.724628){0.045}
\psbezier[linecolor=black, linewidth=0.03](0.017229004,31.019629)(1.017229,30.469627)(2.547229,31.219627)(3.117229,30.919627685546875)
\psbezier[linecolor=black, linewidth=0.03](3.117229,30.929628)(2.887229,29.919628)(2.667229,28.649628)(2.977229,27.979627685546873)
\psbezier[linecolor=black, linewidth=0.03](2.987229,27.989628)(2.017229,27.619627)(0.707229,28.229628)(0.037229005,28.079627685546875)
\psbezier[linecolor=black, linewidth=0.03](0.052229002,28.084627)(0.402229,29.024628)(0.257229,30.129627)(0.027229004,31.019627685546876)
\psframe[linecolor=black, linewidth=0.03, fillstyle=solid,fillcolor=colour1, dimen=outer](8.882229,30.959627)(5.862229,27.969627)
\psbezier[linecolor=black, linewidth=0.01, fillstyle=solid,fillcolor=colour0](6.742229,29.969627)(6.5538526,29.72004)(6.4596586,29.045519)(7.192229,28.769627685546876)(7.9247994,28.493736)(8.7109585,29.768702)(8.312229,30.149628)(7.9135,30.530554)(6.9306054,30.219215)(6.742229,29.969627)
\psbezier[linecolor=black, linewidth=0.04, arrowsize=0.05291667cm 2.0,arrowlength=1.4,arrowinset=0.0]{->}(3.472229,29.879627)(4.062229,30.199627)(4.992229,30.139627)(5.422229,29.819627685546877)
\psbezier[linecolor=black, linewidth=0.04, arrowsize=0.05291667cm 2.0,arrowlength=1.4,arrowinset=0.0]{->}(5.422229,29.229628)(5.062229,28.889627)(4.152229,28.809628)(3.472229,29.339627685546876)
\rput[bl](1.15,29.7){$x$}
\rput[bl](1.8,29.5){$U$}
\rput[bl](7,29.3){$\phi(U)$}
\rput[bl](3.3,30.3){$\phi : U \rightarrow \phi(U)$}
\rput[bl](3.1,28.4){$\phi^{-1} : \phi(U) \rightarrow U$}
\rput[bl](2.5,30.55){$\scrM$}
\rput[bl](8.3,30.5){$\R^n$}
\end{pspicture}
}
\caption{A chart~$(\phi, U)$ around a point~$x \in \scrM$.}
\label{figchart}
\end{figure}%
A collection of charts is an {\em{atlas}}~$\scrA$.
\sindex{atlas!of a topological manifold}%
\sindex{atlas!complete}
We always assume that the atlas is {\em{complete}} in the sense that every point of~$\scrM$ lies in the
domain of a chart of the atlas.

 A chart~$(\phi,U)$ can be seen as an identification of the open subset~$U \subset \scrM$ with the open
 subset~$\phi(U) \subset \R^n$, any~$p \in U$ being identified with the coordinates of its image
\begin{equation}
     \label{eq:local-coordinates}
     \phi(p) = (x^1,\ldots,x^n) \in \R^n \,.
\end{equation}
In this way a chart introduces \emph{local coordinates}~$x^1,\ldots,x^n$ on~$U \subset M$. 
\sindex{coordinates!local}%
Note that by~\eqref{eq:local-coordinates} the coordinates can be understood as the component functions of the map 
$\phi$ and thus as functions on~$U$. Also maps between manifolds can locally be identified with maps between 
open subset of Euclidean spaces. Concretely, if~$F: \scrM \to \scrN$ is a map between two manifolds~$\scrM$ and 
$\scrN$, 
and if~$(\phi,U)$ and~$(\psi,V)$ are charts on~$\scrM$ and~$\scrN$, respectively, with~$F(U) \subset V$, then 
we can consider the map
\begin{equation}
    \label{eq:coordinate-representation}
     \psi \circ F \circ \phi^{-1}: \phi(U) \to \psi(V) \,,
 \end{equation}
 called \emph{coordinate representation} of~$F$.

One would like to use these local identifications to carry over the machinery of differential calculus to maps between manifolds. To this end, the structure of a topological manifold alone is not sufficient. Rather, one needs
additional compatibility conditions on the transition maps between charts:
Given two charts~$(\phi, U)$ and~$(\tilde{\phi}, \tilde{U})$ on a manifold~$\scrM$ with~$U \cap \tilde{U} \neq \varnothing$, the two mappings
\beq \phi|_{U \cap \tilde{U}},\: \tilde{\phi}|_{U \cap U'} \::\: U \cap \tilde{U} \rightarrow \R^n \eeq
are both homeomorphism onto open subsets of~$\R^n$. We define the
{\em{transition map}} by (see Figure~\ref{figtransition})
\beq
    \tilde{\phi}|_{U \cap \tilde{U}}\ \circ \big( \phi|_{U \cap \tilde{U}})^{-1} \::\: \phi \big( U \cap \tilde{U} \big) \rightarrow
    \tilde{\phi} \big( U \cap \tilde{U} \big) \:.
\eeq
Being a mapping between two open subsets of~$\R^n$, it is clear what ``differentiability''
of this mapping means. A {\em{differentiable manifold}} is a topological manifold together with a
\sindex{manifold!differentiable}%
complete atlas with the property that all transition maps are differentiable.
Likewise, a {\em{smooth manifold}} is defined by the requirement that all transition maps are smooth.
\sindex{manifold!smooth}%

The differentiability of all transition maps now allows us to introduce a notion of differentiable maps: A
map~$F : \scrM \rightarrow \scrN$ between two differentiable manifolds~$\scrM$ and~$\scrN$ is called \emph{differentiable map}
\sindex{map!differentiable}% 
\sindex{differentiability!on manifold}%
if all of its coordinate representations~\eqref{eq:coordinate-representation} are differentiable maps between open 
subsets of Euclidean spaces. It turns out that this is actually a local condition, and in order to prove differentiability of 
$F$ in a point~$p \in \scrM$ it suffices to verify that \emph{one} coordinate representation around~$p$ is 
differentiable.

Having introduced the notion of a differentiable map, the next question is how to compute its derivative.
We begin with the case of a real-valued differentiable function~$f: \scrM \to \R$. If~$(\phi,U)$ is a chart of~$\scrM$, 
then by definition the coordinate representation~$f \circ \phi^{-1}: \phi(U) \subset \R^n \to \R$ is differentiable. (Note 
that we take the identity as chart of~$\R$.) We can therefore take partial derivatives~$\frac{\partial (f \circ 
\phi^{-1})}{\partial 
x^i}(x)$ or also the total derivative~$D_x(f \circ \phi^{-1})$ in a point~$x \in \phi(U)$. These, of course, depend 
on the chart~$\phi$, and if~$(\tilde{\phi},U)$ is another chart with the same domain, then by the chain rule we have
 \beq
     D_x(f \circ \phi^{-1})
     = D_x(f \circ \tilde{\phi}^{-1} \circ \tilde{\phi} \circ \phi^{-1})
     = D_{\tilde{\phi}(\phi^{-1}(x))}(f \circ \tilde{\phi}^{-1})
     \cdot D_x(\tilde{\phi} \circ \phi^{-1}) \,,
 \eeq
 or, written in terms of partial derivatives (note that we are using the Einstein summation convention),
 \beq
     \frac{\partial(f \circ \phi^{-1})}{\partial x^i}
     = \frac{\partial(f \circ \tilde{\phi}^{-1} \circ \tilde{\phi} \circ \phi^{-1})}{\partial x^i}
     = \frac{\partial(f \circ \tilde{\phi}^{-1})}{\partial x^j} \cdot
     \frac{\partial (\tilde{\phi} \circ \phi^{-1})^j}{\partial x^i} \,.
 \eeq
 This shows that the partial derivatives with respect to the two charts are linearly related via the Jacobian matrix of the transition map between the two charts.

 It is desirable to introduce a notion of derivative that is independent of the choice of local charts. There are 
different approaches one can take, and here we only sketch the approach that geometrically seems the most 
intuitive. To this end, let again~$f: \scrM \to \R$ be a differentiable function and let~$p \in \scrM$. If~$c: 
(-\varepsilon,\varepsilon) 
\to \scrM$ is a differentiable curve with~$c(0) = p$, then it is easy to check that the composition~$f \circ c: 
(-\varepsilon,\varepsilon) \to \R$ is differentiable, so one can take the derivative~$(f \circ c)'(0)$. For~$\scrM = \R^n$, 
this would yield precisely the directional derivative of~$f$ in~$p$ in direction~$c'(0)$, but in the general case such an 
interpretation is not possible yet (as we have not defined~$c'(0)$ for a curve in a manifold). However, if~$(\phi,U)$ is 
a local chart around~$p$, then we can write
 \begin{equation}
     \label{eq:derivative-in-local-chart}
     (f \circ c)'(0)
     = (f \circ \phi^{-1} \circ \phi \circ c)'(0)
     = D_{\phi(p)}(f \circ \phi^{-1})((\phi \circ c)'(0)) \,.
 \end{equation}
 From this, it follows that for any other differentiable curve~$\tilde{c}: (-\tilde{\varepsilon},\tilde{\varepsilon}) \to \scrM$ 
with~$\tilde{c}(0) = p$ we will have~$(f \circ c)'(0) = (f \circ \tilde{c})(0)$ if we have~$(\phi \circ c)'(0) = (\phi \circ 
\tilde{c})'(0)$. It is easy to check that the latter condition on the two curves~$c$ and~$\tilde{c}$ is actually independent 
of the chart in the sense that if it holds in one chart, then it also holds in any other chart.
Moreover, it is clear that it defines an equivalence relation on the set of differentiable curves passing through~$p$. The corresponding equivalence classes are called \emph{tangent vectors} 
\sindex{tangent vector}%
to~$\scrM$ in~$p$, and the set of all equivalence classes is called the \emph{tangent space} to~$\scrM$ in~$p$ and is denoted by~$T_p\scrM$. 
\sindex{tangent space|textbf}%
\nindex{abu@$T_p\scrM$ -- tangent space of~$\scrM$ at~$p$}%
Any chart~$(\phi,U)$ around~$p$ defines a map
 \beq
     \phi_{*,p}: T_p\scrM \to \R^n
     \,, \quad \phi_{*,p}([c]) := (\phi \circ c)'(0) \,.
 \eeq
 It is not difficult to verify that this map is a bijection, so it can be used to pull back the vector space structure of~$\R^n$ to~$T_p\scrM$ by
 \beq
     [c_1]+[c_2]
     := \phi_{*,p}^{-1} \big( (\phi \circ c_1)'(0) + (\phi \circ c_2)'(0) \big) \,,
 \eeq
 and similarly for scalar multiplication. The induced vector space structure on~$T_p\scrM$ is actually independent of the chart~$\phi$, as follows from the observation that for any other chart~$(\tilde{\phi},U)$ around~$p$, we have
 \Evtl{Exercise}%
\begin{equation}
     \label{eq:transition-map-on-tangent-bundle}
     \tilde{\phi}_{*,p} \circ \phi_{*,p}^{-1}
     = D_{\phi(p)}(\tilde{\phi} \circ \phi^{-1}) \in \mathrm{GL}(\R^n) \:.
 \end{equation}
Having defined the tangent space, we can finally define the \emph{derivative in a point~$p \in \scrM$} of a
\sindex{derivative!on manifolds}%
differentiable map~$f: \scrM \to \R$ as the map
 \begin{equation}
     \label{eq:derivative-of-differentiable-map}
     D_pf: T_p\scrM \to \R \,, \quad D_pf([c]) := (f \circ c)'(0) \,.
 \end{equation}
One easily verifies that this is indeed a linear map.

We note that, for any local chart~$(\phi,U)$, by~\eqref{eq:derivative-in-local-chart} the derivative can be expressed in 
terms of the Jacobian of the coordinate representation~$f \circ \phi^{-1}$. This can be understood a bit more 
systematically by defining the particular tangent vectors
 \begin{equation}
     \label{eq:coordinate-vector-fields}
     \frac{\partial}{\partial x^i}\Big|_p
     := \phi_{*,p}^{-1}(e_i)
     = [t \mapsto \phi^{-1}(\phi(p)+te_i)] \in T_p\scrM \,.
 \end{equation}
They form a basis of~$T_p\scrM$. (As will be considered in more detail later,
they do depend on~$\phi$, even though the notation does not seem to reflect this directly.) Now observe that
 \begin{align}
   D_pf\Big(\frac{\partial}{\partial x^i}\Big|_p\Big)
   &= \frac{\dd}{\dd t}\Big|_{t=0} (f \circ \phi^{-1})(\phi(p)+te_i) \\
   &= D_{\phi(p)}(f \circ \phi^{-1})(e_i)
   = \frac{\partial (f \circ \phi^{-1})}{\partial x^i}(\phi(p)) \,.
 \end{align}
 This shows that the Jacobian of~$f \circ \phi^{-1}$ is precisely the matrix representation of~$D_pf$ in the basis 
$\frac{\partial}{\partial 
x^1}|_p, \ldots, \frac{\partial}{\partial x^n}|_p$. We now come back
to the point that the basis vectors~\eqref{eq:coordinate-vector-fields} depend on the chart~$\phi$, and to how one 
usually resembles this in the notation. To this end, recall that one says that the chart~$\phi$ introduces local 
coordinates~$x^1,\ldots,x^n$ on its domain. A different chart~$\tilde{\phi}$ on the same domain introduces different 
coordinates. One usually denotes these by~$\tilde{x}^1,\ldots,\tilde{x}^n$, and denotes the basis of~$T_p\scrM$ 
induced by~$\tilde{\phi}$ by
 \beq
     \frac{\partial}{\partial \tilde{x}^1}\Big|_p, \ldots,
     \frac{\partial}{\partial \tilde{x}^n}\Big|_p \,.
 \eeq
Thus the reference to the chart in the notation is done via the symbols for the coordinates.  The relation between the two different bases follows from the following computation, which makes use of~\eqref{eq:transition-map-on-tangent-bundle} (note Einstein's summation convention again):
 \begin{align}
   \frac{\partial}{\partial x^i}\Big|_p
   &= \phi_{*,p}^{-1}(e_i)
   = \big(\tilde{\phi}_{*,p}^{-1} \circ \tilde{\phi}_{*,p}^{-1}\big)
     \big( \phi_{*,p}^{-1}(e_i) \big)
   = \tilde{\phi}_{*,p}^{-1}
     \big( (\tilde{\phi}_{*,p}^{-1} \circ \phi_{*,p}^{-1})(e_i) \big) \notag \\
   &\hspace{-0.33cm}\stackrel{~\eqref{eq:transition-map-on-tangent-bundle}}{=} \tilde{\phi}_{*,p}^{-1}\big( D_{\phi(p)}(\tilde{\phi} \circ \phi)(e_i)\big)
   = \tilde{\phi}_{*,p}^{-1} \Big(
     \frac{\partial (\tilde{\phi} \circ \phi)^j}{\partial x^i}(\phi(p)) \cdot e_j
     \Big) \notag \\
   &= \frac{\partial (\tilde{\phi} \circ \phi)^j}{\partial x^i}(\phi(p))
     \cdot \tilde{\phi}_{*,p}^{-1}(e_j)
   = \frac{\partial (\tilde{\phi}^j \circ \phi)}{\partial x^i}(\phi(p))
     \cdot \frac{\partial}{\partial \tilde{x}^j}\Big|_p \,.
 \end{align}
 Recalling that the coordinates induced by a chart can be understood as the component functions of the chart, and 
dropping all reference to the points where one evaluates, this relation can be recast in the shorter and easier to 
memorize the ``reduction rule''
\beq %\label{eq:relation-between-coordinate-vector-fields}
     \frac{\partial}{\partial x^i}
     = \frac{\partial \tilde{x}^j}{\partial x^i}
     \cdot \frac{\partial}{\partial \tilde{x}^j} \:, \eeq
 which might be familiar from the use of ``curvilinear coordinates'' in vector calculus (which are of course nothing else but different local coordinate systems in~$\R^n$).

So far, we only considered real-valued functions; that is, maps~$f: \scrM \to \R$. For a differentiable map~$F: \scrM \to 
\scrN$ between two differentiable manifolds~$\scrM$ and~$\scrN$,
the derivative of~$F$ in a point~$p \in \scrM$ is defined as the map
\beq %\label{eq:derivative-of-map-between-manifolds}
     D_pF: T_p\scrM \to T_{F(p)}\scrN \,, \quad
     D_pF([c]) := [F \circ c] \:. \eeq
 One can check that this is map well-defined and linear. Further, if~$(\phi,U)$ is a chart around~$p \in \scrM$ with 
coordinates~$x^1,\ldots,x^n$, and~$(\psi,V)$ a chart around~$F(p) \in \scrN$ with coordinates~$y^1,\ldots,y^m$, 
then one easily checks that
 \begin{equation}
     \label{eq:derivative-of-smooth-map-in-local-coordinates}
     D_pF\Big(\frac{\partial}{\partial x^i}\Big|_p\Big)
     = \frac{\partial (\psi \circ F \circ \phi^{-1})^j}{\partial x^i}(\phi(p))
     \cdot \frac{\partial}{\partial y^j}\Big|_{F(p)} \,.
 \end{equation}
 This shows that, with respect to the bases induced by the two charts~$\phi$ and~$\psi$, the linear map~$D_pF$ is 
represented by the Jacobian of the coordinate representation~$\psi \circ F \circ \phi^{-1}$. Note that, for a coordinate 
chart~$\phi: U \to \R^n$, we have~$D_p\phi = \phi_{*,p}$. For this reason, some authors denote the derivative of a 
smooth map~$F$ also by~$F_*$ instead of~$DF$.

 We remark that there are different but equivalent ways to define tangent vectors and thus also derivatives of smooth 
maps between smooth manifolds. One commonly used approach is via so-called \emph{derivations}, which is an 
axiomatization of the product rule for (directional) derivatives (for details, see for 
example~\cite[Chapter~3]{lee-smooth}).

The collection of all tangent spaces to a differentiable manifold~$\scrM$ as (disjoint) union~$T\scrM := \bigcup_{p \in 
\scrM} T_p\scrM$ is called the \emph{tangent bundle}
\sindex{tangent bundle}%
\nindex{abv@$T\scrM$ -- tangent bundle of~$\scrM$}%
of~$\scrM$. It turns out that the tangent bundle is again a smooth manifold with certain additional structures. First 
note that we can define a map~$\pi: T\scrM \to \scrM$ by mapping any tangent vector to its \emph{base point}, 
that is, to the point of~$\scrM$ to whose tangent space it belongs. Next, given a chart~$(\phi,U)$ on~$\scrM$,
we can consider the map
 \beq
     T\phi: \pi^{-1}(U) \to \phi(U) \times \R^n \subset \R^{2n} \,, \quad
     v \mapsto (\phi(\pi(v)), \phi_{*,\pi(v)}(v)) \,.
 \eeq
This map is bijective onto the open subset~$\phi(U) \times \R^n \subset \R^{2n}$. One verifies immediately that all 
sets of the form~$T\phi^{-1}(V)$, where~$\phi$ is a chart on~$\scrM$ and~$V \subset \R^{2n}$ is an open subset, form 
what is called a basis for a topology on~$T\scrM$, which is Hausdorff and~$\sigma$-compact. In particular, all these 
sets are open subsets of~$T\scrM$, so the maps~$T \phi$ are local charts. (This is indeed the unique topology on 
$T\scrM$ with these properties; see~\cite[Chapter~3]{lee-smooth} for details.)
Thus~$T\scrM$ is a topological manifold. Furthermore, if~$(\phi,U)$ and~$(\psi,U)$ are two charts on~$\scrM$, then 
by~\eqref{eq:transition-map-on-tangent-bundle} the transition map between the two charts~$T\phi$ and~$T\psi$ on 
$T\scrM$ is given by
 \beq
     T\psi \circ T\phi^{-1}(x,v)
     = \big( x, D_x(\psi \circ \phi^{-1})(v)\big) \,.
 \eeq
 If~$\scrM$ is a smooth manifold, then this is again a smooth map, showing that~$T\scrM$ is also a smooth manifold.

Locally, the tangent bundle has a particular product structure, which is already visible from the construction of the 
charts above. More precisely, note that any chart~$(\phi,U)$ on~$\scrM$ induces a local diffeomorphism
 \beq
     \widehat{\phi}: \pi^{-1}(U) \to U \times \R^n \,, \quad
     v \mapsto (\pi(v),\phi_{*,\pi(v)}(v)) \,,
 \eeq
called a \emph{local trivialization} 
\sindex{trivialization!of the tangent bundle}%
of the tangent bundle. These local trivializations are compatible with the base point projection~$\pi: T\scrM \to \scrM$ 
in the sense that~$\mathrm{pr}_1 \circ \widehat{\phi} = \pi$, where~$\mathrm{pr}_1: U \times \R^n \to U$ denotes the 
projection to the first component. Moreover, they are also compatible with the vector space structures on the tangent 
spaces in the sense that for another chart~$(\psi,V)$ on~$\scrM$ with~$U \cap V \neq \varnothing$, we have
 \beq
     \widehat{\psi} \circ \widehat{\phi}^{-1}(p,v)
     = (p, D_{\phi(p)}(\psi \circ \phi^{-1})(v) ) \,,
 \eeq
 and in the second component we have a smooth map
 \beq
     U \cap V \ni p \mapsto D_{\phi(p)}(\psi \circ \phi^{-1}) \in \mathrm{GL}(\R^n) \,.
 \eeq
 So the transition between two local trivialization simply amounts to a (pointwise fixed) linear transformation on~$\R^n$.

The local structure of the tangent bundle just described can be generalized to the notion of a \emph{vector bundle}.
This notion is helpful for the understanding of causal fermion systems because, under suitable regularity
assumptions, a causal fermion system will give rise to a vector bundle over spacetime with the spin spaces as fibers
(see Section~\ref{sectopvector}).
\Evtl{F\"uge gutes Beispiel ein, z.B.\ komplexes Linienb\"undel mit $\U(1)$-Gruppenwirkung
auf Minkowski-Raum. M\"obiusband? Oder auf Sph\"are? Oder mit \"Ubungsaufgabe?}%
We begin with the notion of a topological vector bundle and explain the differentiable structure afterward.
 \begin{Def} \label{deftvb}
Let~$\B$ and~$\scrM$ be topological spaces and~$\pi : \B \rightarrow \scrM$ a continuous surjective map.
\sindex{vector bundle!topological|textbf}%
\sindex{structure group}%
\sindex{trivialization!local}%
\sindex{chart!bundle}%
Moreover, let~$Y$ be a (real or complex) vector space and~$G \subset \GL(Y)$ a group acting on~$Y$.
Then~$\B$ is a {\bf{topological vector bundle}} with {\bf{fiber}}~$Y$ and {\bf{structure group}}~$G$
if there exists an open covering~$\mathcal{U}$ of~$\scrM$ and for every~$U \in \mathcal{U}$ a 
homeomorphism~$\phi_{U}: \pi^{-1}(U)\to U\times Y$, called a {\bf{local trivialization}} or {\bf{bundle chart}}, such
that the following two properties are satisfied:
\bitem
\item[{\rm{(i)}}] For any~$U \in \mathcal{U}$, the diagram
    \beq \label{commdiag}
    \begin{tikzcd}
        \pi^{-1}(U) \arrow[rd, "\pi"'] \arrow[r, "\phi_U"]
        & U \times Y \arrow[d, "\mathrm{pr}_1"] \\
        & U 
    \end{tikzcd}
    % \begin{array}[c]{cccc}
% \pi^{-1}(U)&\stackrel{\phi_{U}}\longrightarrow&U\times Y\\
% &\searrow&\downarrow \\
% &&U
% \end{array}
\eeq
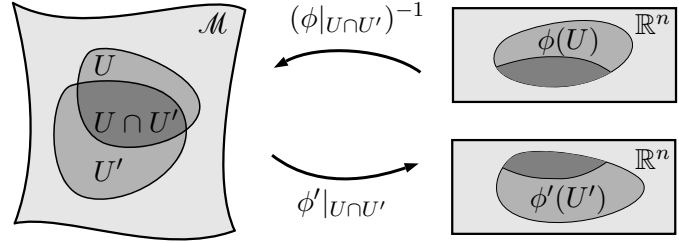
\begin{figure}[tb]
\psscalebox{1.0 1.0} % Change this value to rescale the drawing.
{
\begin{pspicture}(0,27.843513)(8.857229,31.04277)
\definecolor{colour1}{rgb}{0.9019608,0.9019608,0.9019608}
\definecolor{colour0}{rgb}{0.7019608,0.7019608,0.7019608}
\definecolor{colour4}{rgb}{0.5019608,0.5019608,0.5019608}
\psframe[linecolor=black, linewidth=0.03, fillstyle=solid,fillcolor=colour1, dimen=outer](8.857229,30.959627)(5.862229,29.682127)
\psbezier[linecolor=black, linewidth=0.01, fillstyle=solid,fillcolor=colour0](6.592229,30.544085)(6.3488526,30.391603)(6.264659,29.99357)(7.042229,29.907127685546875)(7.8197994,29.820684)(8.535958,30.342434)(8.162229,30.639627)(7.7885,30.936823)(6.835605,30.696564)(6.592229,30.544085)
\pspolygon[linecolor=colour4, linewidth=0.01, fillstyle=solid,fillcolor=colour4](6.459451,30.157406)(6.677229,30.224072)(6.8950067,30.268517)(7.0816736,30.29074)(7.2950068,30.29074)(7.539451,30.268517)(7.70834,30.228516)(7.886118,30.148516)(7.966118,30.090738)(7.837229,30.024073)(7.686118,29.97074)(7.535007,29.930738)(7.3438954,29.908516)(7.0727844,29.904072)(6.86834,29.939629)(6.6905622,29.988516)(6.539451,30.068516)(6.4950066,30.108517)
\psframe[linecolor=black, linewidth=0.03, fillstyle=solid,fillcolor=colour1, dimen=outer](8.847229,29.204628)(5.852229,27.927128)
\psbezier[linecolor=black, linewidth=0.01, fillstyle=solid,fillcolor=colour0](6.697229,28.984085)(6.458853,28.851604)(6.2096586,28.26857)(6.942229,28.122127685546875)(7.6747994,27.975685)(8.730958,28.397432)(8.332229,28.699627)(7.9335,29.001822)(6.9356055,29.116564)(6.697229,28.984085)
\pspolygon[linecolor=colour4, linewidth=0.01, fillstyle=solid,fillcolor=colour4](6.5483403,28.828516)(6.6327844,28.935183)(6.717229,28.988516)(6.863896,29.024073)(7.1216736,29.03296)(7.4416733,29.00185)(7.695007,28.952961)(7.90834,28.899628)(7.846118,28.855183)(7.70834,28.779627)(7.5883403,28.73074)(7.459451,28.695183)(7.2994514,28.677406)(7.1038957,28.686295)(6.957229,28.699627)(6.757229,28.735184)(6.6327844,28.784073)
\pspolygon[linecolor=colour1, linewidth=0.01, fillstyle=solid,fillcolor=colour1](0.042229004,31.004627)(0.212229,30.924627)(0.447229,30.849628)(0.737229,30.824627)(1.277229,30.829628)(1.657229,30.869627)(2.062229,30.929628)(2.442229,30.979628)(2.787229,30.994627)(3.017229,30.964628)(3.117229,30.914627)(3.052229,30.639627)(2.972229,30.264627)(2.912229,29.889627)(2.852229,29.439629)(2.827229,29.099628)(2.822229,28.714628)(2.842229,28.479628)(2.882229,28.249628)(2.967229,27.989628)(2.717229,27.919628)(2.427229,27.874628)(2.102229,27.879627)(1.727229,27.904627)(1.372229,27.964628)(1.002229,28.024628)(0.602229,28.089628)(0.337229,28.104628)(0.15222901,28.109627)(0.052229002,28.099628)(0.172229,28.449627)(0.237229,28.829628)(0.267229,29.314629)(0.257229,29.644629)(0.222229,30.059628)(0.13722901,30.559628)
\pspolygon[linecolor=colour0, linewidth=0.01, fillstyle=solid,fillcolor=colour0](0.89306235,29.947962)(0.8925068,29.823517)(0.8919512,29.685183)(0.9102846,29.572962)(0.9563957,29.417406)(1.1375067,29.204073)(1.4258401,29.084072)(1.669729,29.10574)(1.944729,29.12796)(2.2036178,29.251295)(2.3319511,29.312962)(2.2880623,29.069628)(2.2463956,28.970182)(2.1275067,28.830183)(1.8497291,28.59685)(1.3913957,28.42296)(1.0502845,28.417406)(0.8819512,28.483517)(0.7069512,28.630184)(0.61639565,28.91685)(0.5708401,29.287962)(0.6091735,29.58685)(0.684729,29.798517)(0.7930623,29.92296)
\pspolygon[linecolor=colour0, linewidth=0.01, fillstyle=solid,fillcolor=colour0](1.2880623,30.401295)(1.5080624,30.371294)(1.7030623,30.316294)(1.9980624,30.186295)(2.2180624,30.056293)(2.3380623,29.946295)(2.4580624,29.791294)(2.5030622,29.646294)(2.4930623,29.506294)(2.3386178,29.34685)(2.178618,29.60296)(2.1630623,29.73685)(2.0252845,29.805738)(1.8352846,29.899073)(1.5741735,29.965738)(1.3908402,29.980183)(1.1908401,29.982405)(0.90250677,29.959627)(0.919729,30.163517)(0.9330623,30.111294)(0.97806233,30.266294)(1.0630623,30.371294)(1.1480623,30.406294)
\pspolygon[linecolor=colour4, linewidth=0.01, fillstyle=solid,fillcolor=colour4](0.88417345,29.943516)(0.9241735,29.510183)(1.0780623,29.245184)(1.3469512,29.117962)(1.6930623,29.097406)(2.0875068,29.18296)(2.3380623,29.334627)(2.2591734,29.609072)(2.013618,29.838516)(1.6780623,29.94296)(1.3480624,29.970182)(1.1141734,29.972406)(0.9880623,29.946295)
\psbezier[linecolor=black, linewidth=0.03](0.017229004,31.019629)(1.017229,30.469627)(2.547229,31.219627)(3.117229,30.919627685546875)
\psbezier[linecolor=black, linewidth=0.03](3.117229,30.929628)(2.887229,29.919628)(2.667229,28.649628)(2.977229,27.979627685546873)
\psbezier[linecolor=black, linewidth=0.03](2.987229,27.989628)(2.017229,27.619627)(0.707229,28.229628)(0.037229005,28.079627685546875)
\psbezier[linecolor=black, linewidth=0.03](0.052229002,28.084627)(0.402229,29.024628)(0.257229,30.129627)(0.027229004,31.019627685546876)
\psbezier[linecolor=black, linewidth=0.04, arrowsize=0.05291667cm 2.0,arrowlength=1.4,arrowinset=0.0]{<-}(3.4944513,30.15074)(4.084451,30.47074)(5.014451,30.410738)(5.4444513,30.090738796657934)
\psbezier[linecolor=black, linewidth=0.04, arrowsize=0.05291667cm 2.0,arrowlength=1.4,arrowinset=0.0]{<-}(5.422229,28.878517)(5.0000067,28.738516)(4.152229,28.458517)(3.472229,28.988516574435845)
\psbezier[linecolor=black, linewidth=0.02](1.0300068,30.346294)(0.8316304,30.136707)(0.76965874,29.303297)(1.272229,29.137405463324722)(1.7747992,28.971514)(2.658736,29.295368)(2.4900067,29.746294)(2.3212774,30.197222)(1.2283832,30.555882)(1.0300068,30.346294)
\psbezier[linecolor=black, linewidth=0.02](0.78945124,29.909628)(0.5065304,29.797817)(0.4379701,28.601074)(0.9655924,28.435183241102433)(1.4932146,28.26929)(2.4387112,28.778702)(2.322229,29.425182)(2.205747,30.071665)(1.0723721,30.021439)(0.78945124,29.909628)
\psbezier[linecolor=black, linewidth=0.01](6.454729,30.162128)(6.8492374,30.296852)(7.1112785,30.312778)(7.399729,30.292127685546873)(7.6881795,30.271477)(7.8485155,30.178432)(7.969729,30.097128)
\psbezier[linecolor=black, linewidth=0.01](6.534729,28.822128)(6.8896284,28.658627)(7.075037,28.681772)(7.29084,28.6754610188802)(7.5066433,28.669151)(7.7474756,28.771477)(7.919729,28.897127)
\rput[bl](1.1,29.3){$U \cap U'$}
\rput[bl](1.1,28.65){$U'$}
\rput[bl](1.1,30.05){$U$}
\rput[bl](2.5,30.6){$\scrM$}
\rput[bl](8.3,30.6){$\R^n$}
\rput[bl](8.3,28.8){$\R^n$}
\rput[bl](7,30.3){$\phi(U)$}
\rput[bl](6.9,28.2){$\phi'(U')$}
\rput[bl](3.7,30.6){$(\phi|_{U \cap U'})^{-1}$}
\rput[bl](3.8,28.15){$\phi'|_{U \cap U'}$}
\end{pspicture}
}
\caption{A transition map.}
\label{figtransition}
\end{figure}%
commutes, where~$\mathrm{pr}_1: U \times Y \to U$ denotes the projection onto the first component.
\item[{\rm{(ii)}}] For all~$U,V \in \mathcal{U}$ with~$U\cap V \neq \varnothing$, we have
\beq \label{changechart}
\phi_{U}\circ \phi_{V}^{-1}(x,v)= (x,g_{UV}(x)v) \qquad \text{for all~$x \in U \cap V, v \in Y$}\:,
\eeq
where~$g_{UV}: U\cap V\to G$ is a continuous transition function.
\eitem
\end{Def} \noindent
In the special case~$G = \GL(Y)$, one simply speaks of a vector bundle (i.e., without explicitly mentioning the 
structure group).

Similar to the tangent bundle, by the condition~\eqref{changechart} the local trivializations allow to pull back the 
vector space structure of~$Y$ to the so-called fibers~$\B_x := \pi^{-1}(\{x\}) \subseteq \B$ of a vector bundle. So one 
may think of a vector bundle~$\pi: \B \to \scrM$ as a collection of vector spaces that are parametrized by~$\scrM$.

A simple example of a vector bundle is the Cartesian product
\beq \B = \scrM \times Y \:. \eeq
This vector bundle is globally trivialized, that is, of product form. According to~\eqref{commdiag}, a vector bundle has 
this product structure ``locally''
in~$\pi^{-1}(U)$, but in general not globally. As we have already seen, the tangent bundle~$T\scrM$ of a smooth 
manifold~$\scrM$ is another example. Here, the structure group will in general be~$\mathrm{GL}(\R^n)$, but could 
possibly be taken as a smaller group in special circumstances.

A {\em{differentiable}} (or {\em{smooth}}) vector bundle is a topological vector bundle
where the base~$\scrM$ is a differentiable (or smooth) manifold
together with an atlas of bundle charts such that all transition maps are differentiable
(or smooth, respectively). In this case, also the total space~$\B$ is a differentiable (smooth) manifold and the 
projection map~$\pi$ is differentiable (smooth), as the local trivializations provide a differentiable atlas.

\section{Exercises}
\begin{Exercise} \label{exclosed}  {{(Closed sets)}}  {\em{
\sindex{set!closed}%
Show that the closed sets of a topological space~$E$ (defined as the complements of the open sets) have the following properties:
\bitem
\item[{\rm{(i)}}] The sets~$\varnothing$ and~$E$ are closed.
\item[{\rm{(ii)}}] Closedness under finite unions: For any~$A_1,\ldots,A_n \subset E$,
\beq A_1, \ldots, A_n \text{ closed} \quad \Longrightarrow \quad A_1 \cup \cdots \cup A_n
\text{ closed} \:. \eeq
\item[{\rm{(iii)}}] Closedness under arbitrary intersections: For any family~$\{A_\lambda\}_{\lambda \in \Lambda}$ of subsets of~$E$,
\beq A_\lambda \text{ closed for all } \lambda \in \Lambda \quad \Longrightarrow \quad \bigcap_{\lambda \in \Lambda} A_\lambda \text{ closed} \:. \eeq
\eitem
}} \end{Exercise}

\begin{Exercise} \label{extop1}  {{(Notions of continuity)}}  {\em{
\sindex{continuity!in topological space}%
Let~$E=F=\R$ with the standard topology inherited from the metric~$d(x,y) = |x-y|$.
Show that a real-valued function~$f : E \rightarrow F$ is continuous in the topological sense~\eqref{defcont}
if and only if for every~$x \in \R$ and for every~$\varepsilon>0$ there is~$\delta>0$ such that
\beq \big| f(x) - f(y) \big| < \varepsilon \qquad \text{for all~$y \in \R$ with~$|y-x|<\delta$}\:. \eeq
}} \end{Exercise}

\begin{Exercise} \label{extop2} {{(Extreme value theorem)}\em{
Let~$(E, \O)$ be a non-empty compact topological space and~$f : E \rightarrow \R$ a continuous, real-valued function
on~$E$. Show that~$f$ attains its maximum; that is, there is a point~$x \in E$ with
\beq f(x) \geq \inf_E f \:. \eeq
{\em{Hint:}} First show using basic definitions that a continuous function maps compact sets to compact sets. Then use what you know about compact subsets of~$\R$.

Alternatively, one can use that a compact space is {\em{sequentially compact}} in the sense that every
sequence has a convergent subsequence.
If you are not familiar with these connections, try to prove them starting from the basic definitions.
}} \end{Exercise}

\begin{Exercise} \label{ex:norm} {{(Examples of norms)}}  {\em{
\sindex{normed space}%
\bitem
\item[(a)] Show that the examples~\eqref{lpnorm} and similarly~\eqref{Lpnorm} satisfy all properties of a norm.
{\em{Hint:}} In the case~$p < \infty$, the triangle inequality is also referred to as the {\em{Minkowski inequality}}.
Its proof can be found for example in~\cite[Theorem~3.5]{rudin}.

\item[(b)] We let~$X$ be the vector space of all
complex-valued finite sequences~$(x_n)_{n \in \N}$
(by a finite sequence, we mean a sequences for which only a finite number of
members are non-zero). Show that
\beq
    \big\| (x_n)_{n \in \N} \big\|_p
    := \bigg( \sum_{i=1}^\infty |x_i|^p \bigg)^{\frac{1}{p}}
    \quad\text{for } p < \infty \qquad \text{and} \qquad
    \| x \|_\infty := \sup_{n \in \N} |x_n|
\eeq
defines a norm~$X$. Show that the resulting normed space is infinite-dimensional.
\eitem
}} \end{Exercise}

\begin{Exercise} \label{ex:banach} {{(Examples of Banach spaces)}}  {\em{
\sindex{Banach space}%
\bitem
\item[(a)] Show that~$\C^n$ with the norm~\eqref{lpnorm} is complete.
\item[(b)] Show that the space~$(X, \|.\|_p)$ considered in Exercise~\ref{ex:norm}~(b)
is {\em{not}} complete. Let~$l_p$ be the space of all complex sequences (not necessarily finite)
for which the norm~$\|.\|_p$ is finite. Show that~$l_p$ is a Banach space.
\item[(c)] Show that the space~$C^0_0(\R^n)$ with the norm~\eqref{Lpnorm} is {\em{not}} complete.
{\em{Hint}}: construct a Cauchy sequence that has no limit in~$C^0_0(\R^n)$.
We remark that the completion of these spaces gives the Banach spaces~$L^p(\R^n)$;
details can be found, for example, in~\cite[Chapter~3]{rudin}.
\eitem
}} \end{Exercise}

\begin{Exercise} \label{ex:complete} {{(Completion of a metric space)}}
    {\em{ Let~$(E,d)$ be a metric space.
\sindex{completion!of metric space}%
\bitem
\item[(a)] Show that for two Cauchy sequences~$(x_n)_{n \in \N}, (y_n)_{n \in \N}$ in~$E$, the limit
\beq %\label{distcauchy}
d'\Big( \big( x_n \big)_{n \in \N}, \big(
    y_n \big)_{n \in \N} \Big) := \lim_{n \rightarrow \infty} d(x_n,
    y_n) \eeq
exists. Show further that the function~$d'$ defined
    on the set of all Cauchy sequences of~$E$ in this way has
    the properties~(ii) and~(iii) in Definition~\ref{defmetricspace},
    but will never satisfy property (i) unless~$E$ has only one
    element.
\item[(b)] Verify that
\beq %\label{equivcauchy}
\big( x_n \big)_{n \in
      \N} \simeq \big( y_n \big)_{n \in \N} \qquad \text{if} \qquad
    d'\Big( \big( x_n \big)_{n \in \N}, \big( y_n \big)_{n \in \N}
    \Big) = 0 \eeq
defines an equivalence relation on the set of all Cauchy sequences of~$E$. Show further
    that the function~$d'$ induces a well-defined function~$d''$ on
    the set~$\tilde{E}$ of all equivalence classes. Verify that~$d''$ has property~(i) in Definition~\ref{defmetricspace}
    and still satisfies~(ii) and (iii).
\item[(c)] Show that~$(\tilde{E},d'')$ is a complete metric space.
    Show furthermore that the map~$E \to \tilde{E}$, sending each~$x \in E$ to the constant sequence~$(x)_{n \in \N}$, is
    distance-preserving (and therefore injective).
\item[(d)] Modify the construction in order to form the completion of
    a scalar product space.
\eitem
}} \end{Exercise}

\begin{Exercise} \label{ex:hilbertnorm} {{(Norm of a scalar product space)}}  {\em{
Given a scalar product space~$(V, \la .|. \ra)$, show that~$\| u \| := \sqrt{ \la u | u \ra }$
defines a norm (see Definitions~\ref{defspspace} and~\ref{defnormedspace}).
}} \end{Exercise}

\begin{Exercise} \label{ex:linbound} {{(Boundedness and continuity of linear operators)}}  {\em{
Let~$A : V \rightarrow W$ be a linear operator between normed spaces~$V$ and~$W$.
Show that~$A$ is bounded if and only if it is continuous.
{\em{Hint:}} Write down the usual condition for continuity of~$A$ and try to simplify it using linearity.
}} \end{Exercise}

\begin{Exercise} \label{ex:LVW} {{(Completeness of~$\Lin(V,W)$)}}  {\em{
\bitem
\item[(a)] Show that the operator norm on~$\Lin(V, W)$ is indeed a norm; that is, verify that it has all the properties
in Definition~\ref{defnormedspace}.
\item[(b)] Show that~$\Lin(V,W)$ is complete if and only if~$W$ is complete.
\eitem
}} \end{Exercise}

\begin{Exercise} \label{ex:orthoproject} {{(Orthogonal projection to closed subspaces of a Hilbert space)}}  {\em{
Let~$\H$ be a Hilbert space and~$V \subset \H$ a closed subspace.
\bitem
\item[(a)] Show the parallelogram identity: For all~$u,v \in \H$,
\beq \| u+v \|^2 + \|u-v\|^2 = 2\, \|u\|^2 + 2\, \|v\|^2 \:. \eeq
\item[(b)] Given~$u \in \H$, let~$(v_n)_{n \in \N}$ be a sequence in~$V$ which is a minimizing sequence
of the distance to~$u$; that is,
\beq \|u-v_n\| \rightarrow \inf_{v \in V} \|u-v\| \:. \eeq
Prove that the sequence~$(v_n)_{n \in \N}$ converges.
{\em{Hint:}} Apply the parallelogram identity to show that the sequence is Cauchy. Then
make use of the completeness of the Hilbert space.
\item[(c)] Show that the limit vector~$v := \lim_{n \rightarrow \infty} v_n$ has the property
\beq \la u-v, w \ra = 0 \qquad \text{for all~$w \in V$}\:. \eeq
In view of this equation, the vector~$v$ is also referred to as the orthogonal projection of~$u$ to~$V$.
\eitem
In the finite-dimensional setting, the orthogonal projection can be given more explicitly
as explained in Exercise~\ref{ex:ortho}.
}} \end{Exercise}

\begin{Exercise} \label{ex:FR} {{(Proof of the Fr{\'e}chet-Riesz theorem)}}  {\em{
Let~$\H$ be a Hilbert space and let~$\phi \in \H^*$ be non-zero.
\bitem
\item[(a)] Show that the kernel of~$\phi$ is a closed subspace of~$\H$.
\item[(b)] Apply the result of Exercise~\ref{ex:orthoproject} to construct a nonzero vector~$v$
that is orthogonal to~$\ker \phi$. Show that this vector is unique up to scaling.
\item[(c)] Show that, after a suitable scaling, the vector~$v$ satisfies~\eqref{eq:FR}.
\item[(d)] Show that the vector~$v$ satisfying~\eqref{eq:FR} is unique.
\eitem
}} \end{Exercise}

\begin{Exercise} \label{ex:ortho} {{(Orthogonal complement of a finite-dimensional subspace)}}  {\em{
\bitem
\item[(a)] Let~$I$ be a finite-dimensional subspace of a Hilbert space~$(\H, \la .|. \ra)$.
Show that its orthogonal complement~$I^\perp$ defined by~\eqref{Iperp} is again
a complex vector space.
\item[(b)] Show that restricting the scalar product to~$I$, one again gets a Hilbert space.
In particular, why is it complete again?
\item[(c)] Show that every vector~$u \in \H$ has a unique decomposition of the form~\eqref{udecomp}. \\
{\em{Hint:}} Choosing an orthonormal basis~$e_1, \ldots, e_n$ of~$I$, a good ansatz for~$u^{||}$ is
\beq u^{||} = \sum_{k=1}^n \la e_k | u \ra\: e_k \:. \eeq
\eitem
}} \end{Exercise}

\begin{Exercise} \label{ex:multop1} {{(Multiplication operators)}}  {\em{
\sindex{operator!of multiplication}%
Let~$f \in C^0(\R, \C)$ be a continuous, com\-plex-valued function. Assume that it is bounded, that is,
that~$\sup_\R |f| < \infty$. We consider the multiplication operator~$T_f$ on the Hilbert space~$\H=L^2(\R)$; that is,
\beq T_f \::\: \H \rightarrow \H \:,\qquad (T_f \,\phi)(x) = f(x)\: \phi(x) \:. \eeq
\bitem
\item[(a)] Show that~$T_f$ is a (well-defined) bounded operator whose operator norm is given by
\beq \big\| T_f \| = \sup_{\R} |f| \:. \eeq
\item[(b)] Show that~$T_f$ is symmetric if and only if~$f$ is real-valued.
Under which assumptions on~$f$ is~$T_f$ unitary?
\eitem
}} \end{Exercise}

\begin{Exercise}\label{ex:examples-of-measures}{{(Examples of measures)}} {\em{
Let~$\F$ be a set and~$\mathfrak{M}$ a~$\sigma$-algebra on~$\F$.
\bitem
\item[(a)] Let~$f: \F \to [0,\infty)$ be a function that is nonzero only at finitely many points of~$\F$. Show that~$\rho: \mathfrak{M} \to [0,\infty)$, $A \mapsto \sum_{x \in A} f(x)$ defines a measure. 
\item[(b)] Show that any linear combination of measures on~$\F$ with nonnegative coefficients is again a measure on~$\F$. 
\eitem
}}\end{Exercise}

\begin{Exercise}\label{ex:measure-basic-properties}{{(Basic properties of measures)}} {\em{
Let~$(\F,\mathfrak{M},\rho)$ be a measure space.
\bitem
\item[(a)] Show that~$\rho(A \cup B ) = \rho(A)+\rho(B)-\rho(A\cap B)$ holds for any~$A,B \in \mathfrak{M}$.
\item[(b)] Show that for any~$A,B \in \mathfrak{M}$ with~$A \subset B$, the inequality~$\rho(A) \leq \rho(B)$
holds.
\item[(c)] Show that for any sequence~$(A_n)_{n \in \N}$ of measurable sets (not necessarily pairwise disjoint),
the inequality~$\rho(\bigcup_{n \in \N} A_n) \leq \sum_{n \in \N} \rho(A_n)$ holds.
\eitem
}}\end{Exercise}

\begin{Exercise} \label{ex:borel} {{(Borel algebra)}} {\em{
\sindex{Borel algebra}%
Let~$\F$ be a set.
\bitem
\item[(a)] Show that the power set of~$\F$ (that is, the set of all subsets) forms a $\sigma$-algebra.
\item[(b)] Show that the intersection of $\sigma$-algebras is again a $\sigma$-algebra.
\item[(c)] Assume in addition that~$\F$ is a topological space. Combine~(a) and~(b) to conclude that there is a smallest~$\sigma$-algebra that contains all open subsets of~$\F$.
\eitem
}} \end{Exercise}

\begin{Exercise} \label{exmeasurecomplete} {{(Completion of a measure)}}  {\em{
\sindex{completion!of measure}%
Let~$(\F, {\mathfrak{M}}, \rho)$ be a measure space. We introduce the family of sets
\begin{align}
\tilde{\mathfrak{M}} := \{ A \subset \F \:|\: & \text{there exist~$B, N \in {\mathfrak{M}}$} \notag \\
&\text{with~$(A \setminus B) \cup (B \setminus A) \subset N$ and~$\rho(N)=0$} \} \:.
\end{align}
\bitem
\item[(a)] Show that~$\tilde{\mathfrak{M}}$ is again a $\sigma$-algebra.
\item[(b)] Show that the prescription~$\tilde{\rho}(A) := \rho(B)$ defines a measure on~$\tilde{\mathfrak{M}}$. (Where for~$A \in \tilde{\mathfrak{M}}$ the set~$B \in \mathfrak{M}$ is a set chosen as in the defining property of~$\tilde{\mathfrak{M}}$.)  
\item[(c)] Show that the measure~$\tilde{\rho}$ is complete.
\eitem
}} \end{Exercise}

\begin{Exercise} {{(Understanding the push-forward measure)}} \label{expushforward} {\em{
\sindex{measure!push-forward}%
The purpose of this exercise is to introduce the so-called push-forward measure,
which will be used later for the construction of causal fermion systems.
Let~$\scrM \subset \R^3$ be a smooth surface described by a parametrization~$\Phi$. More precisely, given
an open subset~$\Omega \subset \R^2$, we consider a smooth injective map
\beq \Phi \::\: \Omega \rightarrow \R^3 \:, \eeq
with the property that~$D\Phi|_p : \R^2 \rightarrow \R^3$ has rank two for all~$p \in \Omega$.
Then the surface~$\scrM$ is defined as the image~$\Phi(\Omega) \subset \R^3$.
We now introduce a measure~$\rho$ on~$\R^3$ as the {\em{push-forward measure}} of the
Lebesgue measure on~$\R^2$ through~$\Phi$: Let~$\mu$ be the Lebesgue measure on~$\R^2$.
We define a set~$U \subset \R^3$ to be~$\rho$-measurable if and only if its pre-image~$\Phi^{-1}(U) \subset \R^2$
is~$\mu$-measurable. On the~$\rho$-measurable sets, we define the measure~$\rho$ by
\beq \rho(U) = \mu\big( \Phi^{-1}(U) \big) \:. \eeq
Verify that the $\rho$-measurable sets form a $\sigma$-algebra, and that~$\rho$ is a measure.
What are the sets of $\rho$-measure zero? What is the support of the measure~$\rho$?

Suppose that~$\Phi$ is no longer assumed to be injective. Is~$\rho$ still a well-defined measure?
Is~$\rho$ well-defined if~$\Phi$ is only assumed to be continuous?
What are the minimal regularity assumptions on~$\Phi$ needed for the
push-forward measure to be well-defined? What is the support of~$\rho$ in this general setting?
}} \end{Exercise}

\begin{Exercise} \label{exnofunc} {\em{
\bitem
\item[(a)] Let~$f \in C^0(\R^n, \R)$ be continuous. Show that
\beq \label{exnofunc0}
\lim_{r \searrow 0} \int_{B_r(0)} f(x) \:\Diff^nx = 0 \:.
\eeq
{\em{Hint:}} Use that a continuous function is locally bounded.
\item[(b)] Let~$f : \R^n \rightarrow \R$ be a Lebesgue-integrable function.
Show that~\eqref{exnofunc0} again holds. {\em{Hint:}} Use
Lebesgue's monotone convergence theorem.
\eitem
}} \end{Exercise}

\begin{Exercise} \label{ex:Stop} {{(Topology on the Schwartz space)}} {\em{
\bitem
\item[(a)] Show that the topology on~${\mathcal{S}}(\R^n)$ defined by~\eqref{Stop}
gives rise to the notion of convergence~\eqref{locconvex}.
\item[(b)] Show that a linear functional~$T : {\mathcal{S}}(\R^n) \rightarrow \R$ is
continuous if and only if there are~$p,q \in \N_0$ and~$c>0$ such that the
inequality~\eqref{Tcontinuous} holds.
\item[(c)] Show that for any~$i \in \{1,\ldots, n\}$,
the partial derivative~$D_i : f \mapsto \partial_i f$ is a continuous linear mapping from~${\mathcal{S}}(\R^n)$
to itself.
\eitem
}} \end{Exercise}

\begin{Exercise} \label{ex:fourierinverse} {\em{ Prove the Fourier inversion
formula for tempered distributions
\beq {\mathcal{F}} \circ {\mathcal{F}}^* = {\mathcal{F}}^* \circ {\mathcal{F}} = \1_{{\mathcal{S}}'(\R^n)}
\::\: {\mathcal{S}}'(\R^n) \rightarrow {\mathcal{S}}'(\R^n) \:. \eeq
{\em{Hint:}} Use Lemma~\ref{lemmafourierinverse} together with the definition of the
Fourier transform of a tempered distribution.
}} \end{Exercise}

\begin{Exercise} \label{ex:deltasequence} {{(Dirac sequence)}}  {\em{
Given~$\varepsilon>0$, consider the Gaussian~$\eta_\varepsilon:\R \to \R$ with
\beq \eta_\varepsilon(x) := \frac{1}{\sqrt{4 \pi \varepsilon}}\: \E^{-x^2/(4 \varepsilon)} \:. \eeq
\bitem
\item[(a)] Show that~$\eta_\varepsilon$ is a Schwartz function.
\item[(b)] Show that in the limit~$\varepsilon \searrow 0$, the corresponding regular distribution converges to the $\delta$ distribution
in the sense that for all~$f \in {\mathcal{S}}(\R)$,
\beq \lim_{\varepsilon \searrow 0} T_{\eta_\varepsilon}(f) = \delta(f) \,. \eeq
(We recall that the $\delta$ distribution as introduced informally in Example~\ref{exdelta} is defined
by~$\delta(f)=f(0)$ for all~$f \in {\mathcal{S}}(\R)$).
\eitem
}} \end{Exercise}

\begin{Exercise} \label{ex2} {\em{ This exercise is devoted to a clean proof
of the distributional relation~\eqref{eq:delta-formula} in one dimension. More precisely, we want to prove the slightly
more general statement that for any function~$\eta \in C^1(\R) \cap L^1(\R)$,
\beq \label{deltaint}
\lim_{\varepsilon \searrow 0} \int_\R \eta(x) \left( \frac{1}{x - \ci \varepsilon} - \frac{1}{x + \ci \varepsilon} \right) \Diff x
= 2 \pi \ci\: \eta(0) \:.
\eeq
\bitem
\item[(a)] Let~$\eta \in C^1(\R) \cap L^1(\R)$ with~$\eta(0)=0$. Use Lebesgue's dominated convergence theorem to show that~\eqref{deltaint} holds.
\item[(b)] Use the residue theorem to show that~\eqref{deltaint} holds for the function~$\eta(x)=1/(x^2+1)$.
\item[(c)] Combine the results of~(a) and~(b) to prove~\eqref{deltaint} for general~$\eta \in C^1(\R) \cap L^1(\R)$.
\eitem
}} \end{Exercise}

\begin{Exercise} \label{ex:derivative-of-heaviside} {\em{
Let~$\Theta := \chi_{(0,\infty)}:\R \to \R$ be the {\em{Heaviside function}} defined by
\beq \Theta(x) = \left\{ \begin{array}{cc} 0 & \text{if~$x \leq 0$} \\ 1 & \text{if~$x >0$}\:. \end{array} \right. \eeq
\sindex{Heaviside function}%
\nindex{abv2@$\Theta$ -- Heaviside function}%
Show that~$\Theta' = \delta$ holds in~$\mathcal{S}'(\R)$.
}}\end{Exercise}

\begin{Exercise} \label{ex:fourierdelta} {{(Fourier transform of $\delta$ distribution)}}  {\em{
Let~$T \in {\mathcal{S}}'(\R)$ be the regular distribution corresponding to the constant function with value one; that is,
\beq T(f) := \int_{-\infty}^\infty f(x)\: \Diff x \:. \eeq
\bitem
\item[(a)] Show that~$T(f) = 2 \pi ({\mathcal{F}}^* f)(0)$.
\item[(b)] Use the Fourier inversion theorem for tempered distributions to conclude
that~$\F T = 2 \pi\, \delta$ (where~$\delta$ is again the $\delta$ distribution).
\item[(c)] Apply the Fourier inversion theorem again to compute the Fourier transform
of the $\delta$ distribution.
\item[(d)] Alternatively, one can compute these Fourier transforms directly working with
con\-ver\-gence-generating factors in the style of~\eqref{generate-convergence}.
Do this carefully step by step, making sure that every computation step is mathematically
well-defined.
\eitem
}} \end{Exercise}

\begin{Exercise} {{(Schwartz space)}} {\em{
\sindex{Schwartz space}%
\bitem
\item[(a)] Show that Schwartz functions decay faster than polynomially at infinity together with all their partial derivatives. More precisely, show that for every multi-index~$\alpha$ and for every~$N\in \N$, there exists~$C_{\alpha,N}\ge 0$  such that
\beq
|D^\alpha f(x)|\le \frac{C_{\alpha,N}}{1+|x|^N}\quad \mbox{for all } x\in\R^n.
\eeq
\item[(b)] Which of the following functions belongs to~$\mathcal{S}(\R)$? Motivate your answers!
\beq
f_1(x):=\E^{-x^2},\qquad f_2(x):=\frac{1}{1+x^4},\qquad f_3(x):=\frac{\E^{-x^2}}{2+\sin(\E^{x^2})}.
\eeq
\eitem
}} \end{Exercise}

\begin{Exercise} {{(The principal value integral)}} {\em{
\sindex{principal value integral}%
For every~$f\in\mathcal{S}(\R)$, we define
\beq A(f):=\lim_{\varepsilon\searrow 0}\ \int_{\R\setminus (-\varepsilon,\varepsilon)}\!\frac{f(x)}{x}\,\Diff x \:. \eeq
Does the limit exist? In fact, $A$ is a tempered distribution. Show it.
}} \end{Exercise}

\begin{Exercise} (Multiplication operators) {\em{ Let~$n \in \N$.
\sindex{operator!of multiplication}%
\bitem
\item[(a)] Let~$g$ be a (not necessarily continuous) function~$g: \R^n \rightarrow \R$ 
with the property that the mapping~$x \mapsto g(x) /(1 + \vert x \vert ^r)$ belongs to~$L^1 (\R^n)$
for some~$r >0$. Show that the map~$f \mapsto T_g (f) := \int_{\R^n} f(x) \,g(x)\, \Diff x$ defines a tempered distribution, $T_g \in {\mathcal{S}}(\R^n)'$.
\item[(b)] Find a smooth function~$g\in L^1(\R^n)$ that is not pointwise bounded by any polynomial; that is,
there is no polynomial~$p$ such that
\beq
|g(x)|\le p(|x|)\quad\mbox{for all~$x\in\R^n$} \:.
\eeq
But the corresponding functional 
\begin{equation}\label{integral}
T_g:\mathcal{S}(\R^n)\ni f\mapsto \int_{\R^n}g(x)f(x)\,\Diff^nx
\end{equation}
still yields a well-defined tempered distribution.
\eitem
}} \end{Exercise}

\begin{Exercise} {{(Approximating the $\delta$ distribution)}} {\em{
Not all distributions can be written as in~\eqref{integral}. Nevertheless, it can be shown that every tempered distribution can be approximated by such functionals. 
Let us verify this statement in the concrete example of the $\delta$ distribution. Let~$\varphi\in \mathcal{C}_0^\infty(\R^n)$ fulfill~$\varphi\ge 0$, $\mathrm{supp}\,\varphi\subset B_1(0)$ and~$\|\varphi\|_{L^1}=1$. Define, for every~$\varepsilon>0$,
\beq
\varphi_\varepsilon(x):=\frac{1}{\varepsilon^n}\,\varphi\left(\frac{x}{\varepsilon}\right)\in\mathcal{C}_0^\infty(\R^n)\quad\mbox{and}\quad\delta_\varepsilon:=T_{\varphi_\varepsilon}\ \mbox{(defined as in~\eqref{integral})}
\eeq
Each functional~$\delta_\varepsilon$ is a tempered distribution (why?). Show that, for every~$f\in\mathcal{S}(\R^n)$,
\beq
\delta_\varepsilon(f)\to \delta(f)=f(0)\quad\mbox{as~$\varepsilon\searrow 0$.}
\eeq
{\emph{Remark:} The more general statement that any tempered distribution can be approximated by a sequence of regular distributions can be shown using the method of convolution. We refer to~\cite[Chapter~5]{friedlander2} for details.}
}} \end{Exercise}

\begin{Exercise} {{(Another regular tempered distribution)}} {\em{
Let~$n,k\in\N$ with~$n>k$. Show that the mapping
\beq
\mathcal{S}(\R^n)\ni f\mapsto \int_{\R^n}\frac{f(x)}{|x|^k}\,\Diff^nx\in\C
\eeq
is a well-defined tempered distribution.
}} \end{Exercise}

\begin{Exercise} {{(Fourier transform on~$\mathcal{S}(\R^n)$ and~$\mathcal{S}'(\R^n)$)}} {\em{
\sindex{Fourier transform}%
Compute the Fourier transform of the following functions and tempered distributions.
\bitem
\item[(a)] $f\in\mathcal{S}(\R^n)$ defined by~$f(x):=\E^{-\lambda x^2}$ for~$\lambda\ge 0$;
\item[(b)] $T_g\in\mathcal{S}'(\R^n)$ for the functions~$g\in L^1(\R)$ defined by 
\beq
g(x)=\E^{-|x|}\quad\mbox{and}\quad g(x)=\frac{1}{1+x^2}.
\eeq
\eitem
}} \end{Exercise}

\begin{Exercise} {{(Fourier transform on~$L^1(\R^n)$)}} {\em{
\sindex{Fourier transform}%
The functions in~$L^1(\R^n)$ define tempered distributions by means of the identification~$g\mapsto T_g$. As 
distributions they admit Fourier transform. However, for these functions the Fourier transform can also be defined 
directly via the usual integral form. The goal of this exercise is to show that this is indeed true and that the two 
definitions coincide.
\bitem
\item[(a)] Show that the Fourier transform
\beq
(\mathcal{F}g)(p):=\frac{1}{(2\pi)^{n/2}}\int_{\R^n}\E^{-\cI px}g(x)\,\Diff^n x
\eeq
defines a map
\beq
\mathcal{F}:L^1(\R^n)\rightarrow C^0(\R^n)\cap L^\infty(\R^n).
\eeq
Moreover, show that there exists a constant~$C_n$ such that
\beq
\|\mathcal{F}g\|_\infty\le C_n\|g\|_{L^1}\quad\mbox{for all~$g\in L^1(\R^n)$}.
\eeq
\item[(b)] Let~$g\in L^1(\R^n)$. Show that the Fourier transform of the distribution~$T_g\in\mathcal{S}'(\R^n)$ satisfies the relation
\beq
\mathcal{F}(T_g)=T_{\mathcal{F}g}.
\eeq 
\textit{Hint:} You may use that~$\mathcal{S}(\R^n)$ is dense in~$L^1(\R^n)$.
\eitem
}} \end{Exercise}

\begin{Exercise} {{(On the topology of~$\mathcal{S}(\R^n)$)}} {\em{
\sindex{Schwartz space}%
Consider the Schwartz space~$\mathcal{S}(\R^n)$ equipped with the family of norms~$\|\cdot\|_{p,q}$, where~$p,q\in \N_0$. Show the following statements.
\bitem
\item[(a)] For any~$f,g\in\mathcal{S}(\R^n)$, the following series converges:
\beq
d(f,g):=\sum_{p,q,\in\N_0}\frac{1}{2^{p+q}}\frac{\|f-g\|_{p,q}}{1+\|f-g\|_{p,q}}.
\eeq
Moreover it defines a metric on~$\mathcal{S}(\R^n)$.
\item[(b)] The metric space~$(\mathcal{S}(\R^n),d)$ is complete.
\item[(c)] Show that if a topology on a vector space is induced by a norm, then there is a neighborhood~$U$ of~$0$ such that for every other neighborhood~$V$ of~$0$, there is a positive number~$r$ such that
$U \subset r\cdot V$. Can the topology of~$\mathcal{S}(\R^n)$ be induced by a norm?
\eitem
}} \end{Exercise}

\begin{Exercise} {{(Convolution of Schwartz functions)}} \label{exconv1} {\em{
\bitem
\item[(a)] Show that for two Schwartz functions~$f, g \in \mathcal{S}(\R^n)$,
the integral~\eqref{eq:convolution} defines a smooth function~$(f*g)(x)$.
\item[(b)] Show that this function has rapid decay.
\eitem 
}} \end{Exercise}

\begin{Exercise} {{(Convolution and Fourier transform)}} \label{exconvF} {\em{
Prove the relation~\eqref{convF} on the Fourier transform of a convolution. \\
{\em{Hint:}} Use the definition of the Fourier transform~\eqref{fourier} and rewrite the
resulting double integral.
}} \end{Exercise}

\begin{Exercise} {{(Convolution with the $\delta$ distribution)}} \label{exconvdelta} {\em{
Let~$\delta \in \mathcal{S}'(\R^n)$ be the $\delta$ distribution (at the origin). Show that~$f*\delta = f$ holds for any~$f \in \mathcal{S}(\R^n)$.
}} \end{Exercise}

\begin{Exercise} {{(Smoothing a distribution by convolution)}} \label{exconv2} {\em{ Let~$T \in \mathcal{S}'(\R^n)$ and~$f \in \mathcal{S}(R)^n)$.
\bitem
\item[(a)] Show that the convolution~$f*T$ as defined in~\eqref{fTconv} defines a smooth function.
\item[(b)] Show by a counter-example that the function~$f*T$ in general does {\em{not}} decay at infinity.
\item[(c)]  Let~$n\ge 1$,  $\alpha\in\N^n$, $f\in\mathcal{S}(\R^n)$ and~$T\in\mathcal{S}'(\R^n)$. Prove that~$D^\alpha(f*T)=f*(D^\alpha T)$.
    \eitem
}} \end{Exercise}

\begin{Exercise} {{(A simple manifold)}}  {\em{
Let~$\scrM = S^1 \subset \C$ be the unit circle (considered as a subset of the complex plane).
We choose the charts~$(\phi_1, U_1)$ and~$(\phi_2, U_2)$ with
\begin{align}
U_1 &= \Big\{ \E^{\ci \alpha} \:\Big|\: \alpha \in \Big( -\frac{3 \pi}{2}, \frac{3 \pi}{2} \Big) \Big\} \:,&
\phi_1\big( \E^{\ci \alpha} \big) &= \alpha \\
U_2 &= \Big\{ -\E^{\ci \alpha} \:\Big|\: \alpha \in \Big( -\frac{3 \pi}{2}, \frac{3 \pi}{2} \Big) \Big\} \:,&
\phi_1\big( -\E^{\ci \alpha} \big) &= \alpha \:.
\end{align}
Show that these two charts define a smooth atlas of~$\scrM$.
}} \end{Exercise}

\begin{Exercise}  {\em{
Let~$\scrM, \scrN$ be smooth manifolds and~$F: \scrM \to \scrN$ a map. Show that~$F$ is smooth if and only if for 
any~$p \in\scrM$ there exist charts~$(\phi,U)$ around~$p$ and~$(\psi,V)$ around~$F(p)$ with~$F(U) \subset V$ and 
such that~$\psi \circ F \circ \phi^{-1}$ is smooth (as a map between open subsets of Euclidean spaces).
}} \end{Exercise}

\begin{Exercise}{\em{Unwind the definitions to verify 
formula~\eqref{eq:derivative-of-smooth-map-in-local-coordinates} 
for the derivative of a smooth map~$F: \scrM \to \scrN$ between two smooth manifolds~$\scrM, \scrN$ 
in local coordinates.
}}\end{Exercise}

\begin{Exercise} {{(A simple vector bundle)}}  {\em{
\sindex{vector bundle}%
Let~$\B = \R \times S^1$ be the two-dimensional cylinder, $\scrM=S^1$ and
\beq \pi \::\: \B \rightarrow \scrM \:,\qquad (t, \E^{\ci \alpha}) \mapsto \E^{\ci \alpha} \:. \eeq
Show that~$\B$ is a smooth vector bundle with fiber~$Y = \R$.
}} \end{Exercise}

\begin{Exercise} {{(Another vector bundle)}}  {\em{
\sindex{vector bundle}%
Use the mapping
\beq F \::\: \R \times S^2 \rightarrow \R^3 \setminus \{0\} \:,\qquad
(t, x) \mapsto \E^t\, x \eeq
(where we consider~$S^2$ as the unit sphere embedded in~$\R^3$)
in order to give~$\B = \R^3 \setminus \{0\}$ the structure of a vector bundle on~$S^2$ with fiber~$Y =\R$.
}} \end{Exercise}

\begin{Exercise} {{(The M\"obius bundle)}}  {\em{
Represent the circle~$S^1$ as~$S^1 = [0,1]/(0 \sim 1)$, that is, as the unit interval with its end points being identified. 
Moreover, consider the space~$X := ([0,1] \times \R)/\!\sim$, where~$\sim$ is the equivalence relation generated by 
$(0,v) 
\sim (1,-v)$ for all~$v \in \R$. Show that the map~$\pi: X \to S^1$, $[(x,v)] \mapsto [x]$ is well-defined and 
defines a vector bundle.
}} \end{Exercise}

\chapter{Elements of Operator Theory} \label{chapter:FA}
In this chapter, we introduce some material from functional analysis
that will be needed later in this book.
More precisely, in Section~\ref{secmanop} we explain the concept that
linear operators with certain properties form a submanifold of the
space of all linear operators. This concept will be useful when generalizing
the causal action principle to causal variational principles in Chapter~\ref{chapcvp}.
In Section~\ref{secspectral} we recall the spectral calculus for selfadjoint operators.
Although this material is covered in most functional analysis lectures,
in this book, we do not expect that the reader is already familiar with this topic.
The spectral theorem will be used only when developing functional analytic methods
in spacetime in Chapter~\ref{secFSO}.

%\section{Stone's Formula and Resolvent Methods}

\section{Manifolds of Operators} \label{secmanop}
In this book, it is sometimes useful to observe that certain sets of operators on a Hilbert space
form a smooth manifold. For the purposes in this book, it suffices to work out this concept
in the case of a finite-dimensional Hilbert space (the generalization to the infinite-dimensional case
is a bit more technical; for details see~\cite{banach}).

We begin with a simple example that illustrates the basic concept.

\begin{Example} {\bf{(Grassmann manifold)}} \label{exgrassmann} {\em{
\sindex{Grassmann manifold}%
Given~$f \in \N$, we consider~$\C^f$ with the canonical scalar product~$\la .,. \ra_{\C^f}$.
Let~$\G$ be the set of all orthogonal projection operators to one-dimensional subspaces of~$\C^f$,
\beq \G := \big\{ \pi_V \text{ orthogonal projection to a one-dimensional subspace~$V \subset \C^f$} \big\} \:. \eeq
Let us verify that~$\G$ is a smooth manifold of dimension
\beq \dim \G = 2f - 2 \:. \eeq
To this end, let~$\pi_V$ be such a projection operator. We choose a unit vector~$v$ that spans~$V$.
Next, we let~$V^\perp$ be the orthogonal complement of~$V$
and~$W= B_1(0) \cap V^\perp$ its unit ball. We consider the mapping
\beq F \::\: W \rightarrow \G \:,\qquad u \mapsto \pi_{\text{span}(v+u)} \:. \eeq
It is verified by direct computation that~$F$ is injective and that its image is an open neighborhood of~$\pi_V$
in~$\G$. Also, one easily verifies that it is a local homeomorphism and thus defines a chart
\beq \phi = F|_{F(W)}^{-1} \::\: F(W) \rightarrow W \subset V^\perp \simeq \C^{f-1} \simeq \R^{2f-2} \:. \eeq
Moreover, one verifies directly that all the charts obtained in this way form a smooth atlas
(a more explicit method for constructing charts is explained in Exercise~\ref{exgrasschart}).
This manifold is a special case of a {\em{Grassmann manifold}}.
}} \QEDrem
\end{Example}

This concept can be generalized to so-called {\em{flag manifolds}}
(see for example~\cite[page~142]{helgason}):
\begin{Def} Given~$r \in \N$, we choose~$r$ integers
\beq 0 < d_1 < \cdots < d_r < f \:. \eeq
Consider a sequence~$(L_1, \ldots, L_r)$ of nested subspaces
\sindex{flag manifold}%
\beq L_1 \subset \dots \subset L_r \subset \C^f \qquad \text{with} \qquad
\dim L_i = d_i \eeq
for~$i=1,\ldots, r$. Then, the set of such sequences~$(L_1, \ldots, L_r)$
is referred to as the {\bf{flag manifold}}~$\F^{d_1,\ldots, d_r}$.
\end{Def}
Here we do not need to verify that a flag manifold is indeed a manifold.
Instead, it suffices to consider a specific set of operators that is
related to a flag manifold. Namely, choosing again~$\H=\C^f$ as
well as integers~$p,q$ with~$p+q \leq f$,
we let~$\F^{p,q}$ be the set
\beq \F^{p,q} = \{ A \in \Lin(\H) \,|\, \text{$A$ is symmetric and has~$p$ positive and~$q$ negative eigenvalues} \}\:, \eeq
where we count the eigenvalues with multiplicities.
Taking~$L_1$ as the subspace spanned by the eigenvectors corresponding to the
positive eigenvalues and~$L_2$ as the image of~$A$,
one gets a corresponding flag manifold with~$r=2$ and~$d_1=p$, $d_2=p+q$.
But the operators in~$\F^{p,q}$ contain more information, namely the eigenvalues and the corresponding
eigenspaces. Therefore the set~$\F^{p,q}$ can be regarded as a flag manifold with additional structures.
We now prove that this set is again a smooth manifold, following the method in~\cite[Section~3]{gaugefix}.

\begin{Prp} \label{prppq} The set~$\F^{p,q}$ is a smooth manifold of dimension
\beq \dim \F^{p,q} = 2 f \,(p+q) - (p+q)^2\:. \eeq
\end{Prp}
\Proof Let~$x \in \F^{p,q}$. We denote its image by~$I$ and set~$J=I^\perp$. Thus, using
a block matrix representation in~$\C^f = I \oplus J$, the matrix~$x$ has the representation
\beq \label{xXrep}
x = \begin{pmatrix} X & 0 \\ 0 & 0 \end{pmatrix} \:.
\eeq
The goal is to find a parametrization of operators of~$\F^{p,q}$ in a small neighborhood of~$x$.
We first note that varying~$X$ by a sufficiently small symmetric matrix~$A$, the resulting matrix~$X+A$
has again~$p$ positive and~$q$ negative eigenvalues. In order to also vary the off-diagonal entries
in~\eqref{xXrep}, we make the ansatz
\beq M = (\1+C) \begin{pmatrix} X+A & 0 \\ 0 & 0 \end{pmatrix} (\1+C)^* \eeq
with an $f \times f$-matrix~$C$.
This ansatz has the advantage that~$M$ is obviously symmetric and, for~$C$ sufficiently small, has again~$p$ positive and~$q$ negative eigenvalues (for details see Exercise~\ref{ex:pq}). We want to choose~$C$ such that the upper right block matrix entry of~$M$
has a particularly simple form. This leads us to the parametrization matrix
\begin{align}
M &:= \begin{pmatrix} \1 & 0 \\ B^* (X+A)^{-1} & \1 \end{pmatrix}
\begin{pmatrix} X+A & 0 \\ 0 & 0 \end{pmatrix}
\begin{pmatrix} \1 & (X+A)^{-1} B \\ 0 & \1 \end{pmatrix} \label{Mform0} \\
&\:= \begin{pmatrix} X+A & B \\ B^* & B^* (X+A)^{-1} B \end{pmatrix} \label{Mform1}
\end{align}
with a linear operator~$B : J \rightarrow I$.
We also see that for~$(A,B) = (0,0)$ the parametrization matrix equals~$x$, which is necessary for building a
chart around~$x$.

Thus, for sufficiently small~$\varepsilon$, we obtain the mapping
\beq \label{LambdaM}
\Lambda \::\: \big(\text{Symm}(I) \oplus L(I,J) \big) \cap B_\varepsilon(0) \rightarrow \F^{p,q}\:,\qquad
(A,B) \mapsto M
\eeq
(where~$\text{Symm}(I)$ denotes the symmetric linear operators). Let us verify that (again for
sufficiently small~$\varepsilon$) this mapping is a homeomorphism to an open neighborhood
of~$x \in \F^{p,q}$. It is obvious from~\eqref{Mform1} that~$\Lambda$ is injective. In order to
verify that it maps to an open neighborhood of~$x$, we let~$y \in F^{p,q}$ with~$\|x-y\| < \delta$
(with~$\delta>0$ to be specified below). Diagonalizing~$y$ with a unitary matrix~$U$, we obtain
the block matrix representation
\beq y = \begin{pmatrix} U_{11} & U_{12} \\ U_{21} & U_{22} \end{pmatrix}
\begin{pmatrix} X+C & 0 \\ 0 & 0 \end{pmatrix}
\begin{pmatrix} U_{11}^* & U_{21}^* \\ U_{12}^* & U_{22}^* \end{pmatrix} \:, \eeq
where~$C$ is a symmetric linear operator on~$I$.
In the limit~$y \rightarrow x$, the image of~$y$ converges to the image of~$x$,
implying that the matrix~$U_{11}$ becomes unitary. Therefore, for sufficiently small~$\delta>0$,
the matrix~$U_{11}$ is invertible, giving rise to the representation
\beq y = \begin{pmatrix} \1 & 0 \\ U_{21} \,U_{11}^{-1} & \1 \end{pmatrix}
\begin{pmatrix} U_{11}\, (X+C)\, U_{11}^* & 0 \\ 0 & 0 \end{pmatrix}
\begin{pmatrix} \1 & (U_{11}^*)^{-1}\, U_{21}^* \\ 0 & \1 \end{pmatrix} \:. \eeq
This is of the form~\eqref{Mform0}, and one can even read off~$A$ and~$B$,
\begin{align}
A &= U_{11}\, (X+C)\, U_{11}^* - X \\
B &= \big( U_{11}\, (X+C)\, U_{11}^* \big)\, \big( U_{11}^*)^{-1}\, U_{21}^* \big)\:. 
\end{align}
We conclude that~$\Lambda$ is a bijection to an open neighborhood of~$x \in \F^{p,q}$.
The continuity of~$\Lambda$ and of its inverse are obvious. We have thus constructed a chart
around~$x$.

Performing the above construction around every point~$x \in \F^{p,q}$ gives an atlas. By direct computation,
one verifies that the transition maps are smooth. We conclude that, with the above atlas,
$\F^{p,q}$ is indeed a smooth manifold.

We finally determine its dimension. The linear operator~$B$ is represented by a~$(p+q) \times (f-p-q)$-matrix,
giving rise to~$2 (p+q)(f-p-q)$ real degrees of freedom. The symmetric linear operator~$A$, on the other
hand, is represented by a Hermitian $(p+q) \times (p+q)$-matrix, described by~$(p+q)^2$ real parameters.
Adding these dimensions gives the result.
\QED

We finally remark that the mapping~$ \Lambda$ in~\eqref{LambdaM} gives rise to distinguished
charts, referred to as {\em{symmetric wave charts}}.
\sindex{chart!symmetric wave}%
They have the nice property that they are Gaussian normal coordinates
with respect to the Riemannian metric induced by the Hilbert-Schmidt norm
(for details see~\cite{gaugefix}).

\section{The Spectral Theorem for Selfadjoint Operators} \label{secspectral}
\sindex{spectral theorem!for selfadjoint operators}%
In this book, we will mainly encounter operators of {\em{finite rank}}
(see Definition~\ref{defsymmetric}).
In this case, a symmetric operator~$A$ on a Hilbert space~$(\H, \la .|. \ra_\H)$
(see again Definition~\ref{defsymmetric})
\sindex{operator!of finite rank}%
\sindex{operator!symmetric}%
has real eigenvalues, and there is an orthonormal basis of eigenvectors.
Given an eigenvalue~$\lambda$, we refer to the dimension of the eigenspace as its
{\em{multiplicity}}. 
\sindex{multiplicity of eigenvalue}%
Choosing an orthogonal basis~$e_1, \ldots, e_k$ of this eigenspace, we can
form the {\em{orthogonal projection operator}}~$E_\lambda$ to this eigenspace as
\sindex{orthogonal projection}%
\beq E_\lambda : \H \rightarrow \H \:,\qquad E_\lambda \,u := \sum_{i=1}^k e_i\: \la e_i | u \ra \:. \eeq
Denoting the set of eigenvalues by~$\sigma(A) \subset \R$, the operator~$A$
has the {\em{spectral decomposition}}
\sindex{spectral decomposition}%
\sindex{spectral theorem!for symmetric operators of finite rank}%
\beq \label{specfinite}
A = \sum_{\lambda \in \sigma(A)} \lambda\, E_\lambda \:.
\eeq

We now briefly outline how the spectral decomposition can be generalized to operators
of infinite rank and unbounded operators. In this book, these results will be needed
only in Chapters~\ref{secFSO}--\ref{secperturb} and a few times implicitly in order to
justify that, for example, the unitary time evolution in~\eqref{specdirH} is well-defined.
With this in mind, the reader may skip the remainder of this section in a first reading.
More details on the spectral theorem can be found in textbooks on functional analysis
like, for example, \cite{reed+simon, rudinFA, lax}.
We begin with the case that~$A$ is a {\em{bounded}} and {\em{symmetric}} linear operator
(see Definitions~\ref{defbounded} and~\ref{defsymmetric}).
The {\em{resolvent set}}~$\rho(A) \subset \C$ is defined as the set of all~$\lambda$ for which
the operator~$A-\lambda$ has a bounded inverse, the so-called {\em{resolvent}}~$R_\lambda
:= (A-\lambda)^{-1} \in \Lin(\H)$. 
\sindex{resolvent|textbf}%
The spectrum~$\sigma(A) := \C \setminus \rho(A)$ is
\sindex{spectrum}%
defined as the complement of the resolvent set. The spectrum of a symmetric operator
is always real, $\sigma(A) \subset \R$ (see Exercise~\ref{ex:specreal}). Given a complex polynomial~$p(\lambda)$,
we can form the operator~$p(A)$ by replacing~$\lambda$ with the operator~$A$
(and monomials by powers of the operator). Clearly, this operation is compatible with taking adjoints
and multiplications; that is,
\beq \label{funcrel}
p(A)^* = \overline{p}(A) \qquad \text{and} \qquad p(A)\: q(A) = (pq)(A)
\eeq
for any polynomials~$p$ and~$q$. Moreover, the spectrum and norm of
the operator~$p(A)$ can be expressed easily in terms of the polynomial.
Namely, the {\em{spectral mapping theorem}} 
\sindex{spectral mapping theorem|textbf}%
states that (for details see Exercise~\ref{ex:specmap})
\beq \label{spectralmapping}
\sigma \big( p(A) \big) = p\big( \sigma(A) \big) \:.
\eeq
Moreover, the norm of~$p(A)$ is given by the $C^0$-norm of the polynomial on the spectrum of~$A$,
\beq \label{pA}
\big\| p(A) \big\| = \sup_{\lambda \in \sigma(A)} \big| p(\lambda) \big| \:.
\eeq
\sindex{functional calculus}%
These two properties make it possible to make sense of~$p(A)$ for more general functions~$p$.
Namely, using the Stone-Weierstra{\ss} approximation theorem, the
{\em{continuous functional calculus}} makes it possible to define~$p(A)$ for continuous
functions~$p \in C^0(\R)$. Next, measure-theoretic methods (more precisely, the Riesz
representation theorem to be introduced in Section~\ref{secriesz}) make it possible to
define~$p(A)$ for any bounded Borel function
(a function is called Borel if the pre-image of every open set is a Borel set).
The relations~\eqref{funcrel}, \eqref{spectralmapping} remain valid, whereas~\eqref{pA} becomes an inequality, $\| p(A) \| \leq \sup_{\lambda \in \sigma(A)} | p(\lambda)|$.

The functional calculus for bounded Borel functions makes it possible to choose the function~$p$ in particular
as the characteristic function of a Borel set~$\Omega \subset \R$. We use the notation
\beq E_\Omega := \chi_\Omega(A) \:. \eeq
The mapping that to a Borel set~$\Omega$ associates the operator~$E_\Omega$ is a
{\em{projection-valued measure}}. This means that for every Borel set~$\Omega$,
the operator~$E_\Omega$ is an {\em{orthogonal projection operator}}
in the sense that~$E_\Omega^* = E_\Omega = E_\Omega^2$. 
\sindex{orthogonal projection operator}%
Moreover, for any Borel sets~$U$ and~$V$,
these operators have the properties that~$E_U E_V = E_{U \cap V}$. Finally,
similar to Definition~\ref{defmeasure}, a spectral measure has the property that~$E_\varnothing = 0$
and~$E_\R=\1$, whereas $\sigma$-additivity means that for any sequence of
pairwise disjoint Borel sets~$(\Omega_n)_{n \in \N}$ and every vector~$u \in \H$,
\beq E_{\displaystyle \cup_n \Omega_n} u = \sum_{n=1}^\infty \big( E_{\Omega_n} u \big) \eeq
(where the series converges in the Hilbert space~$\H$; in other words, the
series of operators converges strongly in~$\Lin(\H)$). Being in the measure-theoretic setting has the
major advantage that one can express the functional calculus as an integral; that is,
\sindex{spectral theorem!for bounded symmetric operators}%
\beq \label{speccalcf}
f(A) = \int_\R f(\lambda)\: \dd E_\lambda
\eeq
for any real-valued bounded Borel function~$f$. In particular, the operator~$A$ has the spectral decomposition
\beq \label{specbounded}
A = \int_\R \lambda\: \dd E_\lambda \:.
\eeq
In the case that the spectral measure is supported at a finite number of points,
the integral in~\eqref{specbounded} reduces to a sum, giving us back the spectral decomposition~\ref{specfinite}.
However, in general, the integral~\eqref{specbounded} does not reduce to a sum or series.
Instead, the support of the spectral measure coincides with the spectrum,
\beq \label{sigsupp}
\sigma(A) = \supp E \:,
\eeq
but the spectrum may contain open sets, giving rise to the so-called {\em{continuous spectrum}}.
We remark that the equality in~\eqref{pA} again holds for bounded Borel functions if the supremum is replaced
by the essential supremum (with respect to the spectral measure).
In Exercise~\ref{ex:multop2}, the spectral theorem is illustrated in the example of a multiplication operator.

The above spectral theorem for bounded symmetric operators can be generalized to
{\em{bounded normal}} operators. An operator~$A \in \Lin(\H)$ is called normal if it commutes
with its adjoint,
\sindex{operator!normal}%
\beq [A, A^*] = 0 \:. \eeq
The spectrum of a normal operator is in general complex. The spectral calculus reads
\sindex{spectral theorem!for bounded normal operators}%
\beq \label{specnormal}
f(A) = \int_\C f(\lambda)\: \dd E_\lambda \:,
\eeq
where~$f$ is a complex-valued bounded Borel function on~$\C$ and~$E$ a
spectral measure supported in the complex plane.
The formulas~\eqref{spectralmapping}, \eqref{pA} and~\eqref{sigsupp} continue to hold.
A typical example of a normal operator is a unitary operator (see Definition~\ref{defsymmetric}),
in which case the spectrum lies on the unit circle (see Exercise~\ref{ex:unit}).

Finally, the spectral theorem also applies to {\em{unbounded selfadjoint}} operators, as we now
recall. An unbounded operator~$A$ is not defined on the whole Hilbert space, but only on a dense
subspace~$\D(A) \subset \H$. Thus it is a linear mapping
\beq A \::\: \D(A) \subset \H \rightarrow \H \:. \eeq
The notion of a {\em{symmetric}} operator from Definition~\ref{defsymmetric} extends to
unbounded operators by imposing it only for vectors~$u$ and~$v$ in the domain; that is,
\sindex{operator!symmetric}%
\beq \la A u \,|\, v \ra = \la u \,|\, A v \ra \qquad \text{for all~$u,v \in \D(A)$}\:. \eeq
The operator~$A$ is {\em{selfadjoint}} if the following implication holds,
\sindex{operator!selfadjoint}%
\beq \label{Asa}
\la A u \,|\, v \ra = \la u \,|\, w \ra \quad \text{for all~$u \in \D(A)$}
\qquad \Longrightarrow \qquad v \in \D(A) \quad \text{and} \quad Av=w \:.
\eeq
Clearly, a selfadjoint operator is symmetric. However, the converse is, in general, not true.
Indeed, in order to obtain a selfadjoint operator, one must construct a dense domain
that must be balanced (that is, not too large and not too small) such that the condition on the left
of~\eqref{Asa} implies that~$v$ lies in this domain. In this book, we shall not enter the methods
for the construction of selfadjoint domains. Instead, we assume that a selfadjoint operator~$A$
is given. Then, the spectral theorem yields a spectral measure~$E$ on~$\R$ such that~\eqref{specbounded}
again holds, but with pointwise convergence on the domain; that is,
\sindex{spectral theorem!for unbounded selfadjoint operators}%
\beq %\label{specunbounded}
A u = \int_\R \lambda\: \dd\big( E_\lambda u \big) \qquad \text{for all~$u \in \D(A)$}\:. \eeq
Given a bounded Borel function~$f$ on~$\R$, the spectral calculus is again defined by~\eqref{speccalcf},
giving a bounded operator. If~$f$ is an unbounded Borel function, the spectral calculus~\eqref{speccalcf}
gives an in general unbounded function with dense domain
\beq \D\big(f(A) \big) = \Big\{ u \in \H \:\Big|\: \int_\R |f(\lambda)|^2\: \dd\la u | E_\lambda u \ra < \infty \Big\} \:. \eeq
If~$f$ is real-valued, then the operator~$f(A)$ is again selfadjoint.

\begin{Example} \label{example-multiply} {\bf{(An unbounded multiplication operator)}} {\em{
Generalizing the setting of Exercises~\ref{ex:multop1} and~\ref{ex:multop2}, we let~$g \in C^0(\R, \R)$
be a possibly unbounded, real-valued function. We consider the corresponding
multiplication operator~$A$ on the Hilbert space~$\H=L^2(\R)$; that is,
\beq A:= T_g \::\: \H \rightarrow \H \:,\qquad (A \,\phi)(x) := g(x)\: \phi(x) \:. \eeq
In order to make mathematical sense of this operator, we need to specify a domain. In the first step,
we choose~$\D(A) = C^\infty_0(\R)$ as all smooth test functions.
With this domain, the operator is clearly symmetric, but it is not selfadjoint.
In order to make~$A$ into a selfadjoint operator, we need to choose the domain as
\beq \label{DAmult}
\D(A) := \{ \phi \in L^2(\R)\:|\: g \phi \in L^2(\R) \} \:.
\eeq
Namely, with this choice, the condition on the left of~\eqref{Asa} implies that
$\la u | g v \ra = \la u | w \ra$ for all~$u \in \D(A)$. Using that the domain is dense, it follows
that~$w = gv$, so that the implication in~\eqref{Asa} holds.

Using the relation~$A^2 = T_{g^2}$ inductively, one sees that~$p(A)=T_{p \circ g}$.
Consequently, the functional calculus for the operator~$A$ can be written as~$f(A) = T_{f \circ g}$.
Choosing characteristic functions, one sees that the spectral measure is given by
\beq E_\Omega = T_h \qquad \text{with} \qquad h(x) := \left\{
\begin{array}{ll} 1 & \text{if~$f(x) \in \Omega$} \\ 0 & \text{otherwise}\:. \end{array} \right. \eeq
The support of this spectral measure coincides with the spectrum,
\beq \sigma(A) = \supp E = g(\R)\:. \eeq
If~$g$ is equal to~$\lambda$ on a set of positive Lebesgue-measure, that is, if
\beq \mu\big( g^{-1}(\lambda) \big) > 0 \:, \eeq
then~$\lambda$ is an eigenvalue, and the corresponding eigenspace is given by
\beq \ker (A-\lambda) = \{ \phi \in L^2(R) \:|\: \phi|_{\R \setminus g^{-1}(\lambda)} = 0 \} \:. \eeq
If~$\lambda$ is in the image of~$g$, but the value~$\lambda$ is attained only on a set of measure zero,
then~$\lambda$ lies in the continuous spectrum of~$A$.
More precisely, in this case, the operator~$A-\lambda$ has no kernel (that is, $\lambda$ is not an eigenvalue,
because there are no corresponding eigenvectors in~$\H$), but nevertheless, the operator~$A-\lambda$ has no
bounded inverse.
}} \QEDrem \end{Example}

\begin{Example} {\bf{(The Laplacian in~$\R^3$)}} {\em{
We conclude with an example of a differential operator, namely the Laplacian~$\Delta$ on
the Hilbert space~$L^2(\R^3)$.
\sindex{Laplacian}%
We begin with the simplest domain~$\D(\Delta) = C^\infty_0(\R^3)$ of smooth test functions.
With this domain, the Laplacian is symmetric, but it is not selfadjoint.
One method for obtaining a domain for which~$\Delta$ is selfadjoint is
to take the Fourier transform and use the results for multiplication operators of
the previous Example~\ref{example-multiply}. To this end, it is preferable to choose the
domain as the Schwartz functions, $\D(\Delta) = {\mathcal{S}}(\R^3)$. Taking the Fourier
transform, the Laplacian becomes a multiplication operator; that is, using the notation in Definition~\ref{deffourier},
\beq {\mathcal{F}} \Delta {\mathcal{F}}^* = T_g \qquad \text{with} \qquad g(p) = -|p|^2 \:. \eeq
Similar to~\eqref{DAmult}, the multiplication operator~$T_g$ is selfadjoint with domain
\beq \D\big(T_g \big) = \{ \phi \in L^2(\R^3)\:|\: g \phi \in L^2(\R^3) \} \:. \eeq
Since the Fourier transform is a unitary transformation by Plancherel's theorem
(see Theorem~\ref{thmplancherel}), we obtain a selfadjoint domain of the Laplacian
simply by transforming the domain of~$T_g$ back to position space,
\beq \label{domfourier}
\D(\Delta) = {\mathcal{F}}^* \:\D\big(T_g \big) \:.
\eeq
For the reader familiar with weak derivatives and Sobolev spaces, we remark that this
domain can be expressed more directly as~$\D(\Delta) = L^2(\R^3) \cap W^{2,2}(\R^3)$.
But for the purposes of this book, it suffices to write the domain according to~\eqref{domfourier}. }} \QEDrem
\end{Example}

\section{Exercises}
\begin{Exercise} (Charts of Grassmann manifold) \label{exgrasschart} {\em{
\sindex{Grassmann manifold}%
In this exercise, we want to construct a chart of the Grassmann manifold
of Example~\ref{exgrassmann}. To this end, we consider unit vectors~$u \in \C^f$ with components
\beq u(v) = \big( \sqrt{1-\|v\|^2}, v \big) \qquad \text{with~$v \in \C^{f-1}$ with~$\|v\|<1$} \:. \eeq
We consider the mapping
\beq \label{FGrass}
F \::\: B_1 \subset \C^{f-1} \rightarrow \G \:,\qquad F(v) := |u(v) \ra \la u(v)|
\eeq
(where we used bra/ket-notation; in other words, $F(v)$ is the orthogonal projection to the
span of~$u(v)$). Show that~$F$ is differentiable at the origin and that~$DF|_0$ has maximal rank.
Use this method to construct a differentiable atlas of~$\G$.
Compare the mapping~$\F$ in~\eqref{FGrass} with the mapping~$\Lambda$ in~\eqref{LambdaM}
in the case~$p=1$ and~$q=0$. What are the similarities and differences?
}} \end{Exercise}

\begin{Exercise} \label{ex:pq} {\em{
Let~$x$ be a Hermitian $f \times f$-matrix of rank~$p+q$, which (counting multiplicities)
has~$p$ positive and~$q$ negative eigenvalues. Let~$A$ be another $f \times f$-matrix
(not necessarily Hermitian). Prove the following statements:
\bitem
\item[(a)] The matrix~$A^*xA$ has at most~$p$ positive and at most~$q$ negative eigenvalues. \\
{\em{Hint:}} Consider the maximal positive and negative definite subspaces of the
bilinear forms~$\la .,x. \ra_{\C^f}$ and~$\la .,A^*x A \,. \ra_{\C^f}$. Use that
\beq \la u,A^*x A \,u \ra_{\C^f} = \la (Au) ,x \,(Au) \ra_{\C^f} \:. \eeq
\item[(b)] If~$A$ is invertible, then the matrix~$A^*xA$ has again~$p$ positive and~$q$
negative eigenvalues.
\eitem
}} \end{Exercise}

\begin{Exercise} \label{ex:resolvent} {\em{
Let~$A \in \Lin(H)$. For~$\lambda \in \rho(A)$ in the resolvent set, we define the {\em{resolvent}}~$R_\lambda$
by~$R_\lambda = (A-\lambda)^{-1}$. Prove the so-called {\em{resolvent identity}}
\sindex{resolvent}%
\sindex{resolvent identity}%
\beq R_\lambda \:R_{\lambda'} = \frac{1}{\lambda-\lambda'} \left(
R_\lambda - R_{\lambda'} \right) \:, \eeq
valid for any~$\lambda, \lambda' \in \rho(A)$.
{\em{Hint:}} Multiply by~$A-\lambda$ from the left and by~$A-\lambda'$ from the right.
}} \end{Exercise}

\begin{Exercise} \label{ex:specreal} {\em{
Let~$A \in \Lin(H)$ be a bounded symmetric operator. Show that its spectrum is real,
$\rho(A) \subset \R$. {\em{Hint:}} It might be helpful to prove and make use of
the inequality~$\|(A-\lambda) \,u\|
\geq \im \lambda\,\|u\|$.
}} \end{Exercise}

\begin{Exercise} \label{ex:specmap} {\em{
Prove the spectral mapping theorem~\eqref{spectralmapping} for a polynomial~$p$.
\sindex{spectral mapping theorem}%
{\em{Hint:}} Use that a complex polynomial can be factorized into a product of linear functions.
}} \end{Exercise}

\begin{Exercise} \label{ex:unit} {\em{
Let~$A \in \Lin(\H)$ be unitary.
\sindex{operator!unitary}%
\bitem
\item[(a)] Show that~$A$ is normal.
\item[(b)] Use the spectral calculus~\eqref{specnormal} to show that
\beq \int_\C |\lambda|^2\: \dd E_\lambda = A^* A = \1 \:. \eeq
\item[(c)] Conclude that the spectrum of~$A$ is contained in the unit circle.
\eitem
}} \end{Exercise}

\begin{Exercise} \label{ex:multop2} {{(Spectrum and functional calculus for multiplication operators)}}  {\em{
\sindex{operator!of multiplication}%
We return to the multiplication operators from Exercise~\ref{ex:multop1}.
Let~$f \in C^0(\R, \C)$ be a continuous, complex-valued function. Assume that it is bounded; that is,
that~$\sup_\R |f| < \infty$. We consider the multiplication operator~$T_f$ on the Hilbert space~$\H=L^2(\R)$; that is,
\beq T_f \::\: \H \rightarrow \H \:,\qquad (T_f \,\phi)(x) = f(x)\: \phi(x) \:. \eeq
\bitem
\item[(a)] Show that, if the resolvent exists, it has the form
\beq R_\lambda = (T_f-\lambda)^{-1} = T_g \qquad \text{with} \qquad g(x) = \frac{1}{f(x)-\lambda} \:. \eeq
For which values of~$\lambda$ has the operator~$T_f-\lambda$ a bounded inverse?
What is the spectrum of~$T_f$?
\item[(b)] Are powers of~$T_f$ again multiplication operators? Use your findings
to guess a formula for the continuous spectral calculus~$f(T_a)$.
What is the projection-valued spectral measure~$E$?
\item[(c)] Work out the projection-valued spectral measure explicitly in the example
\beq f(x)=\left\{ \begin{array}{cl} x 
& \text{if~$x \in [0, \frac{1}{2}]$} \\[0.2em]
-x & \text{if~$x \in (\frac{1}{2},1]$} \\[0.2em]
0 & \text{otherwise}\:. \end{array} \right. \eeq
\eitem
}} \end{Exercise}

\chapter{Spinors in Curved Spacetime} \label{Geometry}
This chapter provides a brief introduction to spinors in curved spacetime.
In order to make this book accessible to a broader readership, we mainly restrict
attention to systems in Minkowski space. Nevertheless, many constructions and results
carry over to curved spacetime in a straightforward way. The present section is intended to
provide the necessary background for these generalizations.
The reader not interested in gravity and Lorentzian geometry may skip this section.
More specifically, the results of this chapter will be used only in Sections~\ref{secspincorr},
\ref{seccauchyglobhyp} and~\ref{secwavefront}.
We follow the approach in~\cite{U22}; for other introductions to spinors on manifolds
see. for example, \cite{lawson+michelsohn, friedrich}.

\section{Curved Spacetime and Lorentzian Manifolds}
The starting point for general relativity is the observation that
a physical process involving gravity can be understood in different
ways. Consider for example an observer at rest on earth looking
at a freely falling person (e.g.\ a diver who just jumped from a
diving board). The observer at rest argues that the earth's gravitational
force, which he can feel himself, also acts on the freely falling
person and accelerates him. The person in free fall, on the other hand,
does not feel gravity. He can take the point of view that
he himself is at rest, whereas the earth is accelerated toward him.
He concludes that there are no gravitational fields,
and that the observer on earth merely feels the force of inertia
corresponding to his acceleration.
Einstein postulated that these two points of view should be equivalent
descriptions of the physical process. More generally, it depends on
the observer whether one has a gravitational force or an inertial
force. In other words,
\begin{quote}
{\em{equivalence principle}}: \quad\!\!\!\!
\sindex{equivalence principle|textbf}%
\parbox[t]{8cm}{
no physical experiment can distinguish
between gravitational and inertial forces.} \\[-0em]
\end{quote}
In mathematical language, observers correspond to
coordinate systems, and so the equivalence principle states that the
physical equations should be formulated in general (i.e.\
``curvilinear'') coordinate systems, and should in all these coordinate
systems have the same mathematical structure. This means that the
physical equations should be invariant under diffeomorphisms, and thus
spacetime is to be modeled by a {\em{Lorentzian manifold}}~$(\scrM, g)$.
\sindex{Lorentzian manifold}%
\nindex{abw@$(\scrM, g)$ -- Lorentzian manifold}%

A Lorentzian manifold is ``locally Minkowski space'' in the sense
that at every spacetime point~$p \in \scrM$, the corresponding
{\em{tangent space}}
\sindex{tangent space} %
\nindex{abx@$T_p\scrM$ -- tangent space at~$p$}%
$T_p\scrM$ is a vector space endowed with a scalar product~$\la .,. \ra_p$ of
signature~$(+ \ \!\! - \ \!\! - \ \! - )$. Therefore, we can
distinguish between spacelike, timelike and null tangent vectors.
Defining a non-spacelike curve~$q(\tau)$ by the condition that its
tangent vector~$u(\tau) := \frac{\dd}{\dd \tau} q(\tau) \in T_{q(\tau)}\scrM$ be everywhere
non-spacelike, our definition of the light cone and the notion of
causality given in Section~\ref{SRT} in Minkowski space
immediately carry over to a Lorentzian manifold. In a
coordinate chart, the scalar product~$\la .,. \ra_p$ can be
represented in the form~\eqref{minsp} where~$g_{jk}$ are the components of
the so-called {\em{metric tensor}} (and~$\xi^j$ are~$\eta^k$ are the components of two tangent vectors at~$p$,
again in this coordinate chart). In contrast
\sindex{metric tensor}%
to Minkowski space, the metric tensor is not a constant matrix but
depends on the spacetime point, $g_{jk} = g_{jk}(p)$. Its ten
components can be regarded as the relativistic analogue of Newton's
gravitational potential. For every~$p \in \scrM$ there are coordinate
systems in which the metric tensor coincides with the Minkowski
metric up to second order in a Taylor expansion about the point~$p$,
\beq \label{gnc} g_{jk}(p) =
{\mbox{diag}}(1,-1,-1,-1)\;,\spc
 \partial_j g_{kl}(p) = 0\:.
\eeq
Such {\em{Gaussian normal coordinates}} correspond to the reference frame
\sindex{Gaussian normal coordinates}%
of a ``freely falling observer'' who feels no gravitational forces. However,
it is in general impossible to arrange that also~$\partial_{jk}
g_{lm}(p)=0$. This means that by going into a suitable reference frame,
the gravitational field can be transformed away locally (=in one point),
but not globally. With this in mind, a reference frame corresponding to
Gaussian normal coordinates is also called a {\em{local inertial frame}}.
\sindex{local inertial frame}%

Physical equations can be carried over to 
a Lorentzian manifold by the prescription
that they should in a local inertial frame have the same form as in
Minkowski space; this is referred to as the {\em{strong equivalence
principle}}. 
\sindex{equivalence principle!strong}%
In practice, it amounts to replacing all partial derivatives by
the corresponding {\em{covariant derivatives}}~$\nabla$ of the
Levi-Civita connection of the Lorentzian manifold;
we write symbolically (basics on the covariant derivative can be found, for example, in~\cite{docarmo,
lee-manifold, landau2})
\beq{\textrm{\em{strong equivalence principle}}}: \quad
     \partial \;\longrightarrow\; \nabla \:. \eeq
The gravitational field is described via the curvature of
spacetime. More precisely, the Riemannian {\em{curvature tensor}}
\sindex{curvature!Riemannian}%
is defined by the relations \label{Rijkl}
\beq \label{RT}
R^i_{jkl} \:u^l = \nabla_j \nabla_k u^i - \nabla_k \nabla_j u^i \:.
\eeq
Contracting indices, one obtains the {\em{Ricci tensor}}~$R_{jk} =
R^i_{jik}$ and {\em{scalar curvature}}~$R = R^j_j$.
\sindex{Ricci tensor}%
The relativistic generalization of Newton's gravitational law is
the Ein\-stein equations
\sindex{Einstein equations}%
\beq R_{jk} - \frac{1}{2}\:R\: g_{jk} = 8  \pi \kappa\: T_{jk}\:, \eeq
where~$\kappa$ is the gravitational constant. Here, the {\em{energy-momentum
tensor}}~$T_{jk}$ describes the distribution of matter and energy in
\sindex{energy-momentum tensor}%
spacetime.

\section{The Dirac Equation in Curved Spacetime} \label{secdircurv}
Dirac spinors on a manifold are often formulated using frame bundles, either
an orthonormal frame~\cite{baum, friedrich} or a Newman-Penrose null
frame~\cite{penrose+rindler, chandra}. We here outline an equivalent formulation of
spinors in curved spacetime in the framework of a~$\U(2,2)$ gauge
theory (for details see~\cite{U22}). We begin with constructions
in local coordinates, whereas global issues like topological obstructions
to the existence of spin structures will be discussed in Section~\ref{secspinorbundle}
below. We let~$\scrM$ be a four-dimensional manifold (without Lorentz metric) and
define the {\em{spinor bundle}}~$S\scrM$ as a vector bundle over~$\scrM$ with
\sindex{spinor bundle}%
fiber~$\C^4$. The fibers are endowed with an inner product~$\Sl .|. \Sr$
of signature~$(2,2)$, referred to as the~{\em{spin inner product}}.
\sindex{spin inner product}%
\nindex{aal@$\Sl . \vert . \Sr, \Sl . \vert . \Sr_x$ -- spin inner product}%
\tindex{bb@$\Sl . \vert . \Sr, \Sl . \vert . \Sr_x$ -- spin inner product}%
Sections in the spinor bundle are called {\em{spinors}} or wave functions.
\sindex{spinor!in curved spacetime}%
Choosing local coordinates on~$\scrM$ and local bases~$\{e_\alpha\}_{\alpha=1,\ldots,4}$
of the spin spaces, a wave function is represented by a
four-component complex function in spacetime, usually denoted by~$\psi(x)$.
Choosing at every spacetime point a pseudo-orthonormal basis
$(e_\alpha)_{\alpha=1,\ldots,4}$ in the fibers,
\beq \label{sspgen}
\Sl e_\alpha| e_\beta \Sr = s_\alpha \:\delta_{\alpha \beta}
\:,\qquad s_1=s_2=1,\; s_3=s_4=-1 \:,
\eeq
the spin inner product takes the form~\eqref{f:0c}.
Clearly, such a frame~$\{e_\alpha\}$ is not unique, as one can always perform a transformation according to
\beq e_\alpha \;\longrightarrow\; (U^{-1})^\beta_\alpha\:e_\beta \:, \eeq
where~$U$ is an isometry of the spin inner product, $U \in \U(2,2)$.
\nindex{abz@$\U(2,2)$ -- group of isometries of the spin inner product}%
This basis transformation may depend on the spacetime point.
Under this basis transformation, the spinors behave as follows,
\beq \label{psigauge}
\psi^\alpha(x) \;\longrightarrow\; U^\alpha_\beta(x)\: \psi^\beta(x)\:.
\eeq
In order to simplify the notation in the following computations, we omit the tensor indices and write this
equation with a matrix multiplication,
\beq \label{lgt}
\psi(x) \;\longrightarrow\; U(x)\: \psi(x)
\eeq
(here we implicitly identify the spinor with its concrete realization in a spinor basis).
In view of the analogy to gauge theories, we interpret the transformation law~\eqref{lgt} of the
wave functions as a local gauge transformation with gauge group
${\mathcal{G}}=\U(2,2)$. We refer to a choice of the
spinor basis~$(e_\alpha)$ as a {\em{gauge}}.
\sindex{gauge}%
\sindex{local gauge freedom!$\U(2,2)$}%

Our next goal is to formulate classical Dirac theory in
such a way that the above~$\U(2,2)$ gauge transformations correspond to
a physical symmetry, the {\em{$\U(2,2)$ gauge symmetry}}.
\sindex{gauge symmetry!$\U(2,2)$}%
To this end, we shall consider a Dirac-type operator as the basic
object on~$\scrM$, from which we will then deduce the Lorentz metric and the
gauge potentials. First we define a {\em{differential operator of first order}} as a
linear operator~$\Dir$ on wave functions which in any local chart and gauge takes the form
\beq
\label{4_6}
\Dir = \cI G^j(x) \frac{\partial}{\partial x^j} + B(x)
\eeq
with suitable~$(4 \times 4)$-matrices~$G^j(x)$ and~$B(x)$.
This definition does not depend on a choice of coordinates and gauge, although
the specific form of the matrices~$G^j(x)$ and~$B(x)$ clearly does. More precisely, if we change to other coordinates~$\tilde{x}^i$ while keeping the gauge fixed, a short computation shows that the operator~\eqref{4_6} transforms to
\beq \label{ctrans}
\cI \left( G^k(x)\: \frac{\partial \tilde{x}^j}{\partial x^k} \right)
\frac{\partial}{\partial \tilde{x}^j} + B(\tilde{x})\:.
\eeq
On the other hand, if we perform a gauge transformation~$\psi \to U \psi$ but keep the coordinates~$x^i$ fixed, 
the Dirac operator transforms according to
\begin{align}
\Dir\psi \;\longrightarrow\; U \Dir \psi = U \Dir U^{-1}(U\psi) \:,
\end{align}
and a straightforward computation gives
\beq \label{gtrans}
U \Dir U^{-1} = \cI \left( U G^j U^{-1} \right)
\frac{\partial}{\partial x^j} + \left(
U B U^{-1} + \cI U G^j (\partial_j U^{-1}) \right) .
\eeq
We now define an operator of Dirac type by the requirement that by choosing suitable
coordinates and gauge, one can arrange that the coefficient matrices~$G^j$ of the
partial derivatives ``coincide locally'' with the Dirac matrices of Minkowski space.
\begin{Def}
\label{def1n}
A differential operator~$\Dir$ of first order on a spinor bundle is called {\bf{operator of Dirac type}}
(or Dirac-type operator) if for every~$p \in \scrM$ there is a chart~$(x^i, U)$ around~$p$
\sindex{operator!of Dirac type}%
and a gauge~$(e_\alpha)_{\alpha=1,\ldots, 4}$ on~$U$ such that~$\Dir$ is
of the form~\eqref{4_6} with
\beq
\label{4_21}
G^j(p) = \gamma^j \:,
\eeq
where the~$\gamma^j$ are the Dirac matrices of Minkowski space in the
Dirac representation~\eqref{Dirrep}.
\end{Def}

It may seem unconventional that we defined Dirac-type operators without
having a covariant derivative on the spinor bundle yet.
An advantage of our approach is that the condition~\eqref{4_21} is natural and can be understood
physically as an extension of the equivalence principle which incorporates gauge transformations
(not just coordinate transformations). As we shall see, the above Dirac operator includes both
gravitational fields and electromagnetic potentials, giving a
unified description of electrodynamics and general relativity as a gauge theory.
More concretely, from a Dirac-type operator we shall construct a gauge-covariant
derivative~$D$, also referred to as the corresponding {\em{spin derivative}} or {\em{gauge-covariant derivative}}.
In preparation, we write the transformation law~\eqref{psigauge} in the shorter form
\beq \label{lgtshort}
\psi(x) \;\longrightarrow\; U(x)\: \psi(x)
\eeq
with~$U \in \U(2,2)$.
Clearly, partial derivatives of~$\psi$ do not have a nice behavior under
gauge transformations because we pick up derivatives
of~$U$. This problem disappears if instead of partial derivatives we consider
{\em{gauge-covariant derivatives}}
\sindex{gauge-covariant derivative}
\beq \label{gcd}
D_j = \partial_j - \cI A_j \:,
\eeq
where the gauge potential~$A_j$ is pointwise linear operator on spinor (i.e.\ a~$(4 \times 4)$-matrix valued function if represented in a gauge and coordinate system). Provided that
these gauge potentials~$A_j$ transform under a gauge transformation~$U$ as
\beq \label{Atrans}
A_j \;\longrightarrow\; U A_j U^{-1} + \cI U\:(\partial_j U^{-1})\:,
\eeq
a short calculation shows that then the gauge-covariant derivative behaves
under gauge transformations according to
\beq \label{Dlgt}
D_j \;\longrightarrow\; U\:D_j \:U^{-1} \:,
\eeq
and thus the gauge-covariant derivatives of~$\psi$ obey the simple transformation
rule
\beq D_j \psi \;\longrightarrow\; U\: D_j \psi\:. \eeq
Thus our task is to find a way to construct matrices~$A_j$ that transform under local
gauge transformations according to~\eqref{Atrans}.
This construction will also reveal the structure of the matrix~$B$,
and this will finally lead us to the definition of the
Dirac operator, which involves precisely the gravitational and electromagnetic fields.

Before we come to this construction, we first explain how a Dirac-type operator induces a Lorentzian metric.
In a chart and gauge where~\eqref{4_21} holds in a point~$p$, it is obvious
from~\eqref{f:0b} that the anti-commutator of the matrices~$G^j(p)$ gives
the Minkowski metric. Transforming from this chart to a general coordinate system and gauge
using the transformation rules~\eqref{ctrans} and~\eqref{gtrans}, one sees that in a
general coordinate system and gauge, the anti-commutator of these matrices defines a Lorentzian metric,
\beq
\label{4_4}
g^{jk}(x) \:\1 = \frac{1}{2} \: \{ G^j(x),\: G^k(x) \} \: .
\eeq
Moreover, using that the Dirac matrices in Minkowski space are symmetric
with respect to the spin inner product in Minkowski space (see~\eqref{gammasymmMin}),
one sees (again from the transformation laws~\eqref{ctrans} and~\eqref{gtrans}) that the same is true for the matrices~$G^j$,
\beq \label{gammasymm}
\Sl G^l \psi \,|\, \phi \Sr =
\Sl \psi \,|\, G^l \phi \Sr \qquad {\mbox{for all~$\psi, \phi$}} \:.
\eeq
Thus we see that, via~\eqref{4_4}, a Dirac-type operator induces on the manifold a
Lorentzian structure. We refer to the matrices~$G^j$ as the Dirac matrices in curved spacetime.
Since we can arrange by a choice of coordinate system and gauge that these matrices coincide at any
given spacetime point with the Dirac matrices of Minkowski space, all the familiar relations involving
Dirac matrices generalize in an obvious way to curved space time.
In particular, the pseudo-scalar operator~$\Gamma(x)$ can be introduced in curved spacetime
by writing~\eqref{rhodef} in a coordinate and gauge-covariant form as
\sindex{pseudo-scalar matrix}%
\nindex{aaq@$\pseudo$ -- pseudo-scalar matrix}%
\beq \label{Gammaxdef}
\Gamma(x) = \frac{1}{4!} \:\varepsilon_{jklm}\, G^j(x)\, G^k(x)\, G^l(x)\, G^m(x) \:,
\eeq
where the anti-symmetric tensor~$\varepsilon_{jklm} = \sqrt{|\det g|}\: \epsilon_{jklm}$ differs from
the anti-symmetric symbol~$\epsilon_{jklm}$ by the volume density.
The pseudo-scalar operator gives us again the notion of even and odd
matrices and of chirality~\eqref{chidef}. Furthermore, we introduce the
{\em{bilinear matrices}}~$\sigma^{jk}$ by
\beq \sigma^{jk}(x) = \frac{\cI}{2} \: \big[ G^j(x),\: G^k(x) \big] \: . \eeq
As in Minkowski space, at any spacetime point the matrices
\beq
    G^j\;,\;\;\;\; \pseudo G^j \;,\;\;\;\; \1 \;,\;\;\;\; \cI \pseudo
    \;,\;\;\;\; \sigma^{jk}
    \label{5_10}
\eeq
form a basis of the 16-dimensional (real) vector space
of symmetric matrices (symmetric with respect to the spin inner product~$\Sl .|. \Sr$).
The matrices~$G^j$ and~$\pseudo G^j$ are odd, whereas
$\1$, $\cI\pseudo$ and $\sigma^{jk}$ are even.

We are now ready for the construction of the spin connection from a Dirac-type operator.
The Lorentzian metric~\eqref{4_4} induces the Levi-Civita connection
\sindex{Levi-Civita connection}%
\nindex{acb@$\nabla$ -- Levi-Civita connection}%
$\nabla$ on the tangent bundle, which further induces the Levi-Civita covariant derivative on arbitrary tensor
fields (for basics on the Levi-Civita connection see again~\cite{docarmo, lee-manifold}).
Taking covariant derivatives of the Dirac
matrices by the formula~$\nabla_k G^j = \partial_k G^j + \Gamma^j_{kl} \:G^l$, where~$\Gamma^j_{kl}$ are the Christoffel symbols of the Levi-Civita connection,
we obtain an expression that behaves under coordinate transformations
like a tensor. However, it is not gauge covariant, because
a gauge transformation~\eqref{lgtshort} yields contributions
involving first derivatives of~$U$. More precisely, according to~\eqref{gtrans},
\begin{align}
    \nabla_k G^j \;\longrightarrow\; \nabla_k (U G^j U^{-1}) & =
    U (\nabla_k G^j) U^{-1} + (\partial_k U) G^j U^{-1} +
    U G^j (\partial_k U^{-1}) \nonumber \\
&=U (\nabla_k G^j) U^{-1} - \big[ U (\partial_k U^{-1}), U G^j U^{-1} \big]
    \label{5_5a}
\end{align}
(in the last line, we used the relation~$\partial_j U = - U (\partial_j U^{-1}) U$; for details, see
Exercise~\ref{exUdiff}).

We can use the second summand in~\eqref{5_5a} to partially fix the gauge.
\begin{Lemma}
\label{lemma1}
For every spacetime point~$p \in \scrM$, there is a gauge such that
\beq
    \nabla_k G^j(p) = 0 \qquad \text{for all indices~$j,k$}\:.
    \label{5_5}
\eeq
\end{Lemma}
\Proof
We start with an arbitrary gauge and construct the desired gauge with
two subsequent gauge transformations:
\begin{itemize}[leftmargin=2em]
\item[(i)]
The matrix~$\partial_j \pseudo$ is odd, because
\beq 0 = \partial_j \1 = \partial_j (\pseudo \pseudo) =
(\partial_j \pseudo) \pseudo + \pseudo (\partial_j \pseudo) \: . \eeq
As a consequence, the matrix~$i \pseudo (\partial_j \pseudo)$ is symmetric
(with respect to the spin inner product). Note that for any symmetric matrix~$A$
the family of operators~$U(q) = \exp(-i A (q-p))$ is unitary and~$U(p)=1$, $\partial_j U(p)=-i A$.
Therefore, we can perform a gauge
transformation~$U$ with~$U(p)=\1$ and~$\partial_j U(p)=\frac{1}{2} \pseudo
(\partial_j \pseudo)$. In the new gauge, the matrix~$\partial_j \pseudo(p)$ vanishes, as one checks by the computation
\beq \partial_j \pseudo_{|p} \;\longrightarrow\; \partial_j (U \pseudo U^{-1})_{|p}
= \partial_j \pseudo_{|p} + \frac{1}{2} \left[ \pseudo (\partial_j \pseudo), \:
\pseudo  \right]_{|p} = \partial_j \pseudo_{|p} - \pseudo^2 (\partial_j
\pseudo)_{|p} = 0 \: . \eeq
Differentiating the relation~$\{ \pseudo, G^j \}=0$, one sees that
the matrix~$\nabla_k G^j_{|p}$ is odd (in the sense that it anti-commutes with~$\Gamma$,
exactly as defined in Minkowski space after~\eqref{chiLRrel}). We can thus represent it in the form
\beq
    \nabla_k G^j_{|p} = \Lambda^j_{km} \:G^m_{|p} +
\Theta^j_{km} \: \pseudo_{|p} G^m_{|p}
    \label{5_7}
\eeq
    with suitable coefficients~$\Lambda^j_{km}$ and~$\Theta^j_{km}$.

    This representation can be further simplified: According to Ricci's
    Lemma, $\nabla_n g^{jk}=0$. Expressing the metric via the
    anti-commutation relation~\eqref{4_4} and using the Leibniz rule, we get
\begin{align}
    0 & =  \{ \nabla_n G^j,\:G^k\} + \{ G^j,\:
    \nabla_n G^k\} \nonumber \\
     & = 2 \Lambda^j_{nm} \: g^{mk} - \Theta^j_{nm} \: 2\cI\pseudo
\sigma^{mk}
     + 2 \Lambda^k_{nm} \: g^{mj} - \Theta^k_{nm} \: 2\cI\pseudo
\sigma^{mj} \:, \label{5_9}
\end{align}
    and thus
\beq
    \Lambda^j_{nm} \:g^{mk}_{|p} = -\Lambda^k_{nm} \: g^{mj}_{|p} \:.
    \label{5_8}
\eeq
    Further, in the case~$j=k \neq m$, the relation~\eqref{5_9} yields that~$\Theta^j_{nm}=0$.
    For~$j \neq k$, we obtain~$\Theta^j_{nj} \:\sigma^{jk} +
    \Theta^k_{nk} \:\sigma^{kj} = 0$ and thus~$\Theta^j_{nj} =
\Theta^k_{nk}$
    ($j$ and~$k$ denote fixed indices, no summation is performed).
    We conclude that there are coefficients~$\Theta_n$ with
\beq
    \Theta^j_{nk} = \Theta_n \: \delta^j_k \: .
    \label{5_9a}
\eeq
    \item[(ii)] We perform a gauge transformation~$U$ with~$U(p)=\1$ and
\beq \label{parU}
\partial_k U = -\frac{1}{2} \:\Theta_k \: \pseudo - \frac{\cI}{4} \:
\Lambda^m_{kn} \: g^{nl} \: \sigma_{ml}
\eeq
(the existence of such a gauge transformation follows exactly as explained at the beginning of~(i),
using that the matrix on the right side of~\eqref{parU} is anti-symmetric with respect to the spin inner product).
    Using the representation~\eqref{5_7} together with~\eqref{5_8} and~\eqref{5_9a},
    the matrix~$\nabla_k G^j$ transforms into
\begin{align}
\nabla_k G^j \longrightarrow & \nabla_k G^j + [\partial_k U,\:
G^j] \notag \\
     & =\; \Lambda^j_{km} \:G^m + \Theta_k \: \pseudo G^j
     - \Theta_k \: \pseudo G^j -
     \frac{\cI}{4}\: \Lambda^m_{kn} \:g^{nl} \: [\sigma_{ml},\: G^j] \notag \\
     & =\; \Lambda^j_{km} \:G^m - \frac{\cI}{4}\: \Lambda^m_{kn}
     \:g^{nl} \:2\cI \: (G_m \: \delta^j_l - G_l \:\delta^j_m) \notag \\
     & =\; \Lambda^j_{km} \:G^m + \frac{1}{2}\: \Lambda^m_{kn} \:
     g^{nj} \: G_m - \frac{1}{2}\: \Lambda^j_{km} \: G^m = 0 \: .
\end{align}
\end{itemize}
This concludes the proof.
\QED

We call a gauge satisfying condition~\eqref{5_5} a {\em{normal
gauge}} around~$p$. In order to analyze the remaining gauge freedom, we let~$U$ be a
%%\sindex{gauge!normal}
transformation between two normal gauges. Then according to~\eqref{5_5a}
and~\eqref{5_5}, the commutator
$[U (\partial_k U^{-1}),\: UG^jU^{-1}]$ vanishes at~$p$ or, equivalently,
\beq [\cI (\partial_k U^{-1})\:U,\: G^j]_{|p} = 0 \:. \eeq
As is easily verified in the basis~\eqref{5_10} using the
anti-commutation relations, a matrix that
commutes with all Dirac matrices is a multiple of the identity matrix.
Moreover, the matrix~$\cI (\partial_j U^{-1}) \:U$ is symmetric because
\beq
    (\cI (\partial_j U^{-1}) \:U)^* = -\cI U^{-1}\:(\partial_j U)
= -\cI \partial_j (U^{-1} U) + \cI (\partial_j U^{-1}) \:U
= \cI (\partial_j U^{-1}) \:U \: .
\eeq
We conclude that the matrix
$\cI (\partial_j U^{-1}) \:U$ is a real multiple of the identity matrix.
Transforming it unitarily with~$U$, we see that it also
coincides with the matrix~$\cI U(\partial_j U^{-1})$.
Restricting attention to normal gauges, it is easy to find
expressions with the required behavior~\eqref{Atrans} under
gauge transformations. Namely, setting
\beq \label{adef}
a_j = \frac{1}{4}\: \re \Tr \big(G_j \,B \big) \: \1 \:,
\eeq
where ``Tr'' denotes the trace of a $4\times 4$-matrix,
one sees from~\eqref{gtrans} that
\beq a_j  \;\longrightarrow \;
a_j + \frac{1}{4}\: {\mbox{Re }} \Tr \left( G_j G^k \: i(\partial_k
U^{-1})\:U \right) \1 = a_j + \cI U (\partial_j U^{-1}) \: . \eeq
We can identify the~$a_j$ with the gauge potentials~$A_j$ and
use~\eqref{gcd} as the definition of the spin connection.
\begin{Def} \label{def_sd}
The {\bf{spin derivative}}~$D$ is defined by the condition that
\sindex{spin derivative}%
\nindex{acc@$D$ -- spin derivative}%
it behaves under gauge transformations~\eqref{lgtshort} according to~\eqref{Dlgt},
and that in normal gauges around~$p$ it has the form
\beq
    D_j(p) = \frac{\partial}{\partial x^j} - \cI a_j
    \label{5_11}
\eeq
with the potentials~$a_j$ according to~\eqref{adef}.
\end{Def}
In general gauges, the spin derivative can be written as
\beq
    D_j = \frac{\partial}{\partial x^j} - \cI E_j - \cI a_j
    \label{5_12}
\eeq
with additional matrices~$E_j(x)$, which involve the Dirac matrices and
their first derivatives. The components of~$E_j$ are sometimes referred to as {\em{spin coefficients}}.
\sindex{spin coefficient|textbf}%
The spin coefficients can be regarded in analogy to the Christoffel symbols of the
Levi-Civita connection, but now they act on spinors, not on vector fields.
A short calculation shows that the trace of the matrix~$E_j$ does
not change under gauge transformations, and since it vanishes in
normal gauges, we conclude that the matrices~$E_j$ are trace-free.
It is straightforward to verify that they are given explicitly by
(for details see Appendix~\ref{Anhang_eichfix})
\beq \label{Ejform}
E_j = \frac{\cI}{2}\: \pseudo\: (\partial_j \pseudo) - \frac{\cI}{16}\: \Tr
    (G^m \:\nabla_j G^n) \: G_m G_n + \frac{\cI}{8}\: \Tr (\pseudo G_j \:
    \nabla_m G^m) \:\pseudo \:.
\eeq

In the next two theorems we collect the basic properties of the spin connection.
\begin{Thm}
\label{thm_2}
The spin derivative satisfies for all wave
functions~$\psi, \phi$ the equations
\begin{gather}
\big[ D_k, G^j \big] + \Gamma^j_{kl} \: G^l = 0 \label{5_13} \\
 \partial_j \Sl \psi | \phi \Sr = \Sl D_j \psi \,|\, \phi \Sr + \Sl \psi \,|\, D_j \phi \Sr \:.    \label{5_17}
\end{gather}
\end{Thm}
\Proof
The left side of~\eqref{5_13} behaves under gauge transformations
according to the adjoint representation~$. \to U \:.\: U^{-1}$ of the gauge group.
Therefore, it suffices to check~\eqref{5_13} in a normal gauge, where
\beq [D_k,G^j] + \Gamma^j_{kl} \: G^l = \nabla_k G^j -
\frac{\cI}{4}\: \re \Tr \big(G_j B \big) \: [\1,G^j] = 0 \: . \eeq

Since both sides of~\eqref{5_17} are gauge invariant, it again suffices to
consider a normal gauge. The statement is then an immediate consequence
of the Leibniz rule for partial derivatives and the fact that the spin derivative differs from the partial derivative by
an imaginary multiple of the identity matrix~\eqref{5_11}.
\QED
The identity~\eqref{5_13} means that the coordinate and gauge invariant
derivative of the Dirac matrices vanishes.
The relation~\eqref{5_17} shows that the spin connection is
compatible with the spin inner product.

We define the {\em{curvature}}~${\mathcal{R}}$ of the spin connection
\sindex{curvature!of spin connection}%
as the following two-form,
\beq {\mathcal{R}}_{jk} = \frac{\cI}{2} \: [D_j, D_k] \:. \eeq
\begin{Thm}
\label{thm2}
The spin connection satisfies the relation
\beq [D_j, G_k] = [D_k, G_j] \:. \eeq
Moreover, curvature has the form
\beq {\mathcal{R}}_{jk} = \frac{1}{8} \: R_{mnjk} \:\sigma^{mn} + \frac{1}{2} \:
    (\partial_j a_k - \partial_k a_j) \:, \eeq
where~$R_{mnjk}$ is the Riemannian curvature tensor and
the~$a_j$ are given by~\eqref{adef}.
\end{Thm}
\Proof The identity~\eqref{5_13} yields
\beq \big[ D_j, G_k \big] = \big[D_j, g_{kl} \, G^l \big] = (\partial_j g_{kl})
\: G^l -g_{kl} \: \Gamma^l_{jm} \: G^m = \Gamma^m_{jk} \: G_m \:. \eeq
Thus, using that the Levi-Civita connection is torsion-free, we obtain
\beq [D_j, G_k] = [D_k, G_j] = \big(\Gamma^m_{jk} - \Gamma^m_{kj} \big) \:
G_m = 0 \: . \eeq

Next, again using~\eqref{5_13}, we can rewrite the covariant derivative as a
spin derivative,
\beq G_l\: \nabla_k u^l = \big[ D_k, G_l u^l \big]\:. \eeq
Iterating this relation, we can express the Riemann tensor~\eqref{RT} by
\begin{align}
G_i \: R^i_{jkl} \:u^l &=
\big[ D_j, [D_k, G_l u^l ] \big] - \big[ D_k, [D_j, G_l u^l ] \big] \notag \\
& = \big[ [D_j, D_k], G_l u^l \big] = -2i \: \big[{\mathcal{R}}_{jk}, G_l u^l \big] \: .
\end{align}
This equation determines curvature up to a multiple of the identity 
matrix; that is,
\beq {\mathcal{R}}_{jk}(x) = \frac{1}{8} \: R_{mnjk} \:\sigma^{mn} + \lambda_{jk} \1 \eeq
with unknown parameters~$\lambda_{jk}$. These parameters can be determined by computing
the trace of curvature. Since the matrices~$E_j$ in~\eqref{Ejform} and their partial derivatives are
trace-free, a direct computation starting from~\eqref{5_12} gives
\beq \lambda_{jk} = \frac{1}{4}\: \Tr ({\mathcal{R}}_{jk})\:\1 = \frac{1}{8}\:
\Tr \big( \partial_j A_k - \partial_k A_j \big)\:\1  = \frac{1}{2} \: \big(\partial_j a_k - \partial_k a_j \big) \:, \eeq
concluding the proof.
\QED

We come to the physical interpretation of the above construction.
According to Lemma~\ref{lemma1} we can choose a gauge around~$p$ such that the
covariant derivatives of the Dirac matrices vanish at~$p$.
Moreover, choosing normal coordinates and making a global (=constant) gauge
transformation, we can arrange that~$G(p) = \gamma^j$
and~$\partial_j g_{kl}(p)=0$. Then the covariant derivatives at~$p$
reduce to partial derivatives, and we conclude that
\beq \label{nrf}
G^j(p) = \gamma^j \;,\spc \partial_k G^j(p) = 0 \:.
\eeq
These equations resemble the conditions for normal
coordinates~\eqref{gnc}, except that the role of the metric
is now played by the Dirac matrices. Indeed, by differentiating~\eqref{4_4} one sees
that~\eqref{nrf} implies~\eqref{gnc}. Therefore, \eqref{nrf} is a
stronger condition that poses a condition not only for the coordinates
but also for the gauge. We call a coordinate system and gauge where~\eqref{nrf}
is satisfied a {\em{normal reference frame}} around~$p$.
\sindex{reference frame!normal}%

In a normal reference frame around~$p$, the Dirac matrices, and via~\eqref{4_4} also
the metric, are the same as in Minkowski space up to the order~$\sim(x-p)^2$. 
Combining the equivalence principle with the usual minimal coupling procedure in
physics, it seems a sensible physical assumption that the Dirac equation
at~$p$ should coincide with that in Minkowski space. This implies that there should
be a normal gauge such that all gauge potentials vanish at~$p$, and thus the
Dirac operator at~$p$ should coincide with the vacuum Dirac operator~$i \Pdd$.
This physical argument makes it possible to specify the zero order term in~\eqref{4_6}.

\begin{Def}
\label{def_pdo}
A Dirac-type operator~$\Dir$ is called {\bf{Dirac operator}}
\sindex{Dirac operator!in curved spacetime}%
if for any~$p \in \scrM$ there is a normal reference frame around~$p$ such that~$B(p)=0$.
\end{Def}
Equivalently, the Dirac operator could be defined as a
differential operator of first order~\eqref{4_6} with the additional
structure that for any~$p \in \scrM$ there is a coordinate chart and gauge
such that the following three conditions are satisfied,
\beq G^j(p) = \gamma^j \;,\spc \partial_k G^j(p) = 0 \;,\spc B(p) = 0\:. \eeq
This alternative definition has the disadvantage that it is a-priori not
clear whether the second condition~$\partial_k G^j(p) = 0$
can be satisfied for a general metric. This is the reason why we
preferred to begin with only the first condition
(Definition~\ref{def1n}), then showed that the second condition can be
arranged by choosing suitable coordinates and gauge, and satisfied the
third condition at the end (Definition~\ref{def_pdo}).

In general coordinates and gauge, the Dirac operator can be
written as
\beq \label{Dirphys}
\Dir = \cI G^j D_j = \cI G^j\, \big(\partial_j - \cI E_j - \cI a_j \big) \:,
\eeq
where~$D$ is the spin connection of Definition~\ref{def_sd}.
The matrices~$E_j$ take into account the gravitational field and are
called {\em{spin coefficients}}, whereas the~$a_j$ can be identified with
the {\em{electromagnetic potential}} (compare~\eqref{Dirac2}).
We point out that the gravitational field cannot be introduced into
the Dirac equation by the simple replacement rule~$\partial \rightarrow D$,
because gravity has an effect on both the Dirac matrices and the
spin coefficients. But factorizing the gauge group as
$\U(2,2) = \U(1) \times \SU(2,2)$, the~$\SU(2,2)$-gauge transformations
are linked to the gravitational field because they
influence~$G^j$ and~$E_j$, whereas the~$\U(1)$ can be identified with
the gauge group of electrodynamics. In this sense, we obtain a
unified description of electrodynamics and general relativity as a~$\U(2,2)$
gauge theory. The Dirac equation
\sindex{Dirac equation!in curved spacetime}
\beq \label{diracgrav}
(\Dir-m)\: \psi = 0
\sindex{Dirac equation!in the gravitational field}
\eeq
describes a Dirac particle in the presence of a gravitational and electromagnetic
field. According to Theorem~\ref{thm2}, the curvature of the spin
connection involves both the Riemann tensor and the electromagnetic field
tensor. One can express the classical action in terms of these tensor
fields, so that the corresponding Euler-Lagrange equations give rise to
the classical Einstein-Dirac-Maxwell equations.

For the probabilistic interpretation of the Dirac equation in
curved spacetime, we choose a spacelike hypersurface~$\scrN$
(corresponding to ``space'' for an observer) and consider
in generalization of~\eqref{printMink} on solutions of the Dirac
equation the scalar product
\beq \label{nicsp}
(\psi | \phi)_\scrN = \int_{\scrN} \Sl \psi \,|\, G^j \nu_j\,
\phi \Sr \: \dd\mu_{\scrN} \:,
\eeq
\nindex{acd@$(. \vert . )_\scrN$ -- scalar product on Dirac solutions in curved spacetime}%
\tindex{bb@$(. \vert . )_\scrN$ -- scalar product on Dirac solutions in curved spacetime}%
\sindex{scalar product!on Dirac solutions in curved spacetime}%
where~$\nu$ is the future-directed normal on~${\scrN}$
and~$d\mu_{\scrN}$ is the invariant measure on the Riemannian
manifold~${\scrN}$ (in order to ensure that the integral exists, one can, for example,
restrict attention to solutions of spatially compact support, as will be introduced
after Theorem~\ref{thmcauchy2} below). Then~$(\psi | \psi)_\scrN$ is the {\em{normalization integral}}, which
we again normalize to one.
Its integrand has the interpretation as the {\em{probability density}}.
\sindex{probability density!in curved spacetime}%
\sindex{Dirac current!in curved spacetime}%
In analogy to~\eqref{dc} the {\em{Dirac current}} is introduced
by~$J^k = \Sl \psi \:|\: G^k \psi \Sr$. Using Theorem~\ref{thm_2} one
sees similar as in Minkowski space that the Dirac current is
divergence-free, $\nabla_k J^k = 0$. From the Gau{\ss} divergence
theorem, 
\sindex{Gau{\ss} divergence theorem!for Dirac current}%
\Evtl{Hier Referenz und/oder \"Ubungsaufgabe.}%
one obtains that the scalar product~\eqref{nicsp} does not
depend on the choice of the hypersurface~${\scrN}$ (this follows similarly as explained
in Minkowski space after~\eqref{cc}).

In analogy to~\eqref{stipMin}, we can introduce the {\em{spacetime inner product}}
\sindex{spacetime inner product}%
\nindex{aap@$\bra . \vert . \ket$ -- spacetime inner product}%
\tindex{bb@$\bra . \vert . \ket$ -- spacetime inner product}%
\beq \label{stipintro}
\bra \psi|\phi \ket := \int_\scrM \Sl \psi | \phi \Sr_x\: \dd\mu_\scrM\:.
\eeq
in which the wave functions (which need not satisfy the Dirac
equation but must have a suitable decay at infinity)
are integrated over all of spacetime.
We finally remark that, using Theorem~\ref{thm_2} together with
Gau{\ss}' divergence theorem, one easily verifies that the
Dirac operator is symmetric with respect to this inner product.

\section{Computation of the Dirac Operator} \label{secdiracop}
We now explain how the Dirac operator
can be computed in an efficient way in a given spacetime.
Thus suppose that the Lorentzian metric~$g_{ij}$ is given in a chosen chart.
The general procedure is to first choose matrices~$G^j(x)$ which are symmetric
with respect to the spin inner product~\eqref{gammasymm}
(where in our formulation, the spin inner product is always given by~\eqref{sspgen})
and which satisfy the anti-commutation relations~\eqref{4_4}.
Then, the spin coefficients as given by~\eqref{Ejform} are obtained by
a straightforward computation. Then the spin derivative is given by~\eqref{5_12}
(where~$a_j$ are the components of the electromagnetic potential;
they are set to zero if no electromagnetic field is present).
The Dirac operator is given by~\eqref{Dirphys}; that is,
\beq \label{Dirphys2}
\Dir = \cI G^j D_j = \cI G^j \partial_j + G^j E_j + G^j a_j \:.
\eeq
In this construction, one has a lot of freedom to choose the Dirac matrices~$G^j(x)$
(as described systematically by the $\U(2,2)$-gauge transformations~\eqref{lgtshort}
and~\eqref{Dlgt}).
It is a promising strategy to use this gauge freedom such as to
choose Dirac matrices for which the formulas for the spin coefficients~\eqref{Ejform}
become as simple as possible. Moreover, one should keep in mind that
for the computation of the Dirac operator, one does not need to know all the
matrices~$E_j$, but it suffices to compute the combination~$G^j E_j$ in~\eqref{Dirphys2}.
Indeed, in many spacetimes of physical interest,
making use of the gauge freedom, the combination~$G^j E_j$ can be
computed easily (for details see the computations in black hole geometries in~\cite{FSYperiodic, kerr} or
various examples in~\cite[Section~9]{topology}).
We here illustrate this method by the example of a diagonal metric, in which case it
is even unnecessary to compute the Christoffel symbols:

\begin{Prp} \label{prpDOP} Assume that there is a local chart~$(x^i)$ in which the metric is diagonal; that is,
\beq \label{gdiagonal}
ds^2 = \sum_{i=0}^3 g_{ii}(x)\, \Diff x_i^2 \:.
\eeq
Then there is a gauge in which the Dirac operator (without electromagnetic field) takes the form
\beq \label{Dirac}
\Dir = \cI G^j \frac{\partial}{\partial x^j} + B \:,
\eeq
where
\begin{align}
G^j(x) &= g_{jj}(x)^{-\frac{1}{2}}\: \gamma^j \label{Gchoice} \\
B(x) &= \frac{\cI}{2 \sqrt{|\det g|}}\: \partial_j \left( \sqrt{|\det g|}\, G^j \right) . \label{Bchoice}
\end{align}
(here~$\gamma^j$ are again the Dirac matrices in Minkowski space).
\end{Prp}
\Proof With~\eqref{Gchoice} we have satisfied the anti-commutation relations
\beq \{G^j, G^k \} = 2\, g^{jk}\, \1 \:. \eeq
Moreover, the choice~\eqref{Gchoice} ensures that the pseudo-scalar operator is constant,
and that all derivatives of the~$G^j$ are in the span of~$\gamma^0, \ldots, \sigma^k$.
Therefore, the formula for the zero order term in the Dirac operator~\eqref{Dirphys2}
simplifies to
\beq \label{Bform}
B =  -\frac{\cI}{16}\: \Tr \big( G_m \,(\nabla_j G_n) \big) \: G^j G^m G^n \:,
\eeq
where~$\nabla_j G_n \equiv \partial_j G_n - \Gamma^k_{jn} G_k$ is the covariant derivative
acting on the components of the spinorial matrix.
Using the algebra of the Dirac matrices, one finds that~\eqref{Bform} has a vectorial component
(obtained by using the anti-commutation relations), and an axial component which 
is totally anti-symmetric in the indices~$j$, $m$, and~$n$.
This totally anti-symmetric term vanishes for the following reasons: First, since the Levi-Civita
connection is torsion-free, we may replace the covariant derivative with a partial derivative.
Second, it follows from~\eqref{Gchoice} that the matrix~$\partial_j G_n$ is a multiple of~$G_n$,
implying that the trace~$\Tr(G_m (\partial_j G_n))$ is symmetric in the indices~$m$ and~$n$.

It remains to compute the vectorial component of~\eqref{Bform}. A short computation yields
\beq B = \frac{\cI}{2}\, \nabla_j G^j\:, \eeq
and the usual formula for the covariant divergence of a vector field gives the result.
\QED
The result of this proposition is very useful for applications, as is illustrated in
Exercise~\ref{exschwarzschild}.

\section{Formulation with Vector Bundles, the Spinor Bundle} \label{secspinorbundle}
So far, the Dirac operator was introduced in a local chart.
We intentionally left a large local gauge freedom, having the advantage that
this freedom can be used to simplify the form of the Dirac operator.
The remaining question is whether our constructions in local charts can be
made global to obtain a Dirac operator~$\Dir$ acting on the sections of
the so-called spinor bundle~$S\scrM$.
To this end, we shall consider the Dirac operator in different charts
and patch the Dirac operators in the overlapping regions.

In preparation, we recall the structures introduced so far, using a more abstract notation that
clarifies the dependence on gauge and coordinates.
In our local construction at the beginning of Section~\ref{secdircurv},
the spinor space at a point~$x \in \scrM$ is simply~$\C^4$ with the inner
product~\eqref{sspgen}. Using the same notation as in
Section~\ref{secspinorbundlemink} in Minkowski space,
we now denote the spinor space by~$(S_x\scrM, \Sl .|. \Sr_x)$.
Moreover, we denote the linear operators on~$S_x\scrM$ which are symmetric
with respect to the spin inner product by~$\Symm(S_x\scrM)$.
It is a $16$-dimensional real vector space spanned by the operators in~\eqref{5_10}.
Given a Dirac-type operator~$\Dir$, the Dirac matrices~$G^j(x)$ span a
four-dimensional subspace~$K_x$ of~$\Symm(S_x\scrM)$,
\beq K_x := \text{span} \big\{ G^0(x), \ldots, G^3(x) \big\} \subset \Symm(S_x\scrM) \:, \eeq
referred to as a {\em{Clifford subspace}} at~$x$.
\sindex{Clifford subspace}%
Contracting a tangent vector~$u$ with the Dirac matrices gives rise to
a mapping
\beq \gamma \::\: T_x \scrM \rightarrow K_x \:,\qquad u \mapsto u^j G_j \:. \eeq
Multiplying a spinor by~$\gamma(u)$ is referred to as {\em{Clifford multiplication}}.
\sindex{Clifford multiplication}%
\sindex{anti-commutation relations!in curved spacetime}%
The anti-commutation relations~\eqref{4_4} can be written as
\beq %\label{anticomm}
\frac{1}{2}\: \{ \gamma(u), \gamma(v) \} = g_x(u,v) \:\1_{S_x\scrM} \:, \eeq
showing that Clifford multiplication encodes the Lorentzian metric.

In view of the transformation law~\eqref{ctrans}, the Clifford subspace
does not depend on the choice of coordinates. But it clearly
depends on the gauge. Indeed, in view of~\eqref{gtrans},
it transforms according to
\beq \label{Kgauge}
K_x \rightarrow U \,K_x\,U^{-1} \qquad \text{with~$U \in \U(S_x)$}\:.
\eeq
In order to simplify our problem, it is a good idea to arrange by a gauge transformation
that the Clifford subspace agrees at every spacetime point with the
standard Clifford subspace:

\begin{Lemma} \label{lemmaKfix}
By a gauge transformation~\eqref{Kgauge} we can arrange that
\beq K_x = \text{\rm{span}} \{ \gamma^0, \ldots, \gamma^3 \} \eeq
(where~$\gamma^j$ are again the Dirac matrices in the Dirac representation).
\end{Lemma}
\Proof We consider the pseudo-scalar operator~$\Gamma(x)$
as defined by~\eqref{Gammaxdef}. Working in a coordinate system and gauge where the Dirac matrices
coincide at~$x$ with the usual Dirac matrices in Minkowski space,
one sees immediately that the pseudo-scalar matrix satisfies also in curved spacetime the relations
\beq \Gamma(x)^*=-\Gamma(x) \qquad \text{and} \qquad \Gamma(x)^2 = \1 \:. \eeq
The first relation implies that~$\Gamma(x)$ maps positive definite spinors
to negative definite spinors and vice versa. Therefore, there is a pseudo-orthonormal
basis of the spinor space in which~$\Gamma(x)$ takes the same form as in Minkowski space,
\beq \label{pseudoMink}
\Gamma(x) = \begin{pmatrix} 0 & \1 \\ \1 & 0 \end{pmatrix} \:.
\eeq
Rewriting the change of basis as a gauge transformation,
we have arranged by a transformation of the form~\eqref{Kgauge} that the pseudo-scalar
operator has the same form as in Minkowski space.

It follows from~\eqref{Gammaxdef} and the anti-commutation relations that every vector in~$K$
anti-commutes with~$\Gamma$. Therefore,
\beq K \subset \text{span} \big\{ \gamma^0, \ldots, \gamma^3, \Gamma \gamma^0, \ldots, \Gamma \gamma^3 \big\} \:. \eeq

We next show that the vector space~$K \cap \text{span} \{ \gamma^0, \Gamma \gamma^0\}$
is one-dimensional. To this end, let~$u, v \in T_x\scrM$ with~$\gamma(u) = (a + b \Gamma)\, \gamma^0$
and~$\gamma(v) = (c + d \Gamma)\, \gamma^0$ with real coefficients~$a,b,c,d$. Then
their anti-commutator is computed by
\beq \{ \gamma(u), \gamma(v) \} = 2 \big(a c - b d\big)\:\1 + 2 \big(bc - ad)\: \Gamma \:, \eeq
implying that~$bc-ad=0$. This implies that~$\gamma(u)$ and~$\gamma(v)$ are linearly dependent,
giving the claim.

Repeating the last argument for~$\gamma^1,\ldots, \gamma^3$, we conclude that there is a
(not necessarily pseudo-orthonormal) basis~$u_0,\ldots, u_3$ of~$T_x\scrM$ such that
\beq \gamma(u_j) = (a_j+b_j \Gamma)\: \gamma_j \:. \eeq
Then, for any~$j \neq k$,
\beq \{ \gamma(u_j), \gamma(u_k) \} = \big(b_j a_k - a_j b_k)\: \Gamma \, \big[\gamma_j, \gamma_k \big]\:, \eeq
implying that the four vectors~$(a_j, b_j) \in \R^2$ with~$j=0,\ldots,3$ are all linearly independent.
Therefore, by rescaling the basis vectors~$u_j$ we can arrange that
\beq \gamma(u_j) = (a+b \Gamma)\: \gamma_j \eeq
for real parameters~$a$ and~$b$.

The signature~$(1,3)$ of the Lorentzian metric implies that~$|a|>|b|$. Moreover,
by flipping the sign of the vectors~$u_j$ if necessary we can arrange that~$a>0$.
Therefore, we may represent~$K$ as
\beq K = \text{span} \big\{ (\E^{\alpha \Gamma} \,\gamma^0,
\ldots  \E^{\alpha \Gamma} \, \gamma^3 \big\} \eeq
for some~$\alpha \in \R$ (note that~$\E^{\alpha \Gamma} = \cosh \alpha + \Gamma\, \sinh \alpha$). 
Performing the gauge transformation~\eqref{Kgauge} with~$U$ according to
\beq U = \exp \Big( -\frac{\alpha}{2}\: \Gamma \Big) \eeq
gives the result.
\QED

After these preparations, we are ready to enter the patching construction.
Thus let~$(x,U)$ and~$(\tilde{x}, \tilde{U})$ be two local charts on~$(\scrM, g)$
with non-empty overlap~$U \cap \tilde{U}$.
For technical simplicity, we always restrict attention to the case that spacetime is {\em{time-oriented}}
in the sense that the transition maps between any two charts preserve the time direction.
\sindex{time orientation}%
Moreover, for simplicity we restrict attention to spacetimes which are {\em{oriented}}
in the sense that all transition maps also preserve the orientation
(i.e., in more technical terms, we assume that the determinant of the Jacobian of every transition map
is everywhere positive).
\sindex{orientation}%
We choose the charts such that~$x^0$ and~$\tilde{x}^0$ are time functions that increase to the future.
Then, we can write the Dirac operator in each chart according to~\eqref{Dirphys},
where for clarity we denote the objects in the chart~$\tilde{x}$ with an additional tilde.
We first consider the case without electromagnetic field where the potentials~$a_j$ vanish.
According to Lemma~\ref{lemmaKfix}, there is no loss of generality to restrict
attention to gauges where the Dirac matrices are linear combinations of the Dirac
matrices in Minkowski space; that is,
\beq G^j(x) = h^j_k(x)\: \gamma^k \qquad \text{and} \qquad \tilde{G}^j(\tilde{x})
= \tilde{h}^j_k(\tilde{x})\: \gamma^k \:. \eeq
Since~$x^0$ is a time coordinate, the bilinear form~$\Sl .| G^0(x) . \Sr_x$ is definite
at very spacetime point~$x$, and similar for the tilde coordinates.
We choose the signs of the Dirac matrices such that
the bilinear forms~$\Sl .| G^0(x) . \Sr_x$ and~$\Sl .| \tilde{G}^0(\tilde{x}) . \Sr_y$
are all {\em{positive}} definite.
Moreover, as explained in Lemma~\ref{lemmaKfix}, we always choose the gauge
such that the pseudo-scalar operator~\eqref{Gammaxdef} has 
the same form as in Minkowski space~\eqref{pseudoMink}. Using that spacetime is oriented,
this can be done consistently for all charts (meaning that if~\eqref{pseudoMink} holds in one
chart and gauge, then it also holds in all other charts for the same gauge).

The transformation from the chart~$(x,U)$ to~$(\tilde{x},\tilde{U})$
involves the coordinate transformation as described by~\eqref{ctrans}.
After this transformation, the Dirac matrices
\beq \tilde{G}^j(\tilde{x}) \qquad \text{and} \qquad G^k(x)\: \frac{\partial \tilde{x}^j}{\partial x^k} \eeq
will in general not coincide. But since the matrices are all formed as linear combinations
of the Dirac matrices in Minkowski space, satisfy the same anti-commutation relations,
and have the same time and spatial orientations, they can be obtained from each
other by an orthochronous and proper Lorentz transformation; that is,
\beq \tilde{G}^j(\tilde{x}) = \Lambda^j_l \:G^k(x)\: \frac{\partial \tilde{x}^l}{\partial x^k} \:. \eeq
Now we can proceed just as in the proof of Lorentz invariance of the Dirac equation
in Minkowski space (see Lemma~\ref{Dirtrans}) to conclude that there is
a unitary transformation~$U(x) \in \U(S_x)$ of the form
\beq \label{Uspin}
U := \exp \left( \frac{1}{4}\: \lambda_{lk}\: \gamma^l \,\gamma^k \right)
\eeq
(with an anti-symmetric tensor~$\lambda_{lk}$)
such that the Dirac matrices agree after the gauge transformation; that is,
\beq U(\tilde{x}) \,\tilde{G}^j(\tilde{x})\, U(\tilde{x})^{-1} = G^j(x)\: \frac{\partial \tilde{x}^l}{\partial x^k} \:. \eeq
Since the spin coefficients~$E_j$ in~\eqref{5_12} are given explicitly in terms of the
Dirac matrices and their derivatives (see~\eqref{Ejform}), the lower order terms
in the resulting Dirac operators~\eqref{Dirphys} also agree.
Moreover, using that the only matrices which commute with all Dirac matrices are
multiples of the identity, one sees that the
gauge transformation~$U(\tilde{x})$ of the form~\eqref{Uspin} is uniquely determined 
up to a sign. In this way, to every coordinate transformation, we have found a gauge
transformation, unique up to a sign, such that the Dirac operators agree.

With the above construction, we have found a procedure for matching the
Dirac operators in two overlapping charts. The involved gauge transformations
of the form~\eqref{Uspin} are unique up to signs.
Therefore, once we have decided on the signs, there is a unique way of
identifying the Dirac wave functions in the overlapping region of two charts,
such as to obtain Dirac wave functions in a larger patch of the manifold~$\scrM$.
Proceeding inductively, one can hope to obtain Dirac wave functions on all of~$\scrM$.
The subtle point is whether the signs of the transformations can be chosen
in a compatible way for all charts. In more mathematical terms, one must satisfy
the so-called cocycle conditions, and it turns out that these conditions can be
fulfilled if and only if~$\scrM$ satisfies a topological condition,
the so-called {\em{ spin condition}}
(for details see, for example, \cite[\S~II.1 and \S~II.2]{lawson+michelsohn}).
\sindex{spin condition}%
If the spin condition is satisfied, one can identify the spinor spaces via
the mappings that patch the charts together.
In this way, one obtains a vector bundle
over~$\scrM$, referred to as the {\em{spinor bundle}}~$S\scrM$.
\sindex{spinor bundle! in curved spacetime}%
\nindex{acf@$S\scrM$ -- spinor bundle}%
The fibers of the spinor bundle are the spinor spaces~$S_x \scrM$, which are
four-dimensional complex vector spaces endowed with
an inner product~$\Sl .|. \Sr_x$ of signature~$(2,2)$.
The transformations of the form~\eqref{Uspin} generate a group, the so-called
{\em{spin group}} denoted by
\beq \label{spindef}
\Spin^\uparrow_x \subset \U(S_x)
\eeq
(the reason why we write ``generated by'' is that the operators of the form~\eqref{Uspin}
do {\em{not}} form a group; see Exercise~\ref{ex1m1}).
Elements of the spin group act on vectors of the Clifford subspace by the adjoint representation,
\beq \gamma(v) \rightarrow U\,\gamma(v)\, U^{-1} \:, \eeq
we obtain another vector of the Clifford subspace; that is,
\beq %\label{Oudef}
U\,\gamma(v)\, U^{-1} = \gamma\big( {\mathscr{O}}_U(u) \big) \:. \eeq
Since the anti-commutation relations remain unchanged, the
resulting transformation of the tangent space is an isometry.
Indeed, by Lemma~\ref{Dirtrans} it is a proper orthochronal Lorentz transformation,
\beq \label{Odef}
{\mathscr{O}}_U \in \SO^\uparrow(T_x\scrM) \:.
\eeq
The indices~$\uparrow$ in~\eqref{spindef} and~\eqref{Odef}
indicate that we restrict attention to orthochronous transformations.
We thus obtain the usual commutative diagram
\beq \Z_2 \longrightarrow \Spin^\uparrow_x \overset{{\mathscr{O}}}{\longrightarrow} \SO^\uparrow(T_x\scrM)
\longrightarrow 0 \:. \eeq

The connection to the usual spin group is obtained as follows.
We say that a tangent vector~$u \in T_x\scrM$ is a {\em{unit vector}}
if~$\la u,u \ra = \pm 1$. 
The spin group is defined by
(see, for example, \cite{baum, lawson+michelsohn},
the concise summary in~\cite[Section~2]{baer+gauduchon}
or similarly~\cite{friedrich} in the Riemannian setting)
\beq \label{spingendef}
\Spin_x := \big\{ \text{group generated by~$\gamma(u) \,\gamma(v)$ with unit vectors~$u,v \in T_x\scrM$} \big\} \:.
\eeq
By expanding the exponential in~\eqref{Uspin}, one sees that this matrix is generated
by even products of Dirac matrices, showing that the group~$\Spin^\uparrow_x$
in~\eqref{spindef} is a subgroup of~$\Spin_x$.
The group~$\Spin_x$ also includes operators that are not unitary but satisfy instead the relation~$U^*U=-\1$.
These transformations describe reversals of the time orientation.
Working with the general spin group~\eqref{spingendef} is of advantage in general dimension
or signature. In four-dimensional time-oriented and orientable spacetimes, however,
we can just as well restrict attention to orthochronous proper Lorentz transformations
and the gauge transformations in~\eqref{spindef}.

We finally mention how to treat an electromagnetic field.
Then the starting point is a time-oriented Lorentzian spin manifold~$(\scrM, g)$
together with an anti-symmetric two-tensor~$F$ (the field tensor).
In this situation, after the above coordinate and gauge transformations, the
electromagnetic potentials~$a_j$ and~$\tilde{a}_j$ in the two charts will in
general not coincide. But, since the field tensor is prescribed, they
coincide after a local $U(1)$-gauge transformation.
Identifying the spinor spaces after this gauge transformation
defines the Dirac operator as acting on the spinor bundle~$S\scrM$.
The resulting effective gauge group is~$\U(1) \times \Spin^\uparrow_x$.
We point out that this effective gauge group is obtained under the
condition that the Clifford subspace is fixed at each spacetime
point according to Lemma~\ref{lemmaKfix}.
Dropping this condition gives rise to the larger local gauge group~$\U(2,2)$.

\section{The Dirac Solution Space in Globally Hyperbolic Spacetimes} \label{secdirgh}
We now turn attention to solutions of the Dirac equation.
In Minkowski space, a convenient method for constructing solutions
is the Fourier transformation (see Section~\ref{secfourier}).
However, this method can be used only for PDEs with constant coefficients,
and therefore it does not apply to the Dirac equation in curved spacetime.
Instead, a general method is to solve the Cauchy problem for given
initial data, making it possible to identify the solution space of the Dirac equation
with the space of suitable initial data.
Clearly, for this method to be applicable, one needs to decompose spacetime into
``space'' and ``time,'' because otherwise it would not be clear how to
prescribe initial data at some initial time.
In order to describe the Dirac solutions in all of spacetime by initial data,
this splitting of spacetime into space and time must be performed globally.
Moreover, there must be distinguished notions of the future and past of a spacelike
hypersurface. Intuitively speaking, these requirements are necessary in order to make sense of the
initial value problem in which, given initial data on a spacelike hypersurface, one seeks
global solutions of the Dirac equation to both the future and the past.

In mathematical terms, the necessary assumptions on spacetime needed for a
sensible formulation of the initial value problem, also referred to as the {\em{Cauchy problem}}, are subsumed
\sindex{Cauchy problem}%
in the notion of {\bf{global hyperbolicity}}.
\sindex{global hyperbolicity}%
We first give the formal definition and then explain its consequences.
Let~$(\scrM, g)$ be a Lorentzian manifold.
We again assume that~$\scrM$ is time-oriented (meaning that all transition maps
preserve the time direction).
We consider a parametrized piecewise $C^1$-curve~$\gamma(\tau)$ in~$M$ 
which is {\em{regular}} in the sense that its tangent vector~$\dot{\gamma}(\tau)$ is non-zero
for all~$\tau$ where~$\gamma$ is differentiable.
\sindex{regular curve}%
This spacetime curve is said to be~{\em{causal}}
if its tangent vector is causal (that is, timelike or lightlike)
for all~$\tau$ where~$\gamma$ is differentiable.
\sindex{causal curve}%
Moreover, it is {\em{future-directed}} and {\em{past-directed}}
its tangent vectors are future- and past-directed, respectively.
The manifold~$\scrM$ is said to satisfy the {\em{strong causality condition}} if there are
{\em{no almost closed causal curves}} in the sense that for all~$x \in M$
and for each open neighborhood~$U$ of~$x$ there is an open neighborhood~$V \subset U$
of~$x$ such that every causal curve in~$\scrM$ which is starting and ending in~$V$
is entirely contained in~$U$.
Moreover, in straightforward generalization of the corresponding notions in Minkowski
space as introduced after~\eqref{causalMink}, we let~$J_x^\vee$ (and~$J_x^\wedge$) be the
set of all points~$y \in \scrM$ which can be joined from~$x$ by
a future-directed (respectively past-directed) causal curve.
The manifold~$\scrM$ is said to be {\em{globally hyperbolic}}
if the strong causality condition holds and if the set~$J_x^\vee \cap J_y^\wedge$
is compact for all~$x,y \in \scrM$.
For more details on the abstract definitions we refer to~\cite[Section~6.6]{hawking+ellis},
\cite[Section~1.3]{baer+ginoux}, \cite[Section~3.2]{beem} or~\cite[Chapter~14]{oneillsemi}).

A globally hyperbolic Lorentzian manifold~$(\scrM, g)$ has remarkable properties,
as we now explain. First, global hyperbolicity implies that~$(\scrM, g)$ is
diffeomorphic to a product,
\beq \label{splittop}
\scrM \simeq \R \times \scrN \:,
\eeq
where~$\simeq$ means that there is a smooth diffeomorphism from~$\scrM$
to~$\R \times \scrN$ (with~$\scrN$ a three-dimensional manifold).
Thus every point~$p \in \scrM$ can be written as~$p=(t,x)$ with~$t \in \R$ and~$x \in \scrN$.
One also refers to the above property that~$\scrM$ admits
a {\em{smooth foliation}}~$\scrM=(\scrN_t)_{t \in \R}$, where~$\scrN_t := \{t\} \times \scrN$.
Moreover, the foliation can be chosen such as to have the following properties:
\bitem
\item[(i)] Every surface~$\scrN_t$ is
{\em{spacelike}} (meaning that the metric induced by~$g$ on~$\scrN_t$ is negative definite).
\item[(ii)] Every surface~$\scrN_t$ is a {\em{Cauchy surface}}, meaning that
\sindex{Cauchy surface}%
every inextendible timelike curve in~$\scrM$ intersects~$\scrN_t$ exactly once.
Here the timelike curve is said to be {\em{inextendible}}
if it cannot be extended as a continuous curve.
\eitem
The function~$t$ is also referred to as a {\em{global time function}}.
\sindex{foliation!by Cauchy surfaces}%
\sindex{time function}%
These above properties of globally hyperbolic manifolds were proven in~\cite{bernal+sanchez}
(for more details and more references see again~\cite[Section~1.3]{baer+ginoux}).

The property of~$\scrN_t$ of being a Cauchy surface implies that the Cauchy problem
for the Dirac equation is well-posed, as we now explain.
To this end, let~$(\scrM, g)$ be a four-dimensional globally hyperbolic spacetime.
Then the topological splitting~\eqref{splittop} implies that the
spin condition mentioned before~\eqref{spindef} is satisfied.
Therefore, there is a spinor bundle~$(S\scrM, \Sl .|.\Sr)$,
being a vector bundle with fibers~$S_x \scrM \simeq \C^4$
(there may be different spin structures, but we shall not go into this here).
Moreover, the Dirac operator~$\Dir$ is well-defined; in local coordinates and
local spinor bases it takes the form~\eqref{Dirphys}.
In the Cauchy problem, one seeks for solutions of the Dirac equation
of mass~$m$ for prescribed initial data at time~$t_0$; that is,
\sindex{Cauchy problem!for Dirac equation}%
\beq \label{DirCauchy}
(\Dir - m) \psi = \phi \qquad \text{with} \qquad \psi|_{\scrN_{t_0}} = \psi_0 \:.
\eeq
Then the following result holds:
\begin{Thm} \label{thmcauchy}
For smooth initial data~$\psi_0 \in C^\infty(\scrN_{t_0}, S\scrM)$
and a smooth inhomogeneity~$\phi \in C^\infty(\scrM, S\scrM)$
the Cauchy problem~\eqref{DirCauchy} has a
unique global solution~$\psi \in C^\infty(\scrM, S\scrM)$.
\end{Thm} \noindent
The proof of this theorem uses methods of hyperbolic partial differential equations
and will be given in Section~\ref{seccauchyglobhyp} later in this book.

Having a Cauchy surface is also very useful because we can then define
a scalar product on the solution space as the spatial integral~\eqref{nicsp},
where~$\scrN$ is chosen as a Cauchy surface.
However, for the integral in~\eqref{nicsp} to be well-defined, working with
smooth solutions is not suitable.
Instead, similar as explained in Minkowski space in Section~\ref{secspinorbundlemink},
we better assume that the solution has compact support on the Cauchy surface.
Due to finite propagation speed for solutions of hyperbolic partial differential equations
(as will be made precise in Section~\ref{sec31}), the following result holds:
\begin{Thm} \label{thmcauchy2}
If the initial data and the inhomogeneity have compact support,
\beq \psi_0 \in C^\infty_0(\scrN_{t_0}, S\scrM) \qquad \text{and} \qquad
\phi \in C^\infty_0(\scrM, S\scrM) \:, \eeq
then the solution~$\psi$ of the Cauchy problem~\eqref{DirCauchy}
also has compact support on any other Cauchy surface~$\scrN_t$.
\end{Thm} \noindent
The proof of this theorem will again be given in Section~\ref{seccauchyglobhyp} below.

Using the same notion as in Section~\ref{secspinorbundlemink},
we refer to smooth solutions as in the above theorem as having {\em{spatially compact
support}}. Smooth and spatially compact sections of the spinor bundle are
again denoted by~$\Cisc(\scrM, S\scrM)$.
\sindex{wave function!spatially compact}%
\sindex{function!with spatially compact support}%
\nindex{aau@$\Cisc(\scrM, S\scrM)$ -- spatially compact spinorial wave functions}%
\tindex{ff@$\Cisc(\scrM, S\scrM)$ -- spatially compact spinorial wave functions}%
For Dirac solutions in this class, the scalar product~\eqref{nicsp}
is well-defined. Moreover, due to current conservation, this scalar product does not
depend on the choice of the Cauchy surface (as explained after~\eqref{nicsp}).
Exactly as explained in Section~\ref{secspinorbundlemink},
taking the completion gives the Hilbert space~$(\H_m, (.|.))$ of
weak solutions of the Dirac equation with the property that their restriction
to any Cauchy surface is square integrable (where ``square integrable''
is defined via~\eqref{nicsp}).

\section{Hamiltonian Formulation in Stationary Spacetimes}
Given a foliation~$(\scrN_t)_{t \in \R}$ 
of the globally hyperbolic space\-time~$(\scrM, g)$ by Cauchy surfaces, one can rewrite the Dirac equation
in the Hamiltonian form
\sindex{Dirac equation!in the Hamiltonian form}%
\nindex{aao@$H$ -- Dirac Hamiltonian}%
\beq %\label{hamiltongrav}
\cI \partial_t \psi = H \psi \eeq
with a Hamiltonian~$H$. In order to compute~$H$ in a local chart, one chooses
a coordinate system~$(x^i)$ such that~$x^0=t$ coincides with the time function.
Then, writing the Dirac operator in~\eqref{diracgrav} in the form~\eqref{Dirphys}
and solving for the time derivatives, one obtains in generalization of~\eqref{Hamilton}
\beq H =  -\big(G^0 \big)^{-1} \bigg( \sum \nolimits_{\alpha=1}^3 \cI G^\alpha\, \big(\partial_\alpha - \cI E_\alpha - \cI a_\alpha \big) 
- m \bigg) - E_0 - a_0  \:. \eeq

When analyzing the Dirac equation in the Hamiltonian form, one must be careful
because the Hamiltonian in general is {\em{not symmetric}} with respect to the
Hilbert space scalar product~\eqref{nicsp}. This can be seen as follows.
For the Dirac equation in Minkowski space, the symmetry of the Hamiltonian
is obtained just as for the Schr\"odinger equation by
using that the scalar product is conserved in time~\eqref{symmc}.
In curved spacetime, the scalar product is still conserved (due to current conservation).
But when taking the time derivative, one must take into account that the scalar product
itself is time dependent. More precisely, assuming for notational simplicity that the Cauchy
surfaces admit global charts,
\begin{align}
0 &= \partial_t ( \phi | \psi )
= \frac{\partial}{\partial t} \int_{\scrN_t} \Sl \phi \,|\, G^j \nu_j\, \psi \Sr \: \dd\mu_{\scrN_t} \notag \\
&= (\partial_t \phi \,|\, \psi ) + ( \phi \,|\, \partial_t \psi )
+ \int_{\R^3} \Sl \psi \,|\, \Big( \partial_t \big(G^j \nu_j \sqrt{\det g_{\scrN_t}} \big) \Big) \, \phi \Sr \: \Diff^3x \notag \\
&= -\cI \Big( ( H \phi | \psi ) - ( \phi | H \psi ) \Big)
+ \int_{\R^3} \Sl \psi \,|\, \Big( \partial_t \big(G^j \nu_j \sqrt{\det g_{\scrN_t}} \big) \Big) \, \phi \Sr \: \Diff^3x
\label{nosymm}
\end{align}
(where~$g_{\scrN_t}$ denotes the induced Riemannian metric on the Cauchy surface~$\scrN_t$).
The integral in the last line is in general non-zero. In this case, the Hamiltonian is obviously not symmetric.
At first sight, this might seem surprising because it seems to contradict the axioms of quantum mechanics
(for a detailed account on this issue, see~\cite{arminjon1, arminjon2}).
However, one should keep in mind that the non-symmetric contributions to the Hamiltonian
are needed in order to compensate for the fact that the scalar product itself is time-dependent.

Our interpretation of the above problem is that the Hamiltonian formulation
of the Dirac equation is useful only in situations when the integral in~\eqref{nosymm} vanishes.
This can be arranged if all the coefficients of the metric are time-independent.
In other words, spacetime should be {\em{stationary}} with corresponding Killing field
\sindex{stationary spacetime}%
given by~$\partial_t$. Under these assumptions, the Hamiltonian~$H$ is also time-independent.
Moreover, the computation~\eqref{nosymm} shows that the operator is symmetric.
Using that the time evolution maps smooth and compactly supported
initial data on the Cauchy surface at time~$t_0$ to a smooth and compactly supported solution
at an arbitrary time~$t$, one can use abstract methods to construct a selfadjoint extension of~$H$
(see, for example, \cite{chernoff} for a general situation involving additional boundary conditions).
Then the Cauchy problem can be solved immediately using the spectral theorem for selfadjoint operators,
\beq \label{specdirH}
\psi(t,x) = \big(\E^{-\cI t H}\: \psi_0 \big)(x) \:.
\eeq
This formulation is particularly useful for analyzing the long-time behavior of the solutions
(see, for example, the analysis in the Kerr geometry in~\cite{tkerr, decay}).

\section{Exercises}

\begin{Exercise} {\em{
Verify by elementary integration by parts in a chart that for a diagonal metric~\eqref{gdiagonal},
the Dirac operator~\eqref{Dirac} is symmetric with respect to the inner product~\eqref{stipintro}.
}} \end{Exercise}

\begin{Exercise} \label{exUdiff} {\em{ Let~$U(x)$ with~$x \in \R^4$ be a smooth function of invertible
matrices. Show that
\beq \partial_j U = - U (\partial_j U^{-1}) U \:. \eeq
{\em{Hint:}} Differentiate the relation~$U(x)\, U(x)^{-1}=\1$.
}} \end{Exercise}

\begin{Exercise} \label{ex1m1}
{\em{ The goal of this exercise is to show that the unitary operators
of the form~\eqref{Uspin} do not form a group
(in more mathematical language, the spin group is not exponential;
for details, see~\cite{dokovic} and the references therein).
We proceed in several steps:
\bitem
\item[(a)] Let~$\lambda_{jk}$ be an anti-symmetric tensor. Show using the
anti-commutation relations that
\beq \label{lam2}
\left( \frac{1}{4}\: \lambda_{lk}\: \gamma^l \,\gamma^k \right)^2
= -\frac{1}{16}\: \lambda_{lk}\: \lambda^{lk} \: \1 + \frac{\cI}{16 \cdot 4!}\:
\Gamma \:\varepsilon^{ijkl} \:\lambda_{ij} \lambda_{kl} \:.
\eeq
\item[(b)] Deduce from~(a) that the corresponding unitary transformation~\eqref{Uspin}
is a linear combination of the matrices
\beq \1 \:,\qquad \Gamma\:,\qquad \lambda_{lk}\: \gamma^l \,\gamma^k \qquad \text{and} \qquad
\lambda_{lk}\: \Gamma\, \gamma^l \,\gamma^k \:. \eeq
\item[(c)] Show under the additional assumption~$\varepsilon^{ijkl} \:\lambda_{ij} \lambda_{kl}=0$ that
\beq \label{expexplicit}
\exp \left( \frac{1}{4}\: \lambda_{lk}\: \gamma^l \gamma^k \right)
= \left\{ \begin{array}{cl}
\displaystyle \1 \cos \alpha + \frac{1}{4}\: \lambda_{lk}\: \gamma^l \gamma^k \:\frac{\sin \alpha}{\alpha}
& \text{if~$\lambda_{lk}\: \lambda^{lk} > 0$} \\[0.8em]
\displaystyle \1 \cosh \alpha + \frac{1}{4}\: \lambda_{lk}\: \gamma^l \gamma^k \:\frac{\sinh \alpha}{\alpha}
& \text{if~$\lambda_{lk}\: \lambda^{lk}<0$}\:, \end{array} \right.
\eeq
where~$\alpha := \sqrt{|\lambda_{lk}\: \lambda^{lk}|}/4$.
{\em{Hint:}} Use~\eqref{lam2} in the power series of the exponential.
\item[(d)] Choose a specific tensor~$\lambda_{lk}$ for which the matrix in~\eqref{expexplicit}
is equal to minus the identity.
\item[(e)] We now restrict attention to tensors~$\lambda$ for which the
corresponding unitary transformation~\eqref{Uspin} is a linear combination of
the matrices~$\1$ and~$\gamma^0 \gamma^1$. Infer from~(b) that in this case,
there are real numbers~$\alpha, \beta$ such that
\beq \frac{1}{4}\: \lambda_{lk}\: \gamma^l \,\gamma^k = (\alpha + \cI \Gamma \beta) \: \gamma^0 \gamma^1 \:. \eeq
Deduce that
\beq \exp \left( \frac{1}{4}\: \lambda_{lk}\: \gamma^l \gamma^k \right)
= \cosh \big( \alpha + \cI \Gamma \beta \big)
+ \gamma^0 \gamma^1\: \sinh \big( \alpha + \cI \Gamma \beta \big) \:. \eeq
\item[(f)] Show that the last expression involves no contribution~$\sim \Gamma$ only
if either~$\alpha=0$ or~$\beta=0$
({\em{Hint:}} It might be convenient to work in an eigenvector basis of~$i \Gamma$).
Infer that in the case~$\alpha=0$,
this expression is a linear combination of the matrices~$\1$
and~$\gamma^0 \gamma^1$ only if~$\beta \in \pi \Z$.
Conclude that
\beq \exp \left( \frac{1}{4}\: \lambda_{lk}\: \gamma^l \gamma^k \right)
= a\,\1 + b \, \gamma^0 \gamma^1 \qquad \text{with~$a>0$}\:. \eeq
\item[(g)] Deduce from~(c) and~(e) that the matrices of the form~\eqref{Uspin} do not form a group.
%\item[(a)] Let~$\Lambda \in \SO^\uparrow(1,3)$ be an orthochronous
%proper Lorentz transformation. Show that~$\Lambda$ can be written as
%\beq \Lambda = \exp(\lambda) \:, \eeq
%where~$\lambda$ is a linear operator on Minkowski space which is anti-symmetric
%with respect to the Minkowski inner product (i.e.\ $\la \lambda u, v \ra = - \la u, \lambda v \ra$).
%\item[(b)] Show by explicit computation that the unitary transformation
%\beq U = \exp \Big( -\frac{1}{4}\: \lambda^l_k(\tilde{x})\: \gamma_l \,\gamma^k \Big) \eeq
%satisfies the relation~\eqref{Odef} with~${\mathcal{O}}_U = \Lambda$.
%\item[(c)] Use the relation~$(\gamma^1 \gamma^2)^2 = -\1$ to show that
%\beq \exp \big( t \,\gamma^1 \gamma^2 \big) = \1 \: \cos t + \gamma^1 \gamma^2\: \sin t \:. \eeq
%Conclude that the operator~$-\1$ can be represented in the form~\eqref{Uspin}.
%\item[(d)] Deduce that the unitary operators of the form~\eqref{Uspin} form a group.
\eitem
}} \end{Exercise}

\begin{Exercise} \label{exschwarzschild} {\em{
The {\em{Schwarzschild geometry}} is the simplest mathematical model of a black hole
(for the physical background see, for example, \cite{landau2}).
In so-called Boyer-Lindquist coordinates~$t \in \R$ (the time coordinate),
$r \in (2M, \infty)$ (the radial coordinate) and~$\vartheta \in (0, \pi)$, $\varphi \in (0, 2 \pi)$
(the angular coordinates), the line element takes the form
\beq \dd s^2 = g_{jk}\:\dd x^j \,\dd x^k = \Big( 1 - \frac{2M}{r} \Big) \: \dd t^2
- \Big( 1 - \frac{2M}{r} \Big)^{-1} \:\dd r^2 - r^2\: \dd \vartheta^2 - r^2\: \sin^2 \vartheta\: \dd\varphi^2\:, \eeq
where~$M>0$ is the mass of the black hole.
Compute the Dirac operator in this geometry.
{\em{Hint:}} Use the formulas in Proposition~\ref{prpDOP}. The results can be found in the
literature, for example, in~\cite{FSYperiodic} or~\cite[Section~2.2]{sigbh}.
}} \end{Exercise}

%% Local Variables:
%% mode: latex
%% TeX-master: "intro"
%% End:

%% file: part2.tex
%!TEX root = intro.tex
\part{Causal Fermion Systems: Fun\-da\-men\-tal Structures} \label{parttwo}
%\part{The Theory of Causal Fermion Systems}

\chapter{A Brief Introduction to Causal Fermion Systems} \label{secbrief}
In this chapter, we define and explain the basic objects and structures
of a causal fermion system. Since causal fermion systems introduce a new language
to describe our physical world, we begin with preliminary considerations that explain how the
basic objects of the theory come about and how to think about them. In order to provide different perspectives, the preliminary
considerations motivate causal fermion systems in two somewhat different ways.
In Section~\ref{secencode}, the motivating question is whether and how spacetime structures can be
encoded in quantum mechanical wave functions. In Section~\ref{secex2d}, on the other hand, we begin
with the example of a two-dimensional lattice system and ask the question how one can 
formulate physical equations in this discrete spacetime without making use of specific lattice structures
like the nearest neighbor relations and the lattice spacing.
By extending the setting from the motivating examples (Section~\ref{sectowards})
we are led to the general definition of a causal fermion system (Section~\ref{secgendef}).
Next, as a further example, we explain how the Minkowski vacuum can be described by a
causal fermion system (Section~\ref{seclco}).
In order to formulate equations describing the dynamics of a causal fermion system,
we introduce a variational principle, the so-called causal action principle
(Section~\ref{seccap}). We proceed by explaining how to obtain a spacetime as well as structures
therein (Section~\ref{secinherent}).
We conclude by discussing the form of the causal action principle (Section~\ref{secwhycap})
and by explaining the underlying physical concepts (Section~\ref{secphysicalconcepts}).

\section{Motivation: Encoding Spacetime Structures in Wave Functions} \label{secencode}
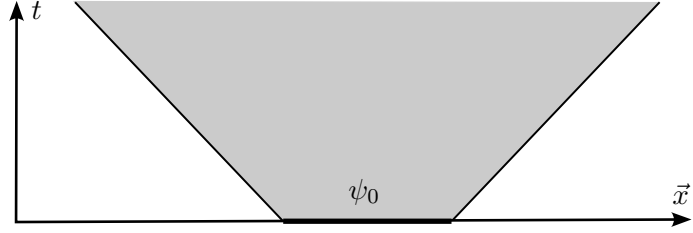
\begin{figure}[t]
\psset{xunit=.5pt,yunit=.5pt,runit=.5pt}
\begin{pspicture}(525.68616972,189.66897799)
{
\newrgbcolor{curcolor}{0.79607844 0.79607844 0.79607844}
\pscustom[linestyle=none,fillstyle=solid,fillcolor=curcolor]
{
\newpath
\moveto(52.52530394,173.45811185)
\lineto(494.46248693,172.64651122)
\lineto(337.82064756,7.14107878)
\lineto(210.50249197,7.14107878)
\closepath
}
}
{
\newrgbcolor{curcolor}{0.80000001 0.80000001 0.80000001}
\pscustom[linewidth=0.99999871,linecolor=curcolor]
{
\newpath
\moveto(52.52530394,173.45811185)
\lineto(494.46248693,172.64651122)
\lineto(337.82064756,7.14107878)
\lineto(210.50249197,7.14107878)
\closepath
}
}
{
\newrgbcolor{curcolor}{0 0 0}
\pscustom[linewidth=2.00000125,linecolor=curcolor]
{
\newpath
\moveto(506.78631685,8.41585295)
\lineto(8.4,6.52802429)
\lineto(8.80885039,161.36545295)
}
}
{
\newrgbcolor{curcolor}{0 0 0}
\pscustom[linestyle=none,fillstyle=solid,fillcolor=curcolor]
{
\newpath
\moveto(503.96512319,14.00521027)
\lineto(519.38623431,8.46357996)
\lineto(504.00754719,2.80528364)
\curveto(506.79517286,6.07786827)(506.77751387,10.73983773)(503.96512319,14.00521027)
\closepath
}
}
{
\newrgbcolor{curcolor}{0 0 0}
\pscustom[linestyle=none,fillstyle=solid,fillcolor=curcolor]
{
\newpath
\moveto(3.20147301,158.58024781)
\lineto(8.84212077,173.96541688)
\lineto(14.40144094,158.55067413)
\curveto(11.1468437,161.35927945)(6.48485705,161.37158949)(3.20147301,158.58024781)
\closepath
}
}
{
\newrgbcolor{curcolor}{0 0 0}
\pscustom[linewidth=1.49999998,linecolor=curcolor]
{
\newpath
\moveto(209.9585348,7.6574076)
\lineto(53.06926866,172.94177547)
}
}
{
\newrgbcolor{curcolor}{0 0 0}
\pscustom[linewidth=1.49999998,linecolor=curcolor]
{
\newpath
\moveto(338.36463496,7.6574076)
\lineto(495.25392,172.94177547)
}
}
{
\newrgbcolor{curcolor}{0 0 0}
\pscustom[linewidth=3.99999979,linecolor=curcolor]
{
\newpath
\moveto(210.50249575,7.14107122)
\lineto(337.82064756,7.14107122)
}
\rput[bl](260,20){$\psi_0$}
\rput[bl](20,160){$t$}
\rput[bl](505,20){$\vec{x}$}
}
\end{pspicture}
\caption{Causal propagation of a wave function.}
\label{figmmt1}
\end{figure}

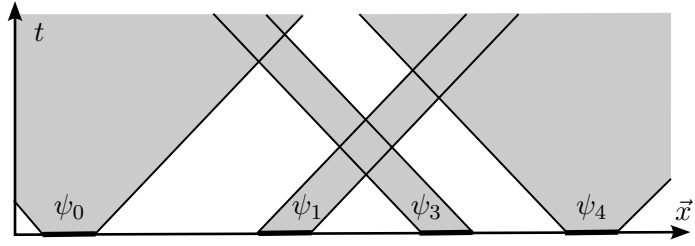
\begin{figure}[t]
\psset{xunit=.5pt,yunit=.5pt,runit=.5pt}
\begin{pspicture}(516.58620279,180.56370532)
{
\newrgbcolor{curcolor}{0.79607844 0.79607844 0.79607844}
\pscustom[linestyle=none,fillstyle=solid,fillcolor=curcolor]
{
\newpath
\moveto(268.69745386,170.15284548)
\lineto(501.43032945,171.2699491)
\lineto(501.34642394,46.53398186)
\lineto(463.48636724,7.66710438)
\lineto(422.41223811,6.25643729)
\closepath
}
}
{
\newrgbcolor{curcolor}{0.80000001 0.80000001 0.80000001}
\pscustom[linewidth=0.99999871,linecolor=curcolor]
{
\newpath
\moveto(268.69745386,170.15284548)
\lineto(501.43032945,171.2699491)
\lineto(501.34642394,46.53398186)
\lineto(463.48636724,7.66710438)
\lineto(422.41223811,6.25643729)
\closepath
}
}
{
\newrgbcolor{curcolor}{0.79607844 0.79607844 0.79607844}
\pscustom[linestyle=none,fillstyle=solid,fillcolor=curcolor]
{
\newpath
\moveto(194.94964535,169.57664139)
\lineto(157.08248315,170.19181619)
\lineto(313.7243263,4.68638375)
\lineto(351.9130885,4.25478438)
\closepath
}
}
{
\newrgbcolor{curcolor}{0.80000001 0.80000001 0.80000001}
\pscustom[linewidth=0.99999871,linecolor=curcolor]
{
\newpath
\moveto(194.94964535,169.57664139)
\lineto(157.08248315,170.19181619)
\lineto(313.7243263,4.68638375)
\lineto(351.9130885,4.25478438)
\closepath
}
}
{
\newrgbcolor{curcolor}{0.79607844 0.79607844 0.79607844}
\pscustom[linestyle=none,fillstyle=solid,fillcolor=curcolor]
{
\newpath
\moveto(6.49712126,170.15284548)
\lineto(223.74078992,170.15284548)
\lineto(66.22308661,5.85330532)
\lineto(27.80870173,2.77807761)
\lineto(6.81657071,28.60848013)
\closepath
}
}
{
\newrgbcolor{curcolor}{0.80000001 0.80000001 0.80000001}
\pscustom[linewidth=0.99999871,linecolor=curcolor]
{
\newpath
\moveto(6.49712126,170.15284548)
\lineto(223.74078992,170.15284548)
\lineto(66.22308661,5.85330532)
\lineto(27.80870173,2.77807761)
\lineto(6.81657071,28.60848013)
\closepath
}
}
{
\newrgbcolor{curcolor}{0.79607844 0.79607844 0.79607844}
\pscustom[linestyle=none,fillstyle=solid,fillcolor=curcolor]
{
\newpath
\moveto(346.49940661,170.67854375)
\lineto(385.93114205,170.76732485)
\lineto(229.28930268,5.26189241)
\lineto(188.52221858,4.3615069)
\closepath
}
}
{
\newrgbcolor{curcolor}{0.80000001 0.80000001 0.80000001}
\pscustom[linewidth=0.99999871,linecolor=curcolor]
{
\newpath
\moveto(346.49940661,170.67854375)
\lineto(385.93114205,170.76732485)
\lineto(229.28930268,5.26189241)
\lineto(188.52221858,4.3615069)
\closepath
}
}
{
\newrgbcolor{curcolor}{0 0 0}
\pscustom[linewidth=2.00000125,linecolor=curcolor]
{
\newpath
\moveto(503.98631055,5.61058028)
\lineto(5.5999937,3.72275162)
\lineto(5.5999937,167.96370532)
}
}
{
\newrgbcolor{curcolor}{0 0 0}
\pscustom[linestyle=none,fillstyle=solid,fillcolor=curcolor]
{
\newpath
\moveto(501.16511689,11.1999376)
\lineto(516.58622801,5.65830729)
\lineto(501.20754089,0.00001097)
\curveto(503.99516656,3.2725956)(503.97750757,7.93456506)(501.16511689,11.1999376)
\closepath
}
}
{
\newrgbcolor{curcolor}{0 0 0}
\pscustom[linestyle=none,fillstyle=solid,fillcolor=curcolor]
{
\newpath
\moveto(-0.00000979,165.16370358)
\lineto(5.5999937,180.56371318)
\lineto(11.19999719,165.16370358)
\curveto(7.93799516,167.96370532)(3.27599225,167.96370532)(-0.00000979,165.16370358)
\closepath
}
}
{
\newrgbcolor{curcolor}{0 0 0}
\pscustom[linewidth=1.49999998,linecolor=curcolor]
{
\newpath
\moveto(312.49049953,5.7360606)
\lineto(155.60122961,171.02042847)
}
}
{
\newrgbcolor{curcolor}{0 0 0}
\pscustom[linewidth=1.49999998,linecolor=curcolor]
{
\newpath
\moveto(66.30753638,4.35214123)
\lineto(223.1968252,169.6365091)
}
}
{
\newrgbcolor{curcolor}{0 0 0}
\pscustom[linewidth=3.99998762,linecolor=curcolor]
{
\newpath
\moveto(26.42963906,3.83580863)
\lineto(66.62276787,4.33579855)
}
}
{
\newrgbcolor{curcolor}{0 0 0}
\pscustom[linewidth=3.99998762,linecolor=curcolor]
{
\newpath
\moveto(189.73374236,4.7029154)
\lineto(229.92687118,5.20290532)
}
}
{
\newrgbcolor{curcolor}{0 0 0}
\pscustom[linewidth=3.99998762,linecolor=curcolor]
{
\newpath
\moveto(421.68123969,5.16727572)
\lineto(461.87439874,5.66726564)
}
}
{
\newrgbcolor{curcolor}{0 0 0}
\pscustom[linewidth=3.99998762,linecolor=curcolor]
{
\newpath
\moveto(311.55567118,4.96973115)
\lineto(351.74879244,5.46972107)
}
}
{
\newrgbcolor{curcolor}{0 0 0}
\pscustom[linewidth=1.49999998,linecolor=curcolor]
{
\newpath
\moveto(27.23871874,3.2655384)
\lineto(4.70387528,29.61521777)
}
}
{
\newrgbcolor{curcolor}{0 0 0}
\pscustom[linewidth=1.5090746,linecolor=curcolor]
{
\newpath
\moveto(188.8727622,4.91506784)
\lineto(347.13878929,170.75044548)
}
}
{
\newrgbcolor{curcolor}{0 0 0}
\pscustom[linewidth=1.49999998,linecolor=curcolor]
{
\newpath
\moveto(229.66205102,5.46924485)
\lineto(386.55132472,170.75361273)
}
}
{
\newrgbcolor{curcolor}{0 0 0}
\pscustom[linewidth=1.49999998,linecolor=curcolor]
{
\newpath
\moveto(461.29514835,5.93360517)
\lineto(501.82676409,46.00060359)
}
}
{
\newrgbcolor{curcolor}{0 0 0}
\pscustom[linewidth=1.49999998,linecolor=curcolor]
{
\newpath
\moveto(423.00097512,5.46924485)
\lineto(266.11172787,170.75361273)
}
}
{
\newrgbcolor{curcolor}{0 0 0}
\pscustom[linewidth=1.49999998,linecolor=curcolor]
{
\newpath
\moveto(352.49812157,4.91034721)
\lineto(195.60889323,170.19471509)
}
\rput[bl](35,15){$\psi_0$}
\rput[bl](215,15){$\psi_1$}
\rput[bl](305,15){$\psi_3$}
\rput[bl](430,15){$\psi_4$}
\rput[bl](20,150){$t$}
\rput[bl](505,15){$\vec{x}$}
}
\end{pspicture}
\caption{Probing with many wave functions.}
\label{figmmt4}
\end{figure}

For the introductory considerations, following~\cite[Section~2.1.1]{mmt-cfs},
we begin with a quantum particle described by a quantum mechanical wave function~$\psi$ satisfying the Klein-Gordon equation~\eqref{KG1}
in Minkowski space or in a curved spacetime.
Suppose that we have access only to the information contained in the
absolute square~$|\psi(x)|^2$ of this wave function. We ask the question: Given this information, what can we infer
on the structure of spacetime? First, let the wave function~$\psi$ be a solution evolved from compactly supported initial data~$\psi_0$ as illustrated in Figure~\ref{figmmt1}.
Then, finite speed of propagation guarantees that the absolute square~$|\psi(x)|^2$ vanishes outside the causal 
future of the support of the initial data. In this way, the support of~$|\psi(x)|^2$ gives us some information on the 
causal structure of our spacetime. But, of course, there is only a limited amount of information that can be 
extracted  from a single wave function. However, if instead we probe with many wave
functions, as illustrated in Figure~\ref{figmmt4}, 
we gain more information.
If we aggregate the information contained in all wave functions evolved from compactly supported initial data,
then we can extract the complete causal structure of our spacetime. We remark that this determines the metric up to a conformal factor~\cite{hawking-king-mccarthy,malament1977class}.

\begin{figure}[t]
\psset{xunit=.5pt,yunit=.5pt,runit=.5pt}
\begin{pspicture}(525.68616972,189.66897799)
{
\newrgbcolor{curcolor}{0.79607844 0.79607844 0.79607844}
\pscustom[linestyle=none,fillstyle=solid,fillcolor=curcolor]
{
\newpath
\moveto(186.78460346,174.39493201)
\lineto(146.95739339,173.52453705)
\lineto(304.91785701,7.20749642)
\lineto(343.14894992,7.72464539)
\closepath
}
}
{
\newrgbcolor{curcolor}{0.80000001 0.80000001 0.80000001}
\pscustom[linewidth=0.99999871,linecolor=curcolor]
{
\newpath
\moveto(186.78460346,174.39493201)
\lineto(146.95739339,173.52453705)
\lineto(304.91785701,7.20749642)
\lineto(343.14894992,7.72464539)
\closepath
}
}
{
\newrgbcolor{curcolor}{0.79607844 0.79607844 0.79607844}
\pscustom[linestyle=none,fillstyle=solid,fillcolor=curcolor]
{
\newpath
\moveto(219.21427654,173.45811185)
\lineto(274.70850898,173.52671028)
\curveto(208.15830803,131.00071343)(158.93693858,62.48147311)(101.05116094,6.95884886)
\lineto(62.84993008,6.78782524)
\closepath
}
}
{
\newrgbcolor{curcolor}{0.80000001 0.80000001 0.80000001}
\pscustom[linewidth=0.99999871,linecolor=curcolor]
{
\newpath
\moveto(219.21427654,173.45811185)
\lineto(274.70850898,173.52671028)
\curveto(208.15830803,131.00071343)(158.93693858,62.48147311)(101.05116094,6.95884886)
\lineto(62.84993008,6.78782524)
\closepath
}
}
{
\newrgbcolor{curcolor}{0 0 0}
\pscustom[linewidth=2.00000125,linecolor=curcolor]
{
\newpath
\moveto(404.79584126,6.9559991)
\lineto(8.4,6.52802429)
\lineto(8.80885039,161.36545295)
}
}
{
\newrgbcolor{curcolor}{0 0 0}
\pscustom[linestyle=none,fillstyle=solid,fillcolor=curcolor]
{
\newpath
\moveto(401.98979502,12.55297626)
\lineto(417.39584177,6.96960288)
\lineto(402.00188727,1.35297581)
\curveto(404.79836552,4.617999)(404.79333212,9.27999919)(401.98979502,12.55297626)
\closepath
}
}
{
\newrgbcolor{curcolor}{0 0 0}
\pscustom[linestyle=none,fillstyle=solid,fillcolor=curcolor]
{
\newpath
\moveto(3.20147301,158.58024781)
\lineto(8.84212077,173.96541688)
\lineto(14.40144094,158.55067413)
\curveto(11.1468437,161.35927945)(6.48485705,161.37158949)(3.20147301,158.58024781)
\closepath
}
}
{
\newrgbcolor{curcolor}{0 0 0}
\pscustom[linewidth=1.49999998,linecolor=curcolor]
{
\newpath
\moveto(304.59402709,7.20749642)
\lineto(145.9438148,173.00820067)
}
}
{
\newrgbcolor{curcolor}{0 0 0}
\pscustom[linewidth=1.49999998,linecolor=curcolor]
{
\newpath
\moveto(101.59512567,7.47518524)
\lineto(181.36025953,90.12086713)
}
}
{
\newrgbcolor{curcolor}{0 0 0}
\pscustom[linewidth=3.99998762,linecolor=curcolor]
{
\newpath
\moveto(61.84010079,7.14107122)
\lineto(101.83100598,7.6413824)
}
}
{
\newrgbcolor{curcolor}{0 0 0}
\pscustom[linewidth=1.50047261,linecolor=curcolor]
{
\newpath
\moveto(62.56143496,8.15771878)
\lineto(219.4507389,173.44208665)
}
}
{
\newrgbcolor{curcolor}{0 0 0}
\pscustom[linewidth=3.99998762,linecolor=curcolor]
{
\newpath
\moveto(303.37705701,7.20749642)
\lineto(344.31969638,7.20749642)
}
}
{
\newrgbcolor{curcolor}{0 0 0}
\pscustom[linewidth=1.49999998,linecolor=curcolor]
{
\newpath
\moveto(345.50616567,6.70458492)
\lineto(188.61680504,173.52453705)
}
}
{
\newrgbcolor{curcolor}{0.60000002 0.60000002 0.60000002}
\pscustom[linestyle=none,fillstyle=solid,fillcolor=curcolor]
{
\newpath
\moveto(194.966689,73.21791798)
\curveto(194.966689,60.10930335)(170.80770487,49.48267155)(141.00608768,49.48267155)
\curveto(111.20447048,49.48267155)(87.04548635,60.10930335)(87.04548635,73.21791798)
\curveto(87.04548635,86.32653261)(111.20447048,96.95316441)(141.00608768,96.95316441)
\curveto(170.80770487,96.95316441)(194.966689,86.32653261)(194.966689,73.21791798)
\closepath
}
}
{
\newrgbcolor{curcolor}{0.60000002 0.60000002 0.60000002}
\pscustom[linewidth=2.4999988,linecolor=curcolor]
{
\newpath
\moveto(194.966689,73.21791798)
\curveto(194.966689,60.10930335)(170.80770487,49.48267155)(141.00608768,49.48267155)
\curveto(111.20447048,49.48267155)(87.04548635,60.10930335)(87.04548635,73.21791798)
\curveto(87.04548635,86.32653261)(111.20447048,96.95316441)(141.00608768,96.95316441)
\curveto(170.80770487,96.95316441)(194.966689,86.32653261)(194.966689,73.21791798)
\closepath
}
}
{
\newrgbcolor{curcolor}{0 0 0}
\pscustom[linewidth=1.49999998,linecolor=curcolor]
{
\newpath
\moveto(180.86804787,90.6417087)
\curveto(183.85080567,94.1623802)(186.8901052,97.63514618)(189.98438173,101.05821894)
\curveto(215.20479496,128.95850618)(244.09594583,153.53668319)(275.67792,173.95910335)
}
\rput[bl](85,15){$\psi_0$}
\rput[bl](300,15){$\phi_0$}
\rput[bl](20,160){$t$}
\rput[bl](405,20){$\vec{x}$}
\rput[bl](102,67){EM field}
}
\end{pspicture}
\caption{Probing an electromagnetic field.}
\label{figmmt3}
\end{figure}
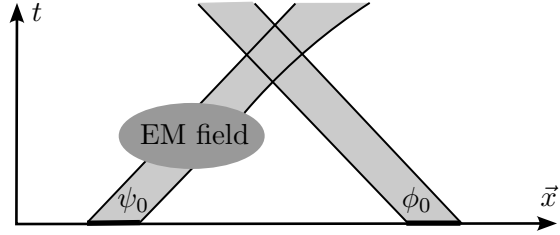

We next consider the situation if an electromagnetic background field is present. 
The coupling of the scalar field to the
electromagnetic field is described by the Klein-Gordon equation~\eqref{KG2}. Now the
wave functions are deflected by the electromagnetic force. Therefore, their absolute square also
encodes information on the electromagnetic field.
In order to retrieve this information, one can use the following procedure.
Suppose that we have access to two wave functions~$\psi$ and~$\phi$ and that we can also
measure the absolute value of superpositions; that is,
\beq \big| \alpha \psi(x) + \beta \phi(x) \big|^2
= \big| \alpha \psi(x) \big|^2 + 2 \re \Big( \overline{\alpha}\, \beta\:\overline{\psi(x)} \:\phi(x) \Big)
+ \big| \beta \phi(x) \big|^2 \eeq
for arbitrary complex coefficients~$\alpha$ and~$\beta$. By varying these coefficients, we can determine
the quantity
\beq %\label{loccorrprelim}
\overline{\psi(x)} \phi(x) \:, \eeq
which tells us about the correlation of the two wave function~$\psi$ and~$\phi$ at the spacetime point~$x$.
This allows us to probe the electromagnetic field, as shown schematically in Figure~\ref{figmmt3}.
Here, we do not need to be specific on what ``probing'' exactly means
(for example, one could determine deflection angles,
recover the Aharanov-Bohm phase shifts of the wave function, etc.).
All that counts is that we can get information also on the electromagnetic field.
Generally speaking, the more wave functions we have to our disposal, the more information on the
electromagnetic field can be retrieved. It seems sensible to expect that,
after suitably increasing the number of wave functions, we can recover both the 
spacetime structures and the matter fields therein from the knowledge of the absolute squares
of all these wave functions alone.

Now we go one step further and formulate the idea of encoding spacetime
structures in a family of wave functions in mathematical terms. To this end, we consider a
(for simplicity finite) number~$f$ of linearly independent wave functions~$\psi_1,\dots,\psi_f: \, \scrM \rightarrow \C$, 
mapping from a classical spacetime~$\scrM$ to the complex numbers. On the complex vector space~$\H$ spanned 
by these wave functions we introduce a scalar product~$\la.|.\ra_\H$ by demanding that the wave functions 
$\psi_1,\ldots,\psi_f$ 
are orthonormal; that is,
\beq \la \psi_k|\psi_l\ra_\H = \delta_{kl} \:. \eeq
We thus obtain an $f$-dimensional Hilbert space~$(\H, \la.|.\ra_\H)$. At any spacetime point~$x \in \scrM$ we can 
now introduce the {\em{local correlation operator}}~$F(x): \H \rightarrow \H$
as the linear operator whose matrix representation in the basis~$\psi_1,\ldots,\psi_f$ is given by
\beq \label{Fnophase}
\left(F(x)\right)^j_{\hphantom{j}k}=\overline{\psi_j(x)}\psi_k(x) \:.
\eeq
The diagonal entries of this matrix are the absolute squares of the wave functions,
whereas the off-diagonal entries tell us about the correlation of two different wave functions at the spacetime point~$x$. This is why we refer to~$F(x)$ as the local correlation operator.
Alternatively, the local correlation operator can be characterized in a basis-invariant form by the identity
\beq \la \psi|F(x)\phi\ra_\H=\overline{\psi(x)}\phi(x) \qquad \text{for all~$\psi, \phi \in \H$} \:. \eeq
By construction, the operator~$F(x)$ is positive semi-definite and has rank at most one
(in order not to distract from the main construction, this will be explained in more detail after~\eqref{Finv} in Section~\ref{secex2d} below). By varying the point~$x$, we obtain a
map~$F: \scrM \to \F$ from the classical spacetime~$\scrM$ to the set~$\F$ of
positive semi-definite linear operators of rank at most one,
\beq \F:= \big\{ y \in \Lin(\H) \,\big|\, y \text{ positive semi-definite of rank at most one} \big\} \:. \eeq
This map encodes all the physical information\footnote{Here by ``physical'' we mean the
information up to local gauge phases, which drop out in~\eqref{Fnophase}. Local gauge
freedom and local gauge transformations will be discussed in Section~\ref{secphysicalconcepts}.}
contained in the wave functions of~$\H$.

Next, we need to formalize the idea that we want to restrict attention to the information encoded
in the wave functions. This entails that we want to disregard all the information contained in the
usual structures of Minkowski space or a curved spacetime
(like the causal structure, the metric, the spinor bundle, and all that).
In order to do so mathematically, we focus on the family of all local correlation operators.
Thus, instead of considering~$F$ as a mapping from our classical spacetime to~$\F$, we restrict attention to
its image~$M:=F(\F)$ as a subset of~$\F$,
\beq \label{Msubset}
M \subset \F \:.
\eeq
In this way, Minkowski space and the corresponding classical spacetime structures no longer enter
our description. Instead, spacetime and all structures therein are encoded in and
must be recovered from the information
contained in the family of wave functions. This point of view of recovering all spacetime structures
from the wave functions will be taken seriously in this book, and we will unravel its consequences step by step.

It turns out that working as in~\eqref{Msubset} merely with a subset of~$\F$ is not quite sufficient.
In order to get into the position to formulate physical equations, we need one more structure: a measure~$\rho$
on spacetime. Here by a ``measure on spacetime'' we mean a mapping which to a subset~$\Omega \subset M$
associates a non-negative number, which can be thought of as the ``volume'' of the spacetime region
corresponding to~$\Omega$.
In non-technical terms, this measure can be obtained by combining the volume measure in Minkowski
space with the map~$F$. More precisely, we take the pre-image~$F^{-1}(\Omega) \subset \F$ and
integrate over it,
\beq \rho (\Omega):= \int_{F^{-1}(\Omega)} \dd\mu \:, \eeq
where~$d\mu=\dd^4x$ is the volume measure on Minkowski space~$\scrM$
(and similarly~$\dd\mu = \sqrt{|\det g|}\: \dd^4x$ in curved spacetime).
In more mathematical terms, the measure~$\rho$ is the {\em{push-forward}} of~$\mu$
under~$F$ (for basics on measure theory and the push-forward measure see Section~\ref{secbasicmeasure}).

This construction leads us to consider a measure~$\rho$ on a set of linear
operators on a Hilbert space as the basic structure describing a physical system in spacetime.
These are indeed all the basic ingredients to define a causal fermion system.
The only modification to be made later is that, instead of complex wave functions, we will
work with sections of a spinor bundle.
One consequence of that is that the local correlation operators will no longer be positive semi-definite.
Instead, they will be of finite rank with a fixed upper bound on the number of positive and negative eigenvalues.

Before coming to these generalizations (Section~\ref{sectowards}), we next explain why
encoding information in the wave functions also has benefits if one wants to formulate
physical equations in a setting that goes beyond a classical continuous spacetime.

\section{Motivating Example: Formulating Equations in Discrete Spacetimes} \label{secex2d}
It is generally believed that for distances as small as the Planck length, spacetime can no longer
be described by Minkowski space or a Lorentzian manifold, but that it should have a different,
possibly discrete structure. There are different approaches to modeling such spacetimes.
The simplest approach is to replace Minkowski space with a discrete lattice.
Indeed, causal fermion systems provide another, more general approach.
In any such approach, one faces the challenge of how to formulate physical 
equations if one gives up the continuous structure of spacetime and thus can no longer work with
partial differential equations like the Klein-Gordon equation or the Dirac equation.

\begin{figure}
% \usepackage[usenames,dvipsnames]{pstricks}
% \usepackage{epsfig}
% \usepackage{pst-grad} % For gradients
% \usepackage{pst-plot} % For axes
% \usepackage[space]{grffile} % For spaces in paths
% \usepackage{etoolbox} % For spaces in paths
% \makeatletter % For spaces in paths
% \patchcmd\Gread@eps{\@inputcheck#1 }{\@inputcheck"#1"\relax}{}{}
% \makeatother
% 
\psscalebox{1.0 1.0} % Change this value to rescale the drawing.
{
\begin{pspicture}(0,-1.6720673)(7.0918274,1.6720673)
\definecolor{colour0}{rgb}{0.0,0.6,0.4}
\definecolor{colour1}{rgb}{0.8,0.2,0.0}
\definecolor{colour2}{rgb}{0.0,0.4,0.0}
\psdots[linecolor=colour0, dotsize=0.12](1.105,-0.78706735)
\psdots[linecolor=colour1, dotsize=0.12](1.105,0.81293267)
\psdots[linecolor=black, dotsize=0.12](1.105,1.6129327)
\psdots[linecolor=black, dotsize=0.12](1.905,1.6129327)
\psdots[linecolor=black, dotsize=0.12](2.705,1.6129327)
\psdots[linecolor=black, dotsize=0.12](3.505,1.6129327)
\psdots[linecolor=black, dotsize=0.12](4.305,1.6129327)
\psdots[linecolor=black, dotsize=0.12](5.105,1.6129327)
\rput[bl](5.375,-0.90206736){\normalsize{$t-\Delta t$}}
\rput[bl](5.385,0.6729327){\normalsize{$t+\Delta t$}}
\rput[bl](5.665,-0.087067336){\normalsize{$t$}}
\psdots[linecolor=colour0, dotsize=0.12](1.905,-0.78706735)
\psdots[linecolor=colour0, dotsize=0.12](2.705,-0.78706735)
\psdots[linecolor=colour0, dotsize=0.12](3.505,-0.78706735)
\psdots[linecolor=colour0, dotsize=0.12](4.305,-0.78706735)
\psdots[linecolor=colour0, dotsize=0.12](5.105,-0.78706735)
\psdots[linecolor=colour0, dotsize=0.12](1.105,0.012932663)
\psdots[linecolor=colour0, dotsize=0.12](1.905,0.012932663)
\psdots[linecolor=colour0, dotsize=0.12](2.705,0.012932663)
\psdots[linecolor=colour0, dotsize=0.12](3.505,0.012932663)
\psdots[linecolor=colour0, dotsize=0.12](4.305,0.012932663)
\psdots[linecolor=colour0, dotsize=0.12](5.105,0.012932663)
\psdots[linecolor=colour1, dotsize=0.12](1.905,0.81293267)
\psdots[linecolor=colour1, dotsize=0.12](2.705,0.81293267)
\psdots[linecolor=colour1, dotsize=0.12](3.505,0.81293267)
\psdots[linecolor=colour1, dotsize=0.12](4.305,0.81293267)
\psdots[linecolor=colour1, dotsize=0.12](5.105,0.81293267)
\psbezier[linecolor=colour1, linewidth=0.04, arrowsize=0.05291667cm 2.0,arrowlength=1.4,arrowinset=0.0]{->}(6.895,0.057932664)(7.18,0.36793268)(7.03,0.53293264)(6.9,0.7579326629638672)
\psbezier[linecolor=colour1, linewidth=0.04, arrowsize=0.05291667cm 2.0,arrowlength=1.4,arrowinset=0.0]{->}(6.89,0.90793264)(7.175,1.2179327)(7.025,1.3829327)(6.895,1.6079326629638673)
\rput[bl](2.405,-0.30706733){\normalsize{$+$}}
\rput[bl](4.395,-0.30206734){\normalsize{$+$}}
%\rput[bl](3.35,-0.39706734){\normalsize{$-$}}
\rput[bl](3.39,-1.0970674){\normalsize{$-$}}
\pspolygon[linecolor=colour2, linewidth=0.02](2.7089655,0.014901032)(3.5057106,-0.7929771)(4.3039584,0.008308901)
\psline[linecolor=colour1, linewidth=0.02, arrowsize=0.05291667cm 2.0,arrowlength=1.4,arrowinset=0.0]{->}(3.495,0.22793266)(3.5,0.6029327)
\psline[linecolor=black, linewidth=0.04](1.095,-1.0920674)(1.195,-1.1920674)(1.395,-1.1920674)(1.495,-1.2920673)(1.595,-1.1920674)(1.795,-1.1920674)(1.895,-1.0920674)
\psline[linecolor=black, linewidth=0.04](0.79,-0.78206736)(0.69,-0.68206733)(0.69,-0.48206735)(0.59,-0.38206732)(0.69,-0.28206733)(0.69,-0.08206734)(0.79,0.017932663)
\rput[bl](0.0,-0.48206735){\normalsize{$\Delta t$}}
\rput[bl](1.215,-1.6720673){\normalsize{$\Delta x$}}
\end{pspicture}
}
\caption{Time evolution of a lattice system~$\scrM \subset \R^{1,1}$.}
\label{lattice2}
\end{figure}
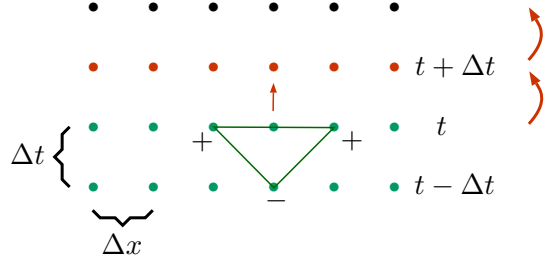

In order to explain the underlying problem more concretely, we now have a closer look at the simple example of a
spacetime lattice (this example was first given in~\cite[Section~1]{dice2018}).
For simplicity, we consider a two-dimensional lattice
(one space and one time dimension), but higher-dimensional lattices could be described similarly.
Thus let~$\scrM \subset \R^{1,1}$ be a rectangular lattice in two-dimensional Minkowski space.
We denote the spacing in time direction by~$\Delta t$ and in spatial direction by~$\Delta x$
(see Figure~\ref{lattice2}).
The usual procedure for setting up equations on a lattice is to replace derivatives
with difference quotients, giving rise to an evolution equation that can be
solved time step by time step according to deterministic rules.
A simple example is the discretization of the
two-dimensional wave equation for a function~$\phi: \scrM \to \C$ on the lattice,
\begin{align} %\label{wavelattice}
0 = \Box \phi(t,x) &:= \frac{1}{(\Delta t)^2} \Big( \phi(t+\Delta t, x) 
- 2 \phi(t, x) + \phi(t-\Delta t, x)\Big) \notag \\
&\quad\; - \frac{1}{(\Delta x)^2} \Big( \phi(t, x+\Delta x) 
- 2 \phi(t, x) + \phi(t, x - \Delta x)\Big) \:.
\end{align}
Solving this equation for~$\phi(t+\Delta t, x)$ gives a deterministic rule
for computing~$\phi(t+\Delta t, x)$ from the values of~$\phi$
at earlier times~$t$ and~$t-\Delta t$ (see again Figure~\ref{lattice2}).

While this method for setting up equations in a discrete spacetime is very simple and yields well-defined evolution equations, it
also has several drawbacks:
\bitem
\itemD The above method of discretizing the continuum equations
is very {\em{ad hoc}}. Why do we choose a regular lattice, why do we work with difference quotients?
There are many other ways of discretizing the wave equation.
\itemD The method is {\em{not background-free}}. In order to speak of the ``lattice spacing,''
the lattice must be thought of as being embedded in a two-dimensional ambient spacetime.
\itemD The concept of a spacetime lattice is not invariant under general coordinate transformations.
In other words, the assumption of a spacetime lattice is
{\em{not compatible with the equivalence principle}}.
\eitem
In view of these shortcomings, the following basic question arises:
\begin{quote}
Can one formulate equations without referring
to the nearest neighbor relation and the lattice spacing?
\end{quote}
The answer to this question is yes, 
and we will now see how this can be done in the example of our two-dimensional lattice system.
Although our example is somewhat oversimplified, this consideration will lead us quite naturally to the setting
of causal fermion systems.

In order to formulate the equations, we consider on our lattice a family of
complex-valued wave functions~$\psi_1, \ldots, \psi_f: \scrM \to \C$
(for simplicity a finite number, that is, $f < \infty$).
At this stage, these wave functions do not need to satisfy any wave equation.
On the complex vector space~$\H$ spanned by these wave functions we introduce
a scalar product~$\la .|. \ra_\H$ by demanding that the wave functions~$\psi_1,
\ldots, \psi_f$ are orthonormal; that is,
\beq %\label{ortho}
\la \psi_k | \psi_l \ra_\H = \delta_{kl} \:. \eeq
We thus obtain an $f$-dimensional Hilbert space~$(\H, \la .|. \ra_\H)$.
Note that the scalar product is given abstractly (meaning that it has no
representation in terms of the wave functions as a sum over lattice points).
Next, for any lattice point~$(t,x) \in \scrM$ we introduce the
so-called {\em{local correlation operator}}~$F(t,x) : \H \rightarrow \H$
\sindex{local correlation operator}%
as the linear operator whose matrix representation in the basis
$\psi_1,\ldots,\psi_f$ is given by
\beq \label{matrix}
(F(t,x))^j_k = \overline{\psi_j(t,x)} \psi_k(t,x) \:.
\eeq
The diagonal elements of this matrix are the absolute squares~$|\psi_k(t,x)|^2$
of the corresponding wave functions. The off-diagonal elements, on the other hand,
tell us about the correlation of the~$j^\text{th}$ and~$k^\text{th}$ wave function
at the lattice point~$(t,x)$. This is the reason for the name ``local correlation operator.''
This operator can also be characterized in a basis-invariant way by the relations
\beq \label{Finv}
\la \psi \,|\, F(t,x) \,\phi \ra_\H = \overline{\psi(t,x)} \phi(t,x) \:,
\eeq
to be satisfied for all~$\psi, \phi \in \H$.

We now analyze some properties of the local correlation operators.
Taking the complex conjugate, one sees immediately that the matrix defined
by~\eqref{matrix} is Hermitian. Stated equivalently independent of bases,
the local correlation operator is a {\em{symmetric}} linear operator on~$\H$
(see Definition~\ref{defsymmetric} in the preliminaries). Moreover, a local correlation operator
has {\em{rank at most one}} and is {\em{positive semi-definite}}. This can be seen
in detail by expressing it in terms of the operator
\beq e(t,x) : \H \rightarrow \C \:,\qquad \psi \mapsto \psi(t,x) \:, \eeq
which to every vector associates the corresponding wave function evaluated at the spacetime
point~$(t,x)$ (this mapping is sometimes referred to as the {\em{evaluation map}}). Indeed, rewriting the right-hand
side of~\eqref{Finv} as
\beq \overline{\psi(t,x)} \phi(t,x) = \overline{ \big(e(t,x) \psi \big) } \big(e(t,x) \phi \big)
= \la \psi \,|\, e(t,x)^* \,e(t,x) \,\phi \ra_\H \:, \eeq
(where~$e(t,x)^*:\C \to \H$ is the adjoint of the operator~$e(t,x)$ as defined by~\eqref{Addef}),
we can compare with the left-hand side of~\eqref{Finv} to conclude that
\beq \label{ees}
F(t,x) = e(t,x)^* \,e(t,x) \:.
\eeq
This shows that~$F(t,x)$ is positive semi-definite. Moreover, being a mapping to~$\C$,
the operator~$e(t,x)$ has rank at most one. As a consequence, also~$F(t,x)$ has rank at most one.

It is useful to denote the set of all operators with the above properties by
\beq \label{Fone}
\begin{split}
\F := \big\{ F \in \Lin(\H) \:\big|\: &\text{$F$ is symmetric,} \\
& \text{positive semi-definite and has rank at most one} \big\} \:.
\end{split}
\eeq
Varying the lattice point, we obtain a mapping (see Figure~\ref{embed})
\beq F : \scrM \rightarrow \F \:,\qquad (t,x) \mapsto F(t,x) \:. \eeq
\begin{figure}
% \usepackage[usenames,dvipsnames]{pstricks}
% \usepackage{epsfig}
% \usepackage{pst-grad} % For gradients
% \usepackage{pst-plot} % For axes
% \usepackage[space]{grffile} % For spaces in paths
% \usepackage{etoolbox} % For spaces in paths
% \makeatletter % For spaces in paths
% \patchcmd\Gread@eps{\@inputcheck#1 }{\@inputcheck"#1"\relax}{}{}
% \makeatother
% 
\psscalebox{1.0 1.0} % Change this value to rescale the drawing.
{
\begin{pspicture}(-1.5,-1.7327906)(10.322246,1.7327906)
\definecolor{colour0}{rgb}{0.0,0.0,0.6}
\pspolygon[linecolor=white, linewidth=0.02, fillstyle=solid](1.1613568,0.65876496)(1.2102457,0.52543163)(1.3169124,0.42765382)(1.4813569,0.40098715)(1.6502457,0.40987605)(1.743579,0.4632094)(1.5791346,0.5832094)(1.3613569,0.65876496)
\rput[bl](5.597801,-1.7327906){$0$}
\rput[bl](6.562246,-0.23501284){$\F \subset \Lin(\H)$}
\pspolygon[linecolor=white, linewidth=0.02, fillstyle=solid,fillcolor=yellow](3.6458013,1.7209872)(5.479579,-1.5207906)(7.2120233,1.722765)
\psline[linecolor=black, linewidth=0.04](3.663579,1.6898761)(5.472468,-1.5385684)(7.201357,1.7103205)
\rput[bl](0.4880235,0.9576538){$\scrM$}
\psbezier[linecolor=colour0, linewidth=0.04, arrowsize=0.05291667cm 2.0,arrowlength=1.4,arrowinset=0.0]{->}(2.4033267,0.116877794)(2.6445446,-0.122063294)(3.155987,-0.51524955)(4.3238316,-0.02157013655146784)
\rput[bl](3.2280235,0.03876494){\textcolor{colour0}{$F$}}
\pspolygon[linecolor=yellow, linewidth=0.02, fillstyle=solid,fillcolor=yellow](5.050246,0.51400304)(5.084468,0.4511459)(5.1581774,0.39114588)(5.2397842,0.3654316)(5.3398185,0.3711459)(5.392468,0.40257445)(5.2950664,0.4797173)(5.163442,0.52543163)
\psdots[linecolor=black, dotsize=0.12](1.6591346,0.6632094)
\psdots[linecolor=black, dotsize=0.12](0.0591346,0.6632094)
\psdots[linecolor=black, dotsize=0.12](0.4591346,0.6632094)
\psdots[linecolor=black, dotsize=0.12](0.8591346,0.6632094)
\psdots[linecolor=black, dotsize=0.12](1.2591347,0.6632094)
\psdots[linecolor=black, dotsize=0.12](1.6591346,0.26320937)
\psdots[linecolor=black, dotsize=0.12](0.0591346,0.26320937)
\psdots[linecolor=black, dotsize=0.12](0.4591346,0.26320937)
\psdots[linecolor=black, dotsize=0.12](0.8591346,0.26320937)
\psdots[linecolor=black, dotsize=0.12](1.2591347,0.26320937)
\psdots[linecolor=black, dotsize=0.12](1.6591346,-0.13679062)
\psdots[linecolor=black, dotsize=0.12](0.0591346,-0.13679062)
\psdots[linecolor=black, dotsize=0.12](0.4591346,-0.13679062)
\psdots[linecolor=black, dotsize=0.12](0.8591346,-0.13679062)
\psdots[linecolor=black, dotsize=0.12](1.2591347,-0.13679062)
\psdots[linecolor=black, dotsize=0.12](1.6591346,-0.5367906)
\psdots[linecolor=black, dotsize=0.12](0.0591346,-0.5367906)
\psdots[linecolor=black, dotsize=0.12](0.4591346,-0.5367906)
\psdots[linecolor=black, dotsize=0.12](0.8591346,-0.5367906)
\psdots[linecolor=black, dotsize=0.12](1.2591347,-0.5367906)
\psdots[linecolor=black, dotsize=0.08](4.7991347,0.90098715)
\psdots[linecolor=black, dotsize=0.08](5.1391344,0.9309872)
\psdots[linecolor=black, dotsize=0.08](5.4591346,0.9909872)
\psdots[linecolor=black, dotsize=0.08](5.7891345,1.0509871)
\psdots[linecolor=black, dotsize=0.08](6.1591344,1.2509872)
\psdots[linecolor=black, dotsize=0.08](6.1491346,0.96098715)
\psdots[linecolor=black, dotsize=0.08](5.8891344,0.77098715)
\psdots[linecolor=black, dotsize=0.08](5.4991345,0.69098717)
\psdots[linecolor=black, dotsize=0.08](5.1191344,0.64098716)
\psdots[linecolor=black, dotsize=0.08](4.8591347,0.54098713)
\psdots[linecolor=black, dotsize=0.08](5.0791345,0.28098717)
\psdots[linecolor=black, dotsize=0.08](5.2791348,0.35098717)
\psdots[linecolor=black, dotsize=0.08](5.509135,0.43098715)
\psdots[linecolor=black, dotsize=0.08](5.6891346,0.45098716)
\psdots[linecolor=black, dotsize=0.08](5.9291344,0.49098715)
\psdots[linecolor=black, dotsize=0.08](5.1791344,0.030987162)
\psdots[linecolor=black, dotsize=0.08](5.3591347,0.08098716)
\psdots[linecolor=black, dotsize=0.08](5.6091347,0.18098716)
\psdots[linecolor=black, dotsize=0.08](5.759135,0.19098715)
\psdots[linecolor=black, dotsize=0.08](6.0291348,0.21098717)
\rput[bl](4.8180237,1.1376538){$F(\scrM)$}
\end{pspicture}
}
\caption{Embedding in~$\F$.}
\label{embed}
\end{figure}%
For clarity, we note that the set~$\F$ is {\em{not}} a vector space,
because a linear combination of operators in~$\F$ in general has
a rank greater than one. But it is a {\em{conical}} set in the sense
that a positive multiple of any operator in~$\F$ is again in~$\F$
(this is why in Figure~\ref{embed} the set~$\F$ is depicted as a cone).

We point out that the local correlation operators do not involve the lattice spacing
or the nearest neighbor relation (as a matter of fact, we did not even use that~$\scrM$ is a lattice);
instead, they contain information only on
the local correlations of the wave functions at each lattice point.
With this in mind, our strategy for formulating equations which do not involve
the specific structures of the lattice is to work exclusively with the local correlation operators, that is,
with the subset~$F(\scrM) \subset \F$. In other words, in Figure~\ref{embed} we
want to disregard the lattice on the left and work exclusively with the objects on the right.

How can one set up equations purely in terms of the local correlation operators?
In order to explain the general procedure, we consider a finite number of
operators~$F_1, \ldots, F_L \in \F$.
Each of these operators can be thought of as encoding information
on the local correlations of the wave functions at a corresponding
spacetime point. However, this ``spacetime point'' is no longer a lattice point,
because the notions of lattice spacing and nearest lattice point have been dropped.
At this stage, spacetime is merely a point set, where each point is an operator on the Hilbert space.
In order to obtain a ``spacetime'' in the usual sense (like Minkowski space, a
Lorentzian manifold or a generalization thereof), one needs additional
structures and relations between the spacetime points.
Such relations can be obtained by multiplying the operators.
Indeed, the operator product~$F_i \,F_j$ tells us about correlations of the wave functions
at different spacetime points. Taking the trace of this operator product gives a
real number. Our method for formulating physical equations is to use the operators~$F_i$
and their products to set up a variational
principle. This variational formulation has the advantage that symmetries give rise to
conservation laws by Noether's theorem (as will be explained in Chapter~\ref{secOSI}).
Therefore, we want to minimize an action~$\Sact$ defined in terms of the operators~$F_1,\ldots,F_L$.
A simple example is to
\beq \label{Ssum}
\text{minimize} \qquad \Sact(F_1, \ldots, F_L) := \sum_{i,j=1}^L \Tr(F_i \,F_j)^2
\eeq
under variations of the points~$F_1,\ldots, F_L \in \F$.
In order to obtain a mathematically sensible variational principle,
one needs to impose certain constraints.
Here we do not enter the details, because the present example is a bit too simple
(see however Exercise~\ref{exexamplesimp}).
Instead, we merely use it as a motivation for the general setting of causal fermion
systems, which we now introduce.

\section{Toward the General Definition of a Causal Fermion System} \label{sectowards}
In order to get from the previous motivating examples to the general setting of causal fermion systems,
we extend the above constructions in several steps:
\bitem
\item[(a)] The previous example works similarly in higher dimensions,
in particular for a lattice~$\scrM \subset \R^{1,3}$ in four-dimensional Minkowski space.
This has no effect on the resulting structure of a finite number of distinguished
operators~$F_1, \ldots, F_L \in \F$.
\item[(b)] Suppose that we consider multi-component wave
functions~$\psi: \scrM \to \C^N$. Then, clearly, we cannot directly multiply two such wave functions
pointwise as was done on the right side of~\eqref{matrix}. However, assuming that we are given an inner product on~$\C^N$, which we denote by~$\Sl .|. \Sr$
(in mathematical terms, this inner product is a non-degenerate sesquilinear form;
we always use the convention that the wave function in the first argument is complex conjugated), we can adapt the  definition of the local correlation operator~\eqref{matrix} to
\beq %\beq \label{matrix2}
(F(t,x))^j_k = -\Sl \psi_j(t,x) | \psi_k(t,x) \Sr \eeq
\sindex{local correlation operator}%
(the minus sign compared to~\eqref{matrix} merely is a useful convention).
The resulting local correlation operator is no longer an operator of rank
at most one, but it has rank at most~$N$ (as can be seen, for example, by writing
it similar to~\eqref{ees} in the form~$F(t,x)=-e(t,x)^* \,e(t,x)$ with the evaluation
map~$e(t,x) : \H \rightarrow \C^N, \psi \mapsto \psi(t,x)$).
If the inner product~$\Sl .|. \Sr$ on~$\C^N$ is positive definite, then the operator~$F(t,x)$
is negative semi-definite.
However, in the physical applications in mind, this inner product will {\em{not}} be
positive definite. Indeed, a typical example in mind is that of four-component Dirac spinors.
The Lorentz invariant inner product~$\overline{\psi} \phi$ on Dirac spinors in Minkowski space
(with the usual adjoint spinor~$\overline{\psi} := \psi^\dagger \gamma^0$)
is indefinite of signature~$(2,2)$.
In order to describe systems involving leptons and quarks, one must take 
direct sums of Dirac spinors, giving the signature~$(n, n)$ with~$n \in 2\N$.
With this in mind, we assume more generally that 
\beq \Sl .|. \Sr \quad \text{has signature~$(n,n)$ with~$n \in \N$}\:. \eeq
Then the resulting local correlation operators are symmetric operators of rank at most~$2n$,
which (counting multiplicities) have at most~$n$ positive and at most~$n$ negative eigenvalues.
\item[(c)] Finally, it is useful to generalize the setting such as to allow for continuous spacetimes
and for spacetimes which may have both continuous and discrete components. In preparation, we note that
the sums over the operators~$F_1,\ldots, F_L$ in~\eqref{Ssum} can be written as integrals,
\beq \label{Scont}
\Sact(\rho) = \int_\F \Diff\rho(x) \int_\F \Diff\rho(y)\: \Tr(xy)^2 \:,
\eeq
if the measure~$\rho$ on~$\F$ is chosen as the sum of Dirac measures
supported at these operators,
\beq \label{deltasum}
\rho = \sum_{i=1}^L \delta_{F_i} \:.
\eeq
Note that, in this formulation, the measure plays a double role: First, it distinguishes the
points~$F_1, \ldots, F_L$
as those points where the measure is non-zero, as is made mathematically precise by the notion
of the {\em{support}} of the measure (for details see Definition~\ref{defsupp}); that is,
\beq \label{FL}
\text{supp}\, \rho = \{F_1, \ldots, F_L \}\:.
\eeq
Second, a measure makes it possible to integrate over its support, an operation which for
the measure~\eqref{deltasum} reduces to the sum over~$F_1, \ldots, F_L$.

Now one can extend the setting simply by considering~\eqref{Scont} for more general measures
on~$\F$ (like, for example, regular Borel measures).
The main advantage of working with measures is that we get into a mathematical framework
in which variational principles like~\eqref{Ssum} can be studied with powerful analytic methods.
\eitem

\section{Basic Definition of a Causal Fermion System} \label{secgendef}
Motivated by the previous considerations we now give the basic definition of a causal fermion system. This definition evolved over several years. Based on preparations in~\cite{pfp},
the present formulation was first given in~\cite{rrev}.
\begin{Def} \label{defcfs} 
\sindex{causal fermion system!basic definition}%
\nindex{ada@$(\H, \la . \vert . \ra_\H)$ -- Hilbert space of causal fermion system}%
\tindex{ff@$(\H, \la . \vert . \ra_\H)$ -- Hilbert space of causal fermion system}%
\sindex{spin dimension}%
\nindex{adb@$n$ -- spin dimension}%
Given a separable complex Hilbert space~$\H$ with scalar product~$\la .|. \ra_\H$
and a parameter~$n \in \N$ (the {\bf{spin dimension}}), we let~$\F \subset \Lin(\H)$ be the set of all
symmetric operators on~$\H$ of finite rank, which (counting multiplicities) have
at most~$n$ positive and at most~$n$ negative eigenvalues. 
\nindex{ade@$\F$ -- set of operators of causal fermion system}%
\nindex{adf@$\rho$ -- measure on~$\F$ of causal fermion system}%
Moreover, let~$\rho$ be a positive measure on~$\F$ (defined on a $\sigma$-algebra of subsets of~$\F$).
We refer to~$(\H, \F, \rho)$ as a {\bf{causal fermion system}}.
\nindex{adg@$(\H, \F, \rho)$ -- causal fermion system}%
\end{Def} \noindent

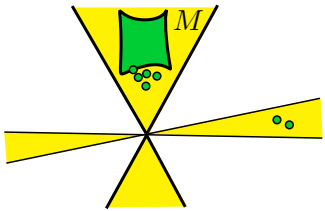
\begin{figure}
% \usepackage[usenames,dvipsnames]{pstricks}
% \usepackage{epsfig}
% \usepackage{pst-grad} % For gradients
% \usepackage{pst-plot} % For axes
% \usepackage[space]{grffile} % For spaces in paths
% \usepackage{etoolbox} % For spaces in paths
% \makeatletter % For spaces in paths
% \patchcmd\Gread@eps{\@inputcheck#1 }{\@inputcheck"#1"\relax}{}{}
% \makeatother
% \psscalebox{1.0 1.0} % Change this value to rescale the drawing.
{
\begin{pspicture}(0,-1.3610485)(4.2347274,1.3610485)
\definecolor{colour0}{rgb}{0.0,0.8,0.2}
\pspolygon[linecolor=white, linewidth=0.02, fillstyle=solid,fillcolor=yellow](0.91751134,1.3317788)(1.9034183,-0.41748416)(2.8348446,1.332738)
\psbezier[linecolor=black, linewidth=0.04, fillstyle=solid,fillcolor=colour0](1.553067,0.89983374)(1.5447986,1.547851)(1.4361526,1.1536276)(1.8759718,1.1576269552442744)(2.315791,1.1616263)(2.0934794,1.5804809)(2.1575115,0.90066475)(2.2215433,0.22084858)(2.306556,0.64546037)(1.8751786,0.54429364)(1.4438012,0.4431269)(1.5613352,0.25181648)(1.553067,0.89983374)
\rput[bl](2.2428446,1.0282936){$M$}
\pscircle[linecolor=black, linewidth=0.02, fillstyle=solid,fillcolor=colour0, dimen=outer](1.7152891,0.4931825){0.05111111}
\pscircle[linecolor=black, linewidth=0.02, fillstyle=solid,fillcolor=colour0, dimen=outer](1.7819558,0.39096028){0.05111111}
\pscircle[linecolor=black, linewidth=0.02, fillstyle=solid,fillcolor=colour0, dimen=outer](1.8930669,0.43984917){0.05111111}
\pscircle[linecolor=black, linewidth=0.02, fillstyle=solid,fillcolor=colour0, dimen=outer](1.884178,0.27540472){0.05111111}
\pscircle[linecolor=black, linewidth=0.02, fillstyle=solid,fillcolor=colour0, dimen=outer](2.0308447,0.40873808){0.05111111}
\pspolygon[linecolor=white, linewidth=0.02, fillstyle=solid,fillcolor=yellow](2.448178,-1.3365248)(1.8933822,-0.34726194)(1.3664002,-1.3508174)
\psline[linecolor=black, linewidth=0.04](0.904178,1.3406092)(1.892704,-0.36415082)(2.8375113,1.3514048)
\pspolygon[linecolor=yellow, linewidth=0.02, fillstyle=solid,fillcolor=yellow](1.953067,-0.35029832)(4.197794,0.09984918)(4.224178,-0.4068175)
\pspolygon[linecolor=yellow, linewidth=0.02, fillstyle=solid,fillcolor=yellow](0.010844693,-0.3490397)(1.8375113,-0.37570637)(0.041955803,-0.7357064)
\psline[linecolor=black, linewidth=0.02](0.0019558037,-0.33570638)(4.2286224,-0.41570637)
\psline[linecolor=black, linewidth=0.02](0.02862247,-0.7490397)(4.197511,0.11318251)
\psline[linecolor=black, linewidth=0.04](2.4064002,-1.331133)(1.8934299,-0.37748414)(1.3575114,-1.3508174)
\pscircle[linecolor=black, linewidth=0.02, fillstyle=solid,fillcolor=colour0, dimen=outer](3.6264002,-0.17348416){0.05111111}
\pscircle[linecolor=black, linewidth=0.02, fillstyle=solid,fillcolor=colour0, dimen=outer](3.7908447,-0.24015082){0.05111111}
\end{pspicture}
}
\caption{A causal fermion system.}
\label{cfs}
\end{figure}

The definition of a causal fermion system is illustrated in Figure~\ref{cfs}.
The set~$\F$ is invariant under the transformation where an operator
is multiplied by a real number, as is indicated in the figure by the double cones.
The support of the measure, denoted by
\sindex{spacetime!of causal fermion system}%
\nindex{adh@$M$ -- spacetime of causal fermion system}%
\beq \label{Mdef}
M := \supp \rho \:,
\eeq
is referred to as {\em{spacetime}} (intuitively speaking, the support of a measure consists
of all points where the measure is non-zero; for mathematical details see Definition~\ref{defsupp}).
In contrast to the example of the lattice system, where
spacetime consisted of discrete points~\eqref{FL}, in general, the measure~$\rho$ can also have
continuous components. For example, $M$ could be a subset of~$\F$ having the
additional structure of being a four-dimensional manifold.
The space~$\F$ should be thought of as a space of very large dimension\footnote{This statement
is made precise in~\cite{gaugefix, banach} as follows.
The operators of~$\F$ of maximal rank~$2n$ form a Banach manifold.
If the Hilbert space~$\H$ is finite-dimensional, then this manifold also has finite dimension
given by~$4n \dim \H - 4n^2$; see also Proposition~\ref{prppq} in the Preliminaries.},
so that~$M$ typically is a low-dimensional subset of~$\F$.
The measure~$\rho(\Omega)$ of a measurable subset~$\Omega \subset M$
can be regarded as the volume of the spacetime region~$\Omega$.
In the example of the lattice system, this volume is simply the number of spacetime points in~$\Omega$,
whereas for a continuous spacetime, it is the four-dimensional Lebesgue measure of~$\Omega$.
It is a specific feature of a causal fermion system that a spacetime point~$x \in M$ is a linear operator on
the Hilbert space~$\H$. This endows spacetime with a lot of additional structure.
In particular, as will be explained in Section~\ref{secinherent}, the spacetime point operators give rise to
a family of spinorial wave functions and to causal and geometric structures.
The general idea is that a causal fermion system describes a spacetime together with all structures therein.
Before entering these structures in more detail, we illustrate the general definition
by the simple and concrete example of Dirac wave functions in Minkowski space.

\section{Example: Dirac Wave Functions in Minkowski Space} \label{seclco}
As a further example, we now explain how to construct a causal fermion system in Minkowski space.
Recall that in Section~\ref{secspinorbundlemink} (and similarly in
curved spacetime in Section~\ref{secdirgh}), for a given parameter~$m \in \R$
we introduced the Hilbert space~$(\H_m, (.|.))$  of
all solutions of the Dirac equation with mass~$m$ (recall that the scalar product is defined as the
spatial integral~\eqref{printMink}). We now choose a closed subspace~$\H$
of this Hilbert space and denote the scalar product~$(.|.)$ restricted to this subspace
by~$\la .|. \ra_\H$ (changing the notation from round to pointed brackets clarifies that
we consider~$\la .|. \ra_\H$ as an abstract scalar product, without referring to its
representation as a spatial integral~\eqref{printMink}). We thus obtain the
\beq \text{Hilbert space} \qquad \big(\H, \la.|. \ra_\H \big) \:. \eeq
By construction, the vectors in this Hilbert space are solutions of the Dirac equation.
They can be thought of as the ``occupied states'' of the system.
We prefer the notion of {\em{physical wave functions}}, where ``physical'' 
means intuitively that these wave functions are realized in our physical system
(whatever this means; we shall not enter philosophical issues here).
The choice of the subspace~$\H \subset \H_m$ is part of the input which characterizes the
physical system. For example, in order to describe the vacuum, one chooses~$\H$ as the
subspace of all negative-energy solutions of the Dirac equation (see Section~\ref{secsea}).
In order to model a system involving electrons, however, the subspace~$\H$ must be chosen to
include the electronic wave functions of positive frequency.
At this stage, we do not need to specify~$\H$, and in order to clarify the concepts,
it seems preferable to keep our considerations on a general abstract level.
Specific choices and explicit computations can be found in~\cite[Section~1.2]{cfs} and in later chapters
of this book (Chapters~\ref{secFSO}-\ref{sechadamard}).

We point out that the functions in~$\H$ do not need to be continuous
(instead, as mentioned at the end of Section~\ref{secspinorbundlemink},
they are weak solutions whose restriction to any Cauchy surface merely is an $L^2$-function).
Therefore, we cannot evaluate the wave functions pointwise at a spacetime point~$x \in \scrM$.
However, for the following constructions it is crucial to do so.
The way out is to introduce so-called {\em{regularization operators}}~$({\mathfrak{R}}_\varepsilon)$
with~$0 < \varepsilon < \varepsilon_{\max}$ as linear operators that
map~$\H$ to the continuous wave functions,
\sindex{regularization operator|textbf}%
\nindex{adi@${\mathfrak{R}}_\varepsilon$ -- regularization operator on scale~$\varepsilon$}%
\beq \label{Repsdef}
{\mathfrak{R}}_\varepsilon \::\: \H \rightarrow C^0(\scrM, S\scrM) \qquad \text{linear} \:.
\eeq
In the limit~$\varepsilon \searrow 0$, these operators should go over to the identity
(in a suitable sense which we do not specify here as it will not be needed).
The physical picture is that on a small length scale, which can be thought of as the
Planck length scale $\varepsilon \approx 10^{-35}$~meters, the structure of spacetime
must be modified. The regularization operators specify this microscopic
structure of spacetime. Different choices of regularization operators are possible.
A simple example of a regularization operator is obtained by mollifying with a test function.
Thus we let~$h \in C^\infty_0(\scrM, \R)$ be a non-negative test function with
\beq \int_\scrM h(x)\: \Diff^4x=1 \:. \eeq
We define the operators~${\mathfrak{R}}_\varepsilon$ for~$\varepsilon>0$ as the
convolution operators (for basics on the convolution see the paragraph after~\eqref{eq:convolution}
in Section~\ref{secfourier})
\beq ({\mathfrak{R}}_\varepsilon u)(x) := \frac{1}{\varepsilon^4}
\int_\scrM h\Big(\frac{x-y}{\varepsilon}\Big)\: u(y)\: \dd^4y \:. \eeq
Another method is to work in Fourier space (for preliminaries, see Sections~\ref{secsea} and~\ref{secfourier})
by setting
\beq u(x) = \int \frac{\dd^4k}{(2 \pi)^4}\: \hat{u}(k)\: \E^{-\cI k x} \:, \eeq
and to regularize by multiplication with an exponentially decaying cutoff function; that is,
\sindex{regularization!$i \varepsilon$-}
\beq \label{iepsreg}
\big( {\mathfrak{R}}_\varepsilon u \big)(x) = \int \frac{\dd^4k}{(2 \pi)^4}\: \hat{u}(k)\: \E^{-\varepsilon \,|\omega|} \:\E^{-\cI k x} 
\qquad \text{with} \qquad \omega=k^0 \:.
\eeq
This so-called {\em{$\cI \varepsilon$-regularization}} is most convenient for explicit computations
(for more details see~\cite[\S2.4.1]{cfs}).
Clearly, these methods of regularizing Dirac solutions are very special
and should be thought of merely as a mathematical tool for 
constructing simple and explicit examples of causal fermion systems.

Before going on, we briefly remark for the reader familiar with quantum field theory (QFT)
how the above regularization is related to the ultraviolet regularization procedures used in
relativistic QFT. Both in QFT and the setting of causal fermion systems,
regularizations are needed in order to make the theory mathematically well-defined.
In the renormalization program in QFT, one shows that the 
UV regularization can be taken out if other parameters of the theory (like masses
and coupling constants) are suitably rescaled. Then, the regularization can be understood
merely as a computational tool. In the causal fermion systems, however, the physical picture behind the regularization is quite different. Namely, in our setting, the {\em{regularized}} objects are to be considered as the
fundamental physical objects. The regularization models the microscopic structure of spacetime
and has therefore a physical significance.

Next, for any~$x \in \scrM$, we consider the bilinear form
\beq %\label{bxdef}
b_x^\varepsilon \::\: \H \times \H \rightarrow \C\:,\quad
b_x^\varepsilon(u, v) = - \Sl ({\mathfrak{R}}_\varepsilon \,u)(x) \,|\, ({\mathfrak{R}}_\varepsilon \,v)(x) \Sr \:. \eeq
This bilinear form is well-defined and bounded because~${\mathfrak{R}}_\varepsilon$ 
is defined pointwise and because evaluation at~$x$ gives a linear operator of finite rank
(see Exercise~\ref{exbounded}).
Thus for any~$v \in \H$, the anti-linear form~$b_x^\varepsilon(.,v) : \H \rightarrow \C$
is continuous. By the Fr{\'e}chet-Riesz theorem (Theorem~\ref{thm-FR}),
there is a unique vector~$w^\varepsilon \in \H$
such that~$b_x^\varepsilon(u,v) = \la u | w^\varepsilon \ra_\H$ for all~$u \in \H$.
The mapping~$v \mapsto w^\varepsilon$ is linear and bounded. We thus obtain a bounded linear
operator~$F^\varepsilon(x)$ on~$\H$ such that
\beq \label{Fepsdef}
b_x^\varepsilon(u, v) = \la u \,|\, F^\varepsilon(x)\, v \ra_\H \qquad \text{for all~$u,v \in \H$}\:,
\eeq
referred to as the {\em{local correlation operator}}.
\sindex{local correlation operator|textbf}%
\nindex{adj@$F^\varepsilon(x)$ -- local correlation operator at~$x \in \scrM$ with regularization on scale~$\varepsilon$}%
Taking into account that the inner product on the Dirac spinors at~$x$ has signature~$(2,2)$,
the local correlation operator~$F^\varepsilon(x)$ is a symmetric operator on~$\H$
of rank at most four, which has at most two positive and at most two negative eigenvalues.

Varying the point~$x \in \scrM$, for any~$\varepsilon$ we obtain a mapping
\beq \label{FeMink}
F^\varepsilon \::\: \scrM \rightarrow \F \:,
\eeq
where~$\F \subset \Lin(\H)$ is the set of all
symmetric operators on~$\H$ of finite rank which (counting multiplicities) have
at most two positive and at most two negative eigenvalues.
We sometimes refer to~$F^\varepsilon$ as the {\em{local correlation map}}.
\sindex{local correlation map|textbf}%
The last step is to drop all other structures (like the metric and causal structures
of Minkowski space, the spinorial structures, etc.).
As mentioned earlier, the basic concept behind causal fermion systems is to work exclusively with the local correlation operators corresponding to the physical wave functions.
In order to formalize this concept, we introduce the measure~$\rho^\varepsilon$ on~$\F$ as the push-forward
of the volume measure on~$\scrM$ (for details see Section~\ref{secbasicmeasure}
or Exercise~\ref{expushforward}),
\sindex{measure!push-forward}%
\nindex{adk@$F_* \mu$ -- push-forward measure}
\beq \label{rhoMink}
\rho^\varepsilon := F^\varepsilon_* \mu \:.
\eeq
We thus obtain a causal fermion system of spin dimension~$n=2$ (see Definition~\ref{defcfs}).
The local correlation operators are encoded in~$\rho$ as the support~$M$ of this measure.
Working exclusively with the structures of a causal fermion system, we no longer have the usual spacetime structures (particles, fields, causal structure, geometry, \ldots).
The underlying idea is that all these spacetime structures are encoded in the local correlation operators.
At this point, it is not obvious that this concept is sensible. But, as we shall see in the later sections in this book,
it is indeed possible to reconstruct all spacetime structures from the local correlation operators.
In this sense, the structures of a causal fermion system give a complete description of the physical system.

\section{The Causal Action Principle} \label{seccap}
Having given the general definition of a causal fermion system (see Definition~\ref{defcfs}),
the question arises how physical equations can be formulated in this setting.
To this end, we now introduce a variational principle, the so-called {\em{causal action principle}}.
In this variational principle, we minimize a functional, the so-called {\em{causal action}}, under
variations of the measure~$\rho$. The minimality property will then impose strong conditions
on the possible form of this measure.
The mathematical structure of the causal action is similar to the action~\eqref{Ssum}
given in our example of the lattice system. Its detailed form, however, 
is the result of many computations and longer considerations, as will be outlined in Section~\ref{secwhycap} below.

For any~$x, y \in \F$, the product~$x y$ is an operator of rank at most~$2n$. 
However, in general, it is no longer a symmetric operator because~$(xy)^* = yx$,
and this is different from~$xy$ unless~$x$ and~$y$ commute.
As a consequence, the eigenvalues of the operator~$xy$ are in general complex.
We denote these eigenvalues counting algebraic multiplicities
by~$\lambda^{xy}_1, \ldots, \lambda^{xy}_{2n} \in \C$
(more specifically,
denoting the rank of~$xy$ by~$k \leq 2n$, we choose~$\lambda^{xy}_1, \ldots, \lambda^{xy}_{k}$ as all
the non-zero eigenvalues and set~$\lambda^{xy}_{k+1}, \ldots, \lambda^{xy}_{2n}=0$).
We introduce the Lagrangian and the causal action by
\begin{align}
\text{\em{causal Lagrangian:}} && \L(x,y) &= \frac{1}{4n} \sum_{i,j=1}^{2n} \Big( \big|\lambda^{xy}_i \big|
- \big|\lambda^{xy}_j \big| \Big)^2 \label{Lagrange} \\
\text{\em{causal action:}} && \Sact(\rho) &= \iint_{\F \times \F} \L(x,y)\: \Diff\rho(x)\, \Diff\rho(y) \:. \label{Sdef}
\end{align}
\sindex{Lagrangian!causal}%
\nindex{adl@$\L(x,y)$ -- causal Lagrangian}%
\sindex{causal action}%
\nindex{adm@$\Sact(\rho)$ -- causal action}%
The {\em{causal action principle}} is to minimize~$\Sact$ by varying the measure~$\rho$
\sindex{causal action principle}%
under the following constraints:
\begin{align}
\text{\em{volume constraint:}} && \rho(\F) = \text{const} \quad\;\; & \label{volconstraint} \\
\text{\em{trace constraint:}} && \int_\F \tr(x)\: \Diff\rho(x) = \text{const}& \label{trconstraint} \\
\text{\em{boundedness constraint:}} && \T(\rho) := \iint_{\F \times \F} 
\bigg( \sum_{j=1}^{2n} \big|\lambda^{xy}_j \big| \bigg)^2
\: \Diff\rho(x)\, \Diff\rho(y) &\leq C \:, \label{Tdef}
\end{align}
\sindex{constraint!volume}%
\sindex{constraint!trace|textbf}%
\sindex{constraint!boundedness|textbf}%
\nindex{adn@$\T(\rho)$ -- functional in boundedness constraint}%
where~$C$ is a given parameter (and~$\tr$ denotes the trace of a linear operator on~$\H$ of finite rank).
\sindex{trace of linear operator}%
\nindex{ado@$\tr$ -- trace of linear operator}%
As already mentioned, we postpone the physical explanation of the 
detailed form of the Lagrangian to Section~\ref{secwhycap}.
The constraints can be understood mathematically as being
needed in order to get a well-posed variational principle
with non-trivial minimizers. This will be explained in Chapter~\ref{secmeasure}
(see, in particular; Section~\ref{seccounter}; also Exercise~\ref{exm3} is related).

Before going on, for clarity we point out that the mathematical structure of the causal action principle is
quite different from other variational principles considered in physics and mathematics.
There does not seem to be a direct way of deriving or even motivating the causal action principle
from other known action principles or Lagrangians. The only way to get the connection to the known physical
equation is by studying suitable limiting cases of the causal action principle and the corresponding
Euler-Lagrange equations (it will be outlined in Chapters~\ref{seccl} and~\ref{secQFT}
how to get a connection to classical field theory and quantum field theory, respectively).

In order to make the causal action principle mathematically well-defined, one needs
to specify the class of measures in which to vary~$\rho$. To this end,
on~$\F$ we consider the topology induced by the operator norm
\beq \label{supnorm}
\|A\| := \sup \big\{ \|A u \|_\H \text{ with } \| u \|_\H = 1 \big\}
\eeq
(for basics see the preliminaries in Sections~\ref{sectopology} and~\ref{sechilbertintro}).
\nindex{abi@$\NORM . \NORM$, $\NORM . \NORM_\H$ -- $\sup$-norm, operator norm}%
\tindex{dd@$\NORM . \NORM$, $\NORM . \NORM_\H$ -- $\sup$-norm, operator norm}%
In this topology, the Lagrangian as well as the integrands in~\eqref{trconstraint}
and~\eqref{Tdef} are continuous.
The $\sigma$-algebra generated by the open sets of~$\F$ consists of the so-called Borel sets.
A {\em{regular Borel measure}}
\sindex{Borel measure!regular}%
is a measure on the Borel sets with the property that
it is continuous under approximations by compact sets from inside and by open sets from outside
(for basics, see the preliminaries in Section~\ref{secbasicmeasure}).
The right prescription is to vary~$\rho$ within the class of regular Borel measures on~$\F$.

One must distinguish two settings:
\bitem
\item[(a)] The {\em{finite-dimensional setting}}: $\dim \H < \infty$ and~$\rho(\F) < \infty$. \\
\sindex{causal action principle!finite-dimensional setting}%
In this case, we will prove the existence of minimizing measures in Chapter~\ref{secmeasure}.
This will also clarify the significance of the constraints
(see in particular the examples in Section~\ref{seccounter}).
\item[(b)] The {\em{infinite-dimensional setting}}: $\dim \H = \infty$ and~$\rho(\F) = \infty$. \\
\sindex{causal action principle!infinite-dimensional setting}%
An obvious complication in this setting is that the volume constraint~\eqref{volconstraint}
is infinite. Likewise, the other constraints as well as the causal action may diverge.
These divergences can be avoided by restricting attention to variations that change the measure
only on a set of finite volume. By doing so, the differences of the actions and the constraints
are well-defined and finite (this method will be introduced in Sections~\ref{secnoncompact}
and~\ref{seccvpnoncompact}).

With this in mind, the remaining problem is to deal with infinite-dimensional Hilbert spaces.
The question whether physics is to be described on the fundamental level
by finite- or infinite-dimensional Hilbert spaces seems of a more philosophical nature, and we shall not
enter this question here. One way of getting along with the finite-dimensional
setting is to take the point of view that, on a fundamental physical level,
the total volume is finite and the Hilbert space~$\H$ is finite-dimensional, whereas the
infinite-dimensional setting merely is a mathematical idealization needed in
order to describe systems in infinite volume involving an infinite number of quantum particles.
Even if this point of view is taken, the infinite-dimensional case is of independent mathematical interest and should
also be the appropriate effective description in many physical situations.
This case also seems to be mathematically sensible.
However, the existence theory has not yet been developed. But at least, it is known that the Euler-Lagrange (EL) equations corresponding to the causal action principle still have a mathematical meaning 
in the infinite-dimensional setting (for details see~\cite{cfs}).
\eitem

We now explain how the spacetime of a causal fermion system is endowed with a topological and causal structure.
Recall that, given a minimizing measure~$\rho$, {\em{spacetime}}~$M \subset \F$ is defined as the
support of~$\rho$ (see~\eqref{Mdef}; this is illustrated in Exercise~\ref{expushforward}).
Thus, the spacetime points are symmetric linear operators on~$\H$.
On~$M$ we consider the topology induced by~$\F$ (generated by the $\sup$-norm~\eqref{supnorm}
on~$\Lin(\H)$). Moreover, the measure~$\rho|_M$ restricted to~$M$ can be regarded as a volume
measure on spacetime. This turns spacetime into a {\em{topological measure space}}.
\sindex{topological measure space}%
Furthermore, one has the following notion of causality:

\begin{Def} {\bf{(causal structure)}} \label{def2}
\sindex{causality!in causal fermion system}%
\sindex{causal structure!of causal fermion system}%
For any~$x, y \in \F$, the product~$x y$ is an operator
of rank at most~$2n$. We denote its non-trivial eigenvalues (counting algebraic multiplicities)
by~$\lambda^{xy}_1, \ldots, \lambda^{xy}_{2n}$.
\nindex{adq@$\lambda^{xy}_1, \ldots, \lambda^{xy}_{2n}$ -- non-trivial eigenvalues of~$xy$}%
The points~$x$ and~$y$ are
called {\bf{spacelike}} separated if all the~$\lambda^{xy}_j$ have the same absolute value.
They are said to be {\bf{timelike}} separated if the~$\lambda^{xy}_j$ are all real and do not all 
have the same absolute value.
In all other cases (that is, if the~$\lambda^{xy}_j$ are not all real and do not all 
have the same absolute value),
the points~$x$ and~$y$ are said to be {\bf{lightlike}} separated.
\end{Def} \noindent
\sindex{timelike separation}%
\sindex{spacelike separation}%
\sindex{lightlike separation}%
Restricting the causal structure of~$\F$ to~$M$, we get causal relations in spacetime. 

Before going on, we point out that it is not obvious whether and in which sense this definition of causality
agrees with the usual notion of causality in Minkowski space
(or, more generally, in a Lorentzian spacetime).
In order to get the connection, one can consider the causal fermion system
constructed in Section~\ref{seclco} with the Hilbert space~$\H \subset \H_m$ chosen
as the subspace of all negative-energy solutions of the Dirac equation
(thereby realizing the concept of the Dirac sea as explained in Section~\ref{secsea}).
Then the above ``spectral definition'' of causality goes over to the causal structure of Minkowski
space in the limiting case~$\varepsilon \searrow 0$.
Since the detailed computations for getting this correspondence are a bit lengthy, we do not
present them here but refer the interested reader instead to~\cite[Section~1.2]{cfs}.

The Lagrangian~\eqref{Lagrange} is compatible with the above notion of causality in the
following sense. Suppose that two points~$x, y \in \F$ are spacelike separated.
Then the eigenvalues~$\lambda^{xy}_i$ all have the same absolute value.
As a consequence, the Lagrangian~\eqref{Lagrange} vanishes.
Thus, pairs of points with spacelike
separation do not enter the action. This can be seen in analogy to the usual notion of causality where
points with spacelike separation cannot influence each other.
This analogy is the reason for the notion ``causal'' in ``causal fermion system''
and ``causal action principle.''

A causal fermion also system distinguishes a {\em{direction of time}}.
In order to see this, for~$x \in \F$, we let~$\pi_x$ be the orthogonal projection in~$\H$ on the subspace~$x(\H) \subset \H$
and introduce the functional
\sindex{time direction!in causal fermion system}%
\nindex{adr@$\mathscr{C}$ -- time direction functional}%
\beq \label{Cform}
{\mathscr{C}} \::\: M \times M \rightarrow \R\:,\qquad
{\mathscr{C}}(x, y) := i \tr \big( y\,x \,\pi_y\, \pi_x - x\,y\,\pi_x \,\pi_y \big) \:.
\eeq
Obviously, this functional is anti-symmetric in its two arguments, making it possible to introduce the notions
\beq %\label{tdir}
\left\{ \begin{array}{cl} \text{$y$ lies in the {\em{future}} of~$x$} &\quad \text{if~${\mathscr{C}}(x, y)>0$} \\[0.2em]
\text{$y$ lies in the {\em{past}} of~$x$} &\quad \text{if~${\mathscr{C}}(x, y)<0$}\:. \end{array} \right. \eeq
We remark that the detailed form of the functional~\eqref{Cform} is not obvious; it must be justified
by working out that it gives back the time direction of Minkowski space in a suitable limiting case
(for details see Exercise~\ref{extimedir} and~\cite[\S1.2.5]{cfs}).

By distinguishing a direction of time, we get a structure similar to
a causal set (see, for example, \cite{sorkin}). However, in contrast to a causal set, our notion of
``lies in the future of'' is not necessarily transitive.
\sindex{future!in causal fermion system}%
\sindex{past!in causal fermion system}%

\section{Basic Inherent Structures} \label{secinherent}
It is the general concept that a causal fermion system describes spacetime as well as all
structures therein (like the causal and metric structures, particles, fields, etc.).
Thus all these structures must be constructed from the basic objects of the theory alone,
using the information already encoded in the causal fermion system.
We refer to these constructed structures as being {\em{inherent}} in the causal fermion system.
\sindex{inherent structures of causal fermion system}%
We now introduce and explain the most important of these structures:
the {\em{spin spaces}}, the {\em{physical wave functions}} and the {\em{kernel of the fermionic projector}}.
Other inherent structures will be introduced later in this book (see Chapters~\ref{secOSI}--\ref{seclqg});
for a more complete account, we also refer to~\cite[Chapter~1]{cfs}.

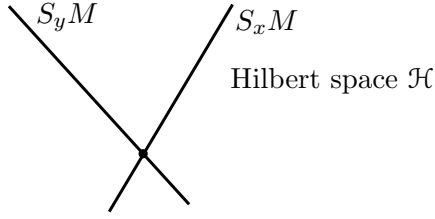
\begin{figure}% spinspace.svg
% \usepackage[usenames,dvipsnames]{pstricks}
% \usepackage{epsfig}
% \usepackage{pst-grad} % For gradients
% \usepackage{pst-plot} % For axes
% \usepackage[space]{grffile} % For spaces in paths
% \usepackage{etoolbox} % For spaces in paths
% \makeatletter % For spaces in paths
% \patchcmd\Gread@eps{\@inputcheck#1 }{\@inputcheck"#1"\relax}{}{}
% \makeatother
% \psscalebox{1.0 1.0} % Change this value to rescale the drawing.
{
\begin{pspicture}(0,-1.3880072)(7.342354,1.3880072)
\rput[bl](2.962354,0.1906667){$\text{Hilbert space}~\Cb{\H}$}
\pscircle[linecolor=black, linewidth=0.02, fillstyle=solid,fillcolor=black, dimen=outer](1.7992429,-0.6066666){0.05111111}
\psline[linecolor=black, linewidth=0.04](0.014798415,1.3555557)(2.410354,-1.28)
\psline[linecolor=black, linewidth=0.04](2.9881318,1.3777778)(1.3481318,-1.3777777)
\rput[bl](3.0245762,0.98622227){$\Cb{S_xM}$}
\rput[bl](0.3756873,1.0217779){$\Cb{S_yM}$}
\end{pspicture}
}
\caption{The spin spaces}
\label{figspinspace}
\end{figure}

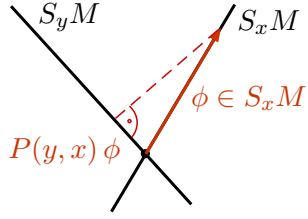
\begin{figure} % kernel.svg
% \usepackage[usenames,dvipsnames]{pstricks}
% \usepackage{epsfig}
% \usepackage{pst-grad} % For gradients
% \usepackage{pst-plot} % For axes
% \usepackage[space]{grffile} % For spaces in paths
% \usepackage{etoolbox} % For spaces in paths
% \makeatletter % For spaces in paths
% \patchcmd\Gread@eps{\@inputcheck#1 }{\@inputcheck"#1"\relax}{}{}
% \makeatother
% \psscalebox{1.0 1.0} % Change this value to rescale the drawing.
{
\begin{pspicture}(0,-1.3880072)(3.8166666,1.3880072)
\definecolor{colour0}{rgb}{0.8,0.2,0.0}
\definecolor{colour1}{rgb}{0.8,0.0,0.0}
\pscircle[linecolor=black, linewidth=0.02, fillstyle=solid,fillcolor=black, dimen=outer](1.8191111,-0.6066666){0.05111111}
\psline[linecolor=black, linewidth=0.04](0.034666665,1.3555557)(2.4302223,-1.28)
\psline[linecolor=black, linewidth=0.04](3.008,1.3777778)(1.368,-1.3777777)
\rput[bl](3.0444446,0.98622227){$S_xM$}
\rput[bl](0.39555556,1.0217779){$S_yM$}
\psline[linecolor=colour0, linewidth=0.04, arrowsize=0.05291667cm 2.0,arrowlength=1.4,arrowinset=0.0]{->}(1.8257778,-0.60888886)(2.8257778,1.0666667)
\psline[linecolor=colour1, linewidth=0.02, linestyle=dashed, dash=0.17638889cm 0.10583334cm](2.7857778,1.0400001)(1.4213333,-0.13777773)
\psarc[linecolor=colour1, linewidth=0.02, dimen=outer](1.4191111,-0.19555551){0.2911111}{-40.0}{50.0}
\pscircle[linecolor=colour1, linewidth=0.02, dimen=outer](1.608,-0.17333329){0.02}
\rput[bl](2.4266667,-0.04488885){\textcolor{colour0}{$\phi \in S_xM$}}
\rput[bl](0.0,-0.756){\textcolor{colour0}{$P(y,x)\, \phi$}}
\end{pspicture}
}
\caption{The kernel of the fermionic projector}
\label{figkernel}
\end{figure}

The causal action principle depends crucially on the eigenvalues of the operator
product~$xy$ with~$x,y \in \F$. For computing these eigenvalues, it is convenient not to
consider this operator product on the (possibly infinite-dimensional) Hilbert space~$\H$, 
but instead to restrict attention to a finite-dimensional subspace of~$\H$, chosen such that
the operator product vanishes on the orthogonal complement of this subspace.
This construction leads us to the spin spaces and to the kernel of the fermionic projector,
which we now introduce.
For every~$x \in \F$ we define the {\em{spin space}}~$S_x$ as the image of the operator~$x$; that is,
\beq \label{Sxdef}
S_x:=x(\H) \:;
\eeq
it is a subspace of~$\H$ of dimension at most~$2n$ (see Figure~\ref{figspinspace}).
\sindex{spin space}%
\nindex{ads@$S_x := x(\H)$ -- spin space}%
\nindex{adt@$(S_xM, \Sl . \vert . \Sr_x)$ -- spin space}%
Moreover, we let
\beq \label{pidef}
\pi_x \::\: \H \rightarrow S_x
\eeq
be the orthogonal projection in~$\H$ on the subspace~$S_x \subset \H$.
For any~$x, y \in M$ we define the
{\em{kernel of the fermionic projector}}~$P(x,y)$ by (see Figure~\ref{figkernel}).
\beq \label{Pxydefintro}
P(x,y) = \pi_x \,y|_{S_y} \::\: S_y \rightarrow S_x
\eeq
(where~$\pi_x$ is again the orthogonal projection on the subspace~$x(\H) \subset \H$).
\sindex{fermionic projector!kernel of}%
\nindex{adu@$\pi_x$ -- orthogonal projection on spin space}%
\nindex{adv@$P(x,y)$ -- kernel of fermionic projector}%
Taking the trace of~\eqref{Pxydefintro} in the case~$x=y$, one
finds that
\beq \tr(x) = \Tr_{S_x}(P(x,x))\:, \eeq
\nindex{adw@$\Tr_{S_x}$ -- trace on spin space}%
making it possible to express
the integrand of the trace constraint~\eqref{trconstraint} in terms of the kernel of the fermionic projector.
In order to also express the eigenvalues of the operator~$xy$ in terms of the kernel of the fermionic projector,
we introduce the {\em{closed chain}}~$A_{xy}$ as the product
\sindex{closed chain}%
\nindex{adx@$A_{xy}$ -- closed chain}%
\beq \label{Axydef}
A_{xy} = P(x,y)\, P(y,x) \::\: S_x \rightarrow S_x\:.
\eeq
The closed chain can be computed in more detail using the formula~\eqref{Pxydefintro}.
In preparation, we note that, from the definition of~$\pi_x$ as the orthogonal projection to the
image of~$x$~\eqref{pidef}, it follows immediately that~$\pi_x x = x$. Taking the adjoint of this relation,
we conclude that
\beq \label{pixx}
\pi_x \,x = x = x \,\pi_x \:.
\eeq
Using these identities, we can compute the closed chain by
\beq A_{xy} = (\pi_x y)(\pi_y x)|_{S_x} = \pi_x\, yx|_{S_x} \:. \eeq
Applying this equation iteratively and using again~\eqref{pixx}, we obtain for the~$p^\text{th}$ power of
the closed chain
\beq (A_{xy})^p = \pi_x\, (yx)^p|_{S_x} \:. \eeq
Taking the trace, one sees in particular that
\beq \label{trid}
\Tr_{S_x}(A_{xy}^p) = \tr \big((yx)^p \big) = \tr \big((xy)^p \big)
\eeq
(where the last identity simply is the invariance of the trace under
cyclic permutations).
Since all our operators have finite rank, for any~$x,y \in \F$ there is a finite-dimensional subspace~$I$
of~$\H$ such that~$xy$ maps~$I$ to itself and vanishes on the orthogonal complement of~$I$.
For example, one can choose~$I$ as the span of the image of~$xy$ and the orthogonal complement of
the kernel of~$xy$,
\beq I = \text{span} \big\{ (xy)(\H), \ker(xy)^\perp \big\} \:. \eeq
Then the non-trivial eigenvalues of the operator product~$xy$ are the non-zero roots
of the characteristic polynomial of the restriction~$xy|_I : I \rightarrow I$.
The coefficients of this characteristic polynomial (like the trace, the determinant, etc.)
are symmetric polynomials in the eigenvalues and can therefore be expressed in terms of traces of
powers of the operator~$xy|_I : I \rightarrow I$ (for details see Exercise~\ref{exchar}).
Using this result similarly for the characteristic polynomial of~$A_{xy}$ and using~\eqref{trid},
we conclude that the eigenvalues of the closed chain coincide with the non-trivial
eigenvalues~$\lambda^{xy}_1, \ldots, \lambda^{xy}_{2n}$ of the operator~$xy$ in Definition~\ref{def2}
(including multiplicities).
\nindex{ady@$\lambda^{xy}_1, \ldots, \lambda^{xy}_{2n}$ -- non-trivial eigenvalues of~$xy$}%
In particular, one sees that kernel of the fermionic projector encodes the causal structure of~$M$.
The above argument also implies that the operator products~$xy$ and~$yx$ are isospectral.
This shows that the causal structure is symmetric in~$x$ and~$y$.
The main advantage of working with the kernel of the fermionic projector is that the closed chain~\eqref{Axydef}
is a linear operator on a vector space of dimension at most~$2n$, making it possible
to compute the~$\lambda^{xy}_1, \ldots, \lambda^{xy}_{2n}$ as the eigenvalues of 
a matrix (in finite dimensions).

Next, it is very convenient to choose inner products on the spin spaces in such a way
that the kernel of the fermionic projector is symmetric in the sense that
\beq \label{Pxysymm}
P(x,y)^* = P(y,x) \:,
\eeq
where the star denotes the adjoint with respect to yet to be specified inner products on the spin spaces.
This identity indeed holds if on the spin space~$S_x$ (and similarly on~$S_y$) one chooses the
{\em{spin inner product}}~$\Sl .|. \Sr_x$ defined by
\beq \label{sspintro}
\Sl u | v \Sr_x := -\la u | x v \ra_\H \qquad \text{(for all~$u,v \in S_x$)}\:.
\eeq
\sindex{spin inner product}%
\nindex{aal@$\Sl . \vert . \Sr, \Sl . \vert . \Sr_x$ -- spin inner product}%
\tindex{bb@$\Sl . \vert . \Sr, \Sl . \vert . \Sr_x$ -- spin inner product}%
\nindex{adt@$(S_xM, \Sl . \vert . \Sr_x)$ -- spin space}%
Due to the factor~$x$ on the right, this definition really makes the kernel of the fermionic projector symmetric,
as is verified by the computation
\begin{align}
\Sl u \,|\, P(x,y) \,v \Sr_x &= - \la u \,|\, x\, P(x,y) \,v \ra_\H = - \la u \,|\, x y \,v \ra_\H \notag \\
&= -\la \pi_y \,x\, u \,|\, y \,v \ra_\H = \Sl P(y,x)\, u \,|\,  v \Sr_y \:,
\end{align}
where we again used~\eqref{pixx} (and~$u \in S_x$, $v \in S_y$).
The spin space~$(S_x, \Sl .|. \Sr_x)$ is an {\em{indefinite}} inner product of signature~$(p,q)$ with~$p,q \leq n$
(for textbooks on indefinite inner product spaces see~\cite{bognar, GLR}).
In this way, indefinite inner product spaces arise naturally when analyzing the
mathematical structure of the causal action principle.

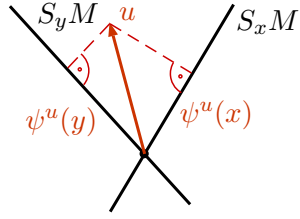
\begin{figure} % wavefunction.svg
% \usepackage[usenames,dvipsnames]{pstricks}
% \usepackage{epsfig}
% \usepackage{pst-grad} % For gradients
% \usepackage{pst-plot} % For axes
% \usepackage[space]{grffile} % For spaces in paths
% \usepackage{etoolbox} % For spaces in paths
% \makeatletter % For spaces in paths
% \patchcmd\Gread@eps{\@inputcheck#1 }{\@inputcheck"#1"\relax}{}{}
% \makeatother
% \psscalebox{1.0 1.0} % Change this value to rescale the drawing.
{
\begin{pspicture}(0,-1.3880072)(3.7745762,1.3880072)
\definecolor{colour0}{rgb}{0.8,0.2,0.0}
\definecolor{colour1}{rgb}{0.8,0.0,0.0}
\pscircle[linecolor=black, linewidth=0.02, fillstyle=solid,fillcolor=black, dimen=outer](1.7992429,-0.6066666){0.05111111}
\psline[linecolor=black, linewidth=0.04](0.014798415,1.3555557)(2.410354,-1.28)
\psline[linecolor=black, linewidth=0.04](2.9881318,1.3777778)(1.3481318,-1.3777777)
\rput[bl](3.0245762,0.98622227){$\Cb{S_xM}$}
\rput[bl](0.3756873,1.0217779){$\Cb{S_yM}$}
\psline[linecolor=colour0, linewidth=0.04, arrowsize=0.05291667cm 2.0,arrowlength=1.4,arrowinset=0.0]{->}(1.8059095,-0.60888886)(1.3392428,1.1422223)
\psline[linecolor=colour1, linewidth=0.02, linestyle=dashed, dash=0.17638889cm 0.10583334cm](1.3481318,1.12)(2.4281318,0.4666667)
\psline[linecolor=colour1, linewidth=0.02, linestyle=dashed, dash=0.17638889cm 0.10583334cm](1.3436873,1.1244445)(0.7836873,0.53333336)
\psarc[linecolor=colour1, linewidth=0.02, dimen=outer](2.4392428,0.45777783){0.2911111}{150.0}{240.0}
\psarc[linecolor=colour1, linewidth=0.02, dimen=outer](0.7725762,0.49777782){0.2911111}{-40.0}{50.0}
\pscircle[linecolor=colour1, linewidth=0.02, dimen=outer](0.9436873,0.5155556){0.02}
\pscircle[linecolor=colour1, linewidth=0.02, dimen=outer](2.281465,0.42222226){0.02}
\rput[bl](1.4690206,1.1684445){\textcolor{colour0}{$u$}}
\rput[bl](2.2734652,-0.2804444){\textcolor{colour0}{$\psi^u(x)$}}
\rput[bl](0.22013175,-0.3693333){\textcolor{colour0}{$\psi^u(y)$}}
\end{pspicture}
}
\caption{The physical wave function}
\label{figwavefunction}
\end{figure}

The kernel of the fermionic projector plays a central role in the analysis for several reasons:
\bitem
\itemD The Lagrangian can be expressed in terms of~$P(x,y)$ (via the closed chain~\eqref{Axydef}
and its eigenvalues).
\itemD Being a mapping from one spin space to another, $P(x,y)$ gives relations between
different spacetime points. In this way, it carries geometric information.
This will be explained in Chapter~\ref{seclqg} (see also~\cite{lqg} or the
introductory survey paper~\cite{nrstg}).
\itemD The kernel of the fermionic projector also encodes all the wave functions of the system.
In order to see the connection, for a vector~$u \in \H$ one introduces the corresponding
{\em{physical wave function}} $\psi^u$ as (see Figure~\ref{figwavefunction})
\sindex{physical wave function}%
\nindex{aeb@$\psi^u$ -- physical wave function of~$u \in \H$}%
\beq \label{psiudef}
\psi^u \::\: M \rightarrow \H\:,\qquad \psi^u(x) = \pi_x u \in S_x \:.
\eeq
Then, choosing an orthonormal basis~$(e_i)$ of~$\H$ and using the completeness relation
as well as~\eqref{sspintro}, one obtains for any~$\phi \in S_y$
\beq P(x,y) \, \phi= \pi_x y|_{S_y} \,\phi = \sum_i \pi_x e_i \: \la e_i | y\,\phi \ra_\H
= -\sum_i \psi^{e_i}(x) \: \Sl \psi^{e_i}(y) \,|\, \phi \Sr_y  \:, \eeq
showing that~$P(x,y)$ is indeed composed of all the physical wave functions; that is,
in bra/ket notation
\beq \label{Prep}
P(x,y) = -\sum_i |\psi^{e_i}(x) \Sr\: \Sl \psi^{e_i}(y)|  \:.
\eeq
\eitem
We remark that knowing the kernel of the fermionic projector in spacetime
makes it possible to reconstruct the causal fermion system
(the detailed construction can be found in~\cite[Section~1.1.2]{rrev}).
We also note that the representation of the kernel of the fermionic projector~\eqref{Prep}
also opens the door to the detailed study of causal fermion systems in Minkowski space
as carried out in~\cite{cfs}; see also Exercises~\ref{exminkfirst}--\ref{exminklast}.

Taking a slightly different perspective, one can say that all structures of the causal fermion
system are encoded in the {\em{family of physical wave functions}}~$\psi^u$ with~$u \in \H$
as defined in~\eqref{psiudef}. In order to make this statement precise, it is most convenient
to introduce the {\em{wave evaluation operator}}~$\Psi(x)$ at the spacetime point~$x \in M$ by
\sindex{wave evaluation operator|textbf}%
\nindex{aed@$\Psi(x)$ -- wave evaluation operator at~$x \in M$}%
\beq \label{weo}
\Psi(x) \::\: \H \rightarrow S_x\:,\qquad u \mapsto \psi^u(x)=\pi_x u \:.
\eeq
Clearly, using~\eqref{psiudef}, the wave evaluation operator can be written simply as
\beq \label{Psix}
\Psi(x) = \pi_x \:.
\eeq
The wave evaluation operator describes the family of all physical wave functions.
Indeed, applying the wave evaluation operator to a vector~$u$ and varying the point~$x$, we get
back the corresponding physical wave function~$\psi^u$.
We next compute the adjoint of~$\Psi(x)$,
\beq \Psi(x)^* : S_x \rightarrow \H \:. \eeq
Taking into account the corresponding inner products, we obtain for any~$\phi \in S_x$ and~$u \in \H$,
\beq \la \Psi(x)^* \phi | u \ra_\H =  \Sl \phi | \Psi(x)\, u \Sr_x \overset{\eqref{sspintro}}{=}
-\la \phi | x\, \Psi(x)\, u \ra_\H \:. \eeq
This shows that
\beq \label{Psixs}
\Psi(x)^* = -x|_{S_x} \:.
\eeq
Combining~\eqref{Psix} and~\eqref{Psixs} and comparing with~\eqref{Pxydefintro}, one sees that
\beq \label{xPdef}
x = -\Psi(x)^* \,\Psi(x) \qquad \text{and} \qquad P(x,y) = -\Psi(x) \,\Psi(y)^* \:.
\eeq
In this way, all the spacetime point operators and the kernel of the fermionic projector can be constructed
from the wave evaluation operator. Moreover, the conclusion after~\eqref{trid} that the
eigenvalues of the closed chain coincide with the nontrivial eigenvalues of the operator product~$xy$
can be seen more directly from the computation
\begin{align}
A_{xy} &= P(x,y)\, P(y,x) = \Psi(x) \Psi(y)^* \Psi(y) \Psi(x)^* \notag \\
&= -\Psi(x) \big( y \,\Psi(x)^* \big) \simeq -\Psi(x)^* \Psi(x) \, y = xy \:,
\end{align}
where by~$\simeq$ we mean that the operators are isospectral
(in the sense that they have the same non-zero eigenvalues with the same algebraic multiplicities).
Here we used that for any two matrices~$A \in \C^{p \times q}$ and~$B \in \C^{q \times p}$,
the matrix product~$AB$ is isospectral to~$BA$ (for details see Exercise~\ref{ex41}).

\section{How Did the Causal Action Principle Come About?} \label{secwhycap}
Causal fermion systems and the causal action principle came to light as a result of
many considerations and computations carried out over several years.
We now give an outline of these developments, also explaining the specific form of the causal action
principle.

The starting point for the considerations leading to causal fermion systems was
the belief that in order to overcome the conceptual problems of quantum field theory,
the structure of spacetime should be modified. Moreover, instead of starting from
differential equations in a spacetime continuum, one should formulate the physical equations
using the new structures of spacetime, which might be non-smooth or discrete.
A more concrete idea in this direction was that the spacetime structures should be encoded
in the family of wave functions which is usually associated to the Dirac sea
(for basics see Section~\ref{secsea}). Thus, instead of disregarding the sea states,
one should take all these wave functions into account. The mutual interaction of all these wave functions
should give rise to the structures of spacetime as we experience them.

The first attempts toward making this idea more precise go back to the early 1990s.
The method was to consider families of Dirac solutions
(the formalism of quantum fields was avoided in order to keep the setting as simple and
non-technical as possible). In order to describe such a family mathematically, the
corresponding two-point kernel~$P(x,y)$ was formed
\beq P(x,y) := -\sum_{l=1}^f |\psi_l(x) \Sr \Sl \psi_l(y) | \eeq
(where~$\psi_1, \ldots, \psi_f$ are suitably normalized solutions of the Dirac equation;
for preliminaries, see Section~\ref{secdirac}). The kernel~$P(x,y)$ is also referred to as
the {\em{kernel of the fermionic projector}}. In the Minkowski vacuum, 
\sindex{fermionic projector!kernel of}%
this kernel is formed of all the states of the Dirac sea. 
\sindex{Dirac sea}%
Then the sum goes over to an integral
over the lower mass shell
\beq \label{Pxyvac}
P^\text{vac}(x,y) = \int \frac{\dd^4k}{(2 \pi)^4}\:(\slashed{k}+m)\: \delta(k^2-m^2)\: \Theta(-k_0)\: \E^{-\cI k(x-y)}
\eeq
(this integral is well-defined as the Fourier transform of a tempered distribution;
see the preliminaries in Section~\ref{secfourier}).
Likewise, a system involving particles and anti-particles is described by
``occupying additional states of positive energy'' and by ``creating holes in the Dirac sea,'' respectively.
Thus, more technically, one sets
\beq \label{Pxyparticles}
P(x,y) = P^\text{vac}(x,y) - \sum_a |\psi_a(x) \Sr \Sl \psi_a(y) | +  \sum_b |\phi_b(x) \Sr \Sl \phi_b(y) | \:,
\eeq
where~$\psi_a$ and~$\phi_b$ are suitably normalized Dirac solutions of positive and negative energy,
respectively. In case a bosonic interaction is present, the kernel of the fermionic projector
should no longer satisfy the vacuum Dirac equation, but the Dirac equation in the presence of
a, say, external potential~$\B$. Clarifying the dependence on the bosonic potential with an additional tilde,
we write the resulting Dirac equation as
\beq \label{DirPtilde}
(\cI \Pdd + \B - m) \, \tilde{P}(x,y) = 0 \:.
\eeq

Analyzing the distribution~$\tilde{P}(x,y)$ in Minkowski space reveals the following facts:
\bitem
\item[(a)] The kernel~$\tilde{P}(x,y)$ contains all the information on the wave functions of the particles
and anti-particles of the system. This statement can be understood from the representation~\eqref{Pxyparticles}
in which all these wave functions appear. Alternatively, the wave functions can be reconstructed
from~$\tilde{P}(x,y)$ as being the image of the corresponding integral operator on~$C^\infty_0(\scrM, S\scrM)$
\beq \phi \mapsto \int_\scrM \tilde{P}(.,y)\: \phi(y)\: \dd^4x \:. \eeq
\item[(b)] The kernel~$\tilde{P}(x,y)$ has singularities on the light cone.
The detailed form of the singularities involves integrals of the potential~$\B$ and its derivatives
along the light cone. In particular, knowing the kernel~$\tilde{P}(x,y)$ makes it possible to
reconstruct the potential~$\B$ at every spacetime point.
These statements follow immediately by looking at the so-called light-cone expansion of~$\tilde{P}(x,y)$
(see Chapter~\ref{sechadamard} in this book or~\cite[Section~2.2 and Appendix~B]{cfs}).
\item[(c)] The singularity structure of~$\tilde{P}(x,y)$ encodes the causal structure of Minkowski space.
This can be seen again from the light-cone expansion of~$\tilde{P}(x,y)$
(see again Chapter~\ref{sechadamard} in this book or~\cite[Section~2.2 and Appendix~B]{cfs}).
\eitem
These findings show that, at least for Dirac systems in the presence of classical bosonic potentials,
the kernel~$\tilde{P}(x,y)$ contains all the information on the physical system.
This led to the concept to regard~$\tilde{P}(x,y)$ as the basic physical object in spacetime.
The more familiar structures and objects like Minkowski space with its causal structure,
the Dirac equation, the classical field equations for the bosonic fields (like the Maxwell or Einstein equations),
however, should no longer be considered as being fundamental.
Consequently, the physical equations should be formulated directly in terms of the kernel of the fermionic projector.

Formalizing this idea in a clean way also made it necessary to disregard
or to prescind from the usual spacetime structures.
This led to the {\em{principle of the fermionic projector}} as formulated around~1990
\sindex{principle of the fermionic projector}%
(see the unpublished preprint~\cite{endlich} and the monograph~\cite{pfp}).
We here present a slightly different but equivalent formulation which is somewhat closer to the setting
of causal fermion systems. Let~$M$ be a discrete set (that is, a point set without additional structures),
the {\em{discrete spacetime}}.
\sindex{spacetime!discrete}%
Moreover, for every~$x \in M$ one chooses an indefinite inner product space~$(S_x, \Sl .|. \Sr_x)$,
referred to as the {\em{spin space}} at~$x$ (usually, one chooses the dimensions and signatures of all
spin spaces to be the same, but this is not crucial for the construction).
Next, we consider a collection of {\em{wave functions}}~$(\psi_a)_a$,
each being a mapping which to every discrete spacetime point~$x \in M$ associates a
vector~$\psi_a(x) \in S_x$ of the corresponding spin space. Out of these wave functions, one can
form the kernel of the fermionic projector
\beq P(x,y) := -\sum_a |\psi_a(x) \Sr \Sl \psi_a(y) | \::\: S_y \rightarrow S_x \:. \eeq
The principle of the fermionic projector asserts that the physical equations should be formulated
purely in terms of the kernel of the fermionic projector in discrete spacetime.

The next question was how precisely these physical equations should look like.
This was a difficult question which took many years to be answered.
Apart from the structural requirements coming from the principle of the fermionic projector,
the following considerations served as guiding principles\footnote{Of course, it is also an important
requirement that our variational principle should give agreement with quantum field theory.
But this connection was not used for finding the causal action principle.
It was worked out more recently; for more details see Chapter~\ref{secQFT}.}:
\bitem
\item[{\rm{(i)}}] In analogy to classical field theory, a {\em{variational approach}} should be used.
One main advantage is the resulting connection between symmetries and conservation laws
(corresponding to the classical Noether theorem), which seems of central importance
in physical applications.
\item[{\rm{(ii)}}] Classical field theory should be obtained in a certain limiting case.
More specifically, the Euler-Lagrange equations coming from our variational principle should reproduce
the {\em{Maxwell}} and {\em{Einstein equations}}.
\item[{\rm{(iii)}}] Also the {\em{Dirac equation}} should be recovered in a certain limiting case.
\eitem
More mathematically, the strategy was to form composite expressions of the kernel of the
fermionic projector. Choosing~$n$ points~$x_1, \ldots, x_n \in M$, one can form
the {\em{closed chain}}
\sindex{closed chain!of $n$-point action}%
\beq \label{Aclosed}
A_{x_1, \ldots, x_n} :=
P(x_1, x_2) \, P(x_2, x_3) \cdots P(x_{n-1}, x_n)\: P(x_n, x_1) \::\: S_{x_1} \rightarrow S_{x_1} \:.
\eeq
Being an endomorphism of the spin space, one can compute the eigenvalues of the closed chain
and form a Lagrangian~$\L[A_{x_1, \ldots, x_n}]$ as a symmetric function of these eigenvalues.
Summing over the spacetime points gives an ansatz for the
\sindex{$n$-point action}%
\beq \text{$n$-point action} \qquad \Sact = \sum_{x_1,\ldots,x_n \in M} \L[A_{x_1, \ldots, x_n}] \:. \eeq

This general ansatz can be made more specific and concrete by considering {\em{gauge phases}}.
\sindex{gauge phase}%
This consideration was motivated by the fact that the kernel of the fermionic projector~$\tilde{P}(x,y)$
formed of Dirac solutions involves gauge phases. More specifically, choosing the potential in
the Dirac equation~\eqref{DirPtilde} as an electromagnetic potential~$\B = \slashed{A}$,
then the leading contribution to the kernel are gauge phases described by line integrals over the
electromagnetic potential,
\beq \label{tilPgauge}
\tilde{P}(x,y) = \E^{-\cI \int_x^y A_j \xi^j}\: P^\text{vac}(x,y) + \cdots\:,
\eeq
where
\beq \int_x^y A_j \xi^j = \int_0^1 A_j(\alpha y+(1-\alpha) x)\: (y-x)^j\: \dd\alpha \eeq
(this can again be seen from the light-cone expansion; more specifically, see~\cite[\S2.2.4]{cfs}).
Here~$\cdots$ stands for many other contributions to~$\tilde{P}(x,y)$ which involve
derivatives of the potential (like the field tensor, the Maxwell current, etc.). All these additional contributions
are small in the sense that they are less singular on the light cone.
These findings will be made precise by the Hadamard and light-cone expansions of the
kernel of the fermionic projector in Chapter~\ref{seccl} of this book.
At this stage, we do not need to be specific. All we need is that gauge phases come into play,
which involves integrals of the potential along the line segment joining the points~$x$ and~$y$.

Let us analyze the effect of the gauge phases on the closed chain~\eqref{Aclosed}.
First of all, the closed chain is {\em{gauge invariant}}. Indeed, if one considers a pure gauge
potential~$A_j = \partial_j \Lambda$, then the gauge phases in~\eqref{tilPgauge} simplify to
\beq \label{Pgauge}
\tilde{P}(x,y) = \E^{-\cI \Lambda(y) + \cI \Lambda(x)}\: P^\text{vac}(x,y) \:,
\eeq
and the phase factors of neighboring factors cancel in~\eqref{Aclosed}.
This consideration also gives a relation between local gauge invariance and
the fact that the adjacent factors in~\eqref{Aclosed} must coincide.
In the case~$n=1$, the kernel of the fermionic projector is evaluated only on the diagonal~$P(x,x)$.
This turns out to be too simple for formulating physical equations,
as can be understood from the fact that no relations between spacetime
points are taken into account. If~$n \geq 3$, the gauge phases in~\eqref{Aclosed}
can be rewritten using Stokes' theorem as {\em{flux integrals}} of the electromagnetic field
through the two-dimensional polygon with vertices~$x_1, \ldots, x_n$.
Analyzing the situation in some more detail, one finds that the resulting Euler-Lagrange equations
will be satisfied only if all fluxes vanish. This implies that the electromagnetic potential must be
a pure gauge potential. In other words, the case~$n \geq 3$ does not allow for an interaction via
gauge potentials. This is the reason why this case was disregarded
(for some more details on this argument, see~\cite[Remark~6.2.5]{pfp}).

After these considerations, we are left with the
\sindex{causal action!two-point}%
\beq %\label{Stwopoint}
\text{two-point action} \qquad \Sact = \sum_{x,y \in M} \L[A_{xy}] \:, \eeq
where~$A_{xy}$ is the closed chain formed of two points,
\beq \label{Axytwopoint}
A_{xy} := P(x,y)\, P(y,x) \:.
\eeq
In this case, the polygon with vertices~$x$ and~$y$ degenerates to a straight line,
implying that the flux through this polygon vanishes as desired.
The starting point for a more quantitative analysis was to choose the Lagrangian
formed by taking products and sums of traces of powers of the closed chain.
A typical example is the Lagrangian
\beq \label{Lpoly}
\L[A_{xy}] := \Tr_{S_x} \big( A_{xy}^2 \big) - c\, \big( \Tr_{S_x} ( A_{xy} ) \big)^2
\eeq
with a real parameter~$c$.
In such examples, the Lagrangian is a symmetric polynomial in the eigenvalues of the closed chain.
The methods and results of this early analysis can be found in the unpublished preprints~\cite{endlich, lightint}.

Generally speaking, the study of such polynomial Lagrangians seemed a promising
strategy toward formulating physically sensible equations.
However, the more detailed analysis revealed the basic problem that {\em{chiral gauge phases}} come into play:
\sindex{gauge phase!chiral}%
As just explained after~\eqref{Pgauge}, the closed chain and therefore also the Lagrangian are
gauge invariant for the electromagnetic potential. However, the situation changes if chiral gauge potentials
are considered. Here chiral gauge potentials are left- or right-handed potentials~$A_L$ and~$A_R$
which can be inserted into the Dirac equation by generalizing~\eqref{Dirac2} to
\beq \big( \ci \Pdd + \chi_R \slashed{A}_L + \chi_L \slashed{A}_R - m \big) \psi = 0 \:, \eeq
where~$\chi_{L\!/\!R}$ are the chiral projection operators~\eqref{chidef}
(for details see, for example, \cite[\S2.2.3]{cfs}).
In physics, the electroweak interaction involves left-handed gauge potentials.
In this case, the left- and right-handed components of~$P(x,y)$ involve phase transformations
by the left- and right-handed gauge potentials, respectively. When forming the closed chain~\eqref{Axytwopoint},
the left- and right-handed components of~$P(x,y)$ are multiplied together. As a consequence,
the closed chain involves relative phases of the left- and right-handed gauge potentials; that is, phase factors
of the form
\beq \E^{\pm \cI \int_x^y (A_L - A_R)_j \xi^j} \:, \eeq
where~$A_L$ and~$A_R$ are the left- and right-handed gauge potentials (here for simplicity
again Abelian).
As a consequence, also the eigenvalues of the closed chain are multiplied by these relative phases.
The traces of powers of the closed chain as in~\eqref{Lpoly} are still real-valued
(this is because the phase factors always come as complex conjugate pairs), but they do not have
fixed signs. Working out the Euler-Lagrange equations, one sees that they also involve the relative gauge phases,
making it difficult to allow for chiral gauge fields.
In order to bypass these difficulties, from around~1999 on Lagrangians were considered which involved
{\em{absolute values}} of the eigenvalues of the closed chain. This had two major advantages:
\bitem
\item[(a)] The {\em{chiral gauge phases drop out}} of the Lagrangian.
\item[(b)] It became natural to formulate non-negative Lagrangians. As a consequence,
in the variational principle one {\em{minimize the action}} instead of merely looking for
critical points.
\item[(c)] A connection to {\em{causality}} was obtained. In order to see how this comes about,
we give a simple computation in the Minkowski vacuum. Suppose that the points~$x$ and~$y$
are either timelike or spacelike separated. Then, the kernel~$P(x,y)$ is well-defined and finite
even without regularization and, due to Lorentz symmetry, it has the form
\beq %\label{Pxyrep}
P(x,y) = \alpha\, \xi_j \gamma^j + \beta\:\1 \eeq
with two complex-valued functions~$\alpha$ and~$\beta$ (where again~$\xi =y-x$, and~$\gamma^j$ are the Dirac matrices).
Taking the adjoint with respect to the spin inner product, we see that
\beq %\label{Pyxrep}
P(y,x) = \overline{\alpha}\, \xi_j \gamma^j + \overline{\beta}\:\1 \:. \eeq
As a consequence,
\beq A_{xy} = P(x,y)\, P(y,x) = a\, \xi_j \gamma^j + b\, \1 \eeq
with two real parameters~$a$ and~$b$ given by
\beq \label{ab}
a = \alpha \overline{\beta} + \beta \overline{\alpha} \:,\qquad
b = |\alpha|^2 \,\xi^2 + |\beta|^2
\eeq
(here~$\xi^2=\xi^i \xi_i$ denotes the Minkowski inner product, which may be negative).
Applying the formula~$(A_{xy} - b \1)^2 = a^2\:\xi^2\,\1$,
the roots of the characteristic polynomial of~$A_{xy}$ are computed by
\beq \label{root}
b \pm \sqrt{a^2\: \xi^2} \:.
\eeq
Therefore, the eigenvalues of the closed chain are either real, or else they
form a complex conjugate pair. Moreover, one gets a connection to causality:
By explicit computation in Minkowski space one sees that~$a$ is non-zero
(the details can be found in~\cite[proof of Lemma~4.3]{lqg}).
Therefore, if~$\xi$ is timelike (i.e.\ $\xi^2>0$) then the relations~\eqref{ab} and~\eqref{root}
show that the eigenvalues are distinct, both real and have the same sign.
If~$\xi$ is spacelike, on the other hand, the eigenvalues are complex and have the same
absolute value. In this way, one gets agreement with the spectral definition of causality in Definition~\ref{def2}.
Moreover, choosing a Lagrangian which depends only on differences of absolute values of the eigenvalues
vanishes for spacelike separation, making it possible to build causality into the action principle.
\eitem
The further analysis led to the class of Lagrangians
\beq \L = \sum_{i,j} \Big( \big|\lambda^{xy}_i \big|^p
- \big|\lambda^{xy}_j \big|^p \Big)^2 \eeq
with a parameter~$p \in \N$, where the~$\lambda^{xy}_i$ are the eigenvalues of~$A_{xy}$
(again counted with algebraic multiplicities). The case~$p=1$ gives the causal
Lagrangian~\eqref{Lagrange} (albeit with the difference of working instead of
the local correlation operators with the kernel of the fermionic projector; the
connection will be explained below). The decision for~$p=1$ was taken
based on the so-called {\em{state stability analysis}}, which revealed that the vacuum Dirac
\sindex{state stability}%
sea configuration~\eqref{Pxyvac} is a local minimizer of the causal action only if~$p=1$
(for details, see~\cite[Section~5.5]{pfp}).
Now that the form of the causal action was fixed, the monograph~\cite{pfp} was completed
and published. The causal action principle is given in this book as an example
of a variational principle in discrete spacetime (see~\cite[Section~3.5]{pfp}).
The boundedness constraint~\eqref{Tdef} already appears, and the causal Lagrangian~\eqref{Lagrange}
arises when combining the Lagrangian with the Lagrange multiplier term corresponding to
the boundedness constraint. The volume constraint~\eqref{volconstraint} is also implemented,
however in discrete spacetime simply as the condition that the number of spacetime points
be fixed (and $\rho$-integrals are replaced by sums over the spacetime points).
The trace constraint, however, was not yet recognized as being necessary and important.

After the publication of the monograph~\cite{pfp}, the causal action principle was analyzed
in more detail and more systematically, starting from simple systems and proceeding to
more realistic physical models, concluding with systems showing all the interactions of the standard model
and gravity (see~\cite[Chapters~3--5]{cfs}).
This detailed study also led to the causal action principle in the form given in Section~\ref{seccap} above.
The path from the monograph~\cite{pfp} to the present formulation in~\cite{cfs}
is outlined in~\cite[Preface to second online edition]{pfp}. We now mention a few points
needed for the basic understanding.

One major conceptual change compared to the setting in indefinite inner product spaces
was to recognize that an underlying Hilbert space structure is needed in order for the
causal variational principle to be mathematically well-defined. This became clear
when working on the existence theory in discrete spacetime~\cite{discrete}.
This Hilbert space structure is built in most conveniently by working instead of the
kernel of the fermionic projector with the local correlation operators
which relate the Hilbert space scalar product to the spin inner product by
\sindex{local correlation operator}%
\beq \label{Finner}
\langle \psi | F(x) \phi \rangle_\H = -\Sl \psi(x) | \phi(x) \Sr_x \:.
\eeq
Using that the operator product~$F(x) F(y)$ has the same non-trivial eigenvalues as the closed chain~$A_{xy}$
given by~\eqref{Axytwopoint} (as we already observed in Section~\ref{secinherent} 
after~\eqref{Axydef}), the causal action principle can also be formulated
in terms of the local correlation operators~$F(x)$ with~$x \in M$.
Moreover, it turned out that measure-theoretic methods can be used to
generalize the setting such as to allow
for the description of not only discrete, but also continuous spacetimes.
In this formulation, the sums over the discrete spacetime points are replaced by
integrals with respect to a measure~$\mu$ on~$M$. 
This setting was first introduced in~\cite{continuum} when working out the existence theory.
In this formulation, the only a-priori structure of spacetime is that of a measure space~$(M, \mu)$.
The local correlation operators give rise to a mapping
\sindex{local correlation map}%
\beq F \::\: M \rightarrow {\mathscr{F}} \:,\quad x \mapsto F(x) \:, \eeq
where~${\mathscr{F}}$ is the subset of finite rank operators on~$\H$ which are symmetric and 
(counting multiplicities) have at most~$n$ positive and at most~$n$ negative eigenvalues
(where~$n$ is introduced via the signature~$(n,n)$ of the indefinite inner product in~\eqref{Finner}).
This analysis also revealed the significance of the trace constraint.
As the final step, instead of working with the measure~$\mu$, the causal action can be expressed
in terms of the push-forward measure~$\rho = F_* \mu$, being a measure on~${\mathscr{F}}$
(defined by~$\rho(\Omega) = \mu(F^{-1}(\Omega))$).
Therefore, it seems natural to leave out the measure space~$(M, \mu)$ and to work instead
directly with the measure~$\rho$ on~${\mathscr{F}}$.

These considerations led to the general definition of causal fermion systems
in Section~\ref{secgendef}, where the physical system is described by a
Hilbert space~$(\H, \langle .|. \rangle_\H)$ and
the measure~$\rho$ on~${\mathscr{F}}$. The causal action principle
takes the form as stated in Section~\ref{seccap}.

\section{Underlying Physical Concepts} \label{secphysicalconcepts}
We now briefly explain a few physical concepts behind causal fermion systems and
the causal action principle. The aim is to convey the reader the correct physical picture
in a non-technical way. Doing so already here makes it necessary to anticipate
some ideas on a qualitative level which will be introduced more systematically and thoroughly
later in this book.

It is a general feature of causal fermion systems that the usual distinction between
the structure of spacetime itself (being modelled by Minkowski space or a Lorentzian manifold)
and structures in spacetime (like wave functions and matter fields) ceases to exist.
Instead, all these structures are described as a whole by {\em{a single object}}: the measure~$\rho$
on~$\F$. Spacetime and all structures therein are different manifestations of this one object.
The dynamics of spacetime and of all objects in spacetime are described in a unified
and holistic manner by the causal action principle.
Clearly, in order to get a connection to the conventional description of physics, one still
needs to construct the familiar physical objects from a causal fermion system. Also, one needs to
rewrite the dynamics as described by the causal action principle in terms of these familiar physical objects.
This study is a main objective of this book. As already exemplified in Section~\ref{secinherent} by
the spin spaces and physical wave functions, the strategy is to identify suitable inherent structures in a causal fermion system, which then may be given suitable names. This must be done carefully in such a way that these
names convey the correct physical picture. Ultimately, the inherent structures serve the
purpose of getting a better understanding of the causal action principle. As we shall see, this will
be achieved by reformulating the Euler-Lagrange equations of the causal action principle in terms
of the inherent structures. When this is done, also the physical names of the inherent structures
will be justified by showing that they agree with the familiar physical objects in specific limiting
cases and generalize these objects in a sensible way.

In view of this unified description of all physical structures by a single mathematical object,
it is difficult to describe the essence of causal fermion systems using the familiar notions
from physics. One simple way of understanding the causal action principle is to focus
on the structure of the {\em{physical wave functions}} and the kernel of the fermionic projector.
Clearly, the resulting picture is a bit oversimplified, because it only captures part of the
structures encoded in a causal fermion system. Nevertheless, it conveys a good and the correct
intuition of what the causal action principle is about.
We saw in Section~\ref{secinherent} that a causal fermion system gives rise to
the family of physical wave functions~$(\psi^u)_{u \in \H}$ (see~\eqref{psiudef}).
The kernel of the fermionic projector~\eqref{Prep} is built up of all the physical wave functions
and thus describes the whole family. It gives rise to the closed chain~\eqref{Axydef}, which in
turn determines the causal action and the constraints.
In this way, the causal action principle becomes a variational principle for the family of
physical wave functions. Thus the interaction described by the causal action principle
can be understood as a direct mutual interaction of all the physical wave functions.
In simple terms, the causal action principle aims at bringing
the family of wave functions into an ``optimal'' configuration. For such optimal configurations,
the family of wave functions gives rise to the spacetime structures as we know them:
the causal and metric structure, the bosonic fields, and all that.

The last step can be understood more concretely starting from Dirac's hole theory
and the picture of the {\em{Dirac sea}} (for basics see again Section~\ref{secsea}).
In our approach, the Dirac sea is taken literally. Thus all the states of the Dirac sea
correspond to physical wave functions. All the information contained in these wave functions 
induces spacetime with the familiar structures. As a simple example, the bosonic potentials~$\B$
are determined via the Dirac equation~\eqref{DirPtilde} from the family of wave functions
as described by~$\tilde{P}(x,y)$. Clearly, in order to make this picture precise, one needs to
verify that, in a certain limiting case, the kernel of the fermionic projector corresponding to
a minimizer of the causal action principle indeed satisfies a Dirac equation of the form~\eqref{DirPtilde}
and thus gives rise to a potential~$\B$. This will be one of the objectives of the later chapters in this book.

We now discuss which {\em{physical principles}} enter the approach, and how they were incorporated.
Causal fermion systems evolved from an attempt to combine several physical principles
in a coherent mathematical setting. As a result, these principles appear in a specific way:
\begin{itemize}[leftmargin=1.3em, itemsep=0.2em]
\itemD The {\bf{principle of causality}}:
\sindex{causality!principle of}%
\sindex{principle of causality}%
A causal fermion system gives rise to a causal structure (see Definition~\ref{def2}).
The causal action principle is compatible with this notion of causality in the sense that the pairs of points with spacelike separation do not enter the Euler-Lagrange equations. In simple terms, points with spacelike separation do not interact.
\itemD The {\bf{local gauge principle}}: Already in the above discussion of how the
\sindex{local gauge invariance}%
causal action principle came about, we mentioned that the Lagrangian is gauge invariant
in the sense that gauge phases drop out of the Lagrangian (see the explanation after~\eqref{Pgauge}
in Section~\ref{secwhycap}). When starting from a general causal fermion system,
{\bf{local gauge invariance}} becomes apparent when representing the physical wave functions in bases of the spin spaces. More precisely, choosing a pseudo-orthonormal basis~$(\mathfrak{e}_\alpha(x))_{\alpha=1,\ldots, \dim S_x}$ of~$S_x$, a physical wave function can be represented as
\beq \label{lgf1}
\psi(x) = \sum_{\alpha=1}^{\dim S_x} \psi^\alpha(x)\: \mathfrak{e}_\alpha(x)
\eeq
with component functions~$\psi^1, \ldots, \psi^{\dim S_x}$.
The freedom in choosing the basis~$(\mathfrak{e}_\alpha)$ is described by the group of unitary transformations with respect to the indefinite spin inner product. This gives rise to the transformations
\beq \label{lgf2}
\mathfrak{e}_\alpha(x) \rightarrow \sum_{\beta} U^{-1}(x)^\beta_\alpha\;
\mathfrak{e}_\beta(x) \qquad \text{and} \qquad
\psi^\alpha(x) \rightarrow  \sum_{\beta} U(x)^\alpha_\beta\: \psi^\beta(x)
\eeq
with~$U \in \U(p,q)$.
As the basis~$(\mathfrak{e}_\alpha)$ can be chosen independently at each spacetime point,
one obtains {\em{local gauge transformations}} of the wave functions,
\sindex{local gauge transformation}%
where the gauge group is determined to be the isometry group of the spin inner product.
The causal action is {\em{gauge invariant}} in the sense that it does not depend on the choice of spinor bases.

\itemD The {\bf{Pauli exclusion principle}} is incorporated in a causal fermion system,
as can be seen in various ways. 
\sindex{Pauli exclusion principle}%
One formulation of the Pauli exclusion principle states that every fermionic one-particle state
can be occupied by at most one particle. In this formulation, the Pauli exclusion principle
is respected because every wave function can either be represented in the form~$\psi^u$
(the state is occupied) with~$u \in \H$ or it cannot be represented as a physical wave function
(the state is not occupied). 
Via these two conditions, the fermionic projector encodes for every state
the occupation numbers~$1$ and~$0$, respectively, but it is
impossible to describe higher occupation numbers.

More technically, one may obtain the connection to the fermionic Fock space formalism
by choosing an orthonormal basis~$u_1, \ldots, u_f$ of~$\H$ and forming the $f$-particle Hartree-Fock state
\sindex{Hartree-Fock state}%
\beq \Psi := \psi^{u_1} \wedge \cdots \wedge \psi^{u_f} \:. \eeq
Clearly, the choice of the orthonormal basis is unique only up to the unitary transformations
\beq u_i \rightarrow \tilde{u}_i = \sum_{j=1}^f U_{ij} \,u_j \quad \text{with} \quad U \in \U(f)\:. \eeq
Due to the anti-symmetrization, this transformation changes the corresponding Hart\-ree-Fock state
only by an irrelevant phase factor,
\beq \psi^{\tilde{u}_1} \wedge \cdots \wedge \psi^{\tilde{u}_f} = \det U \;
\psi^{u_1} \wedge \cdots \wedge \psi^{u_f} \:. \eeq
Thus the configuration of the physical wave functions can be described by a
fermionic multi-particle wave function.
The Pauli exclusion principle becomes apparent
in the total anti-symmetrization of this wave function.

Clearly, the above Hartree-Fock state  does not account for quantum entanglement. Indeed, the description of entanglement requires more general Fock space constructions (this will be described
in more detail in Chapter~\ref{secQFT}).
\itemD The {\bf{equivalence principle}}: Starting from a causal fermion
\sindex{equivalence principle}%
system~$(\H, \F, \rho)$, spacetime~$M$ is given as the support of the measure~$\rho$.
Thus spacetime is a topological space (with the topology on~$M$ induced by the operator norm on~$\Lin(\H)$). In situations when spacetime has a smooth manifold structure, one can describe spacetime by choosing coordinates. However, there is no distinguished coordinate systems, giving rise to the freedom of performing general coordinate transformations. The causal action as well as all the constraints are invariant under such transformations. In this sense, the equivalence principle is implemented in the setting of causal fermion systems.
\end{itemize}
However, other physical principles are missing. For example, the principle of locality is not included. Indeed, the causal action principle is {\em{non-local}}, 
\sindex{principle of locality}%
\sindex{non-locality}%
and locality is recovered only in the continuum limit. Moreover, our concept of causality is quite different from {\em{causation}} 
\sindex{causation}%
(in the sense that the past determines the future) or
{\em{microlocality}}
\sindex{microlocality}%
(stating that the observables of spacelike separated regions must commute).

\newpage
	
\section{A Summary of the Basic Concepts and Objects} \label{secsumm}

\tikz[remember picture,overlay] \draw [fill,blue!20] (current page.north west) rectangle +(0.03\paperwidth,-\paperheight);%

In this section we summarize all important concepts of the preceding sections. You may use this as a reference list for frequently used concepts and objects. 

\vspace{5mm}

\begin{center}
	\begin{tabular*}{\textwidth}{p{0.3\linewidth} p{0.05 \linewidth} p{0.55\linewidth}} 
		
		\textbf{Basic concept} & & \textbf{ Summary and Comments}\tabularnewline \hline 
		
		{\bf{Causal fermion system}} $(\H,\F, \rho)$ & & A separable Hilbert space $\H$, a natural number $n\in \N$, the set $\F$ of symmetric linear operators on $\H$ with at most $n$ positive and $n$ negative eigenvalues as well as a measure $\rho$ defined on a $\sigma$-algebra on $\F$ forms a \textbf{causal fermion system}. \newline \tabularnewline 
		
		\multicolumn{3}{p{0.95\linewidth}}{
		\textit{Remarks:}
		\begin{itemize}[leftmargin=2em]  \setlength\itemsep{1em}
			\item The structure of a causal fermion system provides a general framework for describing generalized spacetimes. Concrete physical systems correspond to specific choices of~$\H$, $n$ and the measure~$\rho$. 
			\item $\H$ should be considered as the Hilbert space spanned by
			all one-particle wave functions realized in our system (the {\em{physical wave functions}}). 
			\item We are mainly interested in the case~$n=2$ (at most two positive and two negative eigenvalues). This case allows for the description of Dirac spinors in four-dimensional spacetimes.
		\end{itemize}
		}	
		\tabularnewline \hline
	Spacetime $M$ & & By definition, we describe spacetime by the support of the measure $M := \text{supp}(\rho)$.\newline \tabularnewline
\multicolumn{3}{p{0.95\linewidth}}{
	\textit{Remarks:}
	\begin{itemize}[leftmargin=2em]  \setlength\itemsep{1em}
		\item All points $x\in M$ are linear operators on $\H$. This fact implies that our spacetime is endowed with more structures and contains additional information. 
		\item In order to describe systems in Minkowski space, we identify spacetime points~$x \in M$ with corresponding points in Minkowski space~$\scrM$ via a mapping~$F^\varepsilon: \scrM \to M$ (for more
details see~\eqref{FeMink}).
	\end{itemize}
}
\tabularnewline \hline 
	\end{tabular*}
\newpage

\tikz[remember picture,overlay] \draw [fill,blue!20] (current page.north east) rectangle +(-0.03\paperwidth,-\paperheight);

\begin{tabular*}{\textwidth}{p{0.2\linewidth} p{0.05\linewidth} p{0.65\linewidth}}  
	\hline

	The measure $\rho$ & &The measure~$\rho$ in Definition~\ref{defcfs} is the most important object of the theory. It describes spacetime as well as all objects therein. \newline
	\tabularnewline 
	\multicolumn{3}{p{0.95\linewidth}}{
	\textit{Remarks:}
	
	\begin{itemize}[leftmargin=2em]  \setlength\itemsep{1em}
		\item A lot of structure is encoded in the measure~$\rho$. In particular, it describes the behavior of spacetime on microscopic scales (Planck scale).
		\item In the example of causal fermion systems describing Minkowski space, the measure is obtained as the push-forward of the Minkowski volume measure $\Diff\mu=\Diff^4x$ under the local correlation map~$F^\varepsilon$; that is, we set~$\rho = F^\varepsilon_*\mu$. 
	\end{itemize}
	}	
	\tabularnewline \hline	
	
	The causal action & & We define a Lagrangian $\mathcal{L}(x,y)$ for two spacetime points $x$ and $y$
	using the eigenvalues $\large(\lambda^{xy}_i\large)_{i = 1,...,2n}$ of the product $xy$, which is an operator of rank at most $2n$. The Lagrangian is given by $\mathcal{L}(x,y) := \frac{1}{4n}\sum_{i,j = 1}^{2n}\large(\vert\lambda^{xy}_i \vert -\vert \lambda^{xy}_j\vert \large)^2$. Finally, the causal action
is defined by taking the double integral $\mathcal{S}(\rho):= \iint_{\F\times \F} \L(x,y) \Diff\rho(x)\Diff\rho(y)$. \newline
	\tabularnewline 
	\multicolumn{3}{p{0.95\linewidth}}{
		\textit{Remarks:}
		\begin{itemize}[leftmargin=2em]
			\item It may happen that the rank of the operator~$xy$ is smaller than~$2n$. In this case, some of the eigenvalues~$\lambda^{xy}_1, \ldots \lambda^{xy}_{2n}$ are zero.
			\item The action depends nonlinearly on the measure $\rho$. Since~$\rho$ describes spacetime and all objects therein, the action also depends on spacetime and on all these object.
	\end{itemize}	}

	\tabularnewline \hline

\end{tabular*}
\newpage

\tikz[remember picture,overlay] \draw [fill,blue!20] (current page.north west) rectangle +(0.03\paperwidth,-\paperheight);

\begin{tabular*}{\textwidth}{p{0.21\linewidth} p{0.04\linewidth} p{0.65\linewidth}}  
	\hline
	The causal action principle & & The causal action principle states that measures describing physical systems
must be minimizers of the causal action under variations of~$\rho$, respecting the constraints~\eqref{volconstraint}, \eqref{trconstraint}
and~\eqref{Tdef}.

	\tabularnewline 
	\multicolumn{3}{p{0.95\linewidth}}{
		\textit{Remarks:}
		\begin{itemize}[leftmargin=2em]
		\item The Euler-Lagrange equations corresponding to the causal action principle are the physical equations of the theory.
			\item 	By varying the measure~$\rho$, we also vary spacetime as well as all structures therein.
	\end{itemize}	}

\tabularnewline \hline	

The physical wave functions & &  Every vector~$u \in \H$ can represented in spacetime by
the physical wave function~$\psi^u$ defined by~$\psi^u(x) = \pi_x u \in S_x$, 
where $\pi_x$ denotes the orthogonal projection in $\H$ onto the subspace $x(\H)\subset\H$.
\tabularnewline \hline	
The kernel of the fermionic projector & &  For any spacetime point operator~$x \in M$ we define the spin space $S_x$ as its image $S_x:= x(\H)$. This gives rise to a mapping between spin spaces at different spacetime points $x,y\in M$ by $P(x,y):= \pi_x y\vert_{S_x}: S_y \to S_x$,
The mapping $P(x,y)$ is the kernel of the fermionic projector. It can be expressed in terms
of all physical wave functions by~$P(x,y) = -\sum_i \vert\psi^{e_i}(x)\Sr \Sl{\psi^{e_i}(y)}\vert$,
where the $(e_i)$ form an orthonormal basis of $\H$.
\newline
\tabularnewline

\multicolumn{3}{p{0.95\linewidth}}{
	\textit{Remarks:}
	\begin{itemize}[leftmargin=2em]
	\item The kernel of the fermionic projector gives relations between spacetime points. In particular, it encodes the causal structure and the geometry of spacetime.
		\item In order to compute the Lagrangian, it is useful to form the \textit{closed chain} $A(x,y):=P(x,y)P(y,x)$. 
%		\item The kernel of the fermionic projector is not a paralleltransport or connection between the two spin spaces. 
%		\item It seems to be useful to choose the inner products on the spin space such that $P^*(x,y) = P(y,x)$. 
\end{itemize}	}

\end{tabular*}

\end{center}

\newpage

\section{Exercises}

\begin{Exercise} \label{exexamplesimp} {\em{
This exercise is devoted to the study of the variational principle~\eqref{Ssum} of the motivating example.
\bitem
\item[(a)] Assume that the operators~$F_1, \ldots, F_L$ are a minimizer of the action~\eqref{Ssum}
under variations of~$F_i \in \F$ with~$\F$ according to~\eqref{Fone}.
Given~$i \in \{1, \ldots f\}$, represent~$F_i$ as
\beq F_i = |\psi_i \ra \la \psi_i | \qquad \text{with~$\psi_i \in \H$} \:. \eeq
Vary the vector~$\psi_i$ to derive the following Euler-Lagrange (EL) equations,
\beq %\label{ELsimp}
\sum_{i,j=1}^f \Tr(F_j F_i)\: F_j \,\psi_i = 0 \:. \eeq
\item[(b)] Deduce that all the matrices~$F_i$ must vanish.
{\em{Hint:}} It is useful to first show that
\beq \label{constraintsimp}
\sum_{i,j=1}^f \big| \Tr(F_j F_i) \big|^2 = 0 \:.
\eeq
\item[(c)] In order to get non-trivial solutions, one can for example impose the constraint
\beq \sum_{i=1}^f \Tr \big(F_i^2 \big) = 1 \:. \eeq
Derive the corresponding EL equations.
\item[(d)] The constraint~\eqref{constraintsimp} also makes it possible to prove existence of minimizer
with a compactness argument. Work out this existence proof in detail.
\eitem
}} \end{Exercise}

\begin{Exercise} {{(A causal fermion system on~$\ell_2$)}} \label{exl2} {\em{
Let~$\mathcal{H}=\ell_2$ be the Hilbert space of square-summable complex-valued sequences, equipped with the scalar product 
\beq
\langle u|v\rangle = \sum_{i=1}^\infty\bar{u}_i\,v_i,\quad u= (u_i)_{i\in\N},\ v= (v_i)_{i\in \N}.
\eeq
For any~$k\in\N$, let~$x_k\in \Lin(\mathcal{H})$ be the operator defined by
\beq
(x_k u)_k:=u_{k+1},\quad (x_k u)_{k+1}:=u_k,\quad (x_k u)_i=0\ \mbox{for~$i\not\in\{k,k+1\}$}.
\eeq
In other words,
\beq
x_k u = \big(\underbrace{0,\dots,0}_{\mbox{\tiny $k-1$ entries}},u_{k+1},u_k,0,\dots\big)
\eeq
Finally, let~$\mu$ the counting measure on~$\N$ (that is, $\mu(X)=|X|$ equals the cardinality of~$X\subset\N$.) 
\bitem
\item[(a)] Show that every operator~$x_k$ has rank two, is symmetric, and has one positive and one negative eigenvalue. Make yourself familiar with the concept that every operator is a point in~$\F$ for spin dimension~$n=1$. 
\item[(b)] Let~$F:\N\rightarrow \F$ be the mapping which to every~$k$ associates the corresponding operator~$x_k$. Show that the push-forward measure~$\rho=F_*\,\mu$ defined by~$\rho(\Omega):=\mu(F^{-1}(\Omega))$ defines a measure on~$\F$. Show that this measure can also be characterized by 
\beq
\rho(\Omega)=|\{k\in\N\:|\: x_k\in\Omega\}|.
\eeq
\item[(c)] Show that~$(\mathcal{H},\F,\rho)$ is a causal fermion system of spin dimension one.
\item[(d)] Show that the support of~$\rho$ consists precisely of all the operators~$x_k$. What is spacetime~$M$? What is the causal structure on~$M$? What is the resulting causal action?
\eitem
}} \end{Exercise}

\begin{Exercise} \label{exbounded} {{(Boundedness of operators of finite rank)}} {\em{
Let~$(\H, \la .|. \ra_\H)$ be a Hilbert space and~$(V, \|.\|)$ a normed space of
finite dimension~$n$. Moreover, let~$A : \H \rightarrow V$ be a linear mapping.
\bitem
\item[(a)] Show that the kernel of~$A$ is a closed subspace of~$\H$.
Show that its orthogonal complement~$(\ker A)^\perp$ has dimension at most~$n$.
\item[(b)] Derive a block matrix representation of~$A$ on~$\H = (\ker A) \oplus (\ker A)^\perp$.
\item[(c)] Deduce that~$A$ is bounded; that is, that there is a constant~$c>0$
with~$\|A u\| \leq c\: \|u\|_\H$ for all~$u \in \H$.
\eitem
}} \end{Exercise}

\begin{Exercise} \label{exm3} {{(On the trace constraint)}} {\em{
\sindex{constraint!trace}%
This exercise shows that the {{trace constraint}} ensures that the action is non-zero.
Let~$(\H, \F, \rho)$ be a causal fermion system of spin dimension~$n$.
\bitem
\item[(a)] Assume that~$\tr(x) \neq 0$. Show that~$\L(x,x)>0$.
(For a quantitative statement of this fact in the setting of discrete spacetimes,
see~\cite[Proposition~4.3]{discrete}.)
\item[(b)] Assume that~$\int_\F \tr(x)\: \Diff\rho \neq 0$.
Show that~$\Sact(\rho) > 0$.
\eitem
}} \end{Exercise}
%
%\begin{Exercise} \label{exm3alt} {{(Well-posedness of the causal action principle)}} {\em{
%This exercise explains why the causal action principle is ill-posed if~$\dim\H=\infty$ and~$\rho(\F)<\infty$.
%\bitem
%\item[(a)] Let~$\H_0$ be a finite-dimensional Hilbert space of dimension~$n$ and~$(\H_0,\F_0,\rho_0)$ 
%be a causal fermion system of finite total volume~$\rho_0(\F_0)<\infty$. Let~$\imath:\H_0\rightarrow\H$ be an
%isometric embedding of Hilbert spaces. Construct a causal fermion system~$(\H,\F,\rho)$ which has
%the same action, the same total volume and the same values for the trace and
%boundedness constraints as the causal fermion system~$(\H_0,\F_0,\rho_0)$.
%\item[(b)] Let~$\H_1:=\H_0\oplus\H_0$. Construct a causal fermion system~$(\H_1,\F_1,\rho_1)$ which has the same total volume and the same value of the trace constraint as~$(\H_0,\F_0,\rho_0)$ but a smaller action and a smaller value of the boundedness constraint.\\[0.3em]
%\textit{Hint:} Let~$F_{1/2}: L(\H_0)\rightarrow L(\H_1)$ be the linear mappings 
%\beq
%F_1(A)(u\oplus v):= (Au)\oplus 0,\quad F_2(A)(u\oplus v):= 0\oplus(Au).
%\eeq
%Show that~$F_{1/2}$ maps~$\F_0$ to~$\F_1$. Define the measure~$\rho_1$ by
%\beq
%\rho_1 = \frac{1}{2}\big((F_1)_*\rho_0+ (F_2)_*\rho_0)\big).
%\eeq
%\item[(c)] Iterate the construction in~$(ii)$ and apply~$(i)$ to obtain a sequence of measures on~$\F$ of fixed total volume and with fixed value of the trace constraint, for which the action and the values of the boundedness constraint tend to zero. Do these measures converge? If yes, what is the limit?
%\eitem
%}} \end{Exercise}

\begin{Exercise} \label{ex41} {{(On the spectrum of the closed chain)}} {\em{
This exercise is devoted to analyzing general properties of the spectrum of
the closed chain.
\bitem
\item[(a)] We let~$x$ and~$y$ be symmetric operators
of finite rank on a Hilbert space~$(\H, \la .|. \ra_\H)$.
Show that there is a finite-dimensional subspace~$I \subset \H$ on which both~$x$
and~$y$ are invariant.
By choosing an orthonormal basis of~$I$ and restricting the operators to~$I$,
we may represent both~$x$ and~$y$ by Hermitian matrices.
Therefore, the remainder of this exercise is formulated for simplicity in terms of Hermitian matrices.
\item[(b)] Show that for any matrix~$Z$, the characteristic polynomials of~$Z$ and of its
adjoint~$Z^*$ (being the transposed complex conjugate matrix) are related by complex
conjugation; that is, $\det(Z^*- \overline{\lambda} \:\1) = \overline{\det(Z-\lambda  \:\1)}$.
\item[(c)] Let~$X$ and~$Y$ be symmetric matrices. Show that the
characteristic polynomials of the matrices~$XY$ and~$YX$ coincide.
\item[(d)] Combine~(b) and~(c) to conclude that the characteristic polynomial
of~$XY$ has real coefficients; that is, $\det(XY- \overline{\lambda}  \:\1) = \overline{\det(XY-\lambda \:\1)}$.
Infer that the spectrum of the matrix product~$XY$ is symmetric about the real axis; that is,
\beq \label{XYsymm}
\det(XY- \lambda  \:\1)=0 \;\;\Longrightarrow\;\; \det(XY- \overline{\lambda} \:\1)=0 \:.
\eeq
\item[(e)] For the closed chain~\eqref{Axydef}, the mathematical setting is somewhat
different, because~$A_{xy}$ is a symmetric operator on the indefinite inner product
space~$(S_x, \Sl .|. \Sr_x)$. On the other hand, we saw after~\eqref{Axydef}
that~$A_{xy}$ is isospectral to~$xy$. Indeed, the symmetry result~\eqref{XYsymm}
can be used to prove a corresponding statement for~$A_{xy}$,
\beq %\label{Axysymm}
\det(A_{xy}- \lambda \:\1)=0 \;\;\Longrightarrow\;\; \det(A_{xy} - \overline{\lambda} \:\1)=0 \:. \eeq
This result is well-known in the theory of indefinite inner product spaces
(see, for example, the textbooks~\cite{bognar, GLR}
or~\cite[Section~3]{discrete}). In order to derive it from~\eqref{XYsymm}, one
can proceed as follows: First, represent the indefinite inner product in the
form~$\Sl .|. \Sr =\la .|S\, x \ra$, where~$\la .|. \ra$ is a scalar product
and~$S$ is an invertible operator which is symmetric (with respect to this scalar product).
Next, show that the operator~$B:=A_{xy} S$ is symmetric (again with
respect to this scalar product). Finally, write the closed chain as~$A_{xy} = B S^{-1}$
and apply~\eqref{XYsymm}.
\eitem
}} \end{Exercise}

\begin{Exercise} {{(Regular spacetime points)}} {\em{
Let~$x\in\F$ have~$p(x)\le n$ negative and~$q(x)\le n$ positive eigenvalues. The
pair~$\mathrm{sign}(x) := (p(x),q(x))$ is
referred to as the \textit{signature} of~$x$. The operator~$x$ is said to be \textit{regular}
\sindex{spacetime point!regular}%
 if~$\mathrm{sign}(x)=(n,n)$. The goal of this exercise is to show that the set~$\F^{\mathrm{reg}}$ of regular points is open in~$\F$. Let us define the \textit{positive} and \textit{negative components} of~$x$ as the operators
\beq
x_\pm :=\frac{x\pm |x|}{2},\quad |x|:=\sqrt{x^2}.
\eeq
From the functional calculus, it follows that~$x\,|x|= |x|\,x$. Prove the following statements.
\bitem
\item[(a)] Let~$\{e_i\:,\:i=1,\dots,m\}$ be an orthogonal set. Show that any vector set~$\{h_i\:,\:i=1,\dots,m\}$ which fulfills the following condition is linearly independent,
\beq
\|e_i-h_i\|<\frac{\inf\{\|e_i\|\:,\:i=1,\dots,m\}}{m}\quad \mbox{for all~$i=1,\dots,m$}.
\eeq 
\item[(b)] For every~$x\in\F$, 
\beq
 x(\mathrm{im}\,x_\pm)\subset \mathrm{im}\,x_\pm\quad\mbox{and}\quad x_+\,x_-=0.
\eeq 
 Moreover, $x|_{\mathrm{im }\, x_-}$ and~$x|_{\mathrm{im }\, x_+}$ are negative and positive definite, respectively.
\item[(c)] Let~$x_0\in\F$. Then there is an orthonormal set~$\{ e_i\:|\: i=1\dots \dim S_{x_0}\}$ of eigenvectors of~$x_0$ such that
\beq
\begin{split}
&\langle e_i|\,x_0\, e_i\rangle < 0 \quad \mbox{ for all~$i\le p(x_0)$}\\
&\langle e_i|\,x_0\, e_i\rangle>0 \quad \mbox{ for all~$p(x_0)<i\le p(x_0)+q(x_0)$}.
\end{split}
\eeq
\item[(d)] The following functions are continuous,
\beq f_i : B_r(x_0)\ni x\mapsto f_i(x):=\begin{cases}
	x_-\, e_i & \qquad \ \ \quad i\le p(x_0)\\
	x_ +\, e_i & p(x_0)<i\le p(x_0)+q(x_0) \:.
\end{cases} \eeq
\textit{Hint: You can use the general inequality~$\||A|-|B|\|\le\|A^2-B^2\|$}
\item[(e)] There is a~$r>0$ such that~$p(x)\ge p(x_0)$ and~$q(x)\ge q(x_0)$ for every~$x\in B_r(x_0)$.\\
\textit{Hint: Use the statements above.}
\item[(f)] Conclude that~$\F^{\mathrm{reg}}$ is an open subset of~$\F$.	
\eitem
}} \end{Exercise}
%
%\begin{Exercise} {{(On the spectrum of the closed chain - part 2)}} {\em{
%Let~$(\H,\F,\rho)$ be a causal fermion system and~$x,y\in\F$. For the closed chain 
%\beq
%A_{xy}:=P(x,y)P(y,x):(S_x,\Sl .|.\Sr_x)\rightarrow (S_x,\Sl .|. \Sr_x),
%\eeq
%the mathematical setting analyzed in Exercise~\ref{ex41} is somewhat different,
%because~$A_{xy}$ is a symmetric operator on an \textit{indefinite inner product space}. On the other hand, we know that~$A_{xy}$ is isospectral to~$xy$.
%Indeed, the symmetry result in Exercise~\ref{ex41}~(iv) can be used to prove a corresponding statement for the closed chain:
%\beq
%\det(A_{xy}-\lambda\,\mathbb{I}) = 0\  \Longleftrightarrow\  \det(A_{xy}-\overline{\lambda}\,\mathbb{I}) = 0.
%\eeq
%This result is well-known in the theory of indefinite inner product spaces. In order to derive it
%from Exercise~\ref{ex41}~(d), one can proceed as follows: First, represent the indefinite inner
%product in the form~$\Sl\,\cdot\,,\,\cdot\,\Sr = \langle\,\cdot\,,S\,\cdot\,\rangle$, where~$\langle\,\cdot\,,\,\cdot\,\rangle$ is a scalar product and~$S$ is an
%invertible operator which is symmetric (with respect to this scalar product). Next,
%show that the operator~$B:= A_{xy}S$ is symmetric (again with respect to this scalar
%product). Finally, write the closed chain as~$A_{xy}=BS^{-1}$ and apply Exercise~\ref{ex41}~(d).
%}} \end{Exercise}
%

\begin{Exercise} {{(A causal causal fermion system on~$\ell_2$ - part 2)}} {\em{
We return to the example of Exercise~\ref{exl2}.. This time we equip it with a \textit{Krein structure}.
\bitem
\item[(a)] For any~$k\in\N$, construct the spin space~$S_{x_k}$ and its spin scalar product.
\item[(b)] Given a vector~$u\in\H$, what is the corresponding wave function~$\psi^u$?
What is the Krein inner product~$<.,.>$?
\item[(c)] What is the topology on the Krein space~$\K$? Does the wave evaluation operator~$\Psi:u\mapsto\psi^u$ give rise to a well-defined and continuous mapping~$\Psi:\H \rightarrow \K$?
If yes, is it an embedding? Is it surjective?
\item[(d)] Repeat part (c) of this exercise for the causal fermion system obtained if the operators~$x_k$ are multiplied by~$k$, i.e.
\beq
x_k u:=(0,\dots,0,k\,u_{k+1},k\,u_k\,0,\dots).
\eeq
\eitem
}} \end{Exercise}

\begin{Exercise} {{(Time direction)}} \label{extimedir} {\em{
\sindex{time direction}%
The ability to distinguish between past and future can be described in mathematical terms by the existence of an anti-symmetric functional~$\mathcal{T}:M\times M\rightarrow \R$. One then says that
\beq
\begin{cases}
\ y\ \mbox{lies in the \emph{future} of}\ x & \mbox{if~$\mathcal{T}(x,y)>0$}\\
\ \ \ y\ \mbox{lies in the \emph{past} of}\ x & \mbox{if~$\mathcal{T}(x,y)<0$}\,.
\end{cases}
\eeq
Can you think of simple non-trivial examples of such a functional which involve only products and linear combinations of the spacetime operators and the orthogonal projections on the corresponding spin spaces?
{\em{Hint:}} One possible functional is
\beq {\mathcal{T}}(x,y) := \tr \big( y\,\pi_x - x\, \pi_y \big) \:; \eeq
this is considered further in~\cite[Exercise~1.22]{cfs}.
}} \end{Exercise}

\begin{Exercise} \label{exchar} {\em{ This exercise is devoted to clarifying the connection between
the characteristic polynomial and traces of powers of a matrix.
We let~$A$ be an~$N \times N$-matrix (not necessarily Hermitian)
and denote the zeros of its characteristic polynomials counting multiplicities by~$\lambda_1, \ldots, \lambda_N \in \C$; that is,
\beq \det (\lambda\,\1 - A) = (\lambda-\lambda_1) \cdots (\lambda-\lambda_N) \:. \eeq
Moreover, we denote the coefficients of the characteristic polynomial by~$a_k$; that is,
\beq \det (\lambda \1 - A) = \lambda^N + a_{1} \lambda^{N-1} + \cdots + a_N \:. \eeq
\bitem
\item[(a)] Show that the coefficients are symmetric polynomials in the eigenvalues of the form
\beq a_n = c_n \mathop{\sum_{B \subset \{1, \ldots, N\}}}_{\text{with~$\# B = n$}}
\;\prod_{k \in B} \lambda_k \:, \eeq
where the sum goes over all subsets of~$\{1, \ldots N\}$ with~$n$ elements, and~$c_n$ are
combinatorial prefactors. Compute the~$c_n$.
\item[(b)] Show that each coefficient~$a_n$ can be written in the form
\begin{align}
a_n &= d_n \Tr(A^n) + d_{1,n-1} \Tr(A)\, \Tr(A^{n-1}) \\
&\quad\: + d_{1,1,n-2} \Tr(A)\, \Tr(A)\,\Tr(A^{n-2}) + \cdots \\
&= \sum_{k=1}^n \;\;\mathop{\sum_{1 \leq j_1 \leq \cdots \leq j_k}}_{\text{with~$j_1+\cdots+ j_k=n$}} \;
d_{j_1, \ldots, j_k}\: \Tr\big( A^{j_1} \big) \cdots \Tr\big( A^{j_k} \big)
\end{align}
with suitable combinatorial factors~$d_n, d_{1,n-1}, \ldots$.
{\em{Hint:}} This formula can be derived in various ways. One method is to proceed inductively in~$n$.
Alternatively, one can use a dimensional argument.
\eitem
}} \end{Exercise}

\begin{Exercise} {{(Embedding of~$S_x\scrM$ into~$S_{F(x)}$)}} \label{ex7} {\em{
The goal of this exercise is to explain how the fibers of the spinor bundle~$S\scrM$ are related to the spin spaces~$S_x$ of the corresponding causal fermion system.
In order to keep the setting as simple as possible, we let~$(\scrM, g)$ be Minkowski space
and~$\H$ a finite-dimensional subspace of the Dirac solution space~$\H_m$, consisting
of smooth wave functions of spatially compact support; that is,
\beq \H \subset \Cisc(\scrM, S\scrM) \cap \H_m \qquad \text{finite-dimensional} \:. \eeq
We again let~$F(x)$ be the local correlation operators; that is,
\beq \la \psi | F(x) \phi \ra = - \Sl \psi(x) | \phi(x) \Sr \qquad \text{for all~$\psi, \phi \in \H$} \eeq
(since~$\H$ consists of smooth functions, we may leave out the regularization operators).
Defining the measure again by~$\Diff\rho = F_*(\Diff^4x)$, we again obtain a causal fermion
system of spin dimension~$n=2$. We next introduce the {\em{evaluation map}}~$e_x$ by
\beq e_x \::\: \H \rightarrow S_x\scrM\:,\qquad e_x(\psi) = \psi(x) \:. \eeq
Restricting the evaluation mapping to the spin space~$S_{F(x)}$ at the spacetime point~$F(x)$
(defined as in~\eqref{Sxdef} as the image of the operator~$F(x)$), we obtain a mapping
\beq e_x|_{S_{F(x)}} \::\: S_{F(x)} \rightarrow S_x\scrM \:. \eeq
\bitem
\item[(a)] Show that~$e_x|_{S_{F(x)}}$ is an isometric embedding.
\item[(b)] Show that for all~$u \in \H$ and~$x \in \scrM$,
\beq e_x|_{S_{F(x)}} \big( \psi^u(F(x)) \big) = u(x) \:. \eeq
\eitem
}} \end{Exercise}

\begin{Exercise} {{(Identification of~$S\scrM$ with~$SM$)}} \label{ex71} {\em{
In the setting of the previous exercise, we now make two additional assumptions:
\bitem
\item[(i)] The mapping~$F : \scrM \rightarrow \F$ is injective and its image is closed in~$\F$.
\item[(ii)] The resulting causal fermion system is {\em{regular}} in the sense that for all~$x \in \scrM$,
the operator~$F(x)$ has rank~$2n$.
\end{itemize}
Using the results of the previous exercise, explain how the following objects can be identified:
\begin{itemize}[leftmargin=2em]
\item[(a)] $x$ with~$F(x)$
\item[(b)] $\scrM$ with~$M$
\item[(c)] The spinor space~$S_x\scrM$ with the corresponding spin space~$S_{F(x)}$
\item[(d)] $u \in \H$ with its corresponding physical wave function~$\psi^u$
\eitem
}} \end{Exercise}

\begin{Exercise} {{(The space~$C^0(M, SM)$)}} {\em{
A {\em{wave function}}~$\psi$ is defined a mapping from~$M$ to~$H$
such that~$\psi(x) \in S_x$ for all~$x \in M$.
It is most convenient to define {\em{continuity}} of a wave function
by the requirement that for all~$x \in M$ and for
every~$\varepsilon>0$ there is~$\delta>0$ such that
\beq \big\| \sqrt{|y|} \,\psi(y) -  \sqrt{|x|}\, \psi(x) \big\|_\H < \varepsilon
\qquad \text{for all~$y \in M$ with~$\|y-x\| \leq \delta$} \:. \eeq
Show that, using this definition, every physical wave function is continuous.
\sindex{continuity!of wave functions in spacetime $M$}%
\nindex{aedb@$C^0(M, SM)$ -- continuous wave functions in spacetime~$M$}%
\tindex{ff@$C^0(M, SM)$ -- continuous wave functions in spacetime~$M$}%
Thus, denoting the space of continuous wave functions by~$C^0(M, SM)$, we obtain an embedding
\beq \H \hookrightarrow C^0(M, SM)\:. \eeq
{\em{Hint:}} You may use the inequality
\beq \Big\| \sqrt{|y|} - \sqrt{|x|} \Big\| \leq \|y-x\|^\frac{1}{4} \:\|y+x\|^\frac{1}{4} \:. \eeq
}} \end{Exercise}

\begin{Exercise} {{(A causal fermion system in~$\R^3$)}} \label{exdiracsphere} {\em{
We choose~$\H=\C^2$ with the canonical scalar product. Moreover, we choose
let~$\scrM=S^2 \subset \R^3$ and~$\Diff\mu$ the Lebesgue measure on~$\scrM$.
Consider the mapping
\beq %\label{Fexdef}
F : \scrM \rightarrow \Lin(\H)\:, \qquad F(p) = 2 \sum_{\alpha=1}^3 p^\alpha \sigma^\alpha + \1 \:, \eeq
where~$\sigma^\alpha$ are the three Pauli matrices~\eqref{pauli}.
\bitem
\item[(a)] Show that for every~$p \in S^2$,
\beq \tr \big( F(p) \big) = 2 \:,\qquad \tr \big( F(p)^2 \big) = 10 \:. \eeq
Conclude that the eigenvalues of~$F(p)$ are equal to~$1 \pm 2$.
\item[(b)] We introduce the measure~$\rho$
as the push-forward measure~$\rho = F_* \mu$ (that is, $\rho(\Omega) := \mu(F^{-1}(\Omega))$).
Show that~$(\H, \F, \rho)$ is a causal fermion system of spin dimension one.
\item[(c)] Show that the support of~$\rho$ coincides with the image of~$F$.
Show that it is is homeomorphic to~$S^2$.
\eitem
This example is also referred to as the {\em{Dirac sphere}}; this and other similar
examples can be found in~\cite[Examples~2.8 and~2.9]{continuum} or~\cite[Example~2.2]{topology}.
}} \end{Exercise}

\begin{Exercise} (The regularized fermionic projector in Minkowski space) \label{exminkfirst} {\em{
The goal of this exercise is to compute the kernel of the fermionic projector
in the Minkowski vacuum for the simplest regularization, the $i \varepsilon$-regularization~\eqref{iepsreg}.
\bitem
\item[(a)] Use the identifications of Exercises~\ref{ex7} and~\ref{ex71} to show that~\eqref{Prep}
holds in the example of Dirac wave functions in Minkowski space
(as constructed in Section~\ref{seclco}) but now with Dirac wave functions and the spin inner product thereon.
\item[(b)] More specifically, we now choose~$\H = \H_m^-$ as the subspace of all negative-frequency
solutions of the Dirac equation. Moreover, we choose the~$i \varepsilon$-regularization~\eqref{iepsreg}.
For clarity, we denote the corresponding kernel of the fermionic projector by~$P^{2\varepsilon}(x,y)$.
Show that
\begin{equation}\label{proj}
P^\varepsilon(x,y)=\int_{\R^4}\frac{\dd^4k}{(2\pi)^4}\,(\slashed{k}+m)\,\delta(k^2-m^2)\,\Theta(-k^0)\,\E^{-\cI k(x-y)}\,\E^{\varepsilon k^0} \:.
\end{equation}
{\em{Hint:}} Work in a suitable orthonormal basis of the Hilbert space.
Without regularization, the computation can be found in~\cite[Lemma~1.2.8]{cfs}.
\item[(c)] Show that~$P^\varepsilon(x,y)$ can be written as~$\slashed{v}^\varepsilon+\beta^\varepsilon$
with~$v^\varepsilon_j,\beta^\varepsilon$ smooth functions of~$\xi=y-x$.
\item[(d)] Compute~$P^\varepsilon(x,x)$. Is this matrix invertible? How does it scale in~$\varepsilon$?
Why does this result show that the resulting causal fermion system is regular?
{\em{Hint:}} The details can also be found in~\cite[Section~2.5]{cfs}.
For an alternative way of proving regularity, see Exercise~\ref{exminklast}.
\item[(e)] For~$\xi$ spacelike or timelike (that is, away from the lightcone) the limit~$\varepsilon\searrow 0$ of~\eqref{proj} is well-defined. More precisely, it can be shown that~$v^\varepsilon_j\to \alpha\,\xi_j$ and~$\beta^\varepsilon\to \beta$ pointwise, for~$\alpha,\beta$ smooth complex functions.
Find smooth real functions~$a,b$ such that
\begin{equation} \label{exclosedchain}
\lim_{\varepsilon \searrow 0} A^\varepsilon_{xy}=a\slashed{\xi}+b.
\end{equation}
\eitem
}} \end{Exercise}

\begin{Exercise} (Correspondence of the causal structure in Minkowski space I) \label{excausalmink1} {\em{
$\quad$ Let~$x,y\in \scrM$ be timelike separated vectors and assume  that~$\xi:=y-x$ is normalized to~$\xi^2=1$. As explained in Exercise~\ref{exminkfirst}, the limit~$\varepsilon\searrow 0$ of the closed chain~$A^\varepsilon_{xy}$ takes the form~$A:=a\,\slashed{\xi}+b$. Consider the matrices
\beq
F_\pm :=\frac{1}{2}\,(\1 \pm\slashed{\xi})\in \Lin(\C^4)\:.
\eeq
Prove the following statements.
\bitem
\item[(a)] The matrices~$F_\pm$ have rank two and map to eigenspaces of~$A$. What are the corresponding eigenvalues? Conclude that the points~$x$ and~$y$ are timelike separated in the sense of Definition~\ref{def2}.
\item[(b)] The matrices~$F_\pm$ are idempotent and symmetric with respect to the spin inner product~$\Sl .|.\Sr$.
\item[(c)] The image of the matrices~$F_\pm$ is positive or negative definite
(again with respect to the spin inner product).
\item[(d)] The image of~$F_+$ is orthogonal to that of~$F_-$ (again with respect to the spin inner product).
\item[(e)] The eigenvalues of~$A$ are strictly positive.
{\em{Hint:}} Use how the functions~$a$ and~$b$ came up in~\eqref{exclosedchain}.
\eitem
The result of~(a)--(d) can be summarized by saying that the~$F_\pm$ are the spectral projection operators of~$A$. We remark that the findings also
mean that the~$x$ and~$y$ are even {\em{properly timelike separated}}
as introduced in~\cite[Definition~1.1.6]{cfs}.
}} \end{Exercise}

\begin{Exercise} (Correspondence of the causal structure in Minkowski space II) {\em{
We now let~$x,y\in \scrM$ be spacelike separated vectors and assume that~$\xi:=y-x$ is normalized to~$\xi^2=-1$.
Consider again the matrix~$A:=a\,\slashed{\xi}+b$ of Exercise~\ref{excausalmink1} and set
\beq
F_\pm :=\frac{1}{2}\,(\1 \pm \cI\,\slashed{\xi})\in \Lin(\C^4)\:.
\eeq
\bitem
\item[(a)] The matrices~$F_\pm$ have rank two and map to eigenspaces of~$A$. What are the corresponding eigenvalues? Conclude that the points~$x$ and~$y$ are spacelike separated in the sense of Definition~\ref{def2}.
\item[(b)] The matrices~$F_\pm$ are idempotent and~$F_+^* = F_-$.
\item[(c)] The image of the matrices~$F_\pm$ is null (in other words, it is a lightlike subspace of the spinor space).
\eitem
These findings illustrate the more general statement that symmetric operators on an indefinite inner
product space may have complex eigenvalues, in which case they form complex conjugate pairs.
}} \end{Exercise}

\begin{Exercise} (Spin spaces for the regularized Dirac sea vacuum) \label{exminklast} {\em{
We consider the causal fermion system constructed in Section~\ref{seclco},
where we choose~$\H=\H_m^-$ as the space of all negative-energy solutions of the Dirac equation.
Moreover, we choose the~$i \varepsilon$-regularization~\eqref{iepsreg}.
For clarity, we denote the corresponding kernel of the fermionic projector by~$P^\varepsilon(x,y)$.
This causal fermion system is also referred to as the {\em{regularized Dirac sea vacuum}}.
\bitem
\item[(a)] Let~$\Sigma_0$ denote the Cauchy surface at time~$t=0$. Show that, for any~$x\in \scrM$
and~$\chi\in\C^4$,
\beq (\cI\slashed{\partial}-m)P^\varepsilon(\,\cdot\,,x)\chi=0\quad\mbox{and}\quad	P^\varepsilon(\,\cdot\,,x)\chi\big|_{\Sigma_0}\in\mathcal{S}(\R^3,\C^4) \:. \eeq
 Conclude that~$P^\varepsilon(\,\cdot\,,x)\chi\in\H_m^-\cap C^\infty(\R^4,\C^4)$.
\item[(b)] Convince yourself that 
\beq \mathfrak{R}_\varepsilon(P^\varepsilon(\,\cdot\,,x)\chi)=P^{2\varepsilon}(\,\cdot\,,x)\chi \:. \eeq
\item[(c)] Let~$\{\mathfrak{e}_1,\dots,\mathfrak{e}_4\}$ denote the canonical basis of~$\C^4$. Using Exercise~\ref{exminkfirst}~(b), show that the wave functions~$P^\varepsilon(\,\cdot\,,x)\mathfrak{e}_\mu$ for~$\mu=1,2,3,4$
 are linearly independent.  
\item[(d)] Let~$S_x:=F^\varepsilon(x)(\H_m^-)$ endowed with~$\Sl u,v\Sr_x := -\langle u|F^\varepsilon(x)v\rangle$ be the \textit{spin space} at~$x\in \scrM$. Show that the following mapping is an isometry of indefinite inner products  (that is, injective and product preserving),
\beq
 \Phi_x:S_x\ni u\mapsto \mathfrak{R}_\varepsilon u(x)\in \C^4.
\eeq
Conclude that the causal fermion system is regular at~$x\in \scrM$ (that is, $\dim S_x=4$) if and only if there exist vectors~$u_\mu\in \H_m^-$, for~$\mu=1,2,3,4$, such that the~$\mathfrak{R}_\varepsilon u_\mu(x)\in\C^4$  are linearly independent.
\item[(e)] Conclude that the causal fermion system is regular at every spacetime point.
\eitem
}} \end{Exercise}

\chapter{Causal Variational Principles} \label{chapcvp}
The causal action principle as introduced in Section~\ref{seccap} has quite a rich structure
and is rather involved. Therefore, it is difficult to analyze it in full generality in one step.
It is preferable to begin with special cases and simplified situations, and to proceed from there step by step.
In fact, doing so leads to a whole class of variational principles, referred to as {\em{causal variational principles}}.
\sindex{causal variational principle}%
These different variational principles capture different features and aspects of the causal action principle.
Proceeding in this way also gives a better understanding of the physical
meaning of the different structures of a causal fermion system and of the interaction as described
by the causal action principle.
We now give an overview of the different settings considered so far.
This has the advantage that in the later chapters of this book, we can always work in the
setting that is most suitable for the particular question in mind.
Moreover, for pedagogical reasons, in this book, we shall sometimes idealize the
setting, for example by assuming, for technical simplicity, that the Lagrangian is smooth.

\section{The Causal Variational Principle on the Sphere} \label{seccvpsphere}
\sindex{causal variational principle!on the sphere|textbf}%
Clearly, the trace constraint~\eqref{trconstraint} and the boundedness constraint~\eqref{Tdef}
complicate the analysis. Therefore, it might be a good idea to consider a simplified setting
where these constraints are not needed. This can be accomplished most easily by {\em{prescribing the
eigenvalues}} of the operators in~$\F$. This method was first proposed in~\cite[Section~2]{continuum}
in a slightly different formulation. We now explain the method in a way that fits best to our setting.
Given~$n \in \N$, we choose real numbers~$\nu_1, \ldots, \nu_{2n}$ with
\beq \label{nudenote}
\nu_1 \leq \cdots \leq \nu_{n} \;\leq 0 \;\leq\; \nu_{n+1} \leq \ldots \leq \nu_{2n} \:.
\eeq
We let~$\F$ be the set of all symmetric operators on~$\H$ of rank~$2n$ whose eigenvalues
(counted with multiplicities) coincide with~$\nu_1, \ldots, \nu_{2n}$.
If~$\H$ is finite-dimensional, the set~$\F$ is compact.
This is the reason why it is sensible to minimize the causal action~\eqref{Sdef}
keeping only the volume constraint~\eqref{volconstraint}, which for simplicity we implement
by restricting attention to normalized measures,
\beq %\label{totovolone}
\rho(\F) = 1 \:. \eeq
Note that, since~$\F$ is compact and the Lagrangian~$\L$ is continuous on~$\F \times \F$, also the action~$S(\rho)$ is finite for any normalized measure~$\rho$. 

The simplest interesting case is obtained by choosing the spin dimension~$n=1$
and the Hilbert space~$\H = \C^2$. In this case, according to~\eqref{nudenote} we have
one non-negative and one non-positive eigenvalue. If these eigenvalues have the same absolute value,
all the operators have trace zero. This case is not of interest because there are trivial minimizers
(for details, see Example~\ref{extraceconstraint} in Section~\ref{seccounter}).
With this in mind, it suffices to consider the case that~$|\nu_1| \neq |\nu_2|$.
Since scaling all the eigenvalues in~\eqref{nudenote} by a real constant does not change the
essence of the variational principle, it is no loss of generality to assume that the two eigenvalues~$\nu_1, \nu_2$
satisfy the relation~$\nu_1+\nu_2=2$, making it possible to parametrize the eigenvalues by
\beq \nu_{1\!/\!2} = 1 \mp \tau \qquad \text{with} \qquad \tau \geq 1 \:. \eeq
Then~$\F$ consists of all Hermitian $2 \times 2$-matrices~$F$ which have
eigenvalues~$\nu_1$ and~$\nu_2$. These matrices can be represented using the
Pauli matrices by
\beq \label{Fspheredef}
\F = \big\{  F = \tau\: \vec{x} \vec{\sigma} + \1 \quad \text{with} \quad \vec{x} \in S^2 \subset \R^3 \big\} \:.
\eeq
Thus the set~$\F$ can be identified with the unit sphere~$S^2$.

The volume constraint~\eqref{volconstraint} can be implemented most easily by
restricting attention to {\em{normalized}} regular Borel measures on~$\F$
(that is, measures with~$\rho(\F)=1$). 
Such a measure can be both continuous, discrete
or a mixture. Examples of {\em{continuous measures}} are obtained by
multiplying the Lebesgue measure on the sphere by a non-negative smooth
function. By a {\em{discrete measure}}, on the other hand, we here
mean a {\em{weighted counting measure}},
\sindex{measure!weighted counting}%
that is, a measure obtained by inserting weight factors into~\eqref{deltasum},
\beq \label{weighted}
\rho = \sum_{i=1}^L c_i \,\delta_{x_i}
\qquad \text{with} \qquad x_i \in \F\:,\quad c_i \geq 0 \quad \text{and} \quad \sum_{i=1}^L c_i = 1 \:.
\eeq

In this setting, a straightforward computation yields for the Lagrangian~\eqref{Lagrange}
(see Exercise~\ref{excomsphere})
\beq \label{Lform}
\begin{split}
\L(x,y) &= \max \big( 0, \D(x,y) \big) 
\qquad \text{with} \\
\D(x,y) &= 2 \tau^2\: \big(1+ \langle x,y \rangle \big) \Big( 2 - \tau^2 \: \big(1 - \langle x,y \rangle \big) \Big) \:,
\end{split}
\eeq
and~$\langle x,y \rangle$ denotes the Euclidean scalar product of two unit
vectors~$x,y \in S^2 \subset \R^3$
(thus~$\langle x,y \rangle = \cos \vartheta$, where~$\vartheta$ is the
angle between~$x$ and~$y$).

The resulting so-called {\em{causal variational principle on the sphere}}
was introduced in~\cite[Chapter~1]{continuum} and analyzed
in~\cite[Sections~2 and~5]{support} and more recently in~\cite{sphere}.
We now explain a few results from these papers.

\begin{figure}
 \centering
\includegraphics[width=10cm]{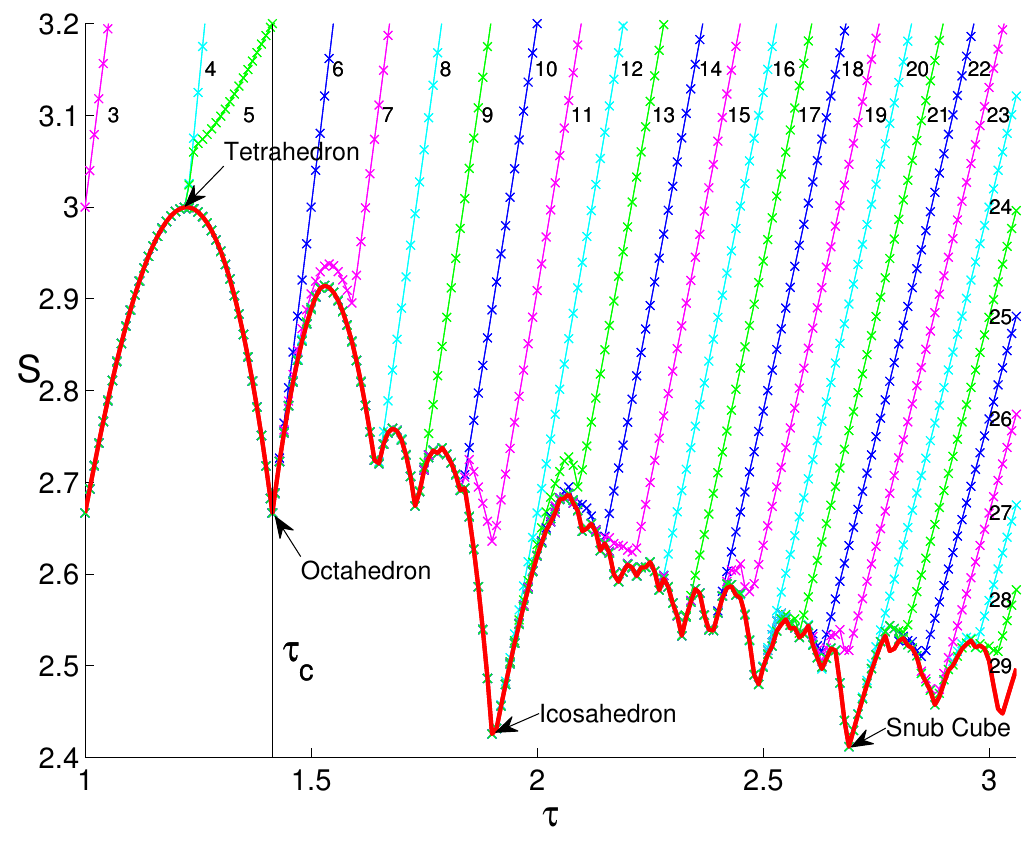}
\caption{Numerical minima for the weighted counting measure on the sphere.}
\label{figvgl_weight}
\end{figure}

First of all, the causal variational principle on the sphere is well-posed,
meaning that the minimum is attained in the class of all normalized regular Borel measures;
(the proof of this statement will be given in Chapter~\ref{secmeasure} using measure-theoretic
methods to be developed later in this book).
Minimizing numerically in the class of weighted counting measures for increasing number~$L$
of points and different values of the parameter~$\tau$, the resulting minimal value of the action
has an interesting non-smooth structure shown in Figure~\ref{figvgl_weight}.
In particular, one finds that the minimizing measure is not unique; indeed, there are typically many minimizers.
From the mathematical perspective, this non-uniqueness can be understood from the fact
that the causal action principle is a {\em{non-convex}} variational principle where one cannot expect
uniqueness. To give a concrete example for the non-uniqueness, we note that in the
case~$\tau=\tau_c = \sqrt{2}$, there is a minimizing measure that is supported on an octahedron
(for details see~\cite[Section~2]{support}).
This minimizing measure is not unique, because every measure obtained from it by
a rotation in~$\SO(3)$ is again a minimizer. But the non-uniqueness goes even further in the sense
that there are pairs of minimizing measures that cannot be obtained from each other by a rotation
or reflection. For example, again in the case~$\tau=\sqrt{2}$, the normalized Lebesgue measure
on the sphere is also a minimizer.

Moreover, the study in~\cite[Section~2]{support} gives the following
\begin{quote}
{\em{numerical result}}: If~$\tau> \sqrt{2}$, every minimizing measure
is a weighted counting measure~\eqref{weighted}.
\end{quote}
Thus, although we minimize over all regular Borel measures
(that is, measures which can have discrete and continuous components),
a minimizing measure always describes a discrete spacetime consisting of a finite
number of spacetime points. This result can be interpreted physically as an indication
that the causal action principle should give rise to discrete spacetime structures.
More details on the numerical findings and the physical interpretation can
be found in the review~\cite[Section~4]{rrev}. A more advanced numerical study of
the causal action principle in low dimensions can be found in~\cite{numerics}.

The above-mentioned numerical findings can be underpinned by analytic results.
We finally mention some of these results, although they will not be needed later on,
and the methods of proof will not be covered in this book.
First, it was proven in~\cite[Section~5.1]{support} that the support has an empty interior:
\begin{Thm}%[F-Schiefeneder 2010]
If~$\tau>\sqrt{2}$, the support of any minimizing measure does not
contain an open subset of~$S^2$.
\end{Thm} \noindent
Intuitively speaking, this result shows that the spacetime points
are a subset of the sphere of dimension strictly smaller than two.
More recently, it was shown in~\cite{sphere} that the dimension
of the support is even strictly smaller than one:
\begin{Thm}
In the case~$\tau > \sqrt{6}$,
the support of any minimizing measure is totally disconnected and has a Hausdorff dimension of at most~$6/7$.
\end{Thm} \noindent
The proof of these theorems uses techniques that will not be covered in this book.
Therefore, we refer the reader interested in more details to the papers cited earlier.

\section{Causal Variational Principles in the Compact and Smooth Settings} \label{seccvp}
Generalizing the causal variational principle on the sphere, one can replace~$\F$
by a smooth compact manifold of dimension~$m \geq 1$.

\begin{Def} \label{defcompactsetting} Let~$\F$ be a smooth compact manifold of dimension~$m \geq 1$
and~$\D \in C^\infty(\F \times \F,\R)$. Define the Lagrangian
$\L \in C(\F \times \F,\R^+_0)$ by
\beq \label{LD}
\L(x,y) := \max \big( 0, \D(x,y) \big) 
\eeq
and assume that~$\L$ has the following properties:
\bitem
\item[{\rm{(i)}}] $\L$ is symmetric: $\L(x,y) = \L(y,x)$ for all~$x,y \in \F$.\label{Cond1}
\item[{\rm{(ii)}}] $\L$ is strictly positive on the diagonal: $\L(x,x)>0$ for all~$x \in \F$. \label{Cond2}
\eitem
The {\bf{causal variational principle in the compact setting}} is to minimize the causal action
\sindex{causal variational principle!in the compact setting|textbf}%
\beq \label{Sactcompact}
\Sact = \int_\F \Diff\rho(x) \int_\F \Diff\rho(y)\: \L(x,y)
\eeq
under variations of measures~$\rho$ in the class of all regular Borel measures on~$\F$ which are normalized; that is,
\beq \label{normonecompact}
\rho(\F) = 1 \:.
\eeq
\end{Def}
This setting was introduced in~\cite[Section~1.2]{support}.
It is the preferable choice for studying phenomena for which the detailed form of the Lagrangian
as well as the constraints of the causal action principle are irrelevant. Note that also in the compact setting the action~$S(\rho)$ is finite for any normalized measure~$\rho$, because~$\F$ is compact and~$\L$ is continuous.

Given a minimizing measure~$\rho$, the Lagrangian induces on spacetime~$M:= \supp \rho$
a causal structure. Namely, two spacetime points~$x,y \in M$
are said to be {\em{timelike}} and {\em{spacelike}} separated if~$\L(x,y)>0$ and~$\L(x,y)=0$, respectively.
But, of course, compared to the causal action principle for causal fermion systems,
spin spaces and physical wave functions (as defined in Section~\ref{secinherent}) are missing in this setting.

We point out that in~\eqref{LD}, we merely assumed that the function~$\D$ is smooth.
The Lagrangian, however, is only Lipschitz continuous. It is in general non-differentiable
on the boundary of the light cone as defined by the level set~$\D(x,y)=0$.
In order to avoid differentiability issues, it is sometimes useful to simplify
the setting
even further by assuming that the Lagrangian itself is smooth,
\sindex{causal variational principle!in the smooth setting|textbf}%
\beq \label{Lsmooth}
\L \in C^\infty(\F \times \F, \R^+_0) \:.
\eeq
This is the so-called {\bf{smooth setting}}.
We point out that the Lagrangian of the causal action~\eqref{Lagrange} is not smooth
if some of the eigenvalues vanish or are degenerate (more precisely, the causal Lagrangian is only
H\"older continuous, as is worked out in detail in~\cite[Section~5]{banach}). Indeed, this
non-smoothness yields interesting effects like the results on the singular
support in~\cite{support, sphere}.
\sindex{minimizing measure!of singular support}%
In view of these results, the smoothness assumption~\eqref{Lsmooth} is a mathematical simplification
that, depending on the application in mind, may or may not be justified.
In this book, we choose the smooth setting mainly for pedagogical reasons, keeping in mind that the
generalizations to non-smooth Lagrangians are rather straightforward. The reader who is interested in or
needs these generalizations will find the details in the research papers.

Before going on, we point out that the assumptions that~$\F$ is a smooth manifold and that
the function~$\D$ in~\eqref{LD} is smooth
are convenient and avoid certain technicalities. But these assumptions are much more than
what is needed for the analysis. More generally, one can choose~$\L$ as a non-negative continuous function,
\beq \label{LC0}
\L \in C^0(\F \times \F, \R^+_0) \:.
\eeq
Going one step further, one may relax the continuity of the Lagrangian
by the condition that~$\L$ be {\em{lower semi-continuous}}; that is, that
for all sequences~$x_n \rightarrow x$ and~$y_{n'} \rightarrow y$,
\sindex{lower semi-continuity}%
\beq \L(x,y) \leq \liminf_{n,n' \rightarrow \infty} \L(x_n, y_{n'})\:. \eeq
Since the Lagrangian of the causal action~\eqref{Lagrange} is continuous, lower semi-continuity
is an unphysical generalization. Nevertheless, this setting is useful for two reasons: First, 
from the point of view of the calculus of variations, it is  a natural generalization to which most methods still apply.
And second, lower semi-continuous Lagrangians are convenient for
formulating explicit examples (like the lattice model in~\cite[Section~5]{jet}).

We finally note that also the assumption of~$\F$ being a smooth manifold can
be relaxed. From the point of view of the calculus of variations, the right setting is
to assume that~$\F$ is a compact topological Hausdorff space.

In this book, for pedagogical reasons, we do not aim for the highest generality and
minimal smoothness and
regularity assumptions. An introduction to a more general and more abstract setting can be found
in~\cite[Section~3]{noncompact}.

\section{Causal Variational Principles in the Non-compact Setting} \label{secnoncompact}
\sindex{causal variational principle!in the non-compact setting|textbf}%
In the compact setting, spacetime~$M$ is a compact subset of~$\F$.
This is not suitable for describing situations when spacetime has an asymptotic future or past
or when spacetime has singularities (like at the big bang or inside a black hole).
For studying such situations, it is preferable to work in the so-called {\em{non-compact setting}}
introduced in~\cite[Section~2.1]{jet}, where~$\F$ is chosen to be a non-compact manifold.

In the non-compact setting, it is not sensible to restrict attention to normalized measures.
Instead, the total volume~$\rho(\F)$ is typically infinite.
In this situation, the causal action~\eqref{Sactcompact} could also be infinite.
Therefore, we need to define in which sense a measure is a minimizer of the action.

\begin{Def} {\bf{(Causal variational principles in the non-compact setting)}}
Let~$\F$ be a non-compact smooth manifold of dimension~$m \geq 1$, and let~$\D \in C^\infty(\F \times \F, \R)$ be such that the Lagrangian~$\L \in C^0(\F \times \F; \R^+_0)$ defined by~\eqref{LD} has the properties~{\rm{(i)}}
and~{\em{(ii)}} in Definition~\ref{defcompactsetting}. Given a regular Borel measure~$\rho$ on~$\F$, another regular Borel measure~$\tilde{\rho}$ on~$\F$ is a {\bf{variation}} of~$\rho$ {\bf{of finite volume}} if it satisfies the conditions
\sindex{variation!of finite volume}%
\beq \label{totvol}
\big| \tilde{\rho} - \rho \big|(\F) < \infty \qquad \text{and} \qquad
\big( \tilde{\rho} - \rho \big) (\F) = 0 \:,
\eeq
where~$|\tilde{\rho}-\rho|$ is the total variation measure
(see Definition~\ref{deftotvar} in Section~\ref{secbasicmeasure}).
For such a variation of finite volume, we consider the (formal) difference of the actions defined by
\begin{align}
\big( \Sact(\tilde{\rho}) - \Sact(\rho) \big)
&:= \int_\F \Diff(\tilde{\rho} - \rho)(x) \int_\F \Diff\rho(y)\: \L(x,y) \notag \\
&\quad\:+ \int_\F \Diff\rho(x) \int_\F \Diff(\tilde{\rho} - \rho)(y)\: \L(x,y) \notag \\
&\quad\:+ \int_\F \Diff(\tilde{\rho} - \rho)(x) \int_\F \Diff(\tilde{\rho} - \rho)(y)\: \L(x,y) \:. \label{integrals}
\end{align}
The measure~$\rho$ is said to be a {\bf{minimizer}} of the causal action
with respect to variations of finite volume
\sindex{minimizing measure!with respect to variations of finite volume}%
if this difference is non-negative for all~$\tilde{\rho}$ satisfying~\eqref{totvol},
\beq \label{Sdiffpos}
\big( \Sact(\tilde{\rho}) - \Sact(\rho) \big) \geq 0 \:.
\eeq
\end{Def} \noindent
We note for clarity that integrals with respect to~$\tilde{\rho}-\rho$ are defined by
\beq \int_\F f(x) \: \Diff(\tilde{\rho} - \rho)(x) := \int_\F f\: \dd\mu^+ - \int_\F f\: \dd\mu^- \eeq
with the finite measures~$\mu_\pm$ as given in Definition~\ref{deftotvar}.
In particular, $( \tilde{\rho} - \rho)(\F) := \mu^+(\F) - \mu^-(\F)$.

Exactly as mentioned at the end of the previous section,
the assumption that~$\F$ is a smooth manifold could be weakened.
From the point of view of calculus of variations, the right setting is
to assume that~$\F$ is a $\sigma$-locally compact topological Hausdorff space
(for details, see again~\cite[Section~3]{noncompact}).

\section{The Local Trace Is Constant}
Causal variational principles as introduced in the previous sections are of interest in their own right
as a novel class of nonlinear variational principles.
Nevertheless, since we are primarily interested in causal fermion systems, it is important
to get a concise mathematical connection to the causal action principle.
In preparation, we now analyze the trace constraint
and derive a first general result on minimizing measures of the causal action principle.
We present this result at such an early stage of this book because this result can be used to simplify
the setup of causal fermion systems, getting the desired connection to causal variational principles
(see Section~\ref{seccfscvp} below). The following result was first obtained in~\cite{lagrange} (albeit with a
different method); see also~\cite[Proposition~1.4.1]{cfs}.
For technical simplicity, we restrict attention to the finite-dimensional setting
(in the infinite-dimensional case, this problem has not yet been studied).
Then the total volume of spacetime as well as the minimal action are finite.

\begin{Prp} \label{prptrxconst}
Consider the causal action principle in the finite-dimensional setting~$\dim \H < \infty$.
Let~$\rho$ be a minimizer of finite total volume, $\rho(\F) < \infty$.
Then there is a real constant~$c$ such that
\beq %\label{trxconst}
\tr(x) = c \qquad \text{for all~$x \in M := \supp \rho$}\:. \eeq
\end{Prp} \noindent
We often refer to~$\tr(x)$ as the {\em{local trace}} at the point~$x$.
\sindex{local trace}%
\Proof[Proof of Proposition~\ref{prptrxconst}.]
We will prove the theorem by contradiction and therefore assume that the local trace is \emph{not} constant. The idea is to use this assumption to construct a suitable variation
\beq (\rho_\tau)_{\tau \in (-\delta,\delta)} \qquad \text{with} \qquad \rho_0 = \rho \eeq
which satisfies the constraints, but makes the action smaller, in contradiction to~$\rho$ being a minimizer.

For the construction of the variation we combine two different general methods. One
method is to multiply the measure~$\rho$ by a positive measurable function~$f_\tau \::\: M \rightarrow \R^+$,
\beq %\label{weight}
\rho_\tau = f_\tau\, \rho \eeq
(alternatively, one can also write this relation as~$\Diff \rho_\tau(x) = f_\tau(x)\: \Diff \rho(x)$).
Clearly, such a variation does not change the support of the measure. In order to change the support,
our second method is to consider a measurable function~$F_\tau \::\: M \rightarrow \F$ and take the push-forward measure,
\beq %\label{pushtau}
\rho_\tau = (F_\tau)_* \rho \:. \eeq
Combining these two methods, we are led to considering variations of the form
\beq \label{rhoFf}
\rho_\tau = (F_\tau)_* \big( f_\tau \,\rho \big) \:.
\eeq
The condition~$\rho_0=\rho$ gives rise to the conditions
\beq \label{fFinit}
f_0 \equiv 1 \qquad \text{and} \qquad F_0 \equiv \1 \:.
\eeq
Finally, we assume that the functions~$f_\tau$ and~$F_\tau$ are smooth in~$\tau$.
The ansatz~\eqref{rhoFf} is particularly convenient for computations. Namely, by definition of the push-forward measure,
\beq \int_\F \L(x,y)\: \Diff\rho_\tau(y) = \int_M \L\big(x, F_\tau(y) \big)\: f_\tau(y)\: \Diff\rho(y) \:, \eeq
and similarly for all other integrals.

Next, for arbitrarily given~$f_\tau$, we want to choose~$F_\tau$ in such a way that the last integral becomes independent of~$\tau$.
To this end, we choose
\beq %\label{Ftauf}
F_\tau(x) := \frac{x}{\sqrt{f_\tau(x)}} \:. \eeq
Using that the causal Lagrangian~$\L(x,y)$ is homogeneous
of degree two in~$y$ (as is obvious from~\eqref{Lagrange} and the fact that the eigenvalues~$\lambda^{xy}_i$
are homogeneous of degree one in~$y$), it follows that
\begin{align}
\int_M &\L\big( x, F_\tau(y) \big)\: f_\tau(y)\: \Diff\rho(y)
= \int_M \L\bigg( x, \frac{y}{\sqrt{f_\tau(y)}} \bigg)\: f_\tau(y)\: \Diff\rho(y) \\
&= \int_M \L(x,y) \frac{1}{f_\tau(y)} \: f_\tau(y)\:\Diff\rho(y) 
= \int_M \L(x,y)\: \Diff\rho(y) \:.
\end{align}
Arguing similarly in the variable~$x$, one sees that our variation does not change the
action. Since the integrand~$|\lambda^{xy}_j|^2$ of the boundedness constraint~\eqref{Tdef} is again
homogeneous of degree two in~$x$ and~$y$,
the above argument applies similarly to the functional~$\T$, showing that the boundedness
constraint is respected by our variations.

Let us analyze the volume and trace constraints. In order to satisfy the volume constraint, we make the ansatz
\beq \label{fgtau}
f_\tau = 1 + \tau g \:,
\eeq
where~$g$ is a bounded function with zero mean,
\beq \label{volc}
\int_M g(x) \: \Diff\rho(x) = 0 \:.
\eeq
This ensures that the volume constraint is satisfied. We finally consider the variation of the trace constraint,\begin{align}
&\int_\F \tr(x)\: \Diff\rho_\tau(x) - \int_\F \tr(x)\: \Diff\rho(x) \notag \\
&= \int_M \tr\big(F_\tau(x)\big)\: f_\tau(x)\: \Diff\rho(x) - \int_M \tr(x)\: \Diff\rho(x) \notag \\
&= \int_M \tr\Big( \frac{x}{\sqrt{f_\tau(x)}} \Big)\: f_\tau(x)\: \Diff\rho(x) 
- \int_M \tr(x)\: \Diff\rho(x) \notag \\
&= \int_M \tr(x)\: \Big( \sqrt{f_\tau(x)}- 1 \Big) \: \Diff\rho(x) \:.
\end{align}
Employing again the ansatz~\eqref{fgtau} and differentiating with respect to~$\tau$,
we obtain for the first variation
\beq \label{intrel}
\frac{\Diff}{\Diff\tau} \int_\F \tr(x)\: \Diff\rho_\tau \bigg|_{\tau=0}
= \frac{1}{2} \int_M \tr(x)\: g(x)\: \Diff\rho(x)
\eeq
(here we may differentiate the integrand using Lebesgue's dominated convergence theorem).
Now we use of our assumption that the local trace is {\em{not}} constant on~$M$,
making it possible to choose~$x_1,x_2 \in M$ with~$\tr(x_1) > \tr(x_2)$.
Moreover, we choose a function~$g$ which supported in a small neighborhood of~$x_1$ and~$x_2$,
has zero mean~\eqref{volc} and is positive at~$x_1$ and negative at~$x_2$.
In this way, we can arrange that the right-hand side of~\eqref{intrel} is strictly positive.
Hence, using~\eqref{volc}, it follows that
\beq \label{vartrpos}
\frac{\Diff}{\Diff\tau} \int_\F \tr(x)\: \Diff\big( \rho_\tau - \rho\big)(x) \bigg|_{\tau=0} >0 \:.
\eeq
To summarize, we have found a variation which respects the boundedness and the volume constraint
and preserves the causal action, but increases the integral of the trace~\eqref{vartrpos}.

The final step is to modify the variation in such a way that the trace and volume constraints are respected,
whereas the action and the boundedness constraints become smaller.
To this end, we transform the measures according to
\beq \rho_\tau \rightarrow (G_\tau)_*(\rho_\tau) \eeq
with
\beq \label{Gscale}
G_\tau(x) = x \:\bigg( \int_M \tr(x)\: \Diff\rho \bigg) \bigg/ \bigg( \int_M \tr(x)\: \Diff\rho_t \bigg) \:.
\eeq
A short computation shows that the trace constraint is respected, and so is the volume constraint.
Moreover, in view of~\eqref{vartrpos}, for small positive~$\tau$ the scaling factor in~\eqref{Gscale} is 
strictly smaller than one, implying that the first variations of the action and of the boundedness constraint
are both strictly negative (here we use again the homogeneity of the Lagrangian and
of the integrand of the boundedness constraint). This is a contradiction to the fact that~$\rho$ is a minimizer
(here we make essential use of the fact that the boundedness constraint~\eqref{Tdef} is an {\em{inequality}} constraint, so that decreasing~$\T$ in the variation is allowed).
We conclude that the local trace must be constant.
% Jan-Hendriks Vorschlag
%At this point we have constructed a variation respecting the boundedness and volume constraint (but not the trace constraint) and keeping the action constant. To obtain the variation we are actually looking for, i.e. one respecting all constraints and reducing the action, we make one more modification. Namely, we transform the measures according to
%% \beq \rho_\tau \rightarrow (G_\tau)_*(\rho_\tau) \eeq
%\beq
%    \widetilde{\rho}_\tau := (G_\tau)_*(\rho_\tau)
%\eeq
%with
%\beq \label{Gscale}
%G_\tau(x) := \frac{\int_M \tr(y) \Diff\rho(y)}{\int_M \tr(y) \Diff\rho_\tau(y)} \;\cdot  x =: s_\tau \cdot x
%%G_\tau(x) = x \:\bigg( \int_M \tr(x)\: \Diff\rho \bigg) \bigg/ \bigg( \int_M \tr(x)\: \Diff\rho_t \bigg) \:.
%\eeq
%A short computation shows that this variation respects the volume and the trace constraint. Moreover, using again the homogeneity properties of~$\L$, one finds
%\beq
%    \Sact(\widetilde{\rho}_\tau) = s_\tau^4 \cdot \Sact(\rho) \quad\text{and}\quad
%    \T(\widetilde{\rho}_\tau) = s_\tau^4 \cdot \T(\rho) \,,
%\eeq
%where~$\T$ is the term in the boundedness constraint~\eqref{Tdef}. In view of~\eqref{vartrpos}, for small positive~$\tau$ we have~$s_\tau < 1$, hence both the action~$\Sact$ and~$\T$ decrease. This shows that firstly the boundedness constraint is respected by the variation (note that it is an {\em{inequality}} constraint, so that decreasing~$\T$ in the variation is allowed), and secondly we finally obtain a contradiction to the fact that~$\rho$ is a minimizer of the action.
%We conclude that the local trace must be constant after all.
\QED

\section{How the Causal Action Principle Fits into the Non-compact Setting} \label{seccfscvp}
Under mild technical assumptions on the minimizing measure, the causal
action principle for causal fermion systems is a special case of the causal variational
principle in the non-compact setting, as we now explain.

Since for minimizers of the causal action principle, all operators in~$M$
have the same trace (see~Proposition~\ref{prptrxconst}),
we can simplify the setting by restricting attention to linear operators in~$\F$
which all have the same trace. Then the trace constraint can be disregarded, as it follows from the volume constraint. We now implement this simplification by modifying our setting.
At the same time, we implement the boundedness constraint by a {\em{Lagrange
multiplier}} term. Here we apply this method naively by modifying the Lagrangian to
\sindex{Lagrange multiplier!for boundedness constraint}%
\beq \label{Lkappadef}
\L_\kappa(x,y) := \frac{1}{4n} \sum_{i,j=1}^{2n} \Big( \big|\lambda^{xy}_i \big| - \big|\lambda^{xy}_j \big| \Big)^2
+ \kappa\: \bigg( \sum_{i=1}^{2n} \big|\lambda^{xy}_i \big| \bigg)^2 \:,
\eeq
where~$\kappa>0$ is the Lagrange multiplier.
We refer to~$\L_\kappa$ as the {\em{$\kappa$-Lagrangian}}.
\sindex{$\kappa$-Lagrangian}%
\nindex{aee@$\L_\kappa$ -- $\kappa$-Lagrangian}%
The justification for this procedure as given in~\cite{lagrange} is a bit subtle, and for brevity,
we shall not enter these constructions here.
It is important to note that, in contrast to the usual Lagrange multiplier, where a minimizer
under constraints in general merely is a critical point of the Lagrangian including the Lagrange multipliers,
here we obtain again a {\em{minimizer}} of the effective action (for details, see~\cite[Theorem~3.13]{lagrange}).

Finally, we make a mild technical simplification. A spacetime point~$x \in M$ is said to be
{\em{regular}} if~$x$ has the maximal possible rank~$2n$. Otherwise, the spacetime point
\sindex{spacetime point!regular|textbf}%
\sindex{spacetime point!singular}%
is {\em{singular}}. In typical physical applications, all spacetimes points are regular, except maybe at
singularities like the center of black holes.
For example, the construction of a causal fermion system in Minkowski space from Dirac wave
functions in Section~\ref{seclco} gives regular spacetime points if~$\H$ is chosen sufficiently
large (in particular for all negative-frequency solutions).
More generally, the interacting systems considered in~\cite[Chapters~3-5]{cfs} all have regular spacetime points.
The same is true for the similar construction in globally hyperbolic spacetimes
(details can be found in~\cite{nrstg}).
With this in mind, in this book, we usually assume that the {\em{causal fermion system}} is {\em{regular}}
in the sense that all spacetime points are regular.
\sindex{causal fermion system!regular}%
This assumption has the advantage that the set of all regular points of~$\F$ is a smooth manifold
(see Proposition~\ref{prppq} in Section~\ref{secmanop}).
We remark that, in the case that~$\H$ is infinite-dimensional, the set of regular points of~$\F$
can be endowed with the structure of a Banach manifold (for details, see~\cite[Section~3]{banach}). These considerations lead us to the following setting:

\begin{Def} Let~$\H$ be a complex Hilbert space. Moreover, suppose we are given
parameters~$n \in \N$ (the spin dimension), $c > 0$ (the constraint for the local trace)
and~$\kappa>0$ (the Lagrange multiplier of the boundedness constraint).
\nindex{aef@$\F^\reg$ -- regular points of~$\F$}%
We then let~$\F^\reg \subset \Lin(\H)$ be the set of all
symmetric operators~$F$ on~$\H$ with the following properties:
\bitem
\item[{\rm{(a)}}] $F$ has finite rank and (counting multiplicities) has exactly~$n$ positive and~$n$ negative eigenvalues. \\[-0.8em]
\item[{\rm{(b)}}] The local trace is equal to~$c$; that is,
\beq \tr(F) = c\:. \eeq
\eitem
On~$\F^\reg$ we again consider the topology induced by the sup-norm on~$\Lin(\H)$. The {\bf{reduced causal action principle for regular systems}} is to minimize the reduced causal action
\beq \Sact_\kappa(\rho) = \iint_{\F \times \F} \L_\kappa(x,y)\: \Diff\rho(x)\, \Diff\rho(y) \eeq
over all regular Borel measures under variations that preserve the total volume.
\sindex{causal action principle!reduced}%
\sindex{causal action principle!reduced for regular systems}%
\end{Def}

In this way, the causal action principle fits into the framework of
causal variational principles in the non-compact setting as introduced in Section~\ref{secnoncompact}.
In agreement with~\eqref{LC0}, the causal Lagrangian is continuous
(in fact, it is even locally H\"older continuous; for details, see~\cite[Section~5.1]{banach}).
Moreover, it has the desired properties~(i) and~(ii) on page~\pageref{Cond1}
(it is strictly positive because the Lagrangian can be estimated from below in terms
of the local trace; see Exercise~\ref{exm3}).

In order to avoid misunderstandings, we point out that the above-mentioned description of causal fermion
systems by measures on~$\F^\reg$ is not a suitable setting for the existence theory
(as will be developed in Chapter~\ref{secmeasure}).
The reason is that~$\F^\reg$ is not closed in~$\F$, because the boundary points in~$\F$ are missing (in a converging sequence, some of the eigenvalues could tend to zero in the limit).
As a consequence, considering a minimizing sequence~$(\rho_n)_{n \in \N}$ of measures in~$\F^\reg$,
the limiting measure might well be supported also on~$\F \setminus \F^\reg$.
For this reason, there is no existence theory for measures on~$\F^\reg$.
But if a minimizing measure is given, it seems sensible to assume that the resulting causal fermion system
is regular. Under this assumption, the analysis of the causal fermion system can be carried out
exclusively in~$\F^\reg$, whereas~$\F$ is no longer needed.
For a convenient and compact notation, in such situations, we shall even omit the superscript~``$\text{reg}$'',
so that~$\F$ denotes the set of all symmetric operators on~$\H$ with the above properties~(a) and~(b).
Moreover, we shall omit the subscript~$\kappa$. Thus, with a slight abuse of notation, we shall denote the
Lagrangian including the Lagrange multiplier term~\eqref{Lkappadef} again by~$\L$.

In this way, assuming that the causal fermion systems under consideration are regular,
we have recovered the causal action principle as a specific causal variational principle.
The connection is summarized schematically as follows:
\begin{gather*}
\hspace*{-5cm} \boxed{\text{causal action principle for causal fermion systems}} \\
\hspace*{4.82cm} \Big\downarrow \;\; \text{ \begin{minipage}[c]{10cm}
implement boundedness constraint \\ by a Lagrange multiplier \end{minipage} } \\
\hspace*{-5cm} \boxed{\text{reduced causal action principle}} \\
\hspace*{4.82cm} \Big\downarrow \;\; \text{ \begin{minipage}[c]{10cm} build in trace constraint, \\
restrict attention to regular causal fermion systems \end{minipage} }  \\
\hspace*{-5cm} \boxed{\text{reduced causal action principle for regular systems}} \\
\hspace*{-3cm} \mathrel{\rotatebox[origin=c]{-90}{$\hookrightarrow$}} \;\; \text{generalize} \\
\hspace*{-5cm} \boxed{\text{causal variational principles}}
\end{gather*}
Whenever the specific form of the causal Lagrangian~\eqref{Lkappadef} is not needed,
we will work in the more general setting of causal variational principles.
Apart from the sake of greater generality, this has the advantage that it becomes clearer
which structures are needed for which results.
Moreover, it is often more convenient not to specify the form of the Lagrangian.
Generally speaking, we can work with causal variational principles unless the physical wave functions
and their induced geometric and analytic structures are invoked.

\section{Exercises}

\begin{Exercise} {{(Derivation of the causal variational principle on the sphere)}} \label{excomsphere}
\sindex{causal variational principle!on the sphere}%
{\em{ We consider the causal fermion systems in the case~$n=1$ and~$f=2$.
For a given parameter~$\tau>1$ we introduce the mapping~$F : S^2 \subset \R^3 \rightarrow \F$ by
\beq \label{FS2}
F(\vec{x}) = \tau\: \vec{x} \vec{\sigma} + \1\:.
\eeq
\bitem
\item[(a)] Compute the eigenvalues of the matrix~$F(\vec{x})$ and verify that it has one positive and one negative
eigenvalue.
\item[(b)] Use the identities between Pauli matrices
\beq %\label{Pauliid}
\sigma^i \sigma^j = \delta^{ij} + \cI \epsilon^{ijk}\: \sigma^k \:, \eeq
in order to compute the matrix product,
\beq F(\vec{x}) \,F(\vec{y}) = \left(1+\tau^2\: \vec{x} \vec{y} \right) \1 
+ \tau\: (\vec{x}+\vec{y}) \vec{\sigma} +\cI \tau^2 \,(\vec{x} \wedge \vec{y}) \vec{\sigma} . \eeq
\item[(c)] Compute the eigenvalues of this matrix product to obtain
\beq %\label{leigen}
\lambda_{1\!/\!2} = 1+\tau^2 \cos \vartheta \pm
 \tau \sqrt{1+\cos \vartheta}\: \sqrt{2 - \tau^2 \:(1-\cos \vartheta)} \:, \eeq
where~$\vartheta$ denotes the angle~$\vartheta$ between~$\vec{x}$ and~$\vec{y}$.
\item[(d)] Verify that if~$\vartheta \leq \vartheta_{\max}$ with
\beq \vartheta_{\max} := \arccos \!\left( 1-\frac{2}{\tau^2} \right), \eeq
then the eigenvalues~$\lambda_{1\!/\!2}$ are both real. Conversely, if~$\vartheta>\vartheta_{\max}$,
then the eigenvalues form a complex conjugate pair.
\item[(e)] Use the formula
\beq \lambda_1 \lambda_2 = \det(F(\vec{x}) F(\vec{y}))
= \det(F(\vec{x})) \, \det(F(\vec{y})) = (1+\tau)^2 (1-\tau)^2 > 0 \eeq
to conclude that if the eigenvalues~$\lambda_{1\!/\!2}$ are both real, then they have the same sign.
\item[(f)] Combine the above findings to conclude that the causal Lagrangian~\eqref{Lagrange}
can be simplified to~\eqref{Lform}.
\eitem
}} \end{Exercise}

\begin{Exercise} {{(The action and boundedness constraint of the Lebesgue measure on the sphere)}} \label{ex28}
{\em{ We consider the causal variational principle on the sphere as introduced in Section~\ref{seccvpsphere}.
We let~$\Diff\mu$ be the surface area measure, normalized such that~$\mu(S^2)=1$. 
\bitem
\item[(a)] Use the formula
for the causal Lagrangian on the sphere~\eqref{Lform} to compute the causal action~\eqref{Sdef}.
Verify that
\beq \label{SFint}
\Sact[F] = \frac{1}{2} \int_0^{\vartheta_{\max}} \L(\cos \vartheta)\: \sin \vartheta\:
\dd \vartheta =  4 - \frac{4}{3 \tau^2}\:.
\eeq
\item[(b)] Show that the functional~$\T$ is given by
\beq \label{TFint}
\T[F] = 4 \tau^2 (\tau^2-2) + 12 -\frac{8}{3 \tau^2} \:.
\eeq
\eitem
Hence the causal action~\eqref{SFint} is bounded uniformly in~$\tau$, although the function~$F$, \eqref{FS2},
as well as the functional~$\T$, \eqref{TFint}, diverge as~$\tau \rightarrow \infty$. }}
\end{Exercise}

\begin{Exercise} \label{excrit} (A minimizer with singular support) {\em{
\sindex{minimizing measure!of singular support}%
We again consider the causal variational principle on the sphere as introduced in Section~\ref{seccvpsphere}.
Verify by direct computation that in the case~$\tau=\sqrt{2}$, the causal action of the
normalized counting measure supported on the
{\em{octahedron}} is smaller than the action of~$\mu$.
{\em{Hint:}} For~$\tau=\sqrt{2}$ the opening angle of the light cone is given by~$\vartheta=90^\circ$,
so that all distinct spacetime points are spacelike separated.
Moreover, the causal action of the normalized Lebesgue measure is given in Exercise~\ref{ex28}~(a).

It turns out that the normalized counting measure supported on the octahedron is indeed a minimizer of the causal action. More details and related considerations can be found in~\cite{continuum, support, rrev}.
}} \end{Exercise}

\begin{Exercise} \label{excvpR} (A causal variational principle on~$\R$) {\em{
We let~$\F=\R$ and consider the Lagrangians
\beq %\label{L24}
\L_2(x,y) = (1+x^2)(1+y^2) \qquad \text{and} \qquad 
\L_4(x,y) = (1+x^4)(1+y^4) \:. \eeq
We minimize the corresponding causal actions~\eqref{Sactcompact} within the class of all
normalized regular Borel measures on~$\R$.
Show with a direct estimate that the Dirac measure~$\delta$ supported at the origin
is the unique minimizer of these causal variational principles.
}} \end{Exercise}

\begin{Exercise} \label{excvpS1} (A causal variational principle on~$S^1$) {\em{
We let~$\F=S^1$ be the unit circle parametrized as~$\E^{\cI \varphi}$ with~$\varphi \in \R \text{ mod } 2 \pi$
and consider the Lagrangian
\beq %\label{LS1}
\L(\varphi, \varphi') = 1 + \sin^2 \big( \varphi - \varphi' \text{ mod } 2 \pi \big) \:. \eeq
We minimize the corresponding causal action~\eqref{Sactcompact} within the class of all
normalized regular Borel measures on~$S^1$.
Show by direct computation and estimates that every minimizer is of the form
\beq \label{rhominS1}
\rho = \tau\, \delta\big( \varphi - \varphi' -\varphi_0 \text{ mod } 2 \pi \big) + (1-\tau) \, \delta \big(
\varphi - \varphi' -\varphi_0 + \pi \text{ mod } 2 \pi \big)
\eeq
for suitable values of the parameters~$\tau \in [0,1]$ and~$\varphi_0 \in \R \text{ mod } 2\pi$.
}} \end{Exercise}

\chapter{The Euler-Lagrange Equations} \label{secEL}
In classical field theory, the dynamics of the physical system is revealed by analyzing the
Euler-Lagrange equations corresponding to the classical action principle.
These Euler-Lagrange equations are the physical equations (like the Maxwell or Einstein equations).
They have the mathematical structure of partial differential equations.
Likewise, for the causal action principle and causal variational principles, the
Euler-Lagrange equations describe the dynamics.
However, they are no longer differential equations but have a quite different form.
In this chapter we shall derive the {\em{Euler-Lagrange equations}} and discuss their general structure.

\section{The Euler-Lagrange Equations}
Let~$\rho$ be a minimizer of the causal variational principle in the non-compact
setting (more precisely, a minimizer with respect to variations of finite volume; see Section~\ref{secnoncompact}).
We now derive the Euler-Lagrange (EL) equations, following the method in the
compact setting~\cite[Lemma~3.4]{support}. We again define spacetime as the support of~$\rho$,
\beq M := \supp \rho \subset \F \:. \eeq
In words, the EL equations state that the causal action is minimal under first variations of the measure.
In order to make mathematical sense of the variations, we need the following assumptions:
\bitem
\item[(i)] The measure~$\rho$ is {\em{locally finite}}
\sindex{measure!locally finite}%
(meaning that any~$x \in \F$ has an open neighborhood~$U$ with~$\rho(U)< \infty$).\label{Cond3}
\item[(ii)] The function~$\L(x,.)$ is~$\rho$-integrable for any~$x \in \F$, and the function
\beq x \mapsto \int_\F \L(x,y) \:\Diff \rho(y) \eeq
is a bounded continuous function on~$\F$. \label{Cond4}
\eitem
These technical assumptions are satisfied in most applications and are sufficiently general
for the purpose of this book (we note that the continuity assumption in~(ii) could be relaxed to
lower semi-continuity; the details are worked out in~\cite{noncompact}).
We introduce the function
\nindex{aeg@$\ell$ -- integrated Lagrangian}%
\beq \label{elldef}
\ell(x) = \int_\F \L(x,y)\: \Diff\rho(y) - \s \::\: \F \rightarrow \R\:,
\eeq
where~$\s \in \R$ is a parameter whose value will be specified below.
\nindex{aeh@$\s$ -- Lagrange parameter in Euler-Lagrange equations}%

\begin{Thm} (The Euler-Lagrange equations) \label{thmEL}
Let~$\rho$ be a minimizer of the causal action with respect to variations of finite volume and assume that~$\rho$ satisfies the conditions~{\rm{(i)}} and~{\rm{(ii)}} above. Then
\sindex{Euler-Lagrange equations!with respect to variations of finite volume}%
\beq \label{EL1}
\ell|_M \equiv \inf_\F \ell \:.
\eeq
\end{Thm}
\Proof Given~$x_0 \in \supp \rho$, we choose an open neighborhood~$U$ with~$0 < \rho(U)<\infty$.
For any~$y \in \F$, we consider the family of measures~$(\tilde{\rho}_\tau)_{\tau \in [0,1)}$ given by
\sindex{first variation of measure|textbf}%
\beq \label{nonlocvary}
\tilde{\rho}_\tau = \chi_{M \setminus U} \,\rho + (1-\tau)\, \chi_U \, \rho + \tau\, \rho(U)\, \delta_{y}
\eeq
(where~$\delta_y$ is the Dirac measure supported at~$y$). Then
\beq \label{tilderho}
\tilde{\rho}_\tau - \rho = -\tau\, \chi_U \, \rho + \tau\, \rho(U)\, \delta_{y} 
= \tau \big( \rho(U)\, \delta_{y} - \chi_U\, \rho \big) \:.
\eeq
Using this formula one readily verifies that~$\tilde{\rho}_\tau$ is a variation of finite volume
satisfying the volume constraint. Hence
\begin{align} 
0 \leq \big(\Sact(\tilde{\rho}_\tau) - \Sact(\rho) \big) &= 2 \tau 
\left( \rho(U)\, \Big( \ell(y) + \s \Big)- \int_U \Big( \ell(x)+ \s \Big)\, \Diff\rho(x) \right) + \O \big(\tau^2 \big) \notag \\
  &= 2\tau\left( \rho(U)\:\ell(y))-\int_U \ell(x) \Diff \rho(x) \right) + \O\big(\tau^2\big)
\end{align}
(here we may carry out the integrals in arbitrary order using Tonelli's theorem for non-negative
integrands).
Since this holds for any~$\tau \in [0,1)$, the linear term must be non-negative, and thus
\begin{align} \label{ELInt}
\ell(y) \geq \frac{1}{\rho(U)} \int_U \ell(x)\, \Diff\rho(x) \:.
\end{align}
Now assume that~\eqref{EL1} is false. Then there is~$x_0 \in \supp \rho$
and~$y \in \F$ such that~$\ell(x_0) > \ell(y)$. Continuity of~$\ell$ implies that there is an open
neighborhood~$U$ of~$x_0$ such that~$\ell(x) > \ell(y)$ for all~$x \in U$. But this contradicts~\eqref{ELInt}.
\QED
It is indeed no loss of generality to restrict attention to first variations within the special class~\eqref{tilderho};
for details see Exercise~\ref{exvar1gen}.

We always choose the parameter~$\s$ such that the infimum of~$\ell$ in~\eqref{EL1} is zero.
Then the EL equations read
\beq \label{EL2}
\ell|_{\supp \rho} \equiv \inf_\F \ell  = 0 \:.
\eeq
The parameter~$\s$ can be understood as the ``action per volume'' (see Exercise~\ref{exsval}).
We finally point out that solutions of the EL equations do not need to be minimizers of the causal
action principle. Similar to the situation for local maxima or saddle points in the finite-dimensional setting,
there may be variations for which~$\Sact$ is stationary, but whose
second or higher variations are negative.

\section{The Restricted Euler-Lagrange Equations in the Smooth Setting} \label{secrestrictedEL}
\begin{figure}
\begin{center}
% \usepackage[usenames,dvipsnames]{pstricks}
% \usepackage{epsfig}
% \usepackage{pst-grad} % For gradients
% \usepackage{pst-plot} % For axes
% \usepackage[space]{grffile} % For spaces in paths
% \usepackage{etoolbox} % For spaces in paths
% \makeatletter % For spaces in paths
% \patchcmd\Gread@eps{\@inputcheck#1 }{\@inputcheck"#1"\relax}{}{}
% \makeatother
% 
\psscalebox{1.0 1.0} % Change this value to rescale the drawing.
{
\begin{pspicture}(0,-0.9933811)(8.918682,0.9933811)
\definecolor{colour0}{rgb}{0.8,0.8,0.8}
\pspolygon[linecolor=colour0, linewidth=0.02, fillstyle=solid,fillcolor=colour0](4.6909046,0.42533422)(4.8159046,0.3653342)(4.9359045,0.3103342)(5.0909047,0.2603342)(5.260905,0.2203342)(5.555905,0.1703342)(5.7959046,0.1503342)(6.115905,0.13533421)(6.3959045,0.1303342)(6.595905,0.1503342)(6.825905,0.1953342)(7.1509047,0.2853342)(7.4609046,0.3803342)(7.7359047,0.4503342)(8.095904,0.5303342)(8.350904,0.5603342)(8.340904,0.055334203)(8.165905,0.020334201)(7.9459047,-0.024665799)(7.7309046,-0.0896658)(7.4959044,-0.1396658)(7.2459044,-0.2096658)(7.0659046,-0.2596658)(6.8559046,-0.3046658)(6.635905,-0.3446658)(6.4609046,-0.3596658)(6.260905,-0.3796658)(6.0159044,-0.3846658)(5.7659044,-0.3746658)(5.535905,-0.3546658)(5.3109045,-0.3096658)(5.1259046,-0.2546658)(4.9359045,-0.1846658)(4.7959046,-0.1196658)(4.6759048,-0.0746658)
\rput[bl](1.1703491,-0.9174436){$M:=\supp \rho$}
\psbezier[linecolor=black, linewidth=0.04](0.009237989,-0.14188802)(0.55699116,-0.42680338)(0.9975413,-0.48874786)(1.6670158,-0.45633246527777943)(2.3364902,-0.42391708)(2.7330673,-0.12844224)(3.709238,-0.021888021)
\psbezier[linecolor=black, linewidth=0.04](4.67146,0.17366754)(5.2192135,-0.11124782)(5.6597633,-0.17319229)(6.329238,-0.14077690972222284)(6.9987125,-0.10836152)(7.3952894,0.18711331)(8.37146,0.29366753)
\rput[bl](8.468127,0.2103342){$M$}
\rput[bl](3.4120157,0.7214453){$\F$}
\rput[bl](8.098682,0.7353342){$\F$}
\pscircle[linecolor=black, linewidth=0.04, fillstyle=solid,fillcolor=black, dimen=outer](1.2559047,0.6403342){0.055}
\rput[bl](1.4203491,0.5325564){$x$}
\psbezier[linecolor=black, linewidth=0.02](4.6614604,-0.07133246)(5.2092133,-0.3562478)(5.6497636,-0.4181923)(6.319238,-0.38577690972222284)(6.9887123,-0.35336152)(7.4752893,-0.12288669)(8.361461,0.048667535)
\psbezier[linecolor=black, linewidth=0.02](4.67146,0.44366753)(5.2192135,0.15875217)(5.7347636,0.1418077)(6.329238,0.12922309027777715)(6.9237123,0.11663848)(7.3952894,0.45711333)(8.37146,0.56366754)
\rput[bl](6.203127,0.2503342){$U$}
\end{pspicture}
}
\end{center}
\caption{Evaluation of~$\ell$ away from and near~$M$.}
\label{figsupp}
\end{figure}
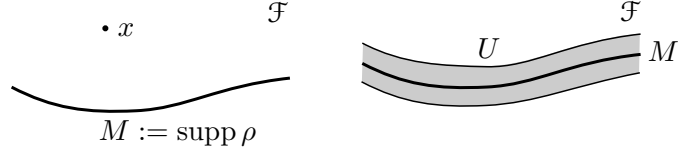

The EL equations~\eqref{EL2} make a statement on the function~$\ell$ even at points~$x \in \F$
which are far away from spacetime~$M$ (see the left of Figure~\ref{figsupp}).
In this way, the EL equations contain much more information than conventional physical equations
formulated in spacetime. At present, it is unclear how this additional information can be used or
interpreted. One way of understanding this situation is to take the point of view that all information
on the physical system must be obtained by performing observations or measurements in spacetime,
which means that the information contained in~$\ell$ away from~$M$ is inaccessible for fundamental reasons.
Here, we shall not take sides or discuss whether or to which extent this point of view is correct.
Instead, we simply note that it seems preferable and physically sensible to restrict attention to the function~$\ell$
in an arbitrarily small neighborhood~$U$ of~$M$ in~$\F$ (see the right of Figure~\ref{figsupp}).
In practice, this means that we shall evaluate~$\ell$ as well as its derivatives only on~$M$.
In this way, the causal action principle gives rise to an interaction described by equations in spacetime.

This concept leads us to the so-called {\em{restricted EL equations}}, which we now introduce.
For technical simplicity, we again restrict attention to the {\em{smooth setting}}
(for a more general derivation, see~\cite[Section~4]{jet}).
\sindex{causal variational principle!in the smooth setting}%
This means that we assume that the Lagrangian is smooth (see~\eqref{Lsmooth} and the discussion thereafter).
To avoid confusion, we point out that this assumption does {\em{not}}
entail that spacetime~$M:=\supp \rho$ is a smooth manifold. Nevertheless, we can speak
of a smooth function or a smooth vector field on~$M$, meaning that the function (or vector field)
\sindex{differentiability!in spacetime $M$}%
\sindex{smoothness in spacetime $M$}%
has a smooth extension to~$\F$.\footnote{We remark that the
question on whether a function or vector field on~$M$ can be
extended smoothly to~$\F$ is rather subtle. The needed conditions are made precise
by Whitney's extension theorem (see, for example, the more recent account in~\cite{fefferman}).
Here, we do not enter the details of these conditions, but use them as implicit
assumptions entering our notion of smoothness.
We remark that these conditions are fulfilled whenever~$M:=\supp \rho$ carries a manifold structure.}
Moreover, for technical simplicity, we assume that also the function~$\ell$ defined by~\eqref{elldef}
is smooth on~$\F$. Under these assumptions, the minimality of~$\ell$ implies that the derivative of~$\ell$
vanishes on~$M$. We thus obtain the equations
\beq \label{ELrestricted}
\ell|_M \equiv 0 \qquad \text{and} \qquad D \ell|_M \equiv 0
\eeq
(where~$D \ell(p) : T_p \F \rightarrow \R$ is the derivative).
In order to combine these two equations in a compact form,
it is convenient to consider a pair~$\u := (a, u)$
consisting of a real-valued function~$a$ on~$M$ and a vector field~$u$
on~$T\F$ along~$M$, and to denote the combination of 
multiplication and directional derivative by
\beq \label{Djet}
\nabla_{\u} \ell(x) := a(x)\, \ell(x) + \big(D_u \ell \big)(x) \:.
\eeq
\sindex{jet derivative|textbf}%
\nindex{aei@$\nabla_\u$ -- jet derivative in direction of~$\u$}%
The pair~$\u=(a,u)$ is referred to as a {\em{jet}}.
\sindex{jet}%
\sindex{jet!for testing}%
\nindex{aej@$\J$ -- space of smooth jets}%
\tindex{ff@$\J$ -- space of smooth jets}%
This jet is a vector in a corresponding {\em{jet space}}~$\J$ defined by
\beq \label{Jinfdef}
\u = (a,u) \;\in\;
\J := C^\infty(M, \R) \oplus \Gamma^\infty(M, T\F) \:,
\eeq
where~$C^\infty(M, \R)$ and~$\Gamma^\infty(M,T\F)$ denote the space of real-valued functions and vector fields
on~$M$, respectively, which admit a smooth extension to~$\F$.
Then the equations~\eqref{ELrestricted} imply that~$\nabla_{\u} \ell(x)$ vanishes for all~$x \in M$,
\beq \label{ELrestricted2}
\nabla_\u \ell|_M = 0 \qquad \text{for all~$\u \in \J$}\:.
\eeq
These are the so-called {\em{restricted EL equations}}.
\sindex{Euler-Lagrange equations!restricted|textbf}%
For brevity, a solution of the restricted EL equations is also referred to as a {\em{critical measure}}.
\sindex{measure!critical}%
We remark that, in the literature, the restricted EL equations are sometimes also referred to as
the {\em{weak}} EL equations. Here we prefer the notion ``restricted'' in order to avoid potential confusion
with weak solutions of these equations (as constructed in~\cite{linhyp}; see also Chapter~\ref{seclinhyp} below).

\section{Symmetries and Symmetric Criticality} \label{secsymmcrit}
In many applications,
variational principles have an underlying symmetry (for example, spherical symmetry or
time independence). Typically, it simplifies the variational problem
to vary within the class of functions that respect this symmetry.
Having found a minimizer within this restricted class, the question arises whether it is also a minimizer
of the full variational problem. The general answer to this question is no, simply because the absolute minimizer
does not necessarily respect the symmetry of the variational principle.
For causal variational principles, the situation is similar,
if we only replace ``function'' by ``measure.'' As a simple example, we saw 
in Section~\ref{seccvpsphere} for the causal variational principle on the sphere that, although
the variational principle is spherically symmetric, minimizing measures are typically weighted counting measures,
thus breaking spherical symmetry.

Nevertheless, one can hope that minimizers within the class of symmetric functions are
critical points of the full variational problem. This statement, referred to as the
{\em{principle of symmetric criticality}}, has been formulated and proven under general
assumptions in~\cite{palais}. 
\sindex{symmetric criticality}%
In this section we explain how the principle of symmetric
criticality can be stated and proved in the setting of causal variational principles.
As we shall see, the proof is quite simple and rather different from that in the classical calculus of variations.
We begin by explaining the basic idea in the simplest possible situation, where we consider
the {\em{compact setting}} and assume that also the {\em{symmetry group is compact}}.
Afterward we explain how to treat a {\em{non-compact symmetry group}}.

As in Section~\ref{seccvp} we let~$\F$ be a compact manifold. Moreover, we again
assume  that the Lagrangian~$\L$ is continuous~\eqref{LC0}, symmetric and
strictly positive on the diagonal (see the assumptions~(i) and~(ii) on page~\pageref{Cond1}).
In order to describe the symmetry, we let~$\G$ be a {\em{compact Lie group}},
which should act as a group of diffeomorphisms on~$\F$
(for basics on Lie groups see, for example, \cite[Chapter~7]{lee-smooth}). More precisely,
we assume the group action~$\Phi : \G \times \F \rightarrow \F$ to be a continuous mapping with the properties that~$\Phi_g:= \Phi(g,.)$
is a diffeomorphism of~$\F$ for any~$g \in \G$, and that
\beq \Phi_g \circ \Phi_h = \Phi_{gh} \qquad \text{for all~$g,h \in \G$}\:. \eeq
Moreover, the symmetry is expressed by the condition that the Lagrangian be invariant under the group action;
that is,
\beq \label{symmLgroup}
\L(\Phi_g x, \Phi_g y) = \L(x,y) \qquad \text{for all~$x,y \in \F$ and~$g \in \G$}\:.
\eeq
\sindex{symmetry!of the causal Lagrangian}%

We denote the set of normalized regular Borel measures on~$\F$ by~${\mathfrak{M}}$.
Taking the push-forward of~$\Phi$, we obtain a group action on~${\mathfrak{M}}$
(for the definition of the push-forward measure see again Section~\ref{secbasicmeasure}).
We denote the measures that are invariant under this group action by~${\mathfrak{M}}^\G$; that is,
\beq \label{MGdef}
{\mathfrak{M}}^\G := \big\{ \rho \in {\mathfrak{M}} \:\big|\: (\Phi_g)_* \rho = \rho \quad\text{for all~$g \in \G$} \big\} \:.
\eeq
We also refer to the measures in~${\mathfrak{M}}^\G$ as being
{\em{equivariant}} (for more details on equivariant causal variational principles see~\cite[Section~4]{lagrange}).
\nindex{aek@${\mathfrak{M}}^\G$ -- equivariant measures under symmetry group~$\G$}
\sindex{minimizing measure!equivariant}%
\sindex{measure!equivariant}%
\sindex{causal variational principle!equivariant|textbf}%
 
\begin{Thm} \label{thmELequi} Let~$\rho$ be a minimizer of the causal action under variations within
the class~${\mathfrak{M}}^\G$ of equivariant measures.
\sindex{symmetric criticality!for causal variational principle}%
Then~$\rho$ is a critical point of the full variational principle in the sense that
the EL equations~\eqref{EL2} hold.
\end{Thm} \noindent
We point out for clarity that minimizers under variations in~${\mathfrak{M}}^\G$
will in general {\em{not}} be minimizers under variations in~${\mathfrak{M}}$.
The reason is that the minimizers in~${\mathfrak{M}}$ are
typically not invariant under the action of the group~$\G$.
A concrete example of this phenomenon is given in Exercise~\ref{excrit2}.

\Proof[Proof of Theorem~\ref{thmELequi}.]
We denote the orbits of the group action by~$\la x \ra := \Phi_\G x$ with~$x \in \F$.
Since~$\G$ is compact, so are the orbits.
On~$\G$ there is a uniquely defined normalized measure that is invariant under the group action
by left multiplication, the so-called {\em{normalized Haar measure}}~$\mu$
(for details on the Haar measure see, for example, \cite[Chapter~16]{lee-smooth}). A particular class of
equivariant measures is obtained by taking the push-forward of~$\mu$ by the mapping~$\Phi(., x)$.
More precisely, for given~$x \in \F$ we define the (also normalized) Borel measure~$\delta_{\la x \ra}$ on~$\F$ by
\beq \label{delHaar}
\delta_{\la x \ra}(\Omega) := \mu \big( \{ g \in \G \:|\: \Phi(g,x) \in \Omega \} \big)
\eeq
for any Borel set~$\Omega \subset \F$. The subscript~$\la x \ra$ indicates that, being equivariant, this measure depends only
on the orbit.

Given~$y \in \F$, we now consider the variation~$(\tilde{\rho}_\tau)_{\tau \in [0,1)}$
within the class of equivariant measures defined by
\beq \label{equivary}
\tilde{\rho}_\tau = (1-\tau)\, \rho + \tau\, \delta_{\la y \ra} \:.
\eeq
Note that, as a convex combination of two normalized measures, also~$\tilde{\rho}_\tau$ is normalized. Using that~$\rho$ is a minimizer within this class, we can proceed similarly as in the proof of Theorem~\ref{thmEL} to obtain
\beq \label{ELprelim}
\int_\F \ell(x)\: \Diff\delta_{\la y \ra}(x) \geq \int_\F \ell(x)\: \Diff\rho(x) \:.
\eeq
Moreover, it follows by symmetry that the function~$\ell$ is constant on the orbits, because
\begin{align}
\ell \big(\Phi_g y \big) &= \int_\F \L\big( \Phi_g y, x \big)\: \Diff\rho(x) - \s
\overset{}{=} \int_\F \L\big( y, \Phi_{g^{-1}} x \big)\: \Diff\rho(x) - \s \notag \\
&= \int_\F \L\big( y, x \big)\: \Diff\rho(x) - \s = \ell(y) \:,
\end{align}
where in the first line we used the symmetry of~$\L$, and in the second line we used that~$\rho$ is equivariant. Hence, integrating over the orbit, we obtain
\beq \ell(y) = \int_\F \ell(x)\: \dd\delta_{\la y \ra} \:. \eeq
Combining this identity with~\eqref{ELprelim}, we conclude that
\beq \ell(y) \geq \int_\F \ell(x)\: \Diff\rho(x) \qquad \text{for all~$y \in \F$}\:. \eeq
Now we can argue exactly as in the proof of Theorem~\ref{thmEL} to obtain the result.
\QED

We next consider the case that the symmetry group~$\G$ is a {\em{non-compact Lie group}}.
A typical example is the translation group, giving rise to the homogeneous causal action
principle as considered in~\cite{elhom}. We again assume that~$\G$ acts on~$\F$ as a group
of diffeomorphisms~$\Phi : \G \times \F \rightarrow \F$. We can again single out the
equivariant measures~${\mathfrak{M}}^\G$ by~\eqref{MGdef}.
Moreover, on~$\G$ one can introduce a left-invariant measure~$\mu$ (again referred to as the Haar measure).
However, in contrast to the case of a compact Lie group, now the measure~$\mu$ has infinite total volume.
As a consequence, it cannot be normalized and, moreover, it is unique only up to a positive prefactor.
It is a basic difficulty that for any non-zero equivariant measure~$\rho$, the integrals
in the causal action~\eqref{Sactcompact} diverge because the integral over the group
elements~$g$ describing the symmetry~\eqref{symmLgroup} is infinite.
In simple terms, this group integral gives an infinite pre\-factor. 
This suggests that the problem could be cured simply by leaving out this integral.
We now explain how this can be done. For simplicity, we restrict attention to the case that~$\G$
{\em{acts freely}} (in the sense that~$g x = x$ with~$g \in \G$ implies that~$g=e$ is the neutral element).
Then, for any~$x \in \F$, the mapping~$g \mapsto \Phi(g,x)$ is a continuous injective mapping from~$\G$ to~$\F$.
In other words, each orbit is homeomorphic to~$\G$. Again denoting the space of orbits by~$\F / \G$,
we can thus identify~$\F \simeq (\F / \G) \times \G$.
Moreover, using this identification, the equivariant measure can be written as
\beq \rho = \rho_{\F/\G} \times \mu \:, \eeq
where~$\rho_{\F/\G}$ is a measure on the orbits. Now we replace the action~\eqref{Sactcompact} by
\beq \label{Sactequi}
\Sact(\rho) = \int_{\F/\G} \Diff\rho_{\F/\G}(x) \int_\F \Diff\rho(y)\: \L(x,y) \qquad \text{with~$\rho \in {\mathfrak{M}}^\G$}\:.
\eeq
The {\em{equivariant causal variational principle}} is to minimize this action under variations in~${\mathfrak{M}}^\G$,
leaving the total volume of~$\rho_{\F/\G}$ fixed.
\sindex{causal variational principle!equivariant}%
If~$\F/\G$ is compact, we can normalize this total volume by demanding that
\beq \label{normFG}
\rho_{\F/\G}(\F/\G) = 1 \:.
\eeq
If~$\F/\G$ is non-compact, the volume constraint can be treated similarly, as explained
for causal variational principles in the non-compact setting in Section~\ref{secnoncompact}.
For more details on this procedure and the resulting existence theory,
we refer to~\cite[Section~4]{lagrange} and~\cite{elhom}.

The justification for considering the equivariant causal variational principle~\eqref{Sactequi} is that
it gives a method for constructing critical points of the full variational principle.
The basis is the following result, which applies in the case that~$\F/\G$ is compact.
\begin{Thm} {\bf{(Symmetric criticality for causal variational principles)}}
\sindex{symmetric criticality!for causal variational principle}%
Let~$\G$ be a non-compact Lie group acting freely on~$\F$ as a group of diffeomorphisms.
Assume that~$\F / \G$ is compact.
Let~$\rho$ be a minimizer of the equivariant causal action principle, which is normalized
on the orbits~\eqref{normFG}.
Then~$\rho$ is a critical point of the full variational principle in the sense that
the EL equations~\eqref{EL2} hold.
\end{Thm}
\Proof The measure~\eqref{delHaar} is normalized on~$\F/\G$. Therefore, the variation~\eqref{equivary}
satisfies the volume constraint~\eqref{normFG}. Computing the first variation of the action, in analogy
to~\eqref{equivary}, we now obtain
\beq \int_{\F/G} \ell(x)\: \dd\delta_{\la y \ra}(x) \geq \int_{\F/\G} \ell(x)\: \Diff\rho_{\F/\G}(x) \eeq
(note that the integrands are constant on the orbits). Carrying out the integral on the left-hand side, we conclude that
\beq \ell(y) = \int_{\F/\G} \ell(x)\: \Diff\rho_{\F/\G}(x) \:, \eeq
giving the claim.
\QED

In the case that~$\F/\G$ is not compact, it is not clear if minimizers exist. One strategy for constructing
minimizers is to exhaust~$\F/\G$ by compact sets, similarly to what is done in~\cite{noncompact}
for causal variational principles in the non-compact setting (see also Section~\ref{seccvpnoncompact}).
If an equivariant minimizer~$\rho$
exists, we know by symmetry that~$\ell$ is constant on the orbits, and moreover the corresponding
EL equations imply that~$\ell$ is minimal on the orbits in the support of~$\rho$.
Combining these facts, we immediately obtain the EL equations~\eqref{EL1}.
In this way, we conclude that symmetric criticality always holds for causal variational principles.

\section{Exercises}

\begin{Exercise} \label{exvar1gen} (More general first variations) {\em{
\sindex{first variation of measure}%
In the proof of Theorem~\ref{thmEL}, we restricted attention to very specific variations~\eqref{nonlocvary}.
In this exercise, we verify that the resulting EL equations~\eqref{EL1} guarantee that the
action is also minimal under more general variations.
To this end, let~$\mu$ be a normalized measure on~$\F$, for technical simplicity with compact support.
Consider variations of the form
\beq \tilde{\rho}_\tau = \chi_{M \setminus U} \,\rho + (1-\tau)\, \chi_U \, \rho + \tau\, \rho(U)\, \mu \:. \eeq
Show that~\eqref{EL1} implies the inequality
\beq \frac{\dd}{\dd\tau} \Sact \big(\tilde{\rho}_\tau \big) \Big|_{\tau=0} \geq 0 \:. \eeq
}} \end{Exercise}

\begin{Exercise} \label{exsval} {\em{ Assume that~$\rho$ is a minimizer of a causal variational principle
with finite total volume. Show that the parameter~$\s$ in~\eqref{elldef} takes the value
\beq \s = \frac{\Sact(\rho)}{\rho(\F)}\:. \eeq
}} \end{Exercise}

\begin{Exercise} (Non-smooth EL equations) {\em{
\sindex{Euler-Lagrange equations!non-smooth}%
We return to the example of the counting measure on the octahedron
as considered in Exercise~\ref{excrit}.
\bitem
\item[(a)] Compute the function~$\ell(x)$. Show that the EL equations~\eqref{EL1} are satisfied.
\item[(b)] Show that the function~$\ell$ is {\em{not}} differentiable
at any point~$x$ of the octahedron. Therefore, it is not possible to formulate the
restricted EL equations~\eqref{ELrestricted2}
\eitem
This example illustrates why in the research papers~\cite{noether, fockbosonic}
one carefully keeps track of differentiability properties by introducing suitable jet spaces.
}} \end{Exercise}

\begin{Exercise} \label{excrit2} (Symmetric criticality on the sphere) {\em{
We consider the causal variational principle on the sphere as introduced in Section~\ref{seccvpsphere}.
\bitem
\item[(a)] Show that the symmetric measure on the sphere
\beq \Diff\mu(\vartheta, \varphi) = \frac{1}{4 \pi}\: \dd\varphi\, \sin \vartheta\: \dd\vartheta \eeq
is critical in the sense that it satisfies the EL equations~\eqref{EL1}.
\item[(b)] Use the minimizer with singular support constructed in Exercise~\ref{excrit}
to argue that minimizers within the class of symmetric measures are, in general, {\em{not}}
minimizers within the class of measures without symmetries.
More details on this effect of symmetry breaking can be found in~\cite{continuum, support, rrev}.
\eitem
}} \end{Exercise}

\chapter{The Linearized Field Equations} \label{seclin}
The EL equations as derived in the previous chapter (see Theorem~\ref{thmEL}
or the restricted EL equations in~\eqref{ELrestricted2}) are {\em{nonlinear}} equations.
This can be seen immediately from the fact that the measure~$\rho$ enters in a twofold way:
It determines the function~$\ell$ via the integration~\eqref{elldef}, and it also determines 
via its support~$M$ where the function~$\ell$ is to be evaluated.
As usual, such nonlinear equations are difficult to analyze. Therefore, it is a good idea
to simplify these equations by linearization. This leads us to the so-called
{\em{linearized field equations}}, which describe linear
perturbations of the system that respect the EL equations.
This chapter is devoted to the derivation of the linearized field equations
and to the construction of explicit examples.
We remark that the linearized field equations are also a suitable starting point for
the analysis of the full EL equations, for example, by using perturbative methods
(see~\cite{perturb} or Chapter~\ref{secperturb}).

\section{Derivation of the Linearized Field Equations in the Smooth Setting} \label{seclinfield}
Linearizations are used frequently in physics and mathematical analysis.
\sindex{linearization}%
Typical physical examples are the anharmonic oscillator for small displacements
or nonlinear waves of small amplitudes. In these examples, the
nonlinearity of the underlying wave equation may be neglected, making it possible to
describe the dynamics by a linear wave equation.
In mathematical terms, the dynamics is usually described by nonlinear equations
that are difficult to solve (like, for example, the Einstein equations).
Linearizing gives linear equations (for example, the equations of linearized gravity),
which are much easier to analyze.
 In order to derive the linearized equations,
one typically considers a family~$G_\tau$ of solutions.
The parameter~$\tau$ can be thought of as the ``amplitude'' of the perturbation,
and~$G_\tau|_{\tau=0}$ describes the unperturbed system (for example, an anharmonic oscillator at rest
or the vacuum). Then the derivative
\beq \frac{\dd}{\dd\tau} G_\tau \big|_{\tau=0} \eeq
is the linearized field; it satisfies a linear equation obtained by differentiating the nonlinear
equation (like Hamilton's equation for the anharmonic oscillator or a nonlinear wave equation)
with respect to~$\tau$.

The concept of linearization is also fruitful in the context of causal variational principles.
Since the system is described by the measure~$\rho$, the above family of solutions
now corresponds to a family of measures~$(\tilde{\rho}_\tau)_{\tau \in [0, \delta)}$
which are all critical points of the causal action principle.
The basic question is how to vary the measure. Indeed, there are many ways of
varying. We begin with a simple method, which we will generalize and discuss afterward.
In order to keep the presentation as simple as possible, we again restrict attention
to the {\em{smooth setting}} (as defined in Section~\ref{seccvp}; this means that the
Lagrangian is smooth~\eqref{Lsmooth}, but~$M$ does not need to have a smooth manifold structure).
Moreover, for technical simplicity, we assume that the Lagrangian has the following property.
\begin{Def} \label{defcompactrange}
\sindex{Lagrangian!of compact range}%
	The Lagrangian has a {\bf{compact range}} if for every compact~$K \subset \F$, there is
	a compact set~$K' \subset \F$ such that
	\beq \L(x,y) = 0 \qquad \text{for all~$x \in K$ and~$y \not \in K'$}\:. \eeq
\end{Def}

We choose a family~$f_\tau$ of positive weight functions and a family~$F_\tau$
of mapping from~$M$ to~$\F$. These functions should all be smooth, also in the parameter~$\tau$; that is,
\beq f \in C^\infty([0, \delta) \times M, \R^+) \qquad \text{and} \qquad
F \in C^\infty([0, \delta) \times M, \F) \eeq
(here, as explained before~\eqref{ELrestricted}, smoothness on~$M$ is defined via the existence of a smooth extension to~$\F$).
We multiply~$\rho$ by~$f_\tau$ and then take the push-forward under~$F_\tau$,
\Evtl{W\"are es eventuell sinnvoll, relativ fr\"uh einen Abschnitt einzuf\"ugen zu ``Wie variiert man ein Ma{\ss}?'' Und dort ggf. auch die technischen Punkte genau zu behandeln? F: Ich sehe ehrlich gesagt nicht, dass das nötig ist.}%
\beq \label{Ffvary}
\tilde{\rho}_\tau := (F_\tau)_* \big( f_\tau \,\rho \big) \:.
\eeq
We assume that for~$\tau=0$, the variation is trivial,
\beq \label{triv0}
f_0 \equiv 1 \qquad \text{and} \qquad F_0 \equiv \1\:.
\eeq
Since multiplying by a positive function leaves the support unchanged,
the support of the measure is transformed only by~$F_\tau$; more precisely,
\beq \label{srt}
\supp \tilde{\rho}_\tau = \overline{F_\tau \big( \supp \rho \big)}
\eeq
(for details, see Exercise~\ref{exsupp1}).

The assumption that all the measures~$\tilde{\rho}_\tau$ are critical means
that they all satisfy the restricted EL equations~\eqref{ELrestricted2}.
Taking into account that the support of the measures changes according to~\eqref{srt}, we know that
for all~$\u \in \J_0$ and all~$x \in M$,
\begin{align}
0 &= \nabla_\u \bigg( \int_\F \L\big( F_\tau(x), y \big)\: \dd\tilde{\rho}_t(y) - \s \bigg) \notag \\
&= \nabla_\u \bigg( \int_\F \L\big( F_\tau(x), F_\tau(y) \big)\: f_\tau(y)\: \Diff\rho(y) - \s \bigg) \:,
\end{align}
where in the last line we used the definition of the push-forward measure.
It is convenient to multiply this equation by~$f_\tau(x)$. We can write this factor inside the brackets,
\beq 0 =  \nabla_\u \bigg( \int_\F f_\tau(x) \:\L\big( F_\tau(x), F_\tau(y) \big)\: f_\tau(y)\: \Diff\rho(y) - 
f_\tau(x) \:\s \bigg) \:, \eeq
because the terms obtained when the derivative~$\nabla_\u$ acts on~$f_\tau(x)$ vanish
in view of the restricted EL equations~\eqref{ELrestricted2}.
Since this equation holds for all~$\tau \in [0, \delta)$, we can differentiate at~$\tau=0$,
\begin{align}
0 &= \frac{\dd}{\dd\tau} \nabla_\u \bigg( \int_\F f_\tau(x) \:\L\big( F_\tau(x), F_\tau(y) \big)\: f_\tau(y)\: \Diff\rho(y) - 
f_\tau(x) \:\s \bigg) \bigg|_{\tau=0} \notag \\
&= \nabla_{\u(x)} \bigg( \int_\F \big( \dot{f}_0(x) + D_{1,\dot{F}_0(x)} + \dot{f}_0(y) + D_{2,\dot{F}_0(y)}
\big) \L(x,y)\: \dd \rho(y) -  \dot{f}_0(x) \:\s \bigg) \:,
\end{align}
where~$D_1$ and~$D_2$ denote partial derivatives acting on the first and second
arguments of the Lagrangian, respectively.
Here, we were allowed to interchange the derivative with the integral because
the integration range is compact by the assumption in Definition~\ref{defcompactrange}.
We write this equation in the shorter form
\sindex{linearization!of the Euler-Lagrange equations}%
\beq \label{limprelim}
\nabla_{\u(x)} \bigg( \int_\F \big( \nabla_{1,\v} + \nabla_{2,\v} \big) \L(x,y)\: \Diff\rho(y) - 
\nabla_\v \:\s \bigg) = 0 \:,
\eeq
where~$\v$ is the jet generated by the functions~$f_\tau$ and~$F_\tau$,
\sindex{jet!for varying}%
\beq \label{vlinear}
\v := \frac{\dd}{\dd\tau} \big( f_\tau, F_\tau \big) \big|_{\tau=0} \;\in\; \J \:.
\eeq

Note that in~\eqref{limprelim}, the $\u$-derivative also acts on the jet~$\v(x)$, giving rise to the terms
\beq \label{extradiff}
\int_\F \nabla_{1,D_u \v} \L(x,y)\: \Diff\rho(y) - \nabla_{D_u \v} \:\s \:.
\eeq
But these terms vanish in view of the restricted EL equations~\eqref{ELrestricted2}.
This observation makes it possible to simplify the formulation of the linearized field equations
by adopting the following computational conventions for partial derivatives and jet derivatives:
\sindex{jet!computational conventions}%
\sindex{jet derivative}%
\bitem
\item[(i)] Partial and jet derivatives with an index~$i \in \{ 1,2 \}$ only act on the respective variable of the function~$\L$.
This implies, for example, that the derivatives commute,
\label{ConventionPartial}
\beq \nabla_{1,\v} \nabla_{1,\u} \L(x,y) = \nabla_{1,\u} \nabla_{1,\v} \L(x,y) \:. \eeq
\item[(ii)] The partial or jet derivatives that do not carry an index act as partial derivatives
on the corresponding argument of the Lagrangian. This implies, for example, that
\beq \nabla_\u \int_\F \nabla_{1,\v} \, \L(x,y) \: \Diff\rho(y) =  \int_\F \nabla_{1,\u} \nabla_{1,\v}\, \L(x,y) \: \Diff\rho(y) \:. \eeq
\eitem
\label{ConventionPartial2}%
We will use these conventions from now on throughout this book.
We point out that, following these conventions, {\em{jets are never differentiated}}.
This is a very convenient convention.
Clearly, we must always verify that this convention may really be used.
As already mentioned above, in~\eqref{limprelim} this convention is justified because the additional terms
obtained if the derivative~$\nabla_\u$ acted on the jet~$\v$ vanish as a consequence of the
restricted EL equations.

We remark that, from a differential geometric perspective, defining
higher derivatives on~$\F$ would make it necessary to introduce a connection on~$\F$.
While this could be done, we here use the simpler method that higher derivatives on~$\F$
are defined as partial derivatives carried out in {\em{distinguished charts}}.
More precisely, around each point~$x \in \F$ we choose a distinguished chart and
carry out all derivatives as partial derivatives acting on each tensor component in this chart.
We remark that, in the setting of causal fermion systems, an atlas of distinguished charts is provided
by the so-called symmetric wave charts (for details, see the remark after Proposition~\ref{prppq}
and~\cite[Section~6.1]{gaugefix} or~\cite[Section~3]{banach}).

\begin{Def} \label{deflinear}
\sindex{linearized field equations}%
\nindex{ael@$\Delta$ -- linearized field operator}%
Let~$\rho$ be a solution of the restricted EL equations~\eqref{ELrestricted2}.
A jet~$\v \in \J$ is referred to as a {\bf{solution of the linearized field equations}} if
\beq \label{linfield}
\la \u, \Delta \v \ra(x) := \nabla_\u \bigg( \int_\F \big( \nabla_{1,\v} + \nabla_{2,\v} \big) \L(x,y)\: \Diff\rho(y) - 
\nabla_\v \:\s \bigg) = 0
\eeq
for all~$\u \in \J$ and all~$x \in M$.
The vector space of all linearized solutions is denoted by~$\Jlin \subset \J$.
\end{Def} \noindent
We often write the linearized field equation in the short form~$\Delta \v = 0$.
For the mathematical analysis of the linearized field equations, it is preferable to
include an {\em{inhomogeneity}}~$\w$
\sindex{inhomogeneity!in linearized field equations}%
\beq \label{Delvw}
\Delta \v = \w \:.
\eeq
In view of the pairing with the jet~$\u$ in~\eqref{linfield}, the inhomogeneity
is a dual jet, that is, $\u(x) \in (\J_x)^*$, making it possible to add on the right
side of~\eqref{linfield} the dual pairing~$\la \u(x), \w(x)\ra_x$.
We remark that, in the analysis of the linearized field equations in Chapter~\ref{seclinhyp}, the
distinction between jets and dual jets will become unnecessary because we will identify
them via a scalar product on the jets given at every spacetime point.

We conclude this section with a brief discussion of our ansatz~\eqref{Ffvary}.
Intuitively speaking, this ansatz means that the support of the
measure is changed smoothly as a whole.
In particular, if~$M$ is a smooth four-dimensional submanifold of~$\F$,
then the varied measure~$M_\tau$ will again have this property.
In physical terms, measures where~$M$ has such a manifold structure
describe {\em{classical spacetimes}}. Consequently, the ansatz~\eqref{Ffvary}
and the corresponding linearization~\eqref{vlinear} correspond to
{\em{classical fields}} in a classical spacetime.
In contrast, if the support~$M:= \supp \rho$ of the measure does {\em{not}} have the structure of
a four-dimensional manifold, then we refer to~$M$ as a {\em{quantum spacetime}}.
\sindex{quantum spacetime}%
The notion ``quantum spacetime'' appears in the literature in different contexts with
rather different meanings. Here we take the above notion as the definition.
In this way, the notion ``quantum spacetime'' gets a precise mathematical meaning.
Our notion is very general. In particular, it allows for the description of non-smooth spacetime structures.
The name ``{\em{quantum}} spacetime'' is justified by the fact, in our setting, all spacetime
structures are encoded in the family of all physical wave functions, being the fundamental quantum objects
of a causal fermion system.

In order to give an idea of how such a quantum spacetime may look like,
let us consider the example where the unperturbed measure~$\rho$
describes a classical spacetime~$M$ (for example, Minkowski space~$M \simeq \R^4$).
As just explained, the ansatz~\eqref{Ffvary} changes the support of the measure smoothly
as a whole (see Figure~\ref{figfragment}~(a)).
\begin{figure}
\begin{center}
% \usepackage[usenames,dvipsnames]{pstricks}
% \usepackage{epsfig}
% \usepackage{pst-grad} % For gradients
% \usepackage{pst-plot} % For axes
% \usepackage[space]{grffile} % For spaces in paths
% \usepackage{etoolbox} % For spaces in paths
% \makeatletter % For spaces in paths
% \patchcmd\Gread@eps{\@inputcheck#1 }{\@inputcheck"#1"\relax}{}{}
% \makeatother
% \psscalebox{1.0 1.0} % Change this value to rescale the drawing.
{
\begin{pspicture}(-0.7,-1.3734679)(14.770368,1.3734679)
\definecolor{colour0}{rgb}{0.8,0.8,0.8}
\pspolygon[linecolor=colour0, linewidth=0.02, fillstyle=solid,fillcolor=colour0](9.339256,-0.2956901)(9.428145,-0.33124566)(9.592589,-0.38902342)(9.788145,-0.42902344)(10.001478,-0.45569012)(10.27259,-0.45124567)(10.543701,-0.44235677)(10.765923,-0.41124567)(11.108145,-0.35346788)(11.339256,-0.30013454)(11.548145,-0.24235678)(11.881478,-0.15791233)(12.094811,-0.10902344)(12.35259,-0.055690106)(12.659256,0.0020876736)(12.877034,0.033198785)(13.041478,0.046532117)(13.045923,0.6287543)(12.721478,0.695421)(12.374812,0.77097654)(12.0148115,0.8331988)(11.748145,0.8643099)(11.3437,0.8643099)(11.050367,0.8331988)(10.717034,0.7443099)(10.388145,0.6287543)(10.103701,0.5531988)(9.725923,0.49097657)(9.4725895,0.53097653)(9.3437,0.57986546)
\rput[bl](1.1703675,-0.8623568){$M:=\supp \rho$}
\psbezier[linecolor=black, linewidth=0.04, linestyle=dashed, dash=0.17638889cm 0.10583334cm](0.0059230137,0.3643099)(0.80589837,0.1160612)(1.3220041,0.2707834)(1.8170341,0.42653211805555347)(2.3120642,0.5822808)(3.01753,0.7522001)(3.705923,0.48430988)
\psbezier[linecolor=black, linewidth=0.04](0.009256347,-0.086801216)(0.5570095,-0.37171656)(0.99755967,-0.43366104)(1.6670341,-0.40124565972222515)(2.3365085,-0.36883026)(2.7330856,-0.07335544)(3.7092564,0.033198785)
\rput[bl](1.6325897,-1.3734679){(a)}
\rput[bl](6.1037006,-1.3690234){(b)}
\rput[bl](10.988145,-1.3690234){(c)}
\rput[bl](0.23703413,0.4665321){$\supp \tilde{\rho}_\tau$}
\psbezier[linecolor=black, linewidth=0.02, arrowsize=0.05291667cm 2.0,arrowlength=1.4,arrowinset=0.0]{->}(2.4488256,-0.16235676)(2.4691994,-0.06597727)(2.505363,0.112867646)(2.494812,0.19444763688016337)(2.4842608,0.27602762)(2.423824,0.42073712)(2.3614786,0.5176432)
\rput[bl](2.5881453,0.039865453){$F$}
\psbezier[linecolor=black, linewidth=0.04](4.6714787,0.22875434)(5.2192316,-0.056161016)(5.659782,-0.11810549)(6.3292565,-0.08569010416666856)(6.9987307,-0.053274717)(7.3953075,0.24220012)(8.371479,0.34875435)
\psbezier[linecolor=black, linewidth=0.04, linestyle=dashed, dash=0.17638889cm 0.10583334cm](4.677034,0.675421)(5.4770093,0.4271723)(5.9931154,0.5818945)(6.4881454,0.7376432291666666)(6.9831753,0.89339197)(7.688641,1.0633112)(8.377034,0.795421)
\psbezier[linecolor=black, linewidth=0.04, linestyle=dashed, dash=0.17638889cm 0.10583334cm](4.690367,-0.59569013)(5.48746,-0.84393877)(6.466691,-0.50699437)(6.9555087,-0.3423567708333326)(7.4443264,-0.17771916)(7.7311215,0.085533455)(8.319257,-0.06680121)
\rput[bl](4.899256,-0.41346788){$F_2$}
\rput[bl](7.5303674,0.395421){$F_1$}
\psbezier[linecolor=black, linewidth=0.02, arrowsize=0.05291667cm 2.0,arrowlength=1.4,arrowinset=0.0]{->}(7.3821588,0.18875434)(7.4025326,0.28513384)(7.4386964,0.46397877)(7.4281454,0.5455587479912742)(7.417594,0.62713873)(7.357157,0.77184826)(7.2948117,0.8687543)
\psbezier[linecolor=black, linewidth=0.02, arrowsize=0.05291667cm 2.0,arrowlength=1.4,arrowinset=0.0]{->}(4.7941318,0.086532116)(4.7737575,-0.009847382)(4.759816,-0.113136746)(4.7703676,-0.1947167341023888)(4.7809186,-0.27629673)(4.8413553,-0.4876729)(4.881479,-0.5934679)
\psbezier[linecolor=black, linewidth=0.04](9.355923,0.22430989)(9.903676,-0.06060546)(10.344226,-0.122549936)(11.0137005,-0.09013454861111199)(11.683175,-0.05771916)(12.079752,0.23775567)(13.055923,0.3443099)
\psbezier[linecolor=white, linewidth=0.04](9.315923,-0.27791232)(9.863676,-0.5628277)(10.570893,-0.46477216)(11.004812,-0.3879123263888903)(11.438731,-0.3110525)(12.070864,-0.060022105)(13.047034,0.046532117)
\psbezier[linecolor=white, linewidth=0.04](9.302589,0.6065321)(9.850343,0.32161677)(10.55756,0.7396723)(11.080367,0.8298654513888895)(11.603175,0.9200586)(12.168641,0.851089)(13.038145,0.6287543)
\rput[bl](11.250367,-0.7601346){$\supp \tilde{\rho}_\tau$}
\psbezier[linecolor=black, linewidth=0.02, arrowsize=0.05291667cm 2.0,arrowlength=1.4,arrowinset=0.0]{->}(11.195492,-0.5934679)(11.004142,-0.572549)(10.904003,-0.54737675)(10.819256,-0.4989638143801977)(10.734508,-0.45055088)(10.681424,-0.40389943)(10.622507,-0.34902343)
\rput[bl](8.628145,0.19542101){$M$}
\rput[bl](3.4370341,1.1065321){$\F$}
\rput[bl](8.143701,1.115421){$\F$}
\rput[bl](12.739256,1.1109766){$\F$}
\end{pspicture}
}
\end{center}
\caption{Illustration of the fragmentation of the measure $\rho$: Smooth variation (a), disintegration into two components (b) and general enlargement of the support (c).}
\label{figfragment}
\end{figure}
More generally, one can consider the situation where the measure~$\rho$ ``disintegrates'' into several
``components'' that are perturbed differently (see Figure~\ref{figfragment}~(b)).
For the mathematical description, we choose a parameter~$L \in \N$ (the ``number of subsystems'')
and consider mappings
%\beq \label{fFadef}
\beq f_\as \in C^\infty\big([0,\delta) \times M,\R^+ \big)\:,\quad
F_\as \in C^\infty\big([0,\delta) \times M, \F \big) \qquad \text{with~$\as = 1, \ldots, L$}\:. \eeq
For the so-called {\em{measure with fragmentation}}, in generalization of~\eqref{Ffvary} we make
\sindex{measure!with fragmentation}%
the ansatz
\beq \label{rhotildea}
\tilde{\rho}_\tau = \frac{1}{L} \sum_{\as=1}^L (F_{\as, \tau})_* \big( f_{\as, \tau} \, \rho \big) \:.
\eeq
The larger~$L$ is chosen, the more freedom we have in perturbing the measure.
We point out that we may choose~$L$ arbitrarily large. In the limit~$L \rightarrow \infty$, one can
even describe situations where the support of the measure~$\rho$ is ``enlarged'' by the perturbation,
as shown in Figure~\ref{figfragment}~(c). In this way, a classical spacetime point may
correspond to many points of the quantum spacetime, making it possible to encode additional
local degrees of freedom. Integrating with respect to the measure~$\rho$ also entails
an integration over these additional degrees of freedom, bearing some similarity to integrating
over field configurations in a path integral.

Assuming that the family~$(\tilde{\rho}_\tau)_{\tau \in [0, \delta)}$ satisfies the
restricted EL equations for all~$\tau$, we can again linearize in~$\tau$ to obtain the corresponding
linearized field equations. They again have the form~\eqref{linfield}, but now with~$\v$ being the
``averaged jet''
\beq \v = \frac{1}{L} \sum_{\as=1}^L \v_\as \qquad \text{with} \qquad \v_\as = 
\frac{\dd}{\dd \tau} \big( f_{\as, \tau}, F_{\as,\tau} \big) \big|_{\tau=0} \:. \eeq
Therefore, for linearized fields the fragmentation does not give anything essentially new.
But on the nonlinear level, fragmentation yields additional effects.
We refer the interested reader for more details to~\cite[Section~5]{positive}
and~\cite[Section~5]{perturb} as well as to the applications to quantum field theory
in~\cite{fockfermionic, fockdynamics} (see also Chapter~\ref{secQFT}).

In view of this consideration, the only restriction in describing linear perturbations of a
measure~$\rho$ by a jet~$\v$ of the form~\eqref{vlinear} is that the support of the measure~$\rho$
is changed continuously in~$\tau$, in the sense that the support~$\supp \tilde{\rho}_\tau$ lies
in a small neighborhood of~$M$ (for details, see Exercise~\ref{exsupp2}).
In particular, we do not cover variations of the form~\eqref{nonlocvary} where part of the measure
is ``transported'' to a point~$y \in \F$ which may be far away from~$M$.
The reason for disregarding such variations is that, similar as explained before introducing the
restricted EL equations in Section~\ref{secrestrictedEL} (see Figure~\ref{figsupp}), analyzing the
EL equations outside a small neighborhood of~$M$ does not seem to be of physical relevance.

\section{Commutator Jets in Causal Fermion Systems} \label{seccomm}
\sindex{jet!commutator|textbf}%
In order to illustrate the linearized field equations, we conclude this chapter by deriving explicit classes of
solutions. These solutions correspond to the inherent symmetries of the system.
In this section, we consider {\em{commutator jets}}, which describe the unitary invariance
of a causal action principle. In the next section (Section~\ref{secinner}),
we shall derive {\em{inner solutions}} by using the
invariance of the measure under diffeomorphisms of~$M$ combined
with a suitable multiplication of~$\rho$ by a smooth weight function.

Let~$(\H, \F, \rho)$ be a causal fermion system.
The causal action principle is {\em{unitarily invariant}} in the following sense. Let~$\scrU \in \U(\H)$ be a
unitary transformation. Given a measure~$\rho$ on~$\F$, we can unitarily transform the measure by
setting
\beq \label{Urhodef}
(\scrU \rho)(\Omega) = \rho \big( \scrU^{-1} \,\Omega\, \scrU \big) 
\qquad \text{for} \qquad \Omega \subset \F \:.
\eeq
By construction of the integral, this also means that
\beq \int_\F \phi(x)\: \Diff(\scrU \rho)(x) = \int_\F \phi \big( \scrU x \scrU^{-1} \big)\: \Diff \rho(x) \eeq
for any integrable function~$\phi$.
Since the eigenvalues of an operator are invariant under unitary transformations,
a measure~$\rho$ is a minimizer or critical point of the causal action principle if and only
if~$\scrU \rho$ is.

Infinitesimally, this unitary invariance gives rise to a special class of solutions of
the linearized field equations, as we now explain. Let~$\rho$ be a critical measure. 
We let~$\scrA$ be a symmetric operator on~$\H$, for technical simplicity of finite rank.
By exponentiating, we obtain a family of unitary operators~$(\scrU_\tau)_{\tau \in \R}$ with
\beq \label{Utaudef}
\scrU_\tau := \exp(\cI \tau \scrA) \:.
\eeq
According to~\eqref{Urhodef}, the support of the measures~$\tilde{\rho}_\tau:= \scrU_\tau \rho$ is
given by
\beq %\label{Munit}
\tilde{M}_\tau := \supp \tilde{\rho}_\tau = \scrU_\tau\, M \, \scrU_\tau^{-1} \:. \eeq
Due to the unitary invariance of the Lagrangian, the measures~$\tilde{\rho}_\tau$ all satisfy the EL equations.
Infinitesimally, the unitary transformations are described by the jet~$\v$ given by
\beq \label{jvdef}
\v := (0,v) \in \Jlin \qquad \text{with} \qquad v(x) = \frac{\dd}{\dd\tau} \big( \scrU_\tau \,x\, \scrU_\tau^{-1} \big) \big|_{\tau=0} 
= \cI \big[\scrA, x \big] \:.
\eeq
Due to the commutator in the last equation, we refer to jets of this form
as {\bf{commutator jets}} (this notion was first introduced in~\cite[Section~3]{dirac}).
The fact that commutator jets generate families of critical measures
implies that they satisfy the linearized field equations:
\begin{Lemma} \label{lemmacommsol} $\;\;\;$ The commutator jet~$\v$ in~\eqref{jvdef} satisfies the linearized field equations~\eqref{linfield}.
\end{Lemma}
\Proof Due to the unitary invariance of the Lagrangian,
\beq \L\big( \scrU_\tau x \scrU_\tau^{-1},\: \scrU_\tau y \scrU_\tau^{-1} \big) = \L(x,y)\:. \eeq
Differentiating with respect to~$\tau$ and applying the product and chain rules gives
\beq %\label{D12zero}
(D_{1,v} + D_{2,v} )\, \L(x,y)\: \Diff\rho(y) = 0 \:. \eeq
Hence the integrand in~\eqref{linfield} vanishes for all~$x,y \in \F$. As a consequence,
the integral in~\eqref{linfield} vanishes for all~$x \in \F$.
Consequently, also its derivative in the direction of~$u$ vanishes.
Using our convention that the
jet derivatives act only on the Lagrangian (see the end of Section~\ref{seclinfield}),
this directional derivative differs from the jet derivative in~\eqref{linfield} by the term~$D_{D_u \v} \ell(x)$
(similar as explained for~\eqref{extradiff}).
This term vanishes in view of the restricted EL equations~\eqref{ELrestricted2}.
\QED

As we shall see in Section~\ref{seccip}, commutator jets are very useful because they give rise to
conserved quantities.

\section{Inner Solutions in Smooth Spacetimes} \label{secinner}
\sindex{inner solution of linearized field equations}%
We now return to causal variational principles in the smooth setting
(thus we again assume that the Lagrangian is smooth~\eqref{Lsmooth}).
We introduce an additional smoothness assumption for the measure~$\rho$ and explain why it
is useful in some applications.
\begin{Def} \label{defsms}
\sindex{smooth manifold structure}%
\sindex{spacetime!with smooth manifold structure}%
Spacetime~$M:= \supp \rho$ has a {\bf{smooth manifold structure}} if
the following conditions hold:
\bitem
\item[{\rm{(i)}}] $M$ is diffeomorphic to a smooth oriented manifold~$\scrM^k$ of dimension~$k$.
\item[{\rm{(ii)}}] In a chart~$(x,U)$ of~$\scrM^k$, the measure~$\rho$ is absolutely continuous with respect
to the Lebesgue measure with a smooth, strictly positive weight function; that is,
\beq \label{hdef}
\Diff\rho = h(x)\: \Diff^kx \qquad \text{with} \quad h \in C^\infty(\scrM^k, \R^+) \:.
\eeq
\eitem
\end{Def} \noindent
Even though there is no reason why physical spacetime should have a smooth manifold
structure on the Planck scale, this assumption is clearly justified on the macroscopic scale
of atomic and gravitational physics. With this in mind, the assumption of a smooth manifold
structure seems admissible in all applications in which the microscopic structure of spacetime
should be irrelevant. Before going on, we point out that one
should carefully distinguish the assumption of a smooth manifold structure from the smooth
setting introduced in Section~\ref{seccvp}. In particular, should keep in mind that the smoothness of~$\L$ does not
imply that~$M$ has a smooth manifold structure, nor vice versa.

The fact that~$\rho$ is defined independent of charts implies that the function~$h$ in~\eqref{hdef}
transform like a tensor density. More precisely, on the overlap of two charts~$(x,U)$ and~$(\tilde{x}, \tilde{U})$,
we know that
\beq h(x)\: \Diff^kx = \tilde{h}(\tilde{x})\: \Diff^k\tilde{x} \eeq
and thus
\beq h(x) = \det \bigg( \frac{\partial x^i}{\partial \tilde{x}^j} \bigg) \: \tilde{h}(\tilde{x}) \:. \eeq
This transformation law makes it possible to define the covariant divergence of a vector field~$v$
on~$M \simeq \scrM^k$ in a local chart by
\nindex{aem@$\div$ -- divergence in smooth spacetime}%
\beq \label{defdiv}
\div v = \frac{1}{h}\: \partial_j \big( h\, v^j \big)
\eeq
(where, following the Einstein summation convention, we sum over~$j=0,\ldots,k$).
Alternatively, the divergence of a vector field~$v \in \Gamma(M, TM)$ 
can be defined independent of charts by the relation
\beq \label{divalt}
\int_M \div v\: \eta(x)\: \Diff\rho = -\int_M D_v \eta(x)\: \Diff\rho(x) \:,
\eeq
to be satisfied for all test functions~$\eta \in C^\infty_0(M, \R)$.
Indeed, integrating partial derivatives by parts and using~\eqref{hdef},
we obtain
\begin{align}
-\int_M D_v \eta(x)\: \Diff\rho(x) &= -\int_M v^j(x)\: \frac{\partial \eta(x)}{\partial x^j} \: h(x)\, \Diff^k x \notag \\
&= \int_M \eta(x)\: \frac{1}{h(x)} \bigg( \frac{\partial}{\partial x^j} \Big( h(x)\: v^j(x) \Big) \bigg)\: h(x)\: \Diff^k x \:,
\end{align}
and using~\eqref{defdiv} and again~\eqref{hdef} gives back the left-hand side of~\eqref{divalt}.
We remark that the right-hand side of~\eqref{divalt} can be understood as a weak formulation of the divergence.
Such a formulation has the advantage that it can be used even in cases where~$v$ is not differentiable.
In what follows, we will always restrict attention to smooth vector fields, so that the weak
and pointwise formulations~\eqref{divalt} and~\eqref{defdiv} are equivalent.
We usually prefer to work with~\eqref{defdiv}, but the weak formulation~\eqref{divalt} still has its value in
being manifestly coordinate independent.

Having vector fields and the divergence at our disposal, we can now introduce a specific class
of linearized solutions. We first define them and explain their significance afterward.

\begin{Def} An {\bf{inner solution}} is a jet~$\v \in \J$ of the form
\beq \v = (\div v, v) \qquad \text{with} \qquad v \in \Gamma(M, TM) \:. \eeq
The vector space of all inner solution is denoted by~$\Jin \subset \J^1$.
\end{Def}

This notion was first introduced in~\cite[Section~3]{fockbosonic}.
The name ``inner {\em{solution}}'' is justified by the following lemma:
\begin{Lemma} \label{lemmaGauss}
Every inner solution~$\v \in \Jin$ of compact support is a solution of the linearized field equations; that is,
\beq \la \u, \Delta \v \ra_M = 0 \qquad \text{for all~$\u \in \J$}\:. \eeq
\end{Lemma}
\Proof Applying the Gau{\ss} divergence theorem, one finds that for every function~$f \in C^1_0(M, \R)$,
\beq \int_M \nabla_\v f\: \Diff\rho = 
\int_M \big( \div v\: f + D_v f \big)\: \Diff\rho 
= \int_M \div \big( f v \big)\: \Diff\rho = 0 \:. \eeq
Likewise, in the linearized field equations, we may integrate by parts in~$y$,
\begin{align}
\la \u, \v \ra_M &= \nabla_\u \bigg( \int_M \big(\nabla_{1,\v} + \nabla_{2,\v} \big) \L(x,y)\: \Diff\rho(y)
- \nabla_\v \s \bigg) \notag \\
&= \nabla_\u \bigg( \int_M \nabla_{1,\v} \L(x,y)\: \Diff\rho(y)
- \nabla_\v \s \bigg) \notag \\
&= \nabla_\u \nabla_\v \ell(x) = \nabla_{\v(x)} \big( \nabla_\u \ell(x) \big) 
- \nabla_{D_v \u} \ell(x) = 0 \:.
\end{align}
The last equality comes about as follows. The summand~$\nabla_{D_v \u} \ell(x)$ vanishes
by the restricted EL equations. Moreover, the restricted EL equations yield that the 
function~$\nabla_\u \ell$ vanishes identically on~$M$.
Therefore, this function is differentiable in the direction of the vector field~$v$ on~$M$,
and this directional derivative is zero. Therefore,
\beq \nabla_{\v(x)} \big( \nabla_\u \ell(x) \big) = D_{v(x)} \big( \nabla_\u \ell(x) \big)
+ \div v(x)\: \nabla_\u \ell(x) = 0  \:, \eeq
giving the result.
\QED
This result also holds for inner solutions~$\v$ of non-compact support, provided that the
vector field~$v$ has suitable decay properties at infinity. For details, we refer to~\cite[Section~3]{fockbosonic}.

We now explain the significance of inner solutions. To this end, we let~$\Phi_t : M \rightarrow M$
with~$t \in (-\delta, \delta)$ and some~$\delta>0$ be the local flow generated by the vector field~$v$; that is,
\beq \Phi_0 = \text{id}_M \qquad \text{and} \qquad \frac{\dd}{\dd t} \Phi_t(x) = v\big|_{\Phi_t(x)}
\quad \text{for all~$t \in (-\delta, \delta)$}\:. \eeq
We consider the corresponding flow of the measure as described by the push-forward
measures~$\rho_t := (\Phi_t)_* \rho$. These measures will in general be different from~$\rho$
(more precisely, $\rho_t = \rho$ for all~$t$ if and only if the vector field is divergence-free).
But one can arrange the measure to remain unchanged by modifying the weight of the measure
with a function~$f_t$; that is,
\beq \label{rhodiffeo}
\rho = (\Phi_t)_* \big( f_t \,\rho) \qquad \text{for all~$t \in (-\delta, \delta)$}
\eeq
for a suitable function~$f_t \in C^\infty(M, \R^+)$ (here we again make use of our assumption
that~$M$ has a smooth manifold structure; see Definition~\ref{defsms}).
One verifies by direct computation that the function~$f_t$ agrees
infinitesimally with the divergence of~$v$; that is,
\beq \dot{f}_0(x) = \div v(x) \:. \eeq
Therefore, the inner solution~$\v=(\div v, v)$ is the infinitesimal generator of the transformation
on the right-hand side of~\eqref{rhodiffeo}.
In other words, inner solutions are infinitesimal generators of transformations of~$M$
which leave the measure~$\rho$ unchanged. Since the causal fermion system is defined in terms of the
measure~$\rho$, inner solutions do not change the causal fermion system,
but they merely describe symmetry transformations of the measure.
For the reader familiar with general relativity, it may be helpful to see inner solutions
as the analog of infinitesimal coordinate transformations, in which case the infinitesimal
change of the metric satisfies the Einstein equations (simply because curvature remains unchanged).

One application of inner solution is that they can be used to simplify the scalar components of jets.
We now explain the general procedure
(explicit examples will be worked out in Chapter~\ref{secexcvp}).
A preparatory question is which scalar components can be realized by inner solutions.
This question can be answered in great generality by applying Moser's theorem
(see, for example, \cite[Section~XVIII, \S2]{langDG} or straightforward generalizations
to non-compact manifolds). For simplicity, we here make additional assumptions that make
it possible to use hyperbolic methods.
\begin{Thm} \label{thmnoscalar} Assume that~$M \simeq \scrM^k$ has a smooth manifold structure.
Moreover, assume that~$\scrM$ is topologically of the form~$\scrM = \R \times \scrN$
with a manifold~$\scrN$.
Let~$a \in \Cisc(\scrM, \R)$ be a smooth function with spatially compact support
(meaning that, for all~$t \in \R$, the function~$a(t,.)$ is compactly supported in~$\scrN$).
Then there is a vector field~$v \in \Cisc(\scrM, T\scrM)$, again with spatially compact support, such
that the jet~$\v := (a, v)$ is an inner solution.
\end{Thm}
\Proof Our task is to solve the equation~$\div v = a$, which can be written equivalently as
\beq \label{diveq}
 \partial_j \big( h\, v^j \big) = h a \:.
\eeq
We first consider the case that~$a$ has compact support.
In order to solve the partial differential equation~\eqref{diveq}, on~$\scrN$ we choose
a complete Riemannian metric~$g_{\scrN}$ (such a metric exists according to~\cite{nomizu-ozeki}).
Moreover, on~$\scrM$ we choose the Lorentzian metric
\beq \label{Lorentz}
\dd s^2 = \dd t^2 - g_{\scrN} \:.
\eeq
Here the choice of the Riemannian metric~$g_{\scrN}$ is irrelevant, and
the arbitrariness in choosing this metric
corresponds to the fact that~\eqref{diveq} is an under-determined equation which admits
many different solutions.

Assume that~$\phi \in \Cisc(\scrM, \R)$ is a spatially compact
solution of the inhomogeneous wave equation
\beq \label{waveinhom}
\bigg( \frac{ha}{\sqrt{|\det g|}} \bigg)(x) = \Box \phi(x) := \frac{1}{\sqrt{|\det g|}} \frac{\partial}{\partial x^j} \bigg( \sqrt{|\det g|}\: g^{jk} \:
\partial_k \phi(x) \bigg)
\eeq
(the existence of such a solution will be shown in the next paragraph).
Then a direct computation shows that the vector field
\beq \label{vconstruct}
v^j := \frac{\sqrt{|\det g|}}{h} \: g^{jk}\: \partial_k \phi
\eeq
indeed satisfies~\eqref{diveq}
(note that, in view of~\eqref{hdef}, we may divide by~$h$ to again obtain a smooth
vector field with spatially compact support).

It remains to show that the inhomogeneous wave equation~\eqref{waveinhom} has solutions
of spatially compact support. Here we must anticipate results for
hyperbolic partial differential equations which will be treated in Chapter~\ref{secshs} later in this book.
A simple method of obtaining the desired solutions uses the 
existence of advanced and retarded Green's operators
denoted by~$S^\vee$ and~$S^\wedge$. In the case that the inhomogeneity~$a$ has
compact support, we can simply set~$\phi = S^\wedge (ha/\sqrt{|\det g|})$.
In the case that~$a$ merely has spatially compact support, we decompose it as
\beq a =a_+ + a_- \:, \eeq
where~$a_+$ is supported in the set~$\{t > 0\}$ and~$a_-$ is supported in the set~$\{t < 1 \}$.
Denoting the advanced and retarded Green's operators of the scalar wave equation
corresponding to the Lorentzian metric~\eqref{Lorentz}
by~$S^\vee$ and~$S^\wedge$, respectively, the function
\beq \phi := S^\wedge \bigg( \frac{h}{\sqrt{|\det g|}}\: a_+ \bigg) + S^\vee\bigg( \frac{h}{\sqrt{|\det g|}}\: a_- \bigg) \eeq
is a well-defined solution of the equation~$\Box \phi = h a$, which is smooth and has spatially compact
support. Therefore, we can again define the vector field~$v$ by~\eqref{vconstruct}.
This gives the result.
\QED

The result of this proposition can be used to change the scalar component
of a linearized solution arbitrarily. As a concrete example, let us consider a causal fermion system
describing an interacting system in Minkowski space which near spatial infinity is the vacuum. In this case, 
all the jets describing the interaction have spatially compact support.
Therefore, we can compensate the scalar components by corresponding inner solutions.
After doing so, the interacting system is described purely in terms of jets without scalar components.
We denote the corresponding jet space similar to~\eqref{Jinfdef} by
\nindex{aen@$\Gamma$ -- space of smooth jets without scalar component}%
\tindex{ff@$\Gamma$ -- space of smooth jets without scalar component}%
\beq \Gamma := \{0\} \oplus \Gamma^\infty(M, T\F) \eeq
(where~$\Gamma^\infty(M,T\F)$ again denotes the space of vector fields on~$M$
which admit a smooth extension to~$\F$). For clarity, we again point out that this simplification can be
made only if spacetime has a smooth manifold structure.

We finally note that, in the above setting, the scalar components of the jets may
be disregarded also for testing. Thus the restricted EL equations~\eqref{ELrestricted2} and the
linearized field equations~\eqref{linfield} can be written equivalently as
\begin{align}
D_u \ell |_M &= 0 \qquad \text{for all~$u \in \Gamma_0$} \label{ELvec} \\
D_u \int_\F \big( D_{1,v} + D_{2,v} \big) \L(x,y)\: \Diff\rho(y) &= 0
\qquad \text{for all~$u \in \Gamma_0$} \label{linvec}
\end{align}
(the subscript zero again denotes the compactly support jets).
In this way, the scalar components of jets can be left out completely for spacetimes
which have a smooth manifold structure and also satisfy the other assumptions of Theorem~\ref{thmnoscalar}.
We now sketch how to reduce to~\eqref{ELvec}
(for the linearized field equation~\eqref{linvec} one argues similarly;
the details of this reduction can be found in~\cite[Section~3]{nonlocal}).
Clearly, \eqref{ELrestricted2} implies~\eqref{ELvec}.
In order to show that also~\eqref{ELvec} implies~\eqref{ELrestricted2}, we assume that~\eqref{ELvec} holds.
Since the vector field~$u$ can be chosen arbitrarily at any given point~$x$,
it follows that the function~$\ell$ is constant on~$M$. After changing the parameter~$\s$
if necessary, the function~$\ell$ vanishes identically, implying~\eqref{ELrestricted2}.

\newpage
\section{Summary of the Linearized Field Equations} \label{secsumm2}

\tikz[remember picture,overlay] \draw [fill,blue!20] (current page.north west) rectangle +(0.03\paperwidth,-\paperheight);%

Following the procedure in Section~\ref{secsumm}, we now summarize the important objects of Chapters~\ref{secEL}
and~\ref{seclin}. One may think of this as a helpful bookmark for the commonly used structures. 
\vspace{2mm}
\begin{center}
	\begin{tabular*}{\textwidth}{p{0.23\linewidth} p{0.02\linewidth} p{0.65\linewidth}}   
		
		\textbf{Basic concept} & & \textbf{ Summary and Comments}\tabularnewline \hline 

		The Euler-La\-grange equations & & For technical simplicity, we here assume that
		\begin{itemize}
			\item[(1)] the measure~$\rho$ is locally finite in the sense that any~$x\in \F$ has a neighborhood~$U$ with~$\rho(U) < \infty$. 
			\item[(2)] the function~$\mathcal{L}(x,.)$ is $\rho$-integrable for all~$x\in \F$.
		\end{itemize}
	
Under these assumptions, we introduce the function~$\ell(x) = \int_\F \mathcal{L}(x,y) \dd \rho(y) - \s.$ The EL equations are formulated in our causal fermion system by~$\ell\vert_{supp\ \rho} \equiv \inf_\F \ell.$ \newline \tabularnewline

	\multicolumn{3}{p{0.95\linewidth}}{
		\textit{Remarks:}
		\begin{itemize}[leftmargin=2em]  \setlength\itemsep{1em}
			\item  The parameter~$\s$ can be chosen arbitrarily. For convenience, we always choose it such that 
			$\inf_\F \ell = 0$. For this choice, the EL equations simplify to~$\ell\vert_{\text{supp} \rho} \equiv 0$.
			\item Every minimizing measure~$\rho$ is a solution of the EL equations. 
\item One should keep in mind that not every solution of the EL equations is also a minimizer of the causal action. It is only a critical point of the variational principle. 
		\end{itemize}
	}
\tabularnewline\hline

Jet space~$\J$ & & It is useful to introduce a pair~$\u:=(a,u)$ with a real-valued function~$a \in C^{\infty}(M,\R)$ and a vector field~$u\in C^{\infty}(M, T\F)$ as a so-called \textit{jet}. The set of all jets is called the jet space~$\J = C^{\infty}(M,\R)\oplus C^{\infty}(M, T\F)$.
\newline \tabularnewline

\multicolumn{3}{p{0.95\linewidth}}{
	\textit{Remarks:}
	\begin{itemize}[leftmargin=2em]  \setlength\itemsep{1em}
		\item We often restrict attention to variations of the measure~$\rho$ in the form~$\rho_{\tau} = F_{\tau}^*(f_{\tau}\rho)$ with families of smooth functions~$(F_\tau) : M\to \F$ and~$(f_\tau) : M \to \R^+$. 
		Infinitesimally, these variations are described by the jets~$\v := \frac{\dd}{\dd\tau}(f_\tau, F_\tau)\vert_{\tau=0}$.
	\end{itemize}
}

	\end{tabular*}	
\end{center}

\newpage
\tikz[remember picture,overlay] \draw [fill,blue!20] (current page.north east) rectangle +(-0.03\paperwidth,-\paperheight);

\begin{center}
\begin{tabular*}{\textwidth}{p{0.175\linewidth} p{0.02\linewidth} p{0.7\linewidth}}   
\hline
The restricted EL equations & & The EL equations also contains information for points on~$\F$
which are far away from our spacetime~$M$. In order to get the connection to usual physical equations
formulated in spacetime, it suffices to restricting attention to a small neighborhood of points in~$M$. In the smooth setting considered in this book, this leads to the following two equations: 
\begin{itemize}
	\item[(1)] $\ell\vert_M \equiv 0$
	\item[(2)] $D\ell\vert_M \equiv 0$
\end{itemize}
These two equations can be combined
in a compact form by using a jet~$\u = (a,u)$ to define~$\nabla_\u \ell(x) := a(x)
\ell(x) + (D_u \ell)(x)$. Therefore, the \textit{restricted EL equations} can be written as~$ \nabla_\u \ell(x)\vert_M = 0$ for all~$\u \in \J$. 
\newline \tabularnewline

\multicolumn{3}{p{0.95\linewidth}}{
	\textit{Remarks:}
	\begin{itemize}[leftmargin=2em]  \setlength\itemsep{1em}
		\item We also refer to solutions of the restricted EL equations as \textit{critical measures}.
	\end{itemize}
}
\tabularnewline\hline

The linearized field equations & &  The linearized field equations read \newline $\la \u, \Delta \v \ra(x) := \nabla_\u \bigg( \int_\F \big( \nabla_{1,\v} + \nabla_{2,\v} \big) \L(x,y)\: \Diff\rho(y) - 
\nabla_\v \:\s \bigg) = 0$
\newline for all~$\u \in \J$ an~$x \in M$. All~$\v \in \J$ that satisfy this equation are called \textit{linearized solutions}. The vector space of all linearized solutions is denoted by~$\Jlin$. 
\newline \tabularnewline

\multicolumn{3}{p{0.95\linewidth}}{
	\textit{Remarks:}
	\begin{itemize}[leftmargin=2em]  \setlength\itemsep{1em}
		\item For simplicity of presentation, in this book, we only consider the {\em{smooth setting}}
		where we assume that the Lagrangian is smooth on~$\F \times \F$.
		\item The EL equations are nonlinear equations for~$\rho$. They are simplified by linearization, giving the linearized field equations.
		\item Often we use the shorthand notation~$\Delta\v = 0$ for the linearized field equation. One also can include an inhomogeneity~$\w$ by writing~$\Delta\v = \w$.
		\item In order to avoid questions of differentiability of the jets, we use a formalism where the jets are never differentiated. 
		\item Solutions of the linearized field equations can be used to obtain solutions for the full EL equations by means of a perturbation expansion.

	\end{itemize}
}

\end{tabular*}

\end{center}

\newpage
\tikz[remember picture,overlay] \draw [fill,blue!20] (current page.north west) rectangle +(0.03\paperwidth,-\paperheight);

\begin{center}
	\begin{tabular*}{\textwidth}{p{0.2\linewidth} p{0.05\linewidth} p{0.65\linewidth}}   
		\hline

		The commutator jet & &   Let~$\scrA$ be a symmetric operator on~$\H$. We get a family of unitary transformations by~$\scrU_\tau := \exp(\cI \tau \scrA)$. Jets of the form~$\v := (0,v) \in \Jlin \qquad \text{with} \qquad v(x) = \frac{\dd}{\dd\tau} \big( \scrU_\tau \,x\, \scrU_\tau^{-1} \big) \big|_{\tau=0} 
		= \cI \big[\scrA, x \big]$ are referred to as commutator jets. They are solutions to the linearized field equations. 
		\newline \tabularnewline
		
		\multicolumn{3}{p{0.95\linewidth}}{
			\textit{Remarks:}
			\begin{itemize}[leftmargin=2em]  \setlength\itemsep{1em}
				\item This is the first special class of solutions of the linearized field equations which we consider. 
				
			\end{itemize}
		}
		
		\tabularnewline\hline
		
		The inner solutions & & If our spacetime has a smooth manifold structure, the measure can be written in terms of the Lebesgue measure by~$\Diff\rho = h(x)\: \Diff^kx$  with~$h \in C^\infty(\scrM^k, \R^+)$. Now we can formulate the covariant divergence of a vector field~$v$ by~$\div v = \frac{1}{h}\: \partial_j \big( h\, v^j \big)$. We refer to the jets of the form~$\v = (\div v, v)$ with~$v \in \Gamma(M, TM)$ as \textbf{inner solutions}. All inner solutions with compact support are solutions of the linearized field equations.  
		\newline \tabularnewline
		
		\multicolumn{3}{p{0.95\linewidth}}{
			\textit{Remarks:}
			\begin{itemize}[leftmargin=2em]  \setlength\itemsep{1em}
				\item This is the second class of special solutions. 
				\item The space of all inner solutions is denoted by~$\Jin$.  
				
			\end{itemize}
		}
	\end{tabular*}

\end{center}

\newpage
\section{Exercises}

\begin{Exercise} \label{exsupp1} {\em{
Let~$F : \F \rightarrow \F$ be continuous and~$\rho$ a measure on~$\F$.
\sindex{measure!push-forward}%
Show that
\beq \supp F_* \rho = \overline{F(\supp \rho)} \:. \eeq
{\em{Hint:}} Recall the definition of the support of a measure~\eqref{suppdef} and use that the
preimage of an open set under a continuous mapping is open.
}} \end{Exercise}

\begin{Exercise} \label{exsupp2} {\em{
\bitem \item[(a)] Assume that~$\F$ is locally compact. Moreover, assume that~$F \in C^0([0, \delta) \times M, \F)$
is continuous and that its preimage of any compact set is compact.
Then, for any~$y \not \in M$ there is~$\tau_0 \in (0, \delta)$ such that
\beq y \not \in \supp \tilde{\rho}_\tau \qquad \text{for all~$\tau \in [0, \tau_0]$} \eeq
(where~$\tilde{\rho}_\tau$ are again the measures~\eqref{Ffvary})
{\em{Hint:}} Use the result of Exercise~\ref{exsupp1}.
\item[(b)] Show that this result remains valid for the variation~\eqref{rhotildea}
with a finite number of subsystems.
\item[(c)] What happens for an infinite number of subsystems?
Also, is the assumption necessary that the preimage of a compact set under~$F$ is compact?
\eitem
}} \end{Exercise}

\begin{Exercise} {{(Linearization of nonlinear partial differential equations)}} {\em{
\sindex{linearization!of nonlinear partial differential equation}%
In this exercise, you are given two \textit{non-linear} partial differential equations with corresponding (soliton) solutions. Check that the functions~$\phi$ do indeed solve the equations. Then try to figure out what it means to \textit{linearize} the equations around the given solutions and do it. 
\bitem
\item[(a)] The \textit{sine-Gordon equation} of velocity~$v\in (-1,1)$:
\beq
 \partial_{tt}\phi-\partial_{xx}\phi+\sin\phi=0,\quad \phi(t,x)=4\arctan\left(\E^{\frac{x-vt}{\sqrt{1-v^2}}}\right).
\eeq
\item[(b)] The \textit{Korteweg-de-Vries equation} of unit speed:
\beq
 \partial_t \phi+6\,\phi\,\partial_x\,\phi+\partial_{xxx}\phi=0,\quad \phi(t,x)=\frac{1}{2}\,\mathrm{sech}^2\left(\frac{x-vt}{{2}}\right).
\eeq
\textit{Hint:} You may use the following identities,
\begin{align}
\sin(4\arctan(x)) &= -4\frac{x^3-x}{(1+x^2)^2} \\
\tanh(x)-\tanh^3(x) &= \mathrm{sech}^2(x)\tanh(x).
\end{align}
\eitem
}} \end{Exercise}

\begin{Exercise} {{(Linearized fields on the sphere)}} {\em{
Let~$\rho$ be a minimizing measure of the causal variational principle of the sphere
as introduced in Section~\ref{seccvpsphere}
(for example, the octahedron in Exercise~\ref{excrit}).
\bitem
\item[(a)] Let~$v$ be the vector field~$\partial/\partial \varphi$ (where~$\varphi$
is the azimuth angle). Show that~$\v = (0, v)$ is a solution of the linearized field equations.
{\em{Hint:}} One can use the fact that the causal variational principle is rotationally symmetric.
\item[(b)] Show that~$\v$ can be written as a commutator jet; that is, in analogy to~\eqref{jvdef},
\beq v(x) = \cI \,\big[c \sigma^3, F(x) \big] \:, \eeq
where~$F : S^2 \subset \R^3 \rightarrow \F$ is the mapping in~\eqref{Fspheredef}.
Compute the constant~$c$.
\eitem
}} \end{Exercise}

\begin{Exercise} {{(Linearized fields for the causal variational principle on~$\R$)}} {\em{
We return to the causal variational principles on~$\R$ introduced in Exercise~\ref{excvpR}.
Let~$\rho=\delta$ be the unique minimizer.
\bitem
\item[(a)] Show that the jet~$\v=(0, v)$ with the vector field~$v=\partial_x$
is a solution of the linearized field equations for the causal variational principle
corresponding to~$\L_4$.
\item[(b)] Show that the jet~$\v=(0, v)$ from~(a) does {\em{not}} satisfy the
linearized field equations for the causal variational principle
corresponding to~$\L_2$.
\eitem
}} \end{Exercise}

\begin{Exercise} \label{excvpS1lin} {{(Linearized fields for the causal variational principle on~$S^1$)}} {\em{
We return to the causal variational principle on~$\R$ introduced in Exercise~\ref{excvpS1}.
Let~$\rho$ be a minimizing measure~\eqref{rhominS1} for~$0 < \tau < 1$.
\bitem
\item[(a)] Show that the jet~$\v=(0,v)$ with the vector field~$v=\partial_\varphi$
satisfies the linearized field equations.
{\em{Hint:}} One can use the fact that the variational principle is rotationally symmetric.
\item[(b)] Show that the jet~$\v=(b,0)$ with~$b(\phi_0)=-b(\phi_0+\pi)$ is a solution
of the linearized field equations.
{\em{Hint:}} Use that the causal action is independent of the parameter~$\tau$.
\item[(c)] Show that every solution of the linearized field equations is
a linear combination of the linearized fields in~(a) and~(b).
\eitem
}} \end{Exercise}

\begin{Exercise} {{(The commutator of commutator jets)}}
\sindex{jet!commutator}%
{\em{ For any symmetric operator~$\scrA$ of finite rank, the
commutator vector field~$\Comm(\scrA)$ is defined by~$\Comm(\scrA)(x) = \cI \big[\scrA, x \big]$
(see~\eqref{jvdef}). It is a vector field on~$M$. For two vector fields~$u,v$ on~$M$, their commutator is defined
by~$[u,v](f) = u(v(f))-v(u(f))$ (with~$f$ any smooth function on~$M$). 
Show that the commutator of two commutator vector fields is again a commutator vector field and
\beq \big[ \Comm(\scrA), \Comm({\mathscr{B}}) \big] = -\Comm\big(\cI [\scrA, {\mathscr{B}}] \big) \:. \eeq
{\em{Hint:}} The proof can be found in~\cite[Lemma~A.2]{dirac}.
}} \end{Exercise}

\chapter{Surface Layer Integrals and Conservation Laws} \label{secOSI}
\sindex{surface layer integral}%
In this chapter, we introduce {\em{surface layer integrals}} as an adaptation
of surface integrals to causal fermion systems and causal variational principles.
The mathematical structure of a surface layer integral fits nicely with the
analytic structures (namely, the EL equations and the
linearized field equations as introduced in Chapters~\ref{secEL} and~\ref{seclin}).
This will become apparent in {\em{conservation laws}}, which generalize Noether's theorem and
the symplectic form to the setting of causal variational principles.
Moreover, we shall introduce a so-called {\em{nonlinear surface layer integral}}, which
makes it possible to compare two measures~$\rho$ and~$\tilde{\rho}$ at a given time.
Finally, we will explain how {\em{two-dimensional}} surface integrals can be described by
surface layer integrals.

\section{The Concept of a Surface Layer Integral} \label{secosiintro}
\sindex{surface layer integral!general concept}%
In daily life, we experience space and objects therein. These objects
are usually described by densities, and integrating these densities over space gives
particle numbers, charges, the total energy, etc.
In mathematical terms, the densities are typically described as the normal components
of vector fields on a Cauchy surface, and conservation laws express that the values of these
integrals do not depend on the choice of the Cauchy surface; that is,
\beq \label{conserve}
\int_{\scrN_1} J^k \nu_k\: \dd\mu_{\scrN_1}(x) = \int_{\scrN_2} J^k \nu_k\: \dd\mu_{\scrN_2}(x) \:,
\eeq
where~$\scrN_1$ and~$\scrN_2$ are two Cauchy surfaces,
$\nu$ is the future-directed normal, and~$\dd\mu_{\scrN_{1\!/\!2}}$ is the induced volume measure.

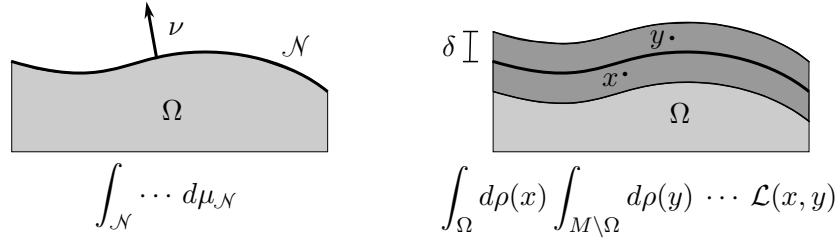
\begin{figure}
\psscalebox{1.0 1.0} % Change this value to rescale the drawing.
{
\begin{pspicture}(0,-1.511712)(10.629875,1.511712)
\definecolor{colour0}{rgb}{0.8,0.8,0.8}
\definecolor{colour1}{rgb}{0.6,0.6,0.6}
\pspolygon[linecolor=black, linewidth=0.002, fillstyle=solid,fillcolor=colour0](6.4146066,0.82162136)(6.739051,0.7238436)(6.98794,0.68384355)(7.312384,0.66162133)(7.54794,0.67939913)(7.912384,0.7593991)(8.299051,0.8705102)(8.676828,0.94162136)(9.010162,0.9549547)(9.312385,0.9371769)(9.690162,0.8571769)(10.036829,0.7371769)(10.365718,0.608288)(10.614607,0.42162135)(10.614607,-0.37837866)(6.4146066,-0.37837866)
\pspolygon[linecolor=black, linewidth=0.002, fillstyle=solid,fillcolor=colour1](6.4146066,1.2216214)(6.579051,1.1616213)(6.770162,1.1127324)(6.921273,1.0905102)(7.103495,1.0816213)(7.339051,1.0549546)(7.530162,1.0638436)(7.721273,1.0993991)(7.8857174,1.1393992)(8.10794,1.2060658)(8.299051,1.2549547)(8.512384,1.3038436)(8.694607,1.3260658)(8.890162,1.3305103)(9.081273,1.3393991)(9.379051,1.3216213)(9.659051,1.2593992)(9.9746065,1.1705103)(10.26794,1.0460658)(10.459051,0.94384354)(10.614607,0.82162136)(10.610162,0.028288014)(10.414606,0.1660658)(10.22794,0.26828802)(10.010162,0.37051025)(9.663495,0.47273245)(9.356829,0.53051025)(9.054606,0.548288)(8.814607,0.54384357)(8.58794,0.5171769)(8.387939,0.48162135)(8.22794,0.44162133)(7.90794,0.34828803)(7.6946063,0.29939914)(7.485718,0.26828802)(7.272384,0.26828802)(7.02794,0.28162134)(6.82794,0.3171769)(6.676829,0.35273245)(6.543495,0.38828802)(6.4146066,0.42162135)
\pspolygon[linecolor=black, linewidth=0.002, fillstyle=solid,fillcolor=colour0](0.014606438,0.82162136)(0.3390509,0.7238436)(0.5879398,0.68384355)(0.9123842,0.66162133)(1.1479398,0.67939913)(1.5123842,0.7593991)(1.8990508,0.8705102)(2.2768288,0.94162136)(2.610162,0.9549547)(2.9123843,0.9371769)(3.290162,0.8571769)(3.6368287,0.7371769)(3.9657176,0.608288)(4.2146063,0.42162135)(4.2146063,-0.37837866)(0.014606438,-0.37837866)
\psbezier[linecolor=black, linewidth=0.04](6.4057174,0.8260658)(7.6346064,0.45939913)(7.8634953,0.8349547)(8.636828,0.92828804)(9.410162,1.0216213)(10.165717,0.7927325)(10.614607,0.42162135)
\psbezier[linecolor=black, linewidth=0.04](0.005717549,0.8260658)(1.2346064,0.45939913)(1.4634954,0.8349547)(2.2368286,0.92828804)(3.0101619,1.0216213)(3.7657175,0.7927325)(4.2146063,0.42162135)
\rput[bl](2.0101619,0.050510235){$\Omega$}
\rput[bl](8.759051,0.0016213481){\normalsize{$\Omega$}}
\psline[linecolor=black, linewidth=0.04, arrowsize=0.09300000000000001cm 1.0,arrowlength=1.7,arrowinset=0.3]{->}(1.9434953,0.85495466)(1.8057176,1.6193991)
\rput[bl](2.0946064,1.1705103){$\nu$}
\psbezier[linecolor=black, linewidth=0.02](6.4146066,0.42384356)(7.6434956,0.057176903)(7.872384,0.43273246)(8.645718,0.52606577)(9.419051,0.61939913)(10.174606,0.39051023)(10.623495,0.019399125)
\psbezier[linecolor=black, linewidth=0.02](6.410162,1.2193991)(7.639051,0.8527325)(7.86794,1.228288)(8.6412735,1.3216213)(9.414606,1.4149547)(10.170162,1.1860658)(10.619051,0.8149547)
\rput[bl](8.499051,0.9993991){\normalsize{$y$}}
\rput[bl](7.8657174,0.49273247){\normalsize{$x$}}
\psdots[linecolor=black, dotsize=0.06](8.170162,0.65273243)
\psdots[linecolor=black, dotsize=0.06](8.796828,1.1327325)
\psline[linecolor=black, linewidth=0.02](6.1146064,1.2216214)(6.103495,0.82162136)
\rput[bl](5.736829,0.8993991){\normalsize{$\delta$}}
\rput[bl](3.6146064,0.888288){$\scrN$}
\rput[bl](1.1146064,-1.4117119){$\displaystyle \int_\scrN \cdots\, \dd\mu_\scrN$}
\rput[bl](5.7146063,-1.511712){$\displaystyle \int_\Omega \dd\rho(x) \int_{M \setminus \Omega} \dd\rho(y)\: \cdots\:\L(x,y)$}
\psline[linecolor=black, linewidth=0.02](6.0146065,1.2216214)(6.2146063,1.2216214)
\psline[linecolor=black, linewidth=0.02](6.0146065,0.82162136)(6.2146063,0.82162136)
\end{pspicture}
}
\caption{A surface integral and a corresponding surface layer integral.}
\label{fignoether1}
\end{figure}

In the setting of causal variational principles, surface integrals like~\eqref{conserve}
are undefined. Instead, one considers so-called {\em{surface layer integrals}}, which we now introduce.
In general terms, a surface layer integral is a double integral of the form
\beq \label{intdouble}
\int_\Omega \bigg( \int_{M \setminus \Omega} (\cdots)\: \L(x,y)\: \dd \rho(y) \bigg)\, \dd \rho(x) \:,
\eeq
where one variable is integrated over a subset~$\Omega \subset M$, and the other
variable is integrated over the complement of~$\Omega$. Here~$(\cdots)$ stands for a
differential operator acting on the Lagrangian to be specified below.
In order to explain the basic idea, we begin with the additional
assumption that the Lagrangian is of {\em{short range}} in the following sense.
We let~$d \in C^0(M \times M, \R^+_0)$ be a suitably chosen distance function on~$M$. Then the
assumption of short range can be quantified by demanding that~$\L$ should vanish
on distances larger than~$\delta$; that is,
\beq \label{shortrange}
d(x,y) > \delta \quad \Longrightarrow \quad \L(x,y) = 0 \:.
\eeq
Under this assumption,
the surface layer integral~\eqref{intdouble} only involves pairs~$(x,y)$ of distance at most~$\delta$,
with~$x$ lying in~$\Omega$ and~$y$ lying in the complement~$M \setminus \Omega$.
As a consequence, the integral only involves points in a layer around the boundary of~$\Omega$
of width~$\delta$; that is,
\beq x, y \in B_\delta \big(\partial \Omega \big) \:. \eeq
Therefore, a double integral of the form~\eqref{intdouble} can be regarded as an approximation
of a surface integral on the length scale~$\delta$, as shown in Figure~\ref{fignoether1}.
In the setting of causal variational principles, such surface layer integrals take the role of surface integrals.

We point out that the causal Lagrangian is {\em{not}} of short range in the sense~\eqref{shortrange}.
But it decays on a length scale that typically coincides with the Compton scale~$1/m$
(where~$m$ denotes the rest mass of the Dirac particles).
With this in mind, the above consideration and the qualitative picture of a surface layer integral
in Figure~\ref{fignoether1} apply to the causal action principle as well.

\section{A Noether-Like Theorem} \label{secnoether}
\sindex{Noether's theorem}
\sindex{Noether-like theorem|textbf}%
In modern physics, the connection between symmetries and conservation laws
is of central importance. For continuous symmetries, this connection is made mathematically precise
by Noether's theorem (see~\cite{noetheroriginal} or the textbooks~\cite[Section~13.7]{goldstein},
\cite[Chapter~III]{barutbook}). As shown in~\cite{noether}, the connection between
symmetries and conservation laws can be extended to the setting of causal variational principles.
As we shall see, both the statement and the proof are quite different from the classical Noether theorem;
this is why we refer to our result as a {\em{Noether-like theorem}}.

The first step is to formulate a symmetry condition for the 
Lagrangian~$\L(x,y)$ of a causal variational principle. Similar to the procedure in
Section~\ref{secsymmcrit}, one could describe the symmetry by a
group of diffeomorphisms. For our purposes, the correct setting
would be to consider a one-parameter group of diffeomorphisms~$\Phi_\tau$ on~$\F$; that is,
\beq \label{try}
\Phi : \R \times \F \rightarrow \F \qquad \text{with} \qquad
\Phi_\tau \Phi_{\tau'} = \Phi_{\tau+\tau'}
\eeq
(we usually write the first argument as a subscript, that is, $\Phi_\tau(x) \equiv \Phi(\tau,x)$).
The symmetry condition could be imposed by demanding that the Lagrangian
be invariant under this one-parameter group in the sense that
\beq \label{symmF1}
\L(x,y) = \L \big( \Phi_\tau(x), \Phi_\tau(y) \big) \qquad \text{for all~$\tau \in \R$
and~$x, y \in \F$\:.}
\eeq
It turns out that this condition is unnecessarily strong for two reasons. First, it suffices to consider families that
are defined locally for~$\tau \in (-\tau_{\max}, \tau_{\max})$. Second, the mapping~$\Phi$ does not
need to be defined on all of~$\F$. Instead, it is more appropriate to impose the symmetry condition
only on spacetime~$M \subset \F$.
This leads us to  consider instead of~\eqref{try} a mapping
\beq \label{Phidef}
\Phi : (-\tau_{\max}, \tau_{\max}) \times M \rightarrow \F
\qquad \text{with} \qquad \Phi_0 = \text{id}_M \:.
\eeq
We refer to~$\Phi_\tau$ as a {\bf{variation}} of~$M$ in~$\F$.
\sindex{variation!of spacetime~$M$ in~$\F$}%
\nindex{aeo@$\Phi_\tau$ -- variation of spacetime}%
Next, we need to specify what we mean by ``smoothness'' of this variation.
This is a subtle point because, as explained in the example of the causal variational principle
on the sphere in Section~\ref{seccvpsphere},
the support of a minimizing measure will in general be singular. Moreover, the
function~$\ell$ defined by~\eqref{elldef} in general will only be Lipschitz continuous.
Our Noether-like theorems only require that this function be differentiable in the
direction of the variations:
\begin{Def} \label{defsymm} A variation~$\Phi_\tau$ of the form~\eqref{Phidef} is
{\bf{continuously differentiable}} if the composition
\beq \ell \circ \Phi \::\: (-\tau_{\max}, \tau_{\max}) \times M \rightarrow \R \eeq
is continuous and if its partial derivative~$\partial_\tau (\ell \circ \Phi)$ exists
and is continuous.
\end{Def} \noindent
The next question is how to adapt the symmetry condition~\eqref{symmF1} to the mapping~$\Phi$
defined only on~$(-\tau_{\max}, \tau_{\max}) \times M$.
This is not obvious because setting~$\tilde{x} = \Phi_\tau(x)$ and using the group property,
the condition~\eqref{symmF1} can be written equivalently as
\beq \label{symmF2}
 \L \big( \Phi_{-\tau}(\tilde{x}), y \big) = \L \big( \tilde{x},\Phi_\tau(y) \big) \qquad \text{for all~$\tau \in \R$
 and~$\tilde{x}, y \in \F$\:.}
\eeq
But if we restrict attention to pairs~$x,y \in M$, the equations in~\eqref{symmF1} and~\eqref{symmF2}
are different. For the general mathematical formulation, it is preferable to
weaken the condition~\eqref{symmF1} starting from the expression in~\eqref{symmF2}.

\begin{Def} \label{defsymmlagr} A variation~$\Phi_\tau$ of the form~\eqref{Phidef} is a {\bf{symmetry
of the Lagrangian}} if
\sindex{symmetry!of the causal Lagrangian|textbf}%
\beq \label{symmlagr}
\L \big( x, \Phi_\tau(y) \big) = \L \big( \Phi_{-\tau}(x), y \big)
\qquad \text{for all~$\tau \in (-\tau_{\max}, \tau_{\max})$
and~$x, y \in M$\:.}
\eeq
\end{Def}

We now state and prove our Noether-like theorem.
\begin{Thm} \label{thmsymmlag} Let~$\rho$ be a critical measure and~$\Phi_\tau$ a 
continuously differentiable symmetry of the Lagrangian. Then, for any compact subset~$\Omega \subset M$,
\sindex{conservation law for surface layer integral}%
\beq
\label{conservationLagrEq}
\frac{\dd}{\dd \tau} \int_\Omega \dd \rho(x) \int_{M \setminus \Omega} \dd \rho(y)\:
\Big( \L \big( \Phi_\tau(x),y \big) - \L \big( \Phi_{-\tau}(x), y \big) \Big) \Big|_{\tau=0} = 0 \:.
\eeq
\end{Thm}
\Proof Integrating~\eqref{symmlagr} 
over~$\Omega \times \Omega$ gives
\begin{align}
0 &= \iint_{\Omega \times \Omega} \Big(
\L \big( x, \Phi_\tau(y) \big) - \L \big( \Phi_{-\tau}(x), y \big) \Big)\, \dd\rho(x)\, \dd\rho(y) \notag \\
&= \iint_{\Omega \times \Omega} \Big(
\L \big( \Phi_\tau(x),y \big) - \L \big( \Phi_{-\tau}(x), y \big) \big) \Big)\, \dd\rho(x)\, \dd\rho(y) \notag \\
&=\int_{\Omega} \dd\rho(x) \int_M \dd\rho(y)\: \chi_\Omega(y)\: \Big(
\L \big( \Phi_\tau(x),y \big) - \L \big( \Phi_{-\tau}(x), y \big) \big) \Big)\:,
\end{align}
where, in the first step, we used the Lagrangian is symmetric in its two arguments and that
the integration range is symmetric in~$x$ and~$y$.
We rewrite the characteristic function~$\chi_\Omega(y)$ as~$1 - (1-\chi_\Omega(y))$, multiply out and use the
definition of~$\ell$, \eqref{elldef}. We thus obtain
\begin{align}
0 &=\int_\Omega
\Big(  \ell \big( \Phi_\tau(x)\big) - \ell\big( \Phi_{-\tau}(x)) \Big) \,\dd\rho(x) \notag \\
&\quad -\int_\Omega \dd\rho(x) \int_M  \dd\rho(y)\:\chi_{M \setminus \Omega}(y)\:
\Big( \L \big( \Phi_\tau(x),y \big) - \L \big( \Phi_{-\tau}(x), y \big) \Big) \:.
\end{align}
We thus obtain the identity
\beq \begin{split}
\int_\Omega &\dd\rho(x) \: \int_{M \setminus \Omega} \dd\rho(y)\:
\Big( \L \big( \Phi_\tau(x), y \big) - \L \big(\Phi_{-\tau}(x),  y \big) \Big) \\
&= \int_\Omega \Big( \ell \big( \Phi_\tau(x) \big)  - \ell \big( \Phi_{-\tau}(x) \big) \Big)\: \dd\rho(x) \:.
\end{split} \label{form}
\eeq
Using that~$\ell(\Phi_\tau(x))$ is continuously differentiable (see Definition~\ref{defsymm})
and that~$\Omega$ is compact, we conclude that
the right-hand side of this equation is differentiable at~$\tau=0$. Moreover,
we are allowed to interchange the $\tau$-differentiation with integration.
The EL equations~\eqref{EL2} imply that
\beq %\label{lfirst}
\frac{\dd}{\dd\tau} \ell \big( \Phi_{\tau}(x) \big) \Big|_{\tau=0} = 0 =
\frac{\dd}{\dd\tau} \ell \big( \Phi_{-\tau}(x) \big) \Big|_{\tau=0} \:. \eeq
Hence, the right-hand side of~\eqref{form} is differentiable at~$\tau=0$,
and the derivative vanishes. This gives the result.
\QED

This theorem requires a detailed explanation. We first clarify the connection to surface layer integrals.
To this end, let us assume for technical simplicity that~$\Phi_\tau$ and the Lagrangian are differentiable in the sense that the derivatives
\beq \label{differentiable}
\frac{\dd}{\dd\tau} \Phi_\tau(x)\big|_{\tau=0} =: u(x) \qquad \text{and} \qquad
\frac{\dd}{\dd\tau} \L\big( \Phi_\tau(x),y \big) \big|_{\tau=0}
\eeq
exist for all~$x, y \in M$ and are continuous on~$M$ and~$M \times M$, respectively.
Then one may exchange differentiation and integration in~\eqref{conservationLagrEq}
and apply the chain rule to obtain
\beq \int_\Omega \dd\rho(x) \int_{M \setminus \Omega} \dd\rho(y)\: D_{1,u} \L(x,y) = 0 \:, \eeq
where~$D_{1,u}$ is the partial derivative at~$x$ in the direction of the vector field~$u(x)$.
This expression is a surface layer integral as in~\eqref{intdouble}.
In general, the derivatives in~\eqref{differentiable} need {\em{not}} exist, because
we merely imposed the weaker differentiability assumption of Definition~\ref{defsymm}.
In this case, the statement of the theorem implies that the derivative 
of the integral in~\eqref{conservationLagrEq} exists and vanishes.

We next explain the connection to conservation laws.
Let us assume that~$M$ admits a sensible notion of ``spatial infinity'' and
that the vector field~$\partial_\tau \Phi \in \Gamma(M, T\F)$ has suitable decay properties at spatial infinity.
Then one can choose a sequence~$\Omega_n \subset M$ of compact sets
that form an exhaustion of a set~$\Omega$ which extends up to spatial infinity
(see Figure~\ref{figjet1}~(a) and~(b)).
\begin{figure}
% \usepackage[usenames,dvipsnames]{pstricks}
% \usepackage{epsfig}
% \usepackage{pst-grad} % For gradients
% \usepackage{pst-plot} % For axes
% \usepackage[space]{grffile} % For spaces in paths
% \usepackage{etoolbox} % For spaces in paths
% \makeatletter % For spaces in paths
% \patchcmd\Gread@eps{\@inputcheck#1 }{\@inputcheck"#1"\relax}{}{}
% \makeatother
% \psscalebox{1.0 1.0} % Change this value to rescale the drawing.
{
\begin{pspicture}(0,-1.0682992)(16.27295,1.0682992)
\definecolor{colour0}{rgb}{0.8,0.8,0.8}
\pspolygon[linecolor=colour0, linewidth=0.02, fillstyle=solid,fillcolor=colour0](9.348506,-0.5327436)(9.348506,1.0361453)(9.63295,0.92947865)(9.926284,0.8805898)(10.339617,0.84058976)(10.837395,0.85836756)(11.326283,0.9072564)(11.668506,0.93836755)(11.988505,0.96058977)(12.184061,0.94281197)(12.801839,0.9961453)(13.241839,1.0050342)(13.259617,-0.537188)(12.948505,-0.50163245)(12.539617,-0.50163245)(12.11295,-0.48385468)(11.748506,-0.497188)(11.459617,-0.47496578)(11.246284,-0.457188)(10.9529505,-0.46163246)(10.717395,-0.4349658)(10.406283,-0.42163244)(9.997395,-0.39941022)(13.259617,-0.5327436)
\pspolygon[linecolor=colour0, linewidth=0.02, fillstyle=solid,fillcolor=colour0](4.5396166,-0.27941024)(4.530728,0.9872564)(4.815172,0.8805898)(5.1085057,0.83170086)(5.530728,0.7961453)(5.9573946,0.7872564)(6.4596167,0.8139231)(6.819617,0.84947866)(7.0773945,0.8761453)(7.3662834,0.8939231)(7.9840612,0.94725645)(8.424061,0.95614535)(8.424061,-0.31052133)(8.03295,-0.33274356)(7.624061,-0.3460769)(7.3040614,-0.38163245)(7.0373945,-0.40385467)(6.779617,-0.4438547)(6.468506,-0.47496578)(6.1885056,-0.50163245)(5.8773947,-0.51052135)(5.55295,-0.497188)(5.179617,-0.4482991)(4.8462834,-0.377188)
\pspolygon[linecolor=colour0, linewidth=0.02, fillstyle=solid,fillcolor=colour0](0.03295012,0.12947866)(0.19739456,0.24503422)(0.35739458,0.3250342)(0.57517236,0.43614534)(0.9485057,0.57836753)(1.3885057,0.6939231)(1.8151723,0.74725646)(2.121839,0.7605898)(2.43295,0.742812)(2.7973945,0.6539231)(3.121839,0.542812)(3.441839,0.3917009)(3.681839,0.25836754)(3.5973945,0.13836755)(3.39295,0.013923102)(3.130728,-0.123854674)(2.8551724,-0.24385467)(2.4773946,-0.31941023)(2.161839,-0.36829913)(1.8151723,-0.38163245)(1.5173945,-0.37274358)(1.081839,-0.30607688)(0.62850565,-0.16385467)(0.281839,-0.0038546752)
\rput[bl](1.8151723,0.025034213){$\Omega_n$}
\rput[bl](6.0973945,0.12281199){\normalsize{$\Omega$}}
\psbezier[linecolor=black, linewidth=0.04](0.02406123,0.12281199)(0.7173699,0.6167855)(1.5001423,0.76261884)(2.0707278,0.7672564358181418)(2.6413136,0.77189404)(3.0712237,0.6218133)(3.7240613,0.242812)
\psbezier[linecolor=black, linewidth=0.04](0.009616784,0.12947866)(0.52181435,-0.15099226)(0.95347565,-0.34627005)(1.5962834,-0.38941023084852644)(2.2390912,-0.4325504)(3.124557,-0.3104089)(3.709617,0.24947865)
\psbezier[linecolor=black, linewidth=0.04](4.525172,-0.297188)(5.046259,-0.4843256)(5.7001424,-0.5373812)(6.1696167,-0.518299119737415)(6.6390915,-0.49921706)(7.315668,-0.33707556)(8.438506,-0.3282991)
\psbezier[linecolor=black, linewidth=0.04](4.520728,0.9917009)(5.0418143,0.8045633)(5.5401425,0.7915077)(5.991839,0.783923102484807)(6.443536,0.7763385)(7.3112235,0.95181334)(8.434061,0.96058977)
\psbezier[linecolor=black, linewidth=0.04](9.338506,1.0494787)(9.859592,0.8623411)(10.375698,0.85372996)(10.751839,0.8683675469292492)(11.12798,0.88300514)(12.129002,1.0095911)(13.251839,1.0183675)
\rput[bl](10.97295,0.10947866){\normalsize{$\Omega_N$}}
\rput[bl](1.7040613,-1.0682992){(a)}
\rput[bl](6.1751723,-1.0638547){(b)}
\rput[bl](11.059617,-1.0638547){(c)}
\rput[bl](7.2085056,-0.23718801){\normalsize{$N_1$}}
\rput[bl](7.2351723,0.482812){\normalsize{$N_2$}}
\rput[bl](12.319616,0.5850342){\normalsize{$N$}}
\end{pspicture}
}
\caption{Choices of spacetime regions: Lens-shaped region (a), region between to Cauchy surfaces (b) and past of a Cauchy surface (c).}
\label{figjet1}
\end{figure}%
Considering the surface layer integrals for~$\Omega_n$ and passing to the limit,
one concludes that also the surface layer integral corresponding to~$\Omega$ vanishes.
Let us assume that the boundary~$\partial \Omega$ has two components~$N_1$
and~$N_2$ (as in Figure~\ref{figjet1}~(b)).
Then, the above theorem implies that the surface layer integrals over~$N_1$ and~$N_2$ coincide
(where the surface layer integral over~$N$ is defined as the surface layer integral
corresponding to a set~$\Omega_N$ with~$\partial \Omega_N = N$ as shown in Figure~\ref{figjet1}~(c)).
In other words, the quantity
\beq \frac{\dd}{\dd\tau} \int_{\Omega_N} \dd\rho(x) \int_{M \setminus {\Omega_N}} \dd\rho(y)\:
\Big( \L \big( \Phi_\tau(x),y \big) - \L \big( \Phi_{-\tau}(x), y \big) \Big) \Big|_{\tau=0} \eeq
is well-defined and independent of the choice of~$N$.
In this setting, the surfaces~$N$ can be interpreted as Cauchy surfaces,
and the conservation law of Theorem~\ref{thmsymmlag} means that the 
surface layer integral is preserved under the time evolution.

As a concrete example, the unitary invariance of the causal action principle gives
rise to a conservation law, which corresponds to current conservation.
This example will be considered in detail in Section~\ref{seccip}.
We finally remark that the conservation laws for {\em{energy-momentum}} can also be
obtained from Theorem~\ref{thmsymmlag}, assuming that the causal fermion system
has symmetries as described by generalized Killing symmetries.
We refer the interested reader to~\cite[Section~6]{noether}.

\section{A Class of Conservation Laws in the Smooth Setting} \label{secconserve}
\sindex{conservation law for surface layer integral}%
In the previous section, we saw that surface layer integrals can be used
to formulate a Noether-like theorem, which relates symmetries to conservation laws.
In this section, we shall derive conservation laws even in
the absence of symmetries. Instead, these conservation laws are closely tied to the structure of the
linearized field equations as derived in Section~\ref{seclinfield}.
In order to focus on the essence of the construction, we again restrict attention to the
smooth setting~\eqref{Lsmooth}.
The basic idea of the construction is explained in the following proposition:
\begin{Prp} \label{prpI1} Let~$\rho$ be a critical measure and~$\Omega \subset M$ be compact. Then for
any solution~$\v \in \Jlin$ of the linearized field equations~\eqref{linfield},
\nindex{aep@$\gamma^\Omega_\rho$ -- conserved one-form}%
\beq
\gamma^\Omega_\rho(\v) := \int_\Omega \dd\rho(x) \int_{M \setminus \Omega} \dd\rho(y)\: 
\big( \nabla_{1,\v} - \nabla_{2,\v} \big) \L(x,y) 
= \int_\Omega \nabla_\v \: \s \: \dd\rho \:. \label{I1osi}
\eeq
\end{Prp}
\Proof In view of the anti-symmetry of the integrand,
\beq \int_\Omega \dd\rho(x) \int_\Omega \dd\rho(y)\: 
\big( \nabla_{1,\v} - \nabla_{2,\v} \big) \L(x,y) = 0 \:. \eeq
Adding this equation to the left-hand side of~\eqref{I1osi}, we obtain
\begin{align}
\gamma^\Omega_\rho(\v) &= \int_\Omega \dd\rho(x) \int_M \dd\rho(y)\: 
\big( \nabla_{1,\v} - \nabla_{2,\v} \big) \L(x,y) \notag \\
&= 2 \int_\Omega \dd\rho(x) \int_M \dd\rho(y)\: 
\big( \nabla_{1,\v} \big) \L(x,y) 
-  \int_\Omega \dd\rho(x) \int_M \dd\rho(y)\: 
\big( \nabla_{1,\v} + \nabla_{2,\v} \big) \L(x,y) \notag \\
&= \int_\Omega \dd\rho(x) \bigg( 2\, \nabla_\v \Big(\ell(x) + \s \Big) 
- \Big( \big( \Delta \v \big)(x) + \nabla_\v \: \s \Big) \bigg) \:,
\end{align}
where in the last line we used the definitions of~$\ell$ and~$\Delta$
(see~\eqref{elldef} and~\eqref{linfield}). Applying the restricted EL equations~\eqref{ELrestricted2}
and the linearized field equations~\eqref{linfield} gives the result.
\QED
Viewing~$\gamma^\Omega_\rho$ as a linear functional on the linearized solutions,
we also refer to~$\gamma^\Omega_\rho(\v)$ as the {\em{conserved one-form}}.
\sindex{conserved one-form|textbf}%
We remark that the identity~\eqref{I1osi} has a similar structure
as the conservation law in the Noether-like theorem~\eqref{conservationLagrEq}.
In order to make the connection precise, one describes the symmetry~$\Phi_\tau$
infinitesimally by a jet~$\v$ with a vanishing scalar component,
\beq \v(x) := \frac{\dd}{\dd\tau} \big(0, \Phi_\tau(x) \big) \Big|_{\tau=0} \:. \eeq
Using the symmetry property~\eqref{symmlagr}, one verifies
similarly as in the proof of Lemma~\ref{lemmacommsol} that this jet
satisfies the linearized field equations~\eqref{linfield}.
Therefore, Proposition~\ref{prpI1} applies, and the right-hand side vanishes
because~$\v$ has no scalar component. We thus recover the identity
obtained by carrying out the $\tau$-derivative in~\eqref{conservationLagrEq}.

We conclude that Proposition~\ref{I1osi} is a generalization of Theorem~\ref{thmsymmlag}.
Instead of imposing symmetries, the identity~\eqref{I1osi} is a consequence of the
linearized field equations.
Again choosing~$\Omega$ as the region between two Cauchy surfaces (see Figure~\ref{figjet1}),
one obtains a relation between the surface layer integrals around~$N_1$ and~$N_2$.
If the scalar component of~$\v$ vanishes, we obtain a conservation law.
Otherwise, the right-hand side of~\eqref{I1osi} tells us how the surface layer integral changes in time.

We now generalize Proposition~\ref{prpI1}. The basic idea is to integrate anti-symmetric
expressions in~$x$ and~$y$ which involve higher derivatives of the Lagrangian.
We again restrict attention to the smooth setting (for the general proof see~\cite{osi}).
Let~$\tilde{\rho}_{s,t}$ with~$s, t \in (-\delta, \delta)$ be a two-parameter family of measures which are solutions of the restricted EL equations. We assume that these
measures are of the form
\beq \label{rhost}
\tilde{\rho}_{s,t} = (F_{s,t})_* \big( f_{s,t} \, \rho \big) \:,
\eeq
where~$f_{s,t}$ and~$F_{s,t}$ are smooth,
\beq \label{fFdef}
f \in C^\infty\big((-\delta, \delta)^2 \times \F,\R^+ \big) \qquad \text{and} \qquad
F \in C^\infty\big((-\delta, \delta)^2 \times \F, \F \big) \:,
\eeq
and are trivial in the case~$s=t=0$~\eqref{fFinit}.
Moreover, we need the following technical assumption:
\bitem
\item[\rm{(ta)}] For all~$x \in M$, $p,q \geq 0$ and~$r \in \{0,1\}$,
the following partial derivatives exist and may be interchanged with integration,
\begin{align}
\int_M &\partial_{s'}^r \partial_s^p \partial_t^q \L\big( F_{s+s',t}(x), F_{s,t}(y) \big) \Big|_{s'=s=t=0} \:\dd\rho(y) \notag \\
&= \partial_{s'}^r \partial_s^p \partial_t^q  \int_M \L\big( F_{s+s',t}(x), F_{s,t}(y) \big) \:\dd\rho(y) \bigg|_{s'=s=t=0} \:. 
\end{align}
%\item[\rm{(r2)}] The jet~$\u(t)$ defined by
%\beq \u = \partial_s \big( f_{s,t}, F_{s,t} \big) \big|_{s=0}
%\qquad \text{is in~$\Jtest_t$ for all~$t \in (-\delta, \delta)$}\:, \eeq
%where (for details see~\cite[Section~4.2]{jet})
%\beq \label{Jtesttau}
%\Jtest_t := \Big\{ \Big( (F_{0,t})_* \big( a + D_u \log f_{0,t} \big), \:(F_{0,t})_* u \Big) \text{ with } \u = (a,u) \in \Jtest \Big\} \:.
%\eeq
\eitem
We now state a general identity between a surface layer integral and a volume integral
which was first obtained in~\cite{osi}. It generalizes the result of Proposition~\ref{prpI1}
and gives rise to additional conservation laws for surface layer integrals, which will be analyzed subsequently
(in Section~\ref{SecSmooth}). The proof of the following theorem also
works out the mathematical essence of our conservation laws.

\begin{Thm} \label{thmosirho}
Let~$f$ and~$F$ be as in~\eqref{fFdef} and~\eqref{fFinit} which satisfy the
above assumption~\text{\em{(ta)}}.
Moreover, assume that the measures~$\tilde\rho_{s,t}$ given by~\eqref{rhost}
satisfy the restricted EL equations for all~$s$ and~$t$. Then
for every compact~$\Omega \subset M$ and every~$k \in \N$,
\sindex{conservation law for surface layer integral}%
\nindex{aeq@$I_{k+1}$ -- general conserved surface layer integral}%
\begin{align}
I_{k+1}^\Omega &:= \int_\Omega \dd\rho(x) \int_{M \setminus \Omega} \dd\rho(y) \notag \\
&\qquad\qquad\qquad \times \big(\partial_{1,s} - \partial_{2,s} \big) \big( \partial_{1,t} + \partial_{2,t} \big)^k f_{s,t}(x)\: \L\big( F_{s,t}(x), F_{s,t}(y) \big)\: f_{s,t}(y) \Big|_{s=t=0}  \notag \\
&\,= \s \int_\Omega\partial_{s} \partial_{t}^k  f_{s,t}(x) \Big|_{s=t=0}\:  \dd\rho(x) \:. \label{Ikdef}
\end{align}
\end{Thm}
\Proof %Siehe Skizzenbuch~12, Seite 75ff.
Introducing the short notation
\beq %\label{Ldefst}
L\big(x_{s,t}, y_{s,t} \big) = f_{s,t}(x)\: \L\big( F_{s,t}(x), F_{s,t}(y) \big)\: f_{s,t}(y) \:, \eeq
the restricted EL equations~\eqref{ELrestricted2} read
\beq \nabla_\u \bigg( \int_M L\big(x_{s,t}, y_{s,t} \big)\: \dd\rho(y) 
- \s\: f_{s,t}(x) \bigg) = 0 \qquad \text{for all~$\u \in \J$} \:. \eeq
In particular for any~$k \geq 0$ and any vector~$v = v^s \partial_s + v^t \partial_s$, we obtain
\begin{align}
\int_M \partial_{1,s} \big( \partial_{1,v} + \partial_{2,v} \big)^k L\big(x_{s,t}, y_{s,t} \big)\: \dd\rho(y)
\Big|_{s=t=0} &=
\s\: \partial_{s} \partial_{v}^k  f_{s,t}(x) \big|_{s=t=0} \label{svk} \\
\int_M \big( \partial_{1,v} + \partial_{2,v} \big)^{k+1} L\big(x_{s,t}, y_{s,t} \big)\: \dd\rho(y)
\Big|_{s=t=0} &= \s\: \partial_{v}^{k+1} f_{s,t}(x) \big|_{s=t=0}
\end{align}
(the derivatives exist and can be exchanged with the integration according to
the above assumption~(ta)).
Differentiating the last equation with respect to~$v^s$ and dividing by~$k+1$, we obtain
\beq \int_M \big( \partial_{1,s} + \partial_{2,s} \big) \big( \partial_{1,v} + \partial_{2,v} \big)^k L\big(x_{s,t}, y_{s,t} \big)\:
\dd\rho(y) = \s\: \partial_{s} \partial_{v}^k  f_{s,t}(x) \:. \eeq
Subtracting twice the identity~\eqref{svk}, we obtain for any~$k \geq 0$ the equation
\beq \int_M \big( \partial_{1,s} - \partial_{2,s} \big) \big( \partial_{1,v} + \partial_{2,v} \big)^k L\big(x_{s,t}, y_{s,t} \big)\:
\dd\rho(y) = \s\: \partial_{s} \partial_{v}^k  f_{s,t}(x) \:. \eeq

Integrating the last equation over~$\Omega$ gives
\beq \label{rel1}
\begin{split}
\int_\Omega & \dd\rho(x) \int_M \dd\rho(y) \: \big( \partial_{1,s} - \partial_{2,s} \big) \big( \partial_{1,v} + \partial_{2,v} \big)^k L\big(x_{s,t}, y_{s,t} \big) \\
&= \s \int_\Omega \partial_{s} \partial_{v}^k  f_{s,t}(x)\: \dd\rho(x) \:.
\end{split}
\eeq
On the other hand, since the integrand is anti-symmetric in its arguments~$x$ and~$y$, we also know that
\beq %\label{rel2}
\int_\Omega \dd\rho(x) \int_\Omega \dd\rho(y) \: \big( \partial_{1,s} - \partial_{2,s} \big) \big( \partial_{1,v} + \partial_{2,v} \big)^k L\big(x_{s,t}, y_{s,t} \big) = 0 \:. \eeq
Subtracting this equation from~\eqref{rel1} and evaluating at~$s=t=0$ gives the result.
\QED

Specializing the statement of this theorem to the case~$k=0$ and setting
\beq \v = \frac{\dd}{\dd s} \big( f_{s,t}, F_{s,t} \big) \Big|_{s=t=0} \:, \eeq
we recover the statement of Proposition~\ref{prpI1}.
The case~$k=1$ will be studied in more detail in Section~\ref{SecSmooth}.

We conclude this section by discussing the conservation law of Proposition~\ref{prpI1} 
for {\em{inner solutions}} as considered in Section~~\ref{secinner}
\sindex{inner solution!conservation law for}%
(commutator jets will be considered afterward in Section~\ref{seccip}).
To this end, we need to assume again that spacetime has a smooth manifold structure.
We first define an integration measure on the boundary of~$\Omega$.
\begin{Def} \label{defmudef}
Let~$\v = (\div v, v) \in \Jin_\rho$ be an inner solution and~$\Omega \subset M$ 
closed with smooth boundary~$\partial \Omega$.
On the boundary, we define the measure~$d\mu(\v,x)$ as the contraction of the volume form on~$M$
with~$v$; that is, in local charts,
\nindex{aer@$\mu(\v,x)$ -- boundary measure induced by inner solution~$\v$}
\beq %\label{mudefmeasure}
\dd\mu(\v,x) = h\: \epsilon_{ijkl} \:v^i\: \dd x^j \dd x^k \dd x^l \:, \eeq
\nindex{aar@$\epsilon_{jklm}$ -- totally anti-symmetric symbol}%
where~$\epsilon_{ijkl}$ is the totally anti-symmetric symbol
(normalized by~$\epsilon_{0123}=1$).
\end{Def} \noindent
We now let~$\v = (\div v, v)$ be an inner solution. Then the integral on the right-hand side of~\eqref{I1osi}
reduces the integral over the divergence of the vector field~$v$,
\beq \label{I1right}
\int_\Omega \nabla_\v \: \s \: \dd\rho = \s \int_\Omega \div v\: \dd\rho \:.
\eeq
On the left side of~\eqref{I1osi}, on the other hand, similar as in Lemma~\ref{lemmaGauss}
we can integrate by parts. But now boundary terms remain,
\begin{align}
\gamma^\Omega_\rho(\v)
&= \int_{\partial \Omega} \dd\mu(\v,x) \int_{M \setminus \Omega} \dd\rho(y)\:
\L(x,y) + \int_{\Omega} \dd\rho(x) \int_{\partial \Omega} \dd\mu(\v,y)\: \L(x,y) \notag \\
&= \int_{\partial \Omega} \dd\mu(\v,x) \int_M \dd\rho(y)\: \L(x,y)
= \s \int_{\partial \Omega} \dd\mu(\v,x) \:, \label{I1left}
\end{align}
where in the last line we used the symmetry of~$\L$ and employed the EL equations.
In this way, the surface layer integral in~\eqref{I1osi} reduces to a
usual surface integral over the hypersurface~$\partial \Omega$.
Moreover, combining~\eqref{I1osi} with~\eqref{I1left} and~\eqref{I1right},
we get back the Gau{\ss} divergence theorem
\sindex{Gauss divergence theorem!for inner solutions}%
\beq \s \int_{\partial \Omega} \dd\mu(\v,x) = \s \int_\Omega \div v\: \dd\rho \:. \eeq
This illustrates that Proposition~\ref{prpI1} is a generalization of the Gau{\ss} divergence theorem
where the vector field is replaced by a general solution of the linearized field equations.
The formulation with surface layer integrals has the further advantage that the result can be
generalized in a straightforward way to non-smooth (for example discrete) spacetimes.

\section{The Commutator Inner Product for Causal Fermion Systems} \label{seccip}
\sindex{commutator inner product|textbf}%
As a concrete example of a conservation law, we now consider {\em{current conservation}}.
To this end, we consider the setting of causal fermion systems. As in Section~\ref{seccomm},
we again let~$\scrA$ be a symmetric operator of finite rank on~$\H$ and~$\scrU_\tau$
be the corresponding one-parameter family of unitary transformations~\eqref{Utaudef}.
Infinitesimally, this one-parameter family is described by the {\em{commutator jet}}~$\v$~\eqref{jvdef}.
\sindex{jet!commutator}%
The unitary invariance of the causal action implies that the commutator jets satisfy the
linearized field equations (see Lemma~\ref{lemmacommsol}). Moreover, using that the scalar
component of commutator jets vanishes, Proposition~\ref{prpI1} gives
for any compact~$\Omega \subset M$ the conservation law
\sindex{conserved one-form}%
\beq \label{osig}
\gamma^\Omega_\rho(\v) := \int_\Omega \dd\rho(x) \int_{M \setminus \Omega} \dd\rho(y)\: 
\big( \nabla_{1,\v} - \nabla_{2,\v} \big) \L(x,y) = 0 \:.
\eeq
In order to understand the significance of this conservation law, it is useful to choose~$\scrA$
more specifically as an operator of rank one. 
More precisely, given a non-zero vector~$\psi \in \H$, we form the symmetric linear operator~$\scrA \in \Lin(\H)$
of rank one by
\beq \label{Adef}
\scrA u := \la u | \psi\ra_\H \: \psi 
\eeq
(thus in bra/ket notation, $\scrA = |\psi\ra \la \psi|$). 
We now form the corresponding commutator jet~\eqref{jvdef}.
Varying the vector~$\psi$, we obtain a mapping
\beq \label{jdef}
\j \::\: \H \rightarrow \Jlin \:,\qquad \psi \mapsto \v \:.
\eeq
Moreover, we choose~$\Omega$ again as the past of a Cauchy surface (as shown in Figure~\ref{figjet1}~(c)).
We write the corresponding conserved surface layer integral in~\eqref{osig} as
\beq \label{IOr}
{\mathscr{C}}^\Omega_\rho(u) := 
\int_\Omega \dd\rho(x) \int_{M \setminus \Omega} \dd\rho(y)\:
\big( D_{1,\j(u)} - D_{2,\j(u)} \big) \L(x,y) \qquad \text{with~$u \in \H$} \:,
\eeq
where for technical simplicity we assume smoothness in order to interchange differentiation with
integration.
Clearly, the mapping~$\j$ in~\eqref{jdef}, and consequently also the mapping~${\mathscr{C}}^\Omega_\rho$,
are homogeneous of degree two; that is,
\beq {\mathscr{C}}^\Omega_\rho \big( \lambda u \big) = |\lambda|^2\: {\mathscr{C}}^\Omega_\rho(u)
\qquad \text{for all~$u \in \H$ and~$\lambda \in \C$}\:. \eeq
Therefore, we can use the polarization formula to define a sesquilinear form on the Hilbert space~$\H$,
\beq \label{commsesqui}
\la u | v \ra^\Omega_\rho :=
\frac{1}{4} \:\Big( {\mathscr{C}}^\Omega_\rho(u+v) - {\mathscr{C}}^\Omega_\rho(u-v) \Big)
-\frac{\cI}{4} \:\Big( {\mathscr{C}}^\Omega_\rho(u+\cI v) - {\mathscr{C}}^\Omega_\rho(u-\cI v) \Big) \:.
\eeq
\nindex{aet@$\la .\vert. \ra^\Omega_\rho$ -- commutator inner product}%
\tindex{bb@$\la .\vert. \ra^\Omega_\rho$ -- commutator inner product}%
This sesquilinear form is referred to as the {\em{commutator inner product}}
(for details see~\cite[Section~3]{dirac}).
In~\cite[Section~5.2]{noether} it is shown that for Dirac systems describing the Minkowski vacuum,
the commutator inner product coincides (up to an irrelevant prefactor) with the scalar product
on Dirac solutions~\eqref{printMink}. In this way, the conservation law for the commutator inner product
gives back the conservation of the Dirac current~\eqref{cc}.
We thus recover current conservation as a special case of a more general conservation
law for causal fermion systems.
\sindex{current conservation!in causal fermion system}%
Since in examples of physical interest, the conserved surface layer integral~$ {\mathscr{C}}^\Omega_\rho(u,v)$ gives back the Hilbert space scalar product, we give this property a name:
\begin{Def}
\sindex{commutator inner product!represents the space scalar product}%
Given a critical measure~$\rho$ and a subset~$\Omega \subset M$, the surface layer integral~${\mathscr{C}}^\Omega_\rho$
is said to {\bf{represent the scalar product}} on the subspace~$\H^\fermi \subset \H$
if there is a non-zero real constant~$c$ such that the sesquilinear form~$\la .|. \ra^\Omega_\rho$
defined by~\eqref{commsesqui} has the property
\beq \label{repc}
\la u | u \ra^\Omega_\rho = c\, \|u\|_\H^2 \qquad \text{for all~$u \in \H^\fermi$}\:.
\eeq
\end{Def} \noindent
In view of the conservation law of Proposition~\ref{prpI1}, this property remains valid if~$\Omega$ is
changed by a compact subset of~$M$.
We point out that the representation~\eqref{repc} {\em{cannot}}
hold on the whole Hilbert space (that is, for all~$u \in \H$); for details, see Exercise~\ref{excommnogo}
and~\cite[Appendix~A]{current}.

At present there is no general argument why
the surface layer integral~${\mathscr{C}}^\Omega_\rho$ should represent the scalar product
on a non-trivial subspace~$\H^\fermi \subset \H$.
Therefore, in this book, we shall not assume that this property holds.
Instead, we make the following weaker assumption.
We assume that the sesquilinear form~${\mathscr{C}}^\Omega_\rho$ is equivalent to
the scalar product in the sense that
\beq \la u | v \ra^\Omega_\rho = \la u \,|\, {\mathscr{C}}_\rho\, v \ra_\H  \qquad \text{for all~$u,v \in \H^\fermi$}\:, \eeq
where~${\mathscr{C}}_\rho$ is a bounded linear operator on~$\H$ with bounded inverse.
Under this assumption, the Hilbert space scalar product can be expressed by
\beq \la u \,|\, v \ra_\H = \la u \,|\, {\mathscr{C}}_\rho^{-1}\, v \ra^\Omega_\rho \qquad \text{for all~$u,v \in \H^\fermi$} \:. \eeq
In this way, the Hilbert space scalar product can be represented by a surface layer
integral involving the physical wave functions in spacetime.

We conclude this section with a remark on the connection between the commutator inner product
and the scalar product on solutions of the Dirac equation. As already mentioned after~\eqref{commsesqui},
for Dirac systems describing the Minkowski vacuum, the commutator inner product~\eqref{commsesqui}
coincides with the scalar product~\eqref{printMink}. Since both inner products are conserved, the same
is true for any Dirac system that evolved from the vacuum (for example, by ``turning on'' an interaction).
The basic shortcoming of this correspondence is that it holds only for the physical wave functions, that is, for
all occupied one-particle states of the system. Thus, in the example of the Minkowski vacuum,
the connection between~\eqref{commsesqui} and~\eqref{printMink} can be made only for the negative-energy
solutions of the Dirac equation. The positive-energy solutions, however, do not correspond to physical
wave functions, so that the commutator inner product is undefined.
In order to improve the situation, one would like to extend the commutator inner product
to more general wave functions, in such a way that it still agrees with~\eqref{printMink}.
This construction is carried out in~\cite{dirac, current}.
Current conservation continues to hold for the extension, provided that the wave functions
satisfy the so-called {\em{dynamical wave equation}}
\sindex{dynamical wave equation}%
\beq \label{dwe}
\int_M Q^\dyn(x,y)\, \psi(y)\: \dd\rho(y) = 0 \:.
\eeq
\nindex{aeu@$Q^\dyn(x,y)$ -- kernel in dynamical wave equation}%
Here, the integral kernel~$Q^\dyn$ is constructed from first variations of the causal Lagrangian.
In this formulation, the commutator inner product takes the form
\beq \label{cli}
\begin{split}
\la \psi | \phi \ra^\Omega_\rho := -2\cI \,\bigg( \int_{\Omega} \!\dd\rho(x) \int_{M \setminus \Omega} \!\!\!\!\!\!\!\dd\rho(y) 
&- \int_{M \setminus \Omega} \!\!\!\!\!\!\!\dd\rho(x) \int_{\Omega} \!\dd\rho(y) \bigg)\: \\
&\times \Sl \psi(x) \:|\: Q^\dyn(x,y)\, \phi(y) \Sr_x \:.
\end{split}
\eeq
For some more details on these connections see Exercises~\ref{exdynwave1} and~\ref{exdynwave2}.

After these extensions have been made, the dynamical wave equation~\eqref{dwe} can be regarded
as the generalization of the Dirac equation to causal fermion systems.
Moreover, the commutator inner product~\eqref{cli} generalizes the scalar product on Dirac
solutions~\eqref{printMink}, thereby also extending current conservation to dynamical waves.

\section{The Symplectic Form and the Surface Layer Inner Product}\label{SecSmooth}
For the applications, the most important surface layer integrals are~$I^\Omega_1$ (also denoted
by~$\gamma^\Omega_\rho$; see Proposition~\ref{prpI1} and Theorem~\ref{thmosirho} in the case~$k=0$)
and~$I^\Omega_2$ (see Theorem~\ref{thmosirho} in the case~$k=1$).
We now have a closer look at the surface layer integral~$I^\Omega_2$.
It is defined by
\beq \label{I2def}
\begin{split}
I_2^\Omega &= \int_\Omega \dd\rho(x) \int_{M \setminus \Omega} \dd\rho(y) \\
&\qquad \times \big(\partial_{1,s} - \partial_{2,s} \big) \big( \partial_{1,t} + \partial_{2,t} \big) f_{s,t}(x)\: \L\big( F_{s,t}(x), F_{s,t}(y) \big)\: f_{s,t}(y) \Big|_{s=t=0}
\end{split}
\eeq
and satisfies for any compact subset~$\Omega \subset M$ the identity
\beq \label{I2cons}
I^\Omega_2 = \s \int_\Omega\partial_{s} \partial_{t}  f_{s,t}(x) \Big|_{s=t=0}\:  \dd\rho(x) \:.
\eeq

These formulas simplify considerably if we {\em{anti-symmetrize}} in the parameters~$s$ and~$t$.
Namely, the formula for~$I_2^\Omega$ reduces to the surface layer integral
\beq \int_\Omega \dd\rho(x) \int_{M \setminus \Omega} \dd\rho(y) 
\big(\partial_{1,s} \partial_{2,t} - \partial_{1,s} \partial_{2,t} \big) f_{s,t}(x)\:
\L\big( F_{s,t}(x), F_{s,t}(y) \big)\: f_{s,t}(y) \Big|_{s=t=0} \:. \eeq
Since this expression involves only first partial derivatives, it can be rewritten with jet derivatives as
\sindex{symplectic form!on linearized solutions}%
\nindex{aev@$\sigma^\Omega_\rho(.,.)$ -- symplectic form on linearized solutions}%
\tindex{bb@$\sigma^\Omega_\rho(.,.)$ -- symplectic form on linearized solutions}%
\beq \label{sigmadef}
\sigma^\Omega_\rho(\u, \v) :=
\int_\Omega \dd\rho(x) \int_{M \setminus \Omega} \dd\rho(y)
\big( \nabla_{1,\u} \nabla_{2,\v} - \nabla_{1,\v} \nabla_{2,\u} \big) \L(x,y) \:,
\eeq
where the jets~$\u$ and~$\v$ are the linearized solutions
\beq \label{uvdef}
\u = \partial_s \big( f_{s,t}, F_{s,t} \big) \big|_{s=t=0} \qquad \text{and} \qquad
\v = \partial_t \big( f_{s,t}, F_{s,t} \big) \big|_{s=t=0} \:.
\eeq
Moreover, the right-hand side of~\eqref{I2cons} vanishes when anti-symmetrizing in~$s$ and~$t$.
We conclude that
\beq \sigma^\Omega_\rho(\u, \v) = 0 \qquad \text{for every compact~$\Omega \subset M$}\:. \eeq
Choosing~$\Omega$ again as explained in Figure~\ref{figjet1},
we obtain a conservation law for a surface layer integral over a neighborhood of
a hypersurface~$N$ which extends to spatial infinity.
We refer to~$\sigma^\Omega_\rho$ as the {\em{symplectic form}}
(the connection to symplectic geometry will be explained after~\eqref{sympex} below).

{\em{Symmetrizing}}~$I^\Omega_2$ in the parameters~$s$ and~$t$ gives the surface layer integral
\beq \label{Isymm}
\int_\Omega \dd\rho(x) \int_{M \setminus \Omega} \dd\rho(y) 
\big(\partial_{1,s} \partial_{1,t} - \partial_{2,s} \partial_{2,t} \big)
f_{s,t}(x)\: \L\big( F_{s,t}(x), F_{s,t}(y) \big)\: f_{s,t}(y) \Big|_{s=t=0} \:.
\eeq
This expression has a more difficult structure because it involves second partial derivatives.
Such second partial derivatives cannot be expressed directly in terms of second
jet derivatives, because the derivatives of the jets also need to be taken into account.
In a differential geometric language, defining second derivatives would make it necessary to
introduce a connection on~$\F$. As explained after~(ii) on page~\pageref{ConventionPartial2},
we here use the simpler method of taking second partial derivatives in distinguished charts
(for example, symmetric wave charts for causal fermion systems; see
the remark after Proposition~\ref{prppq}
and~\cite[Section~6.1]{gaugefix} or~\cite[Section~3]{banach}).
Then it is useful to introduce the {\em{surface layer inner product}}~$(.,.)^\Omega_\rho$
\sindex{surface layer inner product|textbf}%
\nindex{aew@$(.,.)^\Omega_\rho$ -- surface layer inner product}%
\tindex{bb@$(.,.)^\Omega_\rho$ -- surface layer inner product}%
as the contribution to~\eqref{Isymm} involving second derivatives of the Lagrangian; that is,
\beq \label{sliunsoftened}
(\u, \v)^\Omega_\rho :=
\int_\Omega \dd\rho(x) \int_{M \setminus \Omega} \dd\rho(y)
\big( \nabla_{1,\u} \nabla_{1,\v} - \nabla_{2,\u} \nabla_{2,\v} \big) \L(x,y) \:,
\eeq
where the jets~$\u$ and~$\v$ are again the linearized solutions~\eqref{uvdef}.
We point out that, in contrast to the symplectic form, the surface layer inner product does
{\em{not}} correspond to a conservation law. This has two reasons: First because
the right-hand side of~\eqref{I2cons} gives rise to a volume term, and second because
the derivatives of the jets~$\u$ and~$\v$ give additional correction terms.
For the details and the interpretation of these correction terms, we refer to~\cite{osi}.
Here, we only remark that the significance of the surface layer inner product is that
it is an approximate conservation law. In particular, it can be used for estimating solutions
of the linearized field equations and for proving existence results. We will come back to these
applications in Chapter~\ref{seclinhyp}.

We finally comment on the name {\em{symplectic form}}.
\sindex{symplectic form|textbf}%
Clearly, this name is taken from
symplectic geometry, where it refers to a closed and non-degenerate two-form~$\sigma$
on a manifold which we denote by~$\calB$. The connection to the surface layer integral~\eqref{sigmadef}
is obtained if we assume that the set of all critical measures of the form~\eqref{Ffvary}
forms a smooth manifold~$\calB$
(which may be an infinite-dimensional Banach manifold). In this case, a jet~$\v$
describing a first variation of a measure~\eqref{vlinear} is a tangent
vector in~$T_\rho \calB$. Consequently, the jet space~$\J$ can be identified with the
tangent space~$T_\rho \calB$. The surface layer integral~\eqref{sigmadef}
can be regarded as a mapping
\beq \label{sympex}
\sigma^\Omega_\rho \::\: T_\rho \calB \times T_\rho \calB \rightarrow \R \:.
\eeq
Being anti-symmetric, it can be regarded as a two-form.
Similarly, the conserved surface layer integral~$\gamma^\Omega_\rho$ in~\eqref{I1osi}
can be regarded as a one-form.
\sindex{conserved one-form}%
Moreover, the $t$-derivative in~\eqref{I2def} can be regarded as a directional
derivative acting on~$I^\Omega_1=\gamma^\Omega_\rho$. Anti-symmetrizing in~$s$ and~$t$ corresponds to
taking the outer derivative. We thus obtain
\beq \label{sigdgamma}
\sigma^\Omega_\rho = \dd \gamma^\Omega_\rho \:,
\eeq
which also shows again that~$\sigma^\Omega_\rho$ is closed.
Thus, exactly as in symplectic geometry, the symplectic form defined as the
surface layer integral~\eqref{sympex} is a closed two-form.
In contrast to symplectic geometry, it does not need to be non-degenerate.
But this can be arranged by restricting attention to a more specific class
of measures of the form~\eqref{rhost}. We refer to~\cite{jet}
for a more general discussion of this point.

We finally note that the relation~\eqref{sigdgamma} resembles the representation of the symplectic
potential as the derivative of the {\em{symplectic potential}} 
\sindex{symplectic potential}%
(sometimes also referred to as the
tautological one-form or canonical one-form). It is a major difference between classical mechanics
and classical field theory that, in the setting of causal variational principles, the
one-form~$\gamma^\Omega_\rho$ is canonically defined and represented by a conserved
surface layer integral in spacetime.
 
\section{The Nonlinear Surface Layer Integral} \label{secosinonlin}
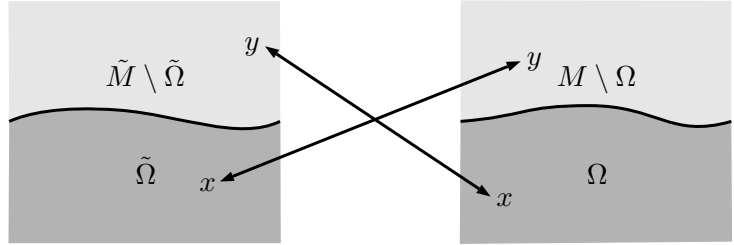
\begin{figure}[tb]
\psscalebox{1.0 1.0} % Change this value to rescale the drawing.
{
\begin{pspicture}(0,26.0)(9.62,29.2)
\definecolor{colour3}{rgb}{0.7019608,0.7019608,0.7019608}
\definecolor{colour4}{rgb}{0.9019608,0.9019608,0.9019608}
\pspolygon[linecolor=colour3, linewidth=0.02, fillstyle=solid,fillcolor=colour3](6.0300636,27.62005)(6.2800636,27.64005)(6.7000637,27.72005)(7.1200633,27.80005)(7.6400633,27.83005)(8.080064,27.810051)(8.570064,27.67005)(8.960064,27.61005)(9.295063,27.60005)(9.420063,27.57005)(9.610064,27.62005)(9.620064,26.04005)(6.0200634,26.01005)
\pspolygon[linecolor=colour4, linewidth=0.02, fillstyle=solid,fillcolor=colour4](6.0300636,29.23005)(9.610064,29.22505)(9.610064,27.64005)(9.395063,27.60005)(9.090063,27.58505)(8.760063,27.62005)(8.330064,27.78005)(7.7700634,27.86005)(7.7700634,27.86005)(7.5400634,27.85005)(7.1000633,27.82005)(6.7700634,27.76005)(6.4100633,27.68005)(6.2000637,27.66005)(6.0300636,27.66005)
\pspolygon[linecolor=colour3, linewidth=0.02, fillstyle=solid,fillcolor=colour3](0.020063477,26.02005)(3.6200635,26.02005)(3.6150634,27.59505)(3.4700634,27.55005)(3.1900635,27.51005)(2.9100635,27.53005)(2.5700636,27.57005)(2.2100635,27.66005)(1.7900635,27.73005)(1.4600635,27.76005)(1.1200634,27.79005)(0.84006345,27.77005)(0.4600635,27.73005)(0.20006348,27.68005)(0.030063476,27.61005)
\pspolygon[linecolor=colour4, linewidth=0.02, fillstyle=solid,fillcolor=colour4](0.020063477,29.22005)(3.6100636,29.22005)(3.6150634,27.63005)(3.4400635,27.560051)(3.1500635,27.53005)(2.6900635,27.58005)(2.1300635,27.70005)(1.6800635,27.77005)(1.2000635,27.83005)(0.64006346,27.78005)(0.24006347,27.72005)(0.030063476,27.64005)
\psline[linecolor=black, linewidth=0.04, arrowsize=0.05291667cm 2.0,arrowlength=1.4,arrowinset=0.0]{<->}(3.4200635,28.63005)(6.4200635,26.63005)
\psline[linecolor=black, linewidth=0.04, arrowsize=0.05291667cm 2.0,arrowlength=1.4,arrowinset=0.0]{<->}(2.8200636,26.83005)(6.8200636,28.43005)
\psbezier[linecolor=black, linewidth=0.04](0.020063477,27.63005)(0.24502629,27.729553)(0.6800635,27.84005)(1.4800634,27.780050048828127)(2.2800634,27.72005)(3.0623415,27.364084)(3.6200635,27.63005)
\psbezier[linecolor=black, linewidth=0.04](6.0200634,27.63005)(6.8838763,27.696981)(6.8200636,27.84005)(7.7300634,27.840050048828125)(8.640063,27.84005)(8.806251,27.39698)(9.620064,27.63005)
\rput[bl](7.3,28){$M \setminus \Omega$}
\rput[bl](7.7,26.8){$\Omega$}
\rput[bl](1.3,28){$\tilde{M} \setminus \tilde{\Omega}$}
\rput[bl](1.7,26.8){$\tilde{\Omega}$}
\rput[bl](2.55,26.7){$x$}
\rput[bl](3.15,28.5){$y$}
\rput[bl](6.5,26.5){$x$}
\rput[bl](6.9,28.3){$y$}
\end{pspicture}
}
\caption{The nonlinear surface layer integral.}
\label{figosinl}
\end{figure}

\sindex{surface layer integral!nonlinear}%
We now introduce a different type of surface layer integral, which can be regarded as a
generalization of the surface layer integrals considered so far.
In order to explain the basic concept, we return to the general structure of
a surface layer integral~\eqref{intdouble}. The differential operator~$(\cdots)$ in
the integrand can be regarded as describing
first or second variations of the measure~$\rho$. As we saw above, the resulting surface layer integrals
give rise to conserved currents, the symplectic form and inner products.
Instead of considering first or second variations of a measure~$\rho$, 
we now consider an additional measure~$\tilde{\rho}$ which can be thought of as a finite
perturbation of the measure~$\rho$. Consequently, we also have two spacetimes
\beq M := \supp \rho \qquad \text{and} \qquad \tilde{M} := \supp \tilde{\rho} \:. \eeq
Choosing two compact subsets~$\Omega \subset M$ and~$\tilde{\Omega} \subset \tilde{M}$
of the corresponding spacetimes, we form the {\em{nonlinear surface layer integral}} by
\beq \label{osinl}
\gamma^{\tilde{\Omega}, \Omega}(\tilde{\rho}, \rho) :=
\int_{\tilde{\Omega}} \dd\tilde{\rho}(x) \int_{M \setminus \Omega} \dd\rho(y)\: \L(x,y)
- \int_{\Omega} \dd\rho(x) \int_{\tilde{M} \setminus \tilde{\Omega}}  \dd\tilde{\rho}(y)\: \L(x,y) \:.
\eeq
\nindex{aex@$\gamma^{\tilde{\Omega}, \Omega}(\tilde{\rho}, \rho)$ -- nonlinear surface layer integral}%
Note that one argument of the Lagrangian is in~$M$, whereas the other is in~$\tilde{M}$.
Moreover, one argument lies inside the set~$\Omega$ respectively~$\tilde{\Omega}$,
whereas the other argument lies outside.
In this way, the nonlinear surface layer integral ``compares'' the two spacetimes
near the boundaries of~$\Omega$ and~$\tilde{\Omega}$,
as is illustrated in Figure~\ref{figosinl}.
If~$\tilde{\rho}$ is a first or second variation of~$\rho$, one recovers surface layer integrals
of the form~\eqref{intdouble}. In this way, the nonlinear surface layer integral can be regarded as a
generating functional for the previous surface layer integrals.
Moreover, it has the advantage that it does not rely on continuous variations or a perturbative
treatment. Instead, it can be used for comparing two arbitrary measures~$\rho$ and~$\tilde{\rho}$.
This nonlinear surface layer integral was introduced in~\cite{fockbosonic}.
It plays a central role in getting the connection to quantum field theory
(as will be outlined in Chapter~\ref{secQFT}).

The nonlinear surface layer integral comes with a corresponding conservation law,
as we now explain. For technical simplicity, we assume that the measure~$\tilde{\rho}$
can be obtained from~$\rho$ by multiplication with a weight function and a push-forward; that is,
\beq %\label{tilrho}
\tilde{\rho} = F_* (f \rho) \eeq
with smooth functions~$f \in C^\infty(M, \R^+)$ and~$F \in C^\infty(M, \F \big)$.
We use the mapping~$F$ in order to identify~$M$ with~$\tilde{M}$. In particular, we choose
\beq \tilde{\Omega} = F(\Omega)\:. \eeq
Then, using the definition of the push-forward measure, the nonlinear surface layer integral can be
written alternatively as
\beq \label{osinl2}
\gamma^{\tilde{\Omega}, \Omega}(\tilde{\rho}, \rho) =
\int_{\Omega} \dd\rho(x) \int_{M \setminus \Omega} \dd\rho(y)\: 
\Big( f(x)\, \L\big(F(x),y \big) -  \L\big(x, F(y) \big)\, f(y) \Big) \:.
\eeq
As explained in Section~\ref{secnoether} in the connection of Noether-like theorems,
by a ``conservation law'' we mean
that the nonlinear surface layer integral should vanish for all compact~$\Omega$.
In preparation for analyzing how to satisfy this condition, we rewrite the nonlinear surface layer integral
as a volume integral by using the anti-symmetry of the integrand in~\eqref{osinl2},
\beq
\gamma^{\tilde{\Omega}, \Omega}(\tilde{\rho}, \rho) =
\int_{\Omega} \dd\rho(x) \int_M \dd\rho(y)\: 
\Big( f(x)\, \L\big(F(x),y \big) -  \L\big(x, F(y) \big)\, f(y) \Big) \:. \label{osinlvol}
\eeq
In order to write this equation in a simpler form, we introduce
a measure~$\nu$ on~$M$ and a measure~$\tilde{\nu}$ on~$\tilde{M}$ by
\begin{align}
\dd\nu(x) &:= \bigg( \int_{\tilde{M}} \L(x,y)\: \dd\tilde{\rho}(y) \bigg) \: \dd\rho(x) \:, \\
\dd\tilde{\nu}(x) &:= \bigg( \int_M \L(x,y)\: \dd\rho(y) \bigg) \: \dd\tilde{\rho}(x) \:.
\end{align}
Intuitively speaking, these measures describe how the measures~$\rho$ and~$\tilde{\rho}$
are connected to each other by the Lagrangian. We refer to them as the {\em{correlation measures}}.
\sindex{measure!correlation}%
\nindex{aey@$\nu$ -- correlation measure}%
Then we can rewrite~\eqref{osinlvol} as
\beq \gamma^{\tilde{\Omega}, \Omega}(\tilde{\rho}, \rho) 
= \tilde{\nu}\big( F(\Omega) \big) - \nu(\Omega) \:. \eeq
In order to obtain a conservation law, this expression should vanish for all compact~$\Omega$.
In other words, the measure~$\nu$ should be the push-forward of the measure~$\tilde{\nu}$
under the mapping~$F$,
\beq  \nu = F_* \tilde{\nu} \:. \eeq
In this way, the task of finding a conservation law is reduced to the following abstract problem:
Given two measures~$\nu$ on~$M$ and~$\tilde{\nu}$ on~$\tilde{M}$, under which assumptions
can one measure be realized as the push-forward of the other?
If~$\nu$ and~$\tilde{\nu}$ are volume forms on compact manifolds, such a push-forward mapping
is obtained from a classical theorem of J\"urgen Moser (see, for example, \cite[Section~XVIII, \S2]{langDG}).
In the non-compact case, the existence of~$F$ has been proven
under general assumptions in~\cite{greene-shiohama}.
In this way, the conservation law for the nonlinear surface layer integral can be arranged by
adjusting the identification of the spacetimes~$M$ and~$\tilde{M}$.

We finally remark how the nonlinear surface layer integral can be used to ``compare'' two
causal fermion systems~$(\H, \F, \rho)$ and~$(\tilde{\H}, \tilde{\F}, \tilde{\rho})$.
In this setting, one must keep in mind that the causal fermion systems are defined on different
Hilbert spaces. Therefore, before forming the nonlinear surface layer integral, we must identify
the Hilbert space~$\H$ and~$\tilde{\H}$ by a unitary transformation~$V : \H \rightarrow \tilde{H}$.
Since this identification is not unique, we are left with the freedom to transform~$V$ according to
\beq V \rightarrow V \scrU \qquad \text{with} \qquad \scrU \in \Lin(\H) \text{ unitary} \:. \eeq
A possible strategy for getting information independent of this freedom is to integrate over the unitary group.
For example, this leads to the so-called {\em{partition function}}
\sindex{partition function}%
\nindex{aez@$Z^{\tilde{\Omega}, \Omega}(\tilde{\rho}, \rho)$ -- partition function}%
\beq Z^{\tilde{\Omega}, \Omega}(\tilde{\rho}, \rho) :=
\int_\G \E^{\beta \gamma^{\tilde{\Omega}, \Omega}(\tilde{\rho}, \scrU \rho)} \:\dd\mu_\G(\scrU) \:, \eeq
where~$\beta$ is a real parameter, and~$\G$ is a compact subgroup of the unitary group on~$\H$
with Haar measure~$d\mu_\G$.
Here the name ``partition function'' stems from an analogy to the path integral formulation of quantum
field theory. For more details, we refer to Chapter~\ref{secQFT} or the research papers~\cite{fockfermionic, fockentangle}.

\section{Two-Dimensional Surface Layer Integrals} \label{secosi2d}
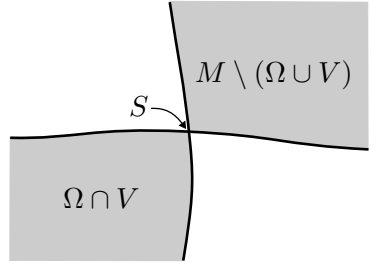
\begin{figure}
\psset{xunit=.5pt,yunit=.5pt,runit=.5pt}
\begin{pspicture}(271.1957508,195.81662156)
{
\newrgbcolor{curcolor}{0.80000001 0.80000001 0.80000001}
\pscustom[linestyle=none,fillstyle=solid,fillcolor=curcolor]
{
\newpath
\moveto(135.05108787,98.74163321)
\lineto(135.05108787,99.64047289)
\lineto(130.78158992,129.30227699)
\lineto(125.38852913,167.50308455)
\lineto(120.44489953,196.49075321)
\lineto(270.55159559,195.81662156)
\lineto(270.7762885,84.80957274)
\lineto(240.66505701,87.05668329)
\lineto(191.22871937,93.34857242)
\lineto(153.2526274,97.16865809)
\closepath
}
}
{
\newrgbcolor{curcolor}{0.80000001 0.80000001 0.80000001}
\pscustom[linewidth=0.37795276,linecolor=curcolor]
{
\newpath
\moveto(135.05108787,98.74163321)
\lineto(135.05108787,99.64047289)
\lineto(130.78158992,129.30227699)
\lineto(125.38852913,167.50308455)
\lineto(120.44489953,196.49075321)
\lineto(270.55159559,195.81662156)
\lineto(270.7762885,84.80957274)
\lineto(240.66505701,87.05668329)
\lineto(191.22871937,93.34857242)
\lineto(153.2526274,97.16865809)
\closepath
}
}
{
\newrgbcolor{curcolor}{0.80000001 0.80000001 0.80000001}
\pscustom[linestyle=none,fillstyle=solid,fillcolor=curcolor]
{
\newpath
\moveto(0.00000378,93.57328423)
\lineto(24.4934589,93.79801494)
\lineto(51.23402079,95.82040612)
\lineto(79.32284976,98.31778187)
\lineto(109.20935811,97.98015667)
\lineto(134.60166803,98.29219447)
\lineto(136.84876724,74.24815919)
\lineto(137.07346016,51.32767919)
\lineto(134.15221795,22.78943037)
\lineto(130.55685165,0.09365463)
\lineto(-0.22473071,0.31834754)
\closepath
}
}
{
\newrgbcolor{curcolor}{0.80000001 0.80000001 0.80000001}
\pscustom[linewidth=0,linecolor=curcolor]
{
\newpath
\moveto(0.00000378,93.57328423)
\lineto(24.4934589,93.79801494)
\lineto(51.23402079,95.82040612)
\lineto(79.32284976,98.31778187)
\lineto(109.20935811,97.98015667)
\lineto(134.60166803,98.29219447)
\lineto(136.84876724,74.24815919)
\lineto(137.07346016,51.32767919)
\lineto(134.15221795,22.78943037)
\lineto(130.55685165,0.09365463)
\lineto(-0.22473071,0.31834754)
\closepath
}
}
{
\newrgbcolor{curcolor}{0 0 0}
\pscustom[linewidth=1.99999997,linecolor=curcolor]
{
\newpath
\moveto(0.02221126,93.13902689)
\curveto(50.07131063,92.0273269)(60.5218105,101.32432678)(132.56010959,97.90832683)
\curveto(204.59840868,94.49232687)(202.72040871,87.97452695)(271.14280784,84.347727)
}
}
{
\newrgbcolor{curcolor}{0 0 0}
\pscustom[linewidth=1.99999997,linecolor=curcolor]
{
\newpath
\moveto(120.77424667,195.8166191)
\curveto(126.77414659,154.49331962)(131.43884654,137.99751983)(136.28644647,87.69692046)
\curveto(139.14604644,58.02412083)(137.84984645,42.66522103)(130.94594654,0.16032156)
}
}
{
\newrgbcolor{curcolor}{0 0 0}
\pscustom[linewidth=1.00157475,linecolor=curcolor]
{
\newpath
\moveto(106.51283906,115.37021652)
\curveto(117.50448756,115.98224675)(124.48166551,111.9394373)(131.00629417,102.78642313)
}
}
{
\newrgbcolor{curcolor}{0 0 0}
\pscustom[linestyle=none,fillstyle=solid,fillcolor=curcolor]
{
\newpath
\moveto(132.68355278,100.43349661)
\lineto(128.13018718,102.50741712)
\lineto(132.20804942,105.41427779)
\closepath
}
}
{
\newrgbcolor{curcolor}{0 0 0}
\pscustom[linewidth=0.66771652,linecolor=curcolor]
{
\newpath
\moveto(132.68355278,100.43349661)
\lineto(128.13018718,102.50741712)
\lineto(132.20804942,105.41427779)
\closepath
}
\rput[bl](40,40){$\Omega \cap V$}
\rput[bl](140,130){$M \setminus (\Omega \cup V)$}
\rput[bl](90,110){$S$}
}
\end{pspicture}
\caption{A two-dimensional surface layer integral.}
\label{figarea}
\end{figure}

\sindex{surface layer integral!two-dimensional}%
The surface layer integrals considered so far were intended to generalize integrals over
hypersurfaces. We now explain how lower-dimensional integrals can be described by surface layer
integrals. We restrict attention to two-dimensional integrals, noting that the methods can be applied
similarly to one-dimensional integrals (that is, integrals along a curve).
It is most convenient to describe a two-dimensional surface~$S \subset M$ as
\beq S = \partial \Omega \cap \partial V \:, \eeq
where~$\Omega$ can be thought of as being the past of a Cauchy surface, and~$V$ describes
a spacetime cylinder. This description has the advantage that the resulting surface layer integrals
will be well-defined even in cases when spacetime is singular or discrete, in which
case the boundaries~$\partial \Omega$ and~$\partial V$ are no longer a sensible concept.
The most obvious way of introducing a surface layer integral localized in a neighborhood of~$S$
is a double integral of the form
\beq \label{osi2d1}
\int_{\Omega \cap V} \bigg( \int_{M \setminus (\Omega \cup V)} (\cdots)\: \L(x,y)\: \dd\rho(y) \bigg)\, \dd\rho(x)
\eeq
(where~$(\cdots)$ again stands for a differential operator acting on the Lagrangian).
If the Lagrangian is of short range, we only get contributions to this surface layer integral if both~$x$
and~$y$ are close to the two-dimensional surface~$S$ (see Figure~\ref{figarea}).

The disadvantage of this method is that the surface layer integral~\eqref{osi2d1}
does not seem to fit together with the EL equations and the linearized field equations.
Therefore, at present there is no corresponding conservation law.
If one considers flows of two-surfaces, it seems preferable to use the following method
introduced in~\cite{jacobson}. We need to assume that~$M$ has a smooth manifold structure
and is four-dimensional (see Definition~\ref{defsms}) and that~$v$ is a vector field which is transverse to the
hypersurface~$\partial \Omega$ and tangential to~$\partial V$.
Following Definition~\ref{defmudef}, the inner solution corresponding to~$v$ gives rise to
a volume measure~$\mu$ on~$\partial \Omega$.
Thus we can introduce a two-dimensional surface layer integral by
\beq A := \int_{\partial \Omega \cap V} \dd\mu(\v, x) \int_{M \setminus V} \dd\rho(y)\: (\cdots) \L(x,y) \:. \eeq
Applying the Gau{\ss} divergence theorem, this surface layer integral can also be written
in the usual way as a double spacetime integral involving jet derivatives of the inner solution,
\begin{align}
A &= \int_{\Omega \cap V} \dd\rho(x) \:\nabla_\v \int_{M \setminus V} \dd\rho(y)\: (\cdots) \L(x,y) \notag \\
&= \int_{\Omega \cap V} \dd\rho(x) \int_{M \setminus V} \dd\rho(y)\: \big( \nabla_{1,\v} \pm \nabla_{2,\v} \big) (\cdots) \L(x,y) \:,
\label{Aform}
\end{align}
where the notation~$\pm$ means that the formula holds for either choice of the sign
(this is because the corresponding term vanishes, as one sees after integrating by parts
as in the proof of Lemma~\ref{lemmaGauss} and using that~$v$ is tangential to~$\partial V$).
The obtained surface layer integral~\eqref{Aform} harmonizes with the structures of the
EL equations and the linearized field equations, as is exemplified in~\cite{jacobson} by 
a simple connection between area change and matter flux.
\Evtl{Insgesamt ein sehr anspruchsvolles Kapitel. Besonders die letzten drei Abschnitte sind sehr formal.
F: Generell mehr Details, mehr Erkl\"arungen.}%

\section{Exercises}

\begin{Exercise} (Noether-like theorems) {\em{
\sindex{Noether-like theorem}%
The goal of this exercise is to illustrate the Noether-like theorems.
In order to simplify the problem as far as possible, we consider the compact setting and assume furthermore that
the Lagrangian is smooth; that is, $\L\in C^\infty(\F\times\F,\R_0^+)$.  Let~$\rho$ be a minimizer of the
action under variations of~$\rho$ in the class of (positive) normalized regular Borel measures. Let~$u\in T\F$ be a vector field on~$\F$. Assume that~$u$ is a symmetry of the Lagrangian in the sense that
\begin{equation}\label{equation1}
\left(u(x)^j\frac{\partial}{\partial x^j}+u(y)^j\frac{\partial}{\partial y^j}\right)\L(x,y)=0\quad\mbox{for all }x,y\in\F.
\end{equation}
Prove that for any measurable set~$\Omega\subset\F$,
\beq
\int_\Omega \dd\rho(x)\int_{\F\setminus\Omega}\dd\rho(y)\,u(x)^j\frac{\partial}{\partial x^j}\L(x,y)=0.
\eeq
\textit{Hint:}  Integrate~\eqref{equation1} over~$\Omega\times\Omega$. Transform the integral using the
symmetry~$\L(x,y)=\L(y,x)$. Finally make use of the EL equations.
}} \end{Exercise}

\begin{Exercise} {{(Commutator jets and conserved surface layer integrals)}} {\em{
\sindex{commutator jet}%
\sindex{conserved one-form}%
\sindex{symplectic form}%
Let~$(\H,\F,\rho)$ be a causal fermion system on a finite-dimensional Hilbert space. For any symmetric
operator~$S\in L(\H)$, we define the corresponding \textit{commutator jet} by 
\beq
\mathfrak{C}_S:=(0,\mathcal{C}_S),\quad\mbox{with}\quad \mathcal{C}_S(x):= \cI[S,x]\quad\mbox{for all~$x\in \F$}.
\eeq
Prove the following identity between the conserved one-form and the conserved symplectic form:
\beq
\gamma^\Omega_\rho\big((0,[\mathcal{C}_A,\mathcal{C}_B])\big)=-\frac{1}{2}\,\sigma^\Omega_\rho(\mathfrak{C}_A,\mathfrak{C}_B),
\eeq
where~$[\mathcal{C}_A,\mathcal{C}_B]$ denotes the commutator of vector fields on~$\F$. 
}} \end{Exercise}

\begin{Exercise} (Representation of the commutator inner product) \label{exdynwave1}
\sindex{commutator inner product}%
{\em{ The goal of this exercise is to represent the commutator inner product in a form similar to~\eqref{cli}.
\bitem
\item[(a)] Show that the first variations of the Lagrangian can be written as
\beq \delta \L(x,y) = 2 \re \Tr_{S_xM} \!\big( Q(x,y)\, \delta P(y,x) \big) \eeq
\nindex{afa@$Q(x,y)$ -- kernel describing first variation of causal Lagrangian}%
with a suitable kernel~$Q(x,y) : S_y \rightarrow S_x$. Show that this kernel can be chosen to be symmetric;
that is, that~$Q(x,y)^* = Q(y,x)$.
\item[(b)] Show that the variation described by the commutator jet in~\eqref{jdef} and~\eqref{Adef}
corresponds to the variation of the integrand in~\eqref{IOr}
\beq \big( D_{1,\j(u)} - D_{2,\j(u)} \big) \L(x,y) = -2\cI \:\big( \Sl \psi(x) \:|\:Q(x,y)\, \psi(y) \Sr_x 
- \Sl \psi(y) \:|\:Q(y,x)\, \psi(x) \Sr_y \big) \:. \eeq
\item[(c)] Use the polarization formula~\eqref{commsesqui} to conclude that~$\la u | v \ra^\Omega_\rho$
has the representation~\eqref{cli} with~$\psi = \psi^u$ and~$\phi = \psi^v$.
\eitem
{\em{Hint:}} Details on this construction can be found in~\cite[Section~3]{dirac}.
}} \end{Exercise}

\begin{Exercise} (Extending the commutator inner product) \label{exdynwave2}
\sindex{commutator inner product}%
{\em{ The goal of this exercise is to illustrate how the commutator inner product can be
extended to more general wave functions. To this end, assume that we are given a
space of wave function~${\mathscr{W}}$ which all satisfy the dynamical wave equation~\eqref{dwe}
with a suitable kernel~$Q^\text{dyn}(x,y)$.
Prove that, under these assumptions, the inner product~\eqref{cli} is conserved
for any~$\psi, \phi \in {\mathscr{W}}$. \\
{\em{Hint:}} In a first step it seem a good idea to choose~$\Omega=\Omega_t$ as the past of
an equal time hypersurface and to differentiate with respect to~$t$.
More generally, one can consider the difference of~\eqref{cli} for two sets~$\Omega$ and~$\Omega'$
which differ by a compact set.
}} \end{Exercise}

\begin{Exercise} (Representing the Hilbert space scalar product in a surface layer) \label{excommnogo}
{\em{ The goal of this exercise is to explain why the sesquilinear form~$\la .|. \ra^\Omega_\rho$
{\em{cannot}} represent the scalar product on the whole Hilbert space.
To this end, let us assume conversely that
\beq \label{Ccondtot}
\la u|u \ra^\Omega_\rho = c\, \la u|u \ra_\H \qquad \text{for all~$u \in \H$ and~$c \neq 0$}
\eeq
and derive a contradiction. For technical simplicity, we assume that~$\H$ is finite-dimensional
and disregard all issues of convergence of integrals.
\bitem
\item[(a)] Show that the surface layer integral can be written as
\beq \label{commstruct}
\la u|u \ra^\Omega_\rho =  \cI \int_M \big\la u \,\big|\, [x,  B(x)]\, u \big\ra_\H\: \dd\rho(x)
\eeq
with~$B(x)$ a suitable family of operators on the Hilbert space.
\item[(b)] Carry out the $x$-integral formally to obtain the representation
\beq \la u|u \ra^\Omega_\rho = \la u \,|\,C u \ra_\H\: \dd\rho(x) \eeq
with a trace-free operator~$C$. {\em{Hint:}} Make use of the commutator structure 
of the integrand in~\eqref{commstruct}.
\item[(c)] Conclude from~\eqref{Ccondtot} that~$C$ is a multiple of the identity operator.
Why is this a contradiction?
\eitem
{\em{Hint:}} More details on this argument can be found in~\cite[Appendix~A]{current}.
}} \end{Exercise}

\begin{Exercise} (On the surface layer inner product) {\em{
\sindex{surface layer inner product}%
The goal of this exercise is to show that, under a suitable restriction of the jet space, the surface-layer inner product is indeed positive. On~$\F=\R^2$ we define the Lagrangian 
\beq
\L(x,y)=\frac{1}{2}\,\eta(x_1-y_1)\,(x_2-y_2)^2,\quad \mbox{where }\ \eta\in C_0^\infty(\R,\R^+).
\eeq
Let~$M=\R\subset\F$ equipped with the canonical one-dimensional Lebesgue measure and consider the set of jets
\begin{align}
\mathfrak{J}:=\bigg\{(0,u)\:\bigg|\: & u=\sum_{i=1}^2 u_i\partial_i\in T\F\ \notag \\
& \mbox{with}\  u_1(t,0)=0\ \mbox{and}\  \partial_1{u_2}(t,0)\le 0 \ \mbox{for all~$t\in\R$}\bigg\} \:.
\end{align}
Let~$\Omega_t:=(-\infty,t)\subset M$. Show that the surface-layer inner product~$(\,\cdot\,,\,\cdot\,)^{\Omega_t}|_{\mathfrak{J}\times\mathfrak{J}}$ is positive semi-definite. \textit{Hint:} Remember that jets are never differentiated in expressions like~$\nabla_{i,\mathfrak{v}}\nabla_{j,\mathfrak{u}}$.
}} \end{Exercise}

\chapter{Positive Functionals} \label{secpositive}
\section{Motivation and Setup} \label{secintro}
Many physical quantities have a definite sign
(for example, positive mass, positive energy, negative charge, etc.).
With this in mind, it is an important question whether the structure of a causal variational
principle gives rise to positive functionals. We now briefly explain the structural results
known at present. These were obtained with two different methods.
One method is to make us of the fact that, given a minimizer of a variational principle, second variations 
are always non-negative. This method was worked out in~\cite{positive}, and we will give an outline
in Sections~\ref{secl} and~\ref{secJ}.
The second method is to use that the action of a given minimizing measure~$\rho$
is smaller than the action of any other test measure~$\tilde{\rho}$. By a suitable choice of~$\tilde{\rho}$,
one gets surface layer integral with a definite sign. This second method is explored in detail in~\cite{matter},
and applications are worked out. Here, we only explain the basic idea in Section~\ref{secposnonlin}.

For technical simplicity, we restrict attention to causal variational principles
in the {\em{non-compact smooth setting}} (see~\eqref{Lsmooth} and Section~\ref{secnoncompact}).

\section{Positivity of the Hessian of~$\ell$} \label{secl}
Let~$\rho$ be a minimizer of the causal action. 
According to the EL equations~\eqref{EL1}, the function~$\ell$ is minimal on~$M$.
This clearly implies that its Hessian (as computed in any chart) is positive semi-definite; that is,
\sindex{Hessian of~$\ell$}%
\beq \label{hessian}
D^2 \ell(x) \geq 0 \qquad \text{for all~$x \in M := \supp \rho$} \:.
\eeq
This is the first non-negative quantity obtained from the fact that~$\rho$ is a minimizer.
In view of the restricted EL equations~\eqref{ELrestricted2}, the zero and first order derivatives 
of~$\ell$ vanish for all~$x \in M$.
Adding such lower derivative terms, we can write~\eqref{hessian} with
jet derivatives as
\beq \nabla^2 \ell|_x(\u,\u) \geq 0 \qquad \text{for all~$x \in M$}\:, \eeq
where, following our conventions~(i) and~(ii) on page~\pageref{ConventionPartial},
\beq \nabla^2 \ell|_x(\u,\u) := a(x)^2\, \ell(x) + 2 \,a(x)\, D_u \ell(x) + D^2 \ell|x(u,u) \:. \eeq
Integrating over~$M$ gives the following result:
\begin{Prp} \label{prppos1} Let~$\rho$ be a minimizer of the causal action. Then
\beq \int_M \nabla^2 \ell|_x(\u,\u)\: \dd\rho(x) \geq 0 \qquad \text{for all~$\u \in \J_0$} \:. \eeq
\end{Prp} \noindent

\section{Positivity of Second Variations Generated by Jets} \label{secJ}
\sindex{second variations!of causal action|textbf}%
We now analyze second variations for a 
special class of variations of the measure~$\rho$ to obtain another positive
functional on jets.
Similar to~\cite[Section~3]{jet}, we consider measures of the form
\beq \label{paramtilrho}
\tilde{\rho}_\tau = (F_\tau)_* \big( f_\tau \,\rho \big) \qquad \text{for~$\tau \in (-\tau_{\max}, \tau_{\max})$}
\eeq
with smooth mappings
\beq f \in C^\infty((-\tau_{\max}, \tau_{\max}) \times M, \R^+) \qquad \text{and} \qquad
F \in C^\infty((-\tau_{\max}, \tau_{\max}) \times M, \F) \:, \eeq
where the star denotes the push-forward measure defined
by~$((F_\tau)_*\mu)(\Omega) = \mu ( F_\tau^{-1} (\Omega))$
(for details, see the preliminaries in Section~\ref{secbasicmeasure} or,
for example, \cite[Section~3.6]{bogachev}).
We assume that for~$\tau=0$ the variation is trivial, \eqref{triv0}.
Moreover, for technical simplicity, we assume that~$F_\tau$ and~$f_\tau$ are
 {\em{trivial outside a compact set}}~$K \subset M$, meaning that
\sindex{jet!compactly supported}%
\beq %\label{triv}
f_\tau|_{M \setminus K} \equiv 1 \qquad \text{and} \qquad F_\tau|_{M \setminus K} \equiv \1\:. \eeq
Finally, in order to satisfy the volume constraint on the right-hand side of~\eqref{totvol}, we assume that
\beq \label{vol0}
\int_K f_\tau(x)\: \dd\rho(x) = \rho(K) \qquad \text{for all~$\tau \in (-\tau_{\max}, \tau_{\max})$}\:.
\eeq
Then the transformation~\eqref{paramtilrho} is described infinitesimally by the smooth
and compactly supported jet
\beq \u =(a,u) := \big( \dot{f}_0, \dot{F}_0 \big) \in \J_0 \:, \eeq
where the dot denotes the $\tau$-derivative.
Moreover, we differentiate the volume con\-straint~\eqref{vol0} to obtain
\beq \label{vol1}
\int_K a(x)\: \dd\rho(x) = 0\:.
\eeq

We now compute the first and second variations of the action. Combining~\eqref{integrals}
with the definition of the push-forward measure, we obtain
\begin{align}
&\Sact\big(\tilde{\rho}_\tau \big) - \Sact(\rho) \notag \\
&= 2 \int_K \dd\rho(x) \int_{M \setminus K} \dd\rho(y)\: 
\Big( f_\tau(x)\: \L\big(F_\tau(x), y \big) - \L(x,y) \Big) \notag \\
&\quad\:+ \int_K \dd\rho(x) \int_K \dd\rho(y)\:\Big( f_\tau(x)\: f_\tau(y) \:\L\big(F_\tau(x), F_\tau(y) \big) - \L(x,y) \Big) \:. \label{Sdiff2}
\end{align}
The first variation vanishes because
\begin{align}
\frac{\dd}{\dd\tau} \Sact\big(\tilde{\rho}_\tau \big) \Big|_{\tau=0} &=
2 \int_K \dd\rho(x) \int_M \dd\rho(y) \:\nabla_{1,\u} \L(x,y) \notag \\
&= 2 \int_K \nabla_\u \Big( \ell(x) + \s \Big)\: \dd\rho(x) = 0 \:,
\end{align}
where in the last step we used~\eqref{ELrestricted2} and~\eqref{vol1}
(and~$\nabla_1$ denotes the partial derivative acting on the first argument of the Lagrangian).
Differentiating~\eqref{Sdiff2} twice, the second variation is computed to be
\begin{align}
&\frac{\dd^2}{\dd\tau^2} \Sact\big(\tilde{\rho}_\tau \big) \Big|_{\tau=0} =
2 \int_K \dd\rho(x) \int_K \dd\rho(y) \:\nabla_{1,\u} \nabla_{2,\u} \L(x,y) \notag \\
&+ 2 \int_K \dd\rho(x) \int_M \dd\rho(y) \:\Big( a(x)\: D_{1,u} \L(x,y) \notag \\
& \qquad \qquad\qquad\qquad\qquad+ D_{1,u} D_{1,u} \L(x,y) + 
\big( \ddot{f}_0(x) + D_{1,\ddot{F}_0} \big) \L(x,y) \Big) \:.
\end{align}
In the last line we can carry out the $y$-integration using~\eqref{elldef}.
Applying the EL equations~\eqref{ELrestricted2}, we obtain
\begin{align}
\int_K \dd\rho(x) \int_M a(x)\: D_{1,u} \L(x,y) \:\dd\rho(y) &= 0 \\
\int_K \dd\rho(x) \int_M D_{1,u} D_{1,u} \L(x,y) \: \dd\rho(y) 
&= D^2 \ell|_x(u,u) = \nabla^2 \ell|_x(\u, \u) \\
\int_K \dd\rho(x) \int_M  \:\big( \ddot{f}_0(x) + D_{1,\ddot{F}_0} \big) \L(x,y) \: \dd\rho(y) 
&= \int_K \ddot{f}_0(x)\: \s\: \dd\rho(x) \overset{\eqref{vol0}} = 0\:.
\end{align}
We thus obtain the simple formula
\beq %\label{var2}
\frac{1}{2}\:\frac{\dd^2}{\dd\tau^2} \Sact\big(\tilde{\rho}_\tau \big) \Big|_{\tau=0} =
\int_K \dd\rho(x) \int_K \dd\rho(y) \:\nabla_{1,\u} \nabla_{2,\u} \L(x,y) 
+ \int_K \nabla^2 \ell|_x(\u,\u)\: \dd\rho(x)\:. \eeq
Since~$\rho$ is a minimizer and the first variation vanishes,
the second variation is necessarily non-negative, giving rise
to the inequality
\beq \label{varint}
\int_M \dd\rho(x) \int_M \dd\rho(y) \:\nabla_{1,\u} \nabla_{2,\u} \L(x,y) 
+ \int_M \nabla^2 \ell|_x(\u,\u)\: \dd\rho(x) \geq 0\:,
\eeq
subject to the condition that the scalar component of the jet~$\u$ must satisfy the volume constraint~\eqref{vol1}.
In the next proposition, we remove this condition with a limiting procedure:
\begin{Prp} \label{prppos2} Let~$\rho$ be a minimizer of the causal action. Then
the inequality~\eqref{varint} holds for all~$\u \in \J_0$.
\end{Prp}
\Proof Let~$\u=(a,u) \in \J_0$ be a jet that violates
the volume constraint~\eqref{vol1}. Then, choosing a compact set~$\Omega \subset M$
with~$\rho(\Omega)>0$, the jet~$\hat{\u} := (\hat{a}, u)$ with
\beq \label{cdef}
\hat{a}(x) = a(x) - c(\Omega) \,\chi_\Omega(x) \qquad \text{and} \qquad
c(\Omega) := \frac{1}{\rho(\Omega)} \:\int_\Omega a(x)\: \dd\rho(x)
\eeq
(where~$\chi_\Omega$ is the characteristic function) does satisfy~\eqref{vol1}.
Choosing the scalar variation~$f_\tau = (1-\tau) + \tau \hat{a}$ and
a family of diffeomorphisms~$F_\tau$ with~$\dot{F}_0 = u$, we obtain a variation
which satisfies the volume constraint~\eqref{vol0} (note that~$\ddot{f}=0$).
Clearly, due to the characteristic function, the jet~$\hat{\u}$ is no longer smooth,
but it has again compact support, and an approximation argument using Lebesgue's dominated convergence
theorem shows that the inequality~\eqref{varint} also holds for~$\hat{\u}$.
Expanding in powers of~$c$, we thus obtain the inequality
\begin{align}
0 &\leq \int_K \dd\rho(x) \int_K \dd\rho(y) \:\nabla_{1,\u} \nabla_{2,\u} \L(x,y) 
+ \int_K \nabla^2 \ell|_x(\u,\u)\: \dd\rho(x) \notag \\
&\qquad - 2c \int_M \dd\rho(x) \int_K \dd\rho(y) \:\chi_\Omega(x) \nabla_{2,\u} \L(x,y) \notag \\
&\qquad + c^2 \int_M \dd\rho(x) \int_M \dd\rho(y) \:\chi_\Omega(x) \, \chi_\Omega(y)\: \L(x,y) \notag \\
&\qquad + \int_M \Big( -2 c \,\chi_\Omega(x)\: \nabla_\u \ell(x) + c^2\: \chi_\Omega(x)^2\: \ell(x) \Big)\: \dd\rho(x)
\end{align}
(the integrand in the last line arises from the contributions to~$\nabla^2 \ell|x(\u,\u)$
involving the scalar components of the jets).
The last line vanishes due to the restricted EL equations~\eqref{ELrestricted2}. Hence
\begin{align}
&\int_K \dd\rho(x) \int_K \dd\rho(y) \:\nabla_{1,\u} \nabla_{2,\u} \L(x,y) + \int_K \nabla^2 \ell|_x(\u,\u)\: \dd\rho(x) \notag \\
&\geq 2c \int_K \dd\rho(x) \int_\Omega \dd\rho(y) \:\nabla_{1,\u} \L(x,y) 
-c^2 \int_K \dd\rho(x) \int_K \dd\rho(y) \: \L(x,y) =: A(\Omega) \:.
\end{align}

We now let~$(\Omega_n)_{n \in \N}$ be an exhaustion of~$M$ by compact sets.
We distinguish the two cases when~$\rho(M)$ is finite and infinite and treat these cases separately.
If the total volume~$\rho(M)$ is finite, one can take the limit~$n \rightarrow \infty$ with
Lebesgue's dominated convergence theorem to obtain
\begin{align}
\lim_{n \rightarrow \infty} \int_K &\dd\rho(x) \int_{\Omega_n} \dd\rho(y) \:\nabla_{1,\u} \L(x,y) =
\int_K \dd\rho(x) \int_M \dd\rho(y) \:\nabla_{1,\u} \L(x,y) \notag \\
&= \int_K \nabla_{\u} \bigg( \ell(x) + \s \bigg) \:\dd\rho(x)
= \s \int_K a(x)\: \dd\rho(x) \\
\lim_{n \rightarrow \infty} A(\Omega_n) &= 2\,c(M) \:\s \int_K a(x)\: \dd\rho(x) - c(M)^2 \, \rho(M)\: \s \notag \\
&= \frac{\nu}{2 \rho(M)} \:\left( \int_K a(x)\: \dd\rho(x) \right)^2 \geq 0\:,
\end{align}
where in the last line we substituted the value of~$c(M)$ in~\eqref{cdef}.

In the remaining case that the volume~$\rho(M)$ is infinite, we estimate the terms as follows,
\begin{align}
&c(\Omega_n)^2 \int_K \dd\rho(x) \int_K \dd\rho(y) \: \L(x,y) \notag \\
&\quad \leq c(\Omega_n)^2 \int_K \dd\rho(x) \int_M \dd\rho(y) \: \L(x,y)
= c(\Omega_n)^2 \: \s\: \rho(K) \rightarrow 0 \\
&\int_K \dd\rho(x) \int_{\Omega_n} \dd\rho(y) \:\nabla_{1,\u} \L(x,y)
\rightarrow \int_K \dd\rho(x) \int_M \dd\rho(y) \:\nabla_{1,\u} \L(x,y) \notag \\
&\quad = \int_K \nabla_{\u} \bigg( \ell(x) + \s \bigg) \:\dd\rho(x)
%= \int_K \Big(\nabla_{\u} \s\Big)\: \dd\rho(x)
= \s \int_K a(x)\: \dd\rho(x) \:.
\end{align}
As a consequence, $A(\Omega_n)$ converges to zero as~$n \rightarrow \infty$.
This concludes the proof.
\QED

We note that, restricting attention to scalar jets (that is, $\u = (a,0)$ with~$a$
a real-valued function on~$M$), the inequality in Proposition~\ref{prppos2} reduces to
\beq \label{scalpos}
\int_M \dd\rho(x) \int_M \dd\rho(y) \:a(x)\: \L(x,y) \: a(y) \geq 0 \qquad \text{for all~$a \in C^\infty_0(M)$}\:.
\eeq
This inequality was first derived in~\cite[Lemma~3.5]{support} and used for the
analysis of minimizing measures. For more details, see also Exercise~\ref{exscalhilbert}.

\subsection{Application: Hilbert Spaces of Jets} \label{sechilbert}
As an application, we now explain how our positive functionals can be used to endow
spaces of jets in spacetime with Hilbert space structures.
These Hilbert space structures should be very useful 
because they make functional analytic tools applicable to the 
analysis of the jet spaces and the causal action principle.

We introduce the following bilinear forms on~$\J_0$,
\begin{align}
\lla \u, \v \rra &:= 
\int_M \dd\rho(x) \int_M \dd\rho(y) \:\nabla_{1,\u} \nabla_{2,\v} \L(x,y) 
+ \int_M \nabla^2 \ell|_x(\u,\v)\: \dd\rho(x) \label{sp1} \\
\llb \u, \v \rrb &:= \la \u, \v \ra + \int_M \nabla^2 \ell|_x(\u,\v)\: \dd\rho(x) \:. \label{sp2}
\end{align}
By Propositions~\ref{prppos1} and~\ref{prppos2}, both bilinear forms are positive semi-definite.
The second bilinear form has the advantage that it is bounded from below by the
bilinear form introduced in Proposition~\ref{prppos1}.
Dividing out the null space and forming the completion gives real
Hilbert spaces of jets denoted by~$\H^{\la.,.\ra}$ and~$\H^{\la\!\la.,.\ra\!\ra}$, respectively.
Obviously,
\beq \lla \u, \u \rra \leq \llb \u, \u \rrb \:, \eeq
giving rise to a norm-decreasing mapping~$\H^{\la\!\la.,.\ra\!\ra} \rightarrow \H^{\la .,. \ra}$.

For the scalar components of the jets, the two scalar products~\eqref{sp1} and~\eqref{sp2} obviously agree.
But they are quite different for the vector components.
In order to understand this difference, it is instructive to consider
a jet~$\u=(0,u)$ which describes a {\em{symmetry}}
of the Lagrangian; that is (for details see~\cite[Section~3.1]{noether}),
\beq \big( D_{1,u} + D_{2,u} \big) \L(x,y) = 0 \qquad \text{for all~$x,y \in M$}\:. \eeq
For this jet, a direct computation shows that
\beq \lla \u, \u \rra = 0 \:. \eeq
Hence, symmetry transformations lie in the kernel of the bilinear form~$\lla ., . \rra$
and thus correspond to the zero vector in~$\H^{\la.,.\ra}$.
Generally speaking, the scalar product~$\lla.,.\rra$ makes it possible to
disregard symmetry transformations of the causal Lagrangian.
However, jets describing symmetry transformations do, in general, correspond to
non-zero vectors of the Hilbert space~$\H^{\la\!\la.,.\ra\!\ra}$.

\subsection{Application: A Positive Surface Layer Integral} \label{secosi}
\sindex{surface layer integral!positive}%
We now derive a surface layer integral that is not necessarily conserved,
but has a definite sign.
Similarly as explained at the beginning of Section~\ref{sechilbert},
this can be used to endow the jet space with a Hilbert structure.
But in contrast to the scalar products in Section~\ref{sechilbert}, where the jets
were integrated over spacetime, here the scalar product is given as a surface
layer integral. This should be useful for analyzing the dynamics of jets
in spacetime.

\begin{Prp} \label{prposi}
Let~$\v$ be a solution of the linearized field equations~\eqref{linfield}. Then
for any compact~$\Omega \subset M$, the following surface layer integral is positive,
\beq -\int_\Omega \dd\rho(x) \int_{M \setminus \Omega} \dd\rho(y) \:\nabla_{1,\v} \nabla_{2,\v} \L(x,y) \geq 0 \:. \eeq 
\end{Prp}
\Proof Denoting the components of~$\v$ by~$\v=(b,v)$,
we evaluate~\eqref{linfield} for~$\u=\v$ and integrate over~$\Omega$.
The resulting integrals can be rewritten as follows,
\begin{align}
0 &= \int_\Omega \dd\rho(x) \int_M \dd\rho(y) \:
\nabla_{1,\v} \big( \nabla_{1,\v} + \nabla_{2,\v} \big) \L(x,y) - \s \int_\Omega b(x)^2\, \dd\rho(x) \notag \\
&= \int_\Omega \nabla^2 \ell|_x(\v,\v) \: \dd\rho(x) + 
\int_\Omega \dd\rho(x) \int_M \dd\rho(y) \:\nabla_{1,\v} \nabla_{2,\v} \L(x,y) \notag \\ 
&= \int_\Omega \nabla^2 \ell|_x(\v,\v) \: \dd\rho(x) + 
\int_\Omega \dd\rho(x) \int_\Omega \dd\rho(y) \:\nabla_{1,\v} \nabla_{2,\v} \L(x,y) \label{t1} \\ 
&\quad\, + \int_\Omega \dd\rho(x) \int_{M \setminus \Omega} \dd\rho(y) \:\nabla_{1,\v} \nabla_{2,\v} \L(x,y) \:. \label{t2}
\end{align}
Using characteristic functions, the expression~\eqref{t1} can be written as
\beq \int_M \nabla^2 \ell|_x(\chi_\Omega \v, \chi_\Omega \v) \: \dd\rho(x) + 
\int_M \dd\rho(x) \int_M \dd\rho(y) \:\nabla_{1,\chi_\Omega \v} \nabla_{2,\chi_\Omega \v} \L(x,y) \:. \eeq
Approximating the jet~$\chi_\Omega \v$ by smooth jets with compact support,
one finds that the integrals in~\eqref{t1} are non-negative by Proposition~\ref{prppos2}.
Therefore, the last summand~\eqref{t2} must be non-positive. This gives the result.
\QED

We finally remark that in~\cite[Section~6]{action}, the surface layer integral in the last proposition
is computed in Minkowski space. 

\section{A Positive Nonlinear Surface Layer Integral} \label{secposnonlin}
\sindex{surface layer integral!positive nonlinear}%
In this section we briefly mention another method for obtaining a positive surface layer integral.
This method and the corresponding positivity results will not be used later in this book.
We refer the reader interested in more explanations and applications of this method to~\cite{matter}.

As in Section~\ref{secosinonlin} we again consider two measures: A measure~$\rho$ that describes the
vacuum spacetime, and another measure~$\tilde{\rho}$ that typically describes an interacting spacetime.
We assume that the vacuum measure is a {\em{minimizer}} of the causal action principle
as defined in Section~\ref{secnoncompact} (see~\eqref{Sdiffpos} and~\eqref{integrals}).
We choose subsets~$\Omega \subset M$
and~$\tilde{\Omega} \subset \tilde{M}$ having the same finite volume,
\beq \rho(\Omega) = \tilde{\rho}(\tilde{\Omega}) < \infty \:. \eeq
In order to construct an admissible test measure~$\hat{\rho}$, we ``cut out''~$\Omega$ from~$\rho$
and ``glue in'' the set~$\tilde{\Omega}$; that is,
\beq \hat{\rho} := \chi_{\tilde{\Omega}}\, \tilde{\rho} + \chi_{M \setminus \Omega}\, \rho \:. \eeq
The measure~$\hat{\rho}$ differs from~$\rho$ only on a set of finite volume
and preserves the volume constraint (see~\eqref{totvol}). Therefore, we obtain from~\eqref{Sdiffpos} and~\eqref{integrals} (with~$\tilde{\rho}$ replaced by~$\hat{\rho}$) that
\begin{align}
0 &\leq \big( \Sact(\hat{\rho}) - \Sact(\rho) \big) \notag \\
&= 2 \int_\F \dd(\hat{\rho} - \rho)(x) \int_M \dd\rho(y)\: \L(x,y)
+ \int_\F \dd(\hat{\rho} - \rho)(x) \int_M \dd(\hat{\rho} - \rho)(y)\: \L(x,y) \notag \\
&= 2 \int_{\tilde{\Omega}} \dd\tilde{\rho}(x) \int_M \dd\rho(y)\: \L(x,y)
- 2 \int_{\Omega} \dd\rho(x) \int_M \dd\rho(y)\: \L(x,y) \notag \\
&\quad\: + \int_{\tilde{\Omega}} \dd\tilde{\rho}(x) \int_{\tilde{\Omega}} \dd\tilde{\rho}(y)\: \L(x,y)
- 2 \int_{\tilde{\Omega}} \dd\tilde{\rho}(x) \int_\Omega \dd\rho(y)\: \L(x,y) \notag \\
&\quad\:+ \int_\Omega \dd\rho(x) \int_\Omega \dd\rho(y)\: \L(x,y) \notag \\
&= 2 \int_{\tilde{\Omega}} \dd\tilde{\rho}(x) \int_{M \setminus \Omega} \dd\rho(y)\: \L(x,y)
- 2 \int_{\Omega} \dd\rho(x) \int_{M \setminus \Omega} \dd\rho(y)\: \L(x,y) \notag \\
&\quad\: + \int_{\tilde{\Omega}} \dd\tilde{\rho}(x) \int_{\tilde{\Omega}} \dd\tilde{\rho}(y)\: \L(x,y)
- \int_\Omega \dd\rho(x) \int_\Omega \dd\rho(y)\: \L(x,y) \:. 
\end{align}
We thus obtain the inequality
\beq \label{pososi}
\begin{split}
& 2 \int_{\tilde{\Omega}} \dd\tilde{\rho}(x) \int_{M \setminus \Omega} \dd\rho(y)\: \L(x,y)
\leq  2 \int_{\Omega} \dd\rho(x) \int_{M \setminus \Omega} \dd\rho(y)\: \L(x,y) \\
&\quad - \int_{\tilde{\Omega}} \dd\tilde{\rho}(x) \int_{\tilde{\Omega}} \dd\tilde{\rho}(y)\: \L(x,y)
- \int_\Omega \dd\rho(x) \int_\Omega \dd\rho(y)\: \L(x,y) \:.
\end{split}
\eeq
The left-hand side of this inequality coincides with the first summand in the nonlinear
surface layer integral as introduced in~\eqref{osinl}.
However, the second summand in~\eqref{osinl} is now missing.
We can regard the left-hand side of~\eqref{pososi} again as a nonlinear surface layer integral,
but with a somewhat different mathematical structure.
It is not conserved, but it satisfies instead an {\em{in}}equality.
The first summand on the right-hand side of~\eqref{pososi} can be interpreted as the
surface area of~$\partial \Omega$. The two other summands in~\eqref{pososi}, on the other hand,
can be regarded as volume integrals over~$\tilde{\Omega}$ and~$\Omega$, respectively.

This method can be generalized and adapted in various ways, also to cases
when~$\tilde{\Omega}$ and~$\Omega$ do not have the same volume.
Moreover, the resulting inequality can be written in a particularly useful form if also the measure~$\tilde{\rho}$
satisfies the EL equations.
We finally remark that, assuming that~$\tilde{\rho}$ is again of the form~\eqref{paramtilrho}
and expanding in powers of~$\tau$, one gets inequalities for
surface layer integrals involving jet derivatives.

\section{Exercises}

\begin{Exercise} {{(Positive functionals for the causal variational principle on~$\R$)}} {\em{
We return to the causal variational principles on~$\R$ corresponding to the Lagrangian~$\L_2$
introduced in Exercise~\ref{excvpR}. Let~$\rho=\delta$ be the unique minimizer.
\bitem
\item[(a)] Compute the function~$\ell(x)$ and verify that its Hessian is positive (see~\eqref{hessian}).
\item[(b)] Compute the functional in Proposition~\ref{prppos2} for the jets~$\u=(0, \partial_x)$
and~$\u=(1,0)$.
\item[(c)] What are the resulting scalar products~\eqref{sp1} and~\eqref{sp2}?
\eitem
}} \end{Exercise}

\begin{Exercise} {{(Positive functionals for the causal variational principle on~$S^1$)}} {\em{
We return to the causal variational principle on~$\R$ introduced in Exercise~\ref{excvpS1}.
Let~$\rho$ be a minimizing measure~\eqref{rhominS1} for~$0 < \tau < 1$.
We choose~$\Jtest_0$ as the four-dimensional vector space generated by the scalar jet~$(1,0)$ and the
vector jet~$(0, \partial_\varphi)$ at the two points.
\bitem
\item[(a)] Compute the function~$\ell(x)$ and verify that its Hessian.
\item[(b)] Compute the bilinear form in Proposition~\ref{prppos2}.
\item[(c)] What are the resulting scalar products~\eqref{sp1} and~\eqref{sp2}?
What are the resulting Hilbert spaces of jets~$\H^{\la .,. \ra}$ and~$\H^{\la\!\la.,.\ra\!\ra}$?
Which dimensions do they have? How can this result be understood in view of the
space of linearized solutions as computed in Exercise~\ref{excvpS1lin}?
\eitem
}} \end{Exercise}

\begin{Exercise} \label{exscalhilbert} {{(A positive operator on scalar jets)}} {\em{
In this exercise we specialize the statement of Proposition~\ref{prppos2} to scalar jets
and work out a few consequences.
\bitem
\item[(a)] Show that for jets of the form~$\u=(a,0)$, the statement of Proposition~\ref{prppos2}
reduces to the inequality~\eqref{scalpos}.
\item[(b)] Let~$\rho$ be a minimizing measure and~$x_0, \ldots, x_N \in M$
be a finite number of spacetime points. Show that the {\em{Gram matrix}}~$L$ defined by
\sindex{Gram matrix}%
\beq L=\Big(\L(x_i,x_j)\Big)_{i,j=0,\ldots,N} \eeq
is symmetric and positive semi-definite.
\item[(c)] Show that the operator~$\L_\rho$ defined by
\sindex{second variations!of causal action}%
\nindex{afb@$L_\rho$ -- positive operator arising from second scalar variations}%
\begin{align}
&\L_\rho \::\: C^\infty_0(M) \subset L^2(M, \dd\rho) \rightarrow L^2(M, \dd\rho)\:,\\
&(\L_\rho \psi)(x) := \int_\F \L(x,y)\: \psi(y)\: \dd\rho(y)
 \end{align}
is a symmetric, densely defined operator on the Hilbert space~$L^2(M, \dd\rho)$.
Prove that this operator is positive semi-definite.
\eitem
}} \end{Exercise}

\begin{Exercise} {\em{
\sindex{surface layer integral!positive nonlinear}%
The goal of this exercise is to explore the positive nonlinear surface layer integral of Section~\ref{secposnonlin}
in the limiting case when the measures of the sets~$\Omega$ and~$\tilde{\Omega}$ tend to zero.
For technical simplicity, let us assume that for given~$x \in M$ and~$y \in \tilde{M}$,
there are sequences of open neighborhoods~$\Omega_k$ of~$x$ and~$\tilde{\Omega}_k$ of~$y$
with~$\rho(\Omega_k) = \tilde{\rho}(\tilde{\Omega}_k)$ for all~$k \in \N$ and~$\lim_{k \rightarrow \infty}
\rho(\Omega_k) = 0$. Show that, in the limit~$k \rightarrow \infty$,
the inequality~\eqref{pososi} reduces to the inequality
\beq %\label{poslin}
\ell(y) \geq \ell(x)\:. \eeq
Thus we get back the EL equation~\eqref{EL2}.

In view of this limiting case, the positive nonlinear surface layer integral in Section~\ref{secposnonlin}
can be regarded as a refined nonlinear version of the EL equations.
}} \end{Exercise}

\chapter{Topological and Geometric Structures} \label{seclqg}
This chapter is devoted to the topological and geometric structures of a causal fermion system.
We closely follow the presentation in~\cite{topology} and~\cite{lqg}.

\section{A Topological Vector Bundle} \label{sectopvector}
For the topological structures, it is not essential that the operators in~$\F$ have at most~$n$
positive and at most~$n$ negative eigenvalues (see Definition~\ref{defcfs}).
Instead, it is preferable for the sake of greater generality and broader applicability to
relax this condition in the following way.

\begin{Def} \label{defcfstop}
Given a complex Hilbert space~$(\H, \la .|. \ra_\H)$ 
and parameters~$\p,\q \in \N_0$ with~$\p \leq \q$, we let~$\F \subset \Lin(\H)$ be the set of all
symmetric operators on~$\H$ of finite rank, which (counting multiplicities) have
at most~$\p$ positive and at most~$\q$ negative eigenvalues. On~$\F$, we are given
a positive measure~$\rho$ (defined on a $\sigma$-algebra of subsets of~$\F$).
We refer to~$(\H, \F, \rho)$ as a {\bf{topological fermion system}} of spin
\sindex{topological fermion system}%
\sindex{Riemannian fermion system|textbf}%
signature~$(\p,\q)$.

If~$\p=0$, we call~$(\H, \F, \rho)$ a {\bf{Riemannian fermion system}} of spin dimension~$n:=\q$.
\end{Def} \noindent
Clearly, the case~$\p=\q$ gives back a causal fermion system (see Definition~\ref{defcfs}).
It should be noted that the assumption~$\p \leq \q$ merely is a convention, because otherwise one may
replace~$\F$ with~$-\F$.
The name {\em{Riemannian}} fermion system comes from the fact that in
examples on Riemannian manifolds, the inner product on the fibers is positive definite,
implying that the resulting local correlation operators are negative semi-definite.
For details, see~\cite{topology} or the examples in Exercises~\ref{exrfs1} and~\ref{exrfs2}.
We also note that for Riemannian fermion systems, the causal structure (according to Definition~\ref{def2})
is trivial; see Exercise~\ref{exrfs3}.

In Section~\ref{secbundle}, the notion of a {\em{topological vector bundle}} was introduced
\sindex{vector bundle!topological}%
(see Definition~\ref{deftvb}). Again setting~$M = \supp \rho$, we want to construct a topological vector bundle
having the spin space~$S_x:=x(\H)$ as the fiber at the point~$x \in M$.
To this end, all the spin spaces must have the same dimension and signature,
making it necessary to impose the following condition:
\begin{Def} \label{defreg} The topological fermion system is called {\bf{regular}} if for all~$x \in M$,
\sindex{topological fermion system!regular}%
the operator~$x$ has the maximal possible rank~${\mathfrak{p+q}}$.
\end{Def} \noindent
We note that most of our constructions can be extended to non-regular topological fermion systems by
decomposing~$M$ into subsets on which~$x$ has fixed rank and a fixed number of positive and
negative eigenvalues (for details see~\cite[Section~7]{topology}).

We define~$\B$ as the set of pairs
\beq \B = \{ (x, \psi) \:|\: x \in M,\: \psi \in S_x \} \eeq
and let~$\pi$ be the projection onto the first component.
Moreover, we let~$(Y, \Sl .|. \Sr)$ be an indefinite inner product space of signature~$({\mathfrak{q,p}})$,
and choose~$G = \U({\mathfrak{q,p}})$ as the group of unitary transformations on~$Y$.
In order to construct the bundle charts, for any given~$x \in M$, we choose a unitary
mapping~$\sigma : S_x \rightarrow Y$. By restricting the projection~$\pi_x$ in~\eqref{pidef}
to~$S_y$, we obtain the mapping
\beq \pi_x|_{S_y} \::\: S_y \rightarrow S_x \:. \eeq
In order to compute its adjoint with respect to the spin inner product~\eqref{sspintro},
for~$\psi \in S_x$ and~$\phi \in S_y$, we make the computation
\begin{align}
\Sl \psi \,|\, \pi_x|_{S_y} \,\phi \Sr_x &= -\la \psi | x \phi \ra_\H
= -\la x \psi | \phi \ra_\H = -\la \pi_y \,x \,\psi | \phi \ra_\H
= -\la y \, (y|_{S_y})^{-1}\,\pi_y \,x \,\psi | \phi \ra_\H \notag \\
&= -\big\la (y|_{S_y})^{-1}\, \pi_y \,x\, \psi \,\big|\, y \phi \big\ra_\H
= \Sl (y|_{S_y})^{-1}\, \pi_y \,x\, \psi \,|\, \phi \Sr_y \:.
\end{align}
Hence,
\beq \big( \pi_x|_{S_y} \big)^* =  (y|_{S_y})^{-1}\, \pi_y \,x |_{S_x} \:. \eeq
We now introduce the operator
\beq T_{xy} = \big( \pi_x|_{S_y} \big) \big( \pi_x|_{S_y} \big)^*
= \pi_x \,(y|_{S_y})^{-1}\, \pi_y \,x|_{S_x} \::\: S_x \rightarrow S_x \:. \eeq
By construction, this operator is symmetric and~$T_{xx} =\1$.
We now form the polar decomposition of~$T_{xy}$ to obtain a unitary operator~$U_{xy}$:
By continuity, there is a neighborhood~$U$ of~$x$ such that for all~$y \in U$,
the operator~$T_{xy}$ is invertible and has a unique square root~$\rho_{xy}$
(defined, for example, by the power series~$\sqrt{T_{xy}} = \sqrt{\1 + (T_{xy}-\1)}
= \1 + \frac{1}{2}\: (T_{xy} - \1) + \cdots$).
Introducing the mapping
\beq U_{x,y} = \rho_{xy}^{-1}\: \pi_x|_{S_y} : S_y \rightarrow S_x \:, \eeq
the calculation
\beq U_{x,y} \,U_{x,y}^* = \rho_{xy}^{-1} \:\pi_x|_{S_y}  \big( \pi_x|_{S_y} \big)^* \rho_{xy}^{-1} 
= \rho_{xy}^{-1} \:T_{xy}\: \rho_{xy}^{-1} = \1_{S_x} \eeq
shows that the mapping~$U_{xy}$ is unitary. Moreover, it clearly depends continuously on~$y \in U$.

We define the bundle chart~$\phi_U$ by
\beq \phi_U(y,v) = \big( y,  (\sigma \circ U_{x,y})(v) \big) \:. \eeq
The commutativity of the diagram~\eqref{commdiag} is obvious because~$\phi$ is the identity map
in the first component. Moreover, the transition functions~$g_{UV}$ in~\eqref{changechart} are in~$G$ because
we are working with unitary mappings of the fibers throughout.
We choose the topology on~$\B$ such that all the bundle charts are homeomorphisms.

\begin{Def} \label{defassoc} The topological vector bundle~$\B \rightarrow M$ is referred to as
the vector bundle associated with the regular topological fermion system~$(\H, \F, \rho)$,
or simply the {\bf{associated vector bundle}}.
\sindex{vector bundle!associated to regular topological fermion system}%
\end{Def}

The next result shows that every vector bundle over a manifold can be
realized as the associated vector bundle of a corresponding topological fermion system.
In other words, working with topological fermion systems poses no topological constraints
for the associated vector bundles.
\begin{Thm} Let~$X \rightarrow \scrM$ be a vector bundle over a $k$-dimensional
topological manifold~$\scrM$, whose fibers are isomorphic to
an indefinite inner product space of signature~$(\mathfrak{q, p})$.
Then there is a regular topological fermion system~$(\H, \F, \rho)$ of signature~$(\mathfrak{p,q})$
such that the associated vector bundle (see Definition~\ref{defassoc})
is isomorphic to~$X$. If~$\scrM$ is compact, the Hilbert space~$\H$ can be chosen to
be finite-dimensional.
\end{Thm} \noindent
The proof can be found in~\cite[Section~3.3]{topology}.

\section{Geometric Structures of a Causal Fermion System}
We now outline constructions from~\cite{lqg} which give general notions
of a connection and curvature (see Theorem~\ref{thmspinconnection}, Definition~\ref{defmconn}
and Definition~\ref{defcurvature}). 
So far, these constructions have been carried out only in the case of spin dimension~$n=2$.
This is the most important case because it allows for the description of
Dirac spinors in a four-di\-men\-sio\-nal spacetime.
%We closely follow the presentation in~\cite[Section~3]{nrstg}.

\subsection{Construction of the Spin Connection} \label{secspin}
\sindex{spin connection!of causal fermion system}%
Let~$(\H, \F, \rho)$ be a causal fermion system of spin dimension~$n=2$.
Moreover, we assume that it is {\em{regular}} (see Definition~\ref{defreg}).

An important structure from spin geometry missing so far is Clifford multiplication.
To this end, we need a Clifford algebra represented by symmetric operators on~$S_x$.
For convenience, we first consider Clifford algebras with the maximal number of five
generators; later, we reduce to four spacetime dimensions (see Definition~\ref{defreduce} below).
We denote the set of symmetric linear endomorphisms of~$(S_x, \Sl .|. \Sr_x)$ by~$\Symm(S_x)$;
it is a $16$-dimensional real vector space.

\begin{Def}  A five-dimensional subspace~$K \subset \Symm(S_x)$ is called
a {\bf{Clifford subspace}} if the following conditions hold:
\sindex{Clifford subspace}%
\bitem
\item[{\rm{(i)}}] For any~$u, v \in K$, the anti-commutator~$\{ u,v \} \equiv u v + v u$ is a multiple
of the identity on~$S_x$.
\item[{\rm{(ii)}}] The bilinear form~$\la .,. \ra$ on~$K$ defined by
\beq \frac{1}{2} \left\{ u,v \right\} = \la u,v \ra \, \1 \qquad {\text{for all~$u,v \in K$}} \eeq
is non-degenerate and has signature~$(1,4)$.
\eitem
\end{Def} \noindent
In view of the situation in spin geometry, we would like to distinguish a specific
Clifford subspace. In order to partially fix the freedom in choosing Clifford subspaces,
it is useful to impose that~$K$ should contain a given operator, as is made precise in the
next definitions.

\begin{Def} An operator~$v \in \Symm(S_x)$ is called a {\bf{sign operator}} if~$v^2 = \1$
\sindex{sign operator}%
and if the inner product~$\Sl .|v \,. \Sr\::\: S_x \times S_x \rightarrow \C$ is positive definite.
\end{Def}

\begin{Def} 
Given a sign operator~$v$, the set of {\bf{Clifford extensions}}~$\T^v$
\sindex{Clifford extension|textbf}%
is defined as the set of all Clifford subspaces containing~$v$,
\beq \T^v = \{K {\text{ Clifford subspace with }} v \in K \}\:. \eeq
\end{Def} \noindent
Considering~$x$ as an operator on~$S_x$, this operator has by definition of the spin
dimension two positive and two negative eigenvalues. Moreover, the calculation
\beq \Sl u | (-x) \,u \Sr_x \overset{\eqref{sspintro}}{=} \la u | x^2 u \ra_\H > 0 \quad
\text{for all~$u \in S_x \setminus \{0\}$} \eeq
shows that the operator~$(-x)$ is positive definite on~$S_x$.
Thus, we can introduce a unique sign operator~$s_x$ by demanding that
the eigenspaces of~$s_x$ corresponding to the
eigenvalues~$\pm 1$ are precisely the positive and negative spectral subspaces of
the operator~$(-x)$. 
More constructively, this operator is obtained by
diagonalizing~$(-x)$ and replacing all positive matrix entries with plus one and all negative
matrix entries with minus one (for details, see Exercise~\ref{exsignop}).
This sign operator is referred to as the {\em{Euclidean sign operator}}~$s_x$.
\sindex{sign operator!Euclidean|textbf}%
\nindex{afb1@$s_x$ -- Euclidean sign operator}%
It is worth noting that for Clifford extensions of the Euclidean sign operator, the bilinear form~$\la .,. \ra$
always has Lorentzian signature (see Exercise~\ref{excliffsign}).

A straightforward calculation shows that for two Clifford extensions~$K, \tilde{K} \in \T^v$,
there is a unitary transformation~$U \in \E^{\cI \R v}$ such that~$\tilde{K} = U K U^{-1}$
(for details, see~\cite[Section~3]{lqg}). By dividing out this group action, we obtain a five-dimensional
vector space, endowed with the inner product~$\la ., \ra$. Taking for~$v$ the Euclidean signature
operator, we regard this vector space as a generalization of the usual tangent space.
\begin{Def} \label{deftangent}
The {\bf{tangent space}}~$T_x$ is defined by
\sindex{tangent space! of causal fermion system}%
\nindex{afc@$T_x$ -- tangent space of causal fermion system}%
\beq T_x = \T_x^{s_x} / \exp(i \R s_x)\:. \eeq
It is endowed with an inner product~$\la .,. \ra$ of signature~$(1,4)$.
\end{Def}
Here, the name ``tangent space'' requires an explanation. Since~$M$ does not need to have
a manifold structure, the geometric tangent space (as introduced in Section~\ref{secbundle}
of the preliminaries) cannot be introduced at this stage. Definition~\ref{deftangent} gives another
notion, which does make sense without assuming that~$M$ is a manifold
(for example, it applies also to discrete spacetimes). The basic reason is that our definition of
a tangent space merely describes Clifford structures at each spacetime point, but without using the relations to neighboring spacetime points. In other words, our definition does not incorporate the usual
notion that spacetime is ``approximated infinitesimally'' by the tangent space.
The reason why our notion nevertheless makes sense is that in the example of a causal fermion
system constructed on a smooth spin manifold, the geometric tangent space~$T_x \scrM$
of a spacetime point~$x \in \scrM$ (defined, for example, as in Section~\ref{secbundle}
as equivalence classes of curves through~$x$) can be identified with a distinguished subspace
of~$\Symm(S_x\scrM)$ via Clifford multiplication~$v \in T_x \scrM \mapsto \gamma(v) \in \Symm(S_x\scrM)$.
After identifying~$S_x \scrM$ with~$S_xM$, we thus obtain a distinguished representative
of the tangent space~$T_x$. With this in mind, Definition~\ref{deftangent} can be regarded
as a generalization of the tangent space of a spin manifold, keeping only those structures that can
be defined on the present level of generality.
We remark that a connection between the tangent space~$T_x$ and the local geometry of~$M$
can be made even in the non-smooth setting by working with so-called tangent cone measures
(see Section~\ref{sectangentcone} for the basic concept and~\cite[Section~6.2]{topology} for
the detailed construction).

We next consider two spacetime points. We need the following assumption.
\begin{Def} \label{defproptl}
Two points~$x,y \in M$ are said to be {\bf{properly timelike}}
\sindex{timelike separation!properly|textbf}%
separated if the closed chain~$A_{xy}$ has a strictly positive spectrum and if the corresponding
eigenspaces are definite subspaces of~$S_x$.
\end{Def} \noindent
This definition clearly implies that~$x$ and~$y$ are timelike separated (see Definition~\ref{def2}
and Exercise~\ref{exproper}).
Moreover, the eigenspaces of~$A_{xy}$ are definite if and only if those of~$A_{yx}$ are,
showing that Definition~\ref{defproptl} is again symmetric in~$x$ and~$y$.
As a consequence, the spin space can be decomposed uniquely
into an orthogonal direct sum~$S_x = I^+ \oplus I^-$ of a positive definite subspace~$I^+$ and a
negative definite subspace~$I^-$ of~$A_{xy}$.
This allows us to introduce a unique sign operator~$v_{xy}$ by demanding
that its eigenspaces corresponding to the eigenvalues~$\pm 1$
are the subspaces~$I^\pm$. This sign operator is referred to as the
{\em{directional sign operator}} of~$A_{xy}$.
Having two sign operators~$s_x$ and~$v_{xy}$ at our disposal, we can distinguish unique
corresponding Clifford extensions, provided that the two sign operators satisfy the following
generic condition.
\begin{Def}
Two sign operators~$v, \tilde{v}$ are said to be {\bf{generically separated}} if
their commutator~$[v, \tilde{v}]$ has rank four.
\end{Def}

\begin{Lemma} \label{lemma3}
Assume that the sign operators~$s_x$ and~$v_{xy}$ are generically separated.
\sindex{sign operator!generically separated}
Then there are unique Clifford extensions~$K_x^{(y)} \in \T^{s_x}$ and~$K_{xy} \in \T^{v_{xy}}$ 
and a unique operator~$\rho \in K_x^{(y)} \cap K_{xy}$ with the following properties:
\bitem
\item[{\rm{(i)}}] The relations~$\{ s_x, \rho \} = 0 = \{ v_{xy}, \rho \}$ hold. \\[-1em]
\item[{\rm{(ii)}}] The operator~$U_{xy} := \E^{\cI \rho}$ transforms one Clifford extension to the other,
\beq K_{xy} = U_{xy} \,K_x^{(y)}\, U_{xy}^{-1}\:. \eeq
\item[{\rm{(iii)}}] If~$\{s_x, v_{xy}\}$ is a multiple of the identity,
then~$\rho=0$.
\eitem
The operator~$\rho$ depends continuously on~$s_x$ and~$v_{xy}$.
\end{Lemma} \noindent
We refer to~$U_{xy}$ as the {\em{synchronization map}}.
Exchanging the roles of~$x$ and~$y$, we also have two sign operators~$s_y$ and~$v_{yx}$
at the point~$y$. Assuming that these sign operators are again generically separated,
we also obtain a unique Clifford extension~$K_{yx} \in \T^{v_{yx}}$.

After these preparations, we can now explain the construction of the spin connection~$D$
(for details, see~\cite[Section~3]{lqg}).
For two spacetime points~$x,y \in M$ with the above properties, we want to introduce
an operator
\beq D_{x,y} \::\: S_y \rightarrow S_x \eeq
(generally speaking, by the subscript~$_{xy}$ we always denote an object at the point~$x$, whereas
the additional comma~$_{x,y}$ denotes an operator which maps an object at~$y$
to an object at~$x$).
It is natural to demand that~$D_{x,y}$ is unitary, that~$D_{y,x}$ is its inverse, and that
these operators map the directional sign operators at~$x$ and~$y$ to each other,
\begin{align}
D_{x,y} &= (D_{y,x})^* = (D_{y,x})^{-1} \label{Dc1} \\
v_{xy} &= D_{x,y}\, v_{yx}\, D_{y,x}\:. \label{Dc2}
\end{align}
The obvious idea for constructing an operator with these properties is to take a polar
decomposition of~$P(x,y)$; this amounts to setting
\beq \label{Dfirst}
D_{x,y} = A_{xy}^{-\frac{1}{2}}\: P(x,y)\:.
\eeq
This definition has the shortcoming that it is not compatible with the chosen Clifford extensions.
In particular, it does not give rise to a connection on the corresponding tangent spaces. In order to
resolve this problem, we modify~\eqref{Dfirst} by the ansatz
\beq \label{Dsecond}
D_{x,y} = \E^{\cI \varphi_{xy}\, v_{xy}}\: A_{xy}^{-\frac{1}{2}}\: P(x,y)
\eeq
with a free real parameter~$\varphi_{xy}$. In order to comply with~\eqref{Dc1}, we need to
demand that
\beq \label{phicond}
\varphi_{xy} = -\varphi_{yx} \!\!\!\mod 2 \pi \:;
\eeq
then~\eqref{Dc2} is again satisfied. We can now use the freedom in choosing~$\varphi_{xy}$
to arrange that the distinguished Clifford subspaces~$K_{xy}$ and~$K_{yx}$ are mapped onto
each other,
\beq \label{Dc3}
K_{xy} = D_{x,y} \:K_{yx}\: D_{y,x}\:.
\eeq
It turns out that this condition determines~$\varphi_{xy}$ up to multiples of~$\frac{\pi}{2}$.
In order to fix~$\varphi_{xy}$ uniquely in agreement with~\eqref{phicond}, we need 
to assume that~$\varphi_{xy}$ is not a multiple of~$\frac{\pi}{4}$. This leads us to the
following definition.
\begin{Def} \label{defspinconnect}
Two points~$x,y \in M$ are called {\bf{spin connectable}} if
\sindex{spin connectable}%
the following conditions hold:
\bitem
\item[{\rm{(a)}}] The points~$x$ and~$y$ are properly timelike separated
(note that this already implies that~$x$ and~$y$ are regular
as defined in Section~\ref{secspin}).
\item[{\rm{(b)}}] The Euclidean sign operators~$s_x$ and~$s_y$ are generically separated
from the directional sign operators~$v_{xy}$ and~$v_{yx}$, respectively.
\item[{\rm{(c)}}] Employing the ansatz~\eqref{Dsecond}, the phases~$\varphi_{xy}$
which satisfy condition~\eqref{Dc3} are not multiples of~$\frac{\pi}{4}$.
\nindex{afc@$D_x$ -- spin connection of causal fermion system}%
\eitem
\end{Def} \noindent
We denote the set of points which are spin connectable to~$x$ by~$\I(x)$.
It is straightforward to verify that~$\I(x)$ is an open subset of~$M$.

Under these assumptions, we can fix~$\varphi_{xy}$ uniquely by imposing that
\beq \label{Dc4}
\varphi_{xy} \in \Big(-\frac{\pi}{2}, -\frac{\pi}{4} \Big) \cup \Big(\frac{\pi}{4}, \frac{\pi}{2} \Big)\:,
\eeq
giving the following result (for the proofs, see~\cite[Section~3.3]{lqg}).

\begin{Thm} \label{thmspinconnection}
Assume that two points~$x,y \in M$ are spin connectable. Then there is a unique
{\bf{spin connection}}~$D_{x,y} : S_y \rightarrow S_x$ of the form~\eqref{Dsecond}
having the properties~\eqref{Dc1}, \eqref{Dc2}, \eqref{Dc3} and~\eqref{Dc4}.
\end{Thm}

\subsection{The Metric Connection and Curvature}
We now outline a few further constructions from~\cite[Section~3]{lqg}.
The spin connection induces a connection on the corresponding tangent spaces, as we now explain.
Suppose that~$u_y \in T_y$. Then, according to Definition~\ref{deftangent} and Lemma~\ref{lemma3},
we can consider~$u_y$ as a vector of the representative~$K_y^{(x)} \in \T^{s_y}$.
By applying the synchronization map, we obtain a vector in~$K_{yx}$,
\beq u_{yx} := U_{yx} \,u_y\, U_{yx}^{-1} \in K_{yx}\:. \eeq
According to~\eqref{Dc3}, we can now ``parallel transport'' the vector to the Clifford
subspace~$K_{xy}$,
\beq u_{xy} := D_{x,y} \, u_{yx}\, D_{y,x} \in K_{xy}\:. \eeq
Finally, we apply the inverse of the synchronization map to obtain the vector
\beq u_x := U_{xy}^{-1} \,u_{xy}\, U_{xy} \in K_x^{(y)} \:. \eeq
As~$K_x^{(y)}$ is a representative of the tangent space~$T_x$ and all transformations were unitary,
we obtain an isometry from~$T_y$ to~$T_x$. 
\begin{Def} \label{defmconn} The isometry between the tangent spaces defined by
\beq \nabla_{x,y} \::\: T_y \rightarrow T_x \::\: u_y \mapsto u_x \eeq
\sindex{metric connection!of causal fermion system}%
\nindex{afd@$\nabla_x$ -- metric connection of causal fermion system}%
is  referred to as the {\bf{metric connection}} corresponding to the spin connection~$D$.
\end{Def} \noindent

We next introduce a notion of curvature.
\begin{Def} \label{defcurvature}
Suppose that three points~$x, y, z \in M$ are pairwise spin connectable. Then the
associated {\bf{metric curvature}}~$R$ is defined by
\sindex{curvature!of causal fermion system}%
\beq \label{Rmdef}
R(x,y,z) = \nabla_{x,y} \,\nabla_{y,z} \,\nabla_{z,x} \::\: T_x \rightarrow T_x\:.
\eeq
\end{Def} \noindent
The metric curvature~$R(x,y,z)$ can be thought of as a discrete analog of the
holonomy of the Levi-Civita connection on a manifold, where a tangent vector is parallel
transported along a loop starting and ending at~$x$. On a manifold, the curvature at~$x$ is immediately
obtained from the holonomy by considering the loops in a small neighborhood of~$x$.
With this in mind, Definition~\ref{defcurvature} indeed generalizes the usual notion of curvature to
causal fermion systems.

The following construction relates directional sign operators to vectors of the tangent space.
Suppose that~$y$ is spin connectable to~$x$. By synchronizing the directional sign operator~$v_{xy}$,
we obtain the vector
\beq \label{yxdef}
\hat{y}_x := U_{xy}^{-1} \,v_{xy}\, U_{xy} \in K_x^{(y)} \:.
\eeq
As~$K_x^{(y)} \in \T^{s_x}$ is a representative of the tangent space, we can regard~$\hat{y}_x$ as a
tangent vector. We thus obtain a mapping
\beq \I(x) \rightarrow T_x \;:\; y \mapsto \hat{y}_x \:. \eeq
We refer to~$\hat{y}_x$ as the {\em{directional tangent vector}} of~$y$ in~$T_x$.
As~$v_{xy}$ is a sign operator and the transformations in~\eqref{yxdef} are unitary,
the directional tangent vector is a timelike unit vector with the additional property that
the inner product~$\Sl .| \hat{y}_x . \Sr_x$ is positive definite.

We finally explain how to reduce the dimension of the tangent space to four, with the
desired Lorentzian signature~$(1,3)$. 
\begin{Def} The fermion system is called {\bf{chirally symmetric}} if to every~$x \in M$
we can associate a spacelike vector~$u(x) \in T_x$ which is orthogonal to all directional tangent vectors,
\beq \la u(x), \hat{y}_x \ra = 0 \qquad \text{for all~$y \in \I(x)$} \:, \eeq
and is parallel with respect to the metric connection; that is,
\beq u(x) = \nabla_{x,y} \,u(y)\, \nabla_{y,x} \qquad \text{for all~$y \in \I(x)$} \:. \eeq
\end{Def}

\begin{Def} \label{defreduce}
For a chirally symmetric fermion system, we introduce the {\bf{reduced tangent
space}}~$T_x^\text{\rm{red}}$ by
\beq T_x^\text{\rm{red}} = \langle u_x \rangle^\perp \subset T_x \:. \eeq
\end{Def} \noindent
Clearly, the reduced tangent space has dimension four and signature~$(1,3)$.
Moreover, the operator~$\nabla_{x,y}$ maps the reduced tangent spaces isometrically to each other.
The local operator~$\pseudo := - \cI u/\sqrt{-u^2}$ takes the role of the {\em{pseudo-scalar matrix}}.

\section{Correspondence to Lorentzian Spin Geometry} \label{secspincorr}
We now explain how the above geometric notions correspond to
the usual objects of differential geometry in Minkowski space (Theorem~\ref{thmminkowski})
and on a globally hyperbolic Lorentzian manifold (Theorem~\ref{thmglobhyp}).
We closely follow the presentation in~\cite[Section~3.3]{rrev}; see also the review~\cite{nrstg}.

We let~$(\scrM, g)$ be a time-oriented Lorentzian spin manifold with spinor bundle~$S \scrM$
(for basic definitions, see Section~\ref{secdirgh}).
In order to obtain a corresponding causal fermion system, we adapt the construction
in Minkowski space given in Section~\ref{seclco}:
First, we choose a closed subspace~$\H$ of the Hilbert space of Dirac solutions~$(\H_m, (.|.))$
(as introduced in Section~\ref{secdirgh}). Endowed with the induced scalar product~$\la .|. \ra_\H := (.|.)|_{\H \times \H}$, we obtain a Hilbert space~$(\H, \la .|. \ra_\H)$.
Next, one introduces a regularization operator~\eqref{Repsdef}, for example by mollifying
the initial data on a Cauchy surface (as is explained in~\cite[Section~4]{finite}).
Introducing the local correlation operator~$F^\varepsilon(x)$ for every~$x \in \scrM$ again by~\eqref{Fepsdef},
we define the measure~$\rho$ on~$\F$ as the push-forward of the volume measure~$\mu$ on~$\scrM$, that is,
\beq \rho = (F^\varepsilon)_* \mu \:, \eeq
where, in local coordinates, the measure~$\mu$ has the form
\beq \dd\mu = \sqrt{|\det g|}\: \dd^4x \:. \eeq
We thus obtain a causal fermion system~$(\H, \F, \rho)$ describing the curved spacetime~$(\scrM, g)$.
The basic inherent structures of this causal fermion system (like the spin space~$S_x$ and the spin inner product
as defined in Section~\ref{secinherent}) can be identified canonically with the corresponding objects
of spin geometry (like the spinor space~$S_x\scrM$ with spin inner product; for details
see~\cite[\S1.2.4]{cfs} or~\cite{topology}). With this identification, Clifford multiplication gives
rise to a canonical identification of the tangent space~$T_x \scrM$ with a distinguished Clifford subspace.

Let~$\gamma(t)$ be a smooth, future-directed and timelike curve, for simplicity parametrized by the arc length, defined
on the interval~$[0, T]$ with~$\gamma(0) = y$ and~$\gamma(T) = x$.
Then the parallel transport of tangent vectors along~$\gamma$ with respect to the
Levi-Civita connection~$\nablaLC$ gives rise to the isometry
\beq \nablaLC_{x,y} \::\: T_y \rightarrow T_x \:. \eeq
In order to compare with the metric connection~$\nabla$ of Definition~\ref{defmconn},
we subdivide~$\gamma$ (for simplicity with equal spacing, although a non-uniform spacing
would work just as well). Thus for any given~$N$, we define the
points~$x_0, \ldots, x_N$ by
\beq x_n = \gamma(t_n) \qquad \text{with} \qquad t_n = \frac{n T}{N}\:. \eeq
We define the parallel transport~$\nabla_{x,y}^N$ by successively composing
the parallel transport between neighboring points,
\beq \nabla^N_{x,y} := \nabla_{x_N, x_{N-1}} \nabla_{x_{N-1}, x_{N-2}} \cdots \nabla_{x_1, x_0}
\::\: T_y \rightarrow T_x \:. \eeq

We first state a result in the {\em{Minkowski vacuum}}. We choose~$\H$ as the subspace of all negative-energy
solutions of the Dirac equation (describing the Dirac sea; see the preliminaries in Section~\ref{secsea}).
For technical simplicity, we choose the {\em{$i \varepsilon$-regularization}}, where the
regularization operator~\eqref{Repsdef} is the multiplication
operator by~$\E^{\varepsilon \omega}$ in momentum space in~\eqref{iepsreg}.

\begin{Thm} \label{thmminkowski}
For every~$\varepsilon>0$ we
consider the causal fermion systems~$(\F, \H, \rho)$ describing the vacuum with $i \varepsilon$-regularization.
Then for a generic curve~$\gamma$ and for every~$N \in \N$, there is~$\varepsilon_0$
such that for all~$\varepsilon \in (0, \varepsilon_0]$ and all~$n=1,\ldots, N$, the
points~$x_{n}$ and~$x_{n-1}$ are spin connectable.
Moreover,
\beq \nablaLC_{x,y} = \lim_{N \rightarrow \infty} \:\lim_{\varepsilon \searrow 0} \nabla^N_{x,y} \:. \eeq
\end{Thm} \noindent
By a {\em{generic curve}}, we mean that the admissible curves are dense in the
$C^\infty$-topology (i.e., for any smooth~$\gamma$ and every~$K \in \N$,
there is a sequence~$\gamma_\ell$ of admissible curves such that~$D^k \gamma_\ell \rightarrow D^k \gamma$
uniformly for all~$k =0, \ldots, K$). The restriction to generic curves is needed in order to
ensure that the Euclidean and directional sign operators are generically separated
(see Definition~\ref{defspinconnect}~(b)).
The proof of the above theorem is given in~\cite[Section~4]{lqg}.

Clearly, in this theorem the connection~$\nablaLC_{x,y}$ is trivial. In order to
show that our connection also coincides with the Levi-Civita connection
in the case with curvature, in~\cite[Section~5]{lqg} a globally hyperbolic Lorentzian manifold
is considered. For technical simplicity, we assume that the manifold is flat Minkowski space
in the past of a given Cauchy hypersurface.

\begin{Thm} \label{thmglobhyp} Let~$(\scrM,g)$ be a globally hyperbolic manifold which is
isometric to Minkowski space in the past of a given Cauchy-hypersurface~$\scrN$.
For given~$\gamma$, for any~$\varepsilon>0$ we consider the causal fermion system~$(\H, \F, \rho)$
which in the past of~$\scrN$ coincides with the causal fermion system in Minkowski space
considered in Theorem~\ref{thmminkowski}. Then for a generic curve~$\gamma$
and for every sufficiently large~$N$, there is~$\varepsilon_0$
such that for all~$\varepsilon \in (0, \varepsilon_0]$ and all~$n=1,\ldots, N$, the
points~$x_{n}$ and~$x_{n-1}$ are spin connectable. Moreover,
\beq \lim_{N \rightarrow \infty} \:\lim_{\varepsilon \searrow 0} \nabla^N_{x,y} - \nablaLC_{x,y} 
= \O \!\left( L(\gamma)\: \frac{\nabla R}{m^2} \right)
\Big(1 + \O \Big( \frac{\text{\rm{scal}}}{m^2} \Big) \Big) \:, \eeq
where~$R$ denotes the Riemann curvature tensor, ${\text{\rm{scal}}}$ is scalar curvature,
and~$L(\gamma)$ is the length of the curve~$\gamma$.
\end{Thm} \noindent
Thus the metric connection of Definition~\ref{defmconn} indeed coincides with the
Levi-Civita connection, up to higher order curvature corrections.
For detailed explanations and the proof, we refer to~\cite[Section~5]{lqg}.

We conclude this section with a few remarks on further constructions~\cite{lqg}.
First, there is the subtle point that the unitary
transformation~$U \in \exp(\cI \R s_x)$, which is used to identify two representatives~$K, \tilde{K}
\in T_x$ via the relation~$\tilde{K} = U K U^{-1}$ (see Definition~\ref{deftangent})
is not unique. More precisely, the operator~$U$ can be transformed according to
\beq U \rightarrow -U \qquad \text{and} \qquad U \rightarrow s_x \,U\:. \eeq
As a consequence, the metric connection (see Definition~\ref{defmconn}) is defined only up
to the transformation
\beq \nabla_{x,y} u \rightarrow s_x \,(\nabla_{x,y} u)\, s_x\:. \eeq
Note that this transformation maps representatives of the same tangent vector into each other,
so that~$\nabla_{x,y} u \in T_x$ is still a well-defined tangent vector.
But we get an ambiguity when composing the metric connection several times
(as, for example, in the expression for the metric curvature in Definition~\ref{defcurvature}).
This ambiguity can be removed by considering {\em{parity-preserving systems}}
as introduced in~\cite[Section~3.4]{lqg}.

At first sight, one might conjecture that Theorem~\ref{thmglobhyp} should also apply to
the spin connection in the sense that
\beq \DLC_{x,y} = \lim_{N \rightarrow \infty} \:\lim_{\varepsilon \searrow 0} D^N_{x,y} \:, \eeq
where~$\DLC$ is the spin connection on~$S\scrM$ induced by the Levi-Civita connection and
\beq \label{spinit}
D^N_{x,y} := D_{x_N, x_{N-1}} D_{x_{N-1}, x_{N-2}} \cdots D_{x_1, x_0}
\::\: S_y \rightarrow S_x
\eeq
(and~$D$ is the spin connection of Theorem~\ref{thmspinconnection}).
It turns out that this conjecture is false. But the conjecture becomes true if we
replace~\eqref{spinit} with the operator product
\beq D^N_{(x,y)} := D_{x_N, x_{N-1}} U_{x_{N-1}}^{(x_N | x_{N-2})}
D_{x_{N-1}, x_{N-2}} U_{x_{N-2}}^{(x_{N-1} | x_{N-3})}
\cdots U_{x_1}^{(x_2 | x_0)} D_{x_1, x_0} \:. \eeq
Here, the intermediate factors~$U_{.}^{(.|.)}$ are the so-called {\em{splice maps}}
\sindex{splice map}%
given by
\beq U_x^{(z|y)} = U_{xz} \,V\, U_{xy}^{-1} \:, \eeq
where~$U_{xz}$ and~$U_{xy}$ are synchronization maps, and~$V \in \exp(i \R s_x)$
is an operator which identifies the representatives~$K_{xy}, K_{xz} \in T_x$
(for details, see~\cite[Section~3.7 and Section~5]{lqg}).
The splice maps also enter the {\em{spin curvature}}~$\mathfrak{R}$, which is defined in
analogy to the metric curvature~\eqref{Rmdef} by
\sindex{curvature!of causal fermion system}%
\beq \mathfrak{R}(x,y,z) = U_x^{(z|y)} \:D_{x,y}\: U_y^{(x|z)} \:D_{y,z}\: U_z^{(y|x)}
\:D_{z,x} \::\: S_x \rightarrow S_x \:. \eeq

\section{Exercises}

\begin{Exercise} \label{exrfs1} {{(Vector fields on a closed Riemannian manifold)}} {\em{
\sindex{Riemannian fermion system}%
Let~$(\scrM,g)$ be a smooth compact Riemannian manifold of dimension~$k$ and~$\Delta$
the covariant Laplacian on smooth vector fields. We complexify the vector fields and endow them
with the $L^2$-scalar product
\beq %\label{L2sp}
\la u | v \ra_{L^2} := \int_{\scrM} g_{jk}\, \overline{u^j} \,v^k \: \dd\mu_{\scrM} \:, \eeq
where~$\dd\mu_\scrM = \sqrt{\det g}\, \dd^k x$ is the volume measure on~$\scrM$.
Show the following:
\bitem
\item[(a)] The operator~$-\Delta$ is essentially selfadjoint and has smooth eigenfunctions.
\item[(b)] We choose a parameter~$L>0$ and choose~$\H$ as the spectral subspace of the Laplacian
\beq \H = \text{rg}\, \chi_{[0,L]}(-\Delta) \:. \eeq
Show that~$\H$ is finite-dimensional.
\item[(c)] For any~$p \in \scrM$ we define the local correlation operator~$F(p) \in \Lin(\H)$ by
\beq -g_{ij} \,\overline{u^i(p)} \,v^j(p) = \la u | F(p) v \ra_{L^2} \qquad \text{for all~$u,v \in \H$} \:. \eeq
Show that this operator is well-defined, negative semi-definite and has rank at most~$k$.
\item[(d)] We again introduce the measure by~$\rho = F_* \mu$.
Show that~$(\H, \F, \rho)$ is a Riemannian fermion system of spin dimension~$k$.
\eitem
{\em{Hint:}} For~(a) and~(b) one can use properties of elliptic operators on compact domains,
as can be found for example in~\cite{evans, taylor1}.
}} \end{Exercise}

\begin{Exercise} \label{exrfs2} {{(Spinors on a closed Riemannian manifold)}} {\em{
\sindex{Riemannian fermion system}%
Let~$(\scrM, g)$ be a compact Riemannian spin manifold of dimension~$k \geq 1$.
Then the spinor bundle~$S\scrM$ is a vector bundle with fiber~$S_p\scrM \simeq \C^n$
with~$n=2^{[k/2]}$ (see, for example, \cite{lawson+michelsohn, friedrich}). Moreover,
the spin inner product~$\Sl .|. \Sr_p : S_p \scrM \times S_p \scrM \rightarrow \C$ is positive definite.
On the smooth sections~$\Gamma(S\scrM)$ of the spinor bundle we can thus introduce the scalar product
\beq %\label{sprod-H}
\la \psi | \phi \ra = \int_{\scrM} \Sl \psi | \phi \Sr_p\: \dd\mu_\scrM(p) \:. \eeq
Forming the completion gives the Hilbert space~$L^2(\scrM, S\scrM)$.
\bitem
\item[(a)] The Dirac operator~$\Dir$ with domain of definition~$\Gamma(S\scrM)$
is an essentially selfadjoint operator on~$L^2(\scrM, S\scrM)$. It has a purely discrete spectrum
and finite-dimensional eigenspaces.
\item[(b)] Given a parameter~$L>0$, we let~$\H$ be the space spanned by
all eigenvectors whose eigenvalues lie in the interval~$[-L, 0]$,
\beq \H = \text{rg}\, \chi_{[-L,0]}(\Dir) \subset L^2(\scrM, S\scrM)\:. \eeq
Denoting the restriction of the $L^2$-scalar product to~$\H$ by~$\la .|. \ra_\H$,
we obtain a finite-dimensional Hilbert space~$(\H, \la .|. \ra_\H)$.
Show that this Hilbert space is finite-di\-men\-sio\-nal and consists of smooth wave functions.
\item[(c)] For every~$p \in \scrM$ we introduce the local correlation operator~$F(p)$ by
\beq -\Sl \psi | \phi \Sr_p = \la \psi | F(p) \phi \ra_\H \qquad \text{for all~$\psi, \phi \in \H$} \:. \eeq
Show that this operator is negative semi-definite and has rank at most~$n$.
\item[(d)] We again introduce the measure by~$\rho = F_* \mu$.
Show that~$(\H, \F, \rho)$ is a Riemannian fermion system of spin dimension~$n$.
\eitem
{\em{Hint:}} The Dirac operator on Riemannian manifolds of general dimension is introduced
in~\cite{lawson+michelsohn, friedrich}. For~(a) and~(b) one can again
use properties of elliptic operators on compact domains,
as can be found for example in~\cite{evans, taylor1} or, more specifically for Dirac operators,
in~\cite[Chapter~20]{taylor2}.
}} \end{Exercise}

\begin{Exercise} \label{exrfs3} {{(Causal structure of a Riemannian fermion system)}} {\em{
\sindex{Riemannian fermion system}%
Let~$(\H, \F, \rho)$ be a Riemannian fermion system of spin dimension~$n$ (see Definition~\ref{defcfstop}).
\bitem
\item[(a)] Show that for every~$y \in \F$, the operator~$-y$ is positive semi-definite.
How can its square root~$\sqrt{-y}$ be defined?
\item[(b)] Show that the operator product~$\sqrt{-y}\;(-x) \sqrt{-y}$ (with~$x \in \F$) is positive semi-definite.
\item[(c)] Show that the eigenvalues of the operator product~$xy$ are all real and non-negative.
{\em{Hint:}} Use the relation~$xy = -x \:\sqrt{-y}\, \sqrt{-y}$ together with the fact that the spectrum
is invariant under cyclic permutations.
\item[(d)] What does this mean for the causal structure of Definition~\ref{def2}?
\eitem
}} \end{Exercise}

\begin{Exercise} \label{exsignop}  {{(The Euclidean sign operator)}} 
\sindex{sign operator!Euclidean}%
{\em{ Let~$x \in \F$ be a regular spacetime point.
For convenience, we choose a basis of the Hilbert space where~$x$ is diagonal; that is,
\beq (-x) = \text{diag} \big( \nu_1, \nu_2, \nu_3, \nu_3, 0, \ldots \big)
\qquad \text{with~$\nu_1, \nu_2>0$ and~$\nu_3, \nu_4<0$}\:. \eeq
\bitem
\item[(a)] Show that the eigenspaces corresponding to the positive (negative) eigenvalues of~$(-x)$
are positive (respectively negative) definite with respect to the spin inner product.
\item[(b)] We define the Euclidean sign operator in the above basis by
\beq s_x = \text{diag} \big( 1,1,-1,-1, 0, \ldots \big) \:. \eeq
Show that this operator is uniquely defined without referring to bases by demanding
that~$s_x$ commutes with~$x$, that its eigenvalues are~$\{1,-1,0\}$
and that the eigenspaces corresponding to~$1$ (and~$-1$) are negative (respectively positive) definite 
with respect to the spin inner product.
\eitem
}} \end{Exercise}

\begin{Exercise} \label{excliffsign} {{(Signature of Clifford extensions)}} {\em{
\sindex{Clifford extension}%
\bitem
\item[(a)] Let~$\T^{s_x}$ be a Clifford extension of the Euclidean sign operator~$s_x$.
Show that the resulting bilinear form~$\la .,. \ra$ on~$\T^{s_x}$
is Lorentzian; that is, that it has signature~$(1,k)$ with~$k \in \N$.
{\em{Hint:}} it is most convenient to work in an orthonormal eigenvector basis of the Euclidean sign operator.
You also find the proof in~\cite[Lemma~4.4]{topology}.
\item[(b)] Now let~$\T^v$ be a the Clifford extension of a general sign operator~$v$.
Is the signature of~$\la .,. \ra$ necessarily Lorentzian?
{\em{Hint:}} It may be helpful to have a look at~\cite[Lemma~3.2]{lqg}.
\eitem
}} \end{Exercise}

\begin{Exercise} \label{exdiracsphere2} {{(Clifford extensions on the Dirac sphere)}} {\em{
We return to the Dirac sphere considered in Exercise~\ref{exdiracsphere}.
Thus we let~$F : S^2 \rightarrow \F$ and~$M:= \supp \rho = F(S^2)$.
\bitem
\item[(a)] Given~$p \in S^2$, we consider the spacetime point~$x = F(p) \in M$.
Construct the Euclidean sign operator~$s_x$ at~$x$.
\item[(b)] What is the maximal dimension of Clifford extensions of the Euclidean sign operator?
Show that the Clifford extension of maximal dimension is unique.
\item[(c)] Give an explicit parametrization of this Clifford extension.
How does the inner product~$\la .,. \ra$ look like in your parametrization?
\eitem
}} \end{Exercise}

\begin{Exercise} {{(Stability of the causal structure)}} \label{exproper} {\em{
\sindex{timelike separation!properly}%
A binary relation~$P$ on~$\F$ is said to be \textit{stable under perturbations} if
%		\beq
%		(x_0,y_0)\in P\  \Longrightarrow\  \exists\  B_r(y_0)\subset M\,:\, (x_0,y) \in P \ \ \forall y\in B_r(y_0).
%		\eeq
\beq
(x_0,y_0)\in P\  \Longrightarrow\  \exists\ r>0\ :\   B_r(x_0)\times B_r(y_0)\subset P.
\eeq
Following Definition~\ref{defproptl}, two points~$x,y\in \F$ are said to be \textit{properly timelike} separated if the closed chain~$A_{xy}$ has a strictly positive spectrum and if all eigenspaces are definite subspaces of~$(S_x, \Sl\,\cdot\,,\,\cdot\,\Sr)$.
\bitem
\item[(a)] Show that proper timelike separation implies timelike separation.
\item[(b)] Show by a counterexample with~$3\times 3$ matrices that the notion of timelike separation is \textit{not} stable under perturbations.
\item[(c)] Show that the notion of properly timelike separation is stable under perturbations.
\eitem
}} \end{Exercise}

%%% Local Variables:
%%% mode: latex
%%% TeX-master: t
%%% End:

%% file: part3.tex
%!TEX root = intro.tex
\part{Mathematical Methods and Analytic Constructions} \label{partthree}

\chapter{Measure-Theoretic Methods} \label{secmeasure}
The main goal of this chapter is to prove the existence of minimizers for the causal action
principle in the case that~$\H$ is {\em{finite dimensional}} and~$\rho$ is {\em{normalized}}; that is,
\beq \label{Hfinitenormed}
\dim \H =: f < \infty \quad \text{and} \qquad \rho(\F)=1 \:.
\eeq
\sindex{causal action principle!finite-dimensional setting}%
After introducing the necessary methods (Sections~\ref{secbanachalaoglu} and~\ref{secriesz}),
we first apply them to prove the existence
of minimizers for causal variational principles in the compact setting (Section~\ref{secweakcompact}).
In preparation for the proof for the causal action principle,
we illustrate the constraints by a few examples (Section~\ref{seccounter}).
The difficulties revealed by these examples can be resolved by working
with the so-called moment measures. After introducing the needed mathematical methods
(Section~\ref{secradonnikodym}), the moment measures are introduced (Section~\ref{secmoment}).
Then the existence proof is completed (Section~\ref{secexistence}).
In order to give a first idea for how to deal with an infinite total volume,
we finally prove the existence
of minimizers for causal variational principles in the non-compact setting (Section~\ref{seccvpnoncompact}).

Our general strategy is to apply the {\em{direct method in the calculus of variations}},
\sindex{direct method in the calculus of variations}%
which can be summarized as follows:
\bitem
\item[(a)] Choose a {\em{minimizing sequence}}, that is, a sequence of measures~$(\rho_k)$
which satisfy the constraints such that
\beq \Sact(\rho_k) \rightarrow \inf_\rho \Sact(\rho) \:. \eeq
Such a minimizing sequence always exists by definition of the infimum (note that the
action and therefore also its infimum are non-negative).
\item[(b)] Show that a subsequence of the measures converges in a suitable sense,
\beq \rho_{k_l} \,\text{``$\longrightarrow$''}\, \rho \:. \eeq
Here the quotation marks indicate that we still need to specify in which sense
the sequence should converge (convergence in which space, strong or weak convergence, etc.).
\item[(c)] Finally, one must show that the action is lower semi-continuous; that is,
\beq \Sact(\rho) \leq \liminf_{l \rightarrow \infty} \Sact(\rho_{k_l}) \:. \eeq
Also, one must prove that the limit measure~$\rho$ satisfies the constraints.
\eitem
Once these three steps have been carried out, the measure~$\rho$ is a desired minimizer.
We point out that this procedure does {\em{not}} give a {\em{unique}} minimizer, simply because
there may be different minimizing sequences, and because the choice of the subsequences may
involve an arbitrariness. Indeed, for the causal action principle we do not expect uniqueness.
There should be many different minimizers, which describe different physical systems
(like the vacuum, a system involving particles and fields, etc.). This intuitive picture is confirmed
by the numerical studies in~\cite{support, numerics}, which show that, even if the dimension of~$\H$ is
small, there are many different minimizers.

\section{The Banach-Alaoglu Theorem} \label{secbanachalaoglu}
\sindex{Banach-Alaoglu theorem}%
For our purposes, it suffices to consider the case that the Banach space is separable,
in which case the theorem was first proved by Banach (Alaoglu proved the generalization to non-separable
Banach spaces; this makes use of Tychonoff's theorem and goes beyond what we need here).
Indeed, the idea of proof of the theorem can be traced back to Eduard Helly's
doctoral thesis in 1912, where the closely related ``Helly's selection theorem'' is proved
(of course without reference to Banach spaces, which were introduced later).
We closely follow the presentation in~\cite[Section~10.3]{lax}.

Let~$(E, \|.\|_E)$ be a separable (real or complex)
Banach space and~$(E^*, \|.\|_{E^*})$ its dual space with the usual $\sup$-norm; that is,
\beq \label{sups}
\| \phi \|_{E^*} = \sup_{u \in E, \|u\|=1} \big| \phi(u) \big| \:.
\eeq
A sequence~$(\phi_n)_{n \in \N}$ in~$E^*$ is said to be {\em{weak*-convergent}}
to~$\phi \in E^*$ if
\beq \lim_{n \rightarrow \infty} \phi_n(u) = \phi(u) \qquad \text{for all~$u \in E$} \:. \eeq

\begin{Thm}  {\bf{(Banach-Alaoglu theorem in the separable case)}} \label{thmbanachalaoglu}
Let~$E$ be a separable Banach space. Then every bounded
sequence in~$E^*$ has a weak*-convergent subsequence.
\end{Thm}
\Proof Let~$\phi_n$ be a bounded sequence in~$E^*$, meaning that there is a constant~$c>0$ with
\beq \label{Esbounded}
\|\phi_n\|_{E^*} \leq c \qquad \text{for all~$n \in \N$}\:.
\eeq
We let~$(u_\ell)_{\ell \in \N}$ be a sequence in~$E$ which is dense in~$E$.
Combining~\eqref{Esbounded} with~\eqref{sups}, the estimate
\beq \label{phinbound}
|\phi_n(u_1)| \leq \|\phi_n\|_{E^*}\: \|u_1\|_E \leq c\, \|u_1\|_E
\eeq
shows that~$(\phi_n(u_1))_{n \in \N}$ is a bounded sequence. Thus we can choose
a convergent subsequence. By inductively choosing subsequences and taking the
diagonal sequence, we obtain a subsequence~$(\phi_{n_j})$ such that the
limit
\beq \lim_{j \rightarrow \infty} \phi_{n_j}(u_\ell) \eeq
exists for all~$\ell \in \N$.
Hence setting
\beq \phi(u_\ell) := \lim_{j \rightarrow \infty} \phi_{n_j}(u_\ell) \:, \eeq
we obtain a densely defined functional.
Taking the limit in~\eqref{phinbound} (and the similar inequalities for~$u_2, u_3, \ldots$),
one sees that this functional is again continuous.
Therefore, it has a unique continuous extension to~$E$.
By continuity, the resulting functional~$\phi \in E^*$ satisfies the relations
\beq \phi(u) = \lim_{j \rightarrow \infty} \phi_{n_j}(u) \qquad \text{for all~$u \in \H$}\:. \eeq
In particular, it is again a linear. This concludes the proof.
\QED

\section{The Riesz Representation Theorem} \label{secriesz}
\sindex{Riesz representation theorem|textbf}%
In this section and Section~\ref{secradonnikodym}, we shall introduce the methods
from measure theory needed for the existence proofs.
Apart from the books already mentioned in the preliminaries (Section~\ref{secbasicmeasure}),
we also recommend the book~\cite{evans+gariepy} (this book is only concerned with
measures in~$\R^n$, but otherwise goes far beyond what we need here).

For our purposes, it suffices to restrict attention to the case
that the base space~$\K$ is a {\em{compact}} topological space.
We always consider {\em{bounded regular Borel measures}} on~$\K$ (for the preliminaries, see Section~\ref{secbasicmeasure}). In order to avoid confusion, we note
that by a measure we always mean a {\em{positive}} measure
(signed measures will not be considered in this book).
A {\em{bounded}} measure is also referred to as a measure of {\em{finite total volume}}.
Often, we {\em{normalize}} the measure such that~$\mu(\K)=1$.

In words, the Riesz representation theorem makes it possible to
represent a linear functional on the Banach space of continuous functions
of a topological space by a regular Borel measure on this topological space.
We remark that we already came across the Riesz representation theorem 
in Section~\ref{secspectral}, where it was needed for the construction of spectral measures.
However, in this context, we only needed the special case that the topological space
was an interval of the real line. We now state the general theorem and outline its proof,
mainly following the presentation in~\cite[\S56]{halmosmt}. More details can be found
in~\cite[\S IV.29]{bauer}.

As a simple example, one can choose~$\K$ as the closed unit ball in~$\R^n$.
Restricting the Lebesgue measure to the Borel subsets of~$\K$ gives a Radon measure.
The Lebesgue measure itself is a completion of this Radon measure obtained by
extending the $\sigma$-algebra of measurable sets by all subsets of Borel sets of measure zero.
Since this completion is a rather trivial extension, in what follows we prefer to work with
Radon measures or, equivalently, with normalized regular Borel measures.

\begin{Thm} {\bf{(Riesz representation theorem)}} \label{thmriesz}
Let~$\K$ be a compact topological space, and~$E=C^0(\K, \R)$ the
Banach space of continuous functions on~$\K$ with the usual sup-norm,
\beq \|f\| = \sup_{x \in \K} |f(x)| \:. \eeq
Let~$\Lambda \in E^*$ be a continuous linear functional which is positive in the sense that
\beq \Lambda(f) \geq 0 \qquad \text{for all nonnegative functions~$f \in C^0(\K, \R)$}\:. \eeq
Then there is a unique regular Borel measure~$\mu$ such that
\beq \Lambda(f) = \int_\K f\, \Diff\mu \qquad \text{for all~$f \in C^0(\K, \R)$}\:. \eeq
\end{Thm}
\Proof[Outline of the Proof.]
We follow the strategy in~\cite[\S56]{halmosmt}. Given a Borel set~$A \subset \K$, we set
\beq \lambda(A) = \inf \big\{ \Lambda(f) \:\big|\: f \in C^0(\K, \R) \text{ and } f \geq \chi_A \big\} \in \R^+_0 \:. \eeq
Intuitively speaking, $\lambda$ gives us the desired ``volume'' of the set~$A$.
But there is the technical problem that~$\lambda$ is in general not a regular Borel measure.
Instead, it merely is a {\em{content}}, meaning that it has the following properties:
\begin{itemize}
\item[(i)] {\em{non-negative}} and {\em{finite}}: $0 \leq \lambda(A) < \infty$
\item[(ii)] {\em{monotone}}: $C$, $D$ compact and~$C \subset D$ $\; \Longrightarrow \;$
$\lambda(C) \leq \lambda(D)$
\item[(iii)] {\em{additive}}: $C$, $D$ compact and disjoint 
$\; \Longrightarrow \;$ $\lambda(C \cup D) = \lambda(C) + \lambda(D)$
\item[(iv)] {\em{subadditive}}: $C$, $D$ compact
$\; \Longrightarrow \;$ $\lambda(C \cup D) \leq \lambda(C) + \lambda(D)$
\end{itemize}
At this stage, we are in a similar situation as in the elementary measure theory course
after saying that a cube of length~$\ell$ in~$\R^3$ should have volume~$\ell^3$.
In order to get from this ``volume measure'' to a measure in the mathematical sense,
one has to proceed in several steps invoking the subtle and clever constructions of measure theory
(due to Lebesgue, Hahn, Carath{\'e}odory and others)
in order to get a
mapping from a $\sigma$-algebra to the non-negative real numbers which is $\sigma$-additive.
In simple terms, repeating these constructions starting from the above content gives
the desired Borel measure~$\mu$. For brevity, we here merely outline the constructions
and refer for details to text books on measure theory (like, for example, \cite[Chapter~X]{halmosmt}).

The first step is to approximate (or exhaust) from inside by compact sets. Thus
one introduces the {\em{inner content}}~$\lambda_*$ by
\beq \lambda_*(U) = \sup \big\{ \lambda(C) \:\big|\: C \subset U \text{ compact} \big\} \:. \eeq
This inner content is monotone and countably additive.
The second step is to exhaust from outside by open sets. This gives the
{\em{outer measure}}~$\mu^*$,
\beq \mu^*(U) = \inf \big\{ \lambda_*(\Omega) \:\big|\: \Omega \supset U \text{ open} \big\} \:. \eeq
The outer measure is defined for any subset of~$\K$. Therefore, it remains to distinguish
the measurable sets. This is accomplished by Carath{\'e}odory's criterion, which defines
a set~$A \subset \K$ to be {\em{measurable}} if
\beq %\label{measurable}
\mu^*(A) = \mu^*(A \cap B) + \mu^*(A \setminus B) \eeq
for every subset~$B \subset \K$. Then Carath{\'e}odory's lemma
(for a concise proof see, for example, \cite[Lemma~2.8]{broeckerana2}) implies
that the measurable sets form a $\sigma$-algebra, and that the restriction of~$\mu^*$
to the measurable sets is indeed a measure, denoted by~$\mu$.

In order to complete the proof, one still needs to verify that every Borel set is $\mu$-measurable.
Moreover, it remains to show that the resulting
Borel measure is regular. To this end, one first needs to show that the content~$\lambda$
is regular in the following sense:
\begin{itemize}
\item[(v)] {\em{regular}}: For every compact~$C$,
\beq \lambda(C) = \inf \Big\{ \lambda(D) \:\big|\: D \text{ compact and } C \subset \overset{\circ}{D} \Big\} \:. \eeq
\end{itemize}
As the proofs of these remaining points are rather straightforward and not very instructive, we 
refer for the details to~\cite[\S54--\S56]{halmosmt}.
\QED

\section[Existence of Minimizers in the Compact Setting]{Existence of Minimizers for Causal Variational Principles in the Compact Setting} \label{secweakcompact}
\sindex{causal variational principle!in the compact setting}%
We now apply the above methods to prove the existence of minimizers for causal variational principles
in the compact setting. Our strategy is to apply the Banach-Alaoglu theorem to a specific Banach space,
namely the continuous functions on a compact metric space.
We first verify that this Banach space is separable.

\begin{Prp} \label{prpsep} %\hspace*{1cm}
Let~$\K$ be a compact metric space. Then~$C^0(\K, \R)$
is a separable Banach space.
\end{Prp}
\Proof The proposition is a consequence of the Stone-Weierstrass
theorem, whose proof can be found, for example, in~\cite[7.3.1]{dieudonne1}
We closely follow the proof given in~\cite[7.4.4]{dieudonne1}.

Covering~$\K$ by a finite number of open balls of radii~$1$, $1/2$, $1/3$, \ldots, one
gets an enumerable basis of the open sets~$(U_n)_{n \in \N}$.
For any~$n \in N$, we let~$g_n$ be the continuous function
\beq g_n(x) = d\big(x, \K \setminus U_n \big) \:. \eeq
Clearly, the algebra generated by these functions (by taking finite products and
finite linear combinations) is again separable.
Therefore, it suffices to show that this algebra is dense in~$C^0(\K,\R)$.
To this end, we need to verify the assumptions of the Stone-Weierstrass theorem.
The only assumption which is not obvious is that the algebra separates the points.
This can be seen as follows: Let~$x$ and~$y$ be two distinct points in~$\K$.
Since the~$(U_n)$ are a basis of the topology, there is~$U_n$ with~$x \in U_n$
and~$y \not\in U_n$. As a consequence, $g_n(x)>0$ but~$g_n(y)=0$.
\QED

We proceed by proving a compactness result for Radon measures.
\begin{Thm} \label{thmcompact}
Let~$\rho_n$ be a series of regular Borel measures
on~$C^0(\K,\R)$ which are bounded in the sense that there is a constant~$c>0$ with
\beq \rho_n(\K) \leq c \qquad \text{for all~$n$} \:. \eeq
Then there is a subsequence~$(\rho_{n_k})$ which converges
as a measure; that is,
\beq \label{measureconverge}
\lim_{k \rightarrow \infty} \int_\K f \,\Diff\rho_{n_k} = \int_\K f\, \Diff\rho \qquad \text{for all~$f \in C^0(\K,\R)$}\:.
\eeq
Moreover, the total volume converges; that is,
\beq \label{rhonorm}
\rho(\K) = \lim_{k \rightarrow \infty} \rho_{n_k}(\K) \:.
\eeq
\end{Thm}
\Proof Via
\beq \phi_n(f) := \int_\K f\, \Diff\rho_n \:, \eeq
every measure can be identified with a positive linear functional on~$E:=C^0(\K,\R)$.
Since~$E$ is separable (Proposition~\ref{prpsep}), we can apply the Banach-Alaoglu
theorem in the separable case (Theorem~\ref{thmbanachalaoglu}) to conclude that
there is a weak*-convergent subsequence; that is,
\beq %\label{eq1abs}
\lim_{k \rightarrow \infty} \phi_{n_k}(f) = \phi(f) \qquad \text{for all~$f \in C^0(\K,\R)$}\:. \eeq
Clearly, since all~$\phi_{n_k}$ are positive, the same is true for the limit~$\phi$.
Therefore, the Riesz representation theorem (Theorem~\ref{thmriesz}) makes it possible
to represent~$\phi$ by a regular Borel measure~$\rho$; that is,
\beq %\label{eq2abs}
\phi(f) = \int_\K f\, \Diff\rho \qquad \text{for all~$f \in C^0(\K,\R)$}\:. \eeq
Choosing~$f$ as the constant function, one obtains~\eqref{rhonorm}.
This concludes the proof.
\QED

\begin{Thm} \label{thmexistcompact}
Assume that~$\F$ is a compact topological space and the Lagrangian is continuous,
\beq \L \in C^0(\F \times \F, \R^+_0) \:. \eeq
Then the causal variational principle where the causal action~\eqref{Sactcompact}
is minimized in the class of regular Borel measures under the volume constraint~\eqref{normonecompact}
is well-posed in the sense that every minimizing sequence~$(\rho_n)_{n \in \N}$ has
a subsequence which converges as a measure to a minimizer~$\rho$.
\end{Thm}
\Proof The existence of a convergent subsequence~$(\rho_{n_k})_{k \in \N}$ 
is proven in Theorem~\ref{thmcompact}. It remains to show that the action is continuous; that is,
\beq \lim_{k \rightarrow \infty} \Sact \big(\rho_{n_k} \big) = \Sact(\rho)\:. \eeq
This is verified in detail as follows. Using that the Lagrangian is continuous in its second argument, we know that
\beq \label{rhopointwise}
\lim_{k \rightarrow \infty} \int_\F \L(x,y)\: \Diff\rho_{n_k}(y) = \int_\F \L(x,y)\: \Diff\rho(y) 
\qquad \text{for all~$x \in \F$} \:.
\eeq
Next, since~$\F$ is compact, the Lagrangian is even uniformly continuous on~$\F \times \F$.
Therefore, given~$\varepsilon>0$, every point~$x \in \F$ has
an open neighborhood~$U(x) \subset \F$ such that
\beq \big| \L(\hat{x},y) - \L(x, y) \big| < \varepsilon \qquad \text{for all~$\hat{x} \in U(x)$ and~$y \in \F$}\:. \eeq
Integrating over~$y$ with respect to any normed regular Borel measure~$\tilde{\rho}$, it follows that
\beq \label{rhouniform}
\bigg| \int_\F \L(\hat{x},y)\: \dd\tilde{\rho}(y) \:-\: \int_\F \L(x,y)\: \dd\tilde{\rho}(y) \bigg| \leq \varepsilon \qquad
\text{for all~$\hat{x} \in U(x)$} \:.
\eeq
Covering~$\F$ by a finite number of such neighborhoods~$U(x_1), \ldots, U(x_N)$,
one can combine the pointwise convergence~\eqref{rhopointwise} for~$x=x_1, \ldots, x_N$
with the estimate~\eqref{rhouniform} to conclude that for any~$\varepsilon>0$
there is~$k_0 \in \N$ such that
\beq \bigg| \int_\F \L(x,y)\: \Diff\rho_{n_k}(y) \:-\: \int_\F \L(x,y)\: \Diff\rho(y) \bigg| \leq 3 \varepsilon \qquad
\text{for all~$x \in \F$ and~$k \geq k_0$} \:. \eeq
Integrating over~$x$ with respect to~$\rho_{n_k}$ and~$\rho$ gives for all~$k \geq k_0$ the respective inequalities
\begin{align}
\bigg| \Sact\big(\rho_{n_k} \big) \:-\: \int_\F \Diff\rho_{n_k}(x) \int_\F \Diff\rho(y)\: \L(x,y) \bigg| &\leq 3 \varepsilon \:, \\
\bigg| \int_\F \Diff\rho(x) \int_\F \Diff\rho_{n_k}(y)\: \L(x,y) \:-\: \Sact(\rho) \bigg| &\leq 3 \varepsilon \:.
\end{align}
Combining these inequalities and using
that the Lagrangian is symmetric in its two arguments, we conclude that
\beq \big| \Sact\big(\rho_{n_k} \big) - \Sact(\rho) \big| \leq 6 \varepsilon \:. \eeq
This gives the result.
\QED
We finally remark that the statement of this theorem also holds if the Lagrangian merely is lower semi-continuous,
as is worked out in~\cite[Section~3.2]{noncompact}.

\section{Examples Illustrating the Constraints} \label{seccounter}
Compared to causal variational principles in the compact setting, the
existence proof for the causal action principle is considerably harder
because we need to handle the constraints~\eqref{volconstraint}--\eqref{Tdef}
and face the difficulty that the set~$\F$ is  unbounded and therefore non-compact. 
We now explain the role of the constraints in a few examples.
The necessity of the volume constraint is quite obvious:
If we dropped the constraint of fixed total volume~\eqref{volconstraint}, the measure~$\rho=0$ would be a trivial minimizer. The role of the trace constraint is already less obvious. It is explained in the next two examples. 

\begin{Example} {\bf{(necessity of the trace constraint)}} \label{extraceconstraint} {\em{
\sindex{constraint!trace}%
Let~$x$ be the operator with the matrix representation
\beq %\label{xtrivial}
x = \text{diag} \big( \underbrace{1, \ldots, 1}_{\text{$n$ times}}, 
\underbrace{-1, \ldots, -1}_{\text{$n$ times}}, 0, 0, \ldots \big) \:. \eeq
Moreover, we choose~$\rho$ as a multiple of the Dirac measure supported at~$x$.
Then the action~$\Sact$ vanishes (see~\eqref{Sdef}), whereas the constraint~$\T$ 
is strictly positive (see~\eqref{Tdef}).
}} 

\QEDrem \end{Example}

\begin{Example} {\bf{(non-triviality of the action with trace constraint)}} {\em{
\sindex{constraint!trace}%
Let~$\rho$ be a normalized measure which satisfies the
trace constraint in a non-trivial way; that is,
\beq \int_\F \tr(x)\: \Diff\rho(x) = \text{const} \neq 0 \:. \eeq
Let us prove that the action is non-zero. This will show that the trace constraint
really avoids trivial minimizers of the causal action principle.

\bitem
\item[(a)] Since the integral over the trace is non-zero, there is a point~$x$ in the support of~$\rho$
with~$\tr(x) \neq 0$. We denote the non-trivial eigenvalues of~$x$ by~$\nu_1,\ldots, \nu_{2n}$
and order them according to
\beq %\label{morder}
\nu_1 \leq \cdots \leq \nu_n \;\leq\; 0 \;\leq\;
\nu_{n+1} \leq \cdots \leq \nu_{2n} \:. \eeq
The fact that the trace of~$x$ is non-zero clearly implies that the~$\nu_i$
do not all have the same absolute value. As a consequence, the
nontrivial eigenvalues of the operator product~$x^2$
given by~$\lambda^{xx}_j = \nu_j^2$ are all non-negative and not all equal.
Using the form of the Lagrangian in~\eqref{Lagrange}, we conclude that~$\L(x,x)>0$.
\item[(b)] Since the Lagrangian is continuous in both arguments, there is
an open neighborhood~$U \subset \F$ of~$x$ such that~$\L(y,z)>0$
for all~$y,z \in U$. Since~$x$ lies in the support of~$\rho$, we know that~$\rho(U)>0$.
As a consequence,
\beq \Sact \geq \int_U \Diff\rho(x) \int_U \Diff\rho(y) \: \L(x,y) > 0 \eeq
(because if the integrals vanished, then the integrand would have to be zero
almost everywhere, a contradiction).
\eitem
We remark that this argument is quantified in~\cite[Proposition~4.3]{discrete}.
}} $\;$ $\;\quad$\QEDrem \end{Example}

We now come to the boundedness constraint. In order to explain
how it comes about, we give an explicit example with $(4 \times 4)$-matrices
(for a similar example with $(2 \times 2)$-matrices, see Exercise~\ref{ex28}).

\begin{Example} {\bf{(Necessity of the boundedness constraint)}} \label{exm1} {\em{
\sindex{constraint!boundedness}%
The following example explains why the {\em{boundedness constraint}}~\eqref{Tdef} is needed
in order to ensure the existence of minimizers.
It was first given in~\cite[Example~2.9]{continuum}.
Let~$\H=\C^4$. We introduce the four $4 \times 4$-matrices acting on~$\H$ by
\beq \gamma^\alpha = \begin{pmatrix} \sigma^\alpha & 0 \\ 0 & -\sigma^\alpha \end{pmatrix},
\quad \alpha=1,2,3  \qquad \text{and} \qquad
\gamma^4 = \begin{pmatrix} 0 & \1 \\ \1 & 0 \end{pmatrix} \eeq
(where the~$\sigma^\alpha$ are again the Pauli matrices~\eqref{pauli}).
Given a parameter~$\tau>1$, we consider the following mapping from
the sphere~$S^3 \subset \R^4$ to the linear operators on~$\H$,
\beq F \::\: S^3 \rightarrow \Lin(\H) \:,\qquad F(x) = \sum_{i=1}^4 \tau\: x^i \gamma^i + \1\:. \eeq
\bitem
\item[(a)] {\em{The matrices~$F(x)$ have two positive and two negative eigenvalues:}} \\
Since the computation of the eigenvalues of $4 \times 4$-matrices is tedious, it is
preferable to proceed as follows. The matrices~$\gamma^j$ are the
Dirac matrices of Euclidean~$\R^4$, satisfying the anti-commutation relations
\beq \{\gamma^i, \gamma^j\} = 2 \delta^{ij}\:\1\qquad (i,j=1,\ldots, 4)\:. \eeq
As a consequence,
\begin{align}
F(x) -\1 &= \sum_{i=1}^4 \tau\: x^i \gamma^i \\
\big(F(x) -\1\big)^2 &= \sum_{i,j=1}^4 \tau^2\: x^i\,x^j \gamma^i \gamma^j
= \frac{\tau^2}{2} \sum_{i,j=1}^4 x^i\,x^j \big\{\gamma^i, \gamma^j \big\} \notag \\
&= \frac{\tau^2}{2} \sum_{i,j=1}^4 x^i\,x^j \:2\, \delta_{ij} \:\1 = \tau^2\, \1 \:.
\end{align}
Hence, the matrix~$F(x)$ satisfies the polynomial equation
\beq \big(F(x) -\1\big)^2 = \tau^2\, \1 \:. \eeq
We conclude that~$F(x)$ has the eigenvalues
\beq \nu_\pm = 1 \pm \tau \:. \eeq
Since~$F(x)-\1$ is trace-free, each eigenvalue must appear with multiplicity two.
Using that~$\tau>1$, we conclude that~$F(x)$ really has two positive and two negative eigenvalues.
\item[(b)] {\em{Construction of a causal fermion system:}} \\
Let~$\mu$ be the normalized Lebesgue measure on~$S^3 \subset \R^4$.
Setting~$\rho = F_* \mu$ defines a causal fermion system of spin dimension two
and total volume one. Since the matrices~$F(x)$ all have trace four, we also know that
\beq \int_\F \tr(x)\: \Diff\rho(x) = \int_{S^3} \tr(F(x))\: \Diff\mu(x) = 4 \:. \eeq
Therefore, the volume constraint~\eqref{volconstraint} and the trace constraint~\eqref{trconstraint}
are satisfied, both with constants independent of~$\tau$.
\item[(c)] {\em{Computation of the eigenvalues of~$F(x)\, F(y)$:}} \\
Again, this can be calculated most conveniently using the Clifford relations.
\begin{align}
F(x) \:F(y) &= \Big( \sum_{i=1}^4 \tau\: x^i \gamma^i + \1 \Big) \Big( \sum_{j=1}^4 \tau\: y^j \gamma^j + \1 \Big) \notag \\
&= \big(1 + \tau^2 \:\la x,y \ra \big) \1 + \tau \sum_{i=1}^4 (x^i+y^i) \gamma^i
+ \frac{\tau^2}{2} \:\sum_{i,j=1}^4 x^i y^j \, \big[\gamma^i, \gamma^j\big] \:. \label{trrel}
\end{align}
Using that
\beq \gamma^i\, \big[\gamma^i, \gamma^j \big] = -\big[\gamma^i, \gamma^j \big]\: \gamma^i \:, \eeq
we conclude that
\begin{align}
\Big(F(x) \:F(y) - \big(1 + \tau^2 \:\la x,y \ra \big) \1 \Big)^2
= \tau^2 \sum_{i=1}^4 (x^i+y^i)^2 + 
\bigg( \frac{\tau^2}{2} \:\sum_{i,j=1}^4 x^i y^j \, \big[\gamma^i, \gamma^j\big] \bigg)^2 \:.
\end{align}
This can be simplified with the help of the relations
\begin{align}
\sum_{i=1}^4 (x^i+y^i)^2 &= 2 + 2 \,\la x,y \ra \\
\bigg( \sum_{i,j=1}^4 x^i y^j \, \big[\gamma^i, \gamma^j\big] \bigg)^2
&= -4 \sin^2 \vartheta = -4\,\big( 1-\la x,y \ra^2 \big)\:, \label{blin}
\end{align}
where~$\vartheta$ is the angle between the vectors~$x,y \in \R^4$.
The relation~\eqref{blin} can be verified in detail as follows. The rotational symmetry of the
Euclidean Dirac operator on~$\R^4$ means that for every rotation~$O \in \text{SO}(4)$ there is
a unitary operator~$U \in \text{SU}(4)$ such that
\beq O^i_j\, \gamma^j = U \gamma^i U^{-1} \:. \eeq
Making use of this rotational symmetry, we can arrange that the vector~$x$ is the basis vector~$e_1$
and that~$y = \cos \vartheta\, e_1 + \sin \vartheta\, e_2$. As a consequence,
\begin{align}
\sum_{i,j=1}^4 x^i y^j \, \big[\gamma^i, \gamma^j \big] &= \sin \vartheta\: [\gamma^1, \gamma^2]
= 2\, \sin \vartheta\: \gamma^1 \gamma^2 \\
\bigg( \sum_{i,j=1}^4 x^i y^j \, \big[\gamma^i, \gamma^j\big] \bigg)^2 &= 
4\, \sin^2 \vartheta\: \gamma^1 \gamma^2 \gamma^1 \gamma^2 \:,
\end{align}
and applying the anti-commutation relations gives~\eqref{blin}.

Combining the above equations, we conclude that the product~$F(x) \:F(y)$ satisfies the polynomial equation
\begin{align}
\Big(F(x) \:F(y) - \big(1 + \tau^2 \:\la x,y \ra \big) \1 \Big)^2
&= 2 \,\tau^2 \big(1+ \la x,y \ra \big) - \tau^4 \:\big(1-\la x,y \ra^2 \big) \notag \\
&= \tau^2 \:\Big(1+ \la x,y \ra \Big) \Big( 2 - \tau^2 \:\big(1-\la x,y \ra \big) \Big) \:.
\end{align}
Taking the square root, the zeros of this polynomial are computed by
\beq \label{leigen4}
\lambda_{1\!/\!2} = 1+\tau^2 \: \la x,y \ra \pm
 \tau \,\sqrt{1+\la x,y \ra}\: \sqrt{2 - \tau^2 \:(1-\la x,y \ra)} \:.
\eeq
Moreover, taking the trace of~\eqref{trrel}, one finds
\beq \tr \big( F(x) \:F(y) \big) = 4\,\big(1 + \tau^2 \:\la x,y \ra \big) \:. \eeq
This implies that each eigenvalue in~\eqref{leigen4} has an algebraic multiplicity of two.
\item[(d)] {\em{Computation of the Lagrangian:}} \\
We again denote the angle between the vectors~$x,y \in \R^4$ by~$\vartheta$.
If~$\vartheta$ is sufficiently small, the term~$(1-\la x,y \ra)$ is close to zero, and thus the
arguments of the square roots in~\eqref{leigen4} are all positive. However, if~$\vartheta$ becomes so large that
\beq \vartheta \:\geq\: \vartheta_{\max} := \arccos \!\left( 1-\frac{2}{\tau^2} \right) , \eeq
then the argument of the last square root in~\eqref{leigen4} becomes negative, so that the~$\lambda_{1\!/\!2}$
form a complex conjugate pair. Moreover, a short calculation shows that
\beq \lambda_1 \lambda_2 = (1+\tau)^2 (1-\tau)^2 > 0 \:, \eeq
implying that if the~$\lambda_{1\!/\!2}$ are both real, then they have the same sign.
Using this information, the Lagrangian simplifies to
\begin{align}
\L(F(x), F(y) \big) &= \frac{1}{8} \sum_{i,j=1}^{4} \Big( \big|\lambda^{xy}_i \big| - \big|\lambda^{xy}_j \big| \Big)^2 
= \frac{1}{2} \sum_{i,j=1}^{2} \Big( \big|\lambda_i \big| - \big|\lambda_j \big| \Big)^2 \notag \\
&= \frac{1}{2}\:\Theta(\vartheta_{\max}-\vartheta)
\sum_{i,j=1}^{2} \Big( \lambda_i -\lambda_j \Big)^2 
= \Theta(\vartheta_{\max}-\vartheta) \big( \lambda_1 -\lambda_2 \big)^2 \notag \\
&= 4 \tau^2\: (1+\cos \vartheta) \left( 2 - \tau^2 \:(1-\cos \vartheta) \right) \:
\Theta(\vartheta_{\max}-\vartheta) \:.
\end{align}
\item[(e)] {\em{Computation of the action:}} \\
Inserting this Lagrangian in~\eqref{Sdef} and using the definition of the push-forward measure, we obtain
\begin{align}
\Sact &= \int_{S^3} \Diff\mu(x) \int_{S^3} \Diff\mu(y) \: \L(F(x), F(y) \big) \notag \\
&= \int_{S^3} \Diff\mu(y) \: \L(F(x), F(y) \big) = \frac{2}{\pi} \int_0^{\vartheta_{\max}}
\L(\cos \vartheta)\: \sin^2 \vartheta\:\dd \vartheta \notag \\
&= \frac{512}{15 \pi}\: \frac{1}{\tau} + {\mathscr{O}}(\tau^{-2})\:.
\end{align}
\eitem
Thus setting~$F_k =F|_{\tau=k}$, we have constructed a divergent minimizing sequence.
However, the integral in the boundedness constraint~\eqref{Tdef} also diverges as~$k \rightarrow \infty$.
This example shows that, leaving out the boundedness constraint, there is no minimizer.
}} \QEDrem 
\end{Example}

We finally remark that this example is not as artificial or academic as it might appear
at first sight. Indeed, as observed in the master thesis~\cite{kilbertus},
when discretizing a Dirac system in~$\R \times S^3$ (where the sphere can be thought of as
a spatial compactification of Minkowski space), then in the simplest case of four occupied Dirac states
(referred to as ``one shell,'' i.e.\ $\dim \H=4$), this system reduces precisely to the
last example. In simple terms, this observation can be summarized by saying that
Clifford structures tend to make the causal action small.

\section{The Radon-Nikodym Theorem} \label{secradonnikodym}
\sindex{Radon-Nikodym theorem|textbf}%
As already mentioned at the beginning of the previous section, one difficulty in the existence proof
for the causal action principle is the fact that the set~$\F$ is unbounded and thus non-compact.
In order to deal with this difficulty, we need one more mathematical tool: the Radon-Nikodym theorem.
We now give the proof of the Radon-Nikodym theorem
by von Neumann following the presentation in~\cite[Chapter~6]{rudin}.
An alternative method of proof is given in~\cite{halmosmt, evans+gariepy}.
As in Section~\ref{secriesz}, it again suffices to consider the case that the base space~$\K$
is a {\em{compact}} topological space.

\begin{Def} A Radon measure~$\lambda$ is {\bf{absolutely continuous}}
with respect to another Radon measure~$\nu$, denoted by
\beq \lambda \ll \nu \:, \eeq
\sindex{measure!absolutely continuous}%
\nindex{aga@$\ll$ -- absolute continuity of measures}%
if the implication
\beq \nu(E)=0 \quad \Longrightarrow \quad \lambda(E)=0 \eeq
holds for any Borel set~$E$.
The measure~$\lambda$ is {\bf{concentrated}} on the Borel set~$A$
\sindex{measure!concentrated on}%
if~$\lambda(E) = \lambda(E \cap A)$ for all Borel sets~$E$.
The measures~$\lambda$ and~$\mu$ are {\bf{mutually singular}}, denoted by
\sindex{measure!mutually singular}%
\beq \lambda \perp \nu \:, \eeq
if there are disjoint Borel sets~$A$ and~$B$ such that~$\lambda$ is concentrated in~$A$
and~$\nu$ is concentrated in~$B$.
\end{Def} \noindent
In order to avoid confusion, we point out that the supports of two mutually singular measures
are not necessarily disjoint, as one sees in the simple example
of the Lebesgue measure on~$(0,1)$ and the Dirac measure supported at the origin,
\beq \lambda := \dd x|_{(0,1)} \qquad \text{and} \qquad \mu = \delta_0 \:. \eeq
Since the support is by definition a closed set (see~\eqref{suppdef}),
the support of~$dx|_{(0,1)}$ contains the origin, which is precisely
the support of the Dirac measure. But clearly, the two measures are
concentrated on the sets~$(0,1)$ and~$\{0\}$, respectively,
and are thus mutually singular.

\begin{Thm} {\bf{(Radon-Nikodym)}} \label{thmRN}
Let~$\mu$ and~$\lambda$ be Radon measures on a compact topological space~$\K$.
\bitem
\item[{\rm{(a)}}] There is a unique pair of Borel measures~$\lambda_a$ and~$\lambda_s$ such that
\beq \lambda = \lambda_a + \lambda_s \qquad \text{and} \qquad \lambda_a \ll \mu \:,\quad
\lambda_s \perp \mu \:. \eeq
\item[{\rm{(b)}}]  There is a unique function~$h \in L^1(\K, \Diff\mu)$ such that
\beq \label{RNrep}
\lambda_a(E) = \int_E h\: \Diff\mu \qquad \text{for every Borel set~$E$}\:.
\eeq
\eitem
\end{Thm} \noindent
The pair~$(\lambda_a, \lambda_s)$ is also referred to as the {\bf{Lebesgue decomposition}} of~$\lambda$
with respect to~$\mu$.
\sindex{measure!Lebesgue decomposition|textbf}%
\Proof[Proof of Theorem~\ref{thmRN}.]
The uniqueness of the decomposition is easily seen as follows: Suppose that~$(\lambda'_a, \lambda'_s)$
is another Lebesgue decomposition. Then
\beq \label{diffmeasure}
\lambda'_a - \lambda_a = \lambda_s - \lambda'_s \:.
\eeq
Since~$\lambda_s \perp \mu$ and~$\lambda'_s \perp \mu$,
the measures~$\lambda_s$ and~$\lambda_{s'}$ are concentrated in a Borel set~$A$
with~$\mu(A)=0$. Evaluating~\eqref{diffmeasure} on Borel subsets of~$A$, the
left side vanishes, because~$\lambda_a$ and~$\lambda'_a$ are both absolutely continuous
with respect to~$\mu$. Hence~$\lambda'_s - \lambda_s=0$.
Using this relation in~\eqref{diffmeasure}, we also conclude that~$\lambda'_a - \lambda_a=0$.
This proves uniqueness.

For the existence proof, we let~$\rho$ be the measure~$\rho=\lambda+\mu$. Then
\beq \label{rlm}
\int_\K f\: \Diff\rho = \int_\K f\: \dd\lambda + \int_\K f\: \Diff\mu
\eeq
for any non-negative Borel function~$f$. If~$f \in L^2(\K, \Diff\rho)$, the Schwarz inequality gives
\beq \bigg| \int_\K f\: \dd\lambda \bigg| \leq \int_\K |f|\: \Diff\rho \leq
\sqrt{\rho(\K)}\; \|f\|_{L^2(\K, \Diff\rho)} \:. \eeq
Therefore, the mapping~$f \mapsto \int_\K f\, d\lambda$ is a bounded linear
functional on~$L^2(\K, \Diff\rho)$. By the Fr{\'e}chet-Riesz theorem, we can represent this
linear functional by a function~$g \in L^2(\K, \Diff\rho)$, i.e.
\beq \label{grep}
\int_\K f\: \dd\lambda = \int_\K g\, f\: \Diff\rho \qquad \text{for all~$f \in L^2(\K, \Diff\rho)$}\:.
\eeq

We next want to show that, by modifying~$g$ on a set of~$\rho$-measure zero,
we can arrange that~$g$ takes values in the interval~$[0,1]$.
To this end, we let~$E$ be any Borel set with~$\rho(E) >0$.
Evaluating~\eqref{grep} for~$f=\chi_E$, we obtain
\beq 0 \leq \frac{1}{\rho(E)} \int_E g\: \Diff\rho = \frac{\lambda(E)}{\rho(E)} \leq 1 \:. \eeq
Now the claim follows from elementary measure theory
(see, for example, \cite[Theorem~1.40]{rudin}).

Using~\eqref{rlm}, we can rewrite~\eqref{grep} as
\beq \label{grep2}
\int_\K (1-g)\: f\: \dd\lambda = \int_\K g\, f\: \Diff\mu \qquad \text{for all non-negative~$f \in L^2(\K, \Diff\rho)$}\:.
\eeq
We introduce the Borel sets
\beq A = \{x \in \K \:|\: 0 \leq g(x) < 1 \} \qquad \text{and} \qquad
B = \{x \in \K \:|\: g(x) = 1 \} \:. \eeq
and define the measures~$\lambda_a$ and~$\lambda_s$ by
\beq \dd\lambda_a = \chi_A\: \dd\lambda \qquad \text{and} \qquad \dd\lambda_s = \chi_B \:\dd\lambda \:. \eeq
Choosing~$f=\chi_B$ in~\eqref{grep2}, one sees that~$\mu(B)=0$, implying that~$\lambda_s \perp \mu$.

Moreover, since~$g$ is bounded, we can evaluate~\eqref{grep2} for
\beq f = \big(1+g+\cdots+g^n \big)\: \chi_E \eeq
for any~$n \in \N$ and any Borel set~$E$. Using the same transformation
with ``telescopic sums'' as in the evaluation of the geometric or Neumann series, we obtain
\beq \label{grep3}
\int_E \big(1-g^{n+1} \big) \:\dd\lambda
= \int_E g\, \big(1+g+\cdots+g^n \big)\: \Diff\mu \:.
\eeq
At every point of~$B$, the factor~$(1-g^{n+1})$ in the integrand on the left
vanishes. At every point of~$A$, on the other hand,
the factor~$(1-g^{n+1})$ is monotone increasing in~$n$ and converges to one. Hence, Lebesgue's
monotone convergence theorem implies that the left-hand side of~\eqref{grep3} converges to
\beq \lim_{n \rightarrow \infty} \int_E \big(1-g^{n+1} \big) \:\dd\lambda = \lambda\big( E \cap A\big) \:. \eeq
The integrand on the right-hand side of~\eqref{grep3}, on the other hand, is monotone increasing in~$n$, so that
the limit
\beq h(x) := \lim_{n \rightarrow \infty} g(x)\, \big(1+g(x)+\cdots+g^n(x) \big)
\quad \text{exists in~$\R^+_0 \cup \{\infty\}$} \:. \eeq
Moreover, the monotone convergence theorem implies that
\beq \lim_{n \rightarrow \infty} \int_E g\, \big(1+g+\cdots+g^n \big)\: \Diff\mu = \int_E h\, \Diff\mu 
\;\in\; \R^+_0 \cup \{\infty\}\:. \eeq
We conclude that, in the limit~$n \rightarrow \infty$, the relation~\eqref{grep3} yields
\beq \lambda_a(E) = \lambda\big( E \cap A\big) = \int_E h\, \Diff\mu \qquad \text{for any Borel set~$E$} \:. \eeq
Choosing~$E=\K$, one sees that~$h \in L^1(\K, \Diff\mu)$. This concludes the proof of~\eqref{RNrep}.
Finally, the representation~\eqref{RNrep} implies that~$\lambda_a \ll \mu$.
\QED

\section{Moment Measures} \label{secmoment}
\sindex{measure!moment}%
\sindex{moment measure|textbf}%
We now introduce an important concept needed for the existence proof: the moment measures.
We again assume that the Hilbert space is finite-dimen\-sio\-nal and that the measure~$\rho$
is normalized~\eqref{Hfinitenormed}.
We consider~$\F$ with the metric induced by the sup-norm on~$\Lin(\H)$; that is,
\beq d(p,q) = \|p-q\| \eeq
(and~$\|.\|$ as in~\eqref{supnorm}).
The basic difficulty in applying the abstract theorems is that~$\F$ is
{\em{not compact}} (indeed, it is a star-shaped in the sense that~$p \in \F$
implies~$\lambda p \in \F$ for all~$\lambda \in \R$).
If the metric space is non-compact, our abstract results no longer apply, as
becomes clear in the following simple example.

\begin{Example} {\em{
Consider the Banach space~$C^0_0(\R,\R)$ of compactly supported continuous
functions. Let~$\rho_n = \delta_n$ be the sequence of Dirac measures supported at~$n \in \N$.
Then for any~$f \in C^0_0(\R,\R)$,
\beq \lim_{n \rightarrow \infty} \int_{-\infty}^\infty f\, \Diff\rho_n = \lim_{n \rightarrow \infty} f(n) = 0 \:. \eeq
Hence the sequence~$(\rho_n)_{n \in \R}$ converges as a measure to zero.
Thus the limiting measure is no longer normalized.
This shows that Theorem~\ref{thmcompact} fails to hold if the base space is non-compact.
}} \QEDrem
\end{Example}

The way out is to make use of the fact that the causal action as well as the constraints
are formed of functionals that are homogeneous under the scaling~$p \rightarrow \lambda p$
of degree zero, one or two. This makes it possible to restrict attention to a compact subset of~$\F$,
and to consider three measures on this compact set. We now give the needed definitions.

\begin{Def} \label{defmm}
Let~$\K$ be the compact metric space
\beq \K = \{ p \in \F \text{ with } \|p\|=1 \} \cup \{0\} \:. \eeq
For a given measure~$\rho$ on~$\F$, we define the measurable sets of~$\K$ by the
requirement that the sets~$\R^+ \Omega = \{ \lambda p \:|\: \lambda \in \R^+, p \in \Omega\}$
and~$\R^- \Omega$ should be $\rho$-measurable in~$\F$. We introduce the measures~$\m^{(0)}$, 
$\m^{(1)}_\pm$ and~$\m^{(2)}$ by
\nindex{agb@$\m^{(\ell)}$ -- $\ell^\text{th}$ moment measure}%
\begin{align}
\m^{(0)}(\Omega) &= \frac{1}{2}\: \rho \big(\R^+ \Omega \setminus \{0\} \big) 
+ \frac{1}{2}\: \rho \big( \R^- \Omega \setminus \{0\} \big)
+ \rho \big( \Omega \cap \{0\} \big) \label{m0def} \\
\m^{(1)}_+(\Omega) &= \frac{1}{2} \int_{\R^+ \Omega} \|p\| \,\Diff\rho(p) \label{m1pdef} \\
\m^{(1)}_-(\Omega) &= \frac{1}{2} \int_{\R^- \Omega} \|p\| \,\Diff\rho(p) \label{m1mdef} \\
\m^{(2)}(\Omega) &= \frac{1}{2} \int_{\R^+ \Omega} \|p\|^2 \,\Diff\rho(p) \:+\:
\frac{1}{2} \int_{\R^- \Omega} \|p\|^2 \,\Diff\rho(p)
\:. \label{m2def}
\end{align}
The measures~$\m^{(l)}$ and~$\m^{(l)}_\pm$ are referred to as the~$l^\text{th}$ {\bf{moment measure}}.
\end{Def} \noindent
As a short notation, it is convenient to abbreviate the difference
of the first moment measures by
\beq
\m^{(1)}(\Omega) := \m^{(1)}_+(\Omega) - \m^{(1)}_-(\Omega) \label{m1def} \:.
\eeq
We remark that~$\m^{(1)}$ can be regarded as a {\em{signed measure}}
(see, for example, \cite[\S28]{halmosmt} or~\cite[Chapter~6]{rudin}).
For simplicity, we here avoid the concept of signed measures by
working instead with the two (positive) measures~$\m^{(1)}_\pm$.
Nevertheless, we introduce an $\m^{(1)}$-integral as a short notation for
the difference of the integrals with respect to~$\m^{(1)}_+$ and~$\m^{(1)}_-$; that is,
\beq \int_\K h\: \dd\m^{(1)} := \int_\K h\: \dd\m^{(1)}_+ - \int_\K h\: \dd\m^{(1)}_- \:, \eeq
where for simplicity we always assume that~$h$ is continuous.

The $\rho$-integrals of homogeneous functions can be rewritten as
integrals over~$\K$ using the moment measures, as we now make precise.
\begin{Def} A function~$h \in C^0(\F)$ is called {\bf{homogeneous of degree}}~$\ell$
with~$\ell \in \{0,1,2\}$ if
\sindex{homogeneous of degree~$\ell$}%
\beq \label{homogeneous}
h(\nu x) = \nu^\ell\, h(x) \qquad \text{for all~$\nu \in \R$ and~$x \in \F$}\:.
\eeq
\end{Def}

\begin{Lemma} Let~$h \in C^0(\F)$ be a function which is homogeneous of degree~$\ell \in \{0,1,2\}$. Then
\beq \int_\F h\: \Diff\rho = \int_\K h\: \dd\m^{(l)}\:. \eeq
\end{Lemma}
\Proof We first note that, using the homogeneity~\eqref{homogeneous}, the function~$h$
is uniquely determined by its restriction to~$\K$.
Moreover, using an approximation argument with Lebesgue's dominated convergence theorem,
it suffices to consider a function~$h$ which is homogeneous of degree~$\ell$
and {\em{simple}} in the sense that its restriction to~$\K$ takes a finite number of values; that is,
\beq h \big|_{\K} = \sum_{i=1}^N c_i\, \chi_{\Omega_i} \eeq
with suitable Borel sets~$\Omega_1, \ldots, \Omega_N \subset \K$
and real coefficients~$c_1, \ldots, c_N$. For such simple functions, the integrals go over to finite sums,
and we obtain
\begin{align}
\int_\F h\: \Diff\rho = \sum_{i=1}^N c_i \int_{\R^+ \Omega_i} \|p\|^\ell\: \Diff\rho(p)
= \sum_{i=1}^N c_i \:\m^{(\ell)} \big( \Omega_i \big) = \int_\K h\: \dd\m^{(l)}\:,
\end{align}
as desired. This concludes the proof.
\QED
Applying this lemma, the normalization~$\rho(\F)=1$ can be expressed in terms
of the moment measures as
\beq \label{m0}
\m^{(0)}(\K) = 1 \:,
\eeq
whereas the action~\eqref{Sdef} as well as the functionals in the
constraints~\eqref{Tdef} and~\eqref{trconstraint} can be written as
\begin{align}
\Sact(\rho) &= \iint_{\K \times \K} \L(p, q)\: \dd\m^{(2)}(p) \,\dd\m^{(2)}(q) \label{Sm2} \\
\T(\rho) &= \iint_{\K \times \K} |p \, q|^2\:  \dd\m^{(2)}(p) \,\dd\m^{(2)}(q) \label{Tm2} \\
\int_\F \tr(x)\: \Diff\rho(x) &= \int_\K \tr(p)\, \dd\m^{(1)}_+(p) - \int_\K \tr(p)\, \dd\m^{(1)}_-(p)  \:. \label{m1}
\end{align}

Working with the moment measures has the advantage that they are measures on the
{\em{compact}} space~$\K$. We also learn that two measures~$\rho$ and~$\tilde{\rho}$
whose moment measures coincide yield the same values for the functionals~$\Sact$ and~$\T$ as well
as for the integral~\eqref{m1} entering the trace constraint.
It is most convenient to work exclusively with the moment measures.
At the end, we shall construct a suitable representative~$\rho$ of the limiting moment
measures. A key step for making this method work is the following a-priori estimate.

\begin{Lemma} \label{lemma212} There is a constant~$\varepsilon=\varepsilon(f,n)>0$ such that for
every measure~$\rho$ on~$\F$ the corresponding moment measures (see Definition~\ref{defmm})
satisfy for all measurable~$\Omega \subset \K$ the following inequalities:
\begin{align}
\Big( \m^{(1)}_+(\Omega) + \m^{(1)}_-(\Omega) \Big)^2 &\leq \m^{(0)}(\Omega)\:\m^{(2)}(\Omega) \label{m1es} \\
\m^{(2)}(\K) &\leq\; \frac{\sqrt{\T(\rho)}}{\varepsilon}\:. \label{m2es}
\end{align}
\end{Lemma}
\Proof The inequality~\eqref{m1es} follows immediately from H\"older's inequality,
\begin{align}
\big| 2 \big(\m^{(1)}_+(\Omega) + \m^{(1)}_-(\Omega) \big) \big|^2 &\leq  \left( \int_{\R \Omega} \|p\| \,\Diff\rho(p) \right)^2 \notag \\
&\leq \rho(\R \Omega) \int_{\R \Omega} \|p\|^2 \,\Diff\rho(p) \leq 4 \m^{(0)}(\Omega)\: \m^{(2)}(\Omega) \:.
\end{align}

In order to prove~\eqref{m2es}, we introduce the mapping
\beq \phi \::\: \K \times \K \rightarrow \R \::\: (p,q) \mapsto |p \, q|\:. \eeq
Clearly, $\phi$ is continuous and
\beq \phi(p,p) = |p^2| = \Tr(p^2) \geq \|p\|^2 = 1 \eeq
(here we used that the Hilbert-Schmidt norm is larger than the
absolute square of each eigenvalue).
Thus every point~$r \in \K$ has a neighborhood~$U(r) \subset \K$ with
\beq \label{Uineq}
\phi(p,q) \geq \frac{1}{2} \qquad \text{for all~$p,q \in U(r)$}\:.
\eeq
Since~$\K$ is compact, there is a finite number of points~$r_1,\ldots, r_N$
such that the corresponding sets~$U_i:=U(r_i)$ cover~$\K$. Due to the additivity property
of measures, there is an index~$i \in \{1, \ldots, N\}$ such that
\beq \label{mui}
\m^{(2)}(U_i) \geq \frac{\m^{(2)}(\K)}{N}\:.
\eeq

We write~$\T$ in the form~\eqref{Tm2} and apply~\eqref{Uineq} as well
as~\eqref{mui} to obtain
\beq \T(\rho) \geq \iint_{U_i \times U_i} |p \, q|^2\: \dd\m^{(2)}(p) \,\dd\m^{(2)}(q)
\geq \frac{1}{2}\: \m^{(2)}(U_i)^2 \geq
\frac{\m^{(2)}(\K)^2}{2 N^2}\:. \eeq
Setting~$\varepsilon=1/(\sqrt{2} N)$, the result follows.
\QED

\section{Existence of Minimizers for the Causal Action Principle} \label{secexistence}
After the above preparations, we can follow the strategy of the direct method in the calculus of
variations described at the beginning of Chapter~\ref{secmeasure}
to obtain the following result.
\begin{Thm} \label{thmexistfinite} Let~$\H$ be a finite-dimensional Hilbert space and~$n \in \N$.
Let~$\rho_k$ be a minimizing sequence of regular Borel measures on~$\F$ satisfying our
constraints~\eqref{volconstraint}, \eqref{trconstraint} and~\eqref{Tdef}
(for fixed and finite constants). Then there is a regular Borel measure~$\rho$ which
also satisfies the constraints (again with the same constants) and
\beq \Sact(\rho) = \liminf_{n \rightarrow \infty} \Sact(\rho_n) \:. \eeq
\end{Thm}
In short, the method for constructing~$\rho$ is to take a limit of the moment measures of the~$\rho_k$
and to realize this limit by the measure~$\rho$. In more detail, we proceed as follows.
In view of Lemma~\ref{lemma212}, we know that the
moment measures are uniformly bounded measures on
the compact metric space~$\K$.
Applying the compactness result of Theorem~\ref{thmcompact}
(based on the Banach-Alaoglu theorem and the Riesz representation theorem),
we conclude that for a suitable subsequence (which we again
denote by~$(\rho_k)$), the moment measures converge in the~$C^0(\K)^*$-topology to
regular Borel measures,
\beq \m_k^{(0)} \rightarrow \m^{(0)} \:,\qquad \m_{k,\pm}^{(1)} \rightarrow \m^{(1)}_\pm \qquad \text{and} \qquad
\m_k^{(2)} \rightarrow \m^{(2)} \:, \eeq
which again have the properties~\eqref{m0}, \eqref{m1es} and~\eqref{m2es}.

We next form the Radon-Nikodym decompositions of~$\m^{(1)}_\pm$ and~$\m^{(2)}$ with
respect to~$\m^{(0)}$. The inequality~\eqref{m1es} shows that every set of $\m^{(0)}$-measure
zero is also a set of measure zero with respect to~$\m^{(1)}_+$ and~$\m^{(1)}_-$.
In other words, the measures~$\m^{(1)}_\pm$ are absolutely continuous with respect
to~$\m^{(0)}$. Hence, applying Theorem~\ref{thmRN}, we obtain the Radon-Nikodym decompositions
\beq \dd\m^{(1)}_\pm = f_\pm \: \dd\m^{(0)} \qquad \text{with} \qquad f_\pm \in L^1(\K, \dd\m^{(0)})\:. \eeq
As a consequence, the difference of measures~$\m^{(1)}$ in~\eqref{m1def} has the decomposition
\beq \label{RNdecomp1}
\dd\m^{(1)} = f^{(1)} \: \dd\m^{(0)} \qquad \text{with} \qquad f^{(1)} := f_+ - f_- \in L^1(\K, \dd\m^{(0)})\:.
\eeq
As we do not know whether also~$\m^{(2)}$ is absolutely continuous with respect to~$m^{(0)}$,
Theorem~\ref{thmRN} gives the decomposition
\beq \label{RNdecomp2}
\dd\m^{(2)} = f^{(2)}\, \dd\m^{(0)} + \dd\m^{(2)}_\text{sing} \qquad \text{with} \qquad f^{(2)} \in L^1(\K, \dd\m^{(0)}) \:,
\eeq
where the measure~$\m^{(2)}_\text{sing}$ is singular with respect to~$\m^{(0)}$.

\begin{Lemma} The two functions~$f^{(1)}$ and~$f^{(2)}$ in the Radon-Nikodym
de\-com\-po\-si\-tions~\eqref{RNdecomp1} and~\eqref{RNdecomp2} can be chosen such that
\beq \big| f^{(1)} \big|^2 \leq f^{(2)} \:. \eeq
\end{Lemma}
\Proof Since~$\m^{(2)}_\text{sing} \perp \m^{(0)}$, there is a Borel set~$V$ such
that
\beq \chi_V\: \dd\m^{(0)} = \dd\m^{(0)} \qquad \text{and} \qquad \chi_V\: \dd\m^{(2)}_\text{sing} = 0 \:. \eeq
Then, using the Radon-Nikodym decompositions~\eqref{RNdecomp1} and~\eqref{RNdecomp2} in~\eqref{m1es},
we obtain for any Borel set~$U \subset V$ the inequality
\beq \bigg| \int_U f^{(1)}\, \dd\m^{(0)} \bigg|^2 \leq \m^{(0)}(U)\: \int_U f^{(2)}\, \dd\m^{(0)} \:. \eeq
If the function~$f^{(1)}$ does not change signs on~$U$, we conclude that
\beq \inf_U \big| f^{(1)} \big|^2 \leq \sup_U f^{(2)} \:. \eeq
By decomposing~$U$ into the two sets where~$f^{(1)}$ is positive
and negative, respectively, one readily sees that this inequality even holds for any Borel set~$U \subset V$.
As a consequence, the inequality~$|f^{(1)}|^2 \leq f^{(2)}$ holds almost everywhere
(with respect to the measure~$\m^{(0)})$, concluding the proof.
\QED
In particular, we conclude that~$f^{(1)}$ even lies in~$L^2(\K, \dd\m^{(0)})$. Setting~$f=f^{(1)}$
and~$\dd\n = (f^{(2)}-|f|^2)\, \dd\m^{(0)} + \dd\m^{(2)}_\text{sing}$,
we obtain the decomposition
\beq \label{RN}
\dd\m^{(1)} = f\, \dd\m^{(0)} \:,\qquad
\dd\m^{(2)} = |f|^2\, \dd\m^{(0)} + \dd\n\:,
\eeq
where~$f \in L^2(\K, \dd\m^{(0)})$, and~$\n$ is a positive measure that need not be
absolutely continuous with respect to~$\m^{(0)}$. From the definition~\eqref{m1def}
it is clear that~$f$ is odd; that is,
\beq %\label{fodd}
f(-p) = -f(p) \quad \text{for all~$p \in \K$}\:. \eeq

The remaining task is to represent the limiting moment measures~$\m^{(l)}$ in~\eqref{RN} by
a measure~$\rho$. Unfortunately, there is the basic problem that such a measure can exist only
if~$\m^{(2)}$ is absolutely continuous with respect to~$\m^{(0)}$, as the following consideration shows:
Assume conversely that~$\m^{(2)}$ is not absolutely continuous with respect to~$\m^{(0)}$.
Then there is a measurable set~$\Omega \subset \K$ with~$\m^{(0)}(\Omega)=0$
and~$\m^{(2)}(\Omega) \neq 0$.
Assume furthermore that there is a measure~$\rho$ on~$\F$ which represents the limiting
moment measures in the sense that~\eqref{m0def}--\eqref{m2def} hold.
From~\eqref{m0def}, we conclude that the set~$\R \Omega \subset \F$
has $\rho$-measure zero. But then the integral~\eqref{m2def} also vanishes, a contradiction.

This problem can also be understood in terms of the limiting sequence~$\rho_k$.
We cannot exclude that there is a star-shaped region~$\R \Omega \subset \F$
such that the measures~$\rho_k(\R \Omega)$ tend to zero, but the corresponding moment
integrals~\eqref{m2def} have a non-zero limit. Using a notion from the calculus of variations
for curvature functionals, we refer to this phenomenon as the possibility of {\em{bubbling}}.
This bubbling effect is illustrated by the following example.

\begin{Example} (Bubbling) \label{ex213} {\em{ We choose~$f=2$ and~$n=1$. 
Furthermore, we let~$\scrM=[0, 1)$ with~$\mu$ the Lebesgue measure.
For any parameters~$\kappa \geq 0$ and~$\varepsilon \in (0, \frac{1}{2})$,
we introduce the mapping~$F_\varepsilon : \scrM \rightarrow \F$ by
\beq F_\varepsilon(x) = \displaystyle \frac{1}{1-2 \varepsilon} \times \left\{ \begin{array}{ll}
-\kappa \,\varepsilon^{-\frac{1}{2}}\: \sigma^3 & \text{if~$x \leq \varepsilon$} \\[.1em]
 \1+ \sigma^1 \,\cos(\nu x) + \sigma^2 \,\sin(\nu x) \quad& \text{if~$\varepsilon < x
\leq 1-\varepsilon$} \\[.1em]
\kappa \,\varepsilon^{-\frac{1}{2}}\: \sigma^3 & \text{if~$x > 1-\varepsilon\:,$}
\end{array} \right. \eeq
where we set~$\nu = 2 \pi/(1-2 \varepsilon)$ (and~$\sigma^j$ are the Pauli matrices).
The corresponding measure~$\rho_\varepsilon$ on~$\F$ has the following properties. On the set
\beq S := \{ \1 + v^1 \sigma^1 + v^2 \sigma^2 \text{ with } (v^1)^2 + (v^2)^2 = 1 \} \:, \eeq
which can be identified with a circle~$S^1$, $\rho_\varepsilon$ is
a multiple of the Lebesgue measure. Moreover, $\rho_\varepsilon$ is supported at the
two points
\beq \label{ppmdef}
p_\pm := \pm \frac{\kappa\, \varepsilon^{-\frac{1}{2}}}{1-2 \varepsilon}\: \sigma^3 \qquad
\text{with} \qquad \rho_\varepsilon(\{p_+\}) = \rho_\varepsilon(\{p_-\}) = \varepsilon \:.
\eeq

A short calculation shows that the trace constraint is satisfied.
Furthermore, the separations of the points~$p_+$ and~$p_-$ from each other and from~$S$
are either spacelike or just in the boundary case between spacelike and timelike. 
Thus for computing the action, we only need to take into account the pairs~$(p_+, p_+)$,
$(p_-, p_-)$ as well as pairs~$(x,y)$ with~$x, y \in S$.
A straightforward computation yields
\beq \label{STform}
\Sact(\rho_\varepsilon) = \frac{3}{(1-2 \varepsilon)^2} \:,\qquad
\T(\rho_\varepsilon) = \frac{6}{(1-2 \varepsilon)^2}  
+ \frac{16 \kappa^2}{(1-2 \varepsilon)^3} + \frac{16 \kappa^4}{(1-2 \varepsilon)^4}\:.
\eeq

Let us consider the limit~$\varepsilon \searrow 0$. From~\eqref{STform} we see that
the functionals~$\Sact$ and~$\T$ converge,
\beq \label{STval}
\lim_{\varepsilon \searrow 0} \Sact = 3\:, \qquad
\lim_{\varepsilon \searrow 0} \T = 6 + 16\, (\kappa^2 + \kappa^4)\:.
\eeq
Moreover, there are clearly no convergence problems on the set~$S$.
Thus it remains to consider the situation at the two points~$p_\pm$, \eqref{ppmdef},
which move to infinity as~$\varepsilon$ tends to zero. These two points enter the moment
measures only at the corresponding normalized points~$\hat{p}_\pm = p_\pm/\|p_\pm\| \in \K$.
A short calculation shows that the limiting moment measures~$\m^{(l)}=\lim_{\varepsilon \searrow 0}
\m^{(l)}_\varepsilon$ satisfy the relations
\beq \m^{(0)}(\{\hat{p}_\pm\}) = 
\m^{(1)}(\{\hat{p}_\pm\}) = 0 \qquad \text{but} \qquad
\m^{(2)}(\{\hat{p}_\pm\}) = \kappa^2 > 0 \:. \eeq
Hence, $\m^{(2)}$ is indeed {\em{not}} absolutely continuous with respect to~$\m^{(0)}$.

In order to avoid misunderstandings, we point out that this example does {\em{not}} show that
bubbling really occurs for minimizing sequences, because we do not know whether the
family~$(\rho_\varepsilon)_{0<\varepsilon<1/2}$ is minimizing. But at least,
our example shows that bubbling makes it possible to arrange arbitrary large values of~$\T$
without increasing the action~$\Sact$ (see~\eqref{STval} for large~$\kappa$).
}} \QEDrem
\end{Example}

In order to handle possible bubbling phenomena, it is important to observe that the second
moment measure does not enter the trace constraint.
Therefore, by taking out the term~$d\n$ in~\eqref{RN} we {\em{decrease}} the
functionals~$\Sact$ and~$\T$ (see~\eqref{Sm2} and~\eqref{Tm2}), without affecting the
trace constraint. It remains to show that the resulting moment measure
can indeed be realized by a measure~$\rho$. This is proven in the next lemma.

\begin{Lemma} \label{lemma214}
For any normalized regular Borel measure~$\m^{(0)}$ on~$\K$ and any function~$f \in L^2(\K, \R)$,
there is a normalized regular Borel measure~$\tilde{\rho}$ whose moment measures~$\tilde{\m}^{(l)}$ are given by
\beq \label{tmmom}
\tilde{\m}^{(0)} = \m^{(0)}\:, \qquad
\dd\tilde{\m}^{(1)} =  f\, \dd\m^{(0)} \:,\qquad
\dd\tilde{\m}^{(2)} = |f|^2\, \dd\m^{(0)}\:.
\eeq
\end{Lemma}
\Proof We introduce the mapping
\beq F \::\: \K \rightarrow \F \:,\qquad F(x) = f(x)\, x \:. \eeq
Choosing~$\tilde{\rho} := F_* \m^{(0)}$, a direct computation shows that the
corresponding moment measures indeed satisfy~\eqref{tmmom}.
\QED

This concludes the proof of Theorem~\ref{thmexistfinite}.
We finally remark that the fact that we dropped the measure~$d\n$ in~\eqref{RN}
implies that the value of~$\T$ might decrease in the limit.
This is the only reason why the boundedness constraint~\eqref{Tdef}
is formulated as an inequality, rather than an equality.
It is not clear if the causal action principle also admits minimizers
if the inequality in~\eqref{Tdef} is replaced by an equality.
We conjecture that the answer is yes. But at present, there is no proof.
We note that, for the physical applications, it makes no difference if~\eqref{Tdef}
is an equality or an inequality, because in this case one works with the
corresponding EL equations, where the constraints
enter only via Lagrange multiplier terms (for details, see~\cite{lagrange}).

\section[Existence of Minimizers in the Non-compact Setting]{Existence of Minimizers for Causal Variational Principles in the Non-compact Setting}\label{seccvpnoncompact}
\sindex{causal variational principle!in the non-compact setting}%
In Theorem~\ref{thmexistfinite}, the existence of minimizers was established in the
case that the Hilbert space~$\H$ is finite-dimensional and the total volume~$\rho(\F)$
of spacetime is finite. From the physical point of view, this existence result is quite satisfying, because
one can take the point of view that it should be possible to describe our universe 
at least approximately by a causal fermion system with~$\dim \H< \infty$ and~$\rho(\F)<\infty$.
From the mathematical point of view, however, it is interesting and important to also study the
cases of an infinite-dimensional Hilbert space and/or infinite total volume.
The case~$\dim \H<\infty$ and~$\rho(\F)=\infty$ is not sensible (see Exercise~\ref{exm2}).
In the case~$\dim \H=\infty$ and~$\rho(\F)<\infty$, on the other hand, there are minimizing
sequences that converge to zero.
Therefore, it remains to study the {\em{infinite-dimensional setting}}~$\dim \H = \infty$ and~$\rho(\F) = \infty$
already mentioned in Section~\ref{seccap}.
In this setting, the existence theory is difficult and has not yet been developed.
Therefore, our strategy is to approach the problem in two steps.
The first step is to deal with infinite total volume; this has been carried out in~\cite{noncompact}.
The second step, which involves the difficulty of dealing with non-locally compact spaces,
is currently under investigation (for first results, see~\cite{langerinfdim} and~\cite{minmov}).

We now outline the basic strategy in the simplest possible case
(more details and a more general treatment can be found in~\cite{noncompact}).
We consider {\em{causal variational principles in the non-compact setting}} as introduced
in Section~\ref{secnoncompact}. Moreover, we consider the {\em{smooth setting}}
by assuming that the Lagrangian is smooth,
\beq \L \in C^\infty(\F \times \F, \R^+_0) \:, \eeq
and has the properties~(i) and~(ii) on page~\pageref{Cond1}.
Moreover, we again assume that the Lagrangian has {\em{compact range}}
(see Definition~\ref{defcompactrange}).
The goal of this section is to prove the following theorem.
\begin{Thm} \label{thmELexist}
Under the above assumptions, there is a regular Borel measure~$\rho$ on~$\F$
(not necessarily bounded) which satisfies the EL equations
\beq \label{ELfinal}
\ell \big|_{\supp \rho} \equiv \inf_M \ell = 0 \qquad \text{with} \qquad
\ell(x) := \int_\F \L(x,y)\: \Diff\rho(y) - 1 \:.
\eeq
\end{Thm}

For the proof, we first exhaust~$\F$ by compact sets~$(K_j)_{j \in \N}$; that is,
\beq K_1 \subset K_2 \subset \cdots \subset \F \qquad \text{and} \qquad
\bigcup_{j=1}^\infty K_j = \F \:. \eeq
On each~$K_j$, we consider the restricted variational principle where we minimize the action
\beq \Sact_{K_j}(\rho) = \int_{K_j} \Diff\rho(x) \int_{K_j} \Diff\rho(y) \: \L(x,y) \eeq
under variations of~$\rho$ within the class of normalized regular Borel measure on~$K_j$.
Using the existence theory in the compact setting (see Theorem~\ref{thmexistcompact}),
each of these restricted variational principles has a minimizer, which we denote by~$\rho_j$.
Each of these measures satisfies the EL equations stated in Theorem~\ref{thmEL}.
Thus, introducing the functions
\beq \ell_j \in C^0(K_j, \R) \:,\qquad \ell(x) := \int_{K_j} \L(x,y)\: \Diff\rho_j(y) - \s_j \:, \eeq
one can choose the parameters~$s_j>0$ such that
\beq \ell_j \big|_{\supp \rho_j} \equiv \inf_{K_j} \ell_j = 0 \:. \eeq

Typically, the support of the measures~$\rho_j$ will be ``spread out'' over larger and larger subsets
of~$\F$. This also means that, working with normalized measures, 
the measures~$\rho_j$ typically converge to the trivial measure~$\rho=0$.
In order to ensure a non-trivial measure, we must perform a suitable {\em{rescaling}}.
To this end, we introduce the measures
\beq \tilde{\rho}_j = \frac{\rho_j}{\s_j} \:. \eeq
These new measures are no longer normalized, but they satisfy the EL equations with~$\tilde{\s}_j=1$; that is,
\beq \label{ELn}
\tilde{\ell}_j \big|_{\supp \tilde{\rho}_j} \equiv \inf_{K_j} \tilde{\ell}_j = 0 \qquad \text{with} \qquad
\tilde{\ell}_j(x) := \int_{K_j} \L(x,y)\: d\tilde{\rho}_j(y) - 1 \:.
\eeq

Our next task is to construct a limit measure~$\rho$ of the measures~$\tilde{\rho}_j$.
We first extend the measures~$\tilde{\rho}_j$ by zero to all of~$\F$ and denote them by~$\rho^{[j]}$,
\beq \rho^{[j]}(U) := \rho_j(U \cap K_j) \qquad \text{for any Borel subset~$U \subset \F$} \:. \eeq
In the next lemma, we show that these measures are bounded on every compact set.
\begin{Lemma} \label{lemmaupper} For every compact subset~$K \subset \F$ there is a constant~$C_K>0$ such that
\beq \label{rhojupper}
\rho^{[j]}(K) \leq C_K \qquad \text{for all~$j \in \N$}\:.
\eeq
\end{Lemma}
\Proof Since~$\L(x,.)$ is continuous and strictly positive at~$x$, there is an open neighborhood~$U(x)$ of~$x$ with
\beq \L(y,z) \geq \frac{\L(x,x)}{2} > 0 \qquad \text{for all~$y,z \in U(x)$}\:. \eeq
Covering~$K$ by a finite number of such neighborhoods~$U(x_1), \ldots, U(x_L)$,
it suffices to show the inequality~\eqref{rhojupper} for the sets~$K \cap U(x_\ell)$ for any~$\ell \in \{1,\ldots, L\}$.
Moreover, we choose~$N$ so large that~$K_N \supset K$ and fix~$k \geq N$.
If~$K \cap \supp \rho^{[k]} = \varnothing$, there is nothing to prove.
Otherwise, there is a point~$z \in K \cap \supp \rho^{[k]}$. Using the EL equations~\eqref{ELn} at~$z$,
it follows that
\beq 1 = \int_\F \L(z,y)\: \Diff\rho^{[k]}(y)
\geq \int_{U(x_\ell)} \L(z,y)\: \Diff\rho^{[k]}(y) \geq 
\frac{\L(x_\ell,x_\ell)}{2}\: \rho^{[k]}(U(x_\ell)) \:. \eeq
Hence
\beq %\label{(lemmaupper)}
	 \rho^{[k]}(U(x_\ell)) \leq \frac{2}{\L(x_\ell,x_\ell)}\:. \eeq
This inequality holds for any~$k \geq N$. We introduce the constants~$c(x_\ell)$ as the maximum of~$2/\L(x_\ell,x_\ell)$
and~$\rho^{[1]}(U(x_\ell)), \ldots, \rho^{[N-1]}(U(x_\ell))$. 
Since the open sets~$U(x_1), \ldots, U(x_L)$ cover~$K$,~we finally introduce the constant~$C_K$ as the sum of the constants~$c(x_1), \ldots, c(x_L)$. 
\QED

Given a compact set~$K$, combining the result of the previous lemma with the compactness
of measures on compact topological spaces (see Theorem~\ref{thmcompact}),
we conclude that there is a subsequence~$(\rho^{[j_n]})$ whose restrictions to~$K$
converge as a measure (that is, in the sense~\eqref{measureconverge}).
Proceeding inductively for the compact sets~$K_1, K_2, \ldots$
and choosing a diagonal sequence, one gets a subsequence of measures on~$\F$,
denoted by~$\rho^{(k)}$, whose restriction to any compact set~$K_j$ converges; that is,
\beq \label{vagueconverge}
\text{$\rho^{(k)} \big|_{K_j}$ converges as~$k \rightarrow \infty$ to~$\rho|_{K_j}$ for all~$j \in \N$}\:,
\eeq
where~$\rho$ is a regular Borel measure on~$\F$ (typically of infinite total volume).
The convergence of measures in~\eqref{vagueconverge} is referred to as {\em{vague convergence}}
(for more details see~\cite[Definition 30.1]{bauer} or~\cite[Section~4.1]{noncompact}).

It remains to show that the obtained measure~$\rho$ is a non-trivial minimizer. In order to
show that it is non-trivial, we make use of the EL equations~\eqref{ELn}.
Let~$x \in \F$. Then~\eqref{ELn} implies that
\beq \int_\F \L(x,y)\: \Diff\rho^{(k)}(y) \geq 1 \:. \eeq
Since~$\L$ has compact range, we may pass to the limit to conclude that
\beq \label{llower}
\int_\F \L(x,y)\: \Diff\rho(y) \geq 1 \:.
\eeq
This shows (in a quantitative way) that the measure~$\rho$ is non-zero.

Our final step for proving the EL equations~\eqref{ELfinal} is to show that the 
EL equations~\eqref{ELn} are preserved in the limit.
In view of the lower bound~\eqref{llower}, it remains to show that~$\ell$ vanishes on the support of~$\rho$.
Thus let~$x \in \supp \rho$. We choose a compact subset~$K \subset \F$ such that~$x$ lies in its interior.
Again using that the Lagrangian has compact range, there is another compact subset~$K' \subset \F$ such that~\eqref{defcompactrange} holds. The fact that~$x$ lies in the support and that the measures~$\rho^{(k)}$
converge vaguely to~$\rho$ implies that there is a sequence~$x_k \in \supp \rho^{(k)}$ which converges to~$x$.
The EL equations for each~$\rho^{(k)}$ imply that, for sufficiently large~$k$,
\beq \int_{K'} \L(x_k, y)\: \Diff\rho^{(k)}(y) = 1 \:. \eeq
Taking the limit is a bit subtle because both the argument~$x_k$ of the Lagrangian and the
integration measure depend on~$k$. Therefore, we begin with the estimate
\begin{align}
\bigg| \int_{K'} &\L(x, y)\: \Diff\rho(y) - \int_{K'} \L(x_k, y)\: \Diff\rho^{(k)}(y)  \bigg| \nonumber \\
&\leq \big| \ell(x) - \ell^{(k)}(x) \big| - \sup_j \big| \ell^{(j)}(x) - \ell^{(j)}(x_k) \big| \:, \label{measintes}
\end{align}
where we set
\beq \ell^{(k)}(z) := \int_{K'} \L(z, y)\: \Diff\rho^{(k)}(y) - 1 \:. \eeq
The first summand on the right side of~\eqref{measintes} tends to zero because
the measures~$\rho^{(k)}$ converge vaguely to~$\rho$. The second summand, on the other hand,
tends to zero because the functions~$\ell^{(j)}$ are equicontinuous
(see~\cite[Section~4.2]{noncompact} for more details on this argument).
This concludes the proof of Theorem~\ref{thmELexist}.

We finally note that, starting from the EL equations~\eqref{ELfinal},
one can also show that~$\rho$ is a {\em{minimizer under variations of finite volume}},
meaning that for every regular Borel measure~$\tilde{\rho}$ satisfying~\eqref{totvol},
the difference of actions~\eqref{integrals} is nonnegative~\eqref{Sdiffpos}.
The proof can be found in~\cite[Section~4.3]{noncompact}.

\section{Tangent Cones and Tangent Cone Measures} \label{sectangentcone}
\sindex{tangent cone}%
\sindex{tangent cone measure}%
In the previous sections of this chapter, measure-theoretic methods have been used in order to prove the
existence of minimizers. But methods of measure theory are also useful for analyzing the structure of
the minimizing measure. Since these methods might be important for the future development
of the theory, we now briefly explain
the concept of a tangent cone measure (more details and applications can be found
in~\cite[Section~6]{topology}). 
We have the situation in mind that spacetime~$M$ does not have a smooth manifold structure,
so that the powerful methods of differential topology and geometry (like the tangent space,
the exponential map, etc.) cannot be used in spacetime.
Nevertheless, the structure of spacetime can be analyzed locally as follows.
Let~$x \in M$ be a spacetime point. We want to analyze a neighborhood of~$x$ in~$M$.
To this end,
it is useful to consider a continuous mapping~${\mathcal{A}}$ from~$M$ to the symmetric operators on the spin
space at~$x$. We always assume that this mapping vanishes at~$x$; that is,
\beq %\label{Afunct}
{\mathcal{A}} : M \rightarrow \Symm(S_x) \qquad \text{with} \qquad {\mathcal{A}}(x)=0 \:. \eeq
There are different possible choices for~${\mathcal{A}}$. The simplest choice is
\beq %\label{cchain}
{\mathcal{A}} \::\: y \mapsto \pi_x \,(y-x)\, x |_{S_x}\:. \eeq
Here the factor~$x$ on the right is needed for the operator to be symmetric, because
\begin{align}
\Sl \psi | {\mathcal{A}} \phi \Sr_x &\overset{\eqref{sspintro}}{=} -\la \psi \,|\, x \, (\pi_x \,(y-x)\, x)\, \phi \ra_\H
= -\la \psi \,|\, x \,\,(y-x)\, x\, \phi \ra_\H \notag \\
&\;\:= -\la \pi_x \,(y-x)\, x\, \psi \,|\, x \, \phi \ra_\H 
= \Sl {\mathcal{A}} \psi | \phi \Sr_x \:.
\end{align}
Alternatively, one can consider mappings involving the operators~$s_y$ or~$\pi_y$, like for example
\begin{align}
{\mathcal{A}} &\::\: y \mapsto \pi_x \,(s_y-s_x)\, x |_{S_x} \\ %\label{As} \\
{\mathcal{A}} &\::\: y \mapsto \pi_x \,(\pi_y-\pi_x)\, x |_{S_x} %\label{Ap}
\end{align}
(where~$\pi_x$ again denotes the orthogonal projection in~$\H$ on~$S_x$).
But, of course, many other choices of~${\mathcal{A}}$ are possible.
The detailed choice of~${\mathcal{A}}$ depends on the application in mind.

A {\em{conical set}} is a set of the form~$\R^+ A$ with~$A \subset \Symm(S_x)$.
We denote the conical sets whose pre-images under~${\mathcal{A}}$
are both $\rho$--measurable by~${\mathfrak{M}}$.
For a conical set~$A \subset \Symm(S_x)$, we consider countable coverings by
measurable conical sets,
\beq A \subset \bigcup_{k=1}^\infty A_k \qquad \text{with} \qquad
\text{$A_k \in {\mathfrak{M}}$}\:. \eeq
We denote the set of such coverings by~${\mathcal{P}}$. We define
\beq %\label{defmus}
\mu^*_\text{\rm{con}}(A) = \inf_{\mathcal{P}} \;\sum_{k=1}^\infty \: \liminf_{\delta \searrow 0}\:
\frac{1}{\rho \big( B_\delta(x) \big)}\: 
\rho \Big( {\mathcal{A}}^{-1} \big(  A_k \big) \cap B_\delta(x) \Big) \eeq
(where~$B_\delta(x) \subset \Lin(\H)$ is the Banach space ball).
We remark for clarity that, since~$x \in M := \supp \rho$, it follows that the
measure~$\rho( B_\delta(x))$ is non-zero for all~$\delta>0$.

The mapping~$\mu^*_\text{\rm{con}}$ defines an outer measure on the conical sets in~$\Symm(S_x)$.
By applying the Carath{\'e}odory extension lemma (see, for example, \cite{bauer, bogachev}), one can construct a corresponding measure denoted by~$\mu_\text{\rm{con}}$.
By restriction one obtains a Borel measure (for details see~\cite[Section~6.1]{topology}).
\begin{Def} We denote the conical Borel sets of~$\Symm(S_x)$ by~${\mathfrak{B}}_\text{\rm{con}}(x)$.
We denote the measure obtained by applying the above construction by
\beq \mu_x \::\: {\mathfrak{B}}_\text{\rm{con}}(x) \rightarrow [0, \infty] \:. \eeq
\nindex{agc@${\mathcal{C}}_x$ -- tangent cone at~$x \in M$}%
\nindex{agd@$\mu_x$ -- tangent cone measure at~$x \in M$}%
It is referred to as the {\bf{tangent cone measure}} corresponding to~${\mathcal{A}}$.
The {\bf{tangent cone}}~${\mathcal{C}}_x$ is defined as the support of the tangent cone measure,
\beq {\mathcal{C}}_x := \supp \mu_x \subset \Symm(S_x)\:. \eeq
\end{Def}

In simple terms, the tangent cone~${\mathcal{C}}_x$ distinguishes directions in which the measure~$\rho$
is non-zero. The tangent cone measure, on the other hand, is a measure supported on the tangent cone.
By integrating functionals on conical subsets of~$\Symm(S_x)$ with respect to this measure, one can
can get fine information on the structure of the measure~$\rho$ in different directions.
For example, one can set up a variational principle by maximizing a suitable integral of this type
under variations of the Clifford section at~$x$.
As a concrete application, this method is used in~\cite[Section~6.2]{topology} in order to choose a
distinguished Clifford section at~$x$.

\section{Exercises}

\begin{Exercise} {\em{
\sindex{Riesz representation theorem}%
Let~$\Lambda$ be the functional
\beq \Lambda \::\: C^0\big([0,1], \R \big) \rightarrow \R\:,\quad \Lambda(f) = \sup_{x \in [0,1]} f(x)\:. \eeq
Can this functional be represented by a measure?
Analyze how your findings are compatible with the Riesz representation theorem.
}} \end{Exercise}

\begin{Exercise} {\em{
\sindex{measure!Lebesgue decomposition}%
Let~$\rho$ be the Borel measure on~$[0, \pi]$ given by
\beq \rho(\Omega) = \int_\Omega \sin x\: \dd x \;+\; \sum_{n=1}^\infty \frac{1}{n^2} \:\chi_\Omega \Big( \frac{1}{n} \Big) \:. \eeq
Compute the Lebesgue decomposition of~$\rho$ with respect to the Lebesgue measure.
}} \end{Exercise}

\begin{Exercise}  {{(Normalized regular Borel measures: compactness results)}} {\em{
\bitem
\item[(a)] Let~$(\rho_n)_{n \in \N}$ be a sequence of normalized regular
Borel measures on~$\R$ with the property that there is a constant~$c>0$ such that
\beq \int_{-\infty}^\infty x^2\: \Diff\rho_n(x) \leq c \qquad \text{for all~$n$\:.} \eeq
Show that a subsequence converges again to a normalized Borel measure on~$\R$. \\
{\em{Hint:}} Apply the compactness result in Theorem~\ref{thmcompact} to the measures
restricted to the interval~$[-L,L]$ and analyze the behavior as~$L \rightarrow \infty$.
\item[(b)] More generally, assume that for a given non-negative function~$f(x)$,
\beq \int_{-\infty}^\infty f(x)\: \Diff\rho_n(x) \leq c \qquad \text{for all~$n$\:.} \eeq
Which condition on~$f$ ensures that a subsequence of the measures converges to
a normalized Borel measure? Justify your answer by a counter example.
\eitem
}} \end{Exercise}

\begin{Exercise} {\em{
Let~$\scrM \subset \R$ be a closed embedded submanifold of~$\R^3$.
We choose a compact set~$K \subset \R^3$ which contains~$\scrM$.
On~$C^0(K, \R)$ we introduce the functional
\beq \Lambda \::\: C^0(K, \R) \rightarrow \R\:,\quad \Lambda(f) = \int_\scrM f(x)\: \Diff\mu_\scrM(x) \eeq
(where~$\Diff\mu_\scrM$ is the volume measure corresponding to the
induced Riemannian metric on~$\scrM$).
Show that this functional is linear, bounded and positive.
Apply the Riesz representation theorem to represent this functional by a Borel measure on~$K$.
What is the support of this measure?
}} \end{Exercise}

\begin{Exercise} \label{exm2} {\em{
This exercise explains why the causal action principle is ill-posed
in the case~$\dim \H=\infty$ and~$\rho(\F)<\infty$.
The underlying estimates were first given in the setting of discrete spacetimes
in~\cite[Lemma~5.1]{discrete}.
\bitem
\item[(a)] Let~$\H_0$ be a finite-dimensional Hilbert space of dimension~$n$
and~$(\H_0, \F_0, \rho_0)$ be a causal fermion system of finite total volume~$\rho_0(\F_0)$.
Let~$\iota : \H_0 \rightarrow \H$ be an isometric embedding.
Construct a causal fermion system~$(\H, \F, \rho)$
which has the same action, the same total volume
and the same values for the trace and boundedness constraints as the causal fermion
system~$(\H_0, \F_0, \rho_0)$.
\item[(b)] Let~$\H_1 = \H_0 \oplus \H_0$. Construct a causal fermion system~$(\H_1, \F_1, \rho_1)$
which has the same total volume and the same value of the trace constraint as~$(\H_0, \F_0, \rho_0)$ but a smaller action
and a smaller value of the boundedness constraint.
{\em{Hint:}} Let~$F_{1\!/\!2} : \Lin(\H_0) \rightarrow \Lin(\H_1)$ be the linear mappings
\beq \big(F_1(A)\big)(u \oplus v) = (Au) \oplus 0 \:,\qquad
\big(F_2(A)\big)(u \oplus v) = 0 \oplus (Av) \:. \eeq
Show that~$F_{1\!/\!2}$ map~$\F_0$ to~$\F_1$. Define~$\rho_1$ by
\beq \rho_1 = \frac{1}{2} \Big( (F_1)_* \rho + (F_2)_* \rho \Big) \:. \eeq
\item[(c)] Iterate the construction in~(b) and apply~(a)
to obtain a series of measures on~$\F$ of fixed total volume
and with fixed value of the trace constraint, for which the action and the values of the boundedness
constraint tend to zero. Do these measures converge? If yes, what is the limit?
\eitem
}} \end{Exercise}

\begin{Exercise} {{(Riesz representation theorem - part 1)}} {\em{
\sindex{Riesz representation theorem}%
Let~$\Lambda$ be the functional
\beq
\Lambda: C^0([0,1],\R)\rightarrow \R,\quad \Lambda(f):=\sup_{x\in [0,1]}f(x).
\eeq
Can this functional be represented by a measure? Analyze how your findings are compatible with the Riesz representation theorem.
}} \end{Exercise}

\begin{Exercise} {{(Riesz representation theorem - part 2)}} {\em{
\sindex{Riesz representation theorem}%
Let~$\M$ be a closed embedded submanifold of~$\R^3$. We choose a compact set~$K\subset \R^3$ which contain~$\M$. On~$C^0(K,\R)$ we introduce the functional 
\beq
\Lambda:C^0(K,\R)\to \R,\quad \Lambda(f)=\int_\M f(x)\,\Diff\mu_\M(x),
\eeq
where~$\Diff\mu_\M$ is the volume measure corresponding to the induced Riemannian metric on~$\M$. Show that this functional is linear, bounded and positive. Apply the Riesz representation theorem to represent this functional by a Borel measure on~$K$. What is the support of this measure?
}} \end{Exercise}

\begin{Exercise} {{(Radon-Nikodym decomposition)}} {\em{
\sindex{Radon-Nikodym theorem}%
Let~$\rho$ be the Borel measure on~$[0,\pi]$ given by
\beq
\rho(\Omega):=\int_\Omega\sin x\,\dd x+\sum_{n=1}^\infty\frac{1}{n^2}\,\chi_\Omega\!\left(\frac{1}{n}\right).
\eeq
Compute the Radon-Nikodym decomposition of~$\rho$ with respect to the Lebesgue measure.
}} \end{Exercise}

\begin{Exercise} {{(Derivative of measures)}} {\em{
Let~$\mu$ be the counting measure on the $\sigma$-algebra~$\mathcal{P}(\N)$. Consider the measure
\beq
\lambda(\varnothing):=0,\quad \lambda(E):=\sum_{n\in E}(1+n)^2,\quad E\in \mathcal{P}(\N).
\eeq
Show that~$\mu$ and~$\lambda$ are equivalent (one is absolutely continuous with respect to the other) and determine the Radon-Nikodym derivative~$\frac{\Diff\mu}{d\lambda}$.
}} \end{Exercise}

\begin{Exercise} {{(Minimizers)}} {\em{
Let~$M$ denote the $2$-sphere~$S^2\subset\R^3$ and let~$\Diff\mu_{M}$ be the normalized canonical surface measure. Consider a Lagrangian  on~$M\times M$ defined by
\beq
\L(x,y):=\frac{1}{1+\|x-y\|_{\R^3}}\quad\mbox{for all~$x,y\in M$}.
\eeq
Show that the action~$\Sact(\mu_M)$ is minimal under variations of the form
\beq
\Diff\rho_{x_0,t}:=(1-t)\Diff\mu_M+t\,\dd\delta_{x_0},\quad \mbox{with~$t\in [0,1)$},
\eeq
where~$\delta_{x_0}$ is the Dirac measure centered at~$x_0\in M$.
}} \end{Exercise}

\begin{Exercise} {{(Moment measures)}} {\em{
\sindex{moment measure}%
Let~$\F=\R^2$ and~$K=S^1\cup\{0\}$ be a compact subset of~$\F$.
Moreover, let~$\rho$ be a Borel measure~$\rho$ on~$\F$.
Compute the moment measures~$\mathfrak{m}^{(0)}$, $\mathfrak{m}^{(1)}$ and~$\mathfrak{m}^{(2)}$
for the following choices of~$\rho$:
\bitem
\item[(a)] $\rho=F_{*}(\mu_{S^1}\!)$, where~$F:S^1\hookrightarrow\R^2$ is the natural injection and~$\mu_{S^1}$ is the normalized Lebesgue measure on~$S^1$.
\item[(b)] $\rho=\delta_{(0,0)}+\delta_{(1,1)}+\delta_{(5,0)}$ (where~$\delta_{(x,y)}$ denote the Dirac measure supported at~$(x,y)\in\R^2$).
\item[(c)] $\rho=F_*(\mu_\R\!)$, where~$\mu_\R$ is the Lebesgue measure on~$\R$ and
\beq
F:\R\rightarrow \R^2,\quad F(x)=(x,2).
\eeq
\eitem
}} \end{Exercise}

\chapter{Methods of Hyperbolic Partial Differential Equations} \label{secshs}
The structures of a causal fermion system are all encoded in the
family of physical wave functions (see Section~\ref{secinherent}). Consequently,
the dynamics of the causal fermion is understood once we know how each physical wave function
propagates in time. In many examples, the physical wave functions satisfy the {\em{Dirac equation}}
(for the simplest example of this type see Section~\ref{seclco}).
More abstractly, the dynamics of the physical wave functions is described by the
{\em{dynamical wave equation}}~\eqref{dwe}. Moreover, we also encountered
the {\em{linearized field equations}} (see Definition~\ref{deflinear}).
We now turn attention to methods for solving these equations.
In this chapter, we begin with linear partial differential equations like the Dirac equation.
Causality is reflected in these equations in the fact that they are {\em{hyperbolic}}.
As we shall see, the methods developed here will also be fruitful for the study of
the linearized field equations, as will be explained in the next chapter (Chapter~\ref{seclinhyp}).
With this in mind, the constructions here can be regarded as a technical preparation
for Chapter~\ref{seclinhyp}.
We remark that the adaptation of the methods to the dynamical wave equation will not be covered
in this book; we refer the interested reader instead to~\cite{dirac}.

\section{The Cauchy Problem, Linear Symmetric Hyperbolic Systems}
\sindex{symmetric hyperbolic system!linear}%
\sindex{Cauchy problem! for linear symmetric hyperbolic system}%
In this section, we shall prove that the Cauchy problem for the Dirac equation in the presence
of an external potential has a unique global solution. Moreover, we will show that the
finite speed of propagation as postulated by special relativity is indeed respected by the solutions
of the Dirac equation. For later purposes, it is preferable to include an inhomogeneity.
Thus we consider the Cauchy problem in Minkowski space
\beq \label{Cauchydirac}
(\cI \Pdd + \B - m) \,\psi = \phi \in C^\infty(\scrM, S\scrM) \:,\qquad \psi|_{t_0} = \psi_0 \in C^\infty(\R^3, S\scrM)
\eeq
for a given inhomogeneity~$\phi$ and initial data~$\psi_0$.
In order to make the standard methods available, we multiply the equation by~$-i \gamma^0$,
\beq \label{diracsymm}
\1_{\C^4}\, \partial_t \psi + \gamma^0 \vec{\gamma} \vec{\nabla} \psi 
-\cI \gamma^0 (\B - m) \psi = -\cI \gamma^0 \phi \:.
\eeq
Now the matrices in front of the derivatives are all Hermitian (with respect to the standard
scalar product on~$\C^4$). Moreover, the matrix in front of the time derivative is
positive definite. Kurt Otto Friedrichs~\cite{friedrichs}
observed that these properties are precisely what is needed in order to get a well-posed Cauchy problem.
He combined these properties in the notion of a {\em{symmetric hyperbolic system}}.
We now give its general definition.
More specifically, we consider a system of~$N$ complex-valued equations
with spatial coordinates~$\vec{x} \in \R^m$ and time~$t$ in an interval~$[0, T]$ with~$T>0$.
The initial data will always be prescribed at time~$t=0$.
For notational clarity, we denote partial derivatives in spatial directions by~$\nabla$.

\begin{Def} \label{defshs} A linear system of differential equations of the form
\beq\label{2c}
A^0(t,\vec{x})\,\partial_t u(t,\vec{x}) +\sum^m_{\alpha=1} A^\alpha(t,\vec{x})\,\nabla_\alpha u(t,\vec{x})+B(t,\vec{x}) \,u(t,\vec{x})=w(t,\vec{x})
\eeq
with
\beq A^0, A^\alpha, B \in C^\infty \big( [0, T] \times \R^m, \Lin(\C^N) \big) \:,\qquad w \in C^\infty
\big( [0, T] \times \R^m, \C^N \big) \:. \eeq
is called {\bf{symmetric hyperbolic}} if
\bitem
\item[{\rm{(i)}}] The matrices~$A^0$ and~$A^\alpha$ are Hermitian,
\beq (A^0)^\dagger = A^0 \qquad \text{and} \qquad (A^\alpha)^\dagger=A^\alpha \eeq
(where~$\dagger$ denotes the adjoint with respect to the canonical scalar product on~$\C^N$).
\item[{\rm{(ii)}}] The matrix~$A^0$ is uniformly positive definite; that is, there is a positive constant~$C$
such that
\beq A^0(t,\vec{x})>C\, \1_{\C^N} \quad \text{for all~$(t,\vec{x}) \in ([0, T] \times \R^m)$}\:. \eeq
\eitem
In the case~$w \equiv 0$, the linear system is called {\bf{homogeneous}}.
\end{Def} \noindent
A good reference for linear symmetric hyperbolic systems
is the book by Fritz John~\cite[Section~5.3]{john} (who was Friedrichs' colleague at the Courant Institute).
Our presentation was also influenced by~\cite[Chapter~8]{rendall}.
We remark that the concept of a symmetric hyperbolic system can be extended to nonlinear
equations of the form
\beq A^0(t,\vec{x},u) \,\partial_t u(t,\vec{x}) + \sum^m_{\alpha=1} A^\alpha(t,\vec{x},u) \,\nabla_\alpha u(t,\vec{x}) +
B(t,\vec{x},u)=0 \:, \eeq
where the matrices~$A^0$ and~$A^\alpha$ should again satisfy the above-mentioned conditions~(i) and~(ii).
For details, we refer to~\cite[Section~16]{taylor3} or~\cite[Section~7]{ringstroem}.
Having the Dirac equation in mind, we always restrict attention to linear systems.
We also note that an alternative method for proving the existence of fundamental solutions is to
work with the so-called Riesz distributions (see~\cite{baer+ginoux} for a good textbook).
Yet another method is to work with estimates in the interaction picture~\cite{deckert+merkl}.
For completeness, we finally note that the concept of symmetric hyperbolic systems was extended by
Friedrichs to so-called symmetric {\em{positive}} systems~\cite{friedrichs2}.

It is a remarkable fact that all partial differential equations in relativistic physics as well as
most wave-type equations can be rewritten as a symmetric hyperbolic system.
As an illustration, we now explain this reformulation in the example of a scalar hyperbolic equation.
\begin{Example} \label{exscalar} {\em{
Consider a scalar hyperbolic equation of the form
\beq \label{scalar}
 \partial_{tt}\phi(t,\vec{x}) = \sum^m_{\alpha, \beta=1} a_{\alpha \beta}(t,\vec{x}) \,\nabla_{\alpha \beta} \phi
+ \sum^m_{\alpha=1} b_\alpha(t,\vec{x}) \,\nabla_\alpha \phi +c(t,\vec{x})\, \partial_t\phi+d(t,\vec{x})\,\phi
\eeq
with~$(a_{\alpha \beta})$ a symmetric, uniformly positive matrix (in the case~$a_{\alpha \beta}=\delta_{\alpha \beta}$ and~$b,c,d=0$, one gets the scalar wave equation). Now the initial data prescribes~$\phi$
and its first time derivatives,
\beq \label{scalinit}
\phi|_{t=0} = \phi_0  \in C^\infty(M) \:,\qquad \partial_t \phi|_{t=0} = \phi_1  \in C^\infty(M) \:.
\eeq
In order to rewrite the equation as a symmetric hyperbolic system, we introduce the vector~$u$ with~$k:=m+2$
components by
\beq \label{udef}
u_1=\nabla_1\phi,\:\dots, u_m=\nabla_m\phi,
\quad u_{m+1}=\partial_t\phi, \quad u_{m+2}=\phi \:.
\eeq
Then the system
\beq \label{system}
\left\{ \begin{array}{cccc}
\displaystyle \sum^m_{\beta=1} a_{\alpha \beta} \,\partial_t u_\beta & \displaystyle - \sum^m_{\beta=1} a_{\alpha\beta} \,\nabla_\beta u_{m+1}
& & =0 \\
\displaystyle -\sum^m_{\alpha,\beta=1} a_{\alpha\beta}\nabla_\beta u_\alpha -\sum^m_{\alpha=1} b_\alpha \,u_\alpha
& \displaystyle +\partial_t u_{m+1} -c \,u_{m+1} & -d\,u_{m+2} &= 0 \\
0 & -u_{m+1} & +\partial_t u_{m+2} &=0
\end{array} \right.
\eeq
is symmetric hyperbolic (as one verifies by direct inspection).
Also, a short calculation shows that if~$\phi$ is a smooth solution of the scalar equation~\eqref{scalar},
then the corresponding vector~$u$ is a solution of the system~\eqref{system}.
Conversely, assume that~$u$ is a smooth solution of~\eqref{system} that
satisfies the initial condition~$u|_{t=0}=u_0$, where~$u_0$
is determined by~$\phi_0$ and~$\phi_1$ via~\eqref{udef}.
Setting~$\phi = u_{m+2}$, the last line in~\eqref{system} shows that~$u_{m+1}=\partial_t \phi$.
Moreover, the first line in~\eqref{system} implies that~$\partial_t u_\beta = \nabla_\beta u_{m+1}
= \partial_t \nabla_\beta \phi$. Integrating over~$t$ and using that the relation~$u_\beta = \nabla_\beta \phi$
holds initially, we conclude that this relation holds for all times.
Finally, the second line in~\eqref{system} yields that~$\phi$ satisfies the scalar hyperbolic
equation~\eqref{scalar}.
In this sense, the Cauchy problem for the system~\eqref{system} is equivalent to that
for the scalar equation~\eqref{scalinit}. }} \QEDrem
\end{Example}

This procedure works similarly for other physical equations like the Klein-Gordon or Maxwell
equations. Exercise~\ref{exmaxwell} is concerned with the example of the homogeneous Maxwell equations.

\section{Finite Propagation Speed and Uniqueness of Solutions} \label{sec31}
\sindex{finite propagation speed!for linear symmetric hyperbolic systems}%
For what follows, it is convenient to combine the time and spatial coordinates
to a spacetime vector~$x = (t, \vec{x}) \in [0, T] \times \R^m$. 
We denote the spacetime dimension by~$n = m+1$. Moreover,
setting~$\partial_0 \equiv \partial_t$, we use Latin spacetime indices~$i \in \{0, \ldots, m\}$
and employ the Einstein summation convention. Then our linear system~\eqref{2c}
can be written in the compact form
\beq \label{Acomp}
A^j(x)\: \partial_j u(x) +B(x)\: u(x) = w(x) \:.
\eeq
Next, a direction in spacetime can be described by a vector~$\xi = (\xi_i)_{i=0,\ldots, m} \in \R^{m+1}$.
Contracting with the matrices~$A^j(x)$, we obtain the Hermitian~$N \times N$-matrix
\beq A(x,\xi) := A^j(x)\: \xi_j \:, \eeq
referred to as the {\em{characteristic matrix}}.
\sindex{characteristic matrix}%
Note that in the example of the Dirac equation~\eqref{diracsymm}, the index~$i$ is a
vector index in Minkowski space, and~$\xi$ should be regarded as a co-vector
(that is, a vector in the cotangent bundle). One should keep in mind that, despite
the suggestive notation, the equation~\eqref{Acomp} should not be considered as being
manifestly covariant, because it corresponds to the Hamiltonian formulation~\eqref{diracsymm}, where
a time direction is distinguished.

The determinant of the characteristic matrix is referred to as the {\em{characteristic polynomial}},
\sindex{characteristic polynomial}%
being a polynomial in the components~$\xi_i$. For our purposes, it is most helpful to consider
whether the characteristic matrix is positive or negative definite.
If the vector~$\xi = (\tau, \vec{0})$ points in the time direction, then~$A(x,\xi) = \tau A^0$, which in
view of Definition~\ref{defshs} is a definite matrix. By continuity, $A(x, \xi)$ is definite if the spatial
components of~$\xi$ are sufficiently small. In the example of the Dirac equation~\eqref{diracsymm},
the fact that
\beq \label{dirsigma}
A(x, \xi) = \1 \xi_0 + \gamma^0 \vec{\gamma} \vec{\xi}
\qquad \text{has eigenvalues} \qquad \xi^0 \pm |\vec{\xi}|
\eeq
shows that~$A(x,\xi)$ is definite if and only if~$\xi$ is a timelike vector.
Moreover, it is positive definite if and only if~$\xi$ is future-directed and timelike.
This suggests that the causal properties of the equation are encoded in the positivity
of the characteristic matrix. We simply use this connection to {\em{define}} the causal
structure for a general symmetric hyperbolic system.

\begin{Def} \label{defshscausal} The vector~$\xi \in \R^{m+1}$ is called {\bf{timelike}}
\sindex{timelike separation!for linear symmetric hyperbolic system}%
\sindex{spacelike separation!for linear symmetric hyperbolic system}%
\sindex{lightlike separation!for linear symmetric hyperbolic system}%
\sindex{causality!for linear symmetric hyperbolic system}%
at the spacetime point~$x$ if the characteristic matrix~$A(x, \xi)$ is definite.
A timelike vector is called {\bf{future-directed}} if~$A(x, \xi)$ is positive definite.
If the characteristic polynomial vanishes, then the vector~$\xi$ is called {\bf{lightlike}}.
A hypersurface~${\mathscr{H}}\subset [0, T]\times\mathbb{R}^m$
with normal~$\nu$ is called {\bf{spacelike}} if the matrix~$A(x, \nu)$ is positive
definite for all~$x \in \H$.
\end{Def} \noindent
The notion of a {\em{normal}} used here requires an explanation. The simplest method is to
represent the hypersurface locally as the zero set of a function~$\phi(x)$. Then
the normal can be defined as the gradient of~$\phi$. In this way, the gradient is
a co-vector, so that the contraction~$A^j \nu_j = A^j \partial_j \phi$ is well-defined
without referring to a scalar product. In particular, the last definition is independent
of the choice of a scalar product on spacetime vectors in~$\R^n$.
We always choose the normal to be future-directed, and we
normalize it with respect to the Euclidean scalar product on~$\R^{m+1}$, 
but these are merely conventions.

We shall now explain why and in which sense the solutions of
symmetric hyperbolic systems comply with this notion of causality.

\begin{Def} Let~$u$ be a smooth solution of the linear symmetric hyperbolic system~\eqref{Acomp}.
A subset~$K$ of the initial value surface~$\{t=0\}$ determines the solution at
a spacetime point~$x \in [0, T]\times\mathbb{R}^m$ if every smooth solution of the system
that coincides on~$K$ with~$u$, also coincides with~$u$ at~$x$.
The {\bf{domain of determination}} of~$K$
\sindex{domain of determination}%
is the set of all spacetime points at which the solution is determined by the initial data on~$K$.
\end{Def}

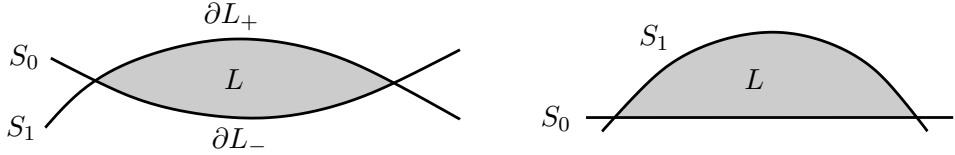
\begin{figure} %
\begin{picture}(0,0)%
\includegraphics{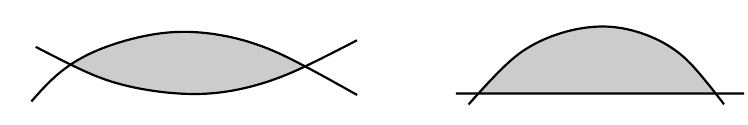}%
\end{picture}%
\setlength{\unitlength}{1533sp}%
\begingroup\makeatletter\ifx\SetFigFont\undefined%
\gdef\SetFigFont#1#2#3#4#5{%
  \reset@font\fontsize{#1}{#2pt}%
  \fontfamily{#3}\fontseries{#4}\fontshape{#5}%
  \selectfont}%
\fi\endgroup%
\begin{picture}(15368,2559)(-1049,-3425)
\put(2289,-3271){\makebox(0,0)[lb]{\smash{{\SetFigFont{11}{13.2}{\rmdefault}{\mddefault}{\updefault}$\partial L_-$}}}}
\put(-989,-1921){\makebox(0,0)[lb]{\smash{{\SetFigFont{11}{13.2}{\rmdefault}{\mddefault}{\updefault}$S_0$}}}}
\put(-1034,-3076){\makebox(0,0)[lb]{\smash{{\SetFigFont{11}{13.2}{\rmdefault}{\mddefault}{\updefault}$S_1$}}}}
\put(2162,-1253){\makebox(0,0)[lb]{\smash{{\SetFigFont{11}{13.2}{\rmdefault}{\mddefault}{\updefault}$\partial L_+$}}}}
\put(2492,-2318){\makebox(0,0)[lb]{\smash{{\SetFigFont{11}{13.2}{\rmdefault}{\mddefault}{\updefault}$L$}}}}
\put(7609,-2919){\makebox(0,0)[lb]{\smash{{\SetFigFont{11}{13.2}{\rmdefault}{\mddefault}{\updefault}$S_0$}}}}
\put(10905,-2318){\makebox(0,0)[lb]{\smash{{\SetFigFont{11}{13.2}{\rmdefault}{\mddefault}{\updefault}$L$}}}}
\put(9201,-1606){\makebox(0,0)[lb]{\smash{{\SetFigFont{11}{13.2}{\rmdefault}{\mddefault}{\updefault}$S_1$}}}}
\end{picture}%
\caption{Lens-shaped regions.}
\label{figlens}
\end{figure}

\begin{Def} An open subset~$L \subset (0, T) \times \R^m$ is called
a {\bf{lens-shaped region}} if~$L$ is relatively compact in~$\R^n$ and if its boundary~$\partial L$
\sindex{lens-shaped region!for linear symmetric hyperbolic system}%
is contained in the union of two smooth hypersurfaces~$S_0$ and~$S_1$
whose intersection with~$\overline{L}$ is spacelike.
We set~$(\partial L)_+ = \partial L \cap S_1$ and
$(\partial L)_-=\partial L \cap S_0$, where we adopt the convention that~$(\partial L)_+$
lies to the future of~$(\partial L)_-$.
\end{Def} \noindent
Figure~\ref{figlens} shows typical examples of lens-shaped regions.
Often, one chooses the initial data surface as~$S_0 = \{t=0\}$.
Moreover, it is often convenient to
write the hypersurface~$S_1$ as a graph~$S_1 = \{(t, \vec{x}) \,|\, t = f(\vec{x})\}$.
In this case, $S_1$ is the zero set of the function~$\phi(t,\vec{x}) = t-f(\vec{x})$,
and the normal~$\nu$ is the gradient of this function; that is,
\beq (\nu_j)_{j=0, \ldots, m} = \left( 1, \nabla_1 f, \ldots, \nabla_m f \right) . \eeq

We first consider the homogeneous equation
\beq\label{2g}
(A^j\partial_j +B)\, u=0\,.
\eeq
The idea for analyzing the domain of determination is to multiply this equation
by a suitable test function and to integrate over a lens-shaped region.
More precisely, we consider the equation
\beq\label{2d}
0=\int_L \E^{-Kt}\: 2\mbox{Re} \langle u,(A^j\partial_j+B)u\rangle \, \dd^nx \:,
\eeq
where~$\langle .,. \rangle$ denotes the canonical scalar product on~$\C^N$,
and~$K>0$ a positive parameter to be determined later.
Since the~$A^j$ are Hermitian, we have
\beq \label{divform}
 \partial_j\langle u,A^ju\rangle = 2 \re \,\langle u,
A^j\partial_ju\rangle +\langle u,(\partial_jA^j)u\rangle\,,
\eeq
and using this identity in~\eqref{2d} gives
\beq \label{312}
0=\int_L \E^{-Kt} \,\Big( \partial_j\langle u,A^ju\rangle + \big\langle
u, \big( B+B^\ast-(\partial_jA^j) \big) u \big\rangle \Big) \,\dd^nx\,.
\eeq
In the first term we integrate by parts with the Gau{\ss} divergence theorem,
\beq\label{2f}
\begin{split}
\int_L &\E^{-Kt}\partial_j\langle u,A^ju\rangle\, \dd^nx =
K \:\int_L\E^{-Kt} \,\langle u,A^0 u\rangle \, \dd^nx \\
& +\int_{(\partial L)_+}\E^{-Kt} \,\langle u,\nu_jA^ju\rangle\,\Diff\mu_{\partial L_+}
- \int_{(\partial L)_-}\E^{-Kt} \,\langle u,\nu_jA^ju\rangle\, \Diff\mu_{\partial L_-} \:.
\end{split}
\eeq 
We now use~\eqref{2f} in~\eqref{312} and solve for the surface integral over~$(\partial L)_+$,
\beq \label{315} \begin{split}
\int_{(\partial L)_+} &\E^{-Kt} \,\langle u,\nu_jA^ju\rangle\,\Diff\mu_{\partial L_+}
= \int_{(\partial L)_-}\E^{-Kt} \,\langle u,\nu_jA^ju\rangle\, \Diff\mu_{\partial L_-} \\
& +\int_L \E^{-Kt} \,\big\langle u, \big( -K-B-B^\ast +(\partial_jA^j) \big) \,u \big\rangle\: \dd^nx\,.
\end{split} \eeq
This identity is the basis for the following uniqueness results.

\begin{Thm} \label{thmsymmunique}
Let~$u_1$ and~$u_2$ be two smooth solutions of the
linear symmetric hyperbolic system~\eqref{2c} that coincide on the past boundary
of a lens-shaped region~$L$,
\beq u_1\vert_{(\partial L)_-}= u_2\vert_{(\partial L)_-}\,. \eeq
Then~$u_1$ and~$u_2$ coincide in the whole set~$L$.
\end{Thm}
\Proof The function~$u:=u_1-u_2$ is a solution of the homogeneous system~\eqref{2g}
with~$u\vert_{(\partial L)_-}=0$. Hence, \eqref{315} simplifies to
\beq \int_{(\partial L)_+} \E^{-Kt} \,\langle u,\nu_jA^ju\rangle\,\Diff\mu_{\partial L_+}
= \int_L \E^{-Kt} \,\big\langle u, \big( -K-B-B^\ast + \partial_j A^j \big) \,u \big\rangle \dd^nx\,. \eeq
Assume that~$u$ does not vanish identically in~$L$. By choosing~$K$ sufficiently
large, we can then arrange that the right-hand side becomes negative.
However, since~$\partial L_+$ is a spacelike hypersurface, the left-hand side is non-negative.
This is a contradiction.
\QED

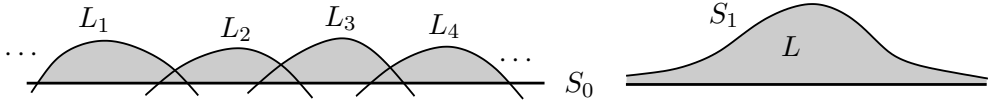
\begin{figure} %
\begin{picture}(0,0)%
\includegraphics{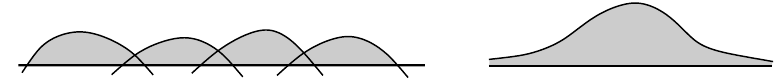}%
\end{picture}%
\setlength{\unitlength}{1533sp}%
\begingroup\makeatletter\ifx\SetFigFont\undefined%
\gdef\SetFigFont#1#2#3#4#5{%
  \reset@font\fontsize{#1}{#2pt}%
  \fontfamily{#3}\fontseries{#4}\fontshape{#5}%
  \selectfont}%
\fi\endgroup%
\begin{picture}(15952,1610)(-1676,-3045)
\put(10905,-2318){\makebox(0,0)[lb]{\smash{{\SetFigFont{11}{13.2}{\rmdefault}{\mddefault}{\updefault}$L$}}}}
\put(7399,-2904){\makebox(0,0)[lb]{\smash{{\SetFigFont{11}{13.2}{\rmdefault}{\mddefault}{\updefault}$S_0$}}}}
\put(-465,-1853){\makebox(0,0)[lb]{\smash{{\SetFigFont{11}{13.2}{\rmdefault}{\mddefault}{\updefault}$L_1$}}}}
\put(1826,-2007){\makebox(0,0)[lb]{\smash{{\SetFigFont{11}{13.2}{\rmdefault}{\mddefault}{\updefault}$L_2$}}}}
\put(3517,-1834){\makebox(0,0)[lb]{\smash{{\SetFigFont{11}{13.2}{\rmdefault}{\mddefault}{\updefault}$L_3$}}}}
\put(5194,-1961){\makebox(0,0)[lb]{\smash{{\SetFigFont{11}{13.2}{\rmdefault}{\mddefault}{\updefault}$L_4$}}}}
\put(-1661,-2405){\makebox(0,0)[lb]{\smash{{\SetFigFont{11}{13.2}{\rmdefault}{\mddefault}{\updefault}$\cdots$}}}}
\put(6338,-2495){\makebox(0,0)[lb]{\smash{{\SetFigFont{11}{13.2}{\rmdefault}{\mddefault}{\updefault}$\cdots$}}}}
\put(9734,-1786){\makebox(0,0)[lb]{\smash{{\SetFigFont{11}{13.2}{\rmdefault}{\mddefault}{\updefault}$S_1$}}}}
\end{picture}%
\caption{Coverings by lens-shaped regions.}
\label{figcover}
\end{figure}

As an immediate consequence, we obtain the following uniqueness result
for solutions of the Cauchy problem.
\begin{Corollary} \label{corunique} Let~$u_1$ and~$u_2$ be two smooth solutions of the
linear symmetric hyperbolic system~\eqref{2c} with the same initial
at time~$t=0$. Then~$u_1 \equiv u_2$ in a neighborhood of the initial data surface.

If the matrices~$A^j$ are uniformly bounded and~$A^0$ is uniformly positive, then~$u_1 \equiv u_2$
in~$[0, T] \times \R^m$.
\end{Corollary}
\Proof The local uniqueness result follows immediately by covering the initial data surface
by lens-shaped regions (see the left of Figure~\ref{figcover}).
For the global uniqueness, for any~$x_0=(t_0, \vec{x}_0) \in [0, T] \times \R^m$ 
our task is to choose a lens-shaped region that contains~$x_0$ and whose past boundary~$S_0$
is contained in the surface~$\{t=0\}$. We need to rule out the situation shown on the right of
Figure~\ref{figcover} that the hypersurface~$S_1$ does not intersect~$S_0$, in which case
we would not get a relatively compact lens-shaped region.
To this end, we must use that the matrices~$A^j$ are uniformly bounded by assumption.
As a consequence, there is~$\varepsilon>0$ such that the inequality~$\|\nabla f \| \leq \varepsilon$ implies that
the hypersurface~$S_1=\{ (t=f(\vec{x}), \vec{x}) \}$ is spacelike.
Possibly after decreasing~$\varepsilon$, we may choose
\beq f(\vec{x})= t_0 + \varepsilon \left( 1 - \sqrt{1+ \| \vec{x} - \vec{x}_0 \|^2} \right) . \eeq
This concludes the proof.
\QED

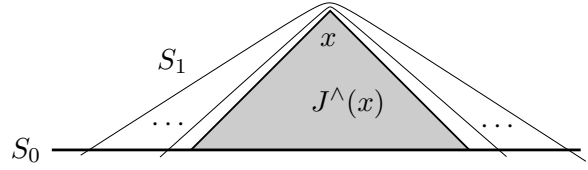
\begin{figure} %
\begin{picture}(0,0)%
\includegraphics{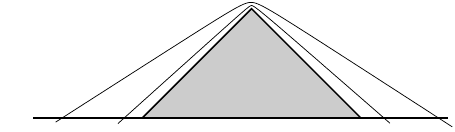}%
\end{picture}%
\setlength{\unitlength}{1533sp}%
\begingroup\makeatletter\ifx\SetFigFont\undefined%
\gdef\SetFigFont#1#2#3#4#5{%
  \reset@font\fontsize{#1}{#2pt}%
  \fontfamily{#3}\fontseries{#4}\fontshape{#5}%
  \selectfont}%
\fi\endgroup%
\begin{picture}(9336,2677)(6071,-3037)
\put(11091,-1073){\makebox(0,0)[lb]{\smash{{\SetFigFont{11}{13.2}{\rmdefault}{\mddefault}{\updefault}$x$}}}}
\put(8456,-1449){\makebox(0,0)[lb]{\smash{{\SetFigFont{11}{13.2}{\rmdefault}{\mddefault}{\updefault}$S_1$}}}}
\put(8359,-2461){\makebox(0,0)[lb]{\smash{{\SetFigFont{11}{13.2}{\rmdefault}{\mddefault}{\updefault}$\cdots$}}}}
\put(13684,-2483){\makebox(0,0)[lb]{\smash{{\SetFigFont{11}{13.2}{\rmdefault}{\mddefault}{\updefault}$\cdots$}}}}
\put(10941,-2130){\makebox(0,0)[lb]{\smash{{\SetFigFont{11}{13.2}{\rmdefault}{\mddefault}{\updefault}$J^\wedge(x)$}}}}
\put(6086,-2896){\makebox(0,0)[lb]{\smash{{\SetFigFont{11}{13.2}{\rmdefault}{\mddefault}{\updefault}$S_0$}}}}
\end{picture}%
\caption{Approximating the light cone by lens-shaped regions.}
\label{figcone}
\end{figure}

By a suitable choice of lens-shaped region, one can get an upper bound
for the maximal propagation speed. For the Dirac equation, where the
causal structure of Definition~\ref{defshscausal} coincides in view of~\eqref{dirsigma} with that
of Minkowski space, one can choose for~$S_1$
a family of spacelike hypersurfaces, which converge to the boundary of a light cone
(see Figure~\ref{figcone}).
This shows that the maximal propagation speed for Dirac waves is indeed the
speed of light (which, according to our conventions, is equal to one).

\section{Global Existence of Smooth Solutions} \label{secshsexist}
In this section, we will show that, by refining the above uniqueness argument, we
even obtain an existence proof. The close connection between existence and uniqueness for
linear equations is a familiar theme in mathematics.
The simplest setting where it occurs is in the study of the linear equation~$Au=v$
with a given vector~$v \in \R^n$ and a quadratic matrix~$A$.
In this case, the uniqueness of the solution implies that the matrix~$A$ is invertible,
which in turn ensures existence. A more interesting example is Fredholm's alternative
for compact operators (see, for example, \cite[Section~VI.5]{reed+simon}).
The procedure for globally hyperbolic systems follows somewhat similar ideas.
Here, the general strategy is to construct a bounded linear functional on a Hilbert space
in such a way that the Fr{\'e}chet-Riesz theorem (see Theorem~\ref{thm-FR}) gives the desired solution.

Before beginning, we point out that, in view of uniqueness and finite propagation speed, it suffices to consider
the problem in a bounded spatial region. Indeed, once we have constructed ``local solutions''
in small lens-shaped regions as shown on the left of Figure~\ref{figcover}, 
uniqueness implies that these solutions agree in the overlap of the lens-shaped regions,
making it possible to ``glue them together'' to obtain the desired solution which is global in space.
We will come back to this construction in more detail in the context of the Dirac equation
in Sections~\ref{secexgreen} and~\ref{seccauchyglobhyp} (see also Figure~\ref{figgluesolution}).
Having this construction in mind, we may start from a local problem and to extend the coefficients of the
symmetric hyperbolic system in an arbitrary way outside. Therefore, it is no loss of generality
to consider a problem in the whole space~$\R^m$. Choosing a bounded time interval~$t \in [0,T]$ (where~$t=0$
is the initial time), we are led to considering the time strip
\beq R_T := [0,T] \times \R^m\:. \eeq
We now write the linear system~\eqref{Acomp} as
\beq \label{Luw}
L u = w \qquad \text{with} \qquad L := A^j\partial_j +B \:,
\eeq
where we again sum over~$j=0,\dots,m$. Again using that the system can be extended
arbitrarily outside a bounded spatial region, we may assume that 
the functions~$A^j$, $B$ and~$w$ are uniformly bounded in~$R_T$ and that~$w$
has spatially compact support (meaning that~$w(t,.) \in C^\infty_0(\R^m)$ for all~$t \in [0, T]$).
Moreover, for convenience, we again assume smoothness of~$A^j$, $B$ and~$w$.
In the {\em{Cauchy problem}}, one seeks a solution to the equation~\eqref{Luw}
with prescribed initial data~$u_0 \in C^\infty(\R^m)$ at time~$t=0$,
\beq \label{cauchyL}
Lu=w\,, \qquad u|_{t=0}=u_0\in C^\infty_0(\R^m)
\eeq
in~$C^s(R_T)$. First of all, we may restrict attention to the case~$u_0 \equiv 0$,
\beq \label{cauchyzero}
Lu=w\,, \qquad u|_{t=0} \equiv 0\:.
\eeq
In order to see this, let~$u$ be a solution of the above Cauchy problem.
Choosing a function~$v \in C^\infty(R_T)$ which at~$t=0$
coincides with~$u_0$. Then the function~$\tilde{u}:=(u-v)$ satisfies
the equation~$L \tilde{u} = \tilde{w}$ with~$\tilde{w} = w+A^j\partial_j v +B v$ and vanishes at~$t=0$.
If conversely~$\tilde{u}$ is a solution to the corresponding Cauchy problem with zero
initial data, then~$u:= \tilde{u} + v$ is a solution to the original problem~\eqref{cauchyL}.

In preparation of the existence proof, we need to introduce the notion of a {\em{weak solution}}.
In order to get into the weak formulation, we multiply the equation~\eqref{Luw} by a test function~$v(t,\vec{x})$
and integrate over~$R_T$, giving rise to the equation
\beq \la v, Lu \ra_{L^2(R_T)} = \la v, w \ra_{L^2(R_T)} \eeq
with the $L^2$-scalar product defined by
\beq \label{L2prod}
\la v, w \ra_{L^2(R_T)} := \int_0^T \dd t \int_{\R^m} \la v(t,\vec{x}), w(t,\vec{x}) \ra\: \dd^mx \:.
\eeq
The next step is to integrate by parts, so that the derivatives act on the test function~$v$.
Before doing so, we need to specify the regularity of the test functions. To this end, for~$\lambda \in [0,T]$
we consider the time strip
\beq R_\lambda := [0,\lambda] \times \R^m\:. \eeq
We denote the $s$-times continuously
differential functions on~$R_\lambda$ with spatially compact support
by~$C^s(R_\lambda)$. The function spaces
\beq \underline{C^s(R_\lambda)} \qquad \text{and} \qquad \overline{C^s(R_\lambda)} \eeq
are defined as the functions which in addition vanish at~$t=0$ and~$t=\lambda$,
respectively. As the space of test functions we choose~$\overline{C^1(R_T)}$; 
this guarantees that integrating by parts does not yield boundary terms at~$t=T$.
For a classical solution~$u\in \underline{C^1(R_T)}$ (that is, a solution with zero Cauchy data~\eqref{cauchyzero}),
also the boundary term at~$t=0$ vanishes. We thus obtain
\beq \label{2m2}
\la v, w \ra_{L^2(R_T)} =  \la \tilde L v,u \ra_{L^2(R_T)} \qquad \text{for all~$v \in \overline{C^1(R_T)}$}\,,
\eeq 
where~$\tilde{L}$ is the formal adjoint of~$L$ with respect to the scalar product~\eqref{L2prod}; that is,
\beq \label{tildeLdef}
\tilde{L} := \tilde{A}^j \partial_j + \tilde{B} \qquad \text{with} \qquad
\tilde{A}^j = -A^j \quad \text{and} \quad
\tilde{B} = B^\dagger - \big( \partial_j A^j \big) \:.
\eeq
Now suppose that a function~$u \in C^1(R_T)$ satisfies~\eqref{2m2}.
Testing with functions~$v \in \overline{C^1(R_T)} \cap \underline{C^1(R_T)}$ which vanish both
at times~$t=0$ and~$t=T$, we can integrate by parts without boundary terms. Using a standard
denseness argument, one finds that~$u$ solves the symmetric hyperbolic system~\eqref{Luw}. Next, testing
with a function~$v \in \overline{C^1(R_T)}$ which does {\em{not}} vanish at~$t=0$, only
the boundary term remains, giving the equation
\beq \int_{\R^m} \la v(0,\vec{x}),u(0,\vec{x}) \ra\: \dd^mx = 0 \qquad \text{for all~$v \in \overline{C^1(R_T)}$} \:, \eeq
which in turn implies that~$u$ vanishes initially. Thus~$u$ is a solution of the Cauchy problem~\eqref{cauchyzero}. To summarize, for functions~$u \in \overline{C^1(R_T)}$, the weak formulation~\eqref{2m2} is equivalent to our Cauchy problem~\eqref{Luw} and~\eqref{cauchyzero}.
Therefore, it is sensible to take~\eqref{2m2} as the definition of a weak solution of the Cauchy problem.
\sindex{weak solution!of symmetric hyperbolic system}%
The main advantage of the weak formulation~\eqref{2m2} is that it is well-defined even for
functions that are not differentiable. 

Our next step is to derive so-called {\em{energy estimates}} for a given solution~$u \in \underline{C^1(R_T)}$.
\sindex{energy estimate!for symmetric hyperbolic system}%
To this end, we return to the formula for the divergence~\eqref{divform}
and using the equation~\eqref{Luw}, we obtain 
\begin{gather}
\label{2k}
 \partial_j \langle u, A^j u \rangle +\langle u,Cu\rangle =
2 \re\, \langle u,w\rangle , \\
C:=B+B^\ast -(\partial_jA^j)\,.
\end{gather}
Next, we integrate~\eqref{2k} over~$R_\lambda$, integrate by parts
and use that the initial values at~$t=0$ vanish. We thus obtain
\beq\label{2L}
E(\lambda) := \int_{t=\lambda}\langle u,A^0u\rangle\:\dd^mx =
\int^\lambda_0 \dd t\int_{\R^m} \Big( 2 \re\, \langle u,w\rangle - \langle u,Cu\rangle \Big) \:\dd^mx\,.
\eeq
Since the matrix~$C$ is uniformly bounded and~$A_0$ is uniformly positive, there
is a constant~$K>1$ such that
\beq |\langle u, Cu\rangle| \leq K\langle u,A^0 u\rangle\,. \eeq
Moreover, the linear term in~$u$ can be estimated with the Schwarz inequality by
\beq 2 \re\, \langle u,w\rangle \leq \mu \:\langle u,u\rangle
+\frac{1}{\mu} \:\langle w,w\rangle \leq \langle u,A^0 u\rangle
+\frac{1}{\mu^2} \:\langle w,A^0 w\rangle \eeq
with a suitable constant~$\mu > 0$. Applying these estimates in~\eqref{2L} gives
\beq E(\lambda) \leq (K+1)\int^\lambda_0 E(t)\:\dd t +\frac{1}{\mu^2}
\int_{R_\lambda}\langle w,A^0w\rangle \:\dd^nx\,. \eeq
Writing this inequality as
\beq \frac{\dd}{\dd\lambda} \E^{-(K+1)\lambda} \int^\lambda_0 E(t)\: \dd t
\leq \E^{-(K+1)\lambda} \frac{1}{\mu^2} \int_{R_T}\langle w,A^0w\rangle \: \dd^nx\,, \eeq
we can integrate over~$\lambda$ to obtain
\beq \int^T_0 E(\lambda) \:\dd\lambda \leq \frac{\E^{(K+1)T}-1}{K+1}\:
\frac{1}{\mu^2}\int_{R_T}\langle w,A^0w\rangle \:\dd^nx \:. \eeq
Finally, we apply the mean value theorem and
use that the exponential function is monotone to conclude that
\beq \label{2m}
\int^T_0 E(\lambda) \:d\lambda
\leq \frac{T}{\mu^2}\:\E^{(K+1)T} \int_{R_T} \langle w,A^0 w \rangle\:\dd^nx \:. 
\eeq
This is the desired energy estimate.

Before going on, we point out that
the notion of ``energy'' used for the quantity~$E(\lambda)$
does in general {\em{not}} coincide with the physical energy.
In fact, for the Dirac equation~\eqref{diracsymm},
$E(\lambda)$ has the interpretation as the electric charge. Following Example~\ref{exscalar},
for the scalar wave equation~$\Box \phi = 0$, we find
\beq \label{Escalar}
E(\lambda) = \int_{\R^m} \Big( |\partial_t \phi|^2 + |\nabla{\phi}|^2 + |\phi|^2 \Big) \dd^mx \:.
\eeq
This differs from the physical energy by the last summand~$|\phi|^2$ (and an overall factor of two).
The name ``energy'' for~$E(\lambda)$ was motivated by the fact, considering only the
highest derivative terms, the expression~\eqref{Escalar} is indeed the physical energy.
We point out that, in contrast to the physical energy, the quantity~$E(\lambda)$
does in general depend on time. The point is that~\eqref{2m} gives an a-priori control
of the energy in terms of the inhomogeneity. The exponential factor in~\eqref{2m}
can be understood in analogy to a Gr\"onwall estimate
\sindex{Gr\"onwall estimate}%
(for the classical Gr\"onwall estimate see, for example, \cite[Lemma~1.15 in Section~VII.1]{amann-escher2}).

For the following construction, it is convenient to introduce on~$C^1(R_T)$ the scalar product
\beq %\label{rprod}
(u,v) = \int_{R_T} \langle u,A^0v\rangle \:\dd^nx \:. \eeq
We denote the corresponding norm by~$\|\cdot\|$. Setting furthermore
\beq \Gamma^2=\frac{T}{\mu^2} \E^{(K+1)T}\,, \eeq
the energy estimate can be written in the compact form
\beq (u,u) \leq \Gamma^2 \:(w,w) \:. \eeq
This inequality holds for every solution~$u$ of the differential equation~$Lu=w$
which vanishes at~$t=0$. Noting that every function~$u \in \underline{C^1(R_T)}$
is a solution of this differential equation with inhomogeneity~$w := L u$,
we obtain
\beq\label{2n2}
\|u\| \leq \Gamma\,\|Lu\| \qquad \text{for all~$u\in\underline{C^1(R_T)}$}\,.
\eeq
This is the form of the energy estimates suitable for an abstract existence proof.

Note that the operator~$\tilde{L}$ in~\eqref{tildeLdef} is also symmetric hyperbolic
and has the same boundedness and positivity properties as~$L$.
Hence, repeating the above arguments, we obtain similar to~\eqref{2n2} the ``dual estimate''
\beq \label{dual}
\| v \| \leq \tilde{\Gamma}\,\|\tilde{L} v\| \qquad \text{for all~$v \in \overline{C^1(R_T)}$}\,.
\eeq

We now want to show the existence of weak solutions with the help of
the Fr{\'e}chet-Riesz theorem (see Theorem~\ref{thm-FR} in the preliminaries or, for
example, \cite{reed+simon, lax}). To this end, we first introduce on~$\overline{C^1(R_T)}$ yet another scalar product
denoted by
\beq \label{spdef}
\langle v,v' \rangle = (\tilde L v,\tilde L v) \:.
\eeq
This scalar product is indeed positive definite, because for any~$v\neq 0$,
\beq \langle v,v \rangle = (\tilde Lv, \tilde Lv) \geq \tilde\Gamma^{-2} \:(v,v)\neq 0\,, \eeq
where in the last step we applied~\eqref{dual}.
Forming the completion, we obtain the Hilbert space~$({\mathscr{H}},\langle .,. \rangle)$.
We denote the corresponding norm by~$\norm . \norm$.
In view of~\eqref{dual} and~\eqref{spdef}, we know that every vector~$v \in \H$
is a function in~$L^2(R_T, \dd^nx)$. Moreover, we know from~\eqref{spdef} that~$\tilde{L} v$
is also in~$L^2(R_T, \dd^nx)$. We remark that, in the language of functional analysis, the space~$\H$
can be identified with the Sobolev space~$W^{1,2}(R_T)$, but we do not need this here.

We now consider for~$w \in C^0(R_T)$ and~$v \in \overline{C^1(R_T)}$ the linear
functional~$\la v,w \ra_{L^2(R_T)}$. In view of the estimate
\beq \big| \la v,w \ra_{L^2(R_T)} \big| \leq \|v\|_{L^2(R_T)}\, \|w\|_{L^2(R_T)}
\leq \frac{\tilde\Gamma}{C} \:\|w\|_{L^2(R_T)}\: \norm v \norm \:, \eeq
this functional is continuous in~$v\in{\mathscr{H}}$. The Fr{\'e}chet-Riesz theorem
shows that there is~$U\in{\mathscr{H}}$ with
\beq \la v,w \ra_{L^2(R_T)} = \la v,U \ra = (\tilde Lv, \tilde L U) \qquad \text{for all~$v \in {\mathscr{H}}$} \,. \eeq
Rewriting the last scalar product as
\beq (\tilde Lv, \tilde L U) = \la Lv, A^0 \tilde L U \ra_{L^2(R_T)} \:, \eeq
one sees that the function~$u:=A^0 \tilde LU\in L^2(R_T, \dd^nx)$ satisfies the equation~\eqref{2m2}
and is thus the desired weak solution. Note that all our methods apply for arbitrarily large~$T$.
We have thus proved the global existence of weak solutions.

We next want to show that the solutions are smooth. Thus our task is to show that
our constructed weak solution~$u$ is of the class~$C^s(R_\lambda)$, where~$s\geq 1$
can be chosen arbitrarily large.
We first show that a linear symmetric hyperbolic system can be ``enlarged'' to
include the partial derivatives of~$\phi$.
\begin{Lemma} \label{lemmaenlarge} Suppose that the system~$A^j \partial_j u + B u = w$ is
symmetric hyperbolic. Then there is a symmetric hyperbolic system of the form
\beq \label{symmlarge}
\tilde{A}^j \partial_j \Psi + \tilde{B} \Psi = \tilde{w}
\eeq
for the vector~$\Psi := (\partial_t u, \nabla_1 u, \ldots, \nabla_m u, u) \in \C^{(n+1) N}$.
\end{Lemma}
\Proof
Let~$i$ be a fixed spacetime index. We differentiate the equation~$Lu=w$,
\beq \partial_i w=\partial_i Lu =
L\partial_i u+(\partial_i A^j) \,\partial_ju + (\partial_i B)\, u \:. \eeq
This equation can be written as
\beq A^j \partial_j \Psi_i + \sum_{j=1}^n \tilde{B}_i^j \,\Psi_j + (\partial_i B)\, u = \tilde{w}_i \:, \eeq
where we set
\beq \tilde{B}_i^j = B\, \delta_i^j + (\partial_i A^j) \qquad \text{and} \qquad
\tilde{w}_i = \partial_i w  \:. \eeq
Combining these equations with the equation~$Lu=w$, we obtain a system of
the form~\eqref{symmlarge}, where the matrices~$A^j$ are block diagonal in the sense that
\beq \tilde{A}^j = \big( (\tilde{A}^j)^\alpha_\beta \big)_{\alpha, \beta=0,\ldots, m+1} \qquad \text{with} \qquad
(\tilde{A}^j)^\alpha_\beta = A^j\: \delta^\alpha_\beta \:. \eeq
Obviously, this system is again symmetric hyperbolic.
\QED
Iterating this lemma, we obtain (at least in principle) a symmetric hyperbolic system
for~$u$ and all its partial derivatives up to any given order~$s$.
Since the corresponding weak solution is in~$L^2(R_T)$, we conclude that~$u$
and all its weak partial derivatives are square integrable.
The next lemma, which is a special case of the general Sobolev embedding theorems
(see, for example, \cite[Section~II.5.]{evans} or~\cite[Section~4]{taylor1}),
gives smoothness of the solution.
\begin{Lemma} \label{lemmasobolev} Let~$s > \frac{m}{2}$ be an integer.
If a function~$g$ on~$\R^m$ is~$s$ times weakly differentiable and
\beq \label{weakder}
\int_{\mathbb{R}^m} |\nabla^\alpha g|^2 \:\dd^mx < C
\eeq
for all multi-indices~$\alpha$ with~$|\alpha| \leq s$, then~$g$ is bounded, $g \in L^\infty(\R^m)$.
Likewise, if~$g$ is~$s+l+1$ times weakly differentiable with~$l \geq 1$
and~\eqref{weakder} holds for all~$\alpha$ with~$|\alpha| \leq s+l+1$, then~$g \in C^l(\R^m)$.
\end{Lemma}
\Proof We apply the Schwarz inequality to the Fourier transform,
\begin{align}
| g(x)|^2 &= \left|\int_{\R^m} \frac{\dd^mk}{(2\pi)^m} \:\hat{g}(k) \:\E^{-\cI kx}\right|^2 \notag \\
&=\left|\int_{\R^m} \frac{\dd^mk}{(2\pi)^m}(1+|k|^2)^{-\frac{s}{2}}\: (1+|k|^2)^{\frac{s}{2}} \:
\hat g(k) \E^{-\cI kx}\right|^2 \notag \\
&\leq c_m\int_{\R^m} \frac{\dd^mk}{(2\pi)^m}(1+|k|^2)^s\: |\hat g(k)|^2\:,
\end{align}
where the constant~$c_m$ is finite due to our choice of~$s$,
\beq c_m=\int_{\R^m} \frac{\dd^mk}{(2\pi)^m} \:(1+|k|^2)^{-s}<\infty\,. \eeq
Using the Plancherel formula together with
the fact that a factor~$k^2$ corresponds to a Laplacian in position space, we obtain
\beq \int_{\R^m} \frac{\dd^mk}{(2\pi)^m} (1+|k|^2)^s \:|\hat g(k)|^2
= \sum^s_{\ell=0} \binom n \ell \: \|\nabla^\ell g\|^2_{L^2(\mathbb{R}^m)}<c\,. \eeq
Hence~$\sqrt{c_m \,c}$ is an $L^\infty$-bound for~$g$.

Next, if~$g$ is~$s+1$ times weakly differentiable, then~$\|Dg\|_{L^\infty}(\R^m) < c$.
As a consequence, the mean value theorem yields~$|g(x)-g(y)| \leq c |x-y|$,
so that~$g$ is Lipschitz continuous.
Finally, if~$g$ is~$s+l+1$ times weakly differentiable, then all partial derivatives~$\nabla^\alpha g$
of order~$|\alpha| \leq l$ are Lipschitz continuous, so that~$g \in C^l(\R^m)$.
\QED

More precisely, in order to apply this lemma, we fix a time~$t$ and consider the
solution~$u(\lambda,.)$. The identity~\eqref{2L} implies that~$E(\lambda)$
is controlled in terms of~$\|w\|$ and~$\|u\|$.
After iteratively applying Lemma~\ref{lemmaenlarge}, we conclude that
the weak derivatives of~$u(\lambda, .)$ exist to any order and are in~$L^2(\R^m)$.
It follows that~$u(\lambda, .)$ is smooth. Finally, one uses the equation
to conclude that~$u$ is also smooth in the time variable.

The results of this section can be summarized as follows.
\begin{Thm} \label{thm10} Consider the Cauchy problem
\beq \big( A^0\partial_t+\sum_{\alpha=1}^m A^\alpha \nabla_\alpha + B \big)u = w \in C^\infty_0([0,T] \times \R^m)
\,,\qquad u\vert_{t=0}=u_0 \in C^\infty_0(\R^m) \:. \eeq
Assume that the matrices~$A^0$, $A^j$
and~$B$ as well as the functions~$w$ and~$u_0$ are smooth. Moreover, assume that all
these functions as well as as all their partial derivatives are uniformly bounded
(where the bound may depend on the order of the derivatives).
Then the Cauchy problem has a smooth solution on~$[0,T] \times \R^m$. \noindent
\end{Thm} \noindent
This theorem also applies in the case~$T=\infty$, giving global existence of a smooth solution.

We finally show that the solutions depend smoothly an parameters.
\begin{Corollary} \label{corpar}
Suppose that the matrices~$A^j, B$ and the functions~$w, u_0$ depend smoothly on
a parameter~$\lambda$. Then the family of solutions~$u(\lambda)$ is also smooth in~$\lambda$.
\end{Corollary}
\Proof First, similar as explained after~\eqref{cauchyzero}, we may restrict attention to the
case~$u_0=0$. 
Differentiating the equation~$Lu=w$ with respect to~$\lambda$, we obtain
\beq L u_\lambda = (\partial_\lambda L) u + \partial_\lambda w =: \tilde{w}\:, \eeq
where~$u_\lambda$ stands for the formal derivative~$\partial_\lambda u$.
This is a symmetric hyperbolic system for~$u_\lambda$.
According to Theorem~\ref{thm10}, we know that~$u$ and therefore~$\tilde{w}$ are smooth.
Hence, applying this theorem again, we conclude that there exists a smooth solution~$u_\lambda$.
Considering the limit of the difference quotients, one verifies that~$u_\lambda$ really
coincides with~$\partial_\lambda u(\lambda)$ for our given family of solutions~$u(\lambda)$.
The higher $\lambda$-derivatives can be treated inductively.
\QED

\section{The Causal Dirac Green's Operators in Minkowski Space} \label{secexgreen}
We now want to apply the previous general existence and uniqueness results
to the Cauchy problem~\eqref{Cauchydirac} for the Dirac equation in Minkowski space in the
presence of an external potential~$\B$.
\begin{Thm} \label{thmexistsdiracmink}
Consider the Cauchy problem for the Dirac equation~\eqref{Cauchydirac}
for smooth initial data~$\psi_0$, a smooth inhomogeneity~$\phi$ and
a smooth matrix-valued potential~$\B \in C^\infty(\scrM, \C^{4 \times 4})$.
Then there is a unique global smooth solution~$\psi \in C^\infty(\scrM, S\scrM)$.
\end{Thm}
\Proof Writing the Dirac equation in the Hamiltonian form~\eqref{diracsymm},
we obtain a symmetric hyperbolic system. In view of the uniqueness result for smooth solutions
of Corollary~\ref{corunique}, it suffices to construct a smooth solution at any given time~$T \in \R$.
It suffices to consider the case~$T>t_0$, because otherwise we reverse the time direction.
Moreover, we can arrange by a time shift that~$t_0=0$.

\begin{figure}[tb]
\psscalebox{1.0 1.0} % Change this value to rescale the drawing.
{
\begin{pspicture}(0,28.37454)(12.130383,30.504921)
\definecolor{colour0}{rgb}{0.8,0.8,0.8}
\definecolor{colour1}{rgb}{0.6,0.6,0.6}
\psframe[linecolor=colour0, linewidth=0.02, fillstyle=solid,fillcolor=colour0, dimen=outer](7.6906495,29.86519)(0.89064944,28.46519)
\pspolygon[linecolor=white, linewidth=0.02, fillstyle=solid](0.110649414,29.690191)(0.24564941,29.71519)(0.40564942,29.725191)(0.5906494,29.67519)(0.7456494,29.555191)(0.9256494,29.35519)(1.1156495,29.13019)(1.2956494,28.90519)(1.4656494,28.690191)(1.5706494,28.565191)(1.6756494,28.46519)(1.1956494,28.47019)(0.76564944,28.48019)(0.095649414,28.485191)
\pspolygon[linecolor=colour1, linewidth=0.02, fillstyle=solid,fillcolor=colour1](3.3256495,29.70519)(3.6356494,29.78519)(3.9756494,29.83519)(4.4706492,29.850191)(5.0606494,29.81019)(5.1956496,29.77519)(5.3706493,29.66519)(5.5056496,29.43519)(5.5256495,29.28019)(5.5156493,29.100191)(5.4706492,28.95519)(5.3706493,28.760191)(5.260649,28.600191)(5.1656494,28.475191)(3.1806495,28.475191)(2.9656494,28.565191)(2.8056495,28.67019)(2.7056494,28.815191)(2.6856494,28.96519)(2.7306495,29.17019)(2.8456495,29.37519)(2.9956493,29.53519)(3.1556494,29.635191)
\psline[linecolor=black, linewidth=0.04, arrowsize=0.05291667cm 2.0,arrowlength=1.4,arrowinset=0.0]{<-}(0.09064941,30.46519)(0.09064941,28.46519)
\psline[linecolor=black, linewidth=0.04, linestyle=dashed, dash=0.17638889cm 0.10583334cm](0.09064941,29.86519)(11.89065,29.86519)
\psbezier[linecolor=black, linewidth=0.02](3.1806495,28.475191)(2.655972,28.620687)(2.5452838,28.931002)(2.8306494,29.3651904296875)(3.116015,29.79938)(3.935972,29.890688)(4.8306494,29.85519)(5.725327,29.819693)(5.686015,29.029379)(5.1706495,28.475191)
\psbezier[linecolor=black, linewidth=0.02, fillstyle=solid](6.1256495,28.46519)(6.8740706,29.088057)(7.2656493,29.84519)(8.17565,29.8551904296875)(9.0856495,29.86519)(9.739432,28.992096)(10.50565,28.46519)
\psline[linecolor=black, linewidth=0.02, linestyle=dashed, dash=0.17638889cm 0.10583334cm](0.89064944,29.86519)(0.89064944,28.46519)
\psline[linecolor=black, linewidth=0.02, linestyle=dashed, dash=0.17638889cm 0.10583334cm](7.6906495,29.86519)(7.6906495,28.46519)
\psbezier[linecolor=black, linewidth=0.02](1.6856494,28.46519)(1.5030638,28.617603)(1.2928774,28.948528)(1.0056494,29.2751904296875)(0.71842134,29.601852)(0.56996405,29.840895)(0.095649414,29.68519)
\psline[linecolor=black, linewidth=0.04, arrowsize=0.05291667cm 2.0,arrowlength=1.4,arrowinset=0.0]{->}(0.09064941,28.46519)(12.09065,28.46519)
\rput[bl](-0.3,29.8){$T$}
\rput[bl](0.25,30.2){$t$}
\rput[bl](12,28.6){$\vec{x}$}
\rput[bl](1,29.35){$[0,T] \times K$}
\rput[bl](8.1,29.2){$L$}
\rput[bl](0.3,29){$L$}
\rput[bl](3.2,28.9){$\supp (\eta_k \tilde{\phi})$}
\end{pspicture}
}
\caption{Construction of local solutions.}
\label{figgluesolution}
\end{figure}
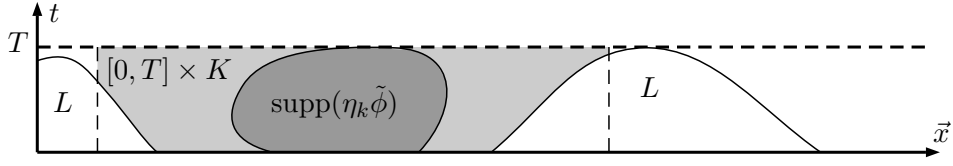

We cannot apply Theorem~\ref{thm10} directly because the coefficient functions in~\eqref{diracsymm}
do not need to be bounded, nor are our initial values compactly supported.
For this reason, we need to construct local solutions and ``glue them together'' using
linearity: We first extend the initial data~$\psi_0$ smoothly to the time strip~$R_T$
and consider the Cauchy problem for~$\tilde{\psi} := \psi - \psi_0$,
\beq %\label{Cauchytil}
(\cI \Pdd + \B - m) \,\tilde{\psi} = \tilde{\phi} \in C^\infty(\scrM, S\scrM) \:,\qquad \tilde{\psi}|_{t_0} = 0 \:. \eeq
We let~$(\eta_k)_{k \in \N}$ be a smooth partition of unity of~$\R^m$ with~$\eta_k \in C^\infty_0(\R^m)$ (for details see, for example, \cite[Theorem~2.13]{rudin}). We extend these functions to static functions on~$R_T$
(that is, $\eta_k(t,\vec{x}):=\eta_k(\vec{x})$. Given~$k \in \N$,
we first solve the Cauchy problem for the inhomogeneity~$\eta_k \tilde{\phi}$.
We choose a compact set~$K \subset \R^m$ such that~$[0,T] \times K$
contains the causal future of the support of~$(\eta_k \tilde{\phi})$ (see Figure~\ref{figgluesolution};
more specifically, we could choose~$K=B_{2T}(\supp \eta_k)$).
Next, we choose a smooth, compactly supported function~$\theta \in C^\infty_0(\R^m)$
with~$\theta|_K\equiv 1$. We again extend~$\theta$ to a static function on~$R_T$.

We now consider the modified Cauchy problem
\beq \Big( \1_{\C^4}\, \partial_t \psi + \gamma^0 \vec{\gamma} \vec{\nabla} 
+ \theta \big( -\cI \gamma^0 (\B - m) \big) \Big) \tilde{\psi}_k = -\cI \gamma^0 \: \eta_k\: \tilde{\phi}
\:,\qquad \tilde{\psi}_k|_{t_0} = 0 \:. \eeq
Now the coefficients in the PDE are uniformly bounded, and the inhomogeneity has compact support.
Therefore, we can apply Theorem~\ref{thm10} to obtain a global smooth solution.
Due to finite propagation speed (see Theorem~\ref{thmsymmunique}, where we choose lens-shaped
regions~$L$ as shown in Figure~\ref{figgluesolution}), this solution vanishes outside~$K$.
Therefore, it is also a solution of the unmodified Dirac equation,
with initial data~$\eta_k \tilde{\phi}$.

Finally, summing over~$k$ gives the desired solution of the original Cauchy problem,
\beq \psi := \sum_{k=1}^\infty \tilde{\psi}_k \:. \eeq
Here the series converges because, again due to finite propagation speed, it is locally finite.
\QED

We next explain how the previous existence and uniqueness results give rise to
the existence of causal Green's operators, being defined as integral operators
with distributional kernels. These kernels are often referred to as {\em{Green's functions}}.
Our main tool is the {\em{Schwartz kernel theorem}}. We do not give a proof of this
more advanced result of distribution theory but refer instead to~\cite[Section~5.2]{hormanderI}
or~\cite[Section~4.6]{taylor1}.
For better consistency with the notation in the perturbative treatment in Chapter~\ref{secperturb}, 
from now on we denote the objects in the presence of an external potential with an additional tilde.
We begin with a representation formula for the solution of the Cauchy problem
in terms of a distribution.

\begin{Thm} \label{thmkext}
Assume that the external potential~$\B$ is smooth and that~$\B$ and all
its partial derivatives are uniformly bounded in Minkowski space.
Then for any~$t, t_0$ there is a unique distribution~$\tilde{k}_m(t,.; t_0, .) \in {\mathcal{D}}'(\R^3 \times \R^3)$
such that the solution of the Cauchy problem~\eqref{cauchyt} has the representation
\beq \label{Bcrep}
\psi(t, \vec{x}) = 2 \pi \int_N \tilde{k}_m(t,\vec{x}; t_0, \vec{y})\, \gamma^0\, \psi_0(\vec{y})\: \dd^3y\:.
\eeq
The integral kernel~$k_m$ is also a distribution in spacetime, $k_m \in {\mathcal{D}}'(M \times M)$ It
is a distributional solution of the Dirac equation,
\beq \label{tkmd}
(\cI \Pdd_x + \B - m) \,\tilde{k}_m(x,y) = 0 \:.
\eeq
\end{Thm}
\Proof The energy estimates combined with the Sobolev embedding of Lemma~\ref{lemmasobolev}
showed that there is~$k \in \N$ and a constant~$C = C(t,t_0, \vec{x}, \B)$ such that the
solution~$\psi(t,.)$ of the Cauchy problem is bounded in terms of the initial data by
\beq \label{pointes}
|\psi(t, \vec{x})| \leq C\, |\psi_0|_{C^k} \:,
\eeq
where~$|\psi|^2 := \Sl \psi | \gamma^0 \psi \Sr$, and the $C^k$-norm is defined by
\beq |\psi_0 |_{C^k} = \max_{|\beta| \leq k} \;\sup_{\vec{x} \in \R^3} |\nabla^\beta \psi_0(\vec{x}) | \:. \eeq
Moreover, this estimate is locally uniform in~$\vec{x}$, meaning that for any compact set~$K \subset \R^3$,
there is a constant~$C$ such that~\eqref{pointes} holds for all~$\vec{x} \in K$.
This makes it possible to apply the Schwartz kernel theorem~\cite[Theorem~5.2.1]{hormanderI},
showing that~$\tilde{k}_m(t,.; t_0, .) \in {\mathcal{D}}'(\R^3 \times \R^3)$.

Next, we note that the constant~$C$ in~\eqref{pointes} can also be chosen locally uniformly in~$t$
and~$t_0$. Thus, after evaluating weakly in~$t$ and~$t_0$, we may again apply the
Schwartz kernel theorem to obtain that~$\tilde{k}_m \in {\mathcal{D}}'(M \times M)$. Finally,
the distributional equation~\eqref{tkmd} follows immediately from the fact that~\eqref{Bcrep} is
satisfies the Dirac equation for any choice of~$\psi_0$.
\QED

The distribution~$\tilde{k}_m$ is referred to as the {\em{causal fundamental solution}}.
Encoding the whole Dirac dynamics, it plays a fundamental role in the analysis of the Dirac equation.
In the next step, we introduce the {\em{causal Green's operators}}
by decomposing~$\tilde{k}_m$ in time. Namely, for any~$t, t_0$ we
introduce the distribution~$\tilde{s}_m^\vee(t,.; t_0, .), \tilde{s}_m^\wedge(t,.; t_0, .)
\in {\mathcal{D}}'(\R^3 \times \R^3)$ by
\beq \label{tsdef}
\left\{ \begin{array}{l} \tilde{s}_m^\vee(t,.; t_0, .) = \;\;\:2 \pi \cI \:\tilde{k}_m(t,.; t_0, .)\: \Theta(t_0-t) \\[0.2em]
\tilde{s}_m^\wedge(t,.; t_0, .) = -2 \pi \cI \:\tilde{k}_m(t,.; t_0, .)\: \Theta(t-t_0) \end{array} \right.
\eeq
(where~$\Theta$ denotes the Heaviside function).
\sindex{causal fundamental solution}%
\nindex{age@$\tilde{k}_m$ -- causal fundamental solution}%
\sindex{Green's operator!causal}%
\nindex{agf@$\tilde{s}_m^\vee, \tilde{s}_m^\wedge$ -- causal Green's operators}%
In this way, we introduce the causal fundamental solutions for any given~$t_0$ and~$t$
as distributions on~$\R^3 \times \R^3$. Alternatively, they can also be introduced as bi-distributions
in spacetime, as is shown in the next lemma.

\begin{Thm} \label{thmgreenhyp}
Assume that the external potential~$\B$ is smooth and that~$\B$ and all
its partial derivatives are uniformly bounded in Minkowski space.
Then there are unique distributions
\beq \tilde{s}_m^\vee, \tilde{s}_m^\wedge \in {\mathcal{D}}'(\scrM \times \scrM) \eeq
which satisfy the distributional equations
\beq \label{diseq}
(\cI \Pdd_x + \B - m) \: \tilde{s}_m(x,y) = \delta^4(x-y)
\eeq
and are supported in the upper respectively lower light cone,
\beq \label{supprop}
\supp \tilde{s}^\lor_m(x,.) \subset J_x^\lor \;,\spc
\supp \tilde{s}^\land_m(x,.) \subset J_x^\land\:.
\eeq
\end{Thm}
\Proof It is clear by construction and the fact that the constant~$C$ in~\eqref{pointes} can be
chosen locally uniformly in~$x$ and~$y$ that the causal Green's operators are
well-defined distributions in~${\mathcal{D}}'(M \times M)$. The support property~\eqref{supprop}
follows immediately from finite propagation speed as explained at the end of Section~\ref{sec31}.
The uniqueness of the causal Green's operators is clear from the uniqueness of solutions of the Cauchy problem.
In order to derive the distributional equations~\eqref{diseq}, we only consider the retarded Green's operator
(the argument for the advanced Green's operator is analogous). Then, according to~\eqref{Bcrep}
and~\eqref{tsdef},
\beq \Theta(t-t_0) \: \psi(t, \vec{x}) = \cI \int_N \tilde{s}^\wedge_m(t,\vec{x}; t_0, \vec{y})\, \gamma^0\,
\psi_0(\vec{y})\: \dd^3y \:, \eeq
where~$\psi$ is the solution of the corresponding Cauchy problem. Applying the Dirac operator
in the distributional sense yields
\beq \cI \gamma^0 \delta(t-t_0) \: \psi_0(t, \vec{x})
= \cI (\Dir_x - m) \int_N \tilde{s}^\wedge_m(t,\vec{x}; t_0, \vec{y})\, \gamma^0\,
\psi_0(\vec{y})\: \dd^3y \:. \eeq
We now choose the initial values as the restriction of a test function in spacetime,
$\psi_0 = \phi|_{t=t_0}$ with~$\phi \in C^\infty_0(M, SM)$. Then we can integrate over~$t_0$ to obtain
\beq \cI \gamma^0 \phi(x) = (\Dir_x - m) \int_M \tilde{s}^\wedge_m(x,y)\: \cI \gamma^0
\phi(y)\: \dd^4y \:. \eeq
This gives the result.
\QED

We remark that, turning the above-mentioned argument around, we can also use the causal Green's operators
in order to define the causal fundamental solution as a bi-distribution in spacetime,
\beq \label{kssdef}
\tilde{k}_m := \frac{1}{2 \pi \cI}\: \big( \tilde{s}^\vee_m - \tilde{s}^\wedge_m \big) \in {\mathcal{D}}'(\scrM \times \scrM) \:.
\eeq

The causal fundamental solution has the remarkable property that it relates the scalar product
with the inner product obtained by integrating the spin inner product over spacetime.
We now explain this relation step by step.
Given two wave functions~$\psi$ and~$\phi$ (not necessarily solutions of the Dirac equation),
we want to integrate their pointwise
inner product~$\Sl \psi | \phi \Sr_x$ over spacetime (similar as already done
in the preliminaries in~\eqref{stipMin} and~\eqref{stipintro}).
In order to ensure that this integral is well-defined, it suffices to assume that one
of the functions is compactly supported. We thus obtain the sesquilinear pairing
\begin{gather}
\bra .|. \ket \::\: C^\infty(\scrM, S\scrM) \times C^\infty_0(\scrM, S\scrM) \rightarrow \C \:, \notag \\
\bra \psi|\phi \ket = \int_\scrM \Sl \psi | \phi \Sr_x \: \Diff\mu_\scrM  \label{stip} 
\end{gather}
(here~$C^\infty(\scrM, S\scrM)$ are again the smooth sections of the spinor bundle,
and~$C^\infty_0(\scrM, S\scrM)$ denotes the smooth sections with compact support).
Restricting the first argument to compactly supported wave functions, we obtain an inner product,
\beq \bra .|. \ket \::\: C^\infty_0(\scrM, S\scrM) \times C^\infty_0(\scrM, S\scrM) \rightarrow \C \:, \eeq
referred to as the {\em{spacetime inner product}}
\sindex{spacetime inner product}%
\nindex{aap@$\bra . \vert . \ket$ -- spacetime inner product}%
\tindex{bb@$\bra . \vert . \ket$ -- spacetime inner product}%
(we remark that this inner product space can be extended to a Krein space; we refer
the interested reader to~\cite[\S1.1.5]{cfs}).
Alternatively, one can also restrict the first argument of~$\bra .|. \ket$ to smooth Dirac solutions
and extend by approximation to the whole Hilbert space~$\H_m$, giving the sesquilinear pairing
\beq \bra .|. \ket \::\: \H_m \times C^\infty_0(\scrM, S\scrM) \rightarrow \C \:. \eeq

The following proposition goes back to John Dimock (see~\cite[Proposition~2.2]{dimock3}).
\begin{Prp} \label{prpdual}
For any~$\psi_m \in \H_m$ and~$\phi \in C^\infty_0(\scrM, S\scrM)$,
\beq \label{dual2}
(\psi_m \,|\, \tilde{k}_m \phi)_m = \bra \psi_m | \phi \ket \:.
\eeq
\end{Prp} \noindent

\Proof We first give the proof under the additional assumption~$\psi_m \in \Cisc(\scrM, S\scrM)$
that the Dirac solution has spatially compact support.
We choose Cauchy surfaces~$\scrN_+$ and~$\scrN_-$ lying in the future and past of~$\supp \phi$,
respectively. Let~$\Omega$ be the spacetime region between these two Cauchy surfaces; that is,
$\partial \Omega = \scrN_+ \cup \scrN_-$. Then, according to~\eqref{kmdef}
and using again the notation~\eqref{nicsp},
\begin{align}
(\psi_m \,|\, \tilde{k}_m \,\phi)_m &= (\psi_m \,|\, \tilde{k}_m \,\phi)_{\scrN_+}
= \frac{\cI}{2 \pi} \:(\psi_m \,|\, \tilde{s}_m^\wedge \,\phi)_{\scrN_+} \notag \\
&= \frac{\cI}{2 \pi} \Big[ (\psi_m \,|\, \tilde{s}_m^\wedge \,\phi)_{\scrN_+} - (\psi_m \,|\, \tilde{s}_m^\wedge \,\phi)_{\scrN_-} \Big] \notag \\
&= \cI \int_\Omega \nabla_j \Sl \psi_m \,|\, \gamma^j\, \tilde{s}_m^\wedge \phi \Sr_x\: \Diff\mu(x)\:,
\end{align}
where in the last line we applied the Gau{\ss} divergence theorem and used~\eqref{print}.
Using that~$\psi_m$ satisfies the Dirac equation, a calculation similar to~\eqref{divfree} yields
\beq (\psi_m \,|\, \tilde{k}_m \,\phi)_m = \int_\Omega \Sl \psi_m \,|\, (\Dir - m) \,\tilde{s}_m^\wedge \phi \Sr_x\: \Diff\mu(x)
\overset{\eqref{Greendefext}}{=} \int_\Omega \Sl \psi_m | \phi \Sr_x\: \Diff\mu(x)\:. \eeq
As~$\phi$ is supported in~$\Omega$, we can extend the last integration to all of~$\scrM$, giving
the result.

In order to extend the result to general~$\psi_m \in \H_m$, we use the following approximation
argument. Let~$\psi_m^{(n)} \in \H_m \cap \Cisc(\scrM, S\scrM)$ be a sequence which converges in~$\H_m$
to~$\psi_m$. Then obviously~$(\psi^{(n)}_m \,|\, \tilde{k}_m \,\phi)_m \rightarrow (\psi_m \,|\, \tilde{k}_m \,\phi)_m$.
In order to show that the right-hand side of~\eqref{dual2} also converges, it suffices to prove
that~$\psi_m^{(n)}$ converges in~$L^2_\text{loc}(\scrM, S\scrM)$ to~$\psi_m$.
Thus let~$K \subset \scrM$ be a
compact set contained in the domain of a chart~$(x, U)$. Using Fubini's theorem, we obtain 
for any~$\psi \in \H_m \cap \Cisc(\scrM, S\scrM)$ the estimate
\beq \int_K \Sl \psi | \nuslsh \psi \Sr \Diff\mu_\scrM = \int \dd x^0 \int \Sl \psi | \nuslsh \psi \Sr \sqrt{|g|} \:\dd^3x
\leq C(K) \,(\psi | \psi)_m \:. \eeq
Applying this estimate to the functions~$\psi = \psi^{(n)}_m - \psi^{(n')}_m$, we see
that~$\psi^{(n)}_m$ converges in~$L^2(K, S\scrM)$ to a function~$\tilde{\psi}$.
This implies that~$\psi^{(n)}_m$ converges to~$\tilde{\psi}$ pointwise almost everywhere
(with respect to the measure~$\Diff\mu_\scrM$).
Moreover, the convergence of~$\psi^{(n)}_m$ in~$\H_m$ to~$\psi_m$ implies that the
restriction of~$\psi^{(n)}_m$ to any Cauchy surface~$\scrN$ converges to~$\psi_m|_\scrN$ 
pointwise almost everywhere (with respect to the measure~$\Diff\mu_\scrN$).
It follows that~$\tilde{\psi} = \psi_m|_K$, concluding the proof.
\QED

\begin{Corollary} \label{corksymm} The operator~$\tilde{k}_m$, \eqref{kmdef},
is symmetric with respect to the inner product~\eqref{stip}.
\end{Corollary}
\Proof Using Proposition~\ref{prpdual}, we obtain for all~$\phi, \psi \in C^\infty_0(\scrM, S\scrM)$,
\beq \bra \tilde{k}_m \phi \,|\, \psi \ket = (\tilde{k}_m \phi \,|\, \tilde{k}_m \psi)_m = \bra \phi \,|\, \tilde{k}_m \psi \ket \:, \eeq
concluding the proof.
\QED

\section{A Polynomial Estimate in Time}
We now derive an estimate which shows that the solutions of the Dirac equation
increase at most polynomially in time. This result will be needed in Section~\ref{secBmop}.
For the proof, we adapt 
standard methods of the theory of partial differential equations to the Dirac equation.
In generalization of~\eqref{sobolev}, we denote the spatial Sobolev norms by
\beq \| \phi \|^2_{W^{a, 2}} = \sum_{\text{$\alpha$ with~$|\alpha| \leq a$}}\:
\int_{\R^3} | \nabla^\alpha \phi(\vec{x}) |^2\: \dd^3x \:. \eeq

\begin{Lemma} \label{lemmaB} We are given two non-negative integers~$a$ and~$b$ as well as
a smooth time-dependent potential~$\B$. In the case~$a>0$ and~$b \geq 0$, we assume furthermore
that the spatial derivatives of~$\B$ decay faster than linearly for large times in the sense that
\beq \label{Bdecay2}
|\nabla \B(t)|_{C^{a-1}} \leq \frac{c}{1 + |t|^{1+\varepsilon}}
\eeq
for suitable constants~$c, \varepsilon>0$. Then there is a constant~$C=C(c, \varepsilon, a,b)$ such
that every family of solutions~$\psi \in \H^\infty$
of the Dirac equation~\eqref{Dout} for varying mass parameter can be estimated for all times in terms of the
boundary values at~$t=0$ by
\beq \big\| \partial^b_m \psi_m|_t \big\|_{W^{a, 2}} \leq C\, \big( 1+|t|^b \big)\: \sum_{p=0}^b \big\| \partial^p_m
\psi_m|_{t=0} \big\|_{W^{a, 2}} \:. \eeq
\end{Lemma}
\Proof We choose a multi-index~$\alpha$ of length~$a:=|\alpha|$
and a non-negative integer~$b$.
Differentiating the Dirac equation~\eqref{Dout} 
with respect to the mass parameter and to the spatial variables gives
\beq (\cI \Pdd + \B - m) \:\nabla^\alpha \partial_m^b \psi_m = b \,\nabla^\alpha \partial_m^{b-1}
\psi_m - \nabla^\alpha \big(\B \,\partial_m^b \psi_m \big)
+ \B \,\nabla^\alpha \partial_m^b \psi_m \:. \eeq
Introducing the abbreviations
\beq \Xi := \nabla^\alpha \partial_m^b \psi_m \qquad \text{and} \qquad
\phi := b\, \nabla^\alpha \partial_m^{b-1} \psi_m - \nabla^\alpha \big(\B \,\partial_m^b \psi_m \big)
+ \B \,\nabla^\alpha \partial_m^b \psi_m \:, \eeq
we rewrite this equation as the inhomogeneous Dirac equation
\beq (\Dir - m) \,\Xi = \phi \:. \eeq
A calculation similar to current conservation yields
\beq -\cI\partial_j \Sl \Xi | \gamma^j \Xi \Sr = \Sl (\Dir -m) \Xi \,|\, \Xi \Sr - \Sl \Xi \,|\, (\Dir -m) \Xi \Sr
= \Sl \phi | \Xi \Sr - \Sl \Xi | \phi \Sr \:. \eeq
Integrating over the equal time hypersurfaces and using the Schwarz inequality, we obtain
\beq \big| \partial_t \big( \Xi|_t \big| \Xi|_t \big)_t \big|
\leq 2\, \big\|\Xi|_t \big\|_t\: \big\|\phi|_t \big\|_t \eeq
and thus
\beq \Big| \partial_t \big\| \Xi|_t \big\| \Big| \leq \big\|\phi|_t \big\|_t \:. \eeq
Substituting the specific forms of~$\Xi$ and~$\phi$ and using the Schwarz and triangle
inequalities, we obtain the estimate
\beq \label{fundes}
\Big| \partial_t \big\| \nabla^\alpha \partial_m^b \psi_m|_t \big\|_t \Big|
\leq b\, \big\| \nabla^\alpha  \partial_m^{b-1} \psi_m|_t \big\|_t + c\, a\,|\nabla \B(t)|_{C^{a-1}}\:
\big\| \partial_m^b \psi_m|_t\big\|_{W^{a-1, 2}}\:,
\eeq
where we used the notation~\eqref{defCk}.

We now proceed inductively in the maximal total order~$a+b$ of the derivatives.
In the case~$a=b=0$, the claim follows immediately from the unitarity of the time evolution.
In order to prove the induction step, we note that in~\eqref{fundes}, the
order of differentiation of the wave function on the right-hand side
is smaller than that on the left-hand side at least by one.
In the case~$a=0$ and~$b \geq 0$, the induction hypothesis yields the inequality
\beq \big| \partial_t \| \partial_m^b \psi_m|_t \| \big| \leq b\, \big\| \partial_m^{b-1} \psi_m|_t \big\| 
\leq b \,C\: \big(1+|t|^{b-1} \big)\: \sum_{p=0}^{b-1} \big\| \partial^p_m \psi_m|_{t=0} \big\| \:, \eeq
and integrating this inequality from~$0$ to~$t$ gives the result.
In the case~$a>0$ and~$b \geq 0$, we apply~\eqref{Bdecay2}
together with the induction hypothesis to obtain
\begin{align}
\Big| \partial_t \big\| \partial_m^b \psi_m|_t \big\|_{W^{a,2}} \Big|
\leq\;& b \,C\: \big(1+|t|^{b-1} \big)\: \sum_{p=0}^{b-1} \big\| \partial^p_m \psi_m|_{t=0} \big\|_{W^{a,2}} \\
&+ c\,C\: \frac{1+|t|^b}{1+|t|^{1+\varepsilon}}\:
\sum_{p=0}^b \big\| \partial_m^p \psi_m|_{t=0} \big\|_{W^{a-1, 2}}\:.
\end{align}
Again integrating over~$t$ gives the result.
\QED

\section{The Cauchy Problem in Globally Hyperbolic Spacetimes} \label{seccauchyglobhyp}
\sindex{Cauchy problem!in globally hyperbolic spacetime}%
We conclude this chapter by extending the global existence and uniqueness result
for the Dirac equation to curved spacetime. These results were already stated
in Section~\ref{secdirgh}. We are now in the position for giving the proof.
The reader not interested in or not familiar with curved spacetime may skip this section.
We note that more details on the geometric properties of globally hyperbolic spacetimes
can be found in~\cite[Section~3.2]{baer+ginoux}.

\Proof[Proof of Theorem~\ref{thmcauchy}.]
Exactly as explained in the proof of Theorem~\ref{thmexistsdiracmink},
by considering the Cauchy problem for~$\psi-\psi_0$ one may reduce the problem
to that of zero initial data zero. Moreover, choosing a partition of unity~$(\eta_k)$
subordinate to the charts of a given atlas, it suffices to consider the
compactly supported inhomogeneity~$\eta_k \phi$
(the sum over~$k$ is again locally finite, similar as explained in the proof of Theorem~\ref{thmexistsdiracmink}).
In view of these constructions, it remains to consider the Cauchy problem
\beq \label{Dircauchycompact}
(\Dir - m) \psi = \phi \in C^\infty_0(\scrM, S\scrM) \:, \qquad \psi|_{\scrN_{t_0}} = 0 \:.
\eeq
We denote the support of~$\phi$ by~$K$.

Clearly, in local charts, the Dirac equation can be written as a symmetric hyperbolic system.
Therefore, the results in Sections~\ref{sec31} and~\ref{secshsexist} yield the existence
and uniqueness of solutions of the Cauchy problem in local charts.
This also yields global uniqueness: Let~$\psi$ and~$\tilde{\psi}$ be two smooth solutions
to the Cauchy problem~\eqref{Dircauchycompact}. Then their difference~$\Xi := \tilde{\psi}-\psi$
is a homogeneous solution that vanishes at time~$t_0$.
In view of a possible time reversal, it suffices to consider the solution in the future of~$t_0$.
Thus let~$x \in \scrM$ be in the future of~$t_0$. Then the past light cone~$J^\wedge(x)$
intersects the future of~$t_0$ in a compact set,
\beq J^\wedge(x) \cap \Big( \bigcup\nolimits_{t \geq t_0} \scrN_t \Big) \qquad \text{is compact} \:. \eeq
Therefore, we can choose~$\delta>0$ such that for every~$\hat{t}$, there is a finite number of
lens-shaped regions which cover the time strip
\beq J^\wedge(x) \cap \Big( \bigcup\nolimits_{t=\hat{t}}^{\hat{t}+\delta} \scrN_t \Big) \:. \eeq
On each lens-shaped regions, the solution for the Cauchy problem with zero initial data
vanishes identically. Therefore, we can proceed inductively to conclude that~$\Xi(x)=0$.
Since~$x$ is arbitrary, the solution~$\Xi$ vanishes identically on~$\scrM$.

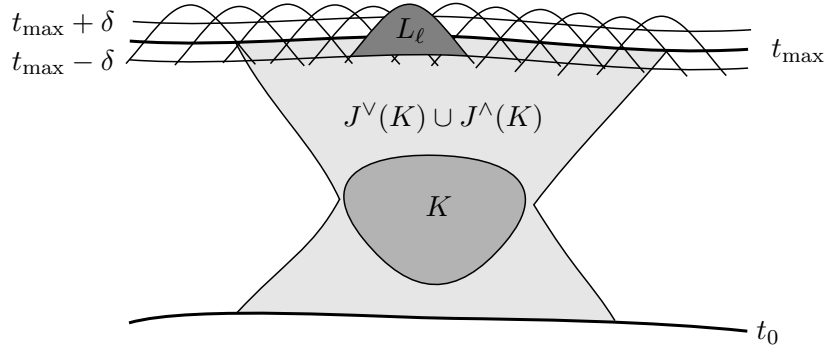
\begin{figure}[tb]
\psscalebox{1.0 1.0} % Change this value to rescale the drawing.
{
\begin{pspicture}(0,27.435057)(8.338042,31.844944)
\definecolor{colour0}{rgb}{0.9019608,0.9019608,0.9019608}
\definecolor{colour1}{rgb}{0.7019608,0.7019608,0.7019608}
\definecolor{colour2}{rgb}{0.5019608,0.5019608,0.5019608}
\pspolygon[linecolor=colour0, linewidth=0.02, fillstyle=solid,fillcolor=colour0](1.4918859,31.298277)(2.162997,31.324944)(3.5007746,31.38272)(4.474108,31.367167)(5.274108,31.322721)(6.6185527,31.222721)(7.2007747,31.196054)(6.814108,30.7605)(6.271886,30.151611)(5.6452193,29.43161)(5.4274416,29.151611)(5.751886,28.633833)(6.2385526,27.99161)(6.522997,27.587166)(5.367441,27.636055)(4.4185524,27.642721)(3.2185526,27.682722)(2.342997,27.709389)(1.5007747,27.722721)(2.0163302,28.216055)(2.4585526,28.633833)(2.734108,28.9805)(2.8718858,29.220499)(2.6896636,29.553833)(2.1763303,30.298277)(1.8296636,30.733833)(1.5585525,31.169388)
\psbezier[linecolor=black, linewidth=0.04](0.057441406,27.574944)(0.88899577,27.81652)(3.3675308,27.648325)(4.367441,27.63494384765625)(5.367352,27.621563)(7.013761,27.599789)(8.277441,27.464943)
\psbezier[linecolor=black, linewidth=0.02, fillstyle=solid,fillcolor=colour1](4.117441,29.804943)(3.1176677,29.826216)(2.547858,29.461876)(3.1774414,28.68494384765625)(3.807025,27.90801)(4.4145703,27.87897)(5.057441,28.644943)(5.7003126,29.410917)(5.117215,29.783672)(4.117441,29.804943)
\psbezier[linecolor=black, linewidth=0.02](2.8607748,29.228277)(2.5007746,28.558277)(2.1174414,28.384943)(1.4874414,27.70494384765625)
\psbezier[linecolor=black, linewidth=0.02](5.4396634,29.146055)(6.0096636,28.216055)(6.1874413,28.134943)(6.5174413,27.60494384765625)
\psbezier[linecolor=black, linewidth=0.04](0.067441404,31.324944)(1.4189957,31.22652)(3.127531,31.438324)(4.3774414,31.38494384765625)(5.6273518,31.331564)(6.873761,31.129787)(8.287441,31.214943)
\psbezier[linecolor=black, linewidth=0.02](5.4341083,29.1405)(6.014108,29.9105)(6.6874413,30.554943)(7.2174416,31.20494384765625)
\psbezier[linecolor=black, linewidth=0.02](2.8585525,29.213833)(2.5285525,29.943832)(1.8852192,30.484943)(1.4652191,31.324943847656247)
\psbezier[linecolor=black, linewidth=0.02](0.1074414,31.584944)(1.4589958,31.486519)(3.167531,31.698324)(4.4174414,31.64494384765625)(5.6673517,31.591562)(6.913761,31.389788)(8.327441,31.474943)
\psbezier[linecolor=black, linewidth=0.02](1.9374413,31.024944)(2.2702944,31.395687)(2.5074415,31.804943)(2.7974415,31.80494384765625)(3.0874414,31.804943)(3.3797596,31.316)(3.6574414,31.014944)
\psbezier[linecolor=black, linewidth=0.02](2.4474413,31.004944)(2.7802944,31.375687)(3.0174415,31.784945)(3.3074415,31.78494384765625)(3.5974414,31.784945)(3.8897595,31.296)(4.1674414,30.994944)
\psbezier[linecolor=black, linewidth=0.02](4.077441,30.994944)(4.4102945,31.365686)(4.6474414,31.774944)(4.9374413,31.77494384765625)(5.2274413,31.774944)(5.5197597,31.286)(5.7974415,30.984943)
\psbezier[linecolor=black, linewidth=0.02](3.5474415,31.044945)(3.8802943,31.415686)(4.117441,31.824944)(4.4074416,31.82494384765625)(4.6974416,31.824944)(4.9897594,31.335999)(5.2674413,31.034945)
\psbezier[linecolor=black, linewidth=0.02](1.3074414,31.034945)(1.6402944,31.405685)(1.8774414,31.814943)(2.1674414,31.81494384765625)(2.4574413,31.814943)(2.7497597,31.325998)(3.0274415,31.024944)
\psbezier[linecolor=black, linewidth=0.02](0.6674414,31.004944)(1.0002944,31.375687)(1.2374414,31.784945)(1.5274414,31.78494384765625)(1.8174415,31.784945)(2.1097596,31.296)(2.3874414,30.994944)
\psbezier[linecolor=black, linewidth=0.02](5.1874413,30.914944)(5.520294,31.285686)(5.7574415,31.694944)(6.0474415,31.69494384765625)(6.3374414,31.694944)(6.62976,31.206)(6.9074416,30.904943)
\psbezier[linecolor=black, linewidth=0.02](4.6674414,30.964943)(5.000294,31.335686)(5.2374415,31.744944)(5.5274415,31.74494384765625)(5.8174415,31.744944)(6.10976,31.255999)(6.3874416,30.954945)
\psbezier[linecolor=black, linewidth=0.02](0.017441407,31.024944)(0.3502944,31.395687)(0.5874414,31.804943)(0.8774414,31.80494384765625)(1.1674414,31.804943)(1.4597597,31.316)(1.7374414,31.014944)
\psbezier[linecolor=black, linewidth=0.02](6.2774415,30.874945)(6.6102943,31.245686)(6.847441,31.654943)(7.1374416,31.65494384765625)(7.4274416,31.654943)(7.7197595,31.165998)(7.9974413,30.864944)
\psbezier[linecolor=black, linewidth=0.02](5.7574415,30.864944)(6.0902944,31.235685)(6.327441,31.644943)(6.617441,31.64494384765625)(6.9074416,31.644943)(7.1997595,31.155998)(7.4774413,30.854944)
\pspolygon[linecolor=colour2, linewidth=0.02, fillstyle=solid,fillcolor=colour2](3.0099413,31.124945)(3.1799414,31.334944)(3.4049413,31.584944)(3.5549414,31.724943)(3.6949415,31.789944)(3.8249414,31.804943)(4.014941,31.724943)(4.1899414,31.564943)(4.3299413,31.394943)(4.4699416,31.219944)(4.5499415,31.129944)(4.159941,31.159945)(3.7249415,31.149944)(3.3299415,31.139944)
\psbezier[linecolor=black, linewidth=0.02](2.9274414,31.034945)(3.2602944,31.405685)(3.4974413,31.814943)(3.7874415,31.81494384765625)(4.077441,31.814943)(4.3697596,31.325998)(4.6474414,31.024944)
\psbezier[linecolor=black, linewidth=0.02](0.067441404,31.074944)(1.4189957,30.97652)(3.127531,31.188324)(4.3774414,31.13494384765625)(5.6273518,31.081564)(6.873761,30.879787)(8.287441,30.964943)
\rput[bl](8.4,27.3){$t_0$}
\rput[bl](8.6,31.05){$t_{\max}$}
\rput[bl](-1.5,30.88){$t_{\max}-\delta$}
\rput[bl](-1.5,31.4){$t_{\max}+\delta$}
\rput[bl](4,29){$K$}
\rput[bl](2.9,30.1){$J^\vee(K) \cup J^\wedge(K)$}
\rput[bl](3.6,31.3){$L_\ell$}
\end{pspicture}
}
\caption{Global solutions in globally hyperbolic spacetimes.}
\label{figglobhyp}
\end{figure}

In order to prove global existence, we proceed indirectly.
In view of a possible time reversal, it suffices to consider the Cauchy problem to the future.
Thus suppose that the solution exists only up to finite time~$t_{\max}$ (see Figure~\ref{figglobhyp}).
Due to finite propagation speed, the solution is supported in the domain of causal dependence
of~$K$,
\beq \supp \psi \subset J^\vee(K) \cup J^\wedge(K) \:. \eeq
By properties of globally hyperbolic spacetimes, the intersection~$D$ of this set with the Cauchy surface~$\scrN_{t_{\max}}$ is compact. Covering~$D$ by a finite number of charts,
we choose~$\delta$ such that the sets~$J^\vee(D) \cup J^\wedge(D) \cap \scrN_t$
lie in the domain of these charts for all~$t \in [t_{\max}-\delta, t_{\max}+\delta]$.
Next we choose a finite number of lens-shaped regions~$L_\ell$ which also cover all these sets
(see again Figure~\ref{figglobhyp}).
In each of these lens-shaped regions we can solve the Cauchy problem with initial data
at time~$t_{\max}-\delta$. In this way, we get a solution up to time~$t_{\max}+\delta$.
This is a contradiction, thereby proving that the solution must exist for all times.
\QED

\Proof[Proof of Theorem~\ref{thmcauchy2}.]
By extending the initial data~$\psi_0$ to a smooth and compactly supported function in spacetime
and considering the Cauchy problem for~$\psi-\psi_0$, it again suffices to consider the case of
zero initial data~\eqref{Dircauchycompact}.
The solution constructed subsequently the proof of Theorem~\ref{thmcauchy}
was supported in~$J^\vee(K) \subset J^\wedge(K)$.
By general properties of globally hyperbolic manifolds, the intersection of this set with every Cauchy surface
is compact. This concludes the proof.
\QED

Similar as explained in Section~\ref{secexgreen} in Minkowski space, 
also in curved spacetime the solution of the Cauchy problem can be expressed
in terms of the causal fundamental solution~$k_m$, as we now explain.
Similar as explained in Minkowski space in Section~\ref{secexgreen}, the
{\em{retarded}} and {\em{advanced Green's operators}}~$s_m^\wedge$ and~$s_m^\vee$ are
linear mappings (for details see, for example, \cite{dimock3, baer+ginoux})
\sindex{Green's operator!causal in globally hyperbolic spacetime}%
\beq s_m^\wedge, s_m^\vee \::\: C^\infty_0(\scrM, S\scrM) \rightarrow \Cisc(\scrM, S\scrM)\:. \eeq
They satisfy the defining equation of the Green's operator
\beq %\label{Greendefdir}
(\Dir - m) \left( s_m^{\wedge, \vee} \phi \right) = \phi \:. \eeq
Moreover, they are uniquely determined by the condition that the support of~$s_m^\wedge \phi$
(or~$s_m^\vee \phi$) lies in the future (respectively the past) of~$\supp \phi$.
The {\em{causal fundamental solution}}~$k_m$ is introduced by
\sindex{causal fundamental solution!in globally hyperbolic spacetime}%
\beq \label{kmdef}
k_m := \frac{1}{2 \pi \cI} \left( s_m^\vee - s_m^\wedge \right) \::\: C^\infty_0(\scrM, S\scrM) \rightarrow \Cisc(\scrM, S\scrM)
\cap \H_m \:.
\eeq
Note that it maps to solutions of the Dirac equation.

\begin{Prp} \label{prp21}
The solution of the Cauchy problem~\eqref{thmcauchy} has the representation
\beq %\label{lemmacauchyk-psi}
 \psi(x) = 2 \pi \int_\scrN k_m(x,y)\, \nuslsh\, \psi_\scrN(y)\: \Diff\mu_\scrN(y)\:, \eeq
where~$k_m(x,y)$ is the causal fundamental solution~\eqref{kmdef}.
\end{Prp}
\Proof Let us consider a point~$x$ in the future of~$\scrN$ (the case for the past is analogous). In this case, due to~\eqref{kmdef}, the lemma simplifies to
\beq
\label{lemmacauchyk-simplified}
\psi(x) = \cI \int_\scrN s^\wedge_m(x,y)\, \nuslsh(y)\, \psi_\scrN(y)\: \Diff\mu_\scrN(y)\:. 
\eeq

In preparation, we want to prove that for any~$\phi \in C^\infty(\scrM, S\scrM)$
which has compact support to the past of~$\scrN$ and with the property
that~$(\Dir-m) \phi$ has compact support the equation
\beq \label{phisformula}
\phi = s^\wedge_m \big( (\Dir-m) \phi \big)
\eeq
holds. To this end, we consider the function
\beq \Xi(x) := \phi - s^\wedge_m \big( (\Dir-m) \phi \big) \:. \eeq
Applying the operator~$(\Dir-m)$ and using the defining equation of the
Green's operators, one sees that~$\Xi$ is a solution of the Dirac equation.
Moreover, $\Xi$ obviously vanishes in the past of the support of~$\phi$.
The uniqueness of the solution of the Cauchy problem implies that~$\Xi$
vanishes identically, proving~\eqref{phisformula}.

In order to derive equation~\eqref{lemmacauchyk-simplified},
we let~$\eta \in C^\infty(\scrM)$ be a function which is identically equal
to one at~$x$ and on~$\scrN$, but such that the function~$\eta \psi$ has compact support to the past.
(For example, in a foliation~$(\scrN_t)_{t \in \R}$ with~$\scrN=\scrN_{t_0}$
one can take~$\eta=\eta(t)$ as a smooth function
with~$\eta|_{[t_0, \infty)} \equiv 1$ which vanishes if~$t<t_0-1$).
Then we can apply~\eqref{phisformula} to the wave function~$\phi = \eta \psi$.
We thus obtain for any~$x$ in the future of~$\scrN$ the relations
\beq \label{seta}
\psi(x) = (\eta \psi)(x) = \Big( s^\wedge_m \big( (\Dir - m) (\eta \psi) \big) \Big)(x)
= \Big( s^\wedge_m \big( \cI \gamma^j (\partial_j \eta)\, \psi \big) \Big)(x) \:,
\eeq
where we have used that~$\psi$ is a solution of the Dirac equation.

To conclude the proof, for~$\eta$ in~\eqref{seta} we choose a
sequence~$\eta_\ell$ which converges in the distributional sense to the
function which in the future~$\scrN$ is equal to one and in the past of~$\scrN$ is equal to zero.
This yields~$\partial_j\eta_\ell \rightarrow \nu$, and thus the right-hand side of~\eqref{seta} 
goes over to the right-hand side of~\eqref{lemmacauchyk-simplified}.
\QED

\section{Exercises}
\begin{Exercise} \label{exmaxwell}  {\em{ The {\em{homogeneous Maxwell equations}} for the
\sindex{Maxwell equations!as symmetric hyperbolic system}%
electric field~$E:\R^3\rightarrow\R^3$ and the magnetic field~$B :\R^3\rightarrow\R^3$ read
\beq \nabla\times B = \partial_t E\,,\quad \nabla\times E=-\partial_t B \:, \eeq
where~$\times$ denotes the cross product in~$\R^3$.
Rewrite these equations as a symmetric hyperbolic system. 
{\em{Remark:}} We here ignore the equations~$\text{div}\, E = \text{div}\, B = 0$.
The reason is that these equations hold automatically if they are satisfied initially.
}} \end{Exercise}

\begin{Exercise} {\em{ Consider the scalar wave equation~$(\partial_{tt}-\Delta_{\R^m})\phi(t,x)=0$.
\sindex{scalar wave equation!as symmetric hyperbolic system}%
\bitem
\item[(a)] Rewrite the equation as a symmetric hyperbolic system
\beq A^0\partial_tu+ \sum_{\alpha=1}^m A^\alpha \nabla_\alpha u+Bu=0. \eeq
\item[(b)] Determine the timelike and future-directed directions. Which directions~$\xi$ are
characteristic (meaning that the characteristic polynomial~$\det A(x,\xi)$ vanishes)?
\item[(c)] Express the ``energy''
\beq E(t)=\int_{\R^m} \langle u,A^0u\rangle\,\dd^mx \eeq
in terms of~$\phi(t,x)$. Compare the resulting expression with the conserved physical energy
\beq \int_{\R^m} (|\partial_t\phi|^2+|\nabla\phi|^2)\,\dd^mx\,. \eeq
\item[(d)] Compute~$\frac{\dd E(t)}{\dd t}$. Prove the inequality
\beq \frac{\dd E(t)}{\dd t}\leq E(t) \eeq
and integrate it (Gr\"onwall's lemma).
\sindex{Gr\"onwall estimate}%
\eitem
}} \end{Exercise}

\begin{Exercise} {\em{ Consider the solution of the homogeneous wave equation
\beq(\partial_{tt}-\Delta_{\R^n})\phi(t,x)=0\eeq
for smooth initial data~$\phi(0,x)=f(x)$ and~$\partial_t\phi(0,x)=g(x)$.\\
Show by a suitable choice of lens-shaped regions that~$\phi(t_0,x_0)$ depends only 
on the initial data in the closed ball~$\{x\in\R^n\,:\,|x-x_0|\leq t_0\}$.
}} \end{Exercise}

\begin{Exercise} {\em{ We consider the system
\begin{align}
	\partial_tu_1(t,x)+\partial_xu_1(t,x)+4\partial_xu_2(t,x)&=0\\
	\partial_tu_2(t,x)+4\partial_xu_1(t,x)+\partial_xu_2(t,x)&=0
\end{align}
\bitem
\item[(a)] Write the system in symmetric hyperbolic form.
\item[(b)] Compute the solution of the Cauchy problem
for the initial data~$u_1(0,x)=\sin x$ and~$u_2(0,x)=\cos x$.
\eitem
}} \end{Exercise}

\begin{Exercise} (The Euler equations)
\sindex{Euler equations equations!as symmetric hyperbolic system}%
{\em{ The evolution equation for an isentropic compressible fluid reads
\begin{equation}
\begin{cases}\partial_t v+\nabla_vv+\frac{1}{\rho} \text{grad}(p)&=0\\
 \partial_t\rho+\nabla_v\rho+ \rho\,\mathrm{div}(v)&=0.\end{cases}
\label{first}
\end{equation}
Here~$v:\R^+\times\R^3\rightarrow\R^3$ is the velocity vector field,
$\rho:\R^+\times\R^3\rightarrow(0,\infty)$ the density and~$p=A\rho^\gamma$ the pressure (where~$A>0$ and~$\gamma>1$).
\bitem
\item[(a)] Show that~\eqref{first} is equivalent to a quasilinear symmetric hyperbolic system, provided that
$\rho$ is bounded away from zero.
\item[(b)] Show that for smooth solutions, the system~\eqref{first} is equivalent to
\begin{equation}
\begin{cases}\partial_t v+\nabla_vv+\text{grad}(h(\rho))&=0\\
 \partial_t\rho+\mathrm{div}(\rho v)&=0,\end{cases}
\label{eight}
\end{equation}
where~$h\in C^\infty(\R)$ satisfies the equation~$h'(\rho)=\rho^{-1}p'(\rho)$.
\item[(c)] Let~$(v,\rho)$ be a solution of~\eqref{eight} with~$v(t,x)=\nabla_x\varphi(t,x)$ for a real-valued
potential~$\varphi$. Prove \textit{Bernoulli's law}: If~$\varphi$ and~$\rho$ decay at infinity sufficiently
fast and if~$h(0)=0$, then
\beq %\label{3}
\partial_t\varphi+\frac{1}{2}|\nabla_x\varphi|^2+h(\rho)=0 \:. \eeq
\item[(d)] Show that~\eqref{first} can also be rewritten as a system for~$(p,v)$,
\beq %\label{nine}
\begin{cases}\partial_t v+\nabla_vv+\rho(p)^{-1}\text{grad}(p)&=0\\
 \partial_tp+\nabla_vp+(\gamma p)\,\mathrm{div}(v)&=0 \:. \end{cases} \eeq
Rewrite this system in symmetric hyperbolic form.
\eitem
}} \end{Exercise}

\begin{Exercise} {\em{ Let~$\lambda>0$. A symmetric hyperbolic system of the form
\beq\partial_t u+A^\alpha(u)\partial_\alpha u+\lambda u=0\,,\eeq
where the matrices~$A^\alpha$ are smooth, uniformly bounded and uniformly positive,
is an example of a so-called {\em{dissipative system}}.
\bitem
\item[(a)] Prove that for spatially compact solutions, the following energy estimate holds:
\beq\frac{\dd}{\dd t}\|u(t)\|_{H^p}^2\leq \Big(-2\lambda+c\|u(t)\|_{C^1}\Big)\|u(t)\|_{H^p}^2\,.\eeq
\item[(b)] Prove: If the initial data~$u_0$ is sufficiently small in the $C^1$-norm, then there exists a global solution. \\ 
\textit{Hint:} Choose~$p$ sufficiently large and use the Sobolev embedding theorem.
\eitem
}} \end{Exercise}

\begin{Exercise} {{(Causality in the setting of symmetric hyperbolic systems)}} {\em{
\sindex{timelike separation!for linear symmetric hyperbolic system}%
\sindex{lightlike separation!for linear symmetric hyperbolic system}%
\sindex{spacelike separation!for linear symmetric hyperbolic system}%
The Dirac equation~$(\cI\slashed{\partial}-m)\psi=0$  can be rewritten as a \emph{symmetric hyperbolic system},
that is, in the form~$(c>0)$
\beq
(A^0(x)\,\partial_0 + A^\alpha(x)\,\partial_\alpha + B(x))\psi = 0,\quad\mbox{with }\ (A^i)^\dagger = A^i\ \mbox{ and }\ A^0(x)\ge c\mathbb{I}. 
\eeq
For such systems a notion of \emph{causality} can be introduced: a vector~$\xi\in \R^4$ is said to be \emph{timelike} or \emph{lightlike} at~$x\in\R^4$, if the matrix~$A(x,\xi):=A^i(x)\,\xi_i$ is definite (either positive or negative) or singular, respectively. 

Find the matrices~$A^i$ and~$B$ for the Dirac equation and show that the above notions of timelike and lightlike vectors coincide with the corresponding notions in Minkowski space.
}} \end{Exercise}

\chapter{Energy Methods for the Linearized Field Equations} \label{seclinhyp}
In the previous chapter, we used energy methods in order to study the Cauchy problem
for linear symmetric hyperbolic systems.
We now briefly explain how these methods can be adapted to the
linearized field equations for causal variational principles as introduced in Chapter~\ref{secEL}.
These constructions are carried out in detail in~\cite{linhyp}; for later developments
see~\cite{dirac, localize}. Here we do not aim for the largest generality, but instead explain
the basic ideas in the simplest possible setting.
We also note that some of the constructions in this section will be illustrated
in Chapter~\ref{secexcvp} with simple concrete examples.

\section{Local Foliations by Surface Layers} \label{seclocfol}
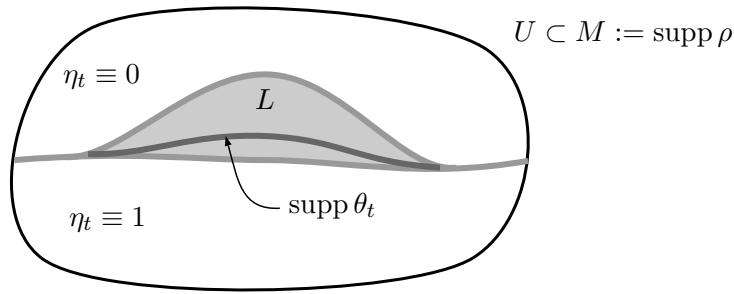
\begin{figure}
% \usepackage[usenames,dvipsnames]{pstricks}
% \usepackage{epsfig}
% \usepackage{pst-grad} % For gradients
% \usepackage{pst-plot} % For axes
% \usepackage[space]{grffile} % For spaces in paths
% \usepackage{etoolbox} % For spaces in paths
% \makeatletter % For spaces in paths
% \patchcmd\Gread@eps{\@inputcheck#1 }{\@inputcheck"#1"\relax}{}{}
% \makeatother
% 
\psscalebox{1.0 1.0} % Change this value to rescale the drawing.
{
\begin{pspicture}(-2.5,-1.9044089)(14.139044,1.9044089)
\definecolor{colour1}{rgb}{0.8,0.8,0.8}
\definecolor{colour0}{rgb}{0.6,0.6,0.6}
\definecolor{colour2}{rgb}{0.4,0.4,0.4}
\pspolygon[linecolor=colour1, linewidth=0.02, fillstyle=solid,fillcolor=colour1](1.1190436,-0.09187103)(1.7090436,-0.101871036)(2.3290436,-0.09187103)(3.1790435,-0.13187103)(3.8890436,-0.15187103)(4.4990435,-0.23187104)(5.1490436,-0.23187104)(5.7290435,-0.28187102)(5.3790436,-0.09187103)(4.8790436,0.23812896)(4.3590436,0.61812896)(3.7590437,0.92812896)(3.2690437,0.97812897)(2.7290435,0.81812894)(2.1090436,0.48812896)
\rput[bl](6.7190437,1.358129){\normalsize{$U \subset M := \supp \rho$}}
\psbezier[linecolor=black, linewidth=0.04](6.389044,1.298129)(5.585711,1.8936595)(2.0168202,2.0884824)(1.0690436,1.5881289672851562)(0.12126687,1.0877756)(-0.35428908,-0.8663405)(0.44904357,-1.461871)(1.2523762,-2.0574017)(5.331267,-1.9822243)(6.179044,-1.451871)(7.02682,-0.92151767)(7.192376,0.70259845)(6.389044,1.298129)
\psbezier[linecolor=colour0, linewidth=0.08](0.07904358,-0.15187103)(1.6590469,-0.024428569)(2.4390447,-0.15040691)(3.4390435,-0.15187103271484376)(4.4390426,-0.15333515)(5.51902,-0.4087202)(6.909044,-0.16187103)
\rput[bl](0.7290436,0.79812896){\normalsize{$\eta_t \equiv 0$}}
\rput[bl](0.8090436,-1.101871){\normalsize{$\eta_t \equiv 1$}}
\psbezier[linecolor=colour0, linewidth=0.08](0.7390436,-0.111871034)(1.4801531,-0.16686673)(2.449648,1.0228906)(3.4490435,0.9881289672851562)(4.448439,0.95336735)(5.2490635,-0.3373064)(5.9590435,-0.25187102)
\psbezier[linecolor=colour2, linewidth=0.08](1.0590435,-0.071871035)(1.870874,-0.09410611)(2.4199824,0.2114487)(3.4190435,0.16812896728515625)(4.4181046,0.124809235)(4.629186,-0.1793354)(5.7390437,-0.25187102)
\rput[bl](3.7290435,-0.93187106){\normalsize{$\supp \theta_t$}}
\psbezier[linecolor=black, linewidth=0.02, arrowsize=0.05291667cm 2.0,arrowlength=1.4,arrowinset=0.0]{->}(3.6090436,-0.791871)(3.0911489,-0.8228388)(3.0490437,-0.41154847)(2.8890436,0.16812896728515625)
\rput[bl](3.2790437,0.528129){\normalsize{$L$}}
\end{pspicture}
}
\caption{A local foliation.}
\label{figlocfol}
\end{figure}

\sindex{local foliation!for causal variational principle}%
We consider {\em{causal variational principles in the compact setting}}
(see Section~\ref{secnoncompact}). Moreover, for technical simplicity, we again
restrict attention to the {\em{smooth setting}} by assuming that
the Lagrangian is smooth~\eqref{Lsmooth}.
\sindex{causal variational principle!in the compact setting}%
\sindex{causal variational principle!in the smooth setting}%
Following our procedure for symmetric hyperbolic systems, we want to analyze
the initial problem ``locally'' in an open subset~$U$ of spacetime~$M$.
In analogy to the time function in a lens-shaped region~$L$ (see Section~\ref{sec31}),
we here choose a foliation of a compact subset~$L \subset U$ by surface layers.
This motivates the following definition.
\begin{Def} \label{deflocfoliate}
Let~$U \subset M$ be an open subset of spacetime and~$I:= [t_{\min}, t_{\max}]$ a compact interval.
Moreover, we let~$\eta \in C^\infty(I \times U, \R)$ be a function with~$0 \leq \eta \leq 1$ which for
all~$t \in I$ has the following properties:
\bitem
\item[{\rm{(i)}}] The function~$\theta(t,.) := \partial_t \eta(t,.)$ is non-negative and compactly supported in~$U$.
\item[{\rm{(ii)}}] For all~$x \in \supp \theta(t,.)$ and all~$y \in M \setminus U$,
the function~$\L(x,y)$ as well as its first and second derivatives vanish.
\eitem
We also write~$\eta(t,x)$ as~$\eta_t(x)$ and~$\theta(t,x)$ as~$\theta_t(x)$.
We refer to~$(\eta_t)_{t \in I}$ as a {\bf{local foliation}} inside~$U$.
\nindex{agh@$(\eta_t)_{t \in I}$ -- local foliation inside~$U \subset M$}%
\end{Def} \noindent
The situation in mind is shown in Figure~\ref{figlocfol}.
The parameter~$t$ can be thought of as the time of a local observer
and will be referred to simply as {\em{time}}.
The support of the function~$\theta_t$ is a {\em{surface layer}}.
The function~$\eta_t$ should be thought of as being equal to one in the past
and equal to zero in the future of this surface layer.
The condition~(i) implies that the set~$L$ defined by
\beq \label{Ldef}
L := \bigcup_{t \in I} \supp \theta_t
\eeq
is compact. It is the region of spacetime described by the local foliation.
The condition~(ii) has the purpose of ensuring that the dynamics in the region~$L$
does not depend on the jets outside~$U$, making it possible to restrict
attention to the spacetime region~$U$.
Sometimes, we refer to this property that~$L$ is {\em{$\L$-localized}} in~$U$.
One way of satisfying~(ii) is to simply choose~$U=M$.
However, in the applications, it may be desirable to ``localize'' the problem,
for example, by choosing~$U$ as the domain of a coordinate chart.

Following the procedure for hyperbolic partial differential equations, our first
goal is to analyze the initial value problem.
For the following constructions, it will be useful to combine the
functions~$\eta_t$ and~$\theta_t$ with the measure~$\rho$ such as to form new measures:
The measure
\beq
\Diff\rho_t(x) := \theta_t(x)\: \Diff\rho(x) \label{rhot}
\eeq
with~$t \in I$ is supported in the surface layer at time~$t$. Likewise, the measures
\beq \eta_t\: \Diff\rho \qquad \text{and} \qquad 
\big(1-\eta_t)\: \Diff\rho \eeq
are supported in the past and future of the surface layer at time~$t$, respectively.
For the measures supported in
a spacetime strip, we use the notation
\beq \label{etabracket}
\eta_{[t_0, t_1]}\: \Diff\rho \qquad \text{with} \qquad 
\eta_{[t_0,t_1]} := \eta_{t_1} - \eta_{t_0} \in C^\infty_0(U) \:,
\eeq
where we always choose~$t_0,t_1 \in I$ with~$t_0 \leq t_1$.
Note that the function~$\eta_{[t_0, t_1]}$ is supported in~$L$.

\section{Energy Estimates and Hyperbolicity Conditions}
\sindex{energy estimate!for causal variational principle|textbf}%
\sindex{hyperbolicity condition!for causal variational principle}%
For the analysis of the linearized field equations, it is helpful to study the surface layer
integrals as introduced in Section~\ref{SecSmooth} for our local foliation~$(\eta_t)_{t \in I}$.
It is useful to ``soften'' these surface layer integrals by 
rewriting the integration domains with characteristic functions
and replacing the characteristic functions with smooth cutoff functions formed of~$\eta_t$; that is, symbolically
\begin{align}
\int_\Omega \Diff\rho(x) \int_{M \setminus \Omega} \Diff\rho(y) \cdots
&= \int_M \Diff\rho(x) \int_M \Diff\rho(y)\: \chi_\Omega(x)\: \big( 1- \chi_\Omega(y) \big) \cdots \notag  \\
& \longrightarrow \int_M \Diff\rho(x) \int_M \Diff\rho(y)\: \eta_t(x)\: \big( 1- \eta_t(y) \big) \cdots \:.
\end{align}
We thus define the {\em{softened symplectic form}} and the {\em{softened surface layer
inner product}} by
\sindex{symplectic form!softened|textbf}%
\sindex{surface layer inner product!softened}%
\nindex{agi@$(.,.)^t$ -- softened surface layer inner product}%
\tindex{bb@$(.,.)^t$ -- softened surface layer inner product}%
\nindex{agj@$\sigma^t(.,.)$ -- softened symplectic form}%
\tindex{bb@$\sigma^t(.,.)$ -- softened symplectic form}%
\begin{align}
(\u, \v)^t &= \int_U \Diff\rho(x)\: \eta_t(x) \int_U \Diff\rho(y)\: \big(1-\eta_t(y)\big)
\: \Big( \nabla_{1,\u} \nabla_{1,\v} - \nabla_{2,\u} \nabla_{2,\v} \Big) \L(x,y) \label{jipdef} \\
\sigma^t(\u,\v) &= \int_U \Diff\rho(x)\: \eta_t(x) \int_U \Diff\rho(y)\: \big(1-\eta_t(y)\big)
\: \Big( \nabla_{1,\u} \nabla_{2,\v} - \nabla_{1,\v} \nabla_{2,\u} \Big) \L(x,y) \label{sympdef} \:.
\end{align}

The quantity~$(\u, \u)^t$ is of central importance for the following constructions,
because it will play the role of the energy used in our energy estimates.
In preparation of these estimates, we derive an energy identity:
\begin{Lemma} {\bf{(energy identity)}} \label{lemmaenid}
For any jet~$\u =(a,u) \in \J$,
\beq \begin{split}
\frac{\dd}{\dd t}\: (\u, \u)^t 
&= 2 \int_U \la \u, \Delta \u \ra(x) \: \Diff\rho_t(x) \\
&\quad\: -2 \int_U \Delta_2[\u, \u] \: \Diff\rho_t(x)
+ \s \int_U a(x)^2\:\Diff\rho_t(x) \:,
\end{split} \label{enid}
\eeq
where the operator~$\Delta_2 \::\: \J \times \J \rightarrow \J^*$ defined by
\begin{align}
\big\la &\u, \Delta_2[\u_1, \u_2] \big\ra(x) \notag \\
&= \frac{1}{2} \:\nabla_{\u} \bigg( \int_{M} \big( \nabla_{1, \u_1} + \nabla_{2, \u_1} \big) \big( \nabla_{1, \u_2} + \nabla_{2, \u_2} \big) \L(x,y)\: \Diff\rho(y)\
-\nabla_{\u_1} \nabla_{\u_2}\, \s\bigg) \:. \label{Lap2def}
\end{align}
\end{Lemma}
\Proof Differentiating~\eqref{jipdef} with respect to~$t$ gives
\begin{align}
\frac{\dd}{\dd t}\: (\u, \u)^t
&= \int_U \Diff\rho(x)\: \theta_t(x) \int_U \Diff\rho(y)\: \big(1-\eta_t(y)\big)
\: \big( \nabla^2_{1,\u} - \nabla^2_{2,\u} \big) \L(x,y) \notag \\
&\qquad -\int_U \Diff\rho(x)\: \eta_t(x) \int_U \Diff\rho(y)\: \theta_t(y) 
\: \big( \nabla^2_{1,\u} - \nabla^2_{2,\u} \big) \L(x,y) \notag \\
&= \int_U \Diff\rho(x)\: \theta_t(x) \int_U \Diff\rho(y)\: \big( \nabla^2_{1,\u} - \nabla^2_{2,\u} \big) \L(x,y) \:. \label{vvt}
\end{align}
Next, for all~$x \in L$ we may use Definition~\ref{deflocfoliate}~(ii) to change the integration range
in~\eqref{linfield} from~$M$ to~$U$,
\beq \la \u, \Delta \u \ra(x) = \int_U \nabla_{1,\u} \big( \nabla_{1,\u} + \nabla_{2,\u} \big)\: \L(x,y)\: \Diff\rho(y)
- \s\:a(x)^2 \:. \eeq
Multiplying by~$\theta_t$ and integrating, we obtain
\begin{align}
0&= \int_U \theta_t(x)\:\la \u, \Delta \u \ra(x) \: \Diff\rho(x) + \s \int_U \theta_t(x) \: a(x)^2\:\Diff\rho(x) \notag \\
&\quad\:
-\int_U \Diff\rho(x)\: \theta_t(x)\: \int_U \Diff\rho(y)\: \big( \nabla_{1,\u}^2 + \nabla_{1,\u} \nabla_{2,\u} \big)\: \L(x,y) \:.
\end{align}
We multiply this equation by two and add~\eqref{vvt}. This gives
\begin{align}
\frac{\dd}{\dd t}\: (\u, \u)^t 
&= -\int_U \Diff\rho(x)\: \theta_t(x) \int_U \Diff\rho(y)\: \big( \nabla_{1,\u} + \nabla_{2,\u} \big)^2 \L(x,y) \notag \\
&\quad\: +2 \int_U \theta_t(x)\:\la \u, \Delta \u \ra(x) \: \Diff\rho(x) + 2 \s \int_U \theta_t(x) \: a(x)^2\:\Diff\rho(x) \:.
\end{align}
Using the property in Definition~\ref{deflocfoliate}~(ii), in the $y$-integral we may
replace the integration range~$U$ by~$M$, making it possible to apply~\eqref{Lap2def}.
Rewriting the obtained integrals using the notation~\eqref{rhot} gives~\eqref{enid}.
\QED

In order to make use of this energy identity, we need to impose a condition
which we call hyperbolicity condition. 
This notion can be understood as follows.
As explained in Chapter~\ref{secshs}, in the theory of hyperbolic partial differential equations,
the hyperbolicity of the equations gives rise to a positive energy
(see~$E(\lambda)$ in~\eqref{2L}). The positivity of this energy was a consequence of the
structure of the equations (more precisely, for symmetric linear hyperbolic systems, it is
a consequence of the positivity statement in Definition~\ref{defshs}~(ii)).
The positivity of the energy is crucial for the analysis of hyperbolic equations
because it gives both uniqueness (see Section~\ref{sec31}) and existence of weak solutions
(see Section~\ref{secshsexist}). With this in mind, our strategy is to express the
hyperbolic nature of the linearized field equations by imposing a positivity condition
for our ``energy''~$(\u, \u)^t$.
As we shall see, this so-called hyperbolicity condition is precisely what is needed in order to
obtain both existence and uniqueness of solutions.
For Dirac systems in Minkowski space, the hyperbolicity conditions can be verified by direct computation
(for details, see~\cite{action}). With this in mind, our hyperbolicity conditions are physically sensible.
But in most situations, imposing the hyperbolicity conditions for all jets in~$\J$ is a too strong assumption.
Instead, these conditions will or can be satisfied only on a suitably chosen subspace of jets,
which we denote by
\nindex{agk@$\Jvary$ -- space of jets used for varying the measure}%
\tindex{ff@$\Jvary$ -- space of jets used for varying the measure}%
\beq \label{Jvarydef}
\Jvary \subset \J \:.
\eeq
Clearly, the smaller the jet space~$\Jvary$ is chosen, the easier it is to satisfy~\eqref{hypcond}.
The drawback is that the Cauchy problem will be solved in a weaker sense.

In order to define the hyperbolicity conditions,
for all~$x \in M$, we choose the subspace of the tangent space spanned by the test jets,
\beq \Gamma_x := \big\{ u(x) \:|\: u \in \Gtest \big\} \;\subset\; T_x\F\:. \eeq
We introduce a Riemannian metric~$g_x$ on~$\Gamma_x$.
The choice of the Riemannian metric is arbitrary; the resulting freedom can be used in order to
satisfy the hyperbolicity conditions below
(note, however, that for causal fermion systems, a canonical Riemannian metric is
obtained from the Hilbert-Schmidt scalar product; see~\cite{gaugefix, banach}).
This Riemannian metric also induces a pointwise scalar product on the jets. Namely, setting
\beq \J_x := \R \oplus \Gamma_x \:, \eeq
we obtain the scalar product on~$\J_x$
\beq
\la .,. \ra_x \,:\, \J_x \times \J_x \rightarrow \R \:,\qquad
\la \u, \tilde{\u} \ra_x := a(x)\, \tilde{a}(x) + g_x \big(u(x),\tilde{u}(x) \big) \label{vsprod}
\eeq
\nindex{agl@$\J_x, \Gamma_x$ -- jets at spacetime point~$x \in M$}%
(where we again denote the scalar and vector components of the jet by~$\u=(a,u)$).
We denote the corresponding norm by~$\|.\|_x$.

\begin{Def} \label{defhypcond}
The local foliation~$(\eta_t)_{t \in I}$ inside~$U$
satisfies the {\bf{hyperbolicity condition}}
\sindex{hyperbolicity condition!for causal variational principle|textbf}%
if there is a constant~$C>0$ such that for all~$t \in I$,
\beq
(\u, \u)^t \geq \frac{1}{C^2} \int_U \Big( \|\u(x)\|_x^2\: + \big|\Delta_2[\u, \u]\big| \Big) \: \Diff\rho_t(x) 
\qquad \text{for all~$\u \in \Jvary$} \:. \label{hypcond}
\eeq
A compact set~$L \subset M$ is a {\bf{lens-shaped region}} inside~$U$
if there is a local foliation~$(\eta_t)_{t \in I}$ inside~$U$
satisfying~\eqref{Ldef} which satisfies the hyperbolicity conditions.
\end{Def}
\sindex{lens-shaped region!for causal variational principle}
\noindent
We point out that these hyperbolicity conditions also pose constraints for the choice of the functions~$\eta_t$;
these constraints can be understood as replacing the condition in the theory of hyperbolic partial differential
equations that the initial data surface be spacelike.
In general situations, the inequality~\eqref{hypcond}
is not obvious and must be arranged and verified in the applications.
More specifically, one can use the freedom in choosing the jet space~$\Jvary$,
the Riemannian metric in the scalar product~\eqref{vsprod} and the functions~$\eta_t$
in Definition~\ref{deflocfoliate} in order to ensure that~\eqref{hypcond} holds.

We now explain how the above hyperbolicity condition can be used to
derive energy estimates. We let~$L$ be a lens-shaped region inside~$U$
with the local foliation~$(\eta_t)_{t \in I}$.
We denote the norm corresponding to the jet scalar product by~$\|\u\|^t := \sqrt{(\u,\u)^t}$.
\nindex{agm@$\NORM . \NORM^t$ -- norm corresponding to softened surface layer inner product}%
\tindex{dd@$\NORM . \NORM^t$ -- norm corresponding to softened surface layer inner product}%
We begin with a simple estimate of the energy identity in Lemma~\ref{lemmaenid}.

\begin{Lemma} \label{lemmaenes}
Assume that the hyperbolicity condition of Definition~\ref{defhypcond} holds.
Then for every~$t \in I$ and all~$\u \in \J$,
\beq \frac{\dd}{\dd t}\: \|\u\|^t 
\leq C\:\|\Delta \u\|_{L^2(U, \Diff\rho_t)}  + c\: \|\u\|^t \label{enes0}
\eeq
with
\beq c := C^2 + \frac{C^2 \,\s}{2} \:. \eeq
\end{Lemma}
\Proof Applying~\eqref{hypcond} in~\eqref{enid}, we obtain
\begin{align}
\frac{\dd}{\dd t}\: (\u, \u)^t 
&\leq 2 \int_U \la \u, \Delta \u \ra_x \: \Diff\rho_t(x)
-2 \int_U \Delta_2[\u, \u] \: \Diff\rho_t(x) + \s \int_U b(x)^2\: \Diff\rho_t(x) \notag \\
&\leq 2 \int_U \la \u, \Delta \u \ra_x \: \Diff\rho_t(x)
+\Big( 2 C^2 + C^2 \,\s \Big)\: (\u,\u)^t \notag \\
&\leq 2 \,\|\u\|_{L^2(U, \Diff\rho_t)} \: \|\Delta \u\|_{L^2(U, \Diff\rho_t)}
+2c\: (\u,\u)^t \notag \\
&\leq 2 C\,\|\u\|^t\: \|\Delta \u\|_{L^2(U, \Diff\rho_t)}
+2c\: (\u,\u)^t \:,
\end{align}
where in the last line, we applied~\eqref{hypcond}.
Using the relation
\beq \partial_t \|\u\|^t = \partial_t (\u, \u)^t / (2 \|\u\|^t) \eeq
gives the result.
\QED

Applying Gr\"onwall-type estimates (see, for example, \cite[Lemma~1.15 in Section~VII.1]{amann-escher2}
\sindex{Gr\"onwall estimate}%
or the proof of Proposition~\ref{prpenes} below), the inequality~\eqref{enes0} shows that~$\|\u\|^t$ grows at most
exponentially in time, provided that~$\Delta \u$ decays in time sufficiently fast.
We here make this statement precise by estimates in Hilbert spaces
of jets with zero initial values.
In the lens-shaped region~$L$, we work with the $L^2$-scalar product
\nindex{agn@$\la .,. \ra_{L^2(L)}$ -- $L^2$-scalar product in lens-shaped region}%
\tindex{bb@$\la .,. \ra_{L^2(L)}$ -- $L^2$-scalar product in lens-shaped region}%
\beq \label{L2prodOSI}
\la \u, \v \ra_{L^2(L)} := \int_L \la \u(x),\v(x) \ra_x\: \eta_I(x) \, \Diff\rho(x) \:,
\eeq
which, according to~\eqref{rhot} and~\eqref{etabracket}, can also be written in terms of a time integral,
\beq \label{L2prodtime}
\la \u, \v \ra_{L^2(L)} = \int_{t_0}^{\tmax} \la\u,\v\ra_{L^2(U, \Diff\rho_t)}\: \dd t \:.
\eeq
The corresponding norm is denoted by~$\| . \|_{L^2(L)}$.
\begin{Prp} {\bf{(energy estimate)}} \label{prpenes}
\sindex{energy estimate!for causal variational principle}%
Assume that the hyperbolicity condition of Definition~\ref{defhypcond} holds.
Then, choosing
\beq \label{Gammachoose}
\Gamma = 2\,C\, \E^{2c\,(\tmax-t_0)}\: (\tmax-t_0)\:,
\eeq
the following estimate holds,
\beq %\beq \label{enes1}
\|\u\|_{L^2(L)} \leq \Gamma\: \|\Delta \u\|_{L^2(L)} \qquad \text{for all~$\u \in \J$ with~$\|\u\|^{t_0}=0$}\:.
\eeq %\eeq
\end{Prp}
\Proof  We write the energy estimate of Lemma~\ref{lemmaenes} as
\beq \frac{\dd}{\dd t}\: \big( \E^{-2c t}\, (\u, \u)^t \big) \leq 2 \: \E^{-2c t} \:C\: \|\u\|^t\: \|\Delta \u\|_{L^2(U, \Diff\rho_t)} \:. \eeq
Integrating over~$t$ from~$t_0$ to some~$t \in I$ and using the hyperbolicity condition~\eqref{hypcond}, we obtain
\begin{align}
\E^{-2c t}\, (\u, \u)^t &=  \int_{t_0}^t \frac{\dd}{\dd t'} \: \big( \E^{-2c t'} (\u, \u)^{t'} \big)\: \dd t' \notag \\
&\leq 2\,C\, \int_{t_0}^{t} \E^{-2c t'} \: \|\u\|^{t'}\: \|\Delta \u\|_{L^2(U, \Diff\rho_{t'})} \: \dd t' \:.
\end{align}
Multiplying by~$\E^{2c t}$ gives the inequality
\begin{align}
(\u, \u)^t &\leq 2\,C\, \int_{t_0}^{t} \E^{2c \,(t-t')} \: \|\u\|^{t'}\: \|\Delta \u\|_{L^2(U, \Diff\rho_{t'})} \: \dd t' \notag \\
&\leq 2\,C\, \E^{2c \,(\tmax-t_0)} \: \int_{t_0}^{\tmax} \|\u\|^{t'}\: \|\Delta \u\|_{L^2(U, \Diff\rho_{t'})} \: \dd t' \notag \\
&\leq 2\,C\, \E^{2c \,(\tmax-t_0)} \: \|\Delta \u\|_{L^2(L)}\:
\bigg( \int_{t_0}^{\tmax} (\u, \u)^{t'} \:\dd t' \bigg)^\frac{1}{2} \:,
\end{align}
where in the last step we used the Schwarz inequality and~\eqref{L2prodtime}.
Integrating once again over~$t$ from~$t_0$ to~$\tmax$ gives
\beq \label{intfinal}
\bigg( \int_{t_0}^{\tmax}(\u, \u)^{t} \:\dd t \bigg)^\frac{1}{2} \leq 2\,C\, \E^{2c \,(\tmax-t_0)}\:(\tmax-t_0) \: \|\Delta \u\|_{L^2(L)} \:.
\eeq

Finally, we apply the hyperbolicity condition~\eqref{hypcond} in~\eqref{L2prodtime},
\beq \|v\|_{L^2(L)} = \bigg( \int_{t_0}^{\tmax}  \|\u \|_{L^2(U, \Diff\rho_{t})}^2 \:\dd t \bigg)^\frac{1}{2}
\leq C \:\bigg( \int_{t_0}^{\tmax}  (\u,\u)^t \:\dd t \bigg)^\frac{1}{2} \:. \eeq
Combining this inequality with~\eqref{intfinal} gives the result.
\QED

\section{Uniqueness of Strong Solutions}
Based on the above energy estimates, we can now prove uniqueness of strong solutions
of the Cauchy problem. The method is quite similar to that employed in Section~\ref{sec31}
for symmetric hyperbolic systems. In preparation of formulating the Cauchy problem,
we need to introduce jets that vanish at the initial time~$t_{\min}$. To this end, we demand
that the jet vanishes in the surface layer and that the
corresponding softened surface layer integrals~\eqref{jipdef} and~\eqref{sympdef} vanish,
\beq %\label{uJtestdef}
\underline{\J}_{t_{\min}} := \big\{ \u \in \J \:\big|\: \eta_{t_{\min}}\, \u \equiv 0 \quad \text{and} \quad
(\u,\v)^{t_{\min}}=0=\sigma^{t_{\min}}(\u, \v) \text{ for all~$\v \in \J$} \big\} \:. \eeq
Similarly, we define the space of jets that vanish at final time~$\tmax$ by
\beq %\label{oJtestdef}
\begin{split}
\overline{\J_U}^\tmax := &\big\{ \u \in \J \:\big|\: \big(1-\eta_{t_{\max}}\big)\, \u \equiv 0  \\
&\quad \text{and}
(\u,\v)^{t_{\max}}=0=\sigma^{t_{\max}}(\u, \v)  \text{ for all~$\v \in \J$} \big\} \:.
\end{split} \eeq
\nindex{ago@$\overline{\J}, \underline{\J}$ -- jet spaces vanishing in future or past}%
\tindex{ff@$\overline{\J}, \underline{\J}$ -- jet spaces vanishing in future or past}%
A {\em{strong solution}} of the Cauchy problem is a jet~$\u \in \J_U$ which satisfies the equations
\sindex{Cauchy problem!strong solution}%
\beq \label{cauchystrong}
\Delta \u = \w \quad \text{in~$L$} \qquad \text{and} \qquad \u-\u_0 \in \underline{\J}_{t_{\min}} \:,
\eeq
where~$\u_0 \in \J$ is the initial data and~$\w$ is the {\em{inhomogeneity}}.
\sindex{inhomogeneity!in linearized field equations}%
More precisely, as explained after~\eqref{Delvw}, the jet~$\w(x)$ can be regarded as a dual jet.
Here, having already introduced a scalar product on the jets at every spacetime point (see~\eqref{vsprod}),
we can identify dual jets with jets. With this in mind, the inhomogeneity simply is a jet~$\w \in \J_U$.

\begin{Prp} \label{prpuniquestrong} {\bf{(uniqueness of strong solutions)}}
Let~$(\eta_t)_{t \in I}$ be a local foliation inside~$U$ which satisfies the hyperbolicity conditions
(as stated in Definitions~\ref{deflocfoliate} and~\ref{defhypcond}).
Then the Cauchy problem~\eqref{cauchystrong} with~$\u_0, \w \in \J_U$ has at most one solution~$\u$ in~$L$.
\end{Prp}
\Proof Let~$\u$ be the difference of two solutions. Then~$\u$ is a solution of the homogeneous equation
with zero initial data. Applying Lemma~\ref{lemmaenes}, we obtain
\beq \Big| \frac{\dd}{\dd t}\: \|\u\|^t \Big| \leq c\: \|\u\|^t \qquad \text{and thus} \qquad
\frac{\dd}{\dd t}\: \big( \E^{-ct} \,\|\u\|^t \big) \leq 0 \:. \eeq
It follows that~$\|\u\|^t$ vanishes for all~$t$ in the respective interval. 
Using~\eqref{hypcond}, we conclude that~$\u$ vanishes identically in~$L$.
This gives the result.
\QED

Similar as explained in Section~\ref{sec31} for symmetric hyperbolic systems,
this uniqueness statement also gives information on the speed of propagation
and the resulting causal structure. For details, we refer to~\cite{linhyp, weyl}.

\section{Existence of Weak Solutions}
Our existence proof is inspired by the method invented by K.O.\ Friedrichs for
symmetric hyperbolic systems in~\cite{friedrichs} as outlined in Section~\ref{secshsexist}.
Our first step is to formulate the linearized field equations weakly. To this end, we need
to ``integrate by parts'' with the help of the following lemma.
\begin{Lemma} {\bf{(Green's formula)}} \label{lemmagreen}
\sindex{Green's formula!for causal variational principle}%
For all~$\u, \v \in \J$,
\beq %\label{green}
\sigma^{\tmax}(\u, \v) - \sigma^{t_{\min}}(\u, \v) = \la \u, \Delta \v \ra_{L^2(L)} - \la \Delta \u, \v \ra_{L^2(L)} \:. \eeq
\end{Lemma}
\Proof Using the definition of the $L^2$-scalar product in~\eqref{L2prodOSI} and
the definition of the linearized field operator~\eqref{linfield}, we obtain
\begin{align}
\la &\u, \Delta \v \ra_{L^2(L)} - \la \Delta \u, \v \ra_{L^2(L)} =
\int_U  \Big( \la \u, \Delta \v \ra - \la \Delta \u, \v \ra \Big)\: \eta_I \,\Diff\rho \notag \\
&= \int_U \Diff\rho(x)\: \eta_I(x)\;
\nabla_{\u} \bigg( \int_M \big( \nabla_{1, \v} + \nabla_{2, \v} \big) \L(x,y)\: \Diff\rho(y) - \nabla_\v \,\s \bigg) \notag \\
&\quad-\int_U \Diff\rho(x)\: \eta_I(x)\;
\nabla_{\v} \bigg( \int_M \big( \nabla_{1, \u} + \nabla_{2, \u} \big) \L(x,y)\: \Diff\rho(y) - \nabla_\u \,\s \bigg) \:.
\end{align}
Here the spacetime point~$x$ is in~$L$. Using Definition~\ref{deflocfoliate}~(ii), we get a contribution to the
integrals only if~$y \in U$. Therefore, we may replace the integration range~$M$ by~$U$.
We thus obtain
\begin{align}
\la &\u, \Delta \v \ra_{L^2(L)} - \la \Delta \u, \v \ra_{L^2(L)} \notag \\
&= \int_U \Diff\rho(x)\: \eta_I(x) \int_U \Diff\rho(y) \big( \nabla_{1,\u} \nabla_{2, \v} 
- \nabla_{2,\u} \nabla_{1, \v} \big) \L(x,y) \:, \label{eq1}
\end{align}
where we used that, following our convention~\eqref{ConventionPartial}, the
second derivatives of the Lagrangian are symmetric.
Using the definition~\eqref{etabracket} as well as
 the anti-symmetry of the integrand, the term~\eqref{eq1} can be rewritten as
\begin{align}
&\int_U \Diff\rho(x)\: \eta_I(x) \int_U \Diff\rho(y) \big( \nabla_{1,\u} \nabla_{2, \v} 
- \nabla_{2,\u} \nabla_{1, \v} \big) \L(x,y) \notag \\
&=\int_U \Diff\rho(x) \int_U \Diff\rho(y) \:\eta_{t}(x)\: \big( \nabla_{1,\u} \nabla_{2, \v}
- \nabla_{2,\u} \nabla_{1, \v} \big) \L(x,y) \Big|_{t_0}^{\tmax} \notag \\
&=\int_U \Diff\rho(x)\ \int_U \Diff\rho(y) \Big( \eta_{t}(x) - \eta_{t}(x)\: \eta_{t}(y) \Big)\,
\big( \nabla_{1,\u} \nabla_{2, \v} - \nabla_{2,\u} \nabla_{1, \v} \big) \L(x,y) \Big|_{t_0}^{\tmax} \notag \\
&=\int_U \Diff\rho(x)\ \int_U \Diff\rho(y) \:\eta_{t}(x)\: \big(1-\eta_{t}(y) \big)\,
\big( \nabla_{1,\u} \nabla_{2, \v} - \nabla_{2,\u} \nabla_{1, \v} \big) \L(x,y) \Big|_{t_0}^{\tmax} \notag \\
&= \sigma^{\tmax}(\u,\v) - \sigma^{t_{\min}}(\u,\v) \:.
\end{align}
This gives the result.
\QED

Assume that~$\u$ is a strong solution to the Cauchy problem~\eqref{cauchystrong}.
As usual, replacing~$\u$ by~$\u-\u_0$ and~$\w$ by~$\w -\Delta \u_0 \in \J$,
it suffices to consider the Cauchy problem for zero initial data; that is,
\beq %\label{cauchyzeroinit}
\Delta \u = \w \quad \text{in~$U$} \qquad \text{and} \qquad \u \in \underline{\J_U}_{t_{\min}}\:. \eeq
Then, applying the above Green's formula, we obtain for any~$\v \in \J$,
\beq \la \v, \w \ra_{L^2(L)} = \la \v, \Delta \u \ra_{L^2(L)} = 
\la \Delta \v, \u \ra_{L^2(L)} - \sigma^{\tmax}(\v, \u) + \sigma^{t_{\min}}(\v, \u) \:. \eeq
Having implemented the vanishing initial data by the
condition~$\u \in \underline{\J}_{t_0}$, the symplectic form vanishes at time~$t_{\min}$.
In order to also get rid of the boundary values at time~$\tmax$, we
restrict attention to test jets that vanish at~$\tmax$. This leads us to the following definition:

\begin{Def} \label{defweak}
\sindex{Cauchy problem!weak solution}%
\sindex{weak solution!of linearized field equations}%
A jet~$\u \in L^2(L)$ is a {\bf{weak solution}} of the Cauchy problem~$\Delta \u = \w$
with zero initial data if
\beq \label{weak}
\la \Delta \v, \u \ra_{L^2(L)} = \la \v, \w \ra_{L^2(L)} \qquad \text{for all~$\v \in \overline{\J}^\tmax$}\:.
\eeq
\end{Def}
Clearly, the energy estimate of Proposition~\ref{prpenes} also holds
if we exchange the roles of~$\tmax$ and~$t_{\min}$; that is,
\beq \label{hyprev}
\|\u\|_{L^2(L)} \leq \Gamma\: \|\Delta \u\|_{L^2(L)} \qquad \text{for all~$\u \in \overline{\J}^\tmax$}
\eeq
(where the constant~$\Gamma$ is again given by~\eqref{Gammachoose}).

We introduce the positive semi-definite bilinear form
\beq \bra .,. \ket \::\: \overline{\J_U}^\tmax \times \overline{\J_U}^\tmax \rightarrow \R\:,\qquad
\bra \u, \v \ket = \la \Delta \u, \Delta \v \ra_{L^2(L)} \:. \eeq
Dividing out the null space and forming the completion, we obtain a Hilbert space
$(\mathcal{H}, \bra .,. \ket)$. The corresponding norm is denoted by~$\norm . \norm$.

We now consider the linear functional~$\la \w, . \ra_{L^2(L)}$ on~$\overline{\J_U}^\tmax$.
Applying the Schwarz inequality and~\eqref{hyprev}, we obtain
\beq \big| \la \w, \u \ra_{L^2(L)} \big| \leq \|\w\|_{L^2(L)} \:  \|\u\|_{L^2(L)} 
\leq \Gamma\:\|\w\|_{L^2(L)} \:  \norm \u \norm \:, \eeq
proving that the linear functional~$\la \w, . \ra_{L^2(L)}$ on~$\overline{\J}^\tmax$
is bounded on~$\mathcal{H}$.
Therefore, it can be extended uniquely to a bounded linear functional on all of~$\mathcal{H}$.
Moreover, by the Fr{\'e}chet-Riesz theorem, there is a unique vector~$U \in \mathcal{H}$ with
\beq %\beq \label{FR}
\la \w, \v \ra_{L^2(L)} = \bra U, \v \ket = 
\la \Delta U, \Delta \v \ra_{L^2(L)}\qquad \text{for all~$\v \in \overline{\J_U}^\tmax$}\:.
\eeq %\eeq
Hence~$\u := \Delta U \in L^2(L)$ is the desired weak solution.
We point out that in the above estimates, the inhomogeneity~$\w$ enters
only via its~$L^2$-norm, making it possible to generalize our methods
to~$\w \in L^2(L)$.
We have obtained the following result:

\begin{Thm} \label{thmexist} Assume that~$(\eta_t)_{t \in I}$ is a local foliation satisfying the
hyperbolicity conditions (see Definitions~\ref{deflocfoliate} and~\ref{defhypcond}).
Then, for every~$\w \in L^2(L)$, there is a weak solution~$\u \in L^2(L)$
of the Cauchy problem~\eqref{weak}.
\end{Thm}

We remark that the construction of weak solutions is the starting point for the more detailed
analysis of linearized fields as carried out in~\cite{linhyp, dirac, weyl, localize}.
One task is to deal with the uniqueness problem for weak solutions
(see Exercise~\ref{exnonunique}). Another issue is to construct global solutions
(for various methods for doing so, see~\cite[Section~4]{linhyp},
\cite[Section~6.3]{dirac}, \cite[Section~3.3]{weyl} and~\cite{localize}).
Once global advanced and retarded solutions have been obtained for a general
class of inhomogeneities, one can also introduce corresponding Green's operators
(see~\cite[Section~5]{linhyp}, \cite[Section~4]{localize} or Exercise~\ref{exgreen}).

\section{Exercises}

\begin{Exercise} (Differentiated form of conservation laws) {\em{
\sindex{conservation law for surface layer integral!differentiated form}%
Conservation laws for causal variational principles are formulated in terms of surface layer integrals
(see, for example, Theorem~\ref{thmsymmlag}, Theorem~\ref{thmosirho}, Proposition~\ref{prpI1}
or Lemma~\ref{lemmagreen}). For the proofs, we rewrote the surface layer integrals
as double volume integrals, making use of anti-symmetry properties
(see, for example, the proof of Theorem~\ref{thmosirho}).
Alternatively, one can prove the conservation laws by computing the time derivatives.
The goal of this exercise is to illustrate this method
(for more details see for example~\cite[Section~2.6]{fockbosonic}).
\bitem
\item[(a)] Given a local foliation~$(\eta_t)_{t \in I}$ we consider the softened symplectic form~\eqref{sympdef}.
Given~$\u, \v \in \J$, compute
\sindex{symplectic form!softened}%
\beq \frac{\dd}{\dd t}\: \sigma^t(\u, \v) \eeq
in a similar style as in the proof of Lemma~\ref{lemmaenid}.
Use this formula to give an alternative proof of Lemma~\ref{lemmagreen}.
\item[(b)] Given a local foliation~$(\eta_t)_{t \in I}$, formulate a softened version of the surface layer integral~$I^\Omega_{k+1}$ in Theorem~\ref{Ikdef}. Differentiate with respect to the time parameter~$t$
to obtain an alternative proof of Theorem~\ref{Ikdef}.
\eitem
}} \end{Exercise}

\begin{Exercise} (Counter example to hyperbolicity conditions) \label{excounterhyp} {\em{
\sindex{hyperbolicity condition!counter example}%
The hyperbolicity conditions in Definition~\ref{defhypcond} were formulated only for jets
in a subspace~$\Jvary \subset \J$. The goal of this exercise is to explain why
it would not be sensible to impose the inequality~\eqref{hypcond} for all~$\u \in \J$.
To this end, consider for simplicity the unsoftened surface layer integral~\eqref{sliunsoftened}.
Show that there is a jet~$\u \in C^0_0(M, \R) \oplus C^0_0(M, T\F)$ with~$(\u, \u)^t<0$.
{\em{Hint:}} Choose points~$x \in \Omega$ and~$y \in M \setminus \Omega$
for which~$\L(x,y) \neq 0$. Choose~$\u$ as a scalar jet that is supported in a small
neighborhood of~$x$ and~$y$.
}} \end{Exercise}

\begin{Exercise} (Non-uniqueness of weak solutions) \label{exnonunique} {\em{
\sindex{weak solution!non-uniqueness of}%
As illustrated in the previous exercise (Exercise~\ref{excounterhyp}),
in order to satisfy the hyperbolicity conditions, the jet space~$\Jvary$ must not be chosen
too large. In particular, in typical examples, the jet space~$\Jvary$ is not dense in~$L^2(L)$.
This leads to a non-uniqueness issue for weak solutions, which will be illustrated in this exercise.
\bitem
\item[(a)] Given an inhomogeneity~$\w \in L ^2(L)$, to which extent are weak solutions unique?
Specify the jet space of all homogeneous solutions.
\item[(b)] On the other hand, the construction before Theorem~\ref{thmexist} gives a
unique solution~$\u = \Delta U$. How does this result fit together with the non-uniqueness
in~(a)? In which sense is the solution~$\u = \Delta U$ distinguished?
\eitem
{\em{Hint:}} Similar questions are analyzed in~\cite[Section~3]{linhyp}.
}} \end{Exercise}

\begin{Exercise} (Causal Green's operators for the linearized field equations) \label{exgreen} {\em{
\sindex{Green's operator!causal}%
In order to avoid the issue of how to ``glue together'' local solutions obtained in different
lens-shaped regions such as to obtain global solutions,
one can simplify the setting by assuming that the spacetime~$M$ 
admits a global foliation~$(\eta_t)_{t \in \R}$.
\bitem
\item[(a)] How can Definition~\ref{deflocfoliate} be modified in order to describe a global foliation?
What is the resulting global analog of Theorem~\ref{thmexist}?
{\em{Hint:}} It might be instructive to compare your definition with Definition~6.4 in~\cite{dirac}.
\item[(b)] Suppose that we know that for any compactly supported jet~$\w \in L^2_0(M)$,
there is a distinguished global weak solution~$\v \in L^2_\loc(M)$; that is,
\beq \la \Delta \v, \u \ra_{L^2(M)} = \la \v, \w \ra_{L^2(M)} \qquad \text{for all~$\v \in \J_0$}\:. \eeq
Then the operator~$S : \w \rightarrow -\v$ is referred to as the Green's operator.
How can one distinguish between the advanced Green's operator~$S^\vee$ and the
retarded Green's operator~$S^\wedge$.
Show that their difference~$G := S^\wedge - S^\vee$ maps to homogeneous weak solutions.
{\em{Hint:}} More details on Green's operators and their causal properties can be found
in~\cite[Section~5]{linhyp}, \cite[Section~3]{weyl} and~\cite[Section~4]{localize}.
\eitem
}} \end{Exercise}

\chapter{Functional Analytic Methods in Spacetime} \label{secFSO}
When constructing a causal fermion system in Minkowski space in Section~\ref{seclco},
we chose~$\H$ as a subspace of the solution space~$\H_m$ of the Dirac equation.
In principle, one can choose~$\H$ as one likes, and different choices give rise to
different causal fermion systems. However, if one wants to describe a given
physical system, one must specify the subspace~$\H \subset \H_m$,
and it important to do it right. It is not obvious what ``right'' and ``wrong'' should be.
Generally speaking, $\H$ can be thought of as the ``occupied states'' of the physical system under
consideration. If we want to describe the vacuum in Minkowski space (that is, no particles and no interaction
is present), then the natural and only physically reasonable choice is to let~$\H$ be the subspace
of all {\em{negative-frequency solutions}} of the Dirac equation.
As already explained in Section~\ref{secsea} in the preliminaries,
this choice corresponds to the
physical concept of the Dirac sea as introduced by Dirac in 1930, which led to the prediction
of anti-matter (discovered shortly afterward in 1932, earning Dirac the Nobel prize in 1933).
Following these physical concepts, it is also clear that if particles and/or anti-particles
(but no interaction of the matter) is present, then~$\H$ is obtained from the
subspace of all negative-frequency solutions by occupying additional particle states
and by creating ``holes'' in the sea corresponding to the anti-particle states.
Once an interaction (for example, an electromagnetic field) is present, it is no longer clear
how~$\H$ is to be chosen. The reason is that, as soon as the fields are time-dependent,
the notion of positive and negative frequency solutions breaks down, so that there is no obvious
decomposition of the solution space into two subspaces.
But for the description of the physical system, a decomposition of the
solution space is needed, and taking the ``wrong'' decomposition leads to artificial
mathematical and physical difficulties.

We now explain a functional analytic method which gives rise to a canonical decomposition of the
solution space into two subspaces, even in the time-dependent situation.
In the static situation, this decomposition reduces to the canonical
frequency splitting. This splitting is ``right'' in the sense that it gives rise to a physically sensible ground state
of the system (a so-called Hadamard state, as we will learn in Chapter~\ref{sechadamard}).
Moreover, when performing our construction perturbatively,
one can compute the singularities of~$P(x,y)$ explicitly working exclusively with bounded line integrals.
These explicit computations are the backbone of the analysis of the continuum limit in~\cite{cfs}.
Before outlining the perturbative treatment (see Chapter~\ref{secperturb}),
we now explain the general functional analytic construction.

\section{General Setting and Basic Ideas} \label{secrecall}
In preparation, we summarize the structures of Chapter~\ref{PhysicalPreliminaries} using a more general
notation, which has the advantage that our setting applies just as well if Minkowski
space is replaced by a globally hyperbolic spacetime. Thus the reader who is familiar
with general relativity and Lorentzian geometry, in what follows may consider~$(\scrM, g)$
as a globally hyperbolic Lorentzian manifold with spinor bundle~$(S\scrM, \Sl .|. \Sr)$.
The Dirac equation is written as
\beq \label{Dir}
(\Dir - m)\, \psi_m = 0
\eeq
(here the subscript~$m$ indicates the mass of the solution; this is of advantage because
later on, we shall consider families of solutions with a varying mass parameter).
In Minkowski space, one chooses~$\Dir = i \Pdd + \B$ such as to get back to~\eqref{Dout},
where~$\B$ is an arbitrary multiplication operator satisfying the symmetry condition~\eqref{Bsymm}.
More generally, in a globally hyperbolic spacetime, the Dirac operator is a first order differential operator,
but the coefficients depend on the metric (for details see Chapter~\ref{Geometry}).
Next, we let~$\scrN$ be any Cauchy surface. 
Then the scalar product~\eqref{printMink} on the solutions can be written more generally as
\nindex{aam@$(.\vert.), (.\vert.)_m$ -- scalar product on Dirac solutions in Minkowski space}%
\tindex{bb@$(.\vert.), (.\vert.)_m$ -- scalar product on Dirac solutions in Minkowski space}%
\beq \label{print}
(\psi_m | \phi_m)_m = 2 \pi \int_\scrN \Sl \psi_m | \nuslsh \phi_m \Sr_x\: \Diff\mu_\scrN(x) \:,
\eeq
where~$\nu$ is the future-directed normal and~$\Diff\mu_\scrN$ the volume measure
given by the induced Riemannian metric on~$\scrN$ (in Minkowski space and~$\scrN = \{t=\text{const}\}$,
the normal has the components~$\nu^i = (1,0,0,0)$ and~$\Diff\mu_\scrN = \dd^3x$,
giving back~\eqref{printMink}).
Similar to the computation~\eqref{divfree}, the vector field~$\Sl \psi_m | \gamma^j \phi_m \Sr_x$
is again divergence-free, implying that the above scalar product is independent of the
choice of the Cauchy surface (for details, see~\cite[Section~2]{finite}).
Forming the completion gives the Hilbert space~$(\H_m, (.|.)_m)$.

For the following constructions, we again need the spacetime inner product~\eqref{stip}.
In order to explain the basic idea of the construction as first given in~\cite{finite},
let us assume for simplicity that the integral in~\eqref{stip} exists for all
solutions~$\psi_m, \phi_m \in \H_m$. This condition is not satisfied in Minkowski space because
the time integral in~\eqref{stip} in general diverges. But it is indeed satisfied in
spacetimes of finite lifetime (for details, see~\cite[Section~3.2]{finite}).
Under this assumption, the spacetime inner product can be extended by
continuity to a sesquilinear form
\beq \bra .|. \ket \::\: \H_m \times \H_m \rightarrow \C\:, \eeq
which is bounded; that is,
\beq \label{stbound}
|\bra \phi_m | \psi_m \ket| \leq c \:\|\phi_m\|_m\: \|\psi_m\|_m
\eeq
(where~$\| . \|_m = (.|.)_m^\frac{1}{2}$ is the norm on~$\H_m$).
Then, applying the Fr{\'e}chet-Riesz theorem (similar as explained in the construction
of the local correlation operator~\eqref{Fepsdef} in Section~\ref{seclco}), we can
uniquely represent this inner product on the Hilbert space~$\H_m$ with a signature operator~$\Sig$,
\beq %\label{Sigdef}
\Sig \::\: \H_m \rightarrow \H_m \qquad \text{with} \qquad
\bra \phi_m | \psi_m \ket = ( \phi_m \,|\, \Sig\, \psi_m)_m \:. \eeq
We refer to~$\Sig$ as the {\bf{fermionic signature operator}}.
\sindex{fermionic signature operator|textbf}%
\nindex{agr@$\Sig$ -- fermionic signature operator}%
It is obviously a symmetric operator. Moreover, it is bounded according to~\eqref{stbound}.
Therefore, the spectral theorem for selfadjoint operators gives the spectral decomposition
\beq \Sig = \int_{\sigma(\Sig)} \lambda\: dE_\lambda \:, \eeq
where~$E_\lambda$ is the spectral measure (see Section~\ref{secspectral}
or, for example, \cite{reed+simon}).
The spectral measure gives rise to the spectral calculus
\beq f(\Sig) = \int_{\sigma(\Sig)} f(\lambda)\: \dd E_\lambda \::\: \H_m \rightarrow \H_m \:, \eeq
where~$f$ is a bounded Borel function on~$\sigma(\Sig) \subset \R$.
Choosing~$f$ as a characteristic function, one obtains the operators~$\chi_{(0,\infty)}(\Sig)$
and~$\chi_{(-\infty,0)}(\Sig)$. Their images are referred to as the {\em{positive}} and {\em{negative
spectral subspace}} of~$\H_m$, respectively. In this way, one obtains the desired
decomposition of the solution space into two subspaces.
We remark that the fermionic signature operator also gives a setting for
doing spectral geometry and index theory with Lorentzian signature. We will not enter these topics
here but refer the interested reader to the papers~\cite{drum, index}.

The basic shortcoming of the above construction is that in many physically interesting
spacetimes (like Minkowski space) the inequality~\eqref{stbound} fails to be true.
The idea to bypass this problem is to make use of the fact that
a typical solution~$\psi \in \Cisc(\scrM, S\scrM) \cap \H_m$
of the Dirac equation oscillates for large times.
If, instead of a single solution, we consider
a {\em{family}} of solutions with a {\em{varying mass parameter}}~$m$,
then the wave functions for different values of~$m$ typically have different phases.
Therefore, integrating over the mass parameter leads to dephasing
(in the physics literature also referred to as destructive interference),
giving rise to decay in time. In order to make this idea mathematically precise,
one considers families of solutions~$(\psi_m)_{m \in I}$ of the family of Dirac equations~\eqref{Dir} with
the mass parameter~$m$ varying in an open interval~$I$.
We need to assume that~$I$ does not contain the origin, because our methods 
for dealing with infinite lifetime do not
apply in the massless case~$m=0$ (this seems no physical restriction because all known
fermions in nature have a non-zero rest mass). By symmetry, it suffices to consider positive masses.
Thus we choose~$I$ as the interval
\beq \label{Idef}
I:=(m_L, m_R) \subset \R \qquad \text{with parameters~$m_L, m_R>0$}\:. 
\eeq
The masses of the Dirac particles of our physical system should be contained in~$I$.
Apart from that, the choice of~$I$ is arbitrary and, as we shall see, all our results will be independent
of the choice of~$m_L$ and~$m_R$.
We always choose the family of solutions~$(\psi_m)_{m \in I}$ in the
class~$\Cisco(\scrM \times I, S\scrM)$ of smooth solutions with spatially compact support
in Minkowski space~$\scrM$ which depend smoothly on~$m$ and vanish identically for~$m$ outside a
compact subset of~$I$. Then the ``decay due to destructive interference'' can be
made precise by demanding that  there is a constant~$c>0$ such that
\beq \label{smopintro}
\bigg| \bra \int_I \phi_m \,\dd m \:|\: \int_I \psi_{m'} \,\dd m'  \ket \bigg|
\leq c \int_I \, \|\phi_m\|_m\, \|\psi_m\|_m\: \dd m
\eeq
for all families of solutions~$(\psi_m)_{m \in I},  (\phi_m)_{m \in I} \in \Cisco(\scrM \times I, S\scrM)$.
The point is that we integrate over the mass parameter {\em{before}} taking the spacetime inner product.
Intuitively speaking, integrating over the mass parameter generates
a decay of the wave function, making sure that the time integral converges.
The inequality~\eqref{smopintro} is one variant of the so-called {\em{mass oscillation property}}.
If~\eqref{smopintro} holds, we shall prove that there is a representation
\beq \label{Sigm}
\bra \int_I \phi_m \,\dd m \,|\, \int_I \psi_{m'} \,\dd m' \ket = \int_I (\phi_m \,|\, \tilde{\Sig}_m \,\psi_m)_m\: \dd m \:,
\eeq
which for every~$m \in I$ uniquely defines the {\em{fermionic signature operator}}~$\tilde{\Sig}_m$.
This operator is bounded and symmetric with respect to the scalar product~\eqref{print}.
Moreover, it does not depend on the choice of the interval~$I$.
Then the positive and negative spectral subspaces
of the operator~$\tilde{\Sig}_m$ again yield the desired splitting of the solution space
into two subspaces.

Before entering the detailed constructions, we explain how the above integrals over the mass parameters
are to be understood. At first sight, integrating over a varying mass parameter~$m \in I$
may look like ``smearing out'' the physical mass in the Dirac equation.
However, this picture is misleading. Instead, one should consider the mass integrals merely as a
technical tool in order to generate decay for large times.
The resulting operators~$\tilde{\Sig}_m$ in~\eqref{Sigm} act on~$\psi_m$ with the corresponding mass~$m \in I$.
Choosing~$m$ again as the physical mass, the operator~$\tilde{\Sig}_m$ acts on standard
Dirac wave functions describing physical particles, without any smearing in the mass parameter.

\section{The Mass Oscillation Properties} \label{secmop}
\sindex{mass oscillation property|textbf}%
In a spacetime of infinite lifetime, the spacetime inner product~$\bra \psi_m | \phi_m \ket$ of two
solutions~$\psi_m, \phi_m \in \H_m$ is in general ill-defined, because the time integral in~\eqref{stip}
may diverge. In order to avoid this difficulty, we shall consider 
families of solutions with a variable mass parameter. The so-called mass oscillation property
will make sense of the spacetime integral in~\eqref{stip} after integrating over the mass parameter.

We consider the mass parameter in a bounded open interval~$I$~\eqref{Idef}.
For a given Cauchy surface~$\scrN$, we consider a function~$\psi_\scrN(x,m) \in S_x\scrM$
with~$x \in \scrN$ and~$m \in I$. We assume that this wave function is smooth and has
compact support in both variables, $\psi_\scrN \in C^\infty_0(\scrN \times I, S\scrM)$.
For every~$m \in I$, we let~$\psi(.,m)$ be the solution of the Cauchy problem for initial data~$\psi_\scrN(.,m)$,
\beq \label{cauchy2}
(\Dir - m) \,\psi(x,m) = 0 \:,\qquad \psi(x,m) = \psi_\scrN(x,m)  \;\; \forall\: x \in \scrN \:.
\eeq
Since the solution of the Cauchy problem is smooth and depends smoothly on parameters,
we know that~$\psi \in C^\infty(\scrM \times I, S\scrM)$.
Moreover, due to finite propagation speed, $\psi(.,m)$ has spatially compact support.
Finally, the solution is clearly compactly supported in the mass parameter~$m$.
We summarize these properties by writing
\beq \label{CscmH}
\psi \in \Cisco(\scrM \times I, S\scrM) \:,
\eeq
\nindex{ags@$\Cisco(\scrM \times I, S\scrM)$ -- space of families of spinorial wave functions}%
\tindex{ff@$\Cisco(\scrM \times I, S\scrM)$ -- space of families of spinorial wave functions}%
where~$\Cisco(\scrM \times I, S\scrM)$ denotes the smooth wave functions with spatially compact support which
are also compactly supported in~$I$. 
We often denote the dependence on~$m$ by a subscript, $\psi_m(x) := \psi(x,m)$.
Then for any fixed~$m$, we can take the scalar product~\eqref{print}. 
On families of solutions~$\psi, \phi \in \Cisco(\scrM \times I, S\scrM)$ of~\eqref{cauchy2},
we introduce a scalar product by integrating over the mass parameter,
\beq \label{spm}
( \psi | \phi)_I := \int_I (\psi_m | \phi_m)_m \: \dd m
\eeq
\nindex{agt@$( . \vert . )_I$ -- scalar product on families of Dirac solutions}%
\tindex{bb@$( . \vert . )_I$ -- scalar product on families of Dirac solutions}%
(where~$\dd m$ is the Lebesgue measure). Forming the completion, we obtain the
Hilbert space~$(\H, (.|.)_I)$. It consists of measurable functions~$\psi(x,m)$
such that for almost all~$m \in I$, the function~$\psi(.,m)$ is a weak solution of the Dirac
equation which is square integrable over any Cauchy surface.
Moreover, this spatial integral is integrable over~$m \in I$, so that
the scalar product~\eqref{spm} is well-defined. We denote the norm on~$\H$
by~$\| . \|_I$.

For the applications, it is useful to introduce
a subspace of the solutions of the form~\eqref{CscmH}:

\begin{Def} \label{defHinf}
We let~$\H^\infty \subset \Cisco(\scrM \times I, S\scrM) \cap \H$ be a subspace
of the smooth solutions with the following properties:
\bitem
\item[{\rm{(i)}}] $\H^\infty$ is invariant under multiplication by smooth functions in the mass parameter,
\beq \eta(m)\, \psi(x,m) \in \H^\infty \qquad \forall\: \psi \in \H^\infty,\;\eta \in C^\infty(I) \:. \eeq
\item[{\rm{(ii)}}] For every~$m \in I$,
the set~$\H^\infty_m := \{ \psi(.,m) \,|\, \psi \in \H^\infty\}$ is a dense subspace of~$\H_m$,
\beq \overline{\H^\infty_m}^{(.|.)_m} = \H_m \qquad \forall \:m \in I \:. \eeq
\eitem
We refer to~$\H^\infty$ as the {\bf{domain}} for the mass oscillation property.
\sindex{mass oscillation property!domain for}%
\nindex{agu@$\H^\infty$ -- domain for mass oscillation property}%
\tindex{ff@$\H^\infty$ -- domain for mass oscillation property}%
\end{Def} \noindent
The simplest choice is to set~$\H^\infty = \Cisco(\scrM \times I, S\scrM) \cap \H$, but in some applications
it is preferable to choose~$\H^\infty$ as a proper subspace of~$\Cisco(\scrM \times I, S\scrM) \cap \H$. 
(for example, in~\cite[Section~6]{infinite}, the space~$\H^\infty$ was chosen as being spanned by a finite number of angular modes, making it unnecessary to prove estimates uniform in the angular mode).

Our motivation for considering a variable mass parameter is that
integrating over the mass parameter should improve the decay properties
of the wave function for large times (similar as explained in the introduction
in the vacuum Minkowski space).
This decay for large times should also make it possible to integrate the Dirac operator
in the inner product~\eqref{stip} by parts without boundary terms,
\beq \bra \Dir \p \psi | \p \phi \ket = \bra \p \psi | \Dir \p \phi \ket \:, \eeq
implying that the solutions for different mass parameters
should be orthogonal with respect to this inner product.
Instead of acting with the Dirac operator, it is technically easier to
work with the operator of multiplication by~$m$, which we denote by
\nindex{agv@$T$ -- operator of multiplication by mass parameter}%
\beq \label{Tmdef}
T \::\: \H \rightarrow \H \:,\qquad (T \psi)_m = m \,\psi_m \:.
\eeq
In view of property~(i) in Definition~\ref{defHinf}, this operator leaves~$\H^\infty$ invariant,
\beq T|_{\H^\infty} \::\: \H^\infty \rightarrow \H^\infty \:. \eeq
Moreover, $T$ is a symmetric operator, and it is bounded because the interval~$I$ is,
\beq %\label{Tsymm}
T^* = T \in \Lin(\H) \:. \eeq
Finally, integrating over~$m$ gives the operation
\beq \label{pdef}
\p \::\: \H^\infty \rightarrow \Cisc(\scrM, S\scrM)\:,\qquad \p \psi = \int_I \psi_m\: \dd m \:.
\eeq
\nindex{agw@$\p$ -- operator of integration over mass parameter}%
We point out for clarity that~$\p \psi$ no longer satisfies a Dirac equation.
The following notions were introduced in~\cite{infinite}, and we refer the reader
to this paper for more details.

\begin{Def} \label{defwmop}
The Dirac operator~$\Dir=i \Pdd + \B$ on Minkowski space~$\scrM$ has the {\bf{weak mass oscillation property}} 
in the interval~$I = (m_L, m_R)$ with domain~$\H^\infty$ if the following conditions hold:
\sindex{mass oscillation property!weak}%
\bitem
\item[{\rm{(a)}}] For every~$\psi, \phi \in \H^\infty$, the
function~$\Sl \p \phi | \p \psi \Sr$ is integrable on~$\scrM$. Moreover, there is
a constant~$c=c(\psi)$ such that
\beq \label{mbound}
|\bra \p \psi | \p \phi \ket| \leq c\, \|\phi\|_I \qquad \textnormal{for all} \: \phi \in \H^\infty \:. 
\eeq
\item[{\rm{(b)}}] For all~$\psi, \phi \in \H^\infty$,
\beq \label{mortho}
\bra \p T \psi | \p \phi \ket = \bra \p \psi | \p T \phi \ket \:.
\eeq
\eitem
\end{Def}

\begin{Def} \label{defsmop}
The Dirac operator~$\Dir=i \Pdd + \B$ on Minkowski space~$\scrM$ has the {\bf{strong mass oscillation property}}
in the interval~$I=(m_L, m_R)$ with domain~$\H^\infty$ if there is a constant~$c>0$ such that
\sindex{mass oscillation property!strong}%
\beq \label{smop}
|\bra \p \psi | \p \phi \ket| \leq c \int_I \, \|\phi_m\|_m\, \|\psi_m\|_m\: \dd m
\qquad \text{for all~$\psi, \phi \in \H^\infty$}\:.
\eeq
\end{Def}

\section{The Fermionic Signature Operator}
\sindex{fermionic signature operator}%
In this section we give abstract constructions based on the mass oscillation property.
We first assume that the {\em{weak mass oscillation property}} of Definition~\ref{defwmop} holds.
Then, in view of the inequality~\eqref{mbound}, every~$\psi \in \H^\infty$
gives rise to a bounded linear functional on~$\H^\infty$.
By continuity, this linear functional can be uniquely extended to~$\H$.
The Fr{\'e}chet-Riesz theorem allows us to represent
this linear functional by a vector~$u \in \H$; that is,
\beq (u | \phi)_I = \bra \p \psi | \p \phi \ket \qquad \forall\: \phi \in \H\:. \eeq
Varying~$\psi$, we obtain the linear mapping
\beq \Sig \::\: \H^\infty \rightarrow \H \:,\qquad
(\Sig \psi | \phi)_I = \bra \p \psi | \p \phi \ket \quad \forall\: \phi \in \H\:. \eeq
This operator is symmetric because
\beq  (\Sig \psi | \phi)_I = \bra \p \psi | \p \phi \ket = (\psi | \Sig \phi)_I \qquad \forall\: \phi, \psi \in \H^\infty\:. \eeq
Moreover, \eqref{mortho} implies that the operators~$\Sig$ and~$T$ commute,
\beq %\label{STcomm}
\Sig \, T = T\, \Sig \::\: \H^\infty \rightarrow \H \:. \eeq
Thus the weak mass oscillation property makes it possible to introduce~$\Sig$ as a densely
defined symmetric operator on~$\H$. It is indeed possible to construct a selfadjoint extension
of the operator~$\Sig^2$ (using the Friedrich's extension), giving rise to a functional calculus
with corresponding spectral measure (for details see~\cite[Section~3]{infinite}).
In this setting the operator~$\Sig$ and the spectral measure are operators
on the Hilbert space~$\H$ which involves an integration over the
mass parameter. In simple terms, this implies that all objects are defined
only for almost all values of~$m$ (with respect to the Lebesgue measure on~$I \subset \R$),
and they can be modified arbitrarily on
subsets of~$I$ of measure zero. But it does not seem possible to ``evaluate pointwise in the mass''
by constructing operators~$\Sig_m$ which act on the Hilbert space~$\H_m$ for fixed mass.

In view of this shortcoming, we shall not enter the spectral calculus based on the
weak mass oscillation operator.
Instead, we move on to the {\em{strong mass oscillation property}}, which
makes life much easier because it implies that~$\Sig$ is a bounded operator.
\begin{Thm} \label{thmsMOP}
The following statements are equivalent:
\begin{itemize}
\item[{\rm{(i)}}] The strong mass oscillation property holds.
\item[{\rm{(ii)}}] There is a constant~$c>0$ such that for all~$\psi, \phi \in \H^\infty$,
the following two relations hold:
\begin{align}
|\bra \p \psi | \p \phi \ket| &\leq c\, \|\psi\|_I\, \|\phi\|_I \label{mb1} \\
\bra \p T \psi | \p \phi \ket &= \bra \p \psi | \p T \phi \ket\:. \label{mb2}
\end{align}
\item[{\em{(iii)}}] There is a family of linear operators~$\Sig_m \in \Lin(\H_m)$ which are uniformly bounded,
\beq \sup_{m \in I} \|\Sig_m\| < \infty\:, \eeq
such that
\beq \label{Smdef}
\bra \p \psi | \p \phi \ket = \int_I (\psi_m \,|\, \Sig_m \,\phi_m)_m\: \dd m \qquad
\forall\: \psi, \phi \in \H^\infty\:.
\eeq
\end{itemize}
\end{Thm}
\Proof The implication (iii)$\Rightarrow$(i) follows immediately from the estimate
\beq |\bra \p \psi | \p \phi \ket| \leq \int_I \big| (\psi_m | \Sig_m \phi_m)_m \big|\: \dd m
\leq \sup_{m \in I} \|\Sig_m\| \int_I \|\psi_m\|_m \:\|\phi\|_m\: \dd m \:. \eeq

In order to prove the implication~(i)$\Rightarrow$(ii), we first apply the Schwarz inequality
to~\eqref{smop} to obtain
\begin{align}
|\bra \p \psi | \p \phi \ket| &\leq c \int_I \, \|\phi_m\|_m\, \|\psi_m\|_m\: \dd m \notag \\
&\leq c \:\Big( \int_I \|\phi_m\|_m^2\: \dd m \Big)^\frac{1}{2}
\Big( \int_I \|\psi_m\|_m^2\: \dd m \Big)^\frac{1}{2} = c\: \|\phi\|_I\, \|\psi\|_I \:,
\end{align}
proving~\eqref{mb1}. Next, given~$N \in \N$ we subdivide the interval~$I=(m_L, m_R)$
by choosing the intermediate points
\beq m_\ell = \frac{\ell}{N}\: (m_R-m_L) + m_L\:,\qquad \ell=0, \ldots, N\:. \eeq
Moreover, we choose non-negative test functions~$\eta_1, \ldots, \eta_N \in C^\infty_0(\R)$ which
form a partition of unity and are supported in small sub-intervals, meaning that
\beq \label{part}
\sum_{\ell=1}^N \eta_\ell \big|_I = 1|_I \qquad \text{and} \qquad \supp \eta_\ell \subset (m_{\ell-1}, m_{\ell+1}) \:,
\eeq
where we set~$m_{-1}=m_L-1$ and~$m_{N+1}=m_R+1$. For any smooth function~$\eta \in C^\infty_0(\R)$ we
define the bounded linear operator~$\eta(T) \::\: \H^\infty \rightarrow \H^\infty$ by
\beq \big( \eta(T) \psi \big)_m = \eta(m)\: \psi_m \:. \eeq
Then, by linearity,
\begin{align}
&\bra \p T \psi | \p \phi \ket - \bra \p \psi | \p T \phi \ket \notag \\
&= \sum_{\ell, \ell'=1}^N \Big( \bra \p \,T \,\eta_{\ell}(T)\, \psi \,|\, \p \,\eta_{\ell'}(T)\, \phi \ket
- \bra \p \,\eta_{\ell}(T)\, \psi \,|\, \p \,T \,\eta_{\ell'}(T)\, \phi \ket \Big) \notag \\
&= \sum_{\ell, \ell'=1}^N 
\Big( \bra \p \,\big(T-m_\ell \big) \,\eta_{\ell}(T)\, \psi \,|\, \p \,\eta_{\ell'}(T)\, \phi \ket \notag \\
&\qquad\qquad
- \bra \p \,\eta_{\ell}(T)\, \psi \,|\, \p \,\big( T-m_\ell \big) \,\eta_{\ell'}(T)\, \phi \ket \big) \:.
\end{align}
Taking the absolute value and applying~\eqref{smop}, we obtain
\beq \big| \bra \p T \psi | \p \phi \ket - \bra \p \psi | \p T \phi \ket \big|
\leq c \sum_{\ell, \ell'=1}^N
\int_I | m-m_\ell | \:\eta_\ell(m)\, \eta_{\ell'}(m)\: \|\phi_m\|_m\, \|\psi_m\|_m\: \dd m \:. \eeq
In view of the second property in~\eqref{part}, we only get a contribution if~$|\ell-\ell'| \leq 1$.
Moreover, we know that~$| m-m_\ell | \leq 2 \,|I|/N$ on the support of~$\eta_\ell$. Thus
\begin{align}
\big| \bra \p T \psi | \p \phi \ket - \bra \p \psi | \p T \phi \ket \big|
&\leq \frac{6 c\,|I|}{N} \sum_{\ell=1}^N
\int_I \eta_\ell(m) \, \|\phi_m\|_m\, \|\psi_m\|_m\: \dd m \notag \\
&= \frac{6 c \,|I|}{N}
\int_I \|\phi_m\|_m\, \|\psi_m\|_m\: \dd m \:.
\end{align}
Since~$N$ is arbitrary, we obtain~\eqref{mb2}.

It remains to prove the implication~(ii)$\Rightarrow$(iii). Combining~\eqref{mb1}
with the Fr{\'e}chet-Riesz theorem, there is a bounded operator~$\Sig \in \Lin(\H)$ with
\beq \label{SdefH}
\bra \p \psi | \p \phi \ket = (\psi | \Sig \phi)_I \qquad \forall\: \psi, \phi \in \H^\infty\:.
\eeq
The relation~\eqref{mb2} implies that the operators~$\Sig$ and~$T$ commute.
Moreover, these two operators are obviously symmetric.
Hence the spectral theorem for commuting selfadjoint operators implies that there is a spectral measure~$F$
on~$\sigma(\Sig) \times I$ such that
\beq \label{dFdef}
\Sig^p\, T^q = \int_{\sigma(\Sig) \times I} \nu^p \,m^q\, \dd F_{\nu, m} \qquad \forall\: p,q \in \N\:.
\eeq
For given~$\psi, \phi \in \H^\infty$, we introduce the Borel measure~$\mu_{\psi, \phi}$
on~$I$ by
\beq \label{mudef}
\mu_{\psi, \phi}(\Omega) = \int_{\sigma(\Sig) \times \Omega} \nu\: \dd(\psi | F_{\nu, m} \phi)_I \:.
\eeq
Then~$\mu_{\psi, \phi}(I) = (\psi | \Sig \phi)_I$ and
\beq \mu_{\psi, \phi}(\Omega) = \int_{\sigma(\Sig) \times I} \nu\: \dd \big( \chi_\Omega(T)\, \psi \,\big|\, 
F_{\nu, m}\, \chi_\Omega(T)\, \phi \big)_I = (\chi_\Omega(T)\, \psi \,|\, \Sig \,\chi_\Omega(T)\, \phi)_I \:. \eeq
Since the operator~$\Sig$ is bounded, we conclude that
\begin{align}
|\mu_{\psi, \phi}(\Omega)| &\leq c \,\|\chi_\Omega(T)\, \psi\|_I\, \|\chi_\Omega(T)\, \phi\|_I 
\overset{\eqref{spm}}{=} c \left( \int_\Omega \|\psi\|_m^2 \:\dd m \; \int_\Omega  \|\phi\|_{m'}^2 \:\dd m' \right)^\frac{1}{2}
\notag \\
&\leq c\, |\Omega| \:\Big( \sup_{m \in \Omega} \|\psi_m\|_m \Big) \Big( \sup_{m' \in \Omega} \|\phi_{m'} \|_{m'} \Big)\:.
\label{RNes}
\end{align}
This shows that the measure~$\mu$ is absolutely continuous with respect to the Lebesgue measure.
The Radon-Nikodym theorem (see Theorem~\ref{thmRN}) %~\cite[Theorem~6.9]{rudin} or~\cite[\S VI.31]{halmosmt})
implies that there is a unique function~$f_{\psi, \phi} \in L^1(I, \dd m)$ such that
\beq \label{fppdef}
\mu_{\psi, \phi}(\Omega) = \int_\Omega f_{\psi, \phi}(m)\: \dd m\:.
\eeq
Using this representation in~\eqref{RNes}, we conclude that for any~$\varphi \in \R$,
\beq \re \left( \E^{\cI \varphi} \int_\Omega f_{\psi, \phi}(m)\: \dd m \right)
\leq \big| \mu_{\psi, \phi}(\Omega) \big| \leq
c\, |\Omega| \:\Big( \sup_{m \in \Omega} \|\psi_m\|_m \Big) \Big( \sup_{m' \in \Omega} \|\phi_{m'} \|_{m'} \Big)\:. \eeq
As a consequence, for almost all~$m \in I$ (with respect to the Lebesgue measure~$\dd m$),
\beq \re \left( \E^{\cI \varphi} \:f_{\psi, \phi}(m) \right) \leq 
c\, \|\psi_m\|_m \:\|\phi_{m} \|_{m} \:. \eeq
Since the phase factor is arbitrary, we obtain the pointwise bound
\beq | f_{\psi, \phi}(m) | \leq c\,\|\psi_m\|_m \: \|\phi_m \|_m \qquad \text{for almost all~$m \in I$}\:. \eeq
Using this inequality, we can apply the Fr{\'e}chet-Riesz theorem to obtain a unique
operator~$\Sig_m \in \Lin(\H_m)$ such that
\beq \label{Smdef2}
f_{\psi, \phi}(m) = (\psi_m | \Sig_m \phi_m)_m \qquad \text{and} \qquad
\|\Sig_m\| \leq c\:.
\eeq
Combining the above results, for any~$\psi, \phi \in \H^\infty$ we obtain
\begin{align}
\bra \p \psi | \p \phi \ket &\overset{\eqref{SdefH}}{=} (\psi | \Sig \phi)_I
\overset{\eqref{dFdef}}{=}\int_{\sigma(\Sig) \times I} \nu\: d( \psi \,|\, F_{\nu, m}\, \phi )_I \notag \\
&\overset{\eqref{mudef}}{=} \int_I \Diff\mu_{\psi, \phi} \overset{\eqref{fppdef}}{=} \int_I f_{\psi, \phi}(m)\: \dd m
\overset{\eqref{Smdef2}}{=} \int_I (\psi_m | \Sig_m \phi_m)_m\: \dd m \:.
\end{align}
This concludes the proof.
\QED

Comparing the statement of Theorem~\ref{thmsMOP}~(ii) with Definition~\ref{defwmop},
we immediately obtain the following result.
\begin{Corollary} The strong mass oscillation property implies the weak mass oscillation property.
\end{Corollary}

We next show uniqueness as well as the independence of the choice of the interval~$I$.
\begin{Prp} {\bf{(uniqueness of~$\Sig_m$)}} \label{prpunique}
The family~$(\Sig_m)_{m \in I}$ in the statement of Theorem~\ref{thmsMOP}
can be chosen such that for all~$\psi, \phi \in \H^\infty$, the expectation
value~$ f_{\psi, \phi}(m) := (\psi_m | \Sig_m \phi_m)_m$ is continuous in~$m$,
\beq \label{flip}
f_{\psi, \phi} \in C^0_0(I) \:.
\eeq
The family~$(\Sig_m)_{m \in I}$ with the properties~\eqref{Smdef} and~\eqref{flip} is unique.
Moreover, choosing two intervals~$\check{I}$ and~$I$ with~$m \in \check{I} \subset I$
and~$0 \not \in \overline{I}$, 
and denoting all the objects constructed in~$\check{I}$ with an additional check,
we have
\beq \label{ScS}
\check{\Sig}_m = \Sig_m \:.
\eeq
\end{Prp}
\Proof Let us show that the function~$f_{\psi, \phi}$ is continuous.
To this end, we choose a function~$\eta \in C^\infty_0(I)$. Then for any~$\varepsilon>0$
which is so small that~$B_\varepsilon(\supp \eta) \subset I$, we obtain
\begin{align}
\int_I & \Big( f_{\psi, \phi}(m+\varepsilon) - f_{\psi, \phi}(m) \Big) \:\eta(m)\: \dd m \notag \\
&= \int_I f_{\psi, \phi}(m) \:\Big( \eta(m-\varepsilon) - \eta(m) \Big)\: \dd m \notag \\
&\!\overset{(\ast)}{=}\, \bra \int_I \Big( \eta(m-\varepsilon) - \eta(m) \Big) \psi_m\: \dd m \:|\: \p \phi \ket \notag \\
&= \bra \int_I \eta(m) \: \Big( \psi_{m+\varepsilon} - \psi_m \Big)\: \dd m \:|\: \p \phi \ket \:,
\end{align}
where in~($\ast$) we used~\eqref{dFdef} and~\eqref{mudef}. Applying~\eqref{mb1}, we obtain
\beq \left| \int_I \Big( f_{\psi, \phi}(m+\varepsilon) - f_{\psi, \phi}(m) \Big) \:\eta(m)\: \dd m \right|
\leq c \, \|\psi_{+\varepsilon} - \psi\|_I\: \|\phi\|_I\: \sup_I |\eta| \:, \eeq
where the vector~$\psi_{+\varepsilon} \in \H^\infty$ is defined by~$(\psi_{+\varepsilon})_m := \psi_{m+\varepsilon}$.
Since
\beq \lim_{\varepsilon \searrow 0} \|\psi_{+\varepsilon} - \psi\|_I = 0 \eeq
and~$\eta$ is
arbitrary, we conclude that~$f_{\psi, \phi}$ is continuous~\eqref{flip}.
This continuity is important because it implies that the function~$f_{\psi, \phi}$ is uniquely defined
pointwise (whereas in~\eqref{fppdef} this function could be modified arbitrarily on sets of measure zero).

In order to prove~\eqref{ScS}, we note that the representation~\eqref{SdefH} implies that
\beq (\psi | \check{\Sig} \phi)_I = (\psi | \Sig \phi)_I \qquad \text{for all~$\psi, \phi \in \check{\H}^\infty$}\:. \eeq
Using~\eqref{mudef} and~\eqref{fppdef}, it follows that
\beq \int_\Omega \check{f}_{\psi, \phi}(m)\:\dd m = \int_\Omega f_{\psi, \phi}(m)\:\dd m
\qquad \text{for all~$\Omega \subset \check{I}$}\:. \eeq
Choosing~$\check{f}_{\psi, \phi}(m)$ and~$f_{\psi, \phi}(m)$ as continuous functions,
we conclude that they coincide for every~$m \in \check{I}$.
It follows from~\eqref{Smdef} that the operators~$\check{\Sig}_m$ and~$\Sig_m$ coincide.
This concludes the proof.
\QED

\section{The Unregularized Kernel of the Fermionic Projector}% as a Distribution}
\label{secPunreg}
\sindex{fermionic projector!unregularized kernel of}%
We now explain how the fermionic signature operator can be used for the construction
of the so-called fermionic projector. This will give a direct connection to the kernel of the fermionic projector
introduced abstractly for causal fermion systems in Chapter~\ref{secbrief} (see~\eqref{Pxydefintro}).
We will explain this connection, which will be elaborated on further in Section~\ref{seccl}.

It follows directly from its defining equation~\eqref{Smdef} that the operator~$\Sig_m$ is symmetric.
Thus the spectral theorem gives rise to the spectral decomposition
\beq \Sig_m = \int_{\sigma(\Sig_m)} \nu\: \dd E_\nu \:, \eeq
where~$E_\nu$ is the spectral measure (see, for example, \cite{reed+simon}).
The spectral measure gives rise to the spectral calculus
\beq f(\Sig_m) = \int_{\sigma(\Sig_m)} f(\nu)\: \dd E_\nu \:, \eeq
where~$f$ is a bounded Borel function.

\begin{Def} \label{ferm_proj_Def}
Assume that the Dirac operator~$\Dir$ on~$(\scrM,g)$ 
satisfies the strong mass oscillation property (see Definition~\ref{defsmop}).
We define the operators
\beq P_\pm \::\: C^\infty_0(\scrM, S\scrM) \rightarrow \H_m \eeq
by
\beq \label{Ppmdef2}
P_+ = \chi_{[0, \infty)}(\Sig_m)\, k_m \qquad \text{and} \qquad P_- = -\chi_{(-\infty, 0)}(\Sig_m)\, k_m
\eeq
(where~$\chi$ denotes the characteristic function).
The {\bf{fermionic projector}}~$P$ is defined by~$P=P_-$.
\end{Def}

\begin{Prp} \label{prpPpm} 
For all~$\phi, \psi \in C^\infty_0(\scrM, S\scrM)$, the operators~$P_\pm$ are symmetric,
\beq \bra P_\pm \phi \,|\, \psi \ket = \bra \phi \,|\, P_\pm \psi \ket \:. \eeq
Moreover, the image of~$P_\pm$ is the positive respectively negative
spectral subspace of~$\Sig_m$; that is,
\beq \label{dense}
\overline{P_+(C^\infty_0(\scrM, S\scrM))} = E_{(0, \infty)}(\H_m) \:,\qquad
\overline{P_-(C^\infty_0(\scrM, S\scrM))} = E_{(-\infty, 0)}(\H_m) \:.
\eeq
\end{Prp}
\Proof According to Proposition~\ref{prpdual},
\begin{align}
\bra P_- \phi \,|\, \psi \ket = (P_- \phi \,|\, k_m \psi )_m
&= - \big( \chi_{(-\infty, 0)}(\Sig_m)\, k_m \phi \,\big|\, k_m \psi \big)_m \notag \\
&= - \big( k_m\,\phi \,\big|\, \chi_{(-\infty, 0)}(\Sig_m)\, k_m \psi \big)_m
= \bra \phi \,|\, P_- \psi \ket \:.
\end{align}
The proof for~$P_+$ is similar.
The relations~\eqref{dense} follow immediately from the fact
that~$k_m(C^\infty_0(\scrM, S\scrM))$ is dense in~$\H_m$.
\QED

Similar as in~\cite[Theorem~3.12]{finite}, the fermionic projector 
can be represented by a two-point distribution on~$\scrM$. As usual,
we denote the space of test functions (with the Fr{\'e}chet topology)
by~${\mathcal{D}}$ and define the space of distributions~${\mathcal{D}}'$ as its dual space.
\begin{Thm} Assume that the strong mass oscillation property holds. Then
there is a unique distribution~${\mathcal{P}} \in {\mathcal{D}}'(\scrM \times \scrM)$ such that 
for all~$\phi, \psi \in C^\infty_0(\scrM, S\scrM)$,
\beq \bra \phi | P \psi \ket = {\mathcal{P}}(\phi \otimes \psi) \:. \eeq
\end{Thm}
\Proof According to Proposition~\ref{prpdual} and Definition~\ref{ferm_proj_Def},
\beq \bra \phi | P \psi \ket = (k_m \phi \,|\, P \psi)_m = - (k_m \phi \,|\, \chi_{(-\infty, 0)}(\Sig_m) \,k_m \psi)_m \:. \eeq
Since the norm of the operator~$\chi_{(-\infty, 0)}(\Sig_m)$ is bounded by one, we conclude that
\beq |\bra \phi | P \psi \ket| \leq \| k_m \phi\|_m \:\|k_m \psi\|_m
= ( \bra \phi | k_m \phi \ket \: \bra \psi | k_m \psi \ket )^\frac{1}{2} \:, \eeq
where in the last step we again applied Proposition~\ref{prpdual}.
As~$k_m \in {\mathcal{D}}'(\scrM \times \scrM)$, the right side is continuous on~${\mathcal{D}}(\scrM \times \scrM)$.
We conclude that also the functional~$\bra \phi | P \psi \ket$ is continuous on~${\mathcal{D}}(\scrM \times \scrM)$.
The result now follows from the Schwartz kernel theorem (see~\cite[Theorem~5.2.1]{hormanderI},
keeping in mind that this theorem applies just as well to bundle-valued distributions on a manifold
simply by working with the components in local coordinates and a local trivialization).
\QED
Exactly as explained in~\cite[Section~3.5]{finite}, it is convenient to use the standard notation with an integral kernel~$P(x,y)$,
\begin{align}
\bra \phi | P \psi \ket &= \iint_{\scrM \times \scrM} \Sl \phi(x) \,|\, P(x,y) \,\psi(y) \Sr_x \: \Diff\mu_\scrM(x)\: \Diff\mu_\scrM(y) \\
(P \psi)(x) &= \int_{\scrM} P(x,y) \,\psi(y) \: \Diff\mu_\scrM(y)
\end{align}
(where~$P(.,.)$ coincides with the distribution~${\mathcal{P}}$ above).
In view of Proposition~\ref{prpPpm}, we know that the last integral is not only a distribution,
but a function which is square integrable over every Cauchy surface.
Moreover, the symmetry of~$P$ shown in Proposition~\ref{prpPpm} implies that
\beq P(x,y)^* = P(y,x) \:, \eeq
where the star denotes the adjoint with respect to the spin inner product.

We next specify the normalization of the fermionic projector. We introduce an operator~$\Pi$ by
\beq \label{Pidef}
\Pi \::\: \H_m \rightarrow \H_m \:, \qquad
(\Pi \,\psi_m)(x) = -2 \pi \int_\scrN P(x,y)\, \nuslsh\, (\psi_m)|_\scrN(y)\: \Diff\mu_\scrN(y)  \:,
\eeq
where~$\scrN$ is any Cauchy surface.
\begin{Prp} {\bf{(spatial normalization)}} \label{prpspatnorm}
\sindex{normalization of fermionic projector!spatial}%
\sindex{fermionic projector!spatial normalization}%
The operator~$\Pi$ is a projection operator on~$\H_m$.
\end{Prp}
\Proof According to Theorem~\ref{thmkext}, %~\ref{prp21},
the spatial integral in~\eqref{Pidef}
can be combined with the factor~$k_m$ in~\eqref{Ppmdef2} to give the solution of the
corresponding Cauchy problem. Thus
\beq \Pi \::\: \H_m \rightarrow \H_m \:, \qquad
(\Pi \,\psi_m)(x) = \chi_{(-\infty, 0)}(\Sig_m)\, \psi_m \:, \eeq
showing that~$\Pi$ is a projection operator.
\QED

Instead of the spatial normalization, one could also consider the {\em{mass normalization}}
\sindex{normalization of fermionic projector!mass}%
\sindex{fermionic projector!mass normalization}%
(for details on the different normalization methods see~\cite{norm}). To this end, one needs
to consider families of fermionic projectors~$P_m$ indexed by the mass parameter. Then
for all~$\phi, \psi \in C^\infty_0(\scrM, S\scrM)$, we can use~\eqref{Smdef} and Proposition~\ref{prpdual}
to obtain
\begin{align}
&\bra \p (P_m \phi) \,|\, \p (P_{m'} \psi) \ket \notag \\
&= \int_I (P_m \phi \,|\, \Sig_m P_m \psi)_m\: \dd m
= \int_I (k_m \phi \,|\, \Sig_m \chi_{(-\infty, 0)}(\Sig_m)\, k_m \psi)_m\: \dd m \notag \\
&= \int_I \bra \phi \,|\, \Sig_m \chi_{(-\infty, 0)}(\Sig_m)\, k_m \psi \ket\: \dd m = -\bra \phi \,|\, \p(\Sig_m P_m \psi) \ket \:,
\end{align}
which can be written in a compact formal notation as
\beq P_m \,P_{m'} = \delta(m-m')\: (-\Sig_m) \, P_m \:. \eeq
Due to the factor~$(-\Sig_m)$ on the right, in general the fermionic projector does {\em{not}} satisfy the
mass normalization condition.
The mass normalization condition could be arranged by modifying the definition~\eqref{Ppmdef2} to
\beq \Sig_m^{-1}\, \chi_{(-\infty, 0)}(\Sig_m)\, k_m \:. \eeq
Here we prefer to work with the spatial normalization. For a detailed discussion of the
different normalization methods we refer to~\cite[Section~2]{norm}.

Finally, the spatial normalization property of Proposition~\ref{prpspatnorm} makes it possible
to obtain a representation of the fermionic projector in terms of one-particle states.
To this end, one chooses an orthonormal basis~$(\psi_j)_{j \in \N}$
of the subspace~$\chi_{(-\infty, 0)}(\Sig_m) \subset \H_m$. Then
\beq \label{Pxygenmink}
P(x,y) = -\sum_{j=1}^\infty |\psi_j(x) \Sr \Sl \psi_j(y)|
\eeq
with convergence in~${\mathcal{D}}'(\scrM \times \scrM)$
(for details see~\cite[Proposition~3.13]{finite}).

This formulas is reminiscent of the decomposition of the kernel of the fermionic projector
into physical wave functions in~\eqref{Prep}. Indeed, these formulas can be understood as
being completely analogous, with the only difference that~\eqref{Pxygenmink} is formed of
wave functions in Minkowski space, whereas in~\eqref{Prep} one works abstractly with
the physical wave functions of a general causal fermion system.
The connection can be made more precise if one identifies the structures of the causal
fermion system with corresponding structures in Minkowski space.
In order to avoid technicalities and too much overlap with~\cite{cfs},
here we shall not enter the details of these identifications (which are worked out in~\cite[Section~1.2]{cfs}).
Instead, we identify~\eqref{Pxygenmink} with~\eqref{Prep} as describing the same object,
on one side in Minkowski space, and on the other side as abstract object of the corresponding causal
fermion system. With this identification, the Hilbert space~$\H$ of the causal fermion system
corresponds to the negative spectral subspace of the fermionic signature operator~$\Sig_m$.
In the vacuum, this gives us back the subspace of all negative frequency solutions as considered
in the example of Exercise~\ref{exminkfirst}.
However, the above identification has one shortcoming: the wave functions in~\eqref{Pxygenmink}
have not yet been regularized. This is why we refer to~$P(x,y)$ as the {\em{unregularized}} kernel.
In order to get complete agreement between~\eqref{Pxygenmink} and~\eqref{exminkfirst},
one needs to introduce an ultraviolet regularization. To this end, one proceeds similar as explained in 
the example in Section~\ref{seclco}: One introduces regularization
operators~$({\mathfrak{R}}_\varepsilon)_{\varepsilon>0}$, computes the local correlation
operators~$F^\varepsilon(x)$ and defines the measure~$\rho$ as the push-forwards~$\Diff\rho=F^\varepsilon_* \Diff\mu_\scrM$. We will come back to this construction in Chapter~\ref{seccl}.

\section{Exercises}

\begin{Exercise} {\em{
Let~$\scrM$ be the ``spacetime strip''
\beq \scrM = \{ (t,\vec{x}) \in \R^{1,3} \text{ with } 0 < t < T \}\:. \eeq
Show that for any solution~$\psi \in \Cisc(\scrM, S \scrM) \cap \H_m$ of the Dirac equation,
the following inequality holds,
\beq \big| \bra \psi | \phi \ket \big| \leq T\: \|\psi\|_m\: \|\phi\|_m \:. \eeq
This estimate illustrates why in spacetimes of finite lifetime, the spacetime inner product is
a bounded sesquilinear form on~$\H_m$.
}} \end{Exercise}

\begin{Exercise} {\em{
Let~$\scrM$ again be the ``spacetime strip'' of the previous exercise.
Let~$\psi, \phi \in \H^\infty := \H \cap \Cisco(\scrM \times I, S\scrM)$ be families of
smooth Dirac solutions of spatially compact support, with compact support in the mass parameter.
Moreover, we again define the operators~$\p$ and~$T$ as in~\eqref{Tmdef} and~\eqref{pdef}.
Does the equation
\beq \bra \p T \psi | \p \phi \ket = \bra \p \psi | \p T \phi \ket \eeq
(which appears in the weak mass oscillation property) in general hold?
Justify your answer by a proof or a counter example.
}} \end{Exercise}

\begin{Exercise} {\em{
Let~$\scrM$ again be the ``spacetime strip'' of the previous exercises.
Moreover, as in Exercise~\ref{ex7} we again let~$\H \subset \H_m$ be a
finite-dimensional subspace of the Dirac solution space~$\H_m$, consisting
of smooth wave functions of spatially compact support; that is,
\beq \H \subset \Cisc(\scrM, S\scrM) \cap H_m \qquad \text{finite-dimensional} \:. \eeq
Show that the fermionic signature operator~$\Sig \in \Lin(\H)$ defined by
\beq \bra \psi | \phi \ket = (\psi | \Sig \phi)_m \qquad \text{for all~$\psi, \phi \in \H$} \eeq
can be expressed within the causal fermion system by
\beq \Sig = -\int_M x \:\Diff\rho(x) \eeq
(where~$\rho$ is again the push-forward of~$\Diff\mu_\scrM$).
}} \end{Exercise}

\begin{Exercise} {\em{
Let~$E$ be the Banach space~$E=C^0([0,1], \C)$
and~$\Lambda : E \times E \rightarrow \C$ be sesquilinear, bounded and positive
semi-definite.
\bitem
\item[(a)] Assume that~$\Lambda$ satisfies for a suitable constant~$c>0$ and all~$f,g \in E$ the inequality
\beq \label{eq1new}
\big| \Lambda(f,g) \big| \leq c \sup_{x \in [0,1]} \big| f(x)\: g(x) \big| \:.
\eeq
Show that there is a regular bounded Borel measure~$\mu$ such that
\beq \Lambda(f,g) = \int_0^1 \overline{f(x)}\: g(x)\: \Diff\mu(x) \:. \eeq
\item[(b)] Now make the stronger assumption that~$\Lambda$ satisfies for a suitable
constant~$\tilde{c}>0$ and all~$f,g \in E$ the inequality
\beq \label{eq2}
\big| \Lambda(f,g) \big| \leq \tilde{c} \int_0^1 \big| f(x)\: g(x) \big| \:\dd x \:.
\eeq
Show that~$\mu$ is absolutely continuous w.r.to the Lebesgue measure.
Show that there is a non-negative function~$h \in L^1([0,1], \dd x)$ such that
\beq \Lambda(f,g) = \int_0^1 \overline{f(x)}\: g(x)\: h(x)\: \dd x \:. \eeq
Show that~$h$ is pointwise bounded by~$c$.
\item[(c)] In order to clarify the different assumptions in this exercise,
give an example for a sesquilinear, bounded and positive semi-definite
functional~$\Lambda$ which violates~\eqref{eq1new}.
Give an example which satisfies~\eqref{eq1new} but violates~\eqref{eq2}.
\eitem
}} \end{Exercise}

\begin{Exercise} {{(Toward the mass oscillation property - part 1)}} {\em{
\sindex{mass oscillation property}%
This exercise illustrates the mass oscillation property.
Let~$0 < m_L < m_R$ and~$\eta \in C^\infty_0((m_L, m_R))$. Show that the
function~$f$ given by
\beq f(t) = \int_{m_L}^{m_R} \eta(m)\: \E^{-\cI \sqrt{1+m^2}\: t} \dd m \eeq
has rapid decay.
Does this result remain valid if~$m_L$ and~$m_R$ are chosen to have opposite signs?
Justify your finding by a proof or a counter example.
}} \end{Exercise}

\begin{Exercise} {{(Toward the mass oscillation property - part 2)}} {\em{
\sindex{mass oscillation property}%
Let~$R_T$ be the ``spacetime strip''
\beq R_T = \{ (t,\vec{x}) \in \R^{1,3} \text{ with } 0 < t < T \}\:. \eeq
Show that for any solutions~$\psi,\phi \in \Cisc(\R^4,\C^4) \cap \H_m$ of the Dirac equation,
the following inequality holds,
\beq 
\big|\bra \psi | \phi \ket_T \big| \leq T\: \|\psi\|_m\: \|\phi\|_m \:,\quad\mbox{where}\quad \bra \psi | \phi \ket_T:=\int_{R_T}\Sl \psi(x)|\phi(x)\Sr\,\Diff^4x.
\eeq
This estimate illustrates how in spacetimes of finite lifetime, the spacetime inner product is
a bounded sesquilinear form on~$\H_m$.
}} \end{Exercise}

\begin{Exercise} {{(Toward the mass oscillation property - part 3)}} {\em{
\sindex{mass oscillation property}%
Let~$R_T$ again be the ``spacetime strip'' of the previous exercises.
Moreover, we again let~$\H \subset \H_m$ be a
finite-dimensional subspace of the Dirac solution space~$\H_m$, consisting
of smooth wave functions of spatially compact support; that is,
\beq \H \subset \Cisc(\R^4,\C^4) \cap \H_m \qquad \text{finite-dimensional} \:. \eeq
Show that the fermionic signature operator~$\Sig \in \Lin(\H)$ defined by
\beq \bra \psi | \phi \ket_T = (\psi | \Sig \phi)_m \qquad \text{for all~$\psi, \phi \in \H$} \eeq
can be expressed within the causal fermion system by
\beq \Sig = -\int_{R_T} x \:\Diff\rho(x) \eeq
(where~$\rho$ is again the push-forward of~$\Diff^4x$).
}} \end{Exercise}

\begin{Exercise} {{(The external field problem)}} {\em{
\sindex{external field problem}%
\noindent In physics, the notion of ``particle'' and ``anti-particle'' is often introduced as follows: Solutions of the Dirac equation with positive frequency are called ``particles'' and solutions with negative frequency ``anti-particles.'' In this exercise, we will check in how far this makes sense.

\noindent To this end, take a look at the Dirac equation in an external field:
\begin{align}\label{DiracExt}
(\cI \slashed{\partial}+\mathcal B-m) \psi = 0.
\end{align}
Assume that~$\mathcal B$ is time-dependent and has the following form:
\begin{align}
\mathcal B(t,x) = V \:\Theta(t-t_0) \Theta(t_1-t),
\end{align}
where~$V \in \R$, $\Theta$ denotes the Heaviside step function and~$t_0 =0, \: t_1=1$. In order to construct a solution thereof, for a given momentum~$\vec k$, we use plane wave solutions of the Dirac equation,
\beq \psi(t,\vec x) = \E^{- \cI \omega t + \cI \vec k \vec x} \chi_{\vec k}, \eeq
where~$\chi_{\vec k}$ is a spinor~$\in \C^4$, and patch them together suitably. (The quantity~$\omega$ is called the ``frequency'' or ``energy'', and~$\vec k$ the``momentum''.) To simplify the calculation, we set~$\vec k =  (k_1,0,0)^T.$ Proceed as follows:
\bitem
\item[(a)] First, take a look at the region~$t<t_0$. Reformulate~\eqref{DiracExt} such that there is only the time derivative on the left hand side. (Hint: Multiply by~$\gamma^0$.)
\item[(b)] Insert the plane wave ansatz with~$\vec k =  (k_1,0,0)^T$ into the equation. Your equation now has the form~$\omega \psi = H(k_1) \psi$. Show that the eigenvalues of~$H(k_1)$ are~$\pm \omega_0$ with~$\omega_0 := \sqrt{(k_1)^2+m^2 }$.
\item[(c)] Show that one eigenvector belonging to~$+ \omega_0$ is
$  \chi_0^+ := ( \frac{m + \omega_0}{k1},0,0,1)^{T}$ and that one eigenvector belonging to~$- \omega_0$ is~$ \chi_0^- := ( \frac{m - \omega_0}{k1},0,0,1)^{T}$. (Both eigenvalues have multiplicity two, but we do not need the other two eigenvectors here.)
\item[(d)] With this, you have constructed plane wave solutions~$\E^{- \cI (\pm \omega_{0}) t + \cI \vec k \vec x} \chi_0^{\pm}$ for~$t < t_0$ and also for~$t > t_1$. By transforming~$m \rightarrow (m-V)$, you immediately obtain plane wave solutions also for~$t_0 < t <t_1$. Denote the respective quantities by~$\omega_1$ and~$\chi_1^{\pm}$.
\item[(e)] Assume that for~$t<t_0$ there is one ``particle'' present with wave function of the form
\beq \psi(t,\vec x) =  \E^{- \cI  \omega_{0} t + \cI \vec k \vec x} \:\chi_0^{+}\: \: \: \textrm{ for } t<t_0 \:. \eeq
Assume that the solution for~$t_0 < t < t_1$ takes the form
\beq A \,\E^{- \cI \omega_{1} t + \cI \vec k \vec x} \chi_1^{+}+ B \,\E^{- \cI (- \omega_{1}) t + \cI \vec k \vec x} \chi_1^{-} \: \: \: \textrm{ with } A, B \in \R.\eeq
Calculate~$A$ and~$B$ for the case~$k_1=1$ and~$V=m$ by demanding continuity of the solution at~$t=t_0$.
\item[(f)] Assume that for~$t > t_1$ the solution  takes the form
\beq C \E^{- \cI \omega_{0} t + \cI \vec k \vec x} \chi_0^{+}+ D \E^{- \cI (- \omega_{0}) t + \cI \vec k \vec x} \chi_0^{-} \: \: \: \textrm{ with } C, D \in \C.\eeq
Calculate~$C$ and~$D$ for~$m=2$ by demanding continuity of the solution at~$t=t_1$ (here you may want to use 
computer algebra).
\item[(g)] Interpret what you have found. Why could this be called the ``external field \textit{problem}''?
\eitem
}} \end{Exercise}

\chapter{Fourier Methods}
In the previous chapter, the fermionic signature operator and the unregularized fermionic projector
were constructed abstractly. We now turn to the question how to compute them in the Minkowski
vacuum. This question can be addressed and answered with Fourier methods.
Since these techniques are frequently used and of independent interest,
we introduce them from a general perspective before entering the
proof of the mass oscillation properties and the construction of the fermionic signature operator.
More details can be found in~\cite{infinite, hadamard}.

\section{The Causal Green's Operators} \label{secgreen}
We already encountered Green's operators in Chapter~\ref{secshs}
when solving the Cauchy problem with methods of hyperbolic partial differential equations (see Theorem~\ref{thmgreenhyp}). In Minkowski space, these Green's operators can be computed
in more detail with Fourier methods.
Our starting point is the definition of the {\em{Green's operator}}~$s_m(x,y)$ of the vacuum Dirac equation
by the distributional equation
\sindex{Green's operator!in Minkowski space}%
\nindex{agx@$s_m(x,y)$ -- Dirac Green's operator in Minkowski space}%
\beq
(\cI \Pdd_x - m) \: s_m(x,y) = \delta^4(x-y) \: , \label{Greendef}
\eeq
where~$\delta^4(x,y)$ denotes the four-dimensional $\delta$ distribution.
Taking the Fourier transform of~\eqref{Greendef},
\beq \label{Ft}
s_m(x,y) = \int \frac{\dd^4k}{(2 \pi)^4} \:s_m(k)\: \E^{-\cI k (x-y)}
\eeq
(where~$x, y \in \scrM$ are spacetime points, $k$ is the four-momentum,
and~$k(x-y)$ denotes the Minkowski inner product)
we obtain the algebraic equation
\beq \label{kalgebra}
(\slashed{k} - m)\: s_m(k) = \1\:.
\eeq
Multiplying by~$\slashed{k}+m$ and using the identity
$(\slashed{k}-m)(\slashed{k}+m)=k^2-m^2$, one sees that if~$k^2 \neq m^2$,
the matrix~$\slashed{k}-m$ is invertible.
If conversely~$k^2=m^2$, we have~$(\slashed{k}-m)^2 = -2 m (\slashed{k}-m)$,
which shows that the matrix~$\slashed{k}-m$ is diagonalizable with eigenvalues~$-2m$ and zero. Since the Dirac 
matrices~\eqref{Dirrep} are trace-free, we have~$\Tr(\slashed{k}-m) =-4m$.
It follows that the matrix~$\slashed{k}-m$ has a two-dimensional kernel if~$k$ is on the mass shell.
This shows that the Green's operator of the Dirac equation is {\em{not unique}}. If we add to it any vector
in the kernel of~$\slashed{k}-m$ (in other words, if we add to it a solution of the homogeneous Dirac equation),
it still satisfies the defining equation~\eqref{Greendef} (for details, see~\cite{bjorken}.)

A convenient method for solving~\eqref{kalgebra} for~$s_m(k)$ is to
use a~$\pm \cI \varepsilon$-regulari\-zation on the mass shell. Common choices are the
{\em{advanced}}
\sindex{Green's operator!advanced}%
\sindex{Green's operator!retarded}%
\sindex{Green's operator!causal}%
and the {\em{retarded}} Green's
functions, which are defined by
\beq
        s^{\lor}_{m}(k) = \lim_{\varepsilon \searrow 0}
            \frac{\slashed{k} + m}{k^{2}-m^{2}-\cI \varepsilon k^{0}}
            \qquad {\mbox{and}} \qquad
           s^{\land}_{m}(k) = \lim_{\varepsilon \searrow 0}
            \frac{\slashed{k} + m}{k^{2}-m^{2}+\cI \varepsilon k^{0}} \:,
        \label{8b}
\eeq
\nindex{agy@$s^\vee_m, s^\wedge_m$ -- causal Green's operators in the vacuum}%
respectively (with the limit~$\varepsilon \searrow 0$ taken
in the distributional sense). Computing their Fourier transform~\eqref{Ft},
one sees that they are {\em{causal}} in the sense that
their supports lie in the upper and lower light cone, respectively,
\sindex{causality!of Green's operator}%
\beq \label{GF}
\supp s^\lor_m(x,.) \subset J_x^\lor \;,\spc
\supp s^\land_m(x,.) \subset J_x^\land\:.
\eeq
Mathematically, the formulas in~\eqref{8b} define the Green's operators
in momentum space as tempered distributions. Taking their Fourier
transform~\eqref{Ft}, the advanced and retarded Green's operators are tempered distributions
in the variable~$\xi := y-x$. We also regard these distributions as integral kernels of
corresponding operators on the wave functions; that is,
\beq (s_m(\psi))(x) := \int_\scrM s_m(x,y)\: \psi(y)\: \dd^4y \:. \eeq
We thus obtain operators
\beq %\label{sop}
s_m^\wedge, s_m^\vee \::\: C^\infty_0(\scrM, S\scrM) \rightarrow \Cisc(\scrM, S\scrM)\:. \eeq
Here~$C^\infty_0(\scrM, S\scrM)$ denote the smooth functions with compact support in~$\scrM$,
\sindex{wave function!spatially compact}%
\nindex{aau@$\Cisc(\scrM, S\scrM)$ -- spatially compact spinorial wave functions}%
\tindex{ff@$\Cisc(\scrM, S\scrM)$ -- spatially compact spinorial wave functions}%
taking values in the spinors, and~$\Cisc$ denotes the smooth functions with
spatially compact support.

\section{The Causal Fundamental Solution and Time Evolution} \label{seckmevolve}
We now state a few properties of the Green's operators and explain why
they are useful. The considerations in this section are valid more generally
in the presence of an external potential. Then the defining equation of the
Green's operator~\eqref{Greendef} is modified similar to~\eqref{Dout} to
\beq
(\cI \Pdd_x + \B - m) \: s_m(x,y) = \delta^4(x-y) \:, \label{Greendefext}
\eeq
where~$\B$ is again a multiplication operator satisfying the symmetry condition~\eqref{Bsymm}.
Then the existence of Green's operators can no longer be proven by Fourier transformation.
Instead, one can use methods of hyperbolic partial differential equations as introduced in Chapter~\ref{secshs}
(see Section~\ref{seccauchyglobhyp}).
Here we do not assume that the reader is familiar with these methods.
Instead, we simply assume that we are given advanced and retarded Green's operators.

The {\em{causal fundamental solution}}~$k_m$ is defined as the difference
between the advanced and the retarded Green's operator,
\sindex{causal fundamental solution!in Minkowski space}%
\nindex{aha@$k_m$ -- causal fundamental solution}%
\beq \label{kmdef2}
k_m(x,y) := \frac{1}{2\pi \cI} \left( s_m^{\vee}(x,y)-s_m^{\wedge}(x,y) \right) .
\eeq
It is a distribution that is causal in the sense that it vanishes if~$x$ and~$y$ have spacelike separation.
Moreover, it is a distributional solution of the homogeneous Dirac equation,
\beq (\cI \Pdd_x +\B - m) \: k_m(x,y) = 0 \:. \eeq

The unique solvability of the Cauchy problem allows us to introduce the time evolution operator
of the Dirac equation as follows. 
Solving the Cauchy problem with initial data at time~$t$ and evaluating the solution
at some other time~$t'$ gives rise to a mapping
\beq U^{t', t} \::\: \H_t \rightarrow \H_{t'}\:, \eeq
\nindex{ahb@$U^{t',t}$ -- time evolution operator}%
referred to as the {\em{time evolution operator}}.
\sindex{time evolution operator}%
Since the scalar product~\eqref{print} is time independent, the operator~$U^{t',t}$
is unitary. Moreover, using that the Cauchy problem can be solved forward and backward in time,
the unitary time evolution operators form a representation of the group~$(\R, +)$. More precisely,
\beq U^{t,t} = \1 \qquad \text{and} \qquad U^{t'',t'}\: U^{t',t} = U^{t'',t}\:. \eeq
Proposition~\ref{prp21} immediately gives the following representation of~$U^{t',t}$:
\begin{align}
\big( U^{t',t} \,\psi|_{t} \big)(\vec{y}) &= \int_{\R^3} U^{t',t}(\vec{y}, \vec{x})\: \psi(t, \vec{x})\: \dd^3x\:, 
\label{Ukern} 
\end{align}
where the kernel~$U^{t',t}(\vec x, \vec y)$ is defined as
\begin{align}
U^{t',t}(\vec{y}, \vec{x}) &= 2 \pi \,k_m \big( (t',\vec{y}), (t, \vec{x}) \big)\, \gamma^0\:.
\label{Ukm}
\end{align}

\section[Proof of the Weak Mass Oscillation Property]{Proof of the Weak Mass Oscillation Property in the Minkowski Vacuum} \label{secmweak}
In the remainder of this chapter, we return to the Dirac equation in Minkowski
space~\eqref{Greendef}. An external potential will be considered in the next chapter
(Chapter~\ref{secscatter}).

The mass oscillation property in the Minkowski vacuum can be proved using Fourier methods.
Here, we shall give two different approaches in detail.
The method of the first proof (in this section) is instructive because it gives an intuitive
understanding of ``mass oscillations.''
However, this method only yields the weak mass oscillation property. The second proof (Section~\ref{secmstrong}) is more abstract but also gives the strong mass oscillation property.

We again consider the foliation~$\scrN_t = \{(t, \vec{x}) \,|\, \vec{x} \in \R^3\}$ of constant time Cauchy hypersurfaces in a fixed reference frame~$(t, \vec{x})$ and a variable mass parameter~$m$
in the interval~$I=(m_L, m_R)$ with~$m_L, m_R > 0$. The families of solutions~$\psi = (\psi_m)_{m \in I}$ of the Dirac equations~$(i \Pdd - m) \psi_m = 0$ are contained in the Hilbert space~$(\H, (.|.))$ with scalar product~\eqref{spm}. 
The subspace~$\H^\infty \subset \H$ in Definition~\ref{defHinf} is chosen as 
\beq \label{Hinfchoice}
\H^\infty = \Cisco(\scrM \times I, S\scrM) \cap \H \:.
\eeq

For what follows, it is convenient to work with the Fourier transform in space; that is,
\beq \hat{\psi}(t, \vec{k}) = \int_{\R^3} \psi(t, \vec{x})\: \E^{-\cI \vec{k} \vec{x}}\: \dd^3x\:,\qquad
 \psi(t, \vec{x}) = \int_{\R^3} \frac{\dd^3k}{(2 \pi)^3} \:\hat{\psi}(t, \vec{k})\: \E^{\cI \vec{k} \vec{x}} 
 \:. \eeq
Then a family of solutions~$\psi \in \H^\infty$ has the representation
\beq \label{Frep}
\hat{\psi}_m(t, \vec{k}) = c_+(\vec{k},m) \: \E^{-\cI \omega(\vec{k}, m)\, t} + c_-(\vec{k}, m)
\: \E^{\cI \omega(\vec{k},m)\, t} \qquad \text{for all~$m \in I$\:,}
\eeq
with suitable spinor-valued coefficients~$c_\pm(\vec{k},m)$ and~$\omega(\vec{k},m):=\sqrt{|\vec{k}|^2+m^2}$.
Integrating over the mass parameter, we obtain a superposition of waves oscillating at different
frequencies. Intuitively speaking, this leads to destructive interference for large~$t$,
giving rise to decay in time. This picture can be made precise using integration by parts in~$m$,
as we now explain. Integrating~\eqref{Frep} over the mass by applying the operator~$\p$, \eqref{pdef},
we obtain
\begin{align}
\p \hat{\psi}(t, \vec{k}) &= \int_I
\left( c_+\: \E^{-\cI \omega t} + c_-\: \E^{\cI \omega t} \right) \dd m \notag \\
&= \int_I \frac{\cI}{t \,\partial_m \omega}
\left( c_+\: \partial_m \E^{-\cI \omega t} - c_-\: \partial_m \E^{\cI \omega t} \right) \dd m \notag \\
&= -\frac{\cI}{t}  \int_I \left[ \partial_m \Big( \frac{c_+}{\partial_m \omega} \Big)\: \E^{-\cI \omega t}
-\partial_m \Big( \frac{c_-}{\partial_m \omega} \Big)\: \E^{\cI \omega t} \right] \dd m \:
\end{align}
\noindent (we do not get boundary terms because~$\psi \in \H^\infty$
has compact support in~$m$).
With~$\partial_m \omega = m/\omega$, we conclude that
\beq \p \hat{\psi}(t, \vec{k}) = -\frac{\cI}{t}  \int_I \left[ \partial_m \Big( \frac{\omega\, c_+}{m} \Big)\:
\E^{-\cI \omega t} -\partial_m \Big( \frac{\omega\, c_-}{m} \Big)\: \E^{\cI \omega t} \right] \dd m \:. \eeq
Since the coefficients~$c_\pm$ depend smoothly on~$m$, the resulting integrand is
bounded uniformly in time, giving a decay at least like~$1/t$, that is, $|\p \hat{\psi}(t, \vec{k})| \lesssim 1/t$.
Iterating this procedure, one even can prove decay rates~$\lesssim 1/t^2, 1/t^3, \ldots$
The price one pays is that higher and higher powers in~$\omega$ come up in the integrand,
which means that in order for the spatial Fourier integral to exist, one needs a faster decay of~$c_\pm$ in~$|\vec{k}|$. Expressed in terms of the initial data, this means that every factor~$1/t$ gives rise to an additional spatial derivative acting on the initial data. This motivates the following basic estimate.
\Evtl{Braucht man denn hier die~$W^{2,2}$-norm? Gen\"ugt es nicht, dass man f\"ur
jedes~$\psi \in \H^\infty$ den quadratischen Abfall hat? Falls ja, vereinfache dies?}%

\begin{Lemma} \label{lemmappsi}
For any~$\psi \in \H^\infty$, there is a constant~$C=C(m_L)$ such that
\beq \label{1es}
\big\| (\p \psi)|_t \big\|_t \leq \frac{C\,|I|}{1+t^2} \:
\sup_{m \in I} \,\sum_{b=0}^2 \big\| (\partial_m^b \psi_m)|_{t=0} \big\|_{W^{2,2}} \:,
\eeq
where~$\|.\|_t$ is the norm corresponding to the scalar product
\beq ( . | .)|_t := 2 \pi \int_{\R^3} \Sl \,.\, | \gamma^0 \,.\, \Sr_{\vec{x}}\: \dd^3x \::\:
L^2(\scrN_t, S\scrM) \times L^2(\scrN_t, S\scrM) \rightarrow \C \eeq
(which is similar to~\eqref{print}, but now applied to wave functions that do not need
to be solutions), and~$\| . \|_{W^{2,2}}$ is the spatial Sobolev norm
\beq \label{sobolev}
\| \phi \|^2_{W^{2,2}} := \sum_{\text{$\alpha$ with~$|\alpha| \leq 2$}}\:
\int_{\R^3} | \nabla^\alpha \phi(\vec{x}) |^2\: \dd^3x \:,
\eeq
where~$\alpha$ is a multi-index.
\end{Lemma} \noindent
The absolute value in~\eqref{sobolev} is the norm~$|\,.\,| := \sqrt{\Sl . | \gamma^0 . \Sr}$
on the spinors. If we again identify all spinor spaces in the Dirac representation with~$\C^4$,
this simply is the standard Euclidean norm on~$\C^4$.

The proof of this lemma will be given later in this section.
Before, we infer the weak mass oscillation property.
\begin{Corollary} \label{corwMOP}
The vacuum Dirac operator~$i \Pdd$ in Minkowski space has the weak mass oscillation property
with domain~\eqref{Hinfchoice}.
\end{Corollary}
\Proof
For every~$\psi, \phi \in \H^\infty$, the Schwarz inequality gives
\beq \label{schwarz}
|\bra \p \psi | \p \phi \ket| = \frac{1}{2 \pi}
\left| \int_{-\infty}^\infty \big( (\p \psi)|_t \,\big|\, \gamma^0 \,(\p \phi)|_t \big)_t \:\dd t \right| 
\leq \int_{-\infty}^\infty \big\| (\p \psi)|_t \big\|_t\: \big\| (\p \phi)|_t \big\|_t\: \dd t \:.
\eeq
Applying Lemma~\ref{lemmappsi} together with the estimate
\begin{align}
\big\| (\p \phi)|_t \big\|^2_t &= \iint_{I \times I} \big( \phi_m|_t \,\big|\, \phi_{m'}|_t \big)_t \:\dd m \,\dd m' \notag \\
&\leq \frac{1}{2} \iint_{I \times I} \Big( \|\phi_m\|_m^2 +  \|\phi_{m'}\|_{m'}^2 \Big)\: \dd m \,\dd m' = |I| \,\|\phi\|^2 \:,
\end{align}
we obtain inequality~\eqref{mbound} with
\beq \label{ces}
c = C\,|I|^\frac{3}{2}\:
\sup_{m \in I} \,\sum_{b=0}^2 \|\partial_m^b (\psi_m)|_{t=0} \|_{W^{2,2}}
\int_{-\infty}^\infty \frac{1}{1+t^2} \: \dd t < \infty \:.
\eeq
 
The identity~\eqref{mortho} follows by integrating the Dirac operator by parts,
\beq \label{intpart}
\begin{split}
\bra \p T \psi | \p \phi \ket &= \bra \p \Dir \psi | \p \phi \ket =
\bra \Dir \p \psi | \p \phi \ket = \int_\scrM \Sl \Dir \p \psi | \p \phi \Sr_x\: \Diff^4x \\
&\overset{(\star)}{=} \int_\scrM \Sl \p \psi | \Dir \p \phi \Sr_x\: \Diff^4x = \bra \p \psi | \Dir \p \phi \ket
= \bra \p \psi | \p T \phi \ket \:.
\end{split}
\eeq
In~$(\star)$, we used that the Dirac operator is symmetric with respect to the inner product~$\bra .|. \ket$.
Moreover, we do not get boundary terms because of the time decay in Lemma~\ref{lemmappsi}.
\QED

The remainder of this section is devoted to the proof of Lemma~\ref{lemmappsi}.
Using the result of Proposition~\ref{prp21},
we can express the solution~$\psi_m$ of the Cauchy problem
in terms of the causal fundamental solution~$k_m$. In order to bring~$k_m$ into a more
explicit form, we use~\eqref{kmdef2} together with formulas for the advanced and retarded Green's operators.
Indeed, these Green's operators are the multiplication operators in momentum space~\eqref{8b}
%\beq s^{\lor}_{m}(p) = \lim_{\varepsilon \searrow 0}
%            \frac{\slashed{p} + m}{k^{2}-m^{2}-i \varepsilon p^{0}}
%            \qquad {\mbox{and}} \qquad
%           s^{\land}_{m}(p) = \lim_{\varepsilon \searrow 0}
%            \frac{\slashed{p} + m}{p^{2}-m^{2}+i \varepsilon k^{0}} \eeq
(with the limit~$\varepsilon \searrow 0$ taken in the distributional sense, and where the vector~$k$ is 
the four-momentum). We thus obtain in momentum space
\begin{align}
k_{m}(p) &=\frac{1}{2 \pi \cI} \:(\slashed{p} + m) \; \lim_{\varepsilon
\searrow 0} \left[ \frac{1}{p^{2}-m^{2}-\cI\varepsilon p^{0}} \:-\:
\frac{1}{p^{2}-m^{2}+\cI\varepsilon p^{0}} \right] \notag \\
&= \frac{1}{2 \pi \cI} \:(\slashed{p} + m) \; \lim_{\varepsilon
\searrow 0} \left[ \frac{1}{p^{2}-m^{2}-\cI\varepsilon} \:-\:
\frac{1}{p^{2}-m^{2}+\cI\varepsilon} \right] \epsilon(p^{0})
\end{align}
(where, for notational clarity, we denoted the momentum variables by~$p$, and
$\epsilon$ is the {\em{sign function}}~$\epsilon(x)=1$ if~$x>0$ and~$\epsilon(x)=-1$ otherwise).
\sindex{sign function}%
\nindex{ahb1@$\epsilon$ -- sign function}%
Employing the distributional equation
\beq \lim_{\varepsilon \searrow 0} \left( \frac{1}{x - \cI \varepsilon} - \frac{1}{x + \cI \varepsilon} \right)
= 2 \pi \cI \: \delta(x) \:, \eeq
we obtain the simple formula
\beq \label{kmp}
k_m(p) = (\slashed{p} + m) \: \delta(p^{2}-m^{2}) \: \epsilon(p^{0}) \:.
\eeq
It is convenient to transform spatial coordinates of the time evolution operator to momentum space. 
First, in the Minkowski vacuum, the time evolution operator can be represented as in~\eqref{Ukern} with
an integral kernel~$U^{t,t'}(\vec{y}, \vec{x})$, which depends only on the difference vector~$\vec{y}-\vec{x}$.
We set
\beq U^{t,t'}(\vec{k}) := \int_{\R^3} U^{t,t'}(\vec{y}, 0) \:\E^{-\cI \vec{k} \vec{y}}\: \dd^3y \:. \eeq
Combining~\eqref{Ukm} with~\eqref{kmp} yields
\beq U^{t,t'}(\vec{k}) = \int_{-\infty}^\infty (\slashed{k} + m)\:\gamma^0 \: \delta(k^{2}-m^{2}) \big|_{k=(\omega, \vec{k})}\;
\epsilon(\omega)\: \E^{-\cI \omega (t-t')}\: \dd\omega \:. \eeq
Carrying out the~$\omega$-integral, we get
\beq \label{Urep}
U^{t,t'}(\vec{k}) = \sum_{\pm} \Pi_\pm(\vec{k}) \:\E^{\mp \cI \omega (t-t')} \:,
\eeq
where we set
\begin{gather}
\Pi_\pm(\vec{k}) := \pm \frac{1}{2 \omega(\vec{k})} \; (\slashed{k}_\pm + m) \,\gamma^0 \label{Pidef2} \\
\text{with} \qquad \omega(\vec{k}) = \sqrt{|\vec{k}|^2 + m^2}
\qquad \text{and} \qquad k_\pm = (\pm \omega(\vec{k}), \vec{k} )\:. \nonumber
\end{gather}
\nindex{ahc@$\Pi_\pm(\vec{k})$ -- projections to Dirac solutions in momentum space}%
Moreover, applying Plancherel's theorem, the scalar product~\eqref{print} can be written in momentum space as
\beq (\psi_m \,|\, \phi_m)_m = (2 \pi)^{-2} \int_{\R^3} \Sl \hat{\psi}_m(t, \vec{k}) \,|\,
\gamma^0 \,\hat{\phi}_m(t, \vec{k}) \Sr\: \dd^3k \:. \eeq
The unitarity of the time evolution operator in position space implies that the matrix~$U^{t,t'}(\vec{k})$ is
unitary (with respect to the scalar product~$\la .\,,. \ra_{\C^2} \equiv \Sl \,.\,| \gamma^0 \,.\, \Sr$), meaning that its eigenvalues are
on the unit circle and the corresponding eigenspaces are orthogonal.
It follows that the operators~$\Pi_\pm(\vec{k})$ in~\eqref{Urep} are the orthogonal projection operators to the
eigenspaces corresponding to the eigenvalues~$\E^{\mp i \omega (t-t')}$; that is,
\beq \label{Piprop}
\gamma^0 \Pi_s^* \gamma^0 = \Pi_s \quad \text{and} \quad \Pi_s(\vec{k})\: \Pi_{s'}(\vec{k}) = \delta_{s, s'}\: \Pi_s(\vec{k}) \qquad \text{for~$s,s' \in \{+,-\}$}
\eeq
(these relations can also be verified by straightforward computations using~\eqref{Pidef2}; see
Exercise~\ref{exPiprop}).

The next two lemmas involve derivatives with respect to the mass parameter~$m$.
For clarity, we again denote the $m$-dependence of the operators by the subscript~$m$.
\begin{Lemma} \label{lemmamint}
The time evolution operator in the vacuum satisfies the relation
\begin{align}
(t-t')\, U_m^{t,t'}(\vec{k}) &= \frac{\partial}{\partial m} V_m^{t,t'}(\vec{k}) + W_m^{t,t'}(\vec{k}) \:, \label{tes}
\intertext{where}
V_m^{t,t'}(\vec{k}) &= \sum_\pm \frac{\cI}{2m}\: (\slashed{k}_\pm + m) \gamma^0  \:\E^{\mp \cI \omega (t-t')}
\label{Vmdef} \\
W_m^{t,t'}(\vec{k}) &= \sum_\pm \frac{\cI}{2} \Big( \frac{\slashed{k}_\pm \gamma^0}{m^2}
\mp \frac{1}{\omega} \Big)  \:\E^{\mp \cI \omega (t-t')} \:. \label{Wmdef}
\end{align}
The operators~$V_m^{t,t'}$ and~$W_m^{t,t'}$ are estimated uniformly by
\beq \label{VWes}
\|V_m^{t,t'}(\vec{k})\| + \|W_m^{t,t'}(\vec{k})\| \leq C \,\bigg( 1 + \frac{|\vec{k}|}{m} \bigg) \:,
\eeq
where the constant~$C$ is independent of~$m$, $\vec{k}$, $t$ and~$t'$
(and~$\| \,.\, \|$ is any norm on the $2\times 2$-matrices).
\end{Lemma}
\Proof First, we generate the factor~$t-t'$ by differentiating the exponential in~\eqref{Urep} with
respect to~$\omega$,
\begin{align}
(t-t')\, U_m^{t,t'}(\vec{k}) &= \sum_\pm
\Pi_\pm(\vec{k}) \Big( \pm \cI \frac{\partial}{\partial \omega}\:\E^{\mp \cI \omega (t-t')} \Big).
\end{align}
Next, we want to rewrite the $\omega$-derivative as a derivative with respect to~$m$.
Taking the total differential of the dispersion relation~$\omega^2 - |\vec{k}|^2 = m^2$ for fixed~$\vec{k}$, one finds that
\beq \label{omegam}
\frac{\partial}{\partial \omega} = \frac{\omega}{m}\: \frac{\partial}{\partial m} \:.
\eeq
Hence
\begin{align}
&(t-t')\,U_m^{t,t'} = \sum_\pm
\Pi_\pm \Big( \pm \cI \:\frac{\omega}{m} \frac{\partial}{\partial m}\:\E^{\mp \cI \omega (t-t')} \Big) \notag \\
&= \frac{\partial}{\partial m} \sum_\pm \left( \pm \cI \:\frac{\omega}{m} \:
\Pi_\pm \:\E^{\mp \cI \omega (t-t')} \right)
- \sum_\pm \left( \frac{\partial}{\partial m} \left[ \pm \cI \:\frac{\omega}{m} \:
\Pi_\pm \right] \right) \E^{\mp \cI \omega (t-t')}\:.
\end{align}
Computing the operators in the round brackets using~\eqref{Pidef2} gives the identities~\eqref{Vmdef}
and~\eqref{Wmdef}. Estimating these formulas, one obtains bounds that are at most linear
in~$|\vec{k}|$, proving~\eqref{VWes}.
\QED

This method can be iterated to generate more factors of~$t-t'$.
In the next lemma, we prove at least quadratic decay in time.
For later use, it is preferable to formulate the result in position space.
\begin{Lemma} \label{lemmaint2}
The time evolution operator in the vacuum has the representation
\beq \label{Udecay}
U_m^{t,t'} = \frac{1}{(t-t')^2} \left( \frac{\partial^2}{\partial m^2} A_m^{t,t'}+ 
\frac{\partial}{\partial m} B_m^{t,t'} + C_m^{t,t'} \right)
\eeq
with operators
\beq A_m^{t,t'}, B_m^{t,t'}, C_m^{t,t'} \::\: W^{2,2}(\scrN_{t'}, S\scrM) \rightarrow L^2(\scrN_t, S\scrM)\:, \eeq
which are bounded uniformly in time by
\beq \label{ABCsob}
\|A_m^{t,t'}(\phi)\|_t + \|B_m^{t,t'}(\phi)\|_t + \|C_m^{t,t'}(\phi)\|_t
\leq c\, \|\phi\|_{W^{2,2}} \:,
\eeq
where~$c$ is a constant that depends only on~$m$.
\end{Lemma}
\Proof A straightforward computation using exactly the same methods as
in Lem\-ma~\ref{lemmamint} yields the representation
\beq \label{t2es}
(t-t')^2\, U_m^{t,t'}(\vec{k}) = \frac{\partial^2}{\partial m^2} A_m^{t,t'}(\vec{k}) + 
\frac{\partial}{\partial m} B_m^{t,t'}(\vec{k}) + C_m^{t,t'}(\vec{k}) \:,
\eeq
where the operators~$A_m^{t,t'}$, $B_m^{t,t'}$ and~$C_m^{t,t'}$ are bounded by
\beq \label{ABCes}
\|A_m^{t,t'}(\vec{k})\| + \|B_m^{t,t'}(\vec{k})\| + \|C_m^{t,t'}(\vec{k})\| 
\leq \frac{C}{m} \:\bigg( 1 + \frac{|\vec{k}|}{m} + \frac{|\vec{k}|^2}{m^2} \bigg) \:,
\eeq
with a numerical constant~$C>0$. We remark that, compared to~\eqref{tes}, the right of~\eqref{ABCes}
involves an additional~$1/m$. This prefactor
is necessary for dimensional reasons, because the additional factor~$t-t'$ in~\eqref{t2es}
(compared to~\eqref{tes}) brings in an additional dimension of length
(and in natural units, the factor~$1/m$ also has the dimension of length).
The additional summand~$|\vec{k}|^2/m^2$ in~\eqref{ABCes} can be understood from the fact that applying~\eqref{omegam} generates a factor of~$\omega/m$, which for large~$|\vec{k}|$
scales like~$|\vec{k}|/m$.

Translating this result to position space and keeping in mind that the vector~$\vec{k}$ corresponds to the derivative~$-\cI\vec{\nabla}$, we obtain the result.
\QED

\Proof[Proof of Lemma~\ref{lemmappsi}.]
First of all, the Schwarz inequality gives
\beq \big\| (\p \psi)|_t \big\|_t \leq \int_I \|\psi_m\|_m\, \dd m \leq \sqrt{|I|} \;\|\psi\| \:. \eeq
Thus, it remains to show the decay for large~$t$; that is,
\beq \label{es1}
\big\| (\p \psi)|_t \big\|_t \leq \frac{C\,|I|}{t^2} \:
\sup_{m \in I} \,\sum_{b=0}^2 \|\partial_m^b (\psi_m)|_{t=0} \|_{W^{2,2}} \:.
\eeq

We apply Lemma~\ref{lemmaint2} and integrate by parts in~$m$ to obtain
\begin{align}
(\p \psi)|_t &= \int_I U_m^{t,0} \:\psi_m|_{t=0}\: \dd m = \frac{1}{t^2} \int_I \big( \partial^2_m A_m^{t,0}
+ \partial_m B_m^{t,0} + C_m^{t,0} \big)\:\psi_m|_{t=0}\: \dd m \notag \\
&= \frac{1}{t^2} \int_I \Big( A_m^{t,0}\:  (\partial^2_m \psi_m|_{t=0})
- B_m^{t,0}\:  (\partial_m \psi_m|_{t=0}) + C_m^{t,0}\: \psi_m|_{t=0} \Big)\: \dd m \:.
\end{align}
Taking the norm and using~\eqref{ABCsob} gives~\eqref{es1}.
\QED

We finally note that the previous estimates are not optimal for two reasons. First, the pointwise
quadratic decay in~\eqref{1es} is more than what is needed for the convergence of the integral in~\eqref{ces}.
Second and more importantly, the Schwarz inequality~\eqref{schwarz} does not catch the optimal
scaling behavior in~$\vec{k}$. This is the reason why the constant in~\eqref{mbound}
involves derivatives of~$\psi_m$ (cf.~\eqref{ces}), making it impossible to prove the inequality~\eqref{smop}
which arises in the strong mass oscillation property.
In order to improve the estimates, one needs to use Fourier methods both in space and time,
as will be explained in the next section.

\section[Proof of the Strong Mass Oscillation Property]{Proof of the Strong Mass Oscillation Property in the Minkowski Vacuum} \label{secmstrong}
\sindex{mass oscillation property!strong in Minkowski space}%
\begin{Thm} \label{thmsMOP2}
The vacuum Dirac operator in Minkowski space has the strong mass oscillation property
with domain~\eqref{Hinfchoice}.
\end{Thm} \noindent
Our proof relies on a Plancherel argument in spacetime. It also provides an
alternative method for establishing the weak mass oscillation property.
\Proof[Proof of Theorem~\ref{thmsMOP2}.]
Let~$\psi=(\psi_m)_{m \in I} \in \H^\infty$ be a family of solutions of the Dirac equation
for a varying mass parameter in the Minkowski vacuum.
Using Proposition~\ref{prp21}, one can express~$\psi_m$ in terms of its values at time~$t=0$ by
\beq \psi_m(x) = 2 \pi \int_{\R^3} k_m(x, (0, \vec{y})) \,\gamma^0\, \psi_m|_{t=0}(\vec{y})\: \dd^3y \:. \eeq
We now take the Fourier transform, denoting the four-momentum by~$k$.
Using~\eqref{kmp}, we obtain
\begin{align}
\psi_m(k) &= 2 \pi k_m(k)\, \gamma^0 \hat{\psi}_m^0(\vec{k}) \notag \\
&= 2 \pi \: \delta(k^2-m^2)\: \epsilon(k^0)\: (\slashed{k}+m) \,\gamma^0 \hat{\psi}_m^0(\vec{k}) \:,
\end{align}
where~$\hat{\psi}_m^0(\vec{k})$ denotes the spatial Fourier transform of~$\psi_m|_{t=0}$
(in order to avoid an ambiguity of notation, the hat of the Fourier transform
in spacetime was omitted).
Obviously, this is a distribution supported on the mass shell. In particular, it is
not square integrable over~$\R^4$.

Integrating over~$m$, we obtain the following function
\beq \label{ppsi}
(\p \psi)(k) = 2 \pi \: \chi_I(m)\: \frac{1}{2m}\: \epsilon(k^0)\: (\slashed{k}+m) \,\gamma^0 
\hat{\psi}_m^0(\vec{k})
\Big|_{m=\sqrt{k^2}} \:,
\eeq
where~$m$ now is a function of the momentum variables.
Since the function~$\psi_m|_{t=0}$ is compactly supported and smooth in the spatial variables,
its Fourier transform~$\hat{\psi}_m^0(\vec{k})$
has rapid decay. This shows that the function~\eqref{ppsi} is indeed square integrable.
Using Plancherel, we see that condition~(a) in Definition~\ref{defwmop} is satisfied.
Moreover, the operator~$T$ is simply the operator of multiplication by~$\sqrt{k^2}$,
so that condition~(b) obviously holds. This again shows the weak mass oscillation property.

In order to prove the strong mass oscillation property, we
need to compute the inner product~$\bra \p \psi | \p \phi \ket$.
To this end, we first write this inner product in momentum space as
\begin{align}
\bra \p \psi | \p \phi \ket &= \int \frac{\dd^4k}{(2 \pi)^4}
4 \pi^2 \: \chi_I(m)\: \frac{1}{4m^2}\: \Sl (\slashed{k}+m) \,\gamma^0 \hat{\psi}_m^0(\vec{k}) \,|\,
(\slashed{k}+m) \,\gamma^0 \hat{\phi}_m^0(\vec{k}) \Sr \Big|_{m=\sqrt{k^2}} \notag \\
&= \int \frac{\dd^4k}{4 \pi^2}\:
\chi_I(m)\: \frac{1}{2m}\: \Sl \gamma^0 \hat{\psi}_m^0(\vec{k}) \,|\,
(\slashed{k}+m) \,\gamma^0 \hat{\phi}_m^0(\vec{k}) \Sr \Big|_{m=\sqrt{k^2}} \:.
\end{align}
Reparametrizing the $k^0$-integral as an integral over~$m$, we obtain
\beq \label{goback}
\bra \p \psi | \p \phi \ket
= \frac{1}{4 \pi^2} \int_I \dd m \int_{\R^3} \frac{\dd^3k}{2\, |k^0|}\: \Sl \gamma^0 \hat{\psi}_m^0(\vec{k}) \,|\,
(\slashed{k}+m) \,\gamma^0 \hat{\phi}_m^0(\vec{k}) \Sr \big|_{k^0 = \pm \sqrt{|\vec{k}|^2+m^2}} \:.
\eeq
Estimating the inner product with the Schwarz inequality and applying Plancherel's theorem, one finds
\beq |\bra \p \psi | \p \phi \ket| \leq \frac{1}{4 \pi^2} \int_I \dd m
\int_{\R^3} \|\hat{\psi}_m^0(\vec{k})\|\, \| \hat{\phi}_m^0(\vec{k})\| \: \dd^3k
\leq 2 \pi \int_I \|\psi_m\|_m\: \|\phi_m\|_m \: \dd m \:. \eeq
Thus the inequality~\eqref{smop} holds.
\QED

Apart from completing the proof of the strong mass oscillation property, the computation in the
above proof also tells us what the fermionic signature operator is.
In order to see this, we return to the formula~\eqref{goback}.
Applying Plancherel's theorem and using~\eqref{print}, we conclude that
\beq \label{pmk}
\bra \p \psi | \p \phi \ket = \int_I ( \psi_m^0 \,|\, \Sig_m\, \phi_m^0)_m \: \dd m \:,
\eeq
where~$\Sig_m$ is the multiplication operator in momentum space
\beq \label{S0def}
\Sig_m(\vec{k}) := \sum_{k^0 = \pm \omega(\vec{k})} \frac{\slashed{k}+m}{2\, \omega(\vec{k})}\: \gamma^0 
= \frac{\vec{k} \vec{\gamma}+m}{\omega(\vec{k})}\: \gamma^0 \:.
\eeq
Comparing~\eqref{pmk} with~\eqref{Smdef}, one sees that the matrix~$\Sig_m(\vec{k})$
is indeed the fermionic signature operator, considered as a multiplication operator in momentum space.
By direct computation, one verifies that the matrix~$\Sig_m(\vec{k})$ has eigenvalues~$\pm 1$
(here one can use that~$\Sig_m = \Pi_+ - \Pi_-$ with~$\Pi_\pm$ as introduced in~\eqref{Pidef2}).

\section{Exercises}

\begin{Exercise} \label{ex201} {\em{ This exercise recalls basics on the principal value
in one dimension
\sindex{principal value integral}%
\beq
\frac{1}{2} \lim_{\varepsilon \searrow 0} \left(
\frac{1}{x - \cI \varepsilon} + \frac{1}{x + \cI \varepsilon} \right) =: \frac{\PP}{x}\:. \label{eq:PP-formula}
\eeq
\bitem
\item[(a)] Repeat the method in Exercise~\ref{ex2} to show that the limit of the left-hand side
of~\eqref{eq:PP-formula} exists for any~$\eta \in C^1(\R) \cap L^1(\R)$.
Derive a corresponding estimate that shows that~$\PP$ is a well-defined tempered distribution.
\item[(b)] Show that for any~$\eta \in C^1(\R) \cap L^1(\R)$,
\beq \PP(\eta) = \lim_{\varepsilon \searrow 0} \left(
\int_{-\infty}^{-\varepsilon} + \int_{\varepsilon}^\infty \right) \frac{\eta(x)}{x} \:dx \:. \eeq
\eitem
}} \end{Exercise}

\begin{Exercise} \label{ex21} {\em{  The goal of this exercise is to justify that the one-dimensional
relations
\begin{align}
\lim_{\varepsilon \searrow 0} \left( \frac{1}{x - \cI \varepsilon} - \frac{1}{x + \cI \varepsilon} \right)
&=\, 2 \pi i \: \delta(x) \label{eq:delta-formula} \\
\frac{1}{2} \lim_{\varepsilon \searrow 0} \left(
\frac{1}{x - \cI \varepsilon} + \frac{1}{x + \cI \varepsilon} \right) &=: \frac{\PP}{x}\:.
\label{eq:PP-formula2}
\end{align}
can be used in the four-dimensional setting to obtain the identity
\beq \label{delta-highdim}
\lim_{\varepsilon \searrow 0} \frac{1}{r^2+(\varepsilon+\cI t)^2}
= \lim_{\varepsilon \searrow 0} \frac{1}{r^2 - t^2 + \cI \varepsilon t }
= -\frac{\PP}{\xi^2} -\cI \pi\, \delta(\xi^2)\,\epsilon(\xi^0)\:,
\eeq
\bitem
\item[(a)] Let~$T$ be a distribution on~$\R$, $\Omega \subset \scrM$ be an open subset of Minkowski space
 and~$f : \Omega \rightarrow \R$ a smooth function with nowhere vanishing gradient. Show that the relation
\beq \big(f^* T)(\eta) := T \big( \phi_f(\eta) \big) \:, \qquad \eta \in C^\infty_0(\Omega) \eeq
with
\beq \phi_f(\eta)(t) := \frac{\partial}{\partial t} \int_{\Omega} \Theta\big( t-f(x) \big)\: \eta(x)\: \Diff^4x \eeq
(where~$\Theta$ is the Heaviside function)
defines~$f^* T$ as a distribution on~$\Omega$ (this is the so-called
{\em{pullback}} of~$T$ under~$f$; for details see~\cite[Section~7.2]{friedlander2}).
\item[(b)] Choosing~$\Omega$ as the half-space in the future, $\Omega = \{ x\in \scrM, x^0>0\}$,
one can rewrite the expression on the left-hand side of~\eqref{delta-highdim} as
\beq \lim_{\varepsilon \searrow 0} \frac{1}{r^2 - t^2 + \cI \varepsilon} \:. \eeq
Use~(a) to conclude that this expression is a well-defined distribution for any~$\varepsilon>0$.
Show that the limit~$\varepsilon \searrow 0$ exists in the distributional sense.
\item[(c)] Repeating the procedure of~(b) for the half-space in the past,
one obtains a distribution on~$\scrM \setminus \{t=0\}$.
Show that this distribution coincides with the limit in~\eqref{delta-highdim}.
{\em{Hint:}} Similar to Exercise~\ref{ex2}, one can estimate the behavior at the origin
with Lebesgue's dominated convergence theorem.
\eitem
}} \end{Exercise}

\begin{Exercise} \label{ex2.1} {\em{
This exercise is devoted to the {\em{advanced Green's operator}}~$s^\vee_m$.
\bitem
\item[(a)] Assume that~$m>0$. Show that the limit~$\nu \searrow 0$
in~\eqref{8b} exists in the distributional sense.
\item[(b)] Show that the limit~$\nu \searrow 0$ in~\eqref{8b} also exists in the
massless case~$m=0$ and that
\beq \lim_{m \searrow 0} s_m^\vee(k) = s_0^\vee(k) \qquad \text{as a distribution}\:. \eeq
{\em{Hint:}} Proceed similarly as in Exercise~\ref{ex21}.
\item[(c)] Consider the Fourier integral in the~$q^0$-variable
\beq \int_{-\infty}^\infty \frac{1}{q^{2}-m^{2}-\cI \nu q^{0}} \:\E^{\cI q^0 t}\: \dd q^0 \:. \eeq
Show with residues that this integral vanishes for sufficiently small~$\nu$ if~$t<0$.
\item[(d)] Argue with Lorentz invariance to prove the left-hand side of~\eqref{GF}.
\eitem
}} \end{Exercise}

\begin{Exercise} \label{ex2.2-1} {\em{
Modifying the location of the poles in~\eqref{8b} gives rise to the distribution
\beq s^F_m(k) := \lim_{\nu \searrow 0}
\frac{\slashed{k} + m}{k^{2}-m^{2}+\cI \nu} \:. \eeq
This is the well-known {\em{Feynman propagator}}, which is often
described intuitively by saying that ``positive frequencies move to the future and
negative frequencies move to the past.'' Make this sentence precise by
a computation similar to that in Exercise~\ref{ex2.1}~(c).
}} \end{Exercise}

\begin{Exercise} \label{ex1d} {\em{ Given~$\omega \in \R$, we
consider the ordinary differential operator~$\Dir = \cI \partial_t + \omega$.
\bitem
\item[(a)] Construct the advanced and retarded Green's operators,
which satisfy in analogy to~\eqref{Greendefext} the equation
\beq \D_t s(t,t') = \delta(t-t') \:. \eeq
\item[(b)] Compute the resulting causal fundamental solution according to~\eqref{kmdef2}. How is it related to the
time evolution operator~$U^{t,t'}$? On which Hilbert space does
the time evolution operator act as a unitary operator?
\eitem
}} \end{Exercise}

\begin{Exercise} \label{ex2d} {\em{ Consider the massless Dirac equation~$\Dir \psi = 0$
in the two-dimensional spacetime cylinder~$\R \times S^1$; that is,
\beq \Dir = \cI \begin{pmatrix} 0 & 1 \\ 1 & 0 \end{pmatrix} \partial_t + 
\cI \begin{pmatrix} 0 & 1 \\ 1 & 0 \end{pmatrix} \partial_\varphi \eeq
with~$t \in \R$ and~$\varphi \in (0, 2 \pi)$.
\bitem
\item[(a)] Choose the spin inner product such that the Dirac matrices become symmetric.
What is the resulting spacetime inner product~$\bra .|. \ket$?
What is the scalar product~$(.|.)$?
\item[(b)] Employ for~$k \in \Z$ the separation ansatz
\beq \psi(t, \varphi) = \E^{-\cI k \varphi} \chi(t) \qquad \text{with} \qquad \chi(t) \in \C^2 \:. \eeq
Derive the resulting ordinary differential equation (ODE) for~$\chi$. Compute the time evolution operator for this
ODE. {\em{Hint:}} Use the result of Exercise~\ref{ex1d}.
\item[(c)] Use a Fourier series decomposition in order to deduce
a series representation of the time evolution operator of the Dirac operator on~$\R \times S^1$.
Try to carry out the infinite series to obtain a closed expression for~$U^{t,t'}$.
How can one see finite propagation speed?
\eitem
}} \end{Exercise}

\begin{Exercise} {\em{ As in Exercise~\ref{ex2d}, we consider the two-dimensional
massless Dirac equation.
\bitem
\item[(a)] Adapt the formulas for the advanced and retarded
Green's operators in momentum space to the two-dimensional massless case.
\item[(b)] Compute the Fourier transform to obtain~$s^\vee(x,y)$ and~$s^\wedge(x,y)$.
\item[(c)] Use the result of~(b) to compute the causal fundamental solution and
the time evolution operator.
\item[(d)] How can one see finite propagation speed?
How is the obtained formula related to the formula in Exercise~\ref{ex2d}~(c)?
\eitem
}} \end{Exercise}

\begin{Exercise} \label{exPiprop} {\em{ Verify the relations~\eqref{Piprop}
by direct computation starting from the definition~\eqref{Pidef2}.
}} \end{Exercise}

\begin{Exercise} \label{exdistrel} {\em{ Verify by formal computation that
in the Minkowski vacuum, the fundamental solution~$k_m$ and the
Green's operator~$s_m$ defined by
\beq %\label{smean2}
s_m := \frac{1}{2}\: \big( s_m^\vee + s_m^\wedge \big) \eeq
satisfy the distributional relations in the mass parameters~$m$ and~$m'$
\begin{align}
k_m\,k_{m'} &= \delta(m-m')\:p_m \\
k_m \,s_{m'}&= s_{m'}\,k_m=\frac{\text{\rm{PP}}}{m-m'}\:k_m\;,
\end{align}
where~${\rm{PP}}$ denotes the principal part, and~$p_m$ is the distribution
\beq %\label{pmdef2}
p_m(k) = (\slashed{k} + m) \: \delta(k^2 - m^2) \:. \eeq
{\em{Hint:}} By a ``formal computation'' we mean that you do not need to
evaluate weakly in the mass with test functions.
}}
\end{Exercise}

\begin{Exercise} \label{exdistrel2} {\em{ Proceed similarly to Exercise~\ref{exdistrel}
to derive a relation for the operator product~$s_m^\vee s_{m'}^\vee$.
Derive the relation
\begin{align}
s_m\,s_{m'}&= \frac{\text{\rm{PP}}}{m-m'}\:(s_m-s_{m'})+\pi^2 \,\delta(m-m')\:p_m\;.
\end{align}
}}
\end{Exercise}

\chapter{Methods of Scattering Theory} \label{secscatter}
\sindex{scattering theory}
In Chapter~\ref{secFSO}, the fermionic signature operator and the unregularized fermionic projector
were introduced abstractly. In the previous chapter, we computed them in the Minkowski vacuum.
It remains to construct them in the presence of an external potential.
In order to prove the mass oscillation properties, our task is to analyze the Dirac solutions
asymptotically for large times and near spatial infinity.
This can be accomplished with methods of scattering theory, which we now briefly introduce.
We follow the presentation in~\cite{hadamard}.

We return to the Cauchy problem in the presence of an external potential,
\beq \label{cauchyt}
(\Dir - m) \,\psi_m = 0 \:,\qquad \psi_m \big|_{t_0} = \psi_0 \in C^\infty(\scrN_{t_0} \simeq \R^3, S\scrM)\:,
\eeq
with~$\Dir$ as in~\eqref{Dout}. For notational clarity, we shall often denote the objects
in the presence of the external potential by a tilde (the ``interacting objects''), whereas the
objects without tilde refer to the Minkowski vacuum.

\section{The Lippmann-Schwinger Equation}
\sindex{Lippmann-Schwinger equation|textbf}%
The Dirac dynamics can be rewritten in terms of a symmetric operator~$\tilde{H}$. To this end, we multiply
the Dirac equation~\eqref{Dout} by~$\gamma^0$ and bring the $t$-derivative separately on one side of the
equation,
\beq \label{Hamilton2}
\cI \partial_t \psi_m = \tilde{H} \psi_m\:, \qquad \text{where} \qquad
\tilde{H} := -\gamma^0 (\cI \vec{\gamma} \vec{\nabla} + \B - m)
\eeq
(note that~$\gamma^j \partial_j = \gamma^0 \partial_t + \vec{\gamma} \vec{\nabla}$).
We refer to~\eqref{Hamilton2} as the Dirac equation in {\em{Hamiltonian form}}.
The fact that the scalar product~\eqref{print} is time independent
implies that for any two solutions~$\phi_m, \psi_m \in \Cisc(\scrM, S\scrM) \cap \H_m$,
\beq \nonumber
0 = \partial_t (\phi_m \,|\, \psi_m)_m\: = \cI \bigl((\tilde{H} \phi_m \,|\, \psi_m)_m - (\phi_m \,|\, \tilde{H} \psi_m)_m\bigr)\:,
\eeq
showing that the Hamiltonian is a symmetric operator on~${\mathscr{H}}_m$.

The Lippmann-Schwinger equation can be used to compare the dynamics in the Minkowski vacuum with the dynamics in the presence of an external potential. We denote the time evolution operator in the Minkowski vacuum by~$U^{t,t_0}_m$.
\begin{Prp} The Cauchy problem~\eqref{cauchyt} has a solution~$\psi_m$ which satisfies the equation
\beq \label{lse}
\psi_m|_t = U^{t,t_0}_m \psi_0 +\cI \int_{t_0}^t U^{t, \tau}_m \,\big( \gamma^0 \B\: \psi_m \big) \big|_\tau\: \dd\tau \:,
\eeq
referred to as the {\bf{Lippmann-Schwinger equation}}.
\end{Prp}
\Proof Obviously, the wave function~$\psi_m|_t$ given by~\eqref{lse} has the correct
initial values at~$t=t_0$. Thus it remains to show that this wave function satisfies the Dirac equation.
To this end, we rewrite the Dirac equation in the Hamiltonian form~\eqref{Hamilton2}, and
separate the vacuum Hamiltonian~$H$ from the term involving the external potential,
\beq \label{hamex}
(\cI \partial_t - H)\, \psi_m = -\gamma^0 \B \,\psi_m \qquad \text{with} \qquad
H = -\cI \gamma^0 \vec{\gamma} \vec{\nabla} + \gamma^0 m \:.
\eeq
Applying the operator~$\cI \partial_t - H$ to~\eqref{lse} and observing that the time evolution
operator maps to solutions of the vacuum Dirac equation, only the derivative of the upper
limit of integration contributes,
\beq (\cI \partial_t - H)\, \psi_m|_t = - U^{t, \tau}_m \:\left( \gamma^0 \B\: \psi_m \right)\big|_{\tau=t}
= -\gamma^0 \B\: \psi_m|_t \:, \eeq
so that~\eqref{hamex} is indeed satisfied.
\QED
We remark that one way of solving the Lippmann-Schwinger equation is to
substitute the left-hand side on the right-hand side to obtain
\begin{align}
\psi_m|_t &= U^{t,t_0}_m \psi_0 + \cI \int_{t_0}^t U^{t, \tau}_m \,\big( \gamma^0 \B\: \psi_m \big) \big|_\tau\: \dd\tau
\notag \\
&= U^{t,t_0}_m \psi_0 + \cI \int_{t_0}^t U^{t, \tau}_m \,\gamma^0 \B|_\tau\: 
U^{\tau,t_0}_m \psi_0 \dd\tau \notag \\
&\quad\:- \int_{t_0}^t U^{t, \tau}_m \,\gamma^0 \B|_\tau \int_{t_0}^\tau
U^{\tau, \tau'}_m \,\big( \gamma^0 \B\: \psi_m \big) \big|_{\tau'}\: \dd\tau' \dd\tau \:.
\end{align}
Iterating this procedure, one gets a series of nested integrals referred to
as the {\em{Dyson series}}, which is commonly used
\sindex{Dyson series}%
in perturbative quantum field theory (see, for example, \cite[Section~3.5]{weinberg}).
The Dyson series can be regarded as an {\em{ordered exponential}}
(see Exercise~\ref{ex2.4-3}).

\section[The Mass Oscillation Property in an External Potential]{The Mass Oscillation Property in the Presence of an External Potential}
The goal of this section is to prove the following result:
\begin{Thm} \label{thmmop}
Assume that the external potential~$\B$ is smooth and for large times
decays faster than quadratically in the sense that
\beq |\B(t)|_{C^2} \leq \frac{c}{1 + |t|^{2+\varepsilon}} \label{Bdecay}
\eeq
for suitable constants~$\varepsilon, c>0$. Then the strong mass oscillation property holds.
\sindex{mass oscillation property!strong in presence of external potential}%
\end{Thm} \noindent
In words, the condition~\eqref{Bdecay} means that the potential and its up to second derivatives
must decay for large times faster than quadratically. This condition does not seem to have any
physical significance; it is needed in order for our methods to apply.

The $C^2$-norm in~\eqref{Bdecay} is defined as follows.
We denote spatial derivatives by~$\nabla$ and use the
notation with multi-indices; that is, for a multi-index~$\alpha = (\alpha_1, \ldots, \alpha_p)$,
we set~$\nabla^\alpha = \partial_{\alpha_1 \cdots \alpha_p}$ and
denote the length of the multi-index by~$|\alpha|=p$.
Then the spatial $C^k$-norms of the potential are defined by
\beq \label{defCk}
|\B(t)|_{C^k} := \max_{|\alpha| \leq k} \;\sup_{\vec{x} \in \R^3} |\nabla^\alpha \B(t, \vec{x})| \:,
\eeq
where~$| \,.\, |$ is the $\sup$-norm corresponding to the norm~$|\phi|^2 := \Sl \phi | \gamma^0 \phi \Sr$
on the spinors.

\subsection{Proof of the Weak Mass Oscillation Property} \label{secBmop}
In this section, we prove the following theorem.
\begin{Thm} \label{thm1}
Assume that the time-dependent external potential~$\B$ is smooth and
decays faster than quadratically for large times in the sense that~\eqref{Bdecay} 
holds for suitable constants~$c, \varepsilon>0$.
Then the Dirac operator~$\Dir = \cI \Pdd + \B$ has the weak mass oscillation property.
\sindex{mass oscillation property!weak in presence of external potential}%
\end{Thm}
\noindent We expect that this theorem could be improved by weakening the decay assumptions
on the potential.
However, this would require refinements of our methods which would go beyond the scope of this paper.
Also, using that Dirac solutions dissipate, the pointwise decay in time could probably be replaced
or partially compensated by suitable spatial decay assumptions.
Moreover, one could probably refine the
result of the above theorem by working with other norms (like weighted~$C^k$- or Sobolev norms).

The main step is the following basic estimate, which is the analog of Lemma~\ref{lemmappsi}
in the presence of an external potential.
\begin{Prp} \label{prpbasic}
Under the decay assumptions~\eqref{Bdecay} on the external potential~$\B$,
there are constants~$c, \varepsilon>0$ such that for every family~$\psi \in \H^\infty$
of solutions of the Dirac equation~\eqref{Dout} with varying mass,
\beq \label{ppsies}
\big\| ( \p \psi \big)|_t \big\|_t \leq \frac{c}{1 + |t|^{1+\varepsilon}} \: \sup_{m \in I}\:
\sum_{b=0}^2  \big\| (\partial_m^b \psi_m)|_{t=0} \big\|_{W^{2,2}} \:.
\eeq
\end{Prp} \noindent

We first show that this proposition implies the weak mass oscillation property.
\Proof[Proof of Theorem~\ref{thm1} assuming that Proposition~\ref{prpbasic} holds.] $\;$
In order to derive the inequality~\eqref{mbound}, we begin with the estimate
\begin{align}
|\bra \p \psi | \p \phi \ket| &\leq  \frac{1}{2 \pi} \int_{-\infty}^\infty \Big| \big( \p \psi|_t \,\big|\, \p \phi|_t \big) \big|_t \Big|\, \dd t
\leq \sup_{t \in \R} \big\|\p \phi|_t \big\|_t \int_{-\infty}^\infty \big\| \p \psi|_t \big\|_t \, \dd t \:.
\end{align}
The last integral is finite by Proposition~\ref{prpbasic}. The supremum can be bounded
by the Hilbert space norm using the H\"older inequality,
\begin{align}
 \|\p \phi|_t\|_t &= \bigg\| \int_I \phi_m|_t\: \dd m \bigg\|_t
\leq \int_I \big\| \phi_m|_t \big\|_t\: \dd m \notag \\
&\leq \sqrt{|I|} \left( \int_I \|\phi_m|_t\|_t^2 \: \dd m \right)^\frac{1}{2}
= \sqrt{|I|} \:\|\phi\| \:,
\end{align}
giving~\eqref{mbound}.

Using~\eqref{Bsymm}, the Dirac operator~$i \Pdd+ \B$ is symmetric with respect
to the inner product~$\bra .| .\ket$. Therefore, the
identity~\eqref{mortho} can be obtained just as in~\eqref{intpart}
by integrating the Dirac operator in spacetime by parts,
noting that we do not get boundary terms in view of the time decay in
Proposition~\ref{prpbasic}.
\QED

The remainder of this section is devoted to the proof of Proposition~\ref{prpbasic}.
We make use of the Lippmann-Schwinger equation~\eqref{lse},
\beq \label{lse2}
\psi_m|_t = U_m^{t,0} \:\psi_m|_{t=0} +\cI \int_0^t U_m^{t, \tau} \big( \gamma^0 \B\: \psi_m \big) \big|_\tau\, \dd\tau \:.
\eeq
Since the first summand of this equation is controlled by Lemma~\ref{lemmappsi},
it remains to estimate the second summand.
Again using~\eqref{Udecay} and integrating by parts with respect to the mass, we obtain
\beq \int_I U_m^{t, \tau} \big( \gamma^0 \B\: \psi_m \big) \big|_\tau\: \dd m =
\frac{1}{(t-\tau)^2} \int_I \big( A_m^{t,\tau}\:\partial^2_m  -B_m^{t,\tau}  \partial_m + C_m^{t,\tau} \big)
\big( \gamma^0 \B\: \psi_m \big) \big|_\tau \:\dd m \eeq
(where~$I$ is again the interval~\eqref{Idef}) and thus
\begin{align}
\bigg\| \int_I U_m^{t, \tau} \big( \gamma^0 \B\: \psi_m \big) \big|_\tau\: \dd m \bigg\|_t
&\leq \frac{c\,|I|}{(t-\tau)^2} \:\sup_{m \in I} \:\sum_{b=0}^2 \big\| \B(\tau)\: (\partial_m^b \psi_m)|_\tau \big\|_{W^{2,2}}
\nonumber \\
&\leq \frac{c\,|I|}{(t-\tau)^2} \:|\B(\tau)|_{C^2}\: \sup_{m \in I}\,
\sum_{b=0}^2 \big\| \partial_m^b \psi_m|_\tau \big\|_{W^{2,2}} \:.
\end{align}
We now bound~$\B(\tau)$ with the help of~\eqref{Bdecay}. Moreover, we estimate the
Sobolev norm $\big\| \partial_m^b \psi_m|_\tau \big\|_{W^{2,2}}$
at time~$\tau$ by means of Lemma~\ref{lemmaB}.
This gives rise to the inequality
\beq \bigg\| \int_I U_m^{t, \tau} \big( \gamma^0 \B\: \psi_m \big) \big|_\tau\: \dd m \bigg\|_t
\leq \frac{c^2 \,C\,|I|}{(t-\tau)^2} \:\frac{1+|\tau|^2}{1+|\tau|^{2+\varepsilon}} \:
\sup_{m \in I}\,\sum_{b=0}^2 \big\| \partial^b_m \psi_m|_{t=0} \big\|_{W^{2, 2}} \:, \eeq
which yields the desired decay provided that~$\tau$ and~$t$ are not too close to each other.
More precisely, we shall apply this inequality in the case~$|\tau| \leq |t|/2$. Then the estimate simplifies to
\begin{align}
\bigg\| \int_I U_m^{t, \tau} \big( &\gamma^0 \B\: \psi_m \big) \big|_\tau\: \dd m \bigg\|_t \notag \\
& \;\;\,\leq \frac{\tilde{C}}{t^{2}\, (1+|\tau|^\varepsilon)}
\sup_{m \in I}\,\sum_{b=0}^2 \big\| \partial^b_m \psi_m|_{t=0} \big\|_{W^{2, 2}} \qquad
\text{if~$|\tau| \leq |t|/2$} \label{es2}
\end{align}
with a new constant~$\tilde{C}>0$.
In the remaining case~$|\tau| > |t|/2$, we use the unitarity of~$U_m^{t, \tau}$ to obtain
\beq \bigg\| \int_I U_m^{t, \tau} \left( \gamma^0 \B\: \psi_m \right) |_\tau\: \dd m \bigg\|_t
\leq |I| \: |\B(\tau)|_{C^0}\: \sup_{m \in I} \|\psi_m\|\:. \eeq
Applying~\eqref{Bdecay} together with the inequality~$|\tau| > |t|/2$, this gives
\beq \label{es3}
\bigg\| \int_I U_m^{t, \tau} \left( \gamma^0 \B\: \psi_m \right) |_\tau\: \dd m \bigg\|_t
\leq \frac{\tilde{C}}{t^{2+\varepsilon}}\: \sup_{m \in I} \|\psi_m\| \qquad\;\; \text{if~$|\tau| > |t|/2$}\:.
\eeq
This again decays for large~$t$ because~$\tau$ is close to~$t$ and because~$|\B(\tau)|_{C^0}$
decays for large~$\tau$.

Comparing~\eqref{es2} and~\eqref{es3}, we find that the inequality in~\eqref{es2}
even holds for all~$\tau$. Thus integrating this inequality over~$\tau \in [0, t]$, we obtain
the following estimate for the second summand in~\eqref{lse2},
\beq \left\| \int_I \dd m \int_0^t U_m^{t, \tau} \left( \gamma^0 \B\: \psi_m \right) |_\tau\: \dd\tau \right\|_t
\leq \frac{C'}{t^{1+\varepsilon}}
\sup_{m \in I}\,\sum_{b=0}^2 \big\| \partial^b_m \psi|_{t=0} \big\|_{W^{2, 2}} \eeq
(where~$C'>0$ is a new constant).
Combining this inequality with the estimate~\eqref{1es} of the first summand in~\eqref{lse2},
we obtain the desired inequality~\eqref{ppsies}. This concludes the proof of Proposition~\ref{prpbasic}.

\subsection{Proof of the Strong Mass Oscillation Property} \label{secproofmass}
In this section, we prove the following result.

\begin{Thm} \label{thmC2} Assume that the weak mass oscillation property holds and
that the external potential~$\B$ satisfies the condition
\beq \label{L1B}
\int_{-\infty}^\infty |\B(\tau)|_{C^0} \:\dd\tau < \infty \:.
\eeq
Then the Dirac operator~$\Dir = \cI \Pdd + \B$ has the strong mass oscillation property.
\end{Thm} \noindent
Combining this theorem with Theorem~\ref{thm1}, one immediately obtains Theorem~\ref{thmmop}.

For the proof, we shall derive an explicit formula for the fermionic signature operator
(Proposition~\ref{prpSrep}). This formula is obtained by 
comparing the dynamics in the presence of the external potential with that in the Minkowski vacuum
using the Lippmann-Schwinger equation, and by employing distributional relations for products of
fundamental solutions and Green's operators (Lemma~\ref{lemmadistrel}).

In order to compare the dynamics in the presence of the external potential with that in the Minkowski vacuum,
we work with the Hamiltonian formulation.
We decompose the Dirac Hamiltonian~\eqref{Hamilton2} into the Hamiltonian in the Minkowski
vacuum~\eqref{hamex} plus a potential,
\beq \tilde{H} = H + \V \qquad \text{with} \qquad \V := -\gamma^0 \B \:. \eeq

\begin{Prp}  \label{prpSrep} Assume that the potential~$\B$ satisfies the condition~\eqref{L1B}. 
Then for every~$\psi, \phi \in \H^\infty$,
\beq \label{Sdef2}
\bra \p \psi | \p \phi \ket = \int_I ( \psi_m \,|\, \tilde{\Sig}_m\, \phi_m)_m \: \dd m\:,
\eeq
where~$\tilde{\Sig}_m : \H_m \rightarrow \H_m$ are bounded linear operators that act
on the wave functions at time~$t_0$ by
\begin{align}
\tilde{\Sig}_m &= \Sig_m -\frac{\cI}{2} \int_{-\infty}^\infty \epsilon(t-t_0)\,
\big[ \Sig_m \,U^{t_0, t}_m\, \V(t)\, \tilde{U}^{t,t_0}_m
-\tilde{U}^{t_0, t}_m\, \V(t)\, \Sig_m \,U^{t,t_0}_m \big] \:\dd t \label{Sigm1} \\
&\quad +\frac{1}{2} \left( \int_{t_0}^\infty \!\!\!\int_{t_0}^\infty + \int_{-\infty}^{t_0} \int_{-\infty}^{t_0} \right)
\tilde{U}^{t_0, t}_m\, \V(t) \,\Sig_m\,U^{t, t'}_m\, \V(t')\, \tilde{U}^{t',t_0}_m\:\dd t\, \dd t' \label{Sigm2}
\end{align}
(and~$\Sig_m$ is again the fermionic signature operator of the vacuum~\eqref{S0def}).
\end{Prp}

\noindent Before entering the proof of this proposition, it is instructive to verify
that the above formula for~$\tilde{\Sig}_m$ does not depend on the choice of~$t_0$.
\begin{Remark} {\bf{(Independence of~$\tilde{\Sig}_m$ on~$t_0$)}} {\em{
Our strategy is to differentiate the above formula for~$\tilde{\Sig}_m$ with respect to~$t_0$
and to verify that we obtain zero.
We first observe that taking a solution~$\phi_m \in \H_m$ of the Dirac equation in the presence of~$\B$,
evaluating at time~$t_0$ and applying the time evolution operator~$\tilde{U}^{t,t_0}_m$
gives~$\phi_m$ at time~$t$; that is, $\tilde{U}_m^{t,t_0} \phi_m|_{t_0} = \phi_m|_t$.
Differentiating with respect to~$t_0$ yields
\beq \partial_{t_0} \tilde{U}_m^{t,t_0} \phi_m|_{t_0} = 0 \:. \eeq
The situation is different when one considers the time evolution operator of the vacuum.
Namely, in the expression~$U^{t,t_0}_m \phi_m|_{t_0}$, the wave function~$\phi_m$ satisfies the
Dirac equation~$(\cI \partial_t - H) \phi_m = \V \phi_m$, whereas the time evolution operator
solves the Dirac equation with~$\V \equiv 0$. As a consequence,
\beq \partial_{t_0} U_m^{t,t_0} \phi_m|_{t_0} =  -\cI U_m^{t,t_0} (\V \phi_m)|_{t_0} \:. \eeq
Using these formulas together with~$U^{t_0, t_0} = \1 = \tilde{U}^{t_0, t_0}$,
a straightforward computation gives
\begin{align}
\partial_{t_0} \big( &\psi_m \,|\, \eqref{Sigm1} \,\phi_m \big) \big|_{t_0} \notag \\
=& -\cI (\psi_m \,|\, [\Sig_m, \V]\,
 \phi_m) |_{t_0} \notag \\
&-\frac{\cI}{2} \, (-2)
\big( \psi_m \,\big|\, \big( \Sig_m \, \V(t_0) -\V(t_0)\, \Sig_m \big) \,\phi_m \big) \big|_{t_0} \notag \\
&-\frac{\cI}{2} \int_{-\infty}^\infty \epsilon(t-t_0)\,
\big( (-\cI \V(t_0)) \,\psi_m \,\big|\, \Sig_m \,U_m^{t_0, t}\, \V(t)\, \tilde{U}_m^{t,t_0} \,\phi_m \big) \big|_{t_0} \dd t \notag \\
&+\frac{\cI}{2} \int_{-\infty}^\infty \epsilon(t-t_0)\,
\big( \psi_m \,\big|\, \tilde{U}_m^{t_0, t}\, \V(t)\, \Sig_m \,U_m^{t,t_0} \,(-i \V(t_0)) \,\phi_m \big) \big|_{t_0} \dd t \\
\partial_{t_0} \big( &\psi_m \,|\, \eqref{Sigm2} \,\phi_m \big) \big|_{t_0} \notag \\
=&-\frac{1}{2}\int_{-\infty}^\infty \epsilon(t'-t_0) \:
\big( \psi_m \,\big|\, \V(t_0) \,\Sig_m\,U_m^{t_0, t'}\, \V(t')\,
\tilde{U}_m^{t',t_0}\, \phi_m \big) \big|_{t_0} \:\dd t' \notag \\
&-\frac{1}{2}\int_{-\infty}^\infty \epsilon(t-t_0) \:
\big( \psi_m \,\big|\, \tilde{U}_m^{t_0, t}\, \V(t) \,\Sig_m\,U_m^{t, t_0}\, \V(t_0)\, \phi_m \big) \big|_{t_0} \:\dd t \:,
\end{align}
where for notational simplicity we here omitted the restrictions~$|_{t_0}$
for the solutions~$\psi_m$ and~$\phi_m$.
Adding the terms gives zero.
 }} \QEDrem \end{Remark}
The remainder of this section is devoted to the proof of Proposition~\ref{prpSrep}.
Our strategy is to combine the Lippmann-Schwinger equation with estimates in momentum space.
We begin with two technical lemmas.

\begin{Lemma} \label{lemmaC1}
Assume that the external potential~$\B$ satisfies condition~\eqref{L1B}.
For any~$t_0 \in \R$, we denote the characteristic functions in the future and past, respectively,
of this hypersurface~$t=t_0$ by~$\chi_{t_0}^\pm(x)$ (i.e.\ $\chi_{t_0}^\pm(x) = \Theta(\pm(x^0-t_0))$,
where~$\Theta$ is the Heaviside function). Then for any~$\psi_m \in \Cisc(\scrM, S\scrM) \cap \H_m$, the
wave function~$k_m(\chi_{t_0}^\pm \B \psi_m)$ is a well-defined vector in~$\H_{t_0}$ and
\beq \|k_m(\chi_{t_0}^\pm \B \psi_m)\|_{t_0} \leq \frac{1}{2 \pi}\, \|\psi_m\|_m
\int_{-\infty}^\infty \chi_{t_0}^\pm(\tau)\:|\B(\tau)|_{C^0} \:\dd\tau \:. \eeq
\end{Lemma}
\Proof Using the integral kernel representation~\eqref{Ukern} and~\eqref{Ukm}
together with the fact that the time evolution in the vacuum is unitary, we obtain
\begin{align}
2 \pi &\left\| \int_{\R^3} k_m \big( (t_0, .), (\tau, \vec{y}) \big)
\: \big(\chi_{t_0}^\pm \B \psi_m \big)(\tau, \vec{y})\: \dd^3y \right\|_{t_0} \notag \\
&= \big\| U^{t_0, \tau}_m \gamma^0 (\chi_{t_0}^\pm \B \psi_m)|_\tau \big\|_{t_0}
= \big\|\gamma^0 (\chi_{t_0}^\pm \B \psi_m)|_\tau \big\|_\tau \leq |\B(\tau)|_{C^0} \:\|\psi_m\|_m \:.
\end{align}
Integrating over~$\tau$ and using~\eqref{L1B} gives the result.
\QED

The following lemma is proved in~\cite[Eqs.~(2.13)--(2.17)]{grotz} (see Exercises~\ref{exdistrel}
and~\ref{exdistrel2}).

\begin{Lemma} \label{lemmadistrel} $\;$
In the Minkowski vacuum, the fundamental solution~$k_m$ and the Green's operator~$s_m$
defined by
\beq \label{smean}
s_m := \frac{1}{2}\: \big( s_m^\vee + s_m^\wedge \big)
\eeq
satisfy the distributional relations in the mass parameters~$m$ and~$m'$
\begin{align}
k_m\,k_{m'} &= \delta(m-m')\:p_m \\
k_m \,s_{m'}&= s_{m'}\,k_m=\frac{\text{\rm{PP}}}{m-m'}\:k_m \\
s_m\,s_{m'}&= \frac{\text{\rm{PP}}}{m-m'}\:(s_m-s_{m'})+\pi^2 \,\delta(m-m')\:p_m\;,
\end{align}
where~${\rm{PP}}$ denotes the principal part, and~$p_m$ is the distribution
\beq
\label{pmdef}
p_m(k) = (\slashed{k} + m) \: \delta(k^2 - m^2) \:.
\eeq
\end{Lemma}

\Proof[Proof of Proposition~\ref{prpSrep}.]
Let~$\psi \in \H^\infty$ be a family of solutions of the Dirac equation for varying mass.
We denote the boundary values at time~$t_0$ by~$\psi^0_m := \psi_m|_{t_0}$.
Then we can write the Lippmann-Schwinger equation~\eqref{lse} as
\beq \psi_m|_t = U_m^{t,t_0} \psi^0_m +\cI \int_{t_0}^t U_m^{t, \tau} \big( \gamma^0 \B\: \psi_m \big)
\big|_\tau\: \dd\tau \:. \eeq
We now bring this equation into a more useful form.
Expressing the time evolution operator with the help of~\eqref{Ukm} in terms of the fundamental solution,
we obtain
\begin{align}
\psi_m(x) &= 2 \pi \int_{\R^3} k_m \big( x, (t_0, \vec{y}) \big)\: \gamma^0 \psi_m^0(t_0, \vec{y}) \:\dd^3y \notag \\
&\quad + 2 \pi \cI \int_{t_0}^{x^0} \dd y^0 \int_{\R^3} \dd^3y \: k_m(x,y)(\B\: \psi_m)(y)\:.
\end{align}
Applying~\eqref{kmdef2} and using that the advanced and retarded Green's operators
are supported in the future and past light cones, respectively,
we can rewrite the last integral
in terms of the advanced and retarded Green's operators,
\beq \psi_m = 2 \pi \,k_m \big( \gamma^0 \delta_{t_0} \psi_m^0 \big)
- s_m^\wedge \big(\chi_{t_0}^+ \B \psi_m \big) - s_m^\vee \big( \chi_{t_0}^- \B \psi_m \big) \:, \eeq
where~$\delta_{t_0}(x) := \delta(t_0-x^0)$ is the $\delta$ distribution supported on the
hypersurface~$x^0=t_0$. Next, we express the advanced and retarded Green's operators in terms
of the Green's operator~\eqref{smean}: According to~\eqref{kmdef2}, we have the relations
\beq s_m=s_m^{\vee}-\cI\pi k_m=s_m^{\wedge}+\cI\pi k_m \eeq
and thus
\beq \label{gdef}
\psi_m = k_m g_m - s_m \B \psi_m \qquad \text{with} \qquad
g_m := 2 \pi \,\gamma^0 \delta_{t_0} \psi_m^0  + \cI \pi\, \epsilon_{t_0} \B \psi_m \:,
\eeq
where~$\epsilon_{t_0}$ is the step function
\beq \epsilon_{t_0}(x) := \epsilon(x^0-t_0) \eeq
(and we omitted the brackets in expressions like~$k_m g_m \equiv k_m(g_m)$).
Note that the expression~$k_m g_m$ is well-defined according to Lemma~\ref{lemmaC1}.
We also remark that by applying the operator~$(i \Pdd - m)$ to the distribution~$g_m$ in~\eqref{gdef},
one immediately verifies that~$\psi_m$ indeed satisfies the Dirac equation~$(\cI \Pdd - m) \psi_m = - \B \psi_m$.

Now we can compute the inner product~$\bra \p \psi | \p \psi \ket$ with the help of
Lemma~\ref{lemmadistrel}. Namely, using~\eqref{gdef},
\begin{align}
\bra \p \psi | \p \psi \ket \:=& \iint_{I \times I} \bra k_m g_m - s_m \B \psi_m \:|\: k_{m'} g_{m'} - s_{m'}
\B \psi_{m'} \ket \: \dd m\: \dd m' \notag \\
=& \int_I \Big( \bra g_m \,|\, p_m g_m \ket + \pi^2\, \bra \B \psi_m \,|\, p_m \B \psi_m \ket \Big) \dd m \notag \\
&+ \iint_{I \times I} \frac{\text{PP}}{m-m'} \Big(
\bra \B \psi_m \,|\, k_{m'} g_{m'} \ket
- \bra k_m g_m  \,|\, \B \psi_{m'} \ket \notag \\
&\qquad\qquad\qquad\qquad + \bra \B \psi_m \,|\, (s_m - s_{m'}) \B \psi_{m'} \ket \Big) \:\dd m\: \dd m' \:.
\end{align}
Note that this computation is mathematically well-defined in the distributional sense
because~$\psi_m$ and~$g_m$ are smooth and compactly supported
in the mass parameter~$m$.
Employing the explicit formula for~$g_m$ in~\eqref{gdef}, we obtain
\beq \bra \p \psi | \p \psi \ket =
\int_I \Big( \bra g_m \,|\, p_m g_m \ket + \pi^2\, \bra \B \psi_m \,|\, p_m \B \psi_m \ket \Big) \dd m \:. \eeq
Comparing~\eqref{kmp} with~\eqref{pmdef} and taking into account that
the operator~$\Sig_m$ defined by~\eqref{S0def} gives a minus sign for the states of negative frequency,
we get
\beq p_m = \Sig_m \,k_m \:. \eeq
Using this identity together with Proposition~\ref{prpdual} in the vacuum yields the relations
\begin{align}
\bra g_m \,|\, p_m g_m \ket  &= (k_m g_m \,|\, \Sig_m \,k_m g_m) |_{t_0} \\
\bra \B \psi_m \,|\, p_m \B \psi_m \ket &= ( k_m \B \psi_m \,|\, \Sig_m \,k_m \B \psi_m) |_{t_0}\:.
\end{align}
We finally apply Proposition~\ref{prp21} to obtain the representation
\beq \label{stiprel}
\bra \p \psi | \p \psi \ket =
\int_I \Big( ( h_m \,|\, \Sig_m \,h_m )|_{t_0} + \pi^2\, ( k_m \B \psi_m \,|\, 
\Sig_m \,k_m \B \psi_m )|_{t_0} \Big) \: \dd m \:,
\eeq
where
\beq h_m := \psi_m  + \cI \pi\, k_m ( \epsilon_{t_0} \B \psi_m) \:. \eeq

Comparing~\eqref{Sdef2} with~\eqref{stiprel}, we get
\beq ( \psi_m \,|\, \tilde{\Sig}_m\, \psi_m)_m = ( h_m \,|\, \Sig_m \,h_m )|_{t_0} + \pi^2\, ( k_m \B \psi_m \,|\, 
\Sig_m \,k_m \B \psi_m )|_{t_0} \:. \eeq
Expressing the operators~$k_m$ according to~\eqref{Ukm} by the time evolution operator
and writing~$\psi_m$ in terms of the initial data as
\beq \psi_m|_t = \tilde{U}^{t,t_0} \psi|_{t_0} \:, \eeq
we obtain
\begin{align}
&( \psi_m \,|\, \tilde{\Sig}_m\, \psi_m)_m  \notag \\
&= ( \psi | \Sig_m \psi)|_{t_0} - \frac{\cI}{2} 
\int_{-\infty}^\infty \epsilon(t-t_0)\, \big( \psi \,\big|\, \Sig_m\, U^{t_0, t} \,\V(t)\, \tilde{U}^{t, t_0} \,
\psi \big) \big|_{t_0}\:\dd t \notag \\
&\quad\: +\frac{\cI}{2} \int_{-\infty}^\infty \epsilon(t-t_0)\: \big(U^{t_0, t} \,\V(t)\, \tilde{U}^{t, t_0}\,
\psi \,\big|\,  \Sig_m\, \psi \big) \big|_{t_0}\:\dd t \notag \\
&\quad\: +\frac{1}{4} \iint_{\R \times \R} \epsilon(t-t_0) \,\epsilon(t'-t_0)\:
\big( U^{t_0, t} \,\V(t)\, \tilde{U}^{t, t_0}\, \psi \,\big|\, \Sig_m\, U^{t_0, t'} \,\V(t')\,
\tilde{U}^{t', t_0} \,\psi \big) \big|_{t_0}\:\dd t\, \dd t' \notag \\
&\quad\:
+\frac{1}{4} \iint_{\R \times \R} \big( U^{t_0, t} \,\V(t)\, \tilde{U}^{t, t_0}\, \psi \,\big|\, \Sig_m\, U^{t_0, t'} \,\V(t')\,
\tilde{U}^{t', t_0} \,\psi \big) \big|_{t_0}\:\dd t\, \dd t' \:.
\end{align}
Rearranging the terms and polarizing gives the result.
\QED

\Proof[Proof of Theorem~\ref{thmC2}.]
Since the time evolution operators are unitary and the operators~$\Sig_m$ have norm one
(see~\eqref{S0def}), the representation~\eqref{Sigm1} and~\eqref{Sigm2} gives rise to
the following estimate for the $\sup$-norm of~$\tilde{S}_m$,
\beq \big\|\tilde{\Sig}_m \big\| \leq 1 + \int_\R |\V(t)|_{C^0}\:\dd t 
+\iint_{\R \times \R} |\V(t)|_{C^0}\; |\V(t')|_{C^0} \:\dd t\, \dd t' \:. \eeq
The decay assumption~\eqref{L1B} implies that the $\sup$-norm of~$\tilde{S}_m$ is
bounded uniformly in~$m$. Using this fact in~\eqref{Sdef2} gives the inequality~\eqref{smop},
thereby establishing the strong mass oscillation property.
\QED

We finally remark that the uniqueness statement in Proposition~\ref{prpunique}
implies that the relations~\eqref{Sigm1} and~\eqref{Sigm2} yield
an explicit representation of the fermionic signature operator
in the presence of a time-dependent external potential.

\section{Exercises}

\begin{Exercise} \label{ex2.4-3}  {\em{ 
For a smooth one-parameter family of matrices~$F(\alpha)$, $\alpha
\in \R$, the {\bf{ordered exponential}}~$\Pexp (\int F(\alpha) \:d\alpha)$
\sindex{ordered exponential}%
\nindex{ahd@$\Pexp$ -- ordered exponential}%
\begin{align}
\Pexp \bigg( \int_a^b F(\alpha) \:d\alpha \bigg)
&= \1 + \int_a^b F(t_0) \:\dd t_0 \:+\: \int_a^b \dd t_0\:F(t_0) \int_{t_0}^b \dd t_1 \: F(t_1) \notag \\
&\qquad \!+ \int_a^b \dd t_0\:F(t_0) \int_{t_0}^b \dd t_1 \: F(t_1) \int_{t_1}^b \dd t_2 \:F(t_2) + \cdots \:. %\label{defPexp}
\end{align}
In this exercise, we collect a few elementary properties of the ordered exponential.
\bitem
\item[(a)] Assume that the matrix-valued function~$F$ is commutative in the sense that
\beq \big[F(\alpha), F(\beta) \big] = 0 \qquad \text{for all~$\alpha, \beta \in [a,b]$}\:. \eeq
Show that the ordered exponential reduces to the ordinary exponential,
\beq \Pexp \bigg( \int_a^b F(\alpha) \:\dd\alpha \bigg) = \exp \bigg( \int_a^b F(\alpha) \:\dd\alpha \bigg) \:. \eeq
{\em{Hint:}} Show inductively that
\beq \int_a^b \dd t_0\:F(t_0) \int_{t_0}^b \dd t_1 \: F(t_1)
\cdots \int_{t_{n-1}}^b \dd t_n \:F(t_n) = \frac{1}{(n+1)!} \bigg(  \int_a^b F(t)\:\dd t \bigg)^{n+1} \:. \eeq
\item[(b)] Assume that~$F$ is continuous on~$[a,b]$. Show that the Dyson series converges
absolutely and that
\beq \bigg\| \Pexp \bigg( \int_a^b F(\alpha) \:\dd\alpha \bigg) \bigg\|
\leq \exp \bigg( \int_a^b \big\|F(\alpha)\big\| \:\dd\alpha \bigg) \:. \eeq
{\em{Hint:}} Estimate the integrals and apply~(a).
\item[(c)] Show by direct computation that the ordered exponential satisfies the equations
\begin{gather}
\frac{\dd}{\dd a} \Pexp \bigg( \int_a^b F(\alpha) \:\dd\alpha \bigg)
= -F(a)\: \Pexp \bigg( \int_a^b F(\alpha) \:\dd\alpha \bigg) \label{Fadiff} \\
\Pexp \bigg( \int_a^a F(\alpha) \:\dd\alpha \bigg) = \1\:. \label{Fgroup2}
\end{gather}
Use the uniqueness theorem for solutions of ordinary differential equations to
give an alternative definition in terms of the solution of an initial-value problem.
Use this reformulation to show the group property
\beq \label{Fgroup}
\Pexp \bigg( \int_a^b F(\alpha) \:\dd\alpha \bigg) \Pexp \bigg( \int_b^c F(\alpha) \:\dd\alpha \bigg)
= \Pexp \bigg( \int_a^c F(\alpha) \:\dd\alpha \bigg) \:.
\eeq
\item[(d)] Show that
\beq \label{Fbdiff}
\frac{\dd}{\dd b} \Pexp \bigg( \int_a^b F(\alpha) \:\dd\alpha \bigg)
= \Pexp \bigg( \int_a^b F(\alpha) \:\dd\alpha \bigg) \: F(b) \:.
\eeq
{\em{Hint:}} Differentiate the identity~\eqref{Fgroup} in the case~$c=a$
and use the group properties~\eqref{Fgroup2} and~\eqref{Fgroup}.
\item[(e)] Show that
\beq \Pexp \bigg( \int_a^b F(\alpha) \:\dd\alpha \bigg)^*
= \Pexp \bigg( \int_b^a \big(-F(\alpha)^*\big) \:\dd\alpha \bigg) \:. \eeq
Deduce that if~$F(\alpha)$ is an anti-Hermitian matrix, then the
ordered exponential is a unitary matrix.
{\em{Hint:}} There are two alternative methods.
One method is to argue using the differential equations~\eqref{Fadiff} and~\eqref{Fbdiff}
or with the group property. A more computational approach is to take the adjoint of the Dyson series
and re-parametrize the integrals.
\eitem
}} \end{Exercise}

\begin{Exercise} {\em{ Given~$\omega \in \R$ and a smooth function~$V(t)$, we
consider the ordinary differential equation
\beq \big( \cI \partial_t + \omega \big) \phi(t) = V(t)\: \phi(t)\:. \eeq
\bitem
\item[(a)] Write down the Lippmann-Schwinger equation, taking the right-hand side of the
equation as the perturbation.
\sindex{Lippmann-Schwinger equation}%
{\em{Hint:}} The free time evolution operator~$U^{t,t'}$ was computed in Exercise~\ref{ex1d}.
\item[(b)] Express the Lippmann-Schwinger equation in the case~$\omega=0$
explicitly as an integral equation. How is it related to the integral equation
used in the Picard iteration (in the proof of the Picard-Lindel\"of theorem)?
\sindex{Picard iteration}%
\eitem
}} \end{Exercise}

\chapter{Methods of Perturbation Theory} \label{secperturb}
\sindex{perturbation theory}%
In Chapter~\ref{secFSO}, the unregularized kernel of the fermionic projector was
constructed abstractly with functional analytic methods. In order to fill these constructions with life,
one can analyze this kernel with methods of perturbation theory.
The resulting explicit formulas give a detailed understanding of the structure of this kernel.
We now outline the perturbative methods; more details can be found in~\cite[Chapter~2]{cfs}
or in the original papers~\cite{sea, grotz, norm}.

As the general setting, we again consider the Dirac equation in Minkowski space~\eqref{Dout} in the presence
of an external potential~$\B$ which we assume to be symmetric~\eqref{Bsymm}.
In preparation, we rewrite the definition of the fermionic signature operator constructed in Chapter~\ref{secFSO}
in a way suitable for the perturbative treatment. Our starting point is the representation~\eqref{Smdef}
of the spacetime inner product in terms of the scalar product,
\beq \label{Smdefpert}
\bra \p \psi | \p \psi' \ket = \int_I (\psi_m \,|\, \Sig_m \,\psi'_m)_m\: \dd m \:,
\eeq
where~$\Sig_m$ is the fermionic signature operator.
Here, $\psi=(\psi_m)_{m \in I}$ and similarly~$\psi'$ are families of solutions of the Dirac equation for a varying
mass parameter. More specifically, we now consider families obtained by acting with the causal
fundamental solution on given test wave functions; that is,
\beq \psi_m = \tilde{k}_m \phi \quad \text{and} \quad \psi'_m = \tilde{k}_m \phi' \qquad \text{with} \qquad
\phi, \phi' \in C^\infty_0(\scrM, S\scrM) \:. \eeq
Using this ansatz in~\eqref{Smdefpert} and pulling the mass integrals outside, we obtain the formula
\beq \label{repform}
\int_I \dd m \int_I \dd m' \bra \tilde{k}_m \phi \,|\, \tilde{k}_{m'} \phi' \ket = \int_I (\tilde{k}_m \phi \,|\, \Sig_m \,\tilde{k}_m \phi')_m\: \dd m \:.
\eeq
Next, we rewrite the integrand on the left-hand side by using that the fundamental solution is symmetric with respect to the spacetime inner product (see Corollary~\ref{corksymm}),
\beq \bra \tilde{k}_m \phi | \tilde{k}_{m'} \phi' \ket = \bra \phi \,|\, \tilde{k}_m \tilde{k}_{m'} \phi' \ket \:. \eeq
Moreover, the integrand on the right-hand side can be rewritten with the help of Proposition~\ref{prpdual} as
\beq (\tilde{k}_m \phi \,|\, \Sig_m \,\tilde{k}_m \phi')_m = \la \phi \,|\, \Sig_m \,\tilde{k}_m \phi' \ra \:. \eeq
Thus~\eqref{repform} becomes
\beq \int_I \dd m \int_I \dd m' \bra \phi \,|\, \tilde{k}_m \,\tilde{k}_{m'} \psi' \ket
= \int_I \la \phi \,|\, \Sig_m \,\tilde{k}_m \phi' \ra\: \dd m \:. \eeq
Here one should keep in mind the product~$\tilde{k}_m \,\tilde{k}_{m'}$ is an operator product
in spacetime,
\beq \label{kkrel}
\big( \tilde{k}_m \,\tilde{k}_{m'} \big)(x,y)
= \int_{\scrM} \tilde{k}_m(x,z) \,\tilde{k}_{m'}(z,y)\: \dd^4z \:,
\eeq
whereas in the product~$\Sig_m \,\tilde{k}_m$, we multiply by an operator on the Hilbert space~$\H_m$
defined, for example, at time~$t$. In order to clarify the notation, we write this product as
\beq \Sig_m \,|_t\,\tilde{k}_m \:. \eeq
Then the relation~\eqref{kkrel} can be written in the short form
\beq \label{kktilde}
\tilde{k}_m \tilde{k}_{m'} = \delta(m-m')\: \Sig_m \,|_t\,\tilde{k}_m \:.
\eeq
In this way, one is led to considering products of operators in spacetime
that involve the mass as a parameter. Carrying out the products gives rise to
$\delta$ distributions in the respective mass parameters.

This computational procedure was introduced in~\cite{sea}.
In the Minkowski vacuum, it can be carried out most conveniently in momentum space.
We begin with the formula for the causal fundamental solution in momentum space~\eqref{kmp},
\beq %\label{kmppert}
k_m(p) = (\slashed{p} + m) \: \delta(p^{2}-m^{2}) \: \epsilon(p^{0}) \:. \eeq
Then, using Plancherel together as well as the anti-commutation relations of the Dirac matrices, we obtain
\begin{align}
\big( k_m\, k_{m'} \big)(p) &= k_m(p)\, k_{m'}(p)
=  (\slashed{p} + m) \: \delta(p^2 - m^2) \: \epsilon(p^0) \;
(\slashed{p} + m') \: \delta(p^2 - (m')^2) \: \epsilon(p^0) \notag \\
&= \big(p^2 + (m+m^\prime) \, \slashed{p} + m m^\prime \big) \: \delta \big(m^2 -
   (m^\prime)^2 \big) \; \delta(k^2 - m^2) \nonumber \\
&= \big(p^2 + (m+m^\prime) \,\slashed{p} + m m^\prime) \: \frac{1}{2m} \:\delta(m -
m^\prime) \; \delta(k^2 - m^2) \nonumber \\
&= \delta(m-m^\prime) \: (\slashed{p} + m) \: \delta(p^2 - m^2)
= \delta(m-m^\prime) \: \epsilon\big( p^0 \big)\: k_m(p) \:.
\end{align}
Comparing with~\eqref{kktilde}, we can read off that the fermionic signature operator
is simply the operator of multiplication operator by the sign of the frequency,
\beq \Sig_m(p) = \epsilon\big( p^0 \big) \:. \eeq
This computation is an efficient way of seeing that, in the Minkowski vacuum, the
fermionic signature operator gives back the frequency splitting.

We proceed by explaining how this computation can be extended to the situation
of the Dirac equation~\eqref{Dout} when an external potential~$\B$ is present.
We want to proceed order by order in 
a perturbation expansion in~$\B$. Before entering the details, we point out that
by a ``perturbation expansion'' we mean a formal expansion in powers of~$\B$.
The resulting formulas will be well-defined and finite to every order.
But it is unknown whether the power series converges. This procedure is
convincing because we already know from our functional analytic constructions
in Chapter~\ref{secFSO} that the fermionic signature operator and the unregularized
fermionic projector are well-defined mathematical objects. With this in mind, the
only purpose of the constructions in this chapter is to compute these objects more
explicitly. For this purpose, a perturbative treatment order by order in perturbation theory
is most suitable.

\section{Perturbation Expansion of the Causal Green's Operators} \label{secperts}
\sindex{perturbation expansion!of causal Green's operators}%
We already encountered the causal Green's operators for the Dirac equation several times
in this book. In Section~\ref{secexgreen}, they were constructed with methods of hyperbolic
partial differential equations (see Theorem~\ref{thmgreenhyp}).
In Section~\ref{secgreen}, on the other hand, we used Fourier methods to derive
explicit formulas for the causal Green's operator in the Minkowski vacuum (see~\eqref{8b}).
Taking these explicit formulas as the starting point, one can also write down closed formulas for the
causal Green's operators in the presence of an external potential.
In order to state these formulas, we consider the Dirac equation~\eqref{Dout} in the
presence of an external potential~$\B$.
We always denote the objects in the presence of the external potential with a tilde,
whereas the objects without a tilde refer to the vacuum.
Then, the advanced and retarded Dirac Green's operators have the perturbation expansions
\beq \tilde{s}_m^{\vee}=\sum_{n=0}^{\infty} \big(-s_m^{\vee}\, \mathscr{B} \big)^n \,s_m^{\vee}\;, \qquad
\tilde{s}_m^{\wedge}=\sum_{n=0}^{\infty} \big( -s_m^{\wedge}\, \mathscr{B} \big)^n \,s_m^{\wedge}\:,
\label{series-scaustilde}
\eeq
as can be verified as follows. First, one sees by direct computation
using the defining equation of the Green's operator~\eqref{Greendef} that they are
formal solutions of~\eqref{Greendefext}. For example, for the advanced Green's operator,
\begin{align}
(\cI &\Pdd_x + \B - m) \bigg( \sum_{n=0}^{\infty} \big(-s_m^{\vee} \mathscr{B} \big)^n s_m^{\vee} \bigg) \notag \\
&= (\cI \Pdd_x - m)\,s_m^{\vee} \bigg( \sum_{n=0}^{\infty} \big(-\mathscr{B} s_m^{\vee} \big)^n \bigg) 
+ \B \bigg( \sum_{n=0}^{\infty} \big( -s_m^{\vee}\mathscr{B} \big)^ns_m^{\vee} \bigg) \notag \\
&= \sum_{n=0}^{\infty}(-\mathscr{B} s_m^{\vee})^n
+ \B \bigg( \sum_{n=0}^{\infty} \big(-s_m^{\vee} \mathscr{B} \big)^n s_m^{\vee} \bigg) = \1 \:.
\end{align}
Second, the fact that the Green's operators in~\eqref{series-scaustilde} are either all advanced
or all retarded implements the causal properties of the respective Green's operators.
Let us consider, for example, the integral kernel of the
first order contribution to the advanced Green's operator
\beq \label{sfirst}
\big( -s_m^\wedge \mathscr{B} s_m^\wedge \big)(x,y)
= -\int_\scrM s_m^\wedge(x,z)\: \mathscr{B}(z)\: s_m^\wedge(z,y)\: \dd^4z \:.
\eeq
The integrand vanishes unless~$z$ lies in the causal future of~$y$ and~$x$ lies in the causal future of~$z$.
Using the transitivity of the causal relations, one concludes that the integral is zero unless~$x$ lies in the
causal future of~$y$. In this sense, the expression~\eqref{sfirst} is again causal and retarded.
The higher orders can be treated similarly by induction.

We finally explain in which sense the perturbation series~\eqref{series-scaustilde} are mathematically
well-defined. To every order in perturbation theory, the operator products are well-defined and finite,
provided that the potential~$\B$ is smooth and decays so fast at infinity that the functions
$\B(x)$, $x^i \B(x)$, and~$x^i x^j \B(x)$ are integrable (for an inductive proof, see~\cite[Lemma~2.1.2]{cfs}).
\Evtl{Hierzu eine \"Ubungsaufgabe?}%
Knowing that the Green's operators are well-defined non-perturbatively (see Chapter~\ref{secshs}),
we disregard the issue of convergence of the perturbation series.

\section{The Causal Perturbation Expansion of the Fermionic Projector} \label{seccpertP}
\sindex{perturbation expansion!of fermionic projector}%
\sindex{causal perturbation expansion}%
Using~\eqref{kssdef}, we also have a unique perturbation expansion for the
causal fundamental solution,
\beq
\tilde{k}_m = \frac{1}{2\pi \cI}(\tilde{s}_m^{\vee}-\tilde{s}_m^{\wedge}) \:. \label{def-ktil}
\eeq
Using the identities
\beq \label{scaus}
s_m^{\vee} = s_m+\cI\pi k_m\:, \qquad s_m^{\wedge} = s_m-\cI\pi k_m \:,
\eeq
where we introduced the {\em{symmetric Green's operator}}
\beq \label{sdef}
s_m := \frac{1}{2}\: \big( s_m^\vee + s_m^\wedge \big) \:,
\eeq
one can write the above perturbation series as operator product expansions. More precisely,
the operator~$\tilde{k}_m$ has the series expansion
\beq \label{kex}
\tilde{k}_m=\sum_{\beta=0}^{\infty}(\cI\pi)^{2\beta} \:b_m^<\, k_m\, (b_m k_m)^{2\beta}\, b_m^>\:,
\eeq
where the factors~$b_m^\bullet$ are defined by
\beq \label{bmdef}
b_m^<=\sum_{n=0}^{\infty}(-s_m\mathscr{B})^n\;,\qquad
b_m=\sum_{n=0}^{\infty}(-\mathscr{B}s_m)^n\mathscr{B}\;,\qquad
b_m^>=\sum_{n=0}^{\infty}(-\mathscr{B}s_m)^n\:.
\eeq

In the following constructions, we need to multiply the operator products in~\eqref{kex}.
These products have a mathematical meaning as distributions in the involved mass parameters,
\begin{align}
p_m\,p_{m'}&=k_m\,k_{m'}=\delta(m-m')\:p_m \\
p_m\,k_{m'}&=k_m\,p_{m'}=\delta(m-m')\:k_m \\
k_m \,b_m^>\, b_{m'}^<\, k_{m'} &= \delta(m-m') \Big( p_m
+ \pi^2\,k_m\, b_m\, p_m\, b_m\, k_m \Big) \:, \label{bbprod}
\end{align}
where
\begin{align}
p_m(q) &= (\slashed{q}+m)\: \delta(q^2-m^2) \label{pmdef3} \\
k_m(q) &= (\slashed{q}+m)\: \delta(q^2-m^2)\: \epsilon(q^0) \:. \label{kmdef3}
\end{align}
Since all these formulas involve a common prefactors~$\delta(m-m')$,
we can introduce a convenient notation by leaving out this factor and omitting the
mass indices. For clarity, we denote this short notation with a dot; that is, symbolically
\beq \label{sdotdef}
A \sdot B = C \qquad \text{stands for} \qquad
A_m \,B_{m'} = \delta(m-m')\: C_m \:.
\eeq
With this short notation, the multiplication rules can be written in the compact form
\beq p \sdot p =k \sdot k =p\:, \qquad p \sdot k=k \sdot p=k\:,\qquad
k \,b^> \sdot b^< k =p+\pi^2 kbpbk \:. \label{rules}
\eeq
In all the subsequent calculations, the operator products are well-defined
provided that the potential~$\B$ is sufficiently smooth and has suitable decay properties
at infinity (for details, see again~\cite[Lemma~2.1.2]{cfs}). But again, all infinite series are to be understood
merely as formal power series in the potential~$\B$.

Using this notation, we can write~\eqref{kex} as
\beq \label{ktildef}
\tilde{k} = k + \Delta k \qquad \text{with} \qquad
\Delta \tilde{k} =\sum_{\beta=0}^{\infty}(\cI\pi)^{2\beta} \:b^<\, k\, (b k)^{2\beta}\, b^> \:-\: k
\eeq
(note that~$\Delta \tilde{k}$ is at least linear in~$\B$).
Powers of the operator~$\tilde{k}$ with the product~\eqref{sdotdef}
are well-defined using the multiplication rules~\eqref{rules}.
This makes it possible to develop a spectral calculus for~$\tilde{k}$, which is
formulated most conveniently with contour integrals. To this end, we introduce the {\em{resolvent}} by
\sindex{resolvent!of causal fundamental solution}%
\nindex{ahe@$R_\lambda$ -- resolvent of causal fundamental solution}%
\beq \label{Resdef}
\tilde{R}_\lambda = (\tilde{k} - \lambda)^{-1}\:.
\eeq
We choose a contour~$\Gamma_+$ that encloses the point~$1$ in counter-clockwise 
direction and does not enclose the points~$-1$ and~$0$.
Likewise, $\Gamma_-$ is chosen as a contour that encloses the point~$-1$ in counter-clockwise 
direction and does not enclose the points~$1$ and~$0$.
Given a holomorphic function~$f$ we define~$f(\tilde{k})$ by
\beq \label{fres}
f \big( \tilde{k} \big) := -\frac{1}{2 \pi \cI} \ointctrclockwise_{\Gamma_+ \cup \Gamma_-} f(\lambda)\:
\tilde{R}_\lambda\: \dd\lambda  \:.
\eeq
Before going on, we need to explain how the resolvent and these contour integrals are to be understood
mathematically. First, the resolvent can be expressed in terms of the vacuum resolvent with a
a perturbation series being a formal Neumann series,
\beq \label{tR}
\tilde{R}_\lambda = (k - \lambda + \Delta k)^{-1}
= (1 + R_\lambda \sdot \Delta k)^{-1} \sdot R_\lambda 
= \sum_{n=0}^\infty (-R_\lambda \sdot \Delta k)^n \sdot R_\lambda \:.
\eeq
In order to define~$R_\lambda$, we note that, according to~\eqref{rules}, the operator~$k$ has the
eigenvalues~$\pm 1$ and~$0$
with corresponding spectral projectors~$(p \pm k)/2$ and~$\1-p$.
Hence, we can write the free resolvent as
\beq \label{Rfree}
R_\lambda = \frac{p+k}{2} \left( \frac{1}{1-\lambda} \right) + \frac{p-k}{2} \left( \frac{1}{-1-\lambda} \right)
- \frac{\1-p}{\lambda} \:.
\eeq
Substituting this formula in~\eqref{tR}, to every order in perturbation theory we obtain
a meromorphic function in~$\lambda$ having poles only at~$\lambda=0$ and~$\lambda= \pm1$.
Therefore, the contour integral in~\eqref{fres} can be computed with residues, and the result
is independent of the choice of the contours~$\Gamma_-$ and~$\Gamma_+$.
In this way, the operator~$f(\tilde{k})$ is uniquely defined as a formal perturbation series.
As explained at the end of the previous section (in the paragraph after~\eqref{sfirst}),
this series is well-defined and finite to every order in perturbation theory.

We now establish the functional calculus.
\begin{Thm} {\bf{(functional calculus)}} \label{thmfcalc}
\sindex{functional calculus!for causal fundamental solution}%
For any functions~$f, g$ which are holomorphic in discs around~$\pm1$
which contain the contours~$\Gamma_\pm$,
\begin{align}
(\cI \Pdd+\B-m)\, f \big( \tilde{k} \big) &= 0 \label{fsol} \\ 
f \big( \tilde{k} \big)^* &= \overline{f}\big( \tilde{k} \big) \label{fstar} \\ 
f \big( \tilde{k} \big) \sdot g \big( \tilde{k} \big) &= (fg)\big( \tilde{k} \big) \:. \label{fg} 
\end{align}
\end{Thm}
\Proof 
Since the image of the operator~$\tilde{k}$ lies in the kernel of the Dirac operator, we know that
\beq (\cI \Pdd + \B - m)\: \tilde{R}_\lambda = (\cI \Pdd + \B - m)\, \big( -\lambda^{-1} \big) . \eeq
Taking the contour integral~\eqref{fres} gives~\eqref{fsol}.

The operators~$p_m$, $k_m$ and~$s_m$ are obviously symmetric (see the 
relations~\eqref{pmdef3}, \eqref{kmdef3}
and~\eqref{sdef}). According to~\eqref{kex}, the operator~$\tilde{k}_m$ is also symmetric. Hence,
the resolvent~$\tilde{R}_\lambda$ defined by~\eqref{Resdef} has the property
\beq \tilde{R}_\lambda^* = \tilde{R}_{\overline{\lambda}}\:. \eeq
The relation~\eqref{fstar} follows by taking the adjoint of~\eqref{fres} and
reparametrizing the integral.

The starting point for proving~\eqref{fg} is the resolvent identity (see Exercise~\ref{ex:resolvent})
\sindex{resolvent identity!for causal fundamental solution}%
\beq \label{rr1}
\tilde{R}_\lambda \sdot \tilde{R}_{\lambda'} = \frac{1}{\lambda-\lambda'} \left(
\tilde{R}_\lambda - \tilde{R}_{\lambda'} \right) \:.
\eeq
We set~$\Gamma=\Gamma_+ \cup \Gamma_-$ and denote the corresponding
contour for~$\lambda'$ by~$\Gamma'$.
Since the integral~\eqref{fres} is independent of the precise choice of the contour,
we may choose
\beq \Gamma = \partial B_{\delta}(1) \cup \partial B_{\delta}(-1) \qquad \text{and} \qquad
\Gamma'=\partial B_{2 \delta}(1) \cup \partial B_{2 \delta}(-1) \eeq
for sufficiently small~$\delta<1/2$.
Then~$\Gamma$ does not enclose any point of~$\Gamma'$, implying that
\beq \label{rr2}
\ointctrclockwise_\Gamma \frac{f(\lambda)}{\lambda-\lambda'} \: \dd\lambda = 0 
\qquad \text{for all~$\lambda' \in \Gamma'$}\:.
\eeq
On the other hand, $\Gamma'$ encloses every point of~$\Gamma$, so that
\beq \label{rr3}
\ointctrclockwise_{\Gamma'} f(\lambda)\, g(\lambda')\: \frac{\tilde{R}_{\lambda}}{\lambda-\lambda'} \: \dd\lambda' =
-2 \pi \cI\,  f(\lambda)\, g(\lambda)\: \tilde{R}_\lambda \qquad \text{for all~$\lambda \in \Gamma$}\:.
\eeq
Combining~\eqref{rr1} with~\eqref{rr2} and~\eqref{rr3}, we obtain
\begin{align}
f \big( \tilde{k} \big) \sdot g \big( \tilde{k} \big) 
&= -\frac{1}{4 \pi^2} \ointctrclockwise_{\Gamma} f(\lambda)\, \dd\lambda \ointctrclockwise_{\Gamma'} g(\lambda')\, \dd\lambda'\;
\frac{1}{\lambda-\lambda'} \left( \tilde{R}_\lambda - \tilde{R}_{\lambda'} \right) \notag \\
&= -\frac{1}{2 \pi \cI} \ointctrclockwise_{\Gamma} f(\lambda)\, g(\lambda)\: \tilde{R}_\lambda \:\dd\lambda
= (fg)\big( \tilde{k} \big) \:. 
\end{align}
This concludes the proof.
\QED

This functional calculus makes it possible to compute the unregularized kernel of the fermionic projector,
as we now explain. Our starting point is the defining equation for the fermionic signature operator~\eqref{kktilde},
which we can now write in short from
\beq \tilde{k} \sdot \tilde{k} = \tilde{\Sig}_m \,|_t\, \tilde{k} \:. \eeq
Iterating this relation, we obtain for any~$p \in \N$
\beq \big( \tilde{k} \sdot \big)^p \,\tilde{k} = \big(\tilde{\Sig}_m \,|_t\, \big)^p \, \tilde{k}
= \big(\tilde{\Sig}_m \big)^p \,\big|_t\, \tilde{k} \qquad \text{for all~$p \in \N$}\:. \eeq
Consequently, this formula also holds for the functional calculus; that is,
\beq f \big(\tilde{k} \big) \sdot \tilde{k} = f\big( \tilde{\Sig}_m \big) \,\big|_t\, \tilde{k} \:. \eeq
This formula makes it possible to express the unregularized kernel~$\tilde{P}_-$ in~\eqref{Ppmdef2} by
\begin{align}
\tilde{P}_- &= -\chi_{(-\infty, 0)}\big( \tilde{\Sig}_m \big)\, \tilde{k}_m
= -\chi_{(-\infty, 0)}\big( \tilde{\Sig}_m \big)\,|_t\, \tilde{k} 
= \chi_{(-\infty, 0)}\big(\tilde{k} \big) \sdot \tilde{k} \notag \\
&= -\Big( \frac{1}{2 \pi \cI} \ointctrclockwise_{\Gamma_-} \: \tilde{R}_\lambda\: \dd\lambda \Big) \sdot \tilde{k}
= -\frac{1}{2 \pi \cI} \ointctrclockwise_{\Gamma_-} (-\lambda)\: \tilde{R}_\lambda\: \dd\lambda \:.
\end{align}
Substituting the perturbation expansion for~$\tilde{R}_\lambda$ in~\eqref{tR}
and writing the vacuum resolvent in the form~\eqref{Rfree}, one can carry out the contour integral
with residues. This gives the desired perturbation expansion for~$P_-$.
More details on this method and explicit formulas can be found in~\cite[Section~3.3 and Appendix~A]{norm}.

\section{Exercises}
\begin{Exercise} (Perturbative description of gauge transformations)
\sindex{gauge transformation!perturbative description}%
{\em{ We consider the perturbation expansion for the Dirac Green's operators~\eqref{series-scaustilde}
for a perturbation by a pure gauge potential; that is,
\beq \mathscr{B}(x) = \Pdd \Lambda(x) \eeq
with a real-valued function~$\Lambda$.
\bitem
\item[(a)] Show that the Dirac operator with interaction can be written as
\beq \cI \Pdd + (\Pdd \Lambda) - m = \E^{\cI \Lambda(x)} \big(\cI \Pdd - m \big)\: \E^{-\cI \Lambda(x)} \:. \eeq
Conclude that the perturbation of the Dirac solutions amounts to multiplication by a phase function; that is,
\beq \tilde{\Psi}(x) = \E^{\cI \Lambda(x)}\: \Psi(x) \:. \eeq
Explain why these findings are a manifestation of the local gauge freedom of electrodynamics.
\item[(b)] Show that the gauge phases also appear in the perturbation expansion~\eqref{series-scaustilde}
in the sense that
\beq \tilde{s}_m^\wedge(x,y) = \E^{\cI \Lambda(x)} \: s_m^\wedge(x,y)\: \E^{- \cI \Lambda(y)} \:. \eeq
{\em{Hint:}} To first order, one needs to show that
\beq - \big( s_m^{\wedge} \,\mathscr{B}\, s_m^{\wedge} \big)(x,y) = \cI \big( \Lambda(x) - \Lambda(y) \big)\, s_m^\wedge(x,y) \:. \eeq
To this end, it is convenient to write the perturbation operator as a commutator,
\beq \mathscr{B} = -\cI \,\big[ (\cI \Pdd - m), \,\Lambda \big] \eeq
and use the defining equation of the Green's operator~\eqref{Greendef}.
To higher order, one can proceed inductively.
\eitem
}} \end{Exercise}

\begin{Exercise} {\em{ Prove the identity~\eqref{bbprod}.
{\em{Hint:}} Use the multiplication rules derived in Exercises~\ref{exdistrel} and~\ref{exdistrel2}.
Make use of the fact that one gets telescopic sums.
}} \end{Exercise}

\begin{Exercise} {\em{ Verify the identity~\eqref{fg} in a perturbation expansion to first and second order.
To this end, compute both sides of this equation using the perturbation expansion of~$\tilde{R}_\lambda$
and carrying out the contour integrals.
{\em{Hint:}} Similar formulas can be found in the appendices of~\cite{grotz} and~\cite{norm}.
}} \end{Exercise}

\chapter{Methods of Microlocal Analysis} \label{sechadamard}
\sindex{microlocal analysis}%
\section{The Hadamard Expansion in Minkowski Space}
\sindex{Hadamard expansion!in Minkowski space}%
In Chapter~\ref{secFSO}, the unregularized kernel of the fermionic projector~$P(x,y)=P_-(x,y)$
was constructed abstractly. In Chapter~\ref{secperturb}, we saw how this kernel can
be expanded in a perturbation series in powers of the external potential.
In order to gain more explicit information on the form of the unregularized kernel,
it is very useful to analyze its singularity structure on the light cone.
It turns out that~$P(x,y)$ has singularities on the light cone, which can be described
by the so-called {\em{Hadamard expansion}} of the form
\beq \label{hadamard1}
P(x,y) = \lim_{\varepsilon \searrow 0} \;\cI \Pdd_x \left( \frac{U(x,y)}{\Gamma_\varepsilon(x,y)}
+ V(x,y)\: \log \Gamma_\varepsilon(x,y) + W(x,y) \right) ,
\eeq
where
\beq \label{Gammadef}
\Gamma_\varepsilon(x,y) := (y-x)^j\, (y-x)_j - \cI \varepsilon \,(y-x)^0 \:,
\eeq
and~$U$, $V$ and~$W$ are smooth functions on~$\scrM \times \scrM$
taking values in the~$4 \times 4$-matrices acting on the spinors
(we always denote spacetime indices by Latin letters running from~$0, \ldots, 3$).
This local expansion is based on the method of {\em{integration along characteristics}},
\sindex{integration along characteristics}%
which will be explained below (see after~\eqref{intchar} or also~\cite{hadamardoriginal, friedlander1}
or~\cite{baer+ginoux}).
In Minkowski space, the {\em{light-cone expansion}}~\cite{firstorder, light} 
(see also~\cite[Section~2.2]{cfs}) gives an efficient
procedure for computing an infinite number of Hadamard coefficients in one step.
The Hadamard form~\eqref{hadamard1} carries over to curved spacetime. Moreover,
there is an interesting connection to the so-called wave front set in microlocal analysis.
These generalizations will be briefly outlined in Section~\ref{secwavefront}.
In all the other sections of this chapter, we restrict attention to Minkowski space.

It turns out that, for an external potential in Minkowski space, the
kernel of the fermionic projector is indeed of Hadamard form.
\begin{Thm} \label{thmHadamard}
Assume that the external potential~$\B$ is smooth, and that its time derivatives decay
at infinity in the sense that~\eqref{Bdecay} holds and in addition that
\beq \int_{-\infty}^\infty |\partial_t^p \B(t)|_{C^0}\, \dd t < \infty \qquad \text{for all~$p \in \N$} \eeq
(with the $C^0$-norm as defined in~\eqref{defCk}).
Moreover, assume that the potential satisfies the bound
\beq
\int_{-\infty}^\infty |\B(t)|_{C^0}\, \dd t < \sqrt{2}-1 \:. \label{Bssmall}
\eeq
Then the fermionic projector~$P(x,y)$ is of Hadamard form.
\end{Thm} \noindent
The proof of this theorem will be given in Section~\ref{sechadamardproof} below.

We conclude this section by explaining how the expansion~\eqref{hadamard1} comes about
and how the involved functions~$U$, $V$ and~$W$, at least in principle, can be computed iteratively
using the method of {\em{integration along characteristics}}.
We begin by computing the unregularized kernel in the Minkowski vacuum.
To this end, one rewrites the factor~$(\slashed{k}+m)$ in~\eqref{Pxyvac}
in terms of a differential operator in position space,
\beq \label{Pdiff}
P(x,y) = (\cI \Pdd_x + m) \,T_{m^2}(x,y) \:,
\eeq
where~$T_{m^2}$ is the scalar bi-distribution
\beq %\label{Tm2def}
T_{m^2}(x,y) := \int \frac{\dd^4k}{(2 \pi)^4}\: \delta(k^2-m^2)\: \Theta(-k^0)\: \E^{-\cI k(x-y)} \:. \eeq
\nindex{ahg@$T_{m^2}(x,y)$ -- Fourier transform of lower mass shell}%
We remark that the distribution~$T_{m^2}$ solves the Klein-Gordon equation~$(-\Box-m^2)\, T_{m^2}=0$;
in quantum field theory, it is sometimes denoted by~$\Delta_-$ (see, for example, \cite[Section~5.2]{weinberg}).
Away from the light cone (i.e.\ for~$\xi^2 \neq 0$),
the distribution~$T_{m^2}(x,y)$ is a smooth function given by
\begin{align} \label{Taway}
T_{m^2}(x,y) = \left\{ \begin{array}{cl} 
\displaystyle \frac{m}{16 \pi^2} \:\frac{Y_1\big(m\sqrt{\xi^2} \,\big)}{\sqrt{\xi^2}}
+\frac{\cI m}{16 \pi^2}\: \frac{J_1 \big(m\sqrt{\xi^2} \,\big)}{\sqrt{\xi^2}}\: \epsilon(\xi^0)
& \text{if~$\xi$ is timelike} \\[1em]
\displaystyle \frac{m}{8 \pi^3} \frac{K_1 \big(m\sqrt{-\xi^2} \,\big)}{\sqrt{-\xi^2}} & \text{if~$\xi$ is spacelike}\:,
\end{array} \right. \hspace*{-0.3em}
\end{align}
where we set
\beq %\label{xidef}
\xi := y-x \:, \eeq
and~$J_1$, $Y_1$ and~$K_1$ are Bessel functions.
Expanding the Bessel functions in~\eqref{Taway} in a power series, one obtains
(see~\cite[(10.2.2), (10.8.1) and~(10.25.2), (10.31.1)]{DLMF})
\begin{align}
T_{m^2}(x,y) &= -\frac{1}{8 \pi^3} \:\bigg( \frac{\PP}{\xi^2} + \cI \pi \delta \big( \xi^2 \big) \: \epsilon \big( \xi^0 \big)
\bigg) \nonumber \\
&\quad +\frac{m^2}{32 \pi^3}\sum_{j=0}^\infty \frac{(-1)^j}{j! \: (j+1)!} \: \frac{\big( m^2 \xi^2 \big)^j}{4^j} \:
 \Big( \log \big|m^2 \xi^2 \big| + c_j
+ \cI\pi \:\Theta \big( \xi^2 \big) \:\epsilon\big( \xi^0 \big) \Big)
\end{align}
\nindex{ahg@$T_{m^2}(x,y)$ -- Fourier transform of lower mass shell}%
with real coefficients~$c_j$ (here~$\Theta$ and~$\epsilon$ are again the Heaviside and the sign function,
respectively). In particular, one sees that~$T_{m^2}$ is a distribution that is singular on the light cone.
These singularities can be written in a shorter form using residues as
\beq \label{Tfinal}
\begin{split}
T_{m^2}(x,y) = &\lim_{\varepsilon \searrow 0}
\bigg( -\frac{1}{8 \pi^3} \:\frac{1}{\Gamma_\varepsilon(x,y)} \\
&+\frac{m^2}{32 \pi^3}\sum_{j=0}^\infty \frac{(-1)^j}{j! \: (j+1)!} \: \frac{\big( m^2 \Gamma_\varepsilon(x,y) \big)^j}{4^j}  \Big( \log \big( m^2 \,\Gamma_\varepsilon(x,y) \big) + c_j  \Big) \bigg)
\end{split}
\eeq
(where~$\Gamma_\varepsilon$ is again defined by~\eqref{Gammadef}; for the proof, one
uses the distributional relation~$\lim_{\varepsilon \searrow 0} (r^2+(\varepsilon+it)^2)^{-1}
= -\PP/\xi^2 -\cI \pi\, \delta(\xi^2)\,\epsilon(\xi^0)$ and similarly the behavior of the logarithm in the complex plane).
Noting that the series converge, one obtains a function of the desired form as in the brackets
in~\eqref{hadamard1}. This shows that the term~$\cI \Pdd_x\,T_{m^2}(x,y)$ in~\eqref{Pdiff}
is of Hadamard form. For the term~$m \,T_{m^2}(x,y)$, this can be shown by pulling out one derivative
and working with matrix-valued kernels. Indeed,
\beq m \,T_{m^2}(x,y) = -\frac{1}{m}\: \Box_x \Big( T_{m^2}(x,y) - T_0(x,y) \Big)
= \Pdd_x \Big\{ -\frac{1}{m}\: \Pdd_x \big( T_{m^2}(x,y) - T_0(x,y) \big) \Big\} \:, \eeq
and computing the curly brackets by differentiating~\eqref{Tfinal} one obtains again
an expression of the Hadamard form~\eqref{hadamard1}.

The summands in~\eqref{Tfinal} can be understood by verifying that~$T_{m^2}$ satisfies the
Klein-Gordon equation. Indeed, using the abbreviation~$\xi := y-x$, we obtain
\begin{align}
\frac{\partial}{\partial x^j} \Big( \frac{1}{\Gamma_\varepsilon(x,y)} \Big) &= 
-\frac{\partial_j \Gamma_\varepsilon(x,y)}{\Gamma_\varepsilon(x,y)^2}
= \frac{1}{\Gamma_\varepsilon(x,y)^2} \Big( 2 \xi_j - \cI \varepsilon \delta_{j,0} \Big) \\
\Box_x\Big( \frac{1}{\Gamma_\varepsilon(x,y)} \Big)
&= \frac{2}{\Gamma_\varepsilon(x,y)^3} \Big( 2 \xi_j - \cI \varepsilon \delta_{j,0} \Big)\Big( 2 \xi_j - \cI \varepsilon \delta^j_0 \Big) - \frac{8}{\Gamma_\varepsilon(x,y)^2} \notag \\
&= \frac{2}{\Gamma_\varepsilon(x,y)^3} \Big(4 \xi^2 - 4 \cI \varepsilon \: \xi_0 -\varepsilon^2 \Big)
\Big) - \frac{8}{\Gamma_\varepsilon(x,y)^2} 
= -\frac{2 \varepsilon^2}{\Gamma_\varepsilon(x,y)^3} \:,
\end{align}
and this tends to zero as~$\varepsilon \searrow 0$. Thus the leading term in~\eqref{Tfinal}
satisfies the scalar wave equation. In the Klein-Gordon equation, however, the term involving the mass remains,
\beq \label{merror}
-m^2\: \frac{1}{\Gamma_\varepsilon(x,y)} \:.
\eeq
This term is compensated by the next term in the expansion~\eqref{Tfinal}, because
\begin{align}
\frac{\partial}{\partial x^j} \log \big( \Gamma_\varepsilon(x,y) \big) &= 
\frac{\partial_j \Gamma_\varepsilon(x,y)}{\Gamma_\varepsilon(x,y)}
= -\frac{1}{\Gamma_\varepsilon(x,y)} \Big( 2 \xi_j - \cI \varepsilon \delta_{j,0} \Big) \\
\Box_x \log \big( \Gamma_\varepsilon(x,y) \big)
&= -\frac{1}{\Gamma_\varepsilon(x,y)^2} \Big( 2 \xi_j - \cI \varepsilon \delta_{j,0} \Big)\Big( 2 \xi_j - \cI \varepsilon \delta^j_0 \Big) + \frac{8}{\Gamma_\varepsilon(x,y)} \notag \\
&= -\frac{1}{\Gamma_\varepsilon(x,y)^2} \Big(4 \xi^2 - 4 \cI \varepsilon \: \xi_0 -\varepsilon^2 \Big)
\Big) + \frac{8}{\Gamma_\varepsilon(x,y)} \notag \\
&= \frac{4}{\Gamma_\varepsilon(x,y)} +\frac{\varepsilon^2}{\Gamma_\varepsilon(x,y)^2} \:.
\end{align}
Now the first summand in the last line cancels the term~\eqref{merror} in the Klein-Gordon equation.
Proceeding order by order in powers of~$\Gamma_\varepsilon(x,y)$, one can verify all the coefficients
in~\eqref{Tfinal}.

This method of applying the wave operator term by term is also useful for computing
the functions~$U$, $V$ and~$W$ in~\eqref{hadamard1} in the case that an external potential
is present. In fact, these functions can be expressed in terms of line integrals along the light cone.
This {\em{method of integration along characteristics}} goes back go Hadamard~\cite{hadamardoriginal}
and is described in the classic textbook~\cite{friedlander1} in curved spacetime.
In order to explain the method in the simplest possible context, let us assume that we consider the
wave equation with an external scalar potential~$a(x)$; that is,
\beq \big( -\Box_x - a(x) \big) \tilde{T}(x,y) = 0 \eeq
(the Dirac equation will be treated more systematically in Section~\ref{seclce}).
In modification of the series in~\eqref{Tfinal}, we make the ansatz
\beq \label{intchar}
\tilde{T}(x,y) =  \lim_{\varepsilon \searrow 0}
\bigg( \frac{1}{\Gamma_\varepsilon(x,y)} + \sum_{n=1}^\infty
f_n(x,y)\: \Gamma_\varepsilon(x,y)^n\: \log \big( \Gamma_\varepsilon(x,y) \big)
\bigg)
\eeq
Compared to~\eqref{merror}, now the error term of the first summand involves the potential~$a(x)$,
\beq \label{aerror}
-\frac{a(x)}{\Gamma_\varepsilon(x,y)} \:.
\eeq
The hope is to compensate this term by a suitable choice of~$f_1(x,y)$. Indeed,
\begin{align}
\frac{\partial}{\partial x^j} &\Big( f_1(x,y)\: \log \Gamma_\varepsilon(x,y) \Big) \\
&=  f_1(x,y) \: \frac{\partial_j \Gamma_\varepsilon(x,y)}{\Gamma_\varepsilon(x,y)}
+ \partial_j f_1(x,y) \: \log \Gamma_\varepsilon(x,y) \\
\Box_x & \Big( f_1(x,y)\: \log \Gamma_\varepsilon(x,y) \Big) \notag \\
&=  f_1(x,y) \: \frac{4}{\Gamma_\varepsilon(x,y)} - 2\: \partial_j f_1(x,y) \:
\frac{2 \xi_j}{\Gamma_\varepsilon(x,y)} + \cdots \:,
\end{align}
where~$\cdots$ stands for all terms that either have a lower order singularity on the light cone
or tend to zero as~$\varepsilon \searrow 0$. In order for this contribution to compensate~\eqref{aerror},
the function~$f_1$ must satisfy the equation
\beq 4\: f_1(x,y) - 4\:\xi^j\, \partial_j f_1(x,y) = a(x) \:. \eeq
Such a differential equation of first order can be solved with the method of characteristics
(see, for example, \cite[Section~I.3.2]{evans}). More specifically, the solution is an integral
along the straight line~$\xi \R$. In order to describe the singular behavior on the light cone,
it suffices to consider the case that~$\xi$ is tangential to the light cone.
Similarly, also to higher order in the expansion parameter~$n$, we obtain transport equations
along the light cone, which can be solved iteratively order by order. \Evtl{Gibt es ein physikalisches Bild dazu?}

\section{The Light-Cone Expansion} \label{seclce}
\sindex{light-cone expansion}%
We first give the basic definition of the light-cone expansion and explain it afterward.
\begin{Def} \label{deflce}
A distribution~$A(x,y)$ on~$\scrM \times \scrM$ is
of the order~$\O((y-x)^{2p})$ for~$p \in \Z$ if the product
\beq (y-x)^{-2p} \: A(x,y) \eeq
\nindex{ahh@$\O((y-x)^{2p})$ -- order on the light cone}%
is a regular distribution (that is, a locally integrable function).
An expansion of the form
\beq
A(x,y) = \sum_{j=g}^{\infty} A^{[j]}(x,y) \label{l:6a}
\eeq
with~$g \in \Z$ is called {\bf{light-cone expansion}}
\sindex{light-cone expansion}%
if the~$A^{[j]}(x,y)$ are distributions of the order
$\O((y-x)^{2j})$ and if~$A$ is approximated by the partial sums
in the sense that for all~$p \geq g$,
\beq \label{l:6b}
A(x,y) - \sum_{j=g}^p A^{[j]}(x,y) \qquad {\text{is of the order~$\O\big( (y-x)^{2p+2} \big)$}}\:.
\eeq
\end{Def} \noindent
The parameter~$g$ gives the leading order of the singularity of~$A(x,y)$ on the light cone.
We point out that we do not demand that the infinite series in~\eqref{l:6a} converges. Thus, similar
to a formal Taylor series, the series in~\eqref{l:6a} is defined only via the approximation by the
partial sums~\eqref{l:6b}. The notion of the light-cone expansion is illustrated in Exercise~\ref{ex2.4-21}.

As a concrete example, due to the factors~$\Gamma_\varepsilon(x,y)$, the series~\eqref{Tfinal} is
a light-cone expansion.
The term with the leading singularity becomes integrable after multiplying by~$(y-x)^2$,
showing that~$g=-1$.

Our task is to perform the light-cone expansion of the unregularized kernel of the fermionic
projector. Schematically, this construction consists of several steps:
\bitem
\item[(1)] Perform the light-cone expansion of the causal Green's operators~$\tilde{s}^\vee_m$
and~$\tilde{s}^\wedge_m$. Here, one proceeds inductively for each summand of the
perturbation series~\eqref{series-scaustilde}.
\item[(2)] Using the relation~\eqref{def-ktil}, one obtains a corresponding light-cone expansion
for the causal fundamental solution~$\tilde{k}_m$.
\item[(3)] Finally, the so-called residual argument relates the sought-after light-cone expansion of~$\tilde{P}(x,y)$
to that of~$\tilde{k}_m$.
\eitem
This procedure is described in detail in~\cite[Chapter~2]{cfs}.
In order to avoid an unnecessary overlap, we here focus on the light-cone expansion
of the causal Green's operators
and only introduce the concepts needed for the basics on the continuum limit in Chapter~\ref{seccl}.
Before doing so, we illustrate the light-cone expansion by a simple example.
\begin{Example} \label{exlight} {\em{ 
\sindex{light-cone expansion!in presence of electromagnetic potential}%
Consider the massless Dirac equation in the presence of an external electromagnetic potential~$A$,
\beq \big( \cI \Pdd + \slashed{A} \big) \tilde{P}(x,y) \:. \eeq
For simplicity, we assume that~$A$ is smooth and compactly supported in spacetime.
Then, to first order in perturbation theory, the light-cone expansion of the unregularized kernel~$\tilde{P}(x,y)$
takes the form
\begin{align}
\tilde{P}(x,y) =\:&
\frac{\cI}{2}\: \exp \bigg( -\cI \int_0^1 A_j \big|_{\alpha y + (1-\alpha) x}\: \xi^j\: \dd\alpha \bigg)\:P(x,y) \label{s:Pgag} \\
&-\frac{1}{2} \:\slashed{\xi}\, \xi_i  \int_0^1 (\alpha-\alpha^2)\: j^i \big|_{\alpha y + (1-\alpha) x}\:
\dd\alpha\;T^{(0)} \label{s:xij} \\
&+\frac{1}{4}\:\slashed{\xi} \int_0^1 F^{ij} \big|_{\alpha y + (1-\alpha) x}\, \gamma_i \gamma_j \:\dd\alpha\; T^{(0)} \label{s:FT1} \\
&- \xi_i \int_0^1 (1-\alpha)\, F^{ij} \big|_{\alpha y + (1-\alpha) x}\, \gamma_j \:\dd\alpha\; T^{(0)} \label{s:FT2} \\
&-\xi_i \int_0^1 (1-\alpha)(\alpha-\alpha^2)\, \Pdd j^i \big|_{\alpha y + (1-\alpha) x}\:\dd\alpha\; T^{(1)} \label{s:dj }\\
&-\int_0^1 (1-\alpha)^2\, j^i \big|_{\alpha y + (1-\alpha) x}\,\gamma_i\: \dd\alpha\;T^{(1)} \label{s:jLi} \\
&+ \slashed{\xi} \, (\deg < 1) + (\deg < 0) + \O(A^2) \:, \nonumber
\end{align}
where~$F^{jk} = \partial^j A^k - \partial^k A^j$ is the field tensor
and~$j^k = \partial^k_{\;j} A^j - \Box A^k$ is the corresponding Maxwell current.
Moreover, the factors~$T^{(0)}$ and~$T^{(1)}$ are the leading summands in~\eqref{Tfinal};
more precisely,
\beq \label{Tidef}
\begin{split}
T^{(0)}(x,y) &= -\frac{1}{8 \pi^3} \: \lim_{\varepsilon \searrow 0} \frac{1}{\Gamma_\varepsilon(x,y)} \\
T^{(1)}(x,y) &= \frac{1}{32 \pi^3} \: \lim_{\varepsilon \searrow 0} \log \Gamma_\varepsilon(x,y) \:.
\end{split}
\eeq
\nindex{ahi@$T^{(n)}(x,y)$ -- singular factors in light-cone expansion}%
Each summand has the general structure of being the product of a smooth function
and a distribution that is singular on the light cone. The smooth factor is an integral along
the straight line segment joining the points~$x$ and~$y$. The integrand involves the
electromagnetic potential and its partial derivatives.
We remark for clarity that the term~\eqref{s:Pgag} involves a gauge phase as needed for
gauge invariance (as already mentioned in~\eqref{tilPgauge} in Section~\ref{secwhycap}).
All the other integrands are gauge invariant, as is obvious from the fact that they
are expressed in terms of the electromagnetic field tensor and the Maxwell current.

To higher order on the light cone or to higher order in the mass or the external potentials,
the formulas of the light-cone expansions have a similar structure.
More detailed formulas can be found in the original papers~\cite{firstorder, light},
in~\cite[Appendix~B]{pfp} and~\cite[Appendix~B]{cfs}.
}} \QEDrem
\end{Example}

We now explain how to perform the light-cone expansion of the causal Green's operators.
In order to get a first idea for how to proceed, we begin by considering the
free advanced Green's operator~$s^\vee_m$ of the Dirac equation of mass~$m$ in position space:
Similar to~\eqref{Pdiff}, it is again convenient to pull the Dirac matrices out of~$s^\vee_m$ by setting
\sindex{Green's operator!retarded}%
\nindex{agy@$s^\vee_m, s^\wedge_m$ -- causal Green's operators in the vacuum}%
\beq
s^\vee_m(x,y) = (\cI \Pdd_x + m) \: S^\vee_{m^2}(x,y) \:, \label{l:10}
\eeq
where~$S^\vee_{m^2}$ is the advanced Green's operator of the 
Klein-Gordon operator,
\beq \label{l:11}
S^\vee_{m^2}(x,y) = \lim_{\nu \searrow 0} \int \frac{\dd^4p}{(2 \pi)^4}
\:\frac{1}{p^2-m^2-\cI \nu p^0} \:\E^{-\cI p(x-y)} \:.
\eeq
\nindex{ahl@$S_{m^2}^\vee, S_{m^2}^\wedge$ -- causal Green's operators of Klein-Gordon equation}%
Computing this Fourier integral and expanding the
resulting Bessel function in a power series gives (for details, see Exercise~\ref{ex2.4-2})
\begin{align}
S^\vee_{m^2}(x,y) &= -\frac{1}{2 \pi} \:\delta \big( \xi^2 \big) \:
\Theta \big( \xi^0 \big) \notag \\
&\quad\:+ \frac{m^2}{4 \pi} \:\frac{J_1 \Big( \sqrt{m^2 
\:\xi^2} \Big)}{\sqrt{m^2 \:\xi^2}} \:\Theta\big( \xi^2 \big) \:\Theta \big(\xi^0 \big) \label{l:121} \\
&= -\frac{1}{2 \pi} \:\delta \big( \xi^2 \big) \: \Theta\big(\xi^0 \big) \notag \\
&\quad\:+ \frac{m^2}{8 \pi}
\sum_{j=0}^\infty \frac{(-1)^j}{j! \:(j+1)!} \: \frac{\big( m^2 \xi^2 \big)^j}{4^j} \:
\Theta \big( \xi^2 \big) \:\Theta \big(\xi^0 \big) \:.
\label{l:12}
\end{align}
This computation shows that~$S^\vee_{m^2}(x,y)$ has a~$\delta(\xi^2)$-like
singularity on the light cone. Furthermore, one sees that~$S^\vee_{m^2}$
is a power series in~$m^2$. The important point for what follows is
that the higher order contributions in~$m^2$ contain more 
factors~$\xi^2$ and are thus of higher order on the light cone. 
More precisely,
\beq \label{l:24b}
\left( \frac{\dd}{\dd a} \right)^n S^\vee_{a }(x,y) \Big|_{a=0} \qquad
\text{is of the order~$\O\big(\xi^{2n-2} \big)$}
\eeq
(here and in what follows, we often use the abbreviation~$a=m^2$).
According to~\eqref{l:10}, the Dirac Green's operator is obtained by 
taking the first partial derivatives of~\eqref{l:12}. Therefore,
$s^\vee_m(x,y)$ has a singularity on the light cone that is even
$\sim \delta^\prime(\xi^2)$.
The higher order contributions in~$m$ are again of increasing order on 
the light cone. This means that we can view the Taylor expansion of~\eqref{l:10} in~$m$,
\beq s^\vee_m(x,y) = \sum_{n=0}^\infty (\cI \Pdd + m) \;\frac{m^{2n}}{n!}
\left( \frac{\dd}{\dd a} \right)^n S^\vee_{a}(x,y)  \Big|_{a=0} \: , \eeq
as a light-cone expansion of the free Green's operator. Our idea is to
generalize this formula to the case with interaction. More precisely, we want
to express the perturbed Green's operator in the form
\beq
        \tilde{s}^\vee(x,y) = \sum_{n=0}^\infty F_n(x,y) \: \left( 
        \frac{\dd}{\dd a} \right)^n S^\vee_{a}(x,y)  \Big|_{a=0} 
        \label{l:14a}
\eeq
with factors~$F_n$ that depend on the external potential.
We will see that this method is very convenient; especially, we can in 
this way avoid working with the rather complicated explicit formula~\eqref{l:12}.
Apart from giving a motivation for the desired form~\eqref{l:14a} of the
formulas of the light-cone expansion, the mass expansion~\eqref{l:12}
\sindex{mass expansion}%
leads to the conjecture
that even the higher order contributions in the mass to the {\em{perturbed}}
Green's operators might be of higher order on the light cone.
If this conjecture were true, it would be a good idea to expand the
perturbation expansion of~$\tilde{s}$ with respect to the parameter~$m$.
Therefore, our strategy is to first expand~\eqref{series-scaustilde} with respect to
the mass and to try to express the contributions to the resulting expansion
in a form similar to~\eqref{l:14a}.

The expansion of~\eqref{series-scaustilde} with respect to~$m$ gives a double 
sum over the orders in the mass parameter and in the external 
potential. It is convenient to combine these two expansions in a single 
perturbation series. To this end, we rewrite the Dirac operator as
\beq %\label{l:18a}
\cI \Pdd + \B - m = \cI \Pdd + B \qquad {\mbox{with}} \qquad B:=\B-m \:. \eeq
\nindex{ahm@$B=\B-m$ -- external potential combined with the mass}%
For the light-cone expansion of the Green's operators, we will always
view~$B$ as the perturbation of
the Dirac operator. This has the advantage that the unperturbed objects
are massless. Expanding in powers of~$B$
gives the mass expansion and the perturbation expansion in one step.
\sindex{mass expansion}%
In order to further simplify the notation, for the massless objects we usually
omit the index~$m$. Thus we write the Green's operator of the
massless Dirac equation in the Minkowski vacuum as
\beq %\label{l:11a}
        s^\vee(x,y) = \cI \Pdd_x \:S^\vee_{m^2}(x,y) \big|_{m=0}\:,\qquad
        s^\wedge(x,y) = \cI \Pdd_x \:S^\wedge_{m^2}(x,y) \big|_{m=0}\:. \eeq
\sindex{Green's operator!advanced}%
\sindex{Green's operator!retarded}%
\nindex{agy@$s^\vee_m, s^\wedge_m$ -- causal Green's operators in the vacuum}%
Then the interacting Green's operators are given by the perturbation series
\beq \tilde{s}^\vee = \sum_{k=0}^\infty (-s^\vee B)^k 
        s^\vee \:,\qquad \tilde{s}^\wedge = \sum_{k=0}^\infty
        (-s^\wedge B)^k s^\wedge \: .
        \label{l:11b}
\eeq
\nindex{agf@$\tilde{s}_m^\vee, \tilde{s}_m^\wedge$ -- causal Green's operators}%
The constructions of the following subsections are exactly the same for
the advanced and retarded Green's operators. In order to treat both 
cases at once, in the remainder of this section, we will omit all 
superscripts `$^\vee$' and `$^\wedge$'. The formulas for the advanced and 
retarded Green's operators are obtained by either adding `$^\vee$' or
`$^\wedge$' to all factors~$s$ and~$S$.

We now explain how each contribution
to the perturbation expansion~\eqref{l:11b} can be written similar to 
the right-hand side of~\eqref{l:14a} as a sum of terms of increasing order 
on the light cone. For the mass expansion of~$S_{m^2}$, we again set~$a=m^2$ and
use the notation
\beq \label{l:23b}
S^{(l)} = \left( \frac{\dd}{\dd a} \right)^l S_a \big|_{a=0} \:. \eeq
\nindex{ahp@$S^{(l)}$ -- mass expansion of $S_a$}%
In preparation, we derive some computation rules for the~$S^{(l)}$:
$S_a$ satisfies the defining equation of a Klein-Gordon Green's operator
\beq (-\Box_x - a) \:S_a(x,y) = \delta^4(x-y) \: . \eeq
Differentiating with respect to~$a$ and setting~$a=0$ gives
\beq
-\Box_x S^{(l)}(x,y) = \delta_{l,0} \:\delta^4(x-y)
+ l \:S^{(l-1)}(x,y) \:,\qquad l \geq 0 . \label{l:5}
\eeq
(For~$l=0$, this formula does not seem to make sense because~$S^{(-1)}$ 
is undefined. The expression is meaningful, however, if one keeps 
in mind that, in this case, the factor~$l$ is zero, and thus the whole 
second summand vanishes. We will also use this convention in the following 
calculations.) Next, we differentiate the formulas for~$S_a$ in momentum space,
\beq %\label{l:21x}
S_a^\vee(p) = \frac{1}{p^2-a-\cI \nu p^0} \:,\qquad
        S_a^\wedge(p)= \frac{1}{p^2-a+\cI \nu p^0} \eeq
with respect to both~$p$ and~$a$. Comparing the results gives the 
relation
\beq \frac{\partial}{\partial p^k} S_{a}(p) = -2p_k \:\frac{\dd}{\dd a} S_a(p) \: , \eeq
or, after expanding in the parameter~$a$,
\beq
\frac{\partial}{\partial p^k} S^{(l)}(p) = -2 p_k \:S^{(l+1)}(p) \:,\qquad l \geq 0 .
        \label{l:21a}
\eeq
This formula also determines the derivatives of~$S^{(l)}$ in position space; namely
\begin{align}
\frac{\partial}{\partial x^k} & S^{(l)}(x,y) =
\int \frac{\dd^4p}{(2 \pi)^4} \:S^{(l)}(p) \:(-\cI p_k) \:\E^{-\cI p(x-y)} \notag \\
&\!\!\!\!\!\stackrel{\eqref{l:21a}}{=} \frac{\cI}{2} \int \frac{\dd^4p}{(2 \pi)^4} \: 
\frac{\partial}{\partial p^k} S^{(l-1)}(p) \; \E^{-\cI p(x-y)} \notag \\
&= -\frac{\cI}{2} \int \frac{\dd^4p}{(2 \pi)^4} \: 
         S^{(l-1)}(p) \;\frac{\partial}{\partial p^k} \E^{-\cI p(x-y)} \notag \\
&=\frac{1}{2} \: (y-x)_k \: S^{(l-1)}(x,y) \:,\qquad l \geq 1 .
        \label{l:7}
\end{align}
We iterate this relation to calculate the Laplacian,
\begin{align}
        -\Box_x S^{(l)}(x,y) &= -\frac{1}{2} \:\frac{\partial}{\partial 
        x^k} \left( (y-x)^k \:S^{(l-1)}(x,y) \right) \notag \\
        &= 2 \:S^{(l-1)}(x,y) + \frac{1}{4} \:(y-x)^2 \:S^{(l-2)}(x,y)
        \:,\qquad l \geq 2
\end{align}
(in the last step we used the product rule and applied~\eqref{l:7} for~$l$ replaced by~$l-1$).
After comparing with~\eqref{l:5}, we conclude that
\beq %\label{l:22a}
(y-x)^2 \:S^{(l)}(x,y) = -4l\: S^{(l+1)}(x,y)\:,\qquad l \geq 0 \:. \eeq
Finally, $S^{(l)}(x,y)$ is only a function of~$(y-x)$, which 
implies that
\beq %\label{l:20a}
\frac{\partial}{\partial x^k} S^{(l)}(x,y) =
-\frac{\partial}{\partial y^k} S^{(l)}(x,y) \:,\qquad l \geq 0 \:. \eeq

The following lemma gives the light-cone expansion of an operator product that
is linear in the external potential. It can be used iteratively to perform the 
light-cone expansion of more complicated operator products;
in this case, the potential is a composite 
expression in~$B$ and its partial derivatives. With this in mind, in the next lemma we
denote the external potential by~$V$.
\sindex{light-cone expansion!of the Green's operator!to first order}%

\begin{Lemma}{\bf{(Light-cone expansion to first order)}}
\sindex{light-cone expansion!to first order}%
\label{l:lemma1} For any~$l,r \geq 0$, the operator product~$S^{(l)} \:V\: S^{(r)}$ has the light-cone expansion
\begin{align}
(&S^{(l)} \:V\: S^{(r)})(x,y)\notag \\
&= \sum_{n=0}^\infty 
        \frac{1}{n!} \int_0^1 \alpha^{l} \:(1-\alpha)^{r} \:
        (\alpha - \alpha^2)^n \: (\Box^n V)_{|\alpha y + (1-\alpha) x} \:\dd\alpha \;
        S^{(n+l+r+1)}(x,y) \:.
        \label{l:4}
\end{align}
\end{Lemma}
Before coming to the proof, we briefly explain this lemma. We first recall that, according to~\eqref{l:24b}, the higher 
$a$-derivatives of~$S_a(x,y)$ are of higher order on the light cone. 
Thus the summands in~\eqref{l:4} are of increasing order on the light 
cone, and the infinite sum is mathematically well-defined in the sense of Definition~\ref{deflce}
via the approximation by the partial sums~\eqref{l:6b}.

The second point that requires an explanation is related to the arbitrariness in choosing the potential~$V$
in the case~$l=0$ (and analogously in the case~$r=0$).
In this case, the distribution~$S^{(l)}=S_0$ is supported on the light cone (see~\eqref{l:121}).
Therefore, the function~$V$ enters the operator product on the left side of~\eqref{l:4} only
evaluated on the light cone~$L_x = \{z \:|\: (x-z)^2=0\}$. This means that we may modify the function~$V$ arbitrarily
outside this light cone. When doing so, the argument~$\Box^n V$ in the integrand on the right-hand side
of~\eqref{l:4} does in general change. Therefore, the individual summands in~\eqref{l:4} do in general change.
But, clearly, in order for the identity~\eqref{l:4} to remain valid, the whole series must remain unchanged.
This is indeed the case due to cancellations in the series (this is illustrated in Exercise~\ref{exlightcancel}).
With this in mind, one can sometimes simplify the application of the previous lemma in the case~$l=0$
by choosing~$V$ outside the light cone~$L_x$ in such a way that the computation of
the right-hand side simplifies.

\Proof[Proof of Lemma~\ref{l:lemma1}] The method of proof is to compute the Laplacian of both 
sides of~\eqref{l:4}. The resulting formulas will have a
similar structure, making it possible to proceed inductively.

On the left-hand side of~\eqref{l:4}, we calculate the Laplacian with the 
help of~\eqref{l:5} to
\beq \label{l:6}
-\Box_x (S^{(l)} \:V\: S^{(r)})(x,y) = \delta_{l,0} 
        \:V(x) \: S^{(r)}(x,y) +
        l \: (S^{(l-1)} \:V\: S^{(r)})(x,y) \:.
\eeq

The Laplacian of the integral on the right-hand side of~\eqref{l:4} can be 
computed with the help of~\eqref{l:7} and~\eqref{l:5},
\begin{align}
-&\Box_x \int_0^1 \alpha^{l} \:(1-\alpha)^{r} \:(\alpha-\alpha^2)^n
\: (\Box^n V)_{|\alpha y + (1-\alpha) x} \:\dd\alpha \;
S^{(n+l+r+1)}(x,y) \label{l:8} \\
         & = -\int_0^1 \alpha^{l} \:(1-\alpha)^{r+2} \:(\alpha-\alpha^2)^n \:
         (\Box^{n+1} V)_{|\alpha y + (1-\alpha) x} \:\dd\alpha \;
S^{(n+l+r+1)}(x,y) \notag \\
&\quad-\int_0^1 \alpha^{l} \:(1-\alpha)^{r+1} \:(\alpha-\alpha^2)^n \:
(\partial_k \Box^{n} V)_{|\alpha y + (1-\alpha) x}
\:\dd\alpha \; (y-x)^k \:
S^{(n+l+r)}(x,y) \notag \\
&\quad +(n+l+r+1) \int_0^1 \alpha^{l} \:(1-\alpha)^{r} \:(\alpha-\alpha^2)^n \:
(\Box^{n} V)_{|\alpha y + (1-\alpha) x}
\:\dd\alpha \notag \\
&\qquad\qquad\qquad\qquad\qquad\qquad\qquad\qquad\qquad\qquad\qquad\qquad \times \;S^{(n+l+r)}(x,y) \: .
\end{align}
In the second summand, we rewrite the partial derivative as a
derivative with respect to~$\alpha$,
\beq  (y-x)^k (\partial_k \Box^{n} V)_{|\alpha y + (1-\alpha) x}
= \frac{\dd}{\dd\alpha}  (\Box^{n} V)_{|\alpha y + (1-\alpha) x} \eeq
(as is verified immediately by computing the right side with the chain rule).
This makes it possible to integrate in~$\alpha$ by parts. We thus obtain
\begin{align}
\int_0^1 &\alpha^{l} \:(1-\alpha)^{r+1} \:(\alpha-\alpha^2)^n \:
(\partial_k \Box^{n} V)_{|\alpha y + (1-\alpha) x}
\:\dd\alpha \; (y-x)^k \notag \\
&= \int_0^1 \alpha^{l} \:(1-\alpha)^{r+1} \:(\alpha-\alpha^2)^n \:
\frac{\dd}{\dd\alpha} \Big( (\Box^{n} V) \big|_{\alpha y + (1-\alpha) x} \Big)
\dd\alpha \notag \\
&= -\delta_{n,0} \:\delta_{l,0}\:V(x) - (n+l)
         \int_0^1 \alpha^{l} \:(1-\alpha)^{r+2} \:(\alpha-\alpha^2)^{n-1} \:
(\Box^{n} V)_{|\alpha y + (1-\alpha) x} \:\dd\alpha \notag \\
&\qquad +(n+r+1) \int_0^1 \alpha^{l} \:(1-\alpha)^{r} \:(\alpha-\alpha^2)^n \:
(\Box^{n} V)_{|\alpha y + (1-\alpha) x} \:\dd\alpha \notag \\
&= -\delta_{n,0} \:\delta_{l,0}\:V(x) \notag \\
&\qquad -n \int_0^1 \alpha^{l} \:(1-\alpha)^{r+2} \:(\alpha-\alpha^2)^{n-1} \:
(\Box^{n} V)_{|\alpha y + (1-\alpha) x} \:\dd\alpha \notag \\
&\qquad+(n+l+r+1) \int_0^1 \alpha^{l} \:(1-\alpha)^{r} \:(\alpha-\alpha^2)^n \:
(\Box^{n} V)_{|\alpha y + (1-\alpha) x} \:\dd\alpha \notag \\
&\qquad-l \int_0^1 \alpha^{l-1} \:(1-\alpha)^{r} \:(\alpha-\alpha^2)^{n} \:
(\Box^{n} V)_{|\alpha y + (1-\alpha) x} \:\dd\alpha \: .
\end{align}
We substitute back into the original equation to obtain
\begin{align}
&\eqref{l:8} = \delta_{n,0} \:\delta_{l,0} \: V(x) \: S^{(r)}(x,y) \notag \\
&\quad+l \int_0^1 \alpha^{l-1} \:(1-\alpha)^{r} \:(\alpha-\alpha^2)^{n} \:
(\Box^{n} V)_{|\alpha y + (1-\alpha) x} 
\:\dd\alpha \; S^{(n+l+r)}(x,y) \notag \\
&\quad-\int_0^1 \alpha^{l} \:(1-\alpha)^{r+2} \:(\alpha-\alpha^2)^{n} \:
(\Box^{n+1} V)_{|\alpha y + (1-\alpha) x} 
\:\dd\alpha \; S^{(n+l+r+1)}(x,y) \notag \\
&\quad+n\int_0^1 \alpha^{l} \:(1-\alpha)^{r+2} \:(\alpha-\alpha^2)^{n-1} \:
(\Box^{n} V)_{|\alpha y + (1-\alpha) x} 
\:\dd\alpha \; S^{(n+l+r)}(x,y) \: .
\end{align}
After dividing by~$n!$ and summation over~$n$, the last two summands 
are telescopic and cancel each other. Thus one gets
\begin{align}
&-\Box \sum_{n=0}^\infty \frac{1}{n!} \int_0^1 
\alpha^{l} \:(1-\alpha)^{r} \:(\alpha-\alpha^2)^n \: (\Box^n 
V)_{|\alpha y + (1-\alpha) x} \:\dd\alpha \; S^{(n+l+r+1)}(x,y) \notag \\
&= \delta_{l, 0} \:V(x)\:S^{(r)}(x,y) \notag \\
&\qquad +l \sum_{n=0}^\infty \frac{1}{n!} \int_0^1 
\alpha^{l-1} \:(1-\alpha)^{r} \:(\alpha-\alpha^2)^n \: (\Box^n 
V)_{|\alpha y + (1-\alpha) x} \:\dd\alpha \; S^{(n+l+r)}(x,y) \:.\label{l:9a}
\end{align}

We now compare the formulas~\eqref{l:6} and~\eqref{l:9a} for the Laplacian of 
both sides of~\eqref{l:4}. In the special case~$l=0$, these formulas
coincide, and we can use a uniqueness argument for the solutions of the 
wave equation to prove~\eqref{l:4}: We assume that
we consider the advanced Green's operator (for the 
retarded Green's operator, the argument is analogous). For given~$y$,
we denote the difference of both sides of~\eqref{l:4} by~$F(x)$.
Since the support of~$F(x)$ is in the past light cone~$x \in
L^\wedge_y$, $F$ vanishes in a neighborhood of the 
hypersurface~${\mathcal{H}}=\{z \in \R^4 \:|\: z^0 = y^0 + 1\}$. 
Moreover, the Laplacian of~$F$ is identically zero according to 
\eqref{l:6} and~\eqref{l:9a}. We conclude that
\beq      \Box F = 0 \qquad {\mbox{and}} \qquad F_{|{\mathcal{H}}}= \partial_k 
        F_{|{\mathcal{H}}} = 0 \: . \eeq
Since the wave equation has a unique solution for given initial data 
on the Cauchy surface~${\mathcal{H}}$, $F$ vanishes identically.

The general case follows by induction in~$l$: Suppose that 
\eqref{l:4} holds for given~$\hat{l}$ (and arbitrary~$r$).
Then, according to~\eqref{l:6}, \eqref{l:9a}, and the induction hypothesis,
the Laplacian of both sides of~\eqref{l:4} coincides for~$l = \hat{l} + 1$.
The above uniqueness argument for the solutions of the wave equation again 
gives~\eqref{l:4}.
\QED

We finally remark that the method of the previous lemma generalizes to other operator products.
In particular, in~\cite[Appendix~C]{nonlocal} light-cone expansions are derived, which
involve unbounded line integrals.

\section{The Hadamard Form in Curved Spacetime and the Wave Front Set} \label{secwavefront}
\sindex{Hadamard form!in curved spacetime}%
\sindex{wave front set}%
The Hadamard expansion~\eqref{hadamard1} can also be formulated in curved spacetime.
To this end, one simply replaces the function~\eqref{Gammadef} by
\beq \Gamma_\varepsilon(x,y) :=\Gamma(x,y) - \cI \varepsilon \,\big(\mathfrak{t}(y)-\mathfrak{t}(x) \big) \:, \eeq
where~$\mathfrak{t}$ is a time function and~$\Gamma(x,y)$ is the geodesic distance squared, with the sign convention that~$\Gamma$ is positive in timelike and negative in spacelike directions.
It turns out that if a bi-distribution is of Hadamard form in one chart, it is also of Hadamard form in any other chart.
More details on the Hadamard expansion for Dirac fields can be found in~\cite{sahlmann2001microlocal, hack}
or~\cite[Appendix~A]{lqg}.

The Hadamard form can be formulated alternatively in terms of the wave front set,
as we now briefly mention.
We work in an open subset~$U \subset \R^n$. We denote the distributions in~$U$
by~${\mathcal{D}}'(U)$ (being the dual space of~$C^\infty(U,\C)$ with the topology induced
by the $C^k$-norms).
An {\em{open conic neighborhood}} of a point~$\xi \in \R^n$ is defined to be an
\sindex{neighborhood!open conic}%
open neighborhood that is invariant under the action of~$\R^+$ by multiplication.
Thus an open conic neighborhood can be written in the form
\beq \{ \lambda x \:|\:  x \in S, \lambda \in \R^+ \} \:, \eeq
where~$S$ is an open subset of~$S^{n-1} \subset \R^n$.

\begin{Def} \label{defwfs} Let~$\phi \in {\mathcal{D}}'(U)$. The {\bf{wave front set}}~$\text{\rm{WF}}(\phi)$ is the
\nindex{ahq@$\text{\rm{WF}}(\phi)$ -- wave front set of~$\phi$}%
complement in~$U \times \R^n \setminus \{0\}$ of all points~$(x,\xi) \in U \times \R^n \setminus \{0\}$
with the following property: There exists a function~$f \in C^\infty(U,\R)$ with~$f(x)=1$
and an open conic neighborhood~$V$ of~$\xi$ such that
\beq \label{nowfs}
\sup_{\zeta \in V} \big(1+|\zeta|\big)^N \, \Big| \big(\widehat{f \phi}\big)(\zeta) \Big| < \infty \qquad
\text{for all~$N \in \N$}\:.
\eeq
\end{Def} \noindent
In simple terms, the wave front set consists of all points~$x \in U$ where the distribution is singular,
together with the directions~$\xi$ into which the singularity points.
More precisely, the above definition can be understood as follows. First, in view of taking the complement,
the condition~\eqref{nowfs} ensures that the point~$(x, \xi)$ does {\em{not}} lie in the wave front set.
With the help of the cutoff function~$f$ one can disregard the behavior of~$\phi$ away from~$x$.
In other words, the condition~\ref{nowfs} only depends on the behavior of~$\phi$
in an arbitrarily small neighborhood of~$x$.
This condition states that the Fourier transform has rapid decay in a cone around~$\xi$.
Since decay properties of the Fourier transform correspond to smoothness properties in position space,
we obtain a smoothness statement for~$\phi$ at~$x$, but only along the ``wave front''
described by~$\xi$.
\Evtl{Eventuell Beispiele oder \"Ubungsaufgaben?
Beispielsweise~$\phi(x) = \delta^2(x)$ for~$U=\R^2$? Oder auch~$P(x,y)$ im Minkowski-Vakuum?}

Definition~\ref{defwfs} readily extends to a distribution~$\phi$ on a manifold~$\scrM$, in which
case the wave front set is a subset of the cotangent bundle,
\beq \text{WF}(\phi) \subset T^*\scrM \setminus 0 \eeq
(where~$0$ is the zero section).
The wave front set can also be defined for bundle-valued distributions by choosing a
local trivialization and taking the wave front sets of the component functions.
The unregularized kernel of the fermionic projector~$P$ is a bi-distribution on~$\scrM \times \scrM$.
Therefore, its wave front set takes values in the product of the cotangent bundles,
\beq \text{WF}(P) \subset \big(T^*\scrM \setminus 0 \big) \times \big( T^*\scrM \setminus 0 \big) \:. \eeq

\begin{Def} The unregularized kernel~$P \in {\mathcal{D}}'(\scrM \times \scrM)$ is said to be of {\bf{Hadamard form}}
if its wave front set has the property
\begin{align}
\text{\rm{WF}}(P) \subset \Big\{ &(x_1, \xi_1, x_2, -\xi_2) \,\Big|\,
\text{there is a null geodesic~$\gamma : I \rightarrow \scrM$ with~$a,b \in I$}, \notag \\[-0.5em]
&\text{$\;\;\gamma(a)=x_1, \gamma(b)=x_2 \quad \text{and} \quad
\xi_1 = \dot{\gamma}(a), \xi_2 = \dot{\gamma}(b)$ past-directed} \Big\} \:.
\end{align}
\end{Def} \noindent
In words, this definition means that there are singularities only on the light cone,
and that these singularities are formed only of negative frequencies.
The equivalence of this definition with the local Hadamard expansion~\eqref{hadamard1}
has been established in~\cite{radzikowski}.
Physically, the Hadamard condition can be understood as a microlocal formulation of
an energy condition, noting that ``frequencies'' can also be interpreted as ``energies.''
Good references on microlocal analysis and the wave front set
are~\cite{hormanderI} and~\cite[Chapter~4]{baer+fredenhagen}.

\section{Proof of the Hadamard Property in an External Potential} \label{sechadamardproof}
In this section, we give the proof of Theorem~\ref{thmHadamard}.
We closely follow the presentation in~\cite{hadamard}.
In preparation, we derive so-called frequency splitting estimates that
give control of the ``mixing'' of the positive and negative frequencies in the
solutions of the Dirac equation as generated by the time-dependent external potential
(Theorem~\ref{thmfreqmix}). Based on these estimates, we will complete
the proof of Theorem~\ref{thmHadamard} at the end of Section~\ref{secproofHadamard}.

\subsection{Frequency Mixing Estimates} \label{secfreqmix}
\sindex{frequency mixing estimate}%
For the following constructions, we again choose the hypersurface~$\scrN := \scrN_{t_0}$ at some given time~$t_0$.
Moreover, we always fix the mass parameter~$m>0$.
Since we are no longer considering families of solutions, for ease of notation, we omit
the index~$m$ at the Dirac wave functions, the scalar products and the corresponding norms.
We also identify the solution space~$\H_m$
with the Hilbert space~$\H_{t_0}$ of square integrable wave functions on~$\scrN$.
On~$\H_{t_0}$, we can act with the Hamiltonian~$H$ of the vacuum, and using the above identification,
the operator~$H$ becomes an operator on~$\H_m$ (which clearly depends on the choice of~$t_0)$.

We work with a so-called {\em{frequency splitting}} with respect to the vacuum
dynamics. To this end, we decompose the Hilbert space~$\H_m$ as
\beq \H_m = \H_m^+ \oplus \H_m^- \qquad \text{with} \qquad
\H^\pm = \chi^\pm(H) \H_m \:, \eeq
where~$\chi^\pm$ are the characteristic functions
\beq \label{chipmdef}
\chi^+ := \chi_{[0, \infty)} \qquad \text{and} \qquad \chi^- := \chi_{(-\infty, 0)} \:.
\eeq
For convenience, we write this decomposition in components and
use a block matrix notation for operators; that is,
\beq \psi = \begin{pmatrix} \psi^+ \\ \psi^- \end{pmatrix} \qquad \text{and} \qquad
A = \begin{pmatrix} A^+_+ & A^+_- \\ A^-_+ & A^-_- \end{pmatrix} \:, \eeq
where~$A^s_{s'} = \chi^s(H) A\, \chi^{s'}(H)$ and~$s,s' \in \{\pm\}$.

The representation in Proposition~\ref{prpSrep} 
makes it possible to let the fermionic signature operator~$\tilde{\Sig}_m$ act on the Hilbert space~$\H_m$ (for fixed~$m$).
We decompose this operator with respect to the above frequency splitting,
\beq \tilde{\Sig}_m = \SD + \Delta \tilde{\Sig}\:, \qquad \text{where} \qquad
\SD := \tilde{\Sig}^+_+ + \tilde{\Sig}^-_- \quad \text{and} \quad
\Delta \tilde{\Sig} := \tilde{\Sig}^+_- + \tilde{\Sig}^-_+ \:. \eeq
Thus the operator~$\SD$ maps positive to positive and negative to negative frequencies.
The operator~$\Delta \tilde{\Sig}$, on the other hand, mixes positive and negative frequencies.
In the next theorem, it is shown under a suitable smallness assumption on~$\B$
that the operators~$\chi^\pm(\tilde{\Sig}_m)$ coincide with the projections~$\chi^\pm(H)$, up to
smooth contributions.
The main task in the proof is to control the ``frequency mixing'' as described by the operator~$\Delta \tilde{\Sig}$.
\begin{Thm} \label{thmfreqmix}
Under the assumptions of Theorem~\ref{thmHadamard},
the operators~$\chi^\pm(\tilde{\Sig}_m)$ have the representations
\beq \label{cD2}
\chi^\pm(\tilde{\Sig}_m) = \chi^\pm(H) + \frac{1}{2 \pi \cI} 
\ointctrclockwise_{\partial B_{\frac{1}{2}}(\pm 1)}
(\tilde{\Sig}_m-\lambda)^{-1}\, \Delta \tilde{\Sig} \: (\SD-\lambda)^{-1}\:\dd\lambda \:,
\eeq
where the contour integral is an integral operator with a smooth integral kernel.
\end{Thm} \noindent
Here~$B_{\frac{1}{2}}$ denotes the open ball of radius~$1/2$.
The operator~$(\tilde{\Sig}_m-\lambda)^{-1}$ is also referred to as the {\em{resolvent}} of~$\tilde{\Sig}_m$.

This theorem will be proved in several steps. We begin with a preparatory lemma.

\begin{Lemma} Under the assumptions~\eqref{Bdecay} and~\eqref{Bssmall},
the spectrum of~$\SD$ is located in the set
\beq \label{SDspec}
\sigma(\SD) \subset \bigg[ -\frac{3}{2}, -\frac{1}{2} \bigg] \cup  \bigg[ \frac{1}{2}, \frac{3}{2} \bigg] \:.
\eeq
Moreover,
\beq
\chi^\pm(\SD) = \chi^\pm(H) \label{cD1} \:,
\eeq
and the operators~$\chi^\pm(\tilde{\Sig}_m)$ have the representations~\eqref{cD2}.
\end{Lemma}
\Proof Since the subspaces~$\H^\pm$ are invariant under the action of~$\SD$, our task is to show
that the spectrum of~$\SD|_{\H^\pm}$ is positive and negative, respectively.
This statement would certainly be true if we replaced~$\SD$ with~$\Sig_m$, because
the operator~$\Sig_m$ has the eigenvalues~$\pm 1$ with~$\H^\pm$ as the corresponding eigenspaces.
Estimating the representation in Proposition~\ref{prpSrep} with the Schwarz inequality,
we obtain
\beq \big| (\psi | \SD \phi) - (\psi | \Sig_m \phi) \big|
\leq \left(c + \frac{c^2}{2} \right) \|\psi\|\, \|\phi\| \quad \text{with} \quad
c := \int_{-\infty}^\infty |\B(\tau)|_{C^0}\, \dd\tau\:. \eeq
Using the assumption~\eqref{Bssmall}, we conclude that
\beq \big| (\psi | \SD \phi) - (\psi | \Sig_m \phi) \big| < \frac{1}{2}\: 
\|\psi\|\, \|\phi\| \qquad \text{for all~$\psi, \phi \in \H_m$}\:. \eeq
Standard estimates on the continuity of the spectrum (see, for example, \cite[\S IV.3]{kato})
yield that the spectrum of~$\SD$ differs by that of the operator~$\Sig_m$ at most by~$1/2$.
This gives~\eqref{SDspec} and~\eqref{cD1}.

In order to prove the representation~\eqref{cD2}, we take the resolvent identity
\beq (\tilde{\Sig}_m-\lambda)^{-1} = (\SD-\lambda)^{-1} - (\tilde{\Sig}_m-\lambda)^{-1}\,
\Delta \tilde{\Sig} \, (\SD-\lambda)^{-1} \:, \eeq
form the contour integral and apply~\eqref{cD1}. This gives the result.
\QED

The next lemma relates the smoothness of an integral kernel to
the boundedness of the product of the operator with powers of the vacuum Hamiltonian.
\begin{Lemma} Let~$A \in \Lin(\H_m)$ be an operator that maps smooth functions to
smooth functions and has the property that for all~$p, q \in \N$, the operator product
\beq \label{Hpq}
H^q \,A\, H^p \::\: C^\infty_0(\scrN, S\scrM) \rightarrow C^\infty(\scrN, S\scrM)
\eeq
extends to a bounded linear operator on~$\H_m$.
Then, considering~$A$ as an operator on~$\H_m$, this operator can be represented
as an integral operator with a smooth integral kernel; that is,
\beq (A \psi)(x) = \int_{\scrN} {\mathcal{A}} \big( x,(t_0, \vec{y}) \big)\, \gamma^0\,
\psi(t_0, \vec{y})\, \dd^3y \qquad \text{with} \qquad
{\mathcal{A}} \in C^\infty(\scrM \times \scrM)\:. \eeq
\end{Lemma}
\Proof Since in momentum space, the square of the Hamiltonian takes the form
\beq H \big(\vec{k} \big)^2 = \left( \gamma^0 \big( \vec{\gamma} \vec{k} + m \big) \right)^2 =
\big( -\vec{\gamma} \vec{k} + m \big) \big( \vec{\gamma} \vec{k} + m \big) = |\vec{k}|^2 + m^2 \:, \eeq
the wave function~$\hat{\psi}$ defined by
\beq \hat{\psi}(\vec{k}) := \frac{1}{|\vec{k}|^2 + m^2}\:\E^{\cI \vec{k} \vec{x}_0}\: \Xi \eeq
for a constant spinor~$\Xi$ and~$\vec{x}_0 \in \R^3$, satisfies the equation
\beq H^2 \,\psi(\vec{x}) = \delta^3(\vec{x}-\vec{x_0}) \,\Xi \:. \eeq
Moreover, one verifies immediately that~$\psi \in \H_{t_0}$ is square-integrable.
Using the last equation together with~\eqref{Hpq}, we conclude that
\beq H^q A \big( \delta^3(\vec{x}-\vec{x_0}) \,\Xi \big) = H^q A H^2 \psi \in \H_{t_0} \:. \eeq
Since~$q$ is arbitrary, it follows that~$A$ has an integral representation in the spatial variables,
\beq (A \phi)(\vec{x}) = \int_{\scrN} {\mathcal{A}}(\vec{x}, \vec{y})\,\gamma^0\, \phi(\vec{y})\: \dd^3y \qquad \text{with} \qquad {\mathcal{A}} \in C^\infty(\scrN \times \scrN)\:. \eeq
We now extend this integral kernel to~$\scrM \times \scrM$ by solving the Cauchy problem
in the variables~$x$ and~$y$. This preserves smoothness by the global existence
and regularity results for linear hyperbolic equations, giving the result.
\QED

\begin{Lemma} \label{lemmacommute}
Under the assumptions of Theorem~\ref{thmHadamard},
for all~$p \in \N$ the iterated commutator
\beq \Sig^{(p)} := \underbrace{ \Big[H, \big[H, \ldots ,[H}_{\text{$p$ factors}}, \tilde{\Sig}_m] \cdots \big] \Big] \eeq
is a bounded operator on~$\H_m$.
\end{Lemma}
\Proof In the vacuum, the Hamiltonian clearly commutes with the time evolution operator,
\beq \label{comm0}
\big[ H, U_m^{t,t'} \big] = 0 \:.
\eeq
In order to derive a corresponding commutator relation in the presence of the external potential,
one must take into account that~$\tilde{H}$ is time-dependent. For ease of notation, we do not
write out this dependence, but instead understand that the Hamiltonian is to be evaluated at the
correct time; that is,
\beq \tilde{U}_m^{t,t'} \, \tilde{H} \equiv \tilde{U}_m^{t,t'} \, \tilde{H}(t') \qquad \text{and} \qquad
\tilde{H}\, \tilde{U}_m^{t,t'} \equiv \tilde{H}(t)\, \tilde{U}_m^{t,t'} \:. \eeq
Then
\beq (\cI \partial_t - \tilde{H}) \big( \tilde{H} \,\tilde{U}_m^{t,t'} - \tilde{U}_m^{t,t'} \tilde{H} \big) = \cI \dot{\tilde{H}} \,\tilde{U}_m^{t,t'} \qquad \text{and}
\qquad \tilde{H} \,\tilde{U}_m^{t,t'} - \tilde{U}_m^{t,t'} \tilde{H} \big|_{t=t'}=0 \eeq
(here and in what follows the dot denotes the partial derivative with respect to~$t$).
Solving the corresponding Cauchy problem gives
\beq
\big[ \tilde{H}, \tilde{U}_m^{t,t'} \big] = \int_{t'}^t \tilde{U}_m^{t,\tau} \dot{\tilde{H}} \,\tilde{U}_m^{\tau,t'}\: \dd\tau
\:. \label{comm1}
\eeq

In order to compute the commutator of~$H$ with the operator products in~\eqref{Sigm1}
and~\eqref{Sigm2},
we first differentiate the expression~$U_m^{t'',t} \,\V\, \tilde{U}_m^{t,t'}$ with respect to~$t$,
\beq \label{prelim}
\cI \partial_t \big( U_m^{t'',t} \,\V\, \tilde{U}_m^{t,t'} \big) =
\cI U_m^{t'',t} \,\dot{\V}\, \tilde{U}_m^{t,t'} + U_m^{t'',t} \,\V\, \tilde{H}\, \tilde{U}_m^{t,t'} - U_m^{t'',t} \,H \,\V\, \tilde{U}_m^{t,t'} \:.
\eeq
Moreover, using the commutation relations~\eqref{comm0} and~\eqref{comm1}, we obtain
\begin{align}
H &\, (U_m^{t'',t} \,\V\, \tilde{U}_m^{t,t'}) - (U_m^{t'',t} \,\V\, \tilde{U}_m^{t,t'}) \,\tilde{H} \notag \\
&=  U_m^{t'',t}\, H \, \V\, \tilde{U}_m^{t,t'} - U_m^{t'',t}\, \V\, \tilde{H}\, \tilde{U}_m^{t,t'}
+  U_m^{t'',t}\, \V \,[\tilde{H}, \tilde{U}_m^{t,t'}] \notag \\
&= \cI U_m^{t'',t} \,\dot{\V}\, \tilde{U}_m^{t,t'} - \cI \partial_t \big( U_m^{t'',t} \,\V\, \tilde{U}_m^{t,t'} \big)
+ \int_{t'}^t  U_m^{t'',t}\, \V\, \tilde{U}_m^{t,\tau} \dot{\tilde{H}} \,\tilde{U}_m^{\tau,t'}\: \dd\tau \:,
\end{align}
where in the last step we applied~\eqref{prelim}. It follows that
\begin{align}
\big[ &H , U_m^{t'',t} \,\V\, \tilde{U}_m^{t,t'} \big] = 
H \, (U_m^{t'',t} \V \tilde{U}_m^{t,t'}) - (U_m^{t'',t} \V \tilde{U}_m^{t,t'}) \,\tilde{H}
+ (U_m^{t'',t} \V \tilde{U}_m^{t,t'}) \,\V \notag \\
&= \cI U_m^{t'',t} \,\dot{\V}\, \tilde{U}_m^{t,t'} + (U_m^{t'',t} \V \tilde{U}_m^{t,t'}) \,\V
- \cI \partial_t \big( U_m^{t'',t} \,\V\, \tilde{U}_m^{t,t'} \big)
+ \int_{t'}^t  U_m^{t'',t}\, \V\, \tilde{U}_m^{t,\tau} \dot{\tilde{H}} \,\tilde{U}_m^{\tau,t'}\: \dd\tau \:.
\end{align}
Proceeding in this way, one can calculate the commutator of~$H$ with
all the terms in~\eqref{Sigm1} and~\eqref{Sigm2}. We write the result symbolically as
\beq [H, \tilde{\Sig}_m] = \Sig^{(1)} \:, \eeq
where~$\Sig^{(1)}$ is a bounded operator.
Higher commutators can be computed inductively, giving the result.
\QED
We point out that this lemma only makes a statement on the iterative commutators.
Expressions like~$[H^p, \tilde{\Sig}_m]$ or~$H^q \,\tilde{\Sig}_m\, H^p$ will not be bounded operators in general.
However, the next lemma shows that the operator~$\Delta \tilde{\Sig}$ has the remarkable property
that multiplying by powers of~$H$ from the left and/or right again gives a bounded operator.
\begin{Lemma} \label{lemmaprod} Under the assumptions of Theorem~\ref{thmHadamard},
for all~$p, q \in \N \cup \{0\}$ the product~$H^q \,\Delta \tilde{\Sig}\, H^p$ is a bounded operator on~$\H_m$.
\end{Lemma}
\Proof We only consider the products~$H^q \,\Sig^-_+\, H^p$ because
the operator~$\Sig^+_-$ can be treated similarly.
Multiplying~\eqref{comm1} from the left and right by the resolvent of~$H$, we obtain
\beq \big[(H - \mu)^{-1}, \tilde{\Sig}_m \big] = - (H - \mu)^{-1} \,\Sig^{(1)}\, (H - \mu)^{-1} \:. \eeq
Writing the result of Lemma~\ref{lemmacommute} as
\beq [H, \Sig^{(p)}] = \Sig^{(p+1)} \qquad \text{with} \qquad \Sig^{(p+1)} \in \Lin(\H) \eeq
yields more generally the commutation relations
\begin{align}
\big[ (H-\mu)^{-1}, \Sig^{(p)} \big] = - (H - \mu)^{-1} \,\Sig^{(p+1)}\, (H - \mu)^{-1}
\qquad \text{for~$p \in \N$}\:. \label{rescomm}
\end{align}

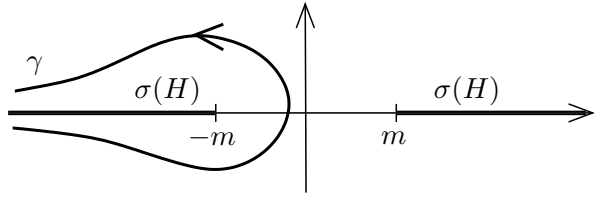
\begin{figure} %
\begin{picture}(0,0)%
\includegraphics{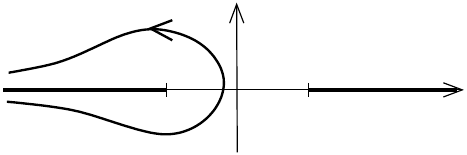}%
\end{picture}%
\setlength{\unitlength}{2486sp}%
\begingroup\makeatletter\ifx\SetFigFont\undefined%
\gdef\SetFigFont#1#2#3#4#5{%
  \reset@font\fontsize{#1}{#2pt}%
  \fontfamily{#3}\fontseries{#4}\fontshape{#5}%
  \selectfont}%
\fi\endgroup%
\begin{picture}(5916,1949)(1037,-6368)
\put(5326,-5401){\makebox(0,0)[lb]{\smash{{\SetFigFont{11}{13.2}{\familydefault}{\mddefault}{\updefault}$\sigma(H)$}}}}
\put(2341,-5401){\makebox(0,0)[lb]{\smash{{\SetFigFont{11}{13.2}{\familydefault}{\mddefault}{\updefault}$\sigma(H)$}}}}
\put(4801,-5866){\makebox(0,0)[lb]{\smash{{\SetFigFont{11}{13.2}{\familydefault}{\mddefault}{\updefault}$m$}}}}
\put(2874,-5873){\makebox(0,0)[lb]{\smash{{\SetFigFont{11}{13.2}{\familydefault}{\mddefault}{\updefault}$-m$}}}}
\put(1262,-5115){\makebox(0,0)[lb]{\smash{{\SetFigFont{11}{13.2}{\familydefault}{\mddefault}{\updefault}$\gamma$}}}}
\end{picture}%

\caption{The contour~$\gamma$.}
\label{figcontour}
\end{figure}

Choosing a contour~$\gamma$ that encloses the interval~$(-\infty, -m]$
as shown in  Figure~\ref{figcontour}, one finds
\begin{align}
H \Sig^-_+ &= -\frac{1}{2 \pi \cI} \int_\gamma \mu\, (H - \mu)^{-1} \,\tilde{\Sig}_m \:\chi^+(H)\:\Diff\mu \notag \\
&= \Sig\: H\, \chi^-(H)\: \chi^+(H) + 
\frac{1}{2 \pi \cI} \int_\gamma \mu\, (H - \mu)^{-1} \,\Sig^{(1)} (H - \mu)^{-1} \:\chi^+(H)\:\Diff\mu \notag \\
&= \frac{1}{2 \pi \cI} \int_\gamma \mu\, (H - \mu)^{-1} \,\Sig^{(1)} (H - \mu)^{-1} \:\chi^+(H)\:\Diff\mu \:,
\end{align}
where in the last step we used that~$\chi^-(H)\, \chi^+(H) = 0$.
In order to show that this operator product is bounded,
it is useful to employ the spectral theorem for~$H$,
which we write as
\beq \label{specthm}
f(H) = \int_{\R \setminus [-m,m]} f(\lambda)\, \dd E_\lambda\:,
\eeq
where~$dE_\lambda$ is the spectral measure of~$H$.
This gives
\begin{align}
H \,\Sig^-_+ &= \iint_{\R \times \R} \bigg(
\frac{1}{2 \pi \cI} \int_\gamma \frac{\mu}{\lambda - \mu}\: \: \frac{1}{\lambda' - \mu} \:\chi^+(\lambda')
\: \dd E_\lambda \bigg)\: \Sig^{(1)}\:\dd E_{\lambda'}\: \:\Diff\mu \nonumber \\
&= -\iint_{\R \times \R} \frac{\lambda}{\lambda - \lambda'}\:
\chi^-(\lambda)\:\chi^+(\lambda') \: \dd E_\lambda\: \Sig^{(1)}\:\dd E_{\lambda'}\:. \label{H0S}
\end{align}
Note that the term~$\lambda-\lambda'$ is bounded away from zero.
Thus the factor~$\lambda/(\lambda-\lambda')$ is bounded, showing that the
operator~$H \Sig^-_+$ is in~$\Lin(\H_m)$.

This method can be iterated. To this end, we first rewrite the product with commutators,
\begin{align}
H^q \,\Sig^-_+ &= \chi^-(H) \:\big( H^- \,\chi^-(H) \big)^p \,\tilde{\Sig}_m\, \chi^+(H) \notag \\
&= \chi^-(H) \Big[ H^-, \big[ H^-, \ldots, [ H^-, \Sig ] \cdots \big] \Big]\:
\chi^+(H) \:,
\end{align}
where we used the abbreviation~$H^- := H \,\chi^-(H)$.
Multiplying from the right by~$H^p$, we can commute factors~$H^+:= H\, \chi^+(H)$ to the left to obtain
\beq H^q \,\Sig^-_+ H^p = (-1)^p\: \chi^-(H) 
\,\underbrace{\big[ H^+, \ldots, \big[ H^+}_{\text{$p$ factors}} 
, \underbrace{\big[H^-, \ldots, \big[ H^-}_{\text{$q$ factors}}, \tilde{\Sig}_m \big] \cdots \big]
\big] \cdots \big] \, \chi^+(H)\:. \eeq
Representing each factor~$H^\pm$ by a contour integral, one can
compute the commutators inductively with the help~\eqref{rescomm}.
Applying the spectral theorem~\eqref{specthm} to the left and right of the resulting factor~$\Sig^{(p+q)}$
yields a constant times the expression
\begin{align} \iint_{\R \times \R} \chi^-(\lambda)\:\chi^+&(\lambda') \: \dd E_\lambda \,\Sig^{(p+q)}\, \dd E_{\lambda'} \notag \\
&\times \ointctrclockwise_{\gamma_1} \frac{\mu_1\: \Diff\mu_1}{(\lambda-\mu_1)(\lambda'-\mu_1)} \cdots
\ointctrclockwise_{\gamma_{p+q}} \frac{\mu_{p+q}\: \Diff\mu_{p+q}}{(\lambda-\mu_{p+q})(\lambda'-\mu_{p+q})} \:
\:.
\end{align}
Carrying out the contour integrals with residues, we obtain similar to~\eqref{H0S}
an expression of the form
\beq H^q \,\Sig^-_+ H^p = \iint_{\R \times \R} f(\lambda, \lambda')\:
\chi^-(\lambda)\:\chi^+(\lambda') \; \dd E_\lambda\: \Sig^{(p+q)}\:\dd E_{\lambda'} \eeq
with a bounded function~$f$. This concludes the proof.
\QED

\Proof[Proof of Theorem~\ref{thmfreqmix}.]
It remains to show that the contour integral in~\eqref{cD2} has a smooth integral kernel.
To this end, we multiply the integrand from the left by~$H^q$ and from the right by~$H^p$
and commute the factors~$H$ iteratively to the inside.
More precisely, we use the formula
\beq H^q (\tilde{\Sig}_m-\lambda)^{-1} = \sum_{a=0}^q \underbrace{ \big[H, \ldots, [H}_{\text{$a$ factors}},
(\tilde{\Sig}_m-\lambda)^{-1} ] \cdots \big] \:H^{q-a} \eeq
(note that the sum is telescopic; here we use the convention that the summand for~$a=0$
is simply~$(\tilde{\Sig}_m-\lambda)^{-1} H^q$). Hence
\begin{align}
&H^q\, (\tilde{\Sig}_m-\lambda)^{-1}\, \Delta \tilde{\Sig} \, (\SD-\lambda)^{-1} \,H^p \notag \\
&= \sum_{a=0}^q \sum_{b=0}^p\underbrace{\Big[H, \ldots, \big[H}_{\text{$a$ factors}},
(\tilde{\Sig}_m-\lambda)^{-1} \big] \cdots \Big] \notag \\
&\qquad\qquad\qquad\qquad \times \:H^{q-a} \, \Delta \tilde{\Sig} \, H^{p-b} \: \Big[ \cdots \big[ (\SD-\lambda)^{-1}, \underbrace{H\big], \ldots, H\Big]}_{\text{$b$ factors}} .
\end{align}
According to Lemma~\ref{lemmaprod}, the intermediate product~$H^{q-a} \, \Delta \tilde{\Sig} \, H^{p-b}$
is a bounded operator. Moreover, the commutators can be computed inductively
by applying Lemma~\ref{lemmacommute} and the formula
\beq \big[H, (\tilde{\Sig}_m-\lambda^{-1}) \big]
= -(\tilde{\Sig}_m-\lambda^{-1})\,\big[ H, \tilde{\Sig}_m \big]\, (\tilde{\Sig}_m-\lambda^{-1}) \eeq
(and similarly for~$\SD$). This gives operators that are all bounded
for~$\lambda \in \partial B_\frac{1}{2}(\pm 1)$. Since the integration contour is compact, the
result follows.
\QED

\subsection{Proof of the Hadamard Form} \label{secproofHadamard}
Relying on the frequency mixing estimates of the previous section, we can now
give the proof of Theorem~\ref{thmHadamard}. Recall that the fermionic projector is given by
(see~\eqref{Ppmdef2})
\beq \label{Pshort}
P = -\chi^-(\tilde{\Sig}_m)\, \tilde{k}_m \:,
\eeq
where we again used the short notation~\eqref{chipmdef}.
Here again the operator~$\chi^-(\tilde{\Sig}_m)$ acts on the solution space~$\H_m$ of the
Dirac equation, which can be identified with the space~$\H_{t_0}$ of square integrable
wave functions at time~$t_0$ (see the beginning of Section~\ref{secfreqmix}). 
For the following arguments, it is important to note
that this identification can be made at any time~$t_0$.

In order to prove that the bi-distribution corresponding to~$P$
is of Hadamard form, we compare the fermionic projectors
for three different Dirac operators
and use the theorem on the propagation of singularities
in~\cite{sahlmann2001microlocal}.
More precisely, we consider the following three fermionic projectors:
\begin{itemize}
\item[(1)] The fermionic projector~$P^\text{vac}$ in the Minkowski vacuum. \\[-0.85em]
\item[(2)] The fermionic projector~$\breve{P}$ in the presence of the external potential
\beq \breve{\B}(x) := \eta \big(x^0 \big)\: \B(x) \:, \eeq
where~$\eta \geq 0$ is a smooth function with~$\eta|_{(-\infty, 0)} \equiv 0$
and~$\eta|_{(1, \infty)} \equiv 1$. \\[-0.85em]
\item[(3)] The fermionic projector~$P$ in the presence of the external potential~$\B(x)$.
\end{itemize}
The potential~$\breve{\B}$ vanishes for negative times, whereas for times~$x^0>1$
it coincides with~$\B$. Thus it smoothly interpolates between the dynamics
with and without external potential. The specific form of the potential~$\breve{\B}$
in the transition region~$0 \leq x^0 \leq 1$ is of no relevance for our arguments.

In the Minkowski vacuum, the relation~\eqref{Pshort} gives the usual two-point function
composed of all negative-frequency solutions of the Dirac equation.
It is therefore obvious that the bi-distribution~$P^\text{vac}(x,y)$ is of Hadamard form.

We now compare~$P^\text{vac}$ with~$\breve{P}$. To this end, we choose an
arbitrary time~$t_0<0$. Then, applying the result of Theorem~\ref{thmfreqmix} to~\eqref{Pshort}, we get
\beq P^\text{vac} = -\chi^-(H)\, k_m \qquad \text{and} \qquad
\breve{P} = -\chi^-(H)\, \breve{k}_m + \text{(smooth)} \:, \eeq
where~$\breve{k}_m$ is the causal fundamental solution in the presence of the potential~$\breve{\B}$.
Since~$\breve{\B}$ vanishes in a neighborhood of the Cauchy surface at time~$t_0$,
we conclude that~$P^\text{vac}$ and~$\breve{P}$ coincide in this neighborhood up to a smooth contribution.
It follows that also~$\breve{P}(x,y)$ is of Hadamard form in this neighborhood.
Using the theorem on the propagation of singularities~\cite[Theorem 5.5]{sahlmann2001microlocal},
we conclude that~$\breve{P}(x,y)$ is of Hadamard form for all~$x,y \in \scrM$.

Next, we compare~$\breve{P}$ with~$P$. Thus we choose an arbitrary time~$t_0>1$.
Using again the result of Theorem~\ref{thmfreqmix} in~\eqref{Pshort}, we obtain
\beq \breve{P} = -\chi^-(H)\, \breve{k}_m + \text{(smooth)} \qquad \text{and} \qquad 
P = -\chi^-(H)\, \tilde{k}_m + \text{(smooth)} \eeq
(where the smooth contributions may of course be different).
Since~$\breve{\B}$ and~$\B$ coincide in a neighborhood of the Cauchy surface at time~$t_0$,
we infer that~$\breve{P}$ and~$P$ coincide in this neighborhood up to a smooth contribution.
As a consequence, $P(x,y)$ is of Hadamard form in this neighborhood.
Again applying~\cite[Theorem 5.5]{sahlmann2001microlocal},
it follows that~$P(x,y)$ is of Hadamard form for all~$x,y \in \scrM$.
This concludes the proof of Theorem~\ref{thmHadamard}.

\section{Exercises}

\begin{Exercise} \label{ex2.4-21} 
\sindex{light-cone expansion!simple examples}%
{\em{ This exercise explains the notion of the {\em{light-cone expansion}} in simple examples.
\bitem
\item[(a)] What is the light-cone expansion of a smooth function on~$\scrM \times \scrM$?
In which sense is it trivial? In which sense is it non-unique?
\item[(b)] Show that~$A(x,y) = \log \big(|y-x|^2 \big)$ is a well-defined distribution on~$\scrM \times \scrM$.
What is the order on the light cone? Write down a light-cone expansion.
\item[(c)] Now consider the distributional derivatives
\beq \Big( \frac{\partial}{\partial x^0} \Big)^p A(x,y) \qquad \text{with} \qquad p \in \N \eeq
and~$A(x,y)$ as in part~(b). What is the order on the light cone? Write down a light-cone expansion.
\item[(d)] Consider the function
\beq E(x,y) = \sin\big( (y-x)^2 \big)\: \log \big(|y-x|^2 \big)\:. \eeq
Determine the order on the light cone and give a light-cone expansion.
\item[(e)] Consider the function
\beq E(x,y) = \left\{ \begin{array}{cl} \displaystyle \E^{-\frac{1}{(y-x)^2}} & \text{if~$(y-x)^2 \geq 0$} \\[0.15em]
0 & \text{otherwise .} \end{array} \right. \eeq
Determine the order on the light cone and give a light-cone expansion.
\item[(f)] Show that the expression
\beq \lim_{\varepsilon \searrow 0} \frac{\log \big(|y-x|^2 \big)}{(y-x)^4 + \cI \varepsilon} \eeq
is a well-defined distribution on~$\scrM \times \scrM$. Derive its light-cone expansion.
\eitem
}} \end{Exercise}

\begin{Exercise} (Understanding the light-cone expansion) {\em{
This exercise aims to familiarize you with some of the particularities of the light-cone expansion.
\bitem
\item[(a)] Let~$A(x,y):= (x-y)^{2k_0}$ with~$k_0 \in \Z$. Which order(s) on the light cone is this? (Prove your answer.) Construct a light-cone expansion of~$A(x,y)$ and prove that it is one.
\item[(b)] \label{Bex} Let~$B(x,y):= (x-y)^{2k_0}+(x-y)^{2k_1}$, where~$k_0, k_1 \in \Z$ and~$k_0 < k_1$. Which order(s) on the light cone is this? (Prove your answer.) Construct a light-cone expansion of~$B(x,y)$ and prove that it is one.
\item[(c)] \label{Cex} Let~$C(x,y):=(x-y)^{2k_0} f(x,y) +(x-y)^{2k_1} g(x,y)$, where~$f$ and~$g$ are smooth functions in~$x$ and~$y$ and~$k0, k1$ as above. Construct a light-cone expansion of~$C(x,y)$ and prove that it is one.
\item[(d)] Let~$D(x,y): = \textrm{sin}\left( (x-y)^2 \right) (x-y)^2$. Use your results from \textrm{(b)} and \textrm{(c)} to construct two different light-cone expansions of~$D(x,y)$. Why might this non-uniqueness not be a problem for the scope of this book?
\item[(e)] Finally, consider the function
\beq E(x,y) = \textrm{sin}\left( (y-x)^2 \right)  + \begin{cases}  \E^{-\frac{1}{(y-x)^2}} & \textrm{if } (y-x)^2 \geq 0 \\ 0 & \textrm{else } \end{cases},
 \eeq
 Determine its order on the light cone and derive a light-cone expansion.
\eitem
{\em{Hint:}} For (d) and (e): Expand the sine function.
}} \end{Exercise}

\begin{Exercise} \label{ex2.4-2} 
\sindex{Green's operator!causal}%
{\em{ This exercise is devoted to computing the Fourier transform of the {\em{advanced Green's operator}}~\eqref{l:11}
and deriving the series expansion~\eqref{l:12}.
\bitem
\item[(a)] We again set~$\xi=y-x$ and~$\xi=(t,\vec{\xi})$ with~$t>0$.
Moreover, we choose polar coordinates~$r=(|\vec{\xi}|, \vartheta, \varphi)$. Carry out the $\omega$-integration
with residues and compute the angular integrals to obtain
\beq S^\vee_{m^2}(x,y) = \frac{\cI}{8 \pi r} \int_0^\infty \frac{p}{\omega(p)} \, \big( \E^{-\cI pr} - \E^{\cI pr}\big) \big(
\E^{\cI \omega(p)\, t} - \E^{-\cI \omega(p)\, t}\big)\, \dd p\:, \eeq
where~$p=|\vec{p}|$ and~$\omega(p):= \sqrt{|\vec{p}^2| + m^2}$.
Justify this integral as the Fourier transform of a distribution and show that
\beq \qquad S^\vee_{m^2}(x,y) =  \frac{\cI}{8 \pi r} \lim_{\varepsilon \searrow 0}\int_0^\infty \E^{-\varepsilon p}\:
\frac{p}{\omega(p)}\, \big( \E^{-\cI pr} - \E^{\cI pr}\big) \big( \E^{\cI \omega(p) \,t} - \E^{-\cI \omega(p) \,t}\big)\, \dd p \:,\eeq
with convergence as a distribution.
\item[(b)] Verify~\eqref{l:121} in the case~$m=0$ by setting~$\omega(p)=p$ and using~\eqref{eq:delta-formula}.
\item[(c)] In order to analyze the behavior away from the light cone, it is most convenient to
take the limit~$r \searrow 0$ and use Lorentz invariance. Show that in this limit,
\begin{align}
S^\vee_{m^2}(x,y) &=  \frac{1}{4 \pi} \lim_{\varepsilon \searrow 0}\int_0^\infty \E^{-\varepsilon p}\:
\frac{p^2}{\omega(p)} \:\big( \E^{\cI \omega(p) \,t} - \E^{-\cI \omega(p) \,t}\big)\, \dd p \label{Spform} \\
&=  \frac{1}{4 \pi} \lim_{\varepsilon \searrow \omega}\int_m^\infty \E^{-\varepsilon p}\:
\sqrt{\omega^2-m^2}\:\big( \E^{\cI \omega t} - \E^{-\cI \omega t}\big)\, \dd \omega\:.
\end{align}
Compute this integral using~\cite[formula (3.961.1)]{gradstein}.
Use the relations between Bessel functions~\cite[(10.27.6), (10.27.11)]{DLMF}
to obtain~\eqref{l:121} away from the light cone.

As an alternative method for computing the Fourier integral, one can begin
from the integral representation for~$J_0$ in~\cite[(10.9.12)]{DLMF}, differentiate with respect to~$x$
and use~\cite[(10.6.3)]{DLMF}.
\item[(d)] Combine the results of~(b) and~(c) to prove~\eqref{l:121}. Why is there no additional
contribution at~$\xi=0$?
\item[(e)] Use the series expansion~\cite[(10.2.2)]{DLMF} to derive~\eqref{l:12}.
\item[(f)] The series expansion~\eqref{l:12} can also be derived without using Bessel functions.
To this end, one expands~\eqref{Spform} in powers of~$m^2$ and computes the
Fourier transform term by term. Verify explicitly that this procedure really gives~\eqref{l:12}.
\eitem
}} \end{Exercise}

\begin{Exercise} \label{exlightcancel}  {\em{
In this exercise, we illustrate the dependence of the light-cone expansion~\eqref{l:4}
on the the function~$V$. We choose~$l=0$, $r \in \N_0$ arbitrary, $x=0$ and~$V(z) = z^2$.
\bitem
\item[(a)] Show that the left-hand side of~\eqref{l:4} vanishes for all~$y \in M$.
{\em{Hint:}} Use the causal structure of~$S_a$ as given in~\eqref{l:121} in the massless case.
\item[(b)] Show that all the summands in~\eqref{l:4} for~$n \geq 2$ vanish.
\item[(c)] Show that the summands in~\eqref{l:4} for~$n=0$ and~$n=1$ are both non-zero
and cancel each other. {\em{Hint:}} Compute~$\Box_z z^2$. Moreover, make use of the
relation between the coefficients of the power series~\eqref{l:12} for~$j=r+1$ and~$j=r+2$.
It might be a good idea to begin with the case~$r=0$.
\eitem
}} \end{Exercise}

%% file: part4.tex
\part{Applications and Outlook} \label{partfour}

\chapter{A Few Explicit Examples of Causal Variational Principles} \label{secexcvp}
In this chapter, we introduce a few examples of causal variational principles and analyze them in detail.
These examples are too simple for being of direct physical interest.
Instead, they are chosen in order to illustrate the different mathematical structures introduced previously.
It is a specific feature of these examples that a minimizing measure can be given in a closed form,
making it possible to analyze the system explicitly. Similar examples were first given in~\cite{static}.

When constructing simple explicit examples, it is often convenient to choose {\em{non-smooth}}
Lagrangians, which involve, for example, characteristic functions or are even distributional.
In order to treat this non-smooth setting in a mathematically convincing way, one needs to
work with additional jet spaces, which we now introduce (for more details, see for example~\cite{jet, fockbosonic}).

Clearly, the fact that a jet~$\u$ is smooth does not imply that the functions~$\ell$
or~$\L$ are differentiable in the direction of~$\u$. This must be ensured by additional
conditions that are satisfied by suitable subspaces of~$\J$, which we now define.
First, we let~$\Gdiff$ be those vector fields for which the
directional derivative of the function~$\ell$ exists,
\beq \Gdiff = \big\{ u \in C^\infty(M, T\F) \;\big|\; \text{$D_{u} \ell(x)$ exists for all~$x \in M$} \big\} \:. \eeq
This gives rise to the jet space
\beq \Jdiff := C^\infty(M, \R) \oplus \Gdiff \;\subset\; \J \:. \eeq
\nindex{aia@$\Jdiff, \Gdiff$ -- Space of jets for which~$\ell$ is differentiable}%
\tindex{ff@$\Jdiff, \Gdiff$ -- Space of jets for which~$\ell$ is differentiable}%
For the jets in~$\Jdiff$, the combination of multiplication and directional derivative
in~\eqref{Djet} is well-defined. 
We choose a linear subspace~$\Jtest \subset \Jdiff$ with the property
that its scalar and vector components are both vector spaces; that is,
%\beq\label{Gammatest}
\beq \Jtest = \Ctest(M, \R) \oplus \Gtest \;\subseteq\; \Jdiff \eeq
\nindex{aib@$\Jtest$ -- space of test jets}%
\tindex{ff@$\Jtest$ -- space of test jets}%
for suitable subspaces~$\Ctest(M, \R) \subset C^\infty(M, \R)$ and~$\Gtest \subset \Gdiff$.
We then write the {\em{restricted EL equations}}~\eqref{ELrestricted2} in the weaker form
\sindex{Euler-Lagrange equations!restricted}%
\beq %\label{ELtest}
\nabla_{\u} \ell|_M = 0 \qquad \text{for all~$\u \in \Jtest$}\:. \eeq
\nindex{agk@$\Jvary$ -- space of jets used for varying the measure}%
\tindex{ff@$\Jvary$ -- space of jets used for varying the measure}%
Finally, when considering weak solutions of the linearized field equations, it is sometimes
useful to restrict attention to jets in
a suitably chosen subspace of~$\Jtest$ which, in agreement with~\eqref{Jvarydef}, we denote by
\beq \Jvary \subset \Jtest \:. \eeq
To summarize, we have the inclusions
\beq \Jvary \subset \Jtest \subset \Jdiff \subset \J \:. \eeq
The compactly supported jets are always denoted by an additional subscript zero.

\section{A One-Dimensional Gaussian} \label{secgaussian}
We let~$\F = \R$ and choose the Lagrangian as the Gaussian
\beq \label{gauss}
\L(x,y) = \frac{1}{\sqrt{\pi}}\: \E^{-(x-y)^2} \:.
\eeq

\begin{Lemma} \label{lemmagauss} The Lebesgue measure
\beq \Diff\rho = \dd x \eeq
is a minimizer of the causal action principle for the Lagrangian~\eqref{gauss}
in the class of variations of finite volume (see~\eqref{integrals} and~\eqref{totvol}).
It is the unique minimizer within this class of variations.
\end{Lemma}
\Proof Writing the difference of the actions as in~\eqref{integrals}, we can carry out the integrals
over~$\rho$ using that the Gaussian is normalized (see Exercise~\ref{exselfsim}),
\beq \int_\F \L(x,y)\: \Diff\rho(y) = 1 \:. \eeq
We thus obtain
\begin{align}
\Sact(\rho) - \Sact(\tilde{\rho})
&= 2\, \int_N \dd(\rho-\tilde{\rho})(x) +  \int_N \dd(\rho-\tilde{\rho})(x) \int_N \dd(\rho-\tilde{\rho})(y) \:\L(x,y) \notag \\
&= \int_N  \dd(\rho-\tilde{\rho})(x) \int_N \dd(\rho-\tilde{\rho})(y)\: \L(x,y) \:,
\end{align}
where in the last line we used the volume constraint~\eqref{totvol}.
In order to show that the last double integral is positive, we take the Fourier transform and
use that the Fourier transform of a Gaussian is again a Gaussian. More precisely,
\beq \label{gaussselfsim}
\int_N \E^{-\cI p(x-y)}\: \L(x,y) \:\dd y = \E^{-\frac{p^2}{4}} =: f(p) \:.
\eeq
Moreover, the estimate
\beq \Big| \int_N \E^{\cI px}\: \dd(\rho-\tilde{\rho})(x) \Big| \leq \big| \tilde{\rho} - \rho \big|(\F) < \infty \eeq
shows that the Fourier transform of the signed measure~$\tilde{\rho} - \rho$ is 
a bounded function~$g \in L^\infty(\R)$.
Approximating this function in~$L^2(\R)$, we can
apply Plancherel's theorem and use the fact that convolution in position space
corresponds to multiplication in momentum space. We thus obtain
\begin{align}
\int_N &\dd(\rho-\tilde{\rho})(x) \int_N \dd(\rho-\tilde{\rho})(y)\: \L(x,y) \notag \\
&= \int_N \big( {\mathcal{F}}^{-1}(f g) \big)(x) \: \dd(\rho-\tilde{\rho})(x)
= \int_{-\infty}^\infty \overline{g(p)} \: \E^{-\frac{p^2}{4}}\: g(p)\:\dd p \geq 0 \:, \label{gausspositive}
\end{align}
and the inequality is strict unless~$\tilde{\rho}=\rho$.
This concludes the proof.
\QED

The EL equations read
\beq \label{ELex1}
\int_\F \L(x,y) \: \Diff\rho(y) = 1 \qquad \text{for all~$x \in \R$} \:.
\eeq
We now specify the jet spaces. Since the Lagrangian is smooth, it is obvious that
\beq \Jdiff = \J = C^\infty(\R) \oplus C^\infty(\R) \eeq
(where we identify a vector field~$a(x)\: \partial_x$ on~$\R$ with the function~$a(x)$).
The choice of~$\Jtest$ is less obvious. For simplicity, we restrict attention to functions
that are bounded together with all their derivatives, denoted by
\beq C^\infty_\text{b} := \big\{ f \in C^\infty(\R) \:\big|\: f^{(n)} \in L^\infty(\R) \text{ for all~$n \in \N_0$} \big\} \:. \eeq
Now different choices are possible. Our first choice is to consider jets whose
scalar components are compactly supported,
\beq \label{Jtestex1}
\Jtest = C^\infty_0(\R) \oplus C^\infty_\text{b}(\R) \:.
\eeq
The linearized field equations~\eqref{linfield} reduce to the scalar equation
\beq \label{lfeex1}
\int_N \big( \nabla_{1, \v} + \nabla_{2, \v} \big) \L(x,y)\: \Diff\rho(y) - \nabla_\v \,1 = 0 \qquad
\text{for all~$x \in \R$} \:,
\eeq
because if this equation holds, then the $x$-derivative of the left side is also zero.
Differentiating the EL equations~\eqref{ELex1} with respect to~$x$, we find that
\beq \int_N \nabla_{1, \v} \L(x,y)\: \Diff\rho(y) - \nabla_\v \,1 = 0 \qquad
\text{for all~$x \in \R$} \:. \eeq
Subtracting this equation from~\eqref{lfeex1}, the linearized field equations simplify to
\beq %\label{Lsimp}
\int_N \nabla_{2,\v} \L(x,y)\: \Diff\rho(y) = 0 \qquad \text{for all~$x \in \R$}\:. \eeq

A specific class of solutions can be given explicitly. Indeed, choosing
\beq \label{inner}
\u = (a, A) \qquad \text{with} \qquad a \in C^\infty_0(\R) \text{ and } A(x) := \int_{\infty}^x a(t)\:\dd t \in C^\infty_{\text{b}}(\R) \:,
\eeq
integration by parts yields
\beq \label{pint}
\int_N \nabla_{2,\u} \L(x,y)\: \Diff\rho(y) = \int_N \big(A'(y) + A(y)\: \partial_y \big) \L(x,y)\: \dd y = 0 \:.
\eeq
These linearized solutions are referred to as {\em{inner solutions}},
as introduced in a more general context in Section~\ref{secinner} and~\cite{fockbosonic}. 
Inner solutions can be regarded as infinitesimal generators of transformations of~$M$
that leave the measure~$\rho$ unchanged. Therefore, inner solutions do not change the
causal fermion system, but merely describe symmetry transformations of the measure.
With this in mind, inner solutions are not of interest by themselves. But they can be used in order
to simplify the form of the jet spaces. For example, by adding suitable inner solutions, one can
arrange that the test jets have vanishing scalar components. Indeed, given a jet~$\v = (b,v)
\in \Jtest$ (with~$\Jtest$ according to~\eqref{Jtestex1}), taking an indefinite integral of~$b$,
\beq B(t) := \int_{-\infty}^t b(\tau)\: \dd\tau \;\in\; C^\infty_\text{b}(\R) \:, \eeq
the resulting jet~$\u := (-b,-B)$ is an inner solution~\eqref{inner}. Adding this jet to~$\v$ gives
\beq \tilde{\v} := \v + \u = (0, v-B) \in \Jtest \:, \eeq
which is physically equivalent to~$\v$ and, as desired, has a vanishing scalar component.

In our example, we can use the inner solutions alternatively in order to eliminate the vector component
of the test jets. To this end, it is preferable to choose the space of test jets as
\beq \label{veccompensate}
\Jtest = C^\infty_\text{b}(\R) \oplus C^\infty_\text{b}(\R) \:.
\eeq
Now the vector component disappears under the transformation
\beq \v = (b,v) \mapsto \tilde{\v} := \v + \u \qquad \text{with} \qquad \u = (-v', -v) \in \Jtest \:. \eeq
Therefore, it remains to consider the scalar components of jets. For technical simplicity, we
restrict attention to compactly supported functions. Thus we choose the jet space~$\Jvary$
as
\beq \Jvary = C^\infty_0(\R) \oplus \{0\} \:. \eeq
Then the linearized field operator in~\eqref{linfield}
reduces to the integral operator with kernel~$\L(x,y)$,
\beq \big( \Delta (b,0) \big)(x) = \int_\F \L(x,y)\: b(y)\: \dd y \:. \eeq

\section{A Minimizing Measure Supported on a Hyperplane} \label{secexhyper}
In the previous example, the support of the minimizing measure was the whole space~$\F$.
In most examples motivated from the physical applications,
however, the minimizing measure will be supported on a low-dimensional
subset of~$\F$ (see, for instance, the minimizers with singular support
for the causal variational principle on the sphere in~\cite{support, sphere}
discussed in Section~\ref{seccvpsphere}).
We now give a simple example where the minimizing measure is supported on a hyperplane of~$\F$.
We let~$\F=\R^2$ and choose the Lagrangian as
\beq \label{gauss2}
\L(x,y; x',y') = \frac{1}{\sqrt{\pi}}\: \E^{-(x-x')^2} \big( 1 + y^2 \big)\big( 1 +y'^2 \big) \:,
\eeq
where~$(x,y), (x',y') \in \F$.
\begin{Lemma} \label{lemmagauss2} The measure
\beq \label{mingauss2}
\Diff\rho = \dd x \times \delta_y
\eeq
(where~$\delta_y$ is the Dirac measure)
is the unique minimizer of the causal action principle for the Lagrangian~\eqref{gauss2} under variations of
finite volume (see~\eqref{integrals} and~\eqref{totvol}).
\end{Lemma} \noindent
Note that this measure is supported on the $x$-axis,
\beq M := \supp \rho = \R \times \{0\} \:. \eeq
\Proof[Proof of Lemma~\ref{lemmagauss2}.]
Let~$\tilde{\rho}$ be a regular Borel measure on~$\F$
satisfying~\eqref{totvol}. Then the difference of actions~\eqref{integrals} is computed by
\begin{align}
&\Sact(\tilde{\rho}) - \Sact(\rho) 
= \frac{2}{\sqrt{\pi}} \int_\F \dd(\tilde{\rho}-\rho)(x,y) \int_N \dd x' \:\E^{-(x-x')^2} \:(1+ y^2) \label{Slin} \\
&\quad\: +  \frac{1}{\sqrt{\pi}}
\int_\F \dd(\tilde{\rho}-\rho)(x,y) \int_\F \dd(\tilde{\rho}-\rho)(x',y') \:\E^{-(x-x')^2}
\:\big( 1 + y^2 \big)\big( 1 +y'^2 \big) \:. \label{Squad}
\end{align}
Using that the negative part of the measure~$\tilde{\rho}-\rho$
is supported on the $x$-axis, the first term~\eqref{Slin} can be estimated by
\begin{align}
&\frac{2}{\sqrt{\pi}} \int_\F \dd(\tilde{\rho}-\rho)(x,y) \int_N \dd x' \:\E^{-(x-x')^2} \:(1+ y^2) \notag \\
&\overset{(*)}{\geq} \frac{2}{\sqrt{\pi}} \int_\F \dd(\tilde{\rho}-\rho)(x,y) \int_N \dd x' \:\E^{-(x-x')^2}
= \int_\F \dd(\tilde{\rho}-\rho)(x,y) = 0 \:,
\end{align}
where in the last step we used the volume constraint.
The second term~\eqref{Squad}, on the other hand, can be rewritten as
\beq \frac{1}{\sqrt{\pi}} \int_\F \Diff\mu(x,y) \int_\F \Diff\mu(x',y') \:\E^{-(x-x')^2} \eeq
with the signed measure~$\rho$ defined by
\beq \Diff\mu(x,y) := \big( 1 + y^2 \big)\: \dd(\tilde{\rho}-\rho)(x,y) \:. \eeq
Now we can proceed as in the proof of Lemma~\ref{lemmagauss} and use
that the Fourier transform of the integral kernel is strictly positive.
For the uniqueness statement one uses that the inequality in~$(*)$ is strict unless~$\tilde{\rho}$
is supported on the $x$-axis. Then one can argue as in the proof of Lemma~\ref{lemmagauss}.
\QED

For the minimizing measure~\eqref{mingauss2}, the function~$\ell$ takes the form
\beq \ell(x,y) = \int_\F \L(x,y; x',y') \: \Diff\rho(x',y') - 1 = y^2 \:, \eeq
showing that the EL equations~\eqref{EL1} are indeed satisfied.
We now specify the jet spaces. Since the Lagrangian is smooth, it is obvious that
\beq %\label{Jdiffex}
\Jdiff = \J = C^\infty(\R) \oplus C^\infty(\R, \R^2) \:, \eeq
where~$C^\infty(\R, \R^2)$ should be regarded as the space of two-dimensional
vector fields along the $x$-axis.
As explained after~\eqref{veccompensate}, we want to use the inner solutions for
simplifying the vector components of the jets. 
To this end, in analogy to~\eqref{veccompensate}, we choose
\beq \label{veccompensate2}
\Jtest = C^\infty_\text{b}(\R) \oplus C^\infty_\text{b}(\R, \R^2) \:.
\eeq
The linearized field equations~\eqref{linfield} read
\begin{align}
0 &= \nabla_{\u} \bigg( \int_{-\infty}^\infty \big( \nabla_{1, \v} + \nabla_{2, \v} \big) 
\E^{-(x-x')^2} \big( 1 + y^2 \big)\big( 1 +y'^2 \big)
\: \dd\rho(x',y') - \nabla_\v \,\sqrt{\pi} \bigg) \bigg|_{y=y'=0} \notag \\
&= \nabla_{\u} \bigg(\big( 1 + y^2 \big) \int_{-\infty}^\infty \big( \nabla_{1, \v} + \nabla_{2, \v} \big) 
\E^{-(x-x')^2} 
\: \dd x' - \nabla_\v \,\sqrt{\pi} \bigg) \bigg|_{y=y'=0} \:. %\label{lininter} 
\end{align}
Now the inner solutions are generated by the vector fields tangential to the~$x$-axis.
More precisely, in analogy to~\eqref{inner}, we consider the jet
\beq %\label{inner2}
\v = \big(b, (B,0) \big) \qquad \text{with} \qquad b \in C^\infty_0(\R) \text{ and } B(x) := \int_{\infty}^x b(t)\:\dd t \in C^\infty_{\text{b}}(\R) \:. \eeq
Exactly as in the example of the one-dimensional Gaussian,
integrating by parts as in~\eqref{pint}, one sees that the jet~$\v$ indeed satisfies the linearized field equations.

By suitably subtracting inner solutions, we can compensate for the tangential components of the
jets. This leads us to choose
\beq %\label{Jvaryex2}
\Jvary = C^\infty_0(\R) \oplus \big( \{0\} \oplus C^\infty_0(\R) \big) \:. \eeq
Then the Laplacian simplifies as follows,
\begin{align}
&\la \u, \Delta \v \ra(x) \notag \\
&= \frac{1}{\sqrt{\pi}}\: \nabla_{\u} \bigg( \int_{-\infty}^\infty \big( \nabla_{1, \v} + \nabla_{2, \v} \big) \E^{-(x-x')^2} \:\big( 1 + y^2 \big)\big( 1 +y'^2 \big)\: \dd x' - \nabla_\v \,\sqrt{\pi} \bigg) \bigg|_{y=y'=0} \notag \\
&= \frac{2}{\sqrt{\pi}} \,u(x)\, v(x) \int_{-\infty}^\infty \E^{-(x-x')^2}\: \dd x' 
+ \frac{1}{\sqrt{\pi}}\: a(x) \int_{-\infty}^\infty  \E^{-(x-x')^2} \: b(x')\: \dd x' \notag \\
&\quad\: + a(x) \bigg( \frac{1}{\sqrt{\pi}} \int_{-\infty}^\infty b(x)\: \E^{-(x-x')^2} 
\: \dd x' - b(x) \bigg) \notag \\
&= 2\,u(x)\, v(x) + \frac{1}{\sqrt{\pi}}\: a(x) \int_{-\infty}^\infty  \E^{-(x-x')^2} \: b(x')\: \dd x' \:,
\end{align}
where~$\u=(a, (0,u))$ and~$\v=(b,(0,v))$. Hence, the inhomogeneous linearized field equations~\eqref{Delvw} with~$\w=(e,w)$
give rise to separate equations for the scalar and vector components,
\beq %\label{linex2}
\frac{1}{\sqrt{\pi}} \int_{-\infty}^\infty  \E^{-(x-x')^2} \: b(x')\: \dd x' = e(x) \:,\qquad v(x) =\frac{w(x)}{2}\:. \eeq

\section{A Non-homogeneous Minimizing Measure}
In the previous examples, the minimizing measure~$\rho$ was translation invariant in the
direction of the $x$-axis. We now give a general procedure for constructing examples
of causal variational principles where the minimizing measure has no translational symmetry.
In order to work in a concrete example, our starting point is again the one-dimensional Gaussian~\eqref{gauss}.
But our method can be adapted to other kernels in a straightforward way.
In view of these generalizations, we begin with the following abstract result.

\begin{Lemma}
\label{MinCrit}
Let~$\mu$ be a measure on the $m$-dimensional manifold~$\F$ whose support is the whole manifold,
\beq \supp \mu = \F \:. \eeq
Moreover, let~$\L(x,y) \in L^1_\loc(\F \times \F, \R^+_0)$ be a symmetric, non-negative kernel on~$\F \times \F$. Next, let~$h \in C^0(\F, \R^+)$ be a strictly positive, continuous function on~$\F$.
Assume that:
\bitem
\item[{\rm{(i)}}] $\displaystyle \int_\F \L(x,y) \: h(y)\: \Diff\mu(y) = 1 \qquad \text{for all~$x \in \F$}$. \\
\item[{\rm{(ii)}}] For all compactly supported bounded functions with zero mean,
\beq g \in L^\infty_0(\F, \R^+) \qquad \text{and} \qquad \int_\F g\: \Diff\mu = 0 \:, \eeq
the following inequality holds,
\beq \label{geqineq}
\int_\F \Diff\mu(x) \int_\F \Diff\mu(y) \:\L(x,y)\: g(x)\: g(y) \geq 0 \:.
\eeq
\eitem
Then the measure~$\Diff\rho := h\, \Diff\mu$ is a minimizer of the causal action principle
under variations of finite volume (see~\eqref{integrals} and~\eqref{totvol}).
If the inequality~\eqref{geqineq} is strict for any non-zero~$g$, then the minimizing measure is unique
within the class of such variations.
\end{Lemma}
\begin{proof} We consider the variation
\beq \label{measvar}
\tilde{\rho}_\tau= \rho + \tau \,g \:\Diff\mu = (h+\tau g)\: \Diff\mu \:.
\eeq
Since~$h$ is continuous and strictly positive and~$g$ is bounded and compactly supported,
the function~$h+\tau g$ is non-negative for sufficiently small~$|\tau|$.
Furthermore, using that~$g$ has mean zero, we conclude that~\eqref{measvar} is an admissible variation of finite volume~\eqref{totvol}. Moreover, the difference of the actions~\eqref{integrals} is well-defined and
computed by
\begin{align}
&\Sact(\tilde{\mu}_\tau) - \Sact(\rho) \notag \\
&= 2\tau \int \Diff\rho(x)\:g(x)\int_N\Diff\rho(y) \:h(y)\:\L(x,y)+ \tau^2 \int_N \Diff\rho(x) \int_N \Diff\rho(x) \:\L(x,y)\:g(x)\:g(y) \notag \\
&\geq  2\tau \int_N g(y) \:\Diff\rho(y)= 0\;, \label{geqineq2}
\end{align}
where in the second step we used the above assumptions~(i) and (ii). The last step follows from the fact that~$g$
has mean zero. If the inequality~\eqref{geqineq} is strict, so is the inequality in~\eqref{geqineq2},
showing that the minimizer~$\rho$ is unique.

We conclude that the measure~$\rho$ is a minimizer under variations of the form~\eqref{measvar}.
In order to treat a general variation of finite volume~\eqref{totvol}, we approximate~$\tilde{\rho}$
by a sequence of functions~$g_n$ with the property that the measures~$g_n \rho$ converge
to~$\tilde{\rho}$ (here one can work with the notion of vague convergence;
for details see~\cite[Definition 30.1]{bauer} or~\cite{elstrodt}).
\end{proof}

Our goal is to apply this lemma to kernels of the form
\beq \label{Lgaussdef}
\L(x,y) = f(x) \: \E^{-(x-y)^2} f(y)
\eeq
with a strictly positive function~$f$, which for convenience we again choose as a Gaussian,
\beq \label{fweight}
f(x) = \E^{\alpha x^2} \qquad \text{with~$\alpha \in \R$}\:.
\eeq
This kernel has the property~(ii) because for all non-trivial~$g \in L^\infty_0(\F, \R^+)$,
\beq \int_\F \Diff\mu(x) \int_\F \Diff\mu(y) \:\L(x,y)\: g(x)\: g(y)
= \int_\F \Diff\mu(x) \int_\F \Diff\mu(y) \:\E^{-(x-y)^2}\: (f g)(x)\: (f g)(y) > 0 \:, \eeq
where the last step is proved exactly as in the example of the Gaussian (see~\eqref{gausspositive}).
In order to arrange~(i), we make an ansatz for~$h$ again with a Gaussian,
\beq \label{gaussweight}
h(x) = c\: \E^{\beta x^2}\:.
\eeq
Then,
\begin{align}
&\int_\F \L(x,y) \: h(y)\: \Diff\mu(y) = c \int_{-\infty}^\infty \E^{\alpha x^2} \: \E^{-(x-y)^2}\: \E^{(\alpha+\beta) y^2}\: \dd y \notag \\
&= c \exp \Big( \alpha x^2 - x^2 - \frac{x^2}{\alpha+\beta-1} \Big)
\int_{-\infty}^\infty \exp \bigg\{ (\alpha+\beta-1) \Big( y - \frac{x}{\alpha+\beta-1} \Big)^2 \bigg\}\:\dd y \notag \\
&= c\: \sqrt{\frac{\pi}{1-\alpha-\beta}}\: \exp \Big( \alpha x^2 - x^2 - \frac{x^2}{\alpha+\beta-1} \Big) \:.
\end{align}
In order to arrange that this function is constant one, we choose
\beq \label{cbetadef}
c = \sqrt{\frac{1-\alpha-\beta}{\pi}} \qquad \text{and} \qquad
\beta = -\frac{\alpha (2-\alpha)}{1-\alpha} \:.
\eeq
For the above Gaussian integral to converge, we need to ensure that~$1-\alpha-\beta > 0$.
In view of the formula
\beq 1 - \alpha - \beta = \frac{1}{1-\alpha} \:, \eeq
this can be arranged simply by choosing~$\alpha < 1$. Our finding is summarized as follows.
\begin{Prp} For any~$\alpha<1$, we let~$f$ and~$h$ be the Gaussians~\eqref{fweight} and~\eqref{gaussweight} with~$c$ and~$\beta$ according to~\eqref{cbetadef}. Then the measure~$\Diff\rho=h\, dx$ is the unique minimizer of the causal action corresponding to the Lagrangian~\eqref{Lgaussdef} within the class
of variations of finite volume.
\end{Prp}

As a concrete example, we consider the well-known {\em{Mehler kernel}}
(see, for example, \cite[Section~1.5]{glimm+jaffe})
\beq E(x,y) = \frac{1}{\sqrt{1-\mu^2}} \:\exp \bigg(-\frac{\mu^2 (x^2+y^2)- 2\mu x y}{(1-\mu^2)} \bigg) \eeq
with~$\mu > 0$. Rescaling~$x$ and~$y$ according to
\beq x,y \rightarrow \sqrt{\frac{1-\mu^2}{\mu}} \: x,y \:, \eeq
the Mehler kernel becomes
\beq E(x,y) = \frac{1}{\sqrt{1-\mu^2}} \:\exp \Big(-\mu (x^2+y^2)- 2 x y \Big) \:. \eeq
This kernel is of the desired form~\eqref{Lgaussdef} if we choose
\beq \alpha = 1-\mu < 1 \:,\qquad \beta = \frac{\mu^2-1}{\mu} \:.  \eeq

We finally remark that this non-homogeneous example can be used as the starting point
for the construction of higher-dimensional examples with minimizing measures supported on
lower-dimensional subsets, exactly as explained for the Gaussian in Section~\ref{secexhyper}.

\section{A Minimizing Measure in Two-Dimensional Minkowski Space} \label{secexmink}
In the previous examples, the Lagrangian was strictly positive
(see~\eqref{gauss}, \eqref{gauss2}, \eqref{Lgaussdef}). Therefore, the causal structure of the
resulting spacetime was trivial, because all pairs or points were timelike separated.
We now give examples where the minimizing measure gives rise to nontrivial causal relations in spacetime.
We let~$\F = \R^2$, denote the coordinates by~$(t,x)$ and choose the Lagrangian
\beq \label{Lmink2}
\L\big( t,x; t',x \big) = \E^{-(t-t')^2}\: \Big( \delta \big((t-t') - (x-x') \big) + \delta \big((t-t') + (x-x') \big) \Big) \:.
\eeq
The Lagrangian is non-negative, and it is strictly positive on the ``light rays''
$(t-t') = \pm (x-x')$.

\begin{Lemma}  \label{lemmaexmink2}
The Lebesgue measure
\beq \Diff\rho = \dd t\, \dd x \eeq
is a minimizer of the causal action principle for the Lagrangian~\eqref{Lmink2}
in the class of variations of finite volume (see~\eqref{integrals} and~\eqref{totvol}).
It is the unique minimizer within this class of variations.
\end{Lemma} 
\Proof Proceeding as in the proof of Lemma~\ref{lemmagauss}, our task is to show that the
Fourier transform of the Lagrangian is strictly positive. To this end, we note that
\beq \int_{\R^2} \delta(t-x)\: \E^{\cI \omega t - \cI k x}\: \dd t\, \dd x = \int_{-\infty}^\infty 
\E^{\cI \omega x - \cI k x}\: dx = 2 \pi\: \delta(\omega-k) \:. \eeq
We thus obtain
\begin{align}
&\int_{\R^2} \Big( \delta \big((t-t') - (x-x') \big) + \delta \big((t-t') + (x-x') \big) \Big)
\: \E^{\cI \omega t - \cI k x}\: \dd t\, \dd x \notag \\
&= 2 \pi\: \big( \delta(\omega+k) + \delta(\omega-k) \big) \:.
\end{align}
Multiplying by the Gaussian in~\eqref{Lmink2} corresponds to a convolution in momentum space
again by a Gaussian. This convolution gives a strictly positive function, as desired.
\QED

The Lagrangian~\eqref{Lmink2} has the shortcoming that it is supported only on the boundary
of the light cone. In order to improve the situation, we next consider the example
\beq \label{Lmink3}
\begin{split}
\L\big( t,x; t',x \big) &= \E^{-(t-t')^2}\: \Big( \delta \big((t-t') - (x-x') \big) + \delta \big((t-t') + (x-x') \big) \Big) \\
&\quad\:+ a\: \E^{-\frac{(t-t')^2}{2}}\: \Theta\big( (t-t')^2 - (x-x')^2 \big) \:.
\end{split}
\eeq
\begin{Lemma}  \label{lemmaexmink3}
Choosing~$|\alpha|<1$, the Lebesgue measure
\beq \Diff\rho = \dd t\, \dd x \eeq
is a minimizer of the causal action principle for the Lagrangian~\eqref{Lmink3}
in the class of variations of finite volume (see~\eqref{integrals} and~\eqref{totvol}).
It is the unique minimizer within this class of variations.
\end{Lemma} 
\Proof We compute the Fourier transform of the Heaviside function. 
\begin{align}
&\int_{\R^2} \Theta\big( t^2-x^2\big)
\: \E^{\cI \omega t - \cI k x}\:\E^{-\varepsilon\,|t|}\: \dd t\, \dd x \notag \\
&= 4 \int_0^\infty \dd x \int_{-\infty}^\infty \dd t \:\Theta( t-x )
\: \cos(\omega t) \: \cos (k x) \:\E^{-\varepsilon t}\: \dd t\, \dd x \notag \\
&= 2 \int_0^\infty \bigg( -\frac{\E^{\cI \omega x - \varepsilon x}}{\cI \omega - \varepsilon}
- \frac{\E^{-\cI \omega x - \varepsilon x}}{-\cI \omega - \varepsilon} \bigg)\: \cos (k x)\:\dd x \notag \\
&= -\frac{1}{\cI \omega - \varepsilon} \bigg( \frac{1}{\cI \omega + k -\varepsilon}
+ \frac{1}{\cI \omega - k -\varepsilon} \bigg) \notag \\
&\quad\: -  \frac{1}{-\cI \omega - \varepsilon} \bigg( \frac{1}{-\cI \omega + k -\varepsilon}
+ \frac{1}{-\cI \omega - k -\varepsilon} \bigg) \:.
\end{align}
In the limit~$\varepsilon \searrow 0$, this converges to a tempered distribution
that is singular on the light cone. Taking the convolution with the Gaussian and choosing~$a$
sufficiently small, the resulting function is dominated near the light cone by the
Fourier transform computed in the proof of Lemma~\ref{Lmink2}.
Moreover, due to its decay properties at infinity, the same is true away from the light cone.
This concludes the proof.
\QED

\section{A Nonlinear Wave Equation in Two-Dimensional Minkowski Space} \label{secexwave}
In the previous examples, the minimizing measures were unique. This means, in particular, that the
systems had no dynamical degrees of freedom, and the linearized field equations only admitted trivial solutions.
We now explain how one can build in dynamical degrees of freedom.
For simplicity, we consider the example of a nonlinear wave equation on a spacetime lattice,
but the method can be generalized to many other situations.
We choose~$\F = \R^2 \times S^1$ and denote the coordinates by~$(t,x) \in \R^2$
and~$\E^{\cI \alpha} \in S^1$. We choose
\begin{align}
\L\big( t,x, \alpha; t',x', \alpha' \big)
&= \E^{-(t-t')^2}\: \E^{-(x-x')^2} + \delta(t-t')\, \delta(x-x')\: (\sin \alpha- \sin \alpha')^2 \notag \\
&\quad\: 
+ g \big( t-t', x-x' \big)\: \sin \alpha\: \sin \alpha' \:,
\end{align}
where~$g$ is the convolution~$g=h*h$ with
\beq h(t,x) := \delta(t-1)\: \delta(x) + \delta(t+1)\: \delta(x) - \delta(t)\: \delta(x+1) - \delta(t)\: \delta(x-1) \eeq
(thus~$h$ is the kernel of a discretized wave operator).
We remark that this Lagrangian violates our usual positivity
assumption~$\L\big( t,x, \alpha; t',x', \alpha' \big) \geq 0$. However, this inequality could be arranged 
without changing the qualitative properties of the example by
mollifying the $\delta$ distributions and adding a constant.

\begin{Prp} \label{prpwave2}
Every minimizing measure~$\rho$ has the form
\beq \Diff\rho(t,x,\alpha) = \dd t\, \dd x\: \delta\big(\alpha - \phi(t,x) \big) \:\dd\alpha \:, \eeq
where~$\phi(t,x)$ solves the nonlinear discrete wave equation
\beq \sin \big( \phi(t+1,x) \big) + \sin \big( \phi(t-1,x) \big)
- \sin \big( \phi(t,x+1) \big) - \sin \big( \phi(t,x-1) \big) = 0 \:. \eeq
\end{Prp}
We begin with a preparatory lemma.
\begin{Lemma} Every minimizing measure has the form
\beq \label{prpwaverho}
\Diff\rho(t,x,\alpha) = \Diff\mu(t,x)\, \delta\big(\alpha - \phi(t,x) \big) \:\dd\alpha \:,
\eeq
with~$\mu$ the push-forward to the first two variables; that is,
\beq \mu = \pi_* \rho \qquad \text{with} \qquad \pi : \R^2 \times S^1 \rightarrow \R^2\:,\;\;\;
(t,x,\alpha) \mapsto (t,x) \:, \eeq
and~$\phi : \R^2 \rightarrow \R$ is a $\mu$-measurable function.
\end{Lemma}
\Proof Let~$\rho$ be a measure on~$\F$. We introduce the function~$\phi(t,x)$ by
\beq \sin \phi(t,x)\: \Diff\mu(t,x) = \int_0^{2 \pi} \sin \alpha\: \Diff\rho(t,x, \alpha) \:. \eeq
In words, $\sin \phi(t,x)$ coincides with the mean of~$\sin \alpha$ integrated over the circle.
The function~$\phi(t,x)$ exists because this mean lies in the interval~$[-1,1]$ and
because the sine takes all values in this interval.
Denoting the resulting measure of the form~\eqref{prpwaverho} by~$\tilde{\rho}$, we obtain
\beq \Sact(\rho) - \Sact(\tilde{\rho}) =
\int_\F \Diff\rho(t,x,\alpha) \int_\F \Diff\rho(t',x',\alpha)\: \delta(t-t')\: \delta(x-x')\:
\big( \sin \alpha - \phi(t,x) \big)^2 \:. \eeq
Therefore, $\rho$ is a minimizer if and only if~$\rho=\tilde{\rho}$.
\QED

\Proof[Proof of Proposition~\ref{prpwave2}.] For measures of the form~\eqref{prpwaverho},
the action takes the form
\begin{align}
\Sact &= \int_{\R^2} \Diff\mu(t,x) \int_{\R^2} \Diff\mu(t',x')\: \E^{-(t-t')^2}\: \E^{-(x-x')^2} \notag \\
&\quad\: + \int_{\R^2} \Diff\mu(t,x) \int_{\R^2} \Diff\mu(t',x')\: 
g\big(t-t',x-x' \big)\: \sin \phi(t,x)\: \sin \phi(t',x') \:.
\end{align}
Using that~$g$ is a convolution,
\beq g\big(t-t',x-x' \big) = \int_{\R^2} h(t-\tau, x-z)\: h(t'-\tau, x'-z)\: \dd\tau\: \dd z \:, \eeq
the action can be rewritten as
\begin{align}
\Sact &= \int_{\R^2} \Diff\mu(t,x) \int_{\R^2} \Diff\mu(t',x')\: \E^{-(t-t')^2}\: \E^{-(x-x')^2} \label{Lmink21} \\
&\quad\: + \int_{\R^2} \bigg( \int_{\R^2} 
h(t-\tau, x-z)\: \sin \phi(t,x) \: \Diff\mu(t,x) \bigg)^2\: \dd\tau\: \dd z \:. \label{Lmink22}
\end{align}
Exactly as shown in Section~\ref{secgaussian}, the minimizer of~\eqref{Lmink21} is given by
the Lebesgue measure. The contribution~\eqref{Lmink22}, on the other hand, is minimal
if~$\sin \phi(t,x)$ satisfies the discrete wave equation.
This concludes the proof.
\QED

\section{Exercises}
\begin{Exercise} {{(Functions with self-similar Fourier transform)}} \label{exselfsim} {\em{
\sindex{Fourier transform!self-similar}%
The example of Lemma~\ref{lemmagauss} was based on the fact that the Fourier transform
of a Gaussian is again Gaussian~\eqref{gaussselfsim}.
\bitem
\item[(a)] Prove~\eqref{gaussselfsim} by direct computation.
\item[(b)] Another example of a function that is self-similar under Fourier transforms is
the distribution in Minkowski space
\beq K_0(p) = \delta(k^2)\: \epsilon(k^0)\:. \eeq
Show that its Fourier transform indeed give, up to a constant, the same distribution back.
{\em{Hint:}} The distribution~$K_0(p)$ is the analog of the causal fundamental
solution~\eqref{kmdef} for the scalar wave equation (see also~\eqref{kmp}).
Using this fact, one can make use of the explicit form of the causal Green's operators for the scalar wave equation.
\item[(c)] Can you think of other functions that are self-similar under the Fourier transform
in the above sense? Is there a systematic way to characterize them all?
\eitem
}} \end{Exercise}

\begin{Exercise} {{(Non-negative functions with non-negative Fourier transforms)}} {\em{
\sindex{Fourier transform!of non-negative function}%
Another specific feature of the Gaussian in~\eqref{gauss}, which was used in
Lemma~\ref{lemmagauss}, is that it is a positive function whose Fourier transform is again
positive.
\bitem
\item[(a)] Show that the same is true for the $\delta$ distribution.
Can you come up with other functions with this property.
\item[(b)] The Lagrangian~\eqref{Lmink2} involves a function of two variables
with the properties that it is non-negative and has a non-negative Fourier transform.
How can this idea be used to construct other Lagrangians with the property that
the Lebesgue measure is a minimizer?
\eitem
}} \end{Exercise}

\chapter{Basics on the Continuum Limit} \label{seccl}
\sindex{continuum limit}%
In the {\em{continuum limit}}, one analyzes the EL equations of the causal action principle
for systems of Dirac seas in the presence of classical bosonic fields.
As worked out in detail in~\cite[Chapters~3-5]{cfs}, this limiting case yields the interactions of
the standard model and gravity on the level of second-quantized fermionic fields
interacting with classical bosonic fields.
In this chapter, we explain schematically how the analysis of the continuum limit works and
give an overview of the obtained results.

\section{Causal Fermion Systems in the Presence of External Potentials}
\sindex{causal fermion system!in presence of external potential}%
In Chapters~\ref{secFSO}--\ref{sechadamard}, it was explained how to construct and
analyze the unregularized kernel of the fermionic projector~$\tilde{P}(x,y)$
in Minkowski space in the presence of an external potential~$\B$.
The general question is whether the causal fermion system corresponding to
this kernel satisfies the EL equations corresponding to the causal action principle.
Thus we would like to evaluate the EL equations as stated abstractly in Theorem~\ref{thmEL}
for~$\tilde{P}(x,y)$.
The basic procedure is to form the closed chain (see~\eqref{Axydef})
and to compute its eigenvalues~$\lambda^{xy}_1, \ldots \lambda^{xy}_{2n} \in \C$.
This, in turn, makes it possible to compute the causal action and the constraints
(see~\eqref{Lagrange}--\eqref{Tdef}). Considering the first variations of~$P(x,y)$,
one then obtains the EL equations.

The main obstacle before one can carry out this program is that, in order to
obtain mathematically well-defined quantities, one needs to introduce an {\em{ultraviolet regularization}}.
As explained in detail in Chapter~\ref{secbrief}, this regularization is not merely a technical procedure,
but it corresponds to implementing a specific microscopic structure of spacetime.
In the vacuum, the regularization was introduced with the help of a regularization operator~${\mathfrak{R}}_\varepsilon$ (see~\eqref{Repsdef}).
Different choices of regularization operators correspond to different microscopic structures of spacetime.
Since the structure of our physical spacetime on the Planck scale is largely unknown,
the strategy is to allow for a general class of regularization operators, making it possible to analyze
later on how the results depend on the regularization
(for more details on this so-called {\em{method of variable regularization}} see~\cite[\S1.2.1]{cfs}).

In more detail, we proceed as follows. In the vacuum, we can follow the procedure
explained in Chapter~\ref{secbrief}, choosing~$\H$ as the subspace of all negative-frequency
solutions of the Dirac equation. In preparation for extending this construction to the interacting situation,
it is useful to note that the causal action principle can be formulated in terms of the kernel
of the fermionic projector given abstractly by~\eqref{Pxydefintro}. Therefore, our task is to compute this kernel.
It can be obtained alternatively by starting from unregularized kernel of the fermionic projector
constructed in Section~\ref{secPunreg} and introducing a regularization.
In the simplest case, working with a regularization that preserves the Dirac equation; that is,
\sindex{regularization operator}%
\nindex{adi@${\mathfrak{R}}_\varepsilon$ -- regularization operator on scale~$\varepsilon$}%
\beq {\mathfrak{R}}_\varepsilon \::\: \H_m \rightarrow \H_m \cap C^0(\scrM, S\scrM) \:, \eeq
the regularization can be introduced similar to~\eqref{Ppmdef2} by
\beq %\label{regPdef}
P^\varepsilon := -{\mathfrak{R}}_\varepsilon\, \pi_\H\, {\mathfrak{R}}_\varepsilon^*
\,k_m \::\: C^\infty_0(\scrM, S\scrM) \rightarrow \H_m \:. \eeq
For more general regularization operators that do {\em{not}} preserve the Dirac equation, one can
introduce the regularization by modifying the right-hand side of~\eqref{xPdef} to
\beq P^\varepsilon(x,y) := - \big( {\mathfrak{R}}_\varepsilon \Psi \big)(x) \,\big( {\mathfrak{R}}_\varepsilon \Psi \big)(y)^* \:, \eeq
where~$\Psi : \H_m \rightarrow L^2_{\text{loc}}(\scrM, S\scrM)$ is the unregularized wave evaluation operator,
and regularization operator~${\mathfrak{R}}_\varepsilon \::\: \H_m \rightarrow C^0(\scrM, S\scrM)$ now
maps more generally to continuous wave functions (not necessarily Dirac solutions).

The latter construction has the advantage that it also applies in the presence of
an external potential. In a perturbative treatment, it gives rise to the causal perturbation
expansion developed in Section~\ref{seccpertP}.
In this way, we obtain the {\em{regularized kernel}}~$\tilde{P}^\varepsilon(x,y)$
in the presence of an external potential.
Following the procedure explained in Chapter~\ref{secbrief}, we obtain a corresponding
causal fermion system. After suitable identifications (as worked out in~\cite[Section~1.2]{cfs}),
this regularized kernel coincides with the kernel of the fermionic projector as defined
abstractly in~\eqref{Pxydefintro}.

The subtle question is whether a chosen regularization of the vacuum also determines
the regularization of the kernel~$\tilde{P}^\varepsilon(x,y)$ in the presence of an external potential.
The general answer to this question is no, simply because the interaction introduces additional
freedoms for regularizing. Moreover, it is not clear a-priori whether the regularized objects
should still satisfy the Dirac equation. But at least, in~\cite[Appendix~F]{cfs} and~\cite[Appendix~D]{pfp}
a canonical procedure is given for {\em{regularizing the light-cone expansion}}
(see~\cite{reghadamard} for related constructions in curved spacetime).
It consists in taking the formulas of the (unregularized) light-cone expansion
(like, for example, \eqref{s:Pgag}--\eqref{s:jLi} in Example~\ref{exlight}) and replacing the singular factors~$T^{(n)}$
(like, for example, \eqref{Tidef}) by corresponding functions where the singularities on the light cone
have been regularized on the scale~$\varepsilon$. The precise procedure will be explained
in the next section.

\section{The Formalism of the Continuum Limit} \label{secclform}
\sindex{continuum limit!formalism of the}
We now give a brief summary of the formalism of the continuum limit. More details
can be found in~\cite[Section~2.4]{cfs}. The reader interested in the derivation of this formalism
is referred to~\cite[Chapter~4]{pfp}.

Having introduced the regularized kernel of the fermionic projector~$\tilde{P}^\varepsilon(x,y)$,
we can form the closed chain
\beq A^\varepsilon_{xy} := \tilde{P}^\varepsilon(x,y)\, \tilde{P}^\varepsilon(y,x) \:, \eeq
compute its eigenvalues and proceed by analyzing the EL equations.
In the continuum limit, one focuses on the limiting case~$\varepsilon \searrow 0$ when the
ultraviolet regularization is removed. This limiting case is comparatively easy to analyze.
This can be understood from the fact that, in the limit~$\varepsilon \searrow 0$,
the closed chain~$A^\varepsilon_{xy}$ becomes singular on the light cone. Therefore, asymptotically for small~$\varepsilon$, it suffices to take into account the contributions to~$A^\varepsilon_{xy}$ on the light cone.
These contributions, on the other hand, are captured precisely by the light-cone expansion
of the unregularized kernel~$\tilde{P}(x,y)$
(see Section~\ref{seclce} or the explicit formulas in Example~\ref{exlight}).
This is the basic reason why, in the continuum limit, the EL equations can be rewritten
as field equations involving fermionic wave functions as well as
derivatives of the bosonic potentials.

More specifically, the asymptotics~$\varepsilon \searrow 0$ is captured by the
{\em{formalism of the continuum limit}}, which we now outline
(for more details see~\cite[Section~2.4]{cfs} or the derivation of the formalism in~\cite[Chapter~4]{pfp}).
In the first step, one regularizes the light-cone expansion symbolically by
leaving all smooth contributions unchanged, whereas to the singular factors~$T^{(n)}$
we employ the replacement rule
\sindex{light-cone expansion!regularized|textbf}%
\beq %\label{contri}
m^p \,T^{(n)} \rightarrow m^p \,T^{(n)}_{[p]}\:. \eeq
Thus for the formulas of Example~\ref{exlight}, the factors~$T^{(n)}$ get an additional index~$[0]$.
If the light-cone expansion involves powers of the rest mass, these powers are taken into account
in the lower index. The resulting factors~$T^{(n)}_{[p]}$ are smooth functions, making all the
subsequent computations well-defined. The detailed form of these functions does not need to be
specified, because all we need are the following computation rules.
In computations, one may treat the factors~$T^{(n)}_{[p]}$
as complex functions. However, one must be careful when tensor indices of factors~$\slashed{\xi}$
are contracted with each other. Naively, this gives a factor~$\xi^2$, which vanishes on the
light cone and thus changes the singular behavior on the light cone. In order to describe this
effect correctly, we first write every summand of the light-cone expansion
such that it involves at most one factor~$\slashed{\xi}$ (this can always be arranged using
the anti-commutation relations of the Dirac matrices).
We now associate every factor~$\slashed{\xi}$ to the corresponding factor~$T^{(n)}_{[p]}$.
In short calculations, this can be indicated by putting brackets around the two factors,
whereas in the general situation, we add corresponding indices
to the factor~$\slashed{\xi}$, giving rise to the replacement rule
\beq %\label{xicontri}
m^p \,\slashed{\xi} T^{(n)} \rightarrow m^p \,\slashed{\xi}^{(n)}_{[p]}\, T^{(n)}_{[p]}\:. \eeq
\nindex{aie@$\xi^{(n)}_{[p]}$ -- ultraviolet regularized factor~$\xi$}%
For example, we write the regularized fermionic projector of the vacuum as
\beq P^\varepsilon = \frac{\cI}{2} \sum_{n=0}^\infty \frac{m^{2n}}{n!}\: \slashed{\xi}^{(-1+n)}_{[2n]}\, T^{(-1+n)}_{[2n]}
+ \sum_{n=0}^\infty \frac{m^{2n+1}}{n!}\: T^{(n)}_{[2n+1]} \:. \eeq
\nindex{aif@$P^\varepsilon(x,y)$ -- regularized kernel of fermionic projector}%
\sindex{fermionic projector!regularized kernel of}%

The kernel~$P(y,x)$ is obtained by taking the conjugate (see~\eqref{Pxysymm}).
The conjugates of the factors~$T^{(n)}_{[p]}$ and~$\xi^{(n)}_{[p]}$ are the complex conjugates,
\beq \overline{T^{(n)}_{[p]}} := \big(T^{(n)}_{[p]} \big)^* \qquad \text{and} \qquad
\overline{\xi^{(n)}_{[p]}} := \big(\xi^{(n)}_{[p]} \big)^* \:. \eeq
One must carefully distinguish between the factors with and without complex conjugation.
In particular, the factors~$\slashed{\xi}^{(n)}_{[p]}$ need not be symmetric; that is, in general,
\beq \big( \slashed{\xi}^{(n)}_{[p]} \big)^* \neq \slashed{\xi}^{(n)}_{[p]} \:. \eeq

When forming composite expressions, the tensor indices of the factors~$\xi$ are 
contracted to other tensor indices.
The factors~$\xi$ that are contracted to other factors~$\xi$ are called {\em{inner factors}}.
\sindex{inner factor}%
The contractions of the inner factors are handled with the so-called {\em{contraction rules}}
\sindex{contraction rule}%
\begin{align}
(\xi^{(n)}_{[p]})^j \, (\xi^{(n')}_{[p']})_j &=
\frac{1}{2} \left( z^{(n)}_{[p]} + z^{(n')}_{[p']} \right) \label{eq52} \\
(\xi^{(n)}_{[p]})^j \, \overline{(\xi^{(n')}_{[p']})_j} &=
\frac{1}{2} \left( z^{(n)}_{[p]} + \overline{z^{(n')}_{[p']}} \right) \label{eq53} \\
z^{(n)}_{[p]} \,T^{(n)}_{[p]} &= -4 \left( n \:T^{(n+1)}_{[p]}
+ T^{(n+2)}_{\{p \}} \right) , \label{eq54}
\end{align}
which are to be complemented by the complex conjugates of these equations.
Here the factors~$z^{(n)}_{[p]}$
\nindex{aig@$z^{(n)}_{[p]}$ -- abbreviation for $(\xi^{(n)}_{[p]})^2$}%
 can be regarded simply as a book-keeping device
to ensure the correct application of the rule~\eqref{eq54}.
The factors~$T^{(n)}_{\{p\}}$
\nindex{aih@$T_{[p]}^{(n)}$ -- ultraviolet regularized $T^{(n)}$ }%
have the same scaling behavior as the~$T^{(n)}_{[p]}$,
but their detailed form is somewhat different; we simply treat them as a new class of symbols.
In cases where the lower index does not need to be specified we write~$T^{(n)}_\circ$.
\nindex{aii@$T^{(n)}_\circ$ -- stands for~$T^{(n)}_{\{p\}}$ or~$T^{(n)}_{[p]}$}%
After applying the contraction rules, all inner factors~$\xi$ have disappeared.
The remaining so-called {\em{outer factors}}~$\xi$
\sindex{outer factor~$\xi$}%
need no special attention and are treated
like smooth functions.

Next, to any factor~$T^{(n)}_\circ$ we associate the {\em{degree}} $\deg T^{(n)}_\circ$ by
\nindex{aij@$\deg$ -- degree on light cone}%
\sindex{degree on the light cone}%
\beq \deg T^{(n)}_\circ = 1-n \:. \eeq
The degree is additive in products, whereas the degree of a quotient is defined as the
difference of the degrees of the numerator and denominator. The degree of an expression
can be thought of as describing the order of its singularity on the light cone, in the sense that a
larger degree corresponds to a stronger singularity (for example, the
contraction rule~\eqref{eq54} increments~$n$ and thus decrements the degree, in
agreement with the naive observation that the function~$z=\xi^2$ vanishes on the light cone).
Using the formal Taylor series, we can expand in the degree. In all our applications, this will
give rise to terms of the form
\beq \label{sfr}
\eta(x,y) \:
\frac{ T^{(a_1)}_\circ \cdots T^{(a_\alpha)}_\circ \:
\overline{T^{(b_1)}_\circ \cdots T^{(b_\beta)}_\circ} }
{ T^{(c_1)}_\circ \cdots T^{(c_\gamma)}_\circ \:
\overline{T^{(d_1)}_\circ \cdots T^{(d_\delta)}_\circ} } \qquad \text{with~$\eta(x,y)$ smooth}\:.
\eeq
The quotient of the two monomials in this equation is referred to as a {\em{simple fraction}}.
\sindex{simple fraction}%

A simple fraction can be given a quantitative meaning by considering one-dimensional integrals
along curves that cross the light cone transversely away from the origin~$\xi=0$.
This procedure is called {\em{weak evaluation on the light cone}}.
\sindex{evaluation on the light cone!weak}%
For our purpose, it suffices to integrate over the time coordinate~$t=\xi^0$ for fixed~$\vec{\xi} \neq 0$.
Moreover, using the symmetry under reflections~$\xi \rightarrow -\xi$, it suffices to consider the upper
light cone~$t \approx |\vec{\xi}|$. The resulting integrals diverge if the regularization
is removed. The leading contribution for small~$\varepsilon$ can be written as
\beq
\int_{|\vec{\xi}|-\varepsilon}^{|\vec{\xi}|+\varepsilon} \dd t \; \eta(t,\vec{\xi}) \:
\frac{ T^{(a_1)}_\circ \cdots T^{(a_\alpha)}_\circ \:
\overline{T^{(b_1)}_\circ \cdots T^{(b_\beta)}_\circ} }
{ T^{(c_1)}_\circ \cdots T^{(c_\gamma)}_\circ \:
\overline{T^{(d_1)}_\circ \cdots T^{(d_\delta)}_\circ} }
\;\approx\; \eta(|\vec{\xi}|,\vec{\xi}) \:\frac{c_\reg}{(\cI |\vec{\xi}|)^L}
\;\frac{\log^r (\varepsilon |\vec{\xi}|)}{\varepsilon^{L-1}}\:, \label{asy}
\eeq
where~$L$ is the degree of the simple fraction and~$c_\reg$, the so-called {\em{regularization parameter}},
\nindex{aik@$L$ -- degree of simple fraction}%
\sindex{regularization parameter}%
\nindex{ail@$c_{\text{reg}}$ -- regularization parameter}%
is a real-valued function of the spatial direction~$\vec{\xi}/|\vec{\xi}|$, which also depends on
the simple fraction and on the regularization details
(the error of the approximation will be specified later). The integer~$r$ describes
a possible logarithmic divergence. Apart from this logarithmic divergence, the
scalings in~\eqref{asy} in both~$\xi$ and~$\varepsilon$ are described by the degree.

When analyzing a sum of expressions of the form~\eqref{sfr}, one must
know if the corresponding regularization parameters are related to each other.
In this respect, the {\em{integration-by-parts rules}}
\sindex{integration-by-parts rule}%
are important, which are described
symbolically as follows. On the factors~$T^{(n)}_\circ$ we introduce a derivation~$\nabla$ by
\beq \nabla T^{(n)}_\circ = T^{(n-1)}_\circ \:. \eeq
\nindex{aim@$\nabla$ -- derivation on the light cone}%
Extending this derivation with the product and quotient rules to simple fractions, the
integration-by-parts rules state that
\beq \label{ipart}
\nabla \left( \frac{ T^{(a_1)}_\circ \cdots T^{(a_\alpha)}_\circ \:
\overline{T^{(b_1)}_\circ \cdots T^{(b_\beta)}_\circ} }
{ T^{(c_1)}_\circ \cdots T^{(c_\gamma)}_\circ \:
\overline{T^{(d_1)}_\circ \cdots T^{(d_\delta)}_\circ} }
\right) = 0 \:.
\eeq
Carrying out the derivative with the product rule, one gets relations between simple fractions.
Simple fractions that are not related
to each other by the integration-by-parts rules are called {\em{basic fractions}}. As shown
in~\cite[Appendix~E]{pfp}, there are no further relations between the basic fractions.
Thus the corresponding {\em{basic regularization parameters}} are independent.
\sindex{regularization parameter!basic}%

The above symbolic computation rules give a convenient procedure to evaluate composite expressions
in the fermionic projector, referred to as the {\em{analysis in the continuum limit}}:
\sindex{continuum limit!analysis in the}%
After applying the contraction rules and expanding in the degree, 
the EL equations can be rewritten as equations involving
a finite number of terms of the form~\eqref{sfr}. By applying the integration-by-parts rules,
we can arrange that all simple fractions are basic fractions.
We evaluate weakly on the light cone~\eqref{asy} and collect the terms according to their
scaling in~$\xi$. Taking for every given scaling in~$\xi$ only the leading pole in~$\varepsilon$,
we obtain equations that involve linear combinations of smooth functions and basic regularization parameters.
We consider the basic regularization parameters as empirical parameters describing the
unknown microscopic structure of spacetime.
\sindex{regularization parameter!basic}%
We thus end up with equations involving
smooth functions and a finite number of free parameters.

We finally specify the error of the above expansions. By not regularizing the bosonic potentials
and fermionic wave functions, we clearly disregard the
\beq \label{ap1}
\text{higher orders in~$\varepsilon/\ell_\text{macro}$}\:.
\eeq
\sindex{error term!higher order in $\varepsilon/\ell_\text{macro}$}%
\nindex{ain@$\ell_\text{macro}$ -- length scale of macroscopic physics}%
Furthermore, in~\eqref{asy} we must stay away from the origin, meaning that we neglect the
\beq \label{ap2}
\text{higher orders in~$\varepsilon/|\vec{\xi}|$}\:.
\eeq
\sindex{error term!higher order in $\varepsilon / \vert \vec{\xi} \vert$}%
The higher order corrections in~$\varepsilon/|\vec{\xi}|$ depend on the fine structure of
the regularization and thus seem unknown for principal reasons. Neglecting the terms in~\eqref{ap1}
and~\eqref{ap2} also justifies the formal Taylor expansion in the degree.
Clearly, leaving out the terms~\eqref{ap2} is justified only if~$|\vec{\xi}| \gg \varepsilon$.
Therefore, whenever using the above formalism, we must always ensure that~$|\vec{\xi}|$
is much larger than~$\varepsilon$.

We finally remark that, when working out the Einstein equations, one must go beyond
error terms of the form~\eqref{ap1} and~\eqref{ap2}. The reason is that the gravitational
scales like~$\kappa \sim \delta^2 \approx \varepsilon^2$. In order not to lose the relevant
terms in the error terms, one must take certain higher order contributions into account.
This is done by using the so-called $\iota$-formalism.
Here, we do not enter the details but merely refer the interested reader to~\cite[\S4.2.7]{cfs}.

\section{Overview of Results of the Continuum Limit Analysis}
The formalism of the continuum limit makes it possible to evaluate the EL equations of the
causal action for the regularized kernel~$\tilde{P}^\varepsilon(x,y)$ in the presence of 
an external potential~$\B$. In order to avoid confusion, we point out that, a-priori, the external
potential can be chosen arbitrarily; in particular, it does need to satisfy any field equations.
We find that the EL equations of the causal action are satisfied in the continuum limit if and only
if the potential~$\B$ has a specific structure and satisfies dynamical equations.
Restricting attention to potentials of this form and complementing the Dirac equation~\eqref{Dout}
by the dynamical equations for~$\B$, the potentials are no longer given as external potentials,
but instead one gets a coupled system of equations describing a mutual interaction
of the Dirac wave functions with classical bosonic fields. The dynamical equations for~$\B$
are referred to as the {\em{classical field equations}}. 
\sindex{classical field equation!derivation in the continuum limit}%
In this way, the classical field equations are {\em{derived}} from the causal action principle.

We now outline the main results of the continuum limit analysis as obtained in~\cite[Chapters~3-5]{cfs}.
The main input is to specify the regularized kernel~$P^\varepsilon(x,y))$ of the {\em{vacuum}}.
This involves:
\bitem
\itemD The fermion configuration in the vacuum, including the masses of the leptons and quarks.
Moreover, it is built in that the neutrinos break the chiral symmetry.
\itemD The vacuum kernel should satisfy the EL equations. This poses a few constraints on
the regularization operator.
\eitem
The output of the continuum limit is the following results:
\bitem
\itemD The structure of the interaction on the level of classical gauge theory.
\itemD The gauge groups and their coupling to the fermions.
\itemD The equations of linearized gravity.
\eitem

In~\cite{cfs} the continuum limit is worked out in three steps for systems of increasing complexity.
In Chapter~3, a system formed of a sum of three Dirac seas is considered.
This configuration, referred to as a {\em{sector}}, can be thought of as a simplified model
describing the three generations of charged leptons~$(e,\mu, \tau)$.
In the continuum limit, we obtain the following results for the interaction as described by the
causal action principle:
\bitem
\itemD The fermions interact via an {\em{axial gauge field}}.
\itemD This axial gauge field is massive, with the mass determined by the masses of the fermions
and the regularization.
\itemD We find that the field equations for the axial gauge field arise in the continuum limit
only if the number of generations equals three. For one or two generations, the resulting
equations are overdetermined, whereas for more than three generations, the equations are
under-determined (which means in particular that there is no well-posed Cauchy problem).
\itemD We obtain nonlocal corrections to the classical field equations described by integral
kernels that decay on the Compton scale. It seems that these nonlocal corrections capture
certain features of the underlying quantum field theory. But the detailed connection has not been
worked out.
\itemD There is no gravitational field and no Higgs field.
\eitem

In~Chapter~4, a system formed as a direct sum of two sectors is considered.
This system is referred to as a {\em{block}}. The first sector looks as in Chapter~3.
In the second sector, however, the chiral symmetry is broken. This system can be
regarded as a model for the leptons, including the three generations of neutrinos.
In the continuum limit, we obtain the following results for the interaction as described by the
causal action principle:
\bitem
\itemD The fermions interact via an~$\SU(2)$ gauge field, which couples only to one chirality
(say, the left-handed fermions).
\itemD The corresponding gauge field is again massive.
\itemD Moreover, the fermions interact linearly via the linearized Einstein equations,
where the coupling constant is related to the regularization length.
\eitem

Finally, in~\cite[Chapter~5]{cfs} a realistic system involving leptons and quarks is considered.
To this end, one considers a direct sum of eight sectors, one of which with broken chiral symmetry
(the neutrino sector).
These eight sectors form pairs, referred to as blocks. The block containing the neutrino sector
describes the leptons, whereas the other three blocks describe the quarks.
Moreover, we obtain the following results:
\bitem
\itemD The fermions interact via the gauge group~$\U(1) \times \SU(2)_L \times \SU(3)$.
\sindex{gauge fields!of the standard model}%
The corresponding gauge fields couple to the fermions as in the standard model.
The $\SU(2)$-field couples only to the left-handed component and is massive.
The other gauge fields are massless.
\itemD Moreover, the fermions interact linearly via the linearized Einstein equations.
\itemD The EL equations corresponding to the causal action principle coincide with those
of the standard model after spontaneous symmetry breaking, plus linearized gravity.
\itemD There are scalar degrees of freedom that can be identified with the Higgs potential.
However, the corresponding dynamical equations have not yet been worked out.
\itemD Again, the fermions interact linearly via the linearized Einstein equations,
where the coupling constant is related to the regularization length.
Taking into account that the causal action principle is diffeomorphism invariant,
we obtain the Einstein equations,
up to possible higher order corrections in curvature (which scale in powers of~$(\delta^2\:
\text{Riem})$, where~$\delta$ is the Planck length and~$\text{Riem}$ is the
curvature tensor). Thus, including error terms, the derived Einstein equations
take the form (see~\cite[Theorems~4.9.3 and~5.4.4]{cfs})
\sindex{Einstein equations}%
\beq %\label{einstein}
R_{jk} - \frac{1}{2}\:R\: g_{jk} + \Lambda\, g_{jk} = G\, T_{jk} 
+ {\mathscr{O}} \big( \delta^4 \,\text{Riem}^2 \big) \eeq
(where~$T_{jk}$ is the energy-momentum tensor and~$G$ is the gravitational coupling constant).
\eitem

We conclude this section by discussing a few aspects of the derivation of these results.
We begin with the system of one sector as considered in~\cite[Chapter~3]{cfs}.
In this case, the kernel of the fermionic projector is the sum of~$g \in \N$ Dirac seas
of masses~$m_1, \ldots, m_g$; that is,
\beq \label{s:A}
P(x,y) = \sum_{\beta=1}^g P_{m_\alpha}(x,y) \:,
\eeq
where again
\beq P_m(x,y) = \int \frac{\dd^4k}{(2 \pi)^4}\: (\slashed{k} +m)\:
\delta \big( k^2-m^2 \big)\: \Theta(-k^0)\: \E^{-\cI k(x-y)}\:. \eeq
In order to perturb the system by gauge potentials, we first introduce the kernel of the
{\em{auxiliary fermionic projector}}~$P^\text{aux}(x,y)$, which is obtained from~$P(x,y)$
by replacing the sums with direct sums,
\beq P^\text{aux}(x,y) = \bigoplus_{\beta=1}^g P_{m_\alpha}(x,y) \eeq
(this means that~$P^\text{aux}(x,y)$ is represented by a $(4g \times 4g)$-matrix).
The auxiliary kernel satisfies the Dirac equation
\beq \Big( \cI \Pdd_x - \begin{pmatrix} m_1 & 0 & 0 \\ 0 & \ddots & 0 \\ 0 & 0 & m_g \end{pmatrix} \Big)
P^\text{aux}(x,y) = 0 \:. \eeq
Therefore, it can be perturbed as usual by inserting a potential~$\B$ into the Dirac equation
\beq %\label{Binsert}
\Big( i \Pdd_x + \B(x) - \begin{pmatrix} m_1 & 0 & 0 \\ 0 & \ddots & 0 \\ 0 & 0 & m_g \end{pmatrix} \Big)
\tilde{P}^\text{aux}(x,y) = 0 \eeq
(where~$\B(x)$ is a matrix potential acting on~$\C^{4g}$). The perturbed kernel~$\tilde{P}^\text{aux}$
can be computed with the methods explained in Chapters~\ref{secperturb} and~\ref{sechadamard}.
Finally, we obtain the perturbed kernel of the fermionic projector by summing over the generation
indices in an operation referred to as the {\em{sectorial projection}},
\beq \tilde{P}(x,y) := \sum_{\alpha, \beta=1}^g \tilde{P}^\alpha_\beta(x,y) \:. \eeq
After introducing an ultraviolet regularization, this kernel can be analyzed in the EL equations of the
causal action principle, exactly as outlined in Section~\ref{secclform} above.

In order to gain the largest possible freedom in perturbing the system, the operator~$\B$
should be chosen as general as possible. For this reason, in~\cite[Chapter~4]{cfs} a general
class of potential was considered, including nonlocal potentials (that is, integral operators).
A general conclusion of the analysis is that, in order to satisfy the EL equations, the
potential~$\B$ must be local. More precisely, it must be a differential operator or a multiplication operator
by a potential  which may involve left- and right-handed potentials, but also bilinear, scalar or pseudo-scalar potentials,
\beq \label{Bchiral}
\B(x) = \chi_L \slashed{A}_R(x) + \chi_R \slashed{A}_L(x) +  \sigma^{ij}\, \Lambda_{ij}(x) + \Phi(x) +
\cI \Gamma \, \Xi(x)
\eeq
(where each of the potentials is a $g \times g$-matrix acting on the generations, and~$\Gamma$
is the pseudo-scalar matrix, which in physics textbooks is often denoted by~$\gamma^5$).
Analyzing the continuum limit for such multiplication operators, one gets the above-mentioned results.

One feature that at first sight might be surprising is that, despite local gauge symmetry,
we get {\em{massive gauge fields}}. 
\sindex{gauge field!massive}%
In order to understand how this comes about,
we need to consider local gauge symmetries in connection with the chiral gauge potentials in~\eqref{Bchiral}.
On the fundamental level of the causal fermion system, {\em{local gauge transformations}}
\sindex{local gauge transformation}%
arise from the freedom in choosing bases of the spin spaces
(see~\eqref{lgf1} and~\eqref{lgf2} in Section~\ref{secphysicalconcepts}).
In the present setting with four-component Dirac spinors, the local gauge transformations take the form
\beq \label{U22gauge}
\psi(x) \rightarrow U(x)\, \psi(x) \qquad \text{with} \qquad U(x) \in \U(2,2) \:,
\eeq
where~$\U(2,2)$ is the group of unitary transformations of the spinors at the spacetime point~$x$.
The causal action principle is gauge invariant in the sense that the causal action
is invariant under such gauge transformations.
The group~$\U(2,2)$ can be used to describe gravity as a gauge theory
(for details, see Section~\ref{secdircurv} or~\cite{U22}). Restricting attention to
flat spacetime, the main interest is that~$\U(2,2)$ contains the gauge group~$\U(1)$ of electrodynamics
as a subgroup. In other words, the causal action principle is gauge invariant under local phase transformations
\beq \psi(x) \rightarrow \E^{-\cI \Lambda(x)}\, \psi(x) \eeq
with a real-valued function~$\Lambda$.

The chiral potentials in~\eqref{Bchiral} also give rise to generalized phase transformations.
This can be seen, for example, by working out the leading term to the light-cone expansion
(similar to~\eqref{s:Pgag} for the electromagnetic potential). One finds that the
chiral gauge potentials lead to phase transformations of the left- and right-handed components
of the wave functions; that is,
\beq \label{nogauge}
\psi(x) \rightarrow U(x)\, \psi(x) \qquad \text{with} \qquad
U(x) := \chi_L \,\E^{-\cI \Lambda_L(x)} + \chi_R \,\E^{-\cI \Lambda_R(x)}
\eeq
\sindex{local gauge transformation!chiral}%
(again with real-valued functions~$\Lambda_L$ and~$\Lambda_R$). The point is that
this transformation is {\em{not unitary}} with respect to the spin inner product, because the
chirality flips when taking the adjoint
\beq U^* = \chi_R \,\E^{\cI \Lambda_L(x)} + \chi_L \,\E^{-\cI \Lambda_R(x)}
\quad \text{but} \quad
U^{-1} = \chi_L \,\E^{\cI \Lambda_L(x)} + \chi_R \,\E^{-\cI \Lambda_R(x)} \eeq
(note that~$\chi_L^*= (1-\Gamma)^*/2 = (1+\Gamma)/2
=\chi_R$ because~$\Gamma^* = -\Gamma$).
\Evtl{\"Ubungsaufgabe zu~$(\Gamma)^*$.}%
Therefore, as soon as~$A_L \neq A_R$,
the generalized phase transformation~$U(x)$ in~\eqref{nogauge} is {\em{not}}
a local transformation of the form~\eqref{U22gauge}.
Consequently, the local transformation in~\eqref{nogauge} does {\em{not}} correspond to
a symmetry of the causal action principle. Therefore, it is not a contradiction if these
gauge potentials arise in the effective field equations as mass terms.

More specifically, the relative phases between left- and right-handed potentials do come up in the
closed chain~$A_{xy} = P(x,y) P(y,x)$, as one sees immediately from the fact that, if~$P(x,y)$
is vectorial, then the chirality flips at the corresponding factor; that is,
\beq \label{chiralphase}
\begin{split}
\chi_L A_{xy} &= \chi_L P(x,y)\: \chi_R P(y,x) \,\\
&\rightarrow\, \exp \Big( -\cI \big( \Lambda_L(x) - \Lambda_R(x) \big) \Big)
\exp \Big( \cI \big( \Lambda_L(y) - \Lambda_R(y) \big) \Big)\: A_{xy} \:.
\end{split}
\eeq
Working out the corresponding contribution to the EL equations in the continuum limit, one
finds that the axial current and a corresponding axial mass term come up in the effective field equations.
The coupling constant and the bosonic mass depend on the detailed form of the regularization.
But they can be computed for specific choices of the regularization, as is exemplified
in~\cite[Chapter~3]{cfs} for a hard cutoff in momentum space and the $i \varepsilon$-regularization.

We now move on to the system of two sectors as analyzed in~\cite[Chapter~4]{cfs}.
The vacuum is described by a kernel of the fermionic projector~$P(x,y)$
being a direct sum of two summands, each of which is of the form~\eqref{s:A},
where we choose the number of generations as~$g=3$.
Hence, $P(x,y)$ is a $8 \times 8$-matrix. Replacing the sums by direct sums, one obtains
the corresponding auxiliary kernel~$P^\text{aux}(x,y)$ (being represented by a $24 \times 24$-matrix).
In order to account for the observational fact that neutrinos are left-handed particles,
one must {\em{break the chiral symmetry}} of one of the sectors (the {\em{neutrino sector}}).
\sindex{sector!charged}%
\sindex{sector!neutrino}%
\sindex{chiral symmetry breaking}%
To this end, we assume that the regularization of the neutrino sectors is different from that of the other sector
(the {\em{charged sector}}) by contributions which are not left-right invariant.
The relevant length scale is denoted by~$\delta \gtrsim \varepsilon$.
This procedure is very general and seems the right thing to do, because the regularization
effects on the scale~$\delta$ are also needed in order to obtain the correct form of the
curvature term in the Einstein equations. In fact, the obtained linearized Einstein equations involve
the coupling constant~$G \sim \delta^2$. As briefly mentioned at the end of Section~\ref{secl},
the derivation of the Einstein equations uses the $\iota$-formalism, which goes beyond the standard
formalism of the continuum limit.

The system analyzed in~\cite[Chapter~5]{cfs} is obtained similarly by adding direct summands
to~$P(x,y)$ describing the three generations of quarks.
We begin with eight sectors. These eight sectors form pairs, giving rise to four blocks.
We conclude by outlining how this mechanism of {\em{spontaneous block formation}} comes about.
\sindex{spontaneous block formation}%
For this purpose, we return to the gauge phases as already mentioned in~\eqref{U22gauge}
and~\eqref{nogauge}. We already saw in~\eqref{chiralphase} that, if the kernel of the fermionic projector
is vectorial, then the relative phases (i.e.\ the difference of left- and right-handed gauge phases) show up in the
eigenvalues of the closed chain. Such phase factors drop out of the causal Lagrangian because
of the absolute values in~\eqref{Lagrange}. However, the situation becomes more involved
if the kernel of the fermionic projector is {\em{not}} vectorial. Indeed, expanding 
the vacuum kernel in powers of the rest
mass, the zero order contribution to~$P(x,y)$ is vectorial, whereas the first order contribution
is scalar (more generally, one sees from~\eqref{hadamard1} that the even orders in the mass
are vectorial, whereas the odd orders are scalar).
As a consequence, the absolute values of the eigenvalues~$|\lambda^{xy}_i|$ depend
in a rather complicated way on the chiral gauge phases.
Moreover, considering a direct sum of Dirac seas, one must keep into account that
the gauge phases in the formulas for $P(x,y)$ (and similarly in composite expressions) must be replaced by generalized phases,
which can be described in terms of ordered exponentials of the gauge potentials.
Evaluating the causal Lagrangian~\eqref{Lagrange}, one gets conditions
for the chiral gauge phases. In simple terms, these conditions can be stated by demanding
that matrices formed of ordered exponentials of the gauge potentials must have degeneracies.
Qualitatively speaking these degeneracies mean that the left-handed gauge potential must be the same
in each block, and this condition even makes it possible to explain why such blocks form.
A more detailed and more precise explanation can be found in~\cite[Chapter~5]{cfs}.

\section{Exercises}

\begin{Exercise} \label{exreghadamard} {\em{ This exercise 
explains in a simple example how the {\em{regularization of the Hadamard expansion}} works.
\sindex{light-cone expansion!regularized}%
\bitem
\item[(a)] Consider the singular term of the first summand of the Hadamard expansion~\eqref{hadamard1}
in Minkowski space,
\beq \label{poledis}
\lim_{\nu \searrow 0} \frac{1}{\xi^2 - \cI \nu \,\xi^0}
\eeq
(where again~$\xi := y-x$). A simple method to remove the pole is not to take the limit~$\nu \searrow 0$,
but instead to set~$\nu = 2 \varepsilon$,
\beq \label{ieps}
\frac{1}{\xi^2 - 2 \cI \varepsilon \,\xi^0} \:.
\eeq
Show that this regularization  can be realized by the replacement
\beq \xi^0 \rightarrow \xi^0 - \cI \varepsilon \:, \eeq
up to a multiplicative error of the order
\beq \label{hadamarderror}
\bigg(1 + \O\Big( \frac{\varepsilon^2}{\xi^2} \Big) \bigg) \:.
\eeq
The basic concept behind the regularized Hadamard expansion is to regularize all singular
terms in this way, leaving all smooth functions unchanged. This gives a consistent
formalism is one works throughout with error terms of the form~\eqref{hadamarderror}.
{\em{Hint:}} This is the so-called $\cI \varepsilon$-regularization introduced in~\cite[Section~2.4]{cfs}.
For details in curved spacetime, see~\cite{reghadamard}.
\item[(b)] Show that for kernels written as Fourier transforms
\beq K(x,y) = \int_M \frac{\dd^4p}{(2 \pi)^4}\: \hat{K}(p)\: \E^{-\cI p(y-x)} \eeq
(with~$\hat{K}$ supported in say the lower half plane~$\{p^0 < 0\}$),
the replacement rule~\eqref{ieps} amounts to inserting a convergence-generating factor~$\E^{\varepsilon p^0}$
into the integrand.
\eitem
}} \end{Exercise}

\begin{Exercise} \label{ex2.9} 
\sindex{weak evaluation on the light cone}%
{\em{The goal of this exercise is to
explore {\em{weak evaluation on the light cone}} in a simple example.
\bitem
\item[(a)] Show that, setting~$t=\xi^0$ and choosing polar coordinates with~$r=|\vec{\xi}|$,
regularizing the pole in~\eqref{poledis} according to~\eqref{ieps} gives the function
\beq \frac{1}{(t- \cI \varepsilon)^2 - r^2} \:. \eeq
\item[(b)] As a simple example of a composite expression, we take the absolute square
of the regularized function
\beq \label{exweakint}
\frac{1}{\big| (t- \cI \varepsilon)^2 - r^2 \big|^2} \:.
\eeq
Show that this expression is ill-defined in the limit~$\varepsilon \searrow 0$
even as a distribution.
\item[(c)] Use the identity
\beq \frac{1}{(t-\cI \varepsilon)^2 - r^2} =
\frac{1}{(t-\cI \varepsilon-r)(t-i \varepsilon+r)} = \frac{1}{2r} \left(
\frac{1}{t-\cI \varepsilon-r} - \frac{1}{t-\cI \varepsilon+r} \right) \eeq
to rewrite the integrand in~\eqref{exweakint} in the form
\beq \sum_{p,q=0}^1 \frac{\eta_{p,q}(t,r, \varepsilon)}{(t-\cI \varepsilon-r)^p\, (t+\cI \varepsilon-r)^q} \:, \eeq
with functions~$\eta_{p,q}(t,r, \varepsilon)$, which in the limit~$\varepsilon \searrow 0$
converge to smooth functions. Compute the functions~$\eta_{p,q}$.
\item[(d)] We now compute the leading contributions and specify what we mean by ``leading.''
First compute the following integrals with residues:
\begin{align}
I_0(\varepsilon) &:= \int_{-\infty}^\infty \frac{1}{(t-\cI \varepsilon-r)\, (t+\cI \varepsilon-r)}\: \dd t \:.
\end{align}
Show that
\beq \int_{-\infty}^\infty \frac{\eta_{1,1}(t,r)}{(t-\cI \varepsilon-r)^2\, (t+\cI \varepsilon-r)^2}\: \dd t \\
= I_0(\varepsilon)\: \eta_{2,2}(r,r) +\O(\varepsilon) \:. \eeq
Explain in which sense this formula is a special case of the weak evaluation formula~\eqref{asy}.
\eitem
}} \end{Exercise}

\chapter{Connection to Quantum Field Theory} \label{secQFT}
\sindex{quantum field theory}%
In this chapter, we give an outlook on how to get a connection between the causal action principle
and the dynamics of quantum fields. Since this direction of research is very recent and partly work in progress,
we do not enter any details but instead try to explain a few basic concepts and ideas.
Our presentation is based on the recent research papers~\cite{fockbosonic, fockfermionic, fockentangle, fockdynamics}. Partly, our methods were already explored in the alternative approach in~\cite{qft}, which is
more closely tied to the analysis of the continuum limit (as outlined in Chapter~\ref{seccl}).

\section{Convex Combinations of Measures and Dephasing Effects}
Before beginning, we point out that in most examples of causal fermion systems considered
in this book, the measure~$\rho$ was the push-forward of the volume measure
on Minkowski space or a Lorentzian manifold. Thus we first constructed a local correlation map
(see~\eqref{FeMink})
\beq F^\varepsilon \::\: \scrM \rightarrow \F \:, \eeq
and then introduced the measure~$\rho$ on~$\F$ by (see~\eqref{rhoMink})
\beq \label{rhopush}
\rho = (F^\varepsilon)_* \mu_{\scrM} \:,
\eeq
where~$\mu_\scrM$ is the four-dimensional volume measure on~$\scrM$.
In all these examples, the measure~$\rho$ had the special property that it was supported
on a smooth four-dimensional subset of~$\F$ given by (for details, see Exercise~\ref{exsupp1})
\beq M := \supp \rho = \overline{F^\varepsilon(\scrM)} \:. \eeq
Also, when varying the measure in the derivation of the linearized field equations
or in the study of interacting systems in the continuum limit, we always
restricted attention to measures having this property
(see~\eqref{srt} in Section~\ref{seclinfield} or Chapter~\ref{seccl}).
While this procedure seems a good starting point for the analysis of the causal action principle
and gives good approximate solutions of the EL equations, we
cannot expect that true minimizers are of this particular form.

With this in mind, our strategy is to allow for more general measures on~$\F$
and to analyze the causal action principle for these general measures.
As we will see, this analysis gives rise to close connections to quantum field theory.
We proceed step by step and begin by explaining a construction that explains why
going beyond push-forward measures of the form~\eqref{rhopush} makes it possible
to further decrease the causal action. In other words, the following argument
shows that minimizers of the causal action will {\em{not}} have the form of a push-forward measures~\eqref{rhopush}, but will have a more complicated structure.
This argument is given in more detail in~\cite[\S1.5.3]{cfs}.
Assume that we are given~$L$ measures~$\rho_1, \ldots, \rho_L$ on~$\F$.
Then their {\em{convex combination}}~$\tilde{\rho}$ given by
\sindex{measure!convex combination}%
\beq \label{convexcombi}
\tilde{\rho} := \frac{1}{L} \sum_{\as=1}^L \rho_\as
\eeq
is again a positive measure on~$\F$. Moreover, if the~$\rho_\as$ satisfy the
linear constraints (that is, the volume constraint~\eqref{volconstraint}
and the trace constraint~\eqref{trconstraint}), then these constraints
are again respected by~$\tilde{\rho}$.

Next, we let~$\rho$ be a minimizing measure
(describing, for example, the vacuum).
Choosing unitary transformations~$U_1, \ldots, U_L$, we introduce the measures~$\rho_\as$
in~\eqref{convexcombi} as
\beq \rho_\as(\Omega) := \rho \big(U^{-1} \Omega U \big) \:. \eeq
Thus, in words, the measures~$\rho_\as$ are obtained from~$\rho$ by taking the unitary
transformations by~$U_\as$. Since the causal action and the constraints are unitarily invariant,
each of the measures~$\rho_\as$ is again a minimizer.
Let us compute the action of the convex combination~\eqref{convexcombi}.
First, by~\eqref{Sdef},
\beq \Sact(\tilde{\rho}) = \frac{1}{L^2} \sum_{\as, \bs=1}^L \iint_{\F \times \F} \L(x,y)\: \Diff\rho_\as(x)\:
\Diff\rho_\bs(y) \:. \eeq
If~$\as=\bs$, we obtain the action of the measure~$\rho_\as$, which, due to unitary invariance,
is equal to the action of~$\rho$. We thus obtain
\beq \label{Smix}
\Sact(\tilde{\rho}) = \frac{\Sact(\rho)}{L} + \frac{1}{L^2} \sum_{\as \neq \bs} \iint_{\F \times \F} \L(x,y)\: \Diff\rho_\as(x)\:
\Diff\rho_\bs(y) \:.
\eeq

Let us consider the contributions for~$\as \neq \bs$ in more detail. In order to simplify the explanations,
it is convenient to assume that the measures~$\rho_\as$ have mutually disjoint supports
(this can typically be arranged by a suitable choice of the unitary transformations~$U_\as$).
Then the spacetime~$\tilde{M} := \supp \tilde{\rho}$ can be decomposed into~$L$ ``sub-spacetimes''
$M_\as := \supp \rho_\as$,
\beq \tilde{M} = M_1 \cup \cdots \cup M_L \qquad \text{and} \qquad M_\as \cap M_\bs = \varnothing \quad
\text{if~$\as \neq \bs$}\:. \eeq
The Lagrangian of the last summand in~\eqref{Smix} is computed from the fermionic projector~$P_{\as,\bs}(x,y)$,
where~$x \in M_\as$ and~$y \in M_\bs$ are in different sub-spacetimes.
Similar to~\eqref{Prep}, it can be expressed in terms of the physical wave functions by
(for details, see~\cite[Lemma~1.5.2]{cfs})
\beq \label{Prepmix}
P_{\as, \bs}(x,y) = -\sum_{i,j} |\psi^{e_i}(x) \Sr\: (U_\as \,U_\bs^*)^i_j \: \Sl \psi^{e_j}(y)|  \:.
\eeq
The point is that this fermionic projector involves the operator product~$U_\as U_\bs^*$.
By choosing the unitary operators~$U_\as$ and~$U_\bs$ suitably, one can arrange that this
operator product involves many phase factors. Moreover, one can arrange that,
when carrying out the sums in~\eqref{Prepmix}, these phases cancel each other due to
destructive interference. In this way, the kernel~$P(x,y)$ can be made small if~$x$ and~$y$
lie in different sub-spacetimes. As a consequence, the last summand in~\eqref{Smix} can be
arranged to be very small. Taking into account the factor~$1/N$ in the
first summand in~\eqref{Smix}, also the causal action of~$\tilde{\rho}$ becomes small.
Clearly, this argument applies only if the number~$L$ of sub-spacetimes is not too large,
because otherwise it becomes more and more difficult to arrange destructive interference
for all summands of the sum in~\eqref{Smix} (estimating the optimal number~$L$ of
subsystems is a difficult problem, which we do not enter here).
Also, we cannot expect that the simple
ansatz~\eqref{convexcombi} will already give a minimizer.
But at least, our argument explains why it is too naive to think of a minimizing measure
as being the push-forward measure of a volume measure under a smooth local correlation map.
Instead, a minimizing measure could be composed of a large number of sub-spacetimes. 

In the above consideration, it is crucial that the kernels~$P_{\as, \bs}(x,y)$ for~$\as \neq \bs$ are very small 
due to decoherence effects. It is a subtle point how small these kernels are.
If they are so small that we may assume that they vanish, then this means that the
sub-spacetimes do not interact with each other.
Therefore, one can take the point of view that, in order to describe all physical phenomena,
it suffices to restrict attention to one sub-spacetime. The appearance of many sub-spacetimes
that are completely decoherent is
an intriguing mathematical effect, which may have interesting philosophical implications,
but it is of no relevance as far as physical predictions are concerned.
For this reason, here we shall not discuss these decoherent sub-spacetimes further.
Also, we leave the question open whether they really occur for minimizing measures.
Instead, we take the point of view that, in case our minimizing measure consists of
several decoherent sub-spacetimes, we restrict it to one sub-spacetime
and denote the resulting measure by~$\rho$.

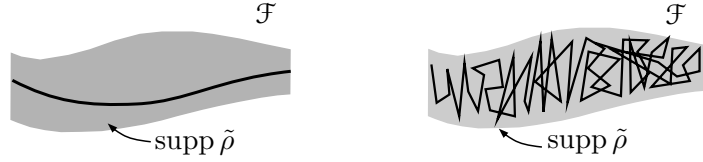
\begin{figure}
\begin{center}
\psscalebox{1.0 1.0} % Change this value to rescale the drawing.
{
\begin{pspicture}(0,28.643261)(9.296764,30.690586)
\definecolor{colour0}{rgb}{0.8,0.8,0.8}
\definecolor{colour1}{rgb}{0.7019608,0.7019608,0.7019608}
\pspolygon[linecolor=colour0, linewidth=0.02, fillstyle=solid,fillcolor=colour0](5.570035,29.212551)(5.658924,29.176996)(5.8233685,29.119217)(6.0189238,29.079218)(6.2322574,29.052551)(6.5033684,29.056995)(6.7744794,29.065884)(6.9967017,29.096996)(7.338924,29.154774)(7.570035,29.208107)(7.778924,29.265884)(8.112257,29.350328)(8.32559,29.399218)(8.583368,29.45255)(8.890035,29.510328)(9.107813,29.54144)(9.272257,29.554773)(9.276702,30.136995)(8.952257,30.203663)(8.605591,30.279217)(8.24559,30.34144)(7.978924,30.372551)(7.5744796,30.372551)(7.281146,30.34144)(6.9478126,30.25255)(6.6189237,30.136995)(6.3344793,30.06144)(5.9567018,29.999218)(5.703368,30.039217)(5.5744796,30.088106)
\pspolygon[linecolor=colour1, linewidth=0.02, fillstyle=solid,fillcolor=colour1](0.020034993,29.162552)(0.10892388,29.126995)(0.27336833,29.069218)(0.4689239,29.029217)(0.68225724,29.00255)(0.9533683,29.006996)(1.2244794,29.015884)(1.4467016,29.046995)(1.7889239,29.104773)(2.020035,29.158106)(2.2289238,29.215885)(2.5622573,29.30033)(2.7755907,29.349218)(3.0333683,29.402552)(3.340035,29.46033)(3.5578127,29.49144)(3.7222571,29.504774)(3.7267017,30.086996)(3.4022572,30.153662)(3.0555906,30.229218)(2.6955905,30.29144)(2.4289238,30.322552)(2.0244794,30.322552)(1.7311461,30.29144)(1.3978127,30.20255)(1.0689238,30.086996)(0.78447944,30.01144)(0.40670165,29.949217)(0.15336832,29.989218)(0.024479438,30.038107)
\psbezier[linecolor=black, linewidth=0.04](0.03670166,29.68255)(0.5844548,29.397636)(1.025005,29.335691)(1.6944795,29.368106553819445)(2.3639538,29.400522)(2.760531,29.695997)(3.7367017,29.802551)
\rput[bl](1.9311461,28.698107){$\supp \tilde{\rho}$}
\psbezier[linecolor=black, linewidth=0.02, arrowsize=0.05291667cm 2.0,arrowlength=1.4,arrowinset=0.0]{->}(1.8762708,28.864773)(1.6849209,28.885693)(1.584782,28.910864)(1.5000342,28.959277288050362)(1.4152865,29.00769)(1.3622024,29.054342)(1.3032857,29.109219)
\rput[bl](3.270035,30.439219){$\F$}
\psline[linecolor=black, linewidth=0.03](5.606702,29.893661)(5.6767015,29.383661)(5.8467016,29.493662)(5.816702,29.833662)(5.9767017,29.203663)(6.0467014,29.913662)(6.0967016,29.533663)(6.2267017,29.273663)(6.3067017,29.753662)(6.1567016,29.843662)(6.5067015,29.953663)(6.5567017,29.593662)(6.3267016,29.563662)(6.3867016,29.283663)(6.7167015,29.203663)(6.6967015,29.993662)(6.4867015,29.163662)(6.936702,29.993662)(6.8467016,29.403662)(6.896702,29.273663)(7.0567017,30.203663)(7.0967016,29.263662)(7.2067018,30.233662)(7.3267016,29.403662)(6.9867015,29.663662)(7.4767017,30.193663)(7.416702,29.383661)(7.666702,30.243662)(7.9067016,30.103662)(7.686702,29.843662)(7.9767017,29.613663)(7.5967016,29.413662)(7.6967015,30.063662)(8.106702,29.963661)(7.7367015,29.593662)(8.126701,29.473661)(8.156702,30.223661)(8.486702,30.213661)(8.346702,30.063662)(8.496701,29.773663)(8.256701,29.523663)(8.246701,30.193663)(8.436702,29.593662)(8.536701,30.053661)(8.756701,30.243662)(8.686702,29.983662)(8.016702,30.083662)(8.786701,29.633661)(8.811702,30.103662)(9.001701,30.123663)(8.531702,29.593662)(9.041701,29.673662)(8.951702,29.883661)(9.171701,30.103662)(9.171701,29.803661)(8.871701,29.733662)(8.621701,29.923662)(7.771702,30.183662)
\rput[bl](7.151146,28.708107){$\supp \tilde{\rho}$}
\psbezier[linecolor=black, linewidth=0.02, arrowsize=0.05291667cm 2.0,arrowlength=1.4,arrowinset=0.0]{->}(7.076271,28.794773)(6.884921,28.815693)(6.784782,28.840864)(6.700034,28.889277288050362)(6.6152864,28.93769)(6.5622025,28.984343)(6.503286,29.039217)
\rput[bl](8.720035,30.429218){$\F$}
\end{pspicture}
}
\end{center}
\caption{A measure obtained by fragmentation (left) and by holographic mixing with fluctuations
(right).}
\label{figholo}
\end{figure}

In order to understand the dynamics of a causal fermion system, it is more interesting
to consider convex combinations of measures that are {\em{not}} completely decoherent.
In order to explain the resulting effects in a simple example, suppose we choose electromagnetic
potentials~$A_1, \ldots, A_L$ in Minkowski space (which do not need to satisfy Maxwell's equations).
Constructing the regularized kernels~$P^\varepsilon_\as(x,y)$ (as explained in Chapters~\ref{secperturb} 
and~\ref{seccl}), one gets corresponding causal fermion systems described by measures~$\rho_\as$.
Abstractly, these measures can be written similarly as explained in the context of
the linearized field equations (see~\eqref{rhotildea} in Section~\ref{seclinfield}) as
\beq \label{tilrhocc}
\tilde{\rho} = \sum_{\as=1}^L (F_\as)_* \big( f_\as \, \rho \big) \:,
\eeq
where~$F_\as$ is the corresponding local correlation map, and~$f_\as$ is a weight function.
Since these measures are obtained from each other by small perturbations, it seems a good idea
to depict the corresponding supports~$M_\as := \supp \rho_\as$ as being close together
(see Figure~\ref{figfragment}~(b)). The convex combination of these measures~\eqref{convexcombi}
is referred to as a measure with {\em{fragmentation}}
\sindex{measure!with fragmentation}%
(see~\cite[Sections~1 and~5]{positive} or~\cite[Section~5]{perturb}).
The reason why we consider convex combinations (rather than general linear combinations)
is that we need to preserve the positivity of the measure.
In the limit when~$N$ gets large, the fragmented measure~$\tilde{\rho}$
goes over to a measure with enlarged support (see Figure~\ref{figfragment}~(c)).
Integrating with respect to this measure also involves an integration over the ``internal degrees of freedom''
corresponding to the directions that are transverse to~$M:= \supp \rho$
(see the left of Figure~\ref{figholo}).
This integral with respect to~$\tilde{\rho}$
bears similarity to the path integral formulation of quantum field theory
if one identifies the above-mentioned ``internal degrees of freedom'' with field configurations.

\section{The Mechanism of Holographic Mixing}
\sindex{holographic mixing}%
For the mathematical description of the interacting measure~$\tilde{\rho}$,
working with fragmented measures as introduced in the previous section does not seem to be the best method.
One difficulty is that it is a-priori not clear how large the number~$L$ of fragments is to be chosen. Moreover,
mechanisms where~$L$ changes dynamically are difficult to implement, at least perturbatively.
For these reasons, it seems preferable to describe~$\tilde{\rho}$ with a different method,
referred to as {\em{holographic mixing}}. At first sight, this method seems very different from fragmentation.
However, as we will explain at the very end of this section, fragmentation also allows for the description
of fragmented measures, at least if the construction is carried out in sufficiently large generality.
We now explain the general idea and a few related constructions.

Let~$(\H, \F, \rho)$ be a causal fermion system (for example, describing the Minkowski vacuum).
The {\em{wave evaluation operator}}~$\Psi$
\sindex{wave evaluation operator}%
\nindex{aio@$\Psi(x)$ -- wave evaluation operator at~$x \in M$}%
introduced in~\eqref{weo} is a mapping that to every vector in~$\H$ associates
the corresponding physical wave function (more details can be found, for example, in~\cite[\S1.1.4]{cfs})
\beq %\label{weo}
\Psi \::\: \H \rightarrow C^0(M, SM)\:, \qquad u \mapsto \psi^u \:, \eeq
where the physical wave function~$\psi^u$ is again given by~\eqref{psiudef}.
Evaluating at a fixed spacetime point gives the mapping
\beq \Psi(x) \::\: \H \rightarrow S_xM\:, \qquad u \mapsto \psi^u(x) \:. \eeq
Working with the wave evaluation operator makes it possible to write the kernel of the fermionic
projector~\eqref{Prep} in the short form (for a detailed proof, see~\cite[Lemma~1.1.3]{cfs})
\beq %\label{Pid}
P(x,y) = -\Psi(x)\, \Psi(y)^*\:. \eeq
The general procedure of holographic mixing is to replace the wave evaluation operator with
a linear combination of wave evaluation operators~$\Psi_\as$,
\beq \label{Psimix}
\tilde{\Psi} := \sum_{\as=1}^L \tilde{\Psi}_\as \:,
\eeq
which in turn are all obtained by perturbing~$\Psi$  (as will be described in more detail in~\eqref{DirBas}
below). Now we form the corresponding local correlation map,
\beq \tilde{F} \::\: M \rightarrow \F\:,\qquad \tilde{F}(x) := -\tilde{\Psi}(x)^*\, \tilde{\Psi}(x) \:, \eeq
and take the corresponding push-forward measure,
\beq \label{pushstar}
\tilde{\rho} := \tilde{F}_* \rho \:.
\eeq
In this way, we have constructed a new measure~$\rho$ that incorporates the
perturbations described all the wave evaluation operators~$\tilde{\Psi}_1, \ldots, \tilde{\Psi}_L$.
However, in contrast to the convex combination of measures~\eqref{tilrhocc}, the
support of the measure~\eqref{pushstar}, in general, does not decompose into several
fragments. In fact, if the mapping~$\tilde{F}$ is continuous, injective and closed,
the support of~$\tilde{\rho}$ will again be homeomorphic to~$M$.
In other words, the topological structure of spacetime remains unchanged
by the above procedure.

More concretely, the perturbed wave evaluation operators~$\Psi_\as$ can be obtained
as follows. Suppose that the causal fermion system~$(\H, \F, \rho)$ was constructed
similarly as in Section~\ref{seclco} from a system of Dirac wave functions satisfying,
for example, the Dirac equation
\beq \big( \Dir - m \big) \psi = 0 \:. \eeq
Then one can perturb the system by considering the Dirac equation
in the presence of classical potentials~$\B_1, \ldots, \B_L$,
\beq \label{DirBas}
\big( \Dir +\B_\as - m \big) \tilde{\psi}_\as = 0 \:.
\eeq
The corresponding wave evaluation operators~$\tilde{\Psi}_\as$ are built up of
all these Dirac solutions.
In this way, the resulting wave evaluation operator~\eqref{Psimix} involves
all the classical potentials~$\B_\as$. Qualitatively speaking, the resulting spacetime~$\tilde{M}$
can be thought of as being in a ``superposition'' of all these potentials. But this analogy
does not carry over to a more technical level.

As already mentioned after~\eqref{pushstar}, taking the push-forward with respect to a mapping~$F$
does not change the topological structure of spacetime. Even more, if~$F$ is
smooth and varies only on macroscopic scales, then
all microscopic structures of spacetime remain unchanged.
This does not account for the
picture of a measure~$\tilde{\rho}$ which accounts for additional ``internal degrees of freedom''
as shown in Figure~\ref{figfragment}~(c) and the left of Figure~\ref{figholo}.
In order to allow the description of such measures, one needs to consider mappings~$F$
that are {\em{not smooth}} but instead ``fluctuate'' on a microscopic scale
(as is shown symbolically on the right of Figure~\ref{figholo}).
If we allow for such fluctuations even on the Planck scale, then the procedure~\eqref{Psimix}
does allow for the description of all measures described previously with fragmentation~\eqref{tilrhocc}.
This considerations explains why the wave evaluation operators~$\tilde{\Psi}_\as$ should be constructed
not only by introducing classical potentials~\eqref{DirBas}, but, in addition, by introducing
small-scale fluctuations. This can be realized as follows. We choose operators~$A_\as$ on~$\H$ which add up to the identity,
\beq \sum_{\as=1}^N A_\as = \1 \:, \eeq
and then decompose the local correlation operator by multiplying from the right by~$A_\as$,
\beq \label{PsiAs}
\Psi_\as := \Psi A_\as \:.
\eeq
In the second step, the physical wave functions in~$\Psi_\as$ are perturbed by classical potentials~$A_\as$,
again by considering the Dirac equation~\eqref{DirBas}.
In the last step, we again take the sum of the wave evaluation operators~\eqref{Psimix}
and form the push-forward measure~\eqref{pushstar}. This procedure is referred to as
{\em{holographic mixing}}.

The resulting wave evaluation operator~$\tilde{\Psi}$ involves both the operators~$A_\as$
and the potentials~$\B_\as$. As explained in~\eqref{Prepmix} in the context of fragmentation,
the operators~$A_\as$ enter the kernel of the fermionic projector,
\beq %\label{Prepholo}
P(x,y) = -\sum_{\as, \bs=1}^N |\psi^{e_i}(x) \Sr\: (A_\as \,A_\bs^*)^i_j \: \Sl \psi^{e_j}(y)|  \:. \eeq
In this way, one can build phase factors into this kernel, possibly giving rise to destructive interference.
In other words, the wave evaluation operator~$\tilde{\Psi}$ is a sum of many, partly decoherent components.
The name ``holographic mixing'' is inspired by
the similarity to a hologram in which several pictures are stored, each of which becomes visible only
when looking at the hologram in the corresponding coherent light.

The above-mentioned ideas and constructions are implemented in the recent paper~\cite{fockdynamics}
in an enhanced way. The main improvement compared to the above description
is to build in {\em{current conservation}}. Indeed, forming the wave evaluation as a sum of
terms~\eqref{Psimix}, each being a solution of a different Dirac equation~\eqref{DirBas}
has the disadvantage that the conservation of the Dirac current
(which holds for each wave function~$\psi_a$) no longer holds for the sum. This is not satisfying, also
because we know from our general setup
that, even in the setting of general quantum spacetimes, there should be a conserved inner product
(namely the commutator inner product introduced in Section~\ref{seccip}).
In order to resolve this shortcoming, it is preferable to work with a single Dirac equation of the form
\beq \big( \Dir +\B - m \big) \tilde{\Psi} = 0 \:. \eeq
This is indeed possible if the operator~$\B$ is chosen as an integral operator
with integral kernel of the form
\beq \label{Bnonloc}
\B(x,y) = \sum_{\as=1}^N \B_\as \Big( \frac{x+y}{2} \Big) \: L_\as(y-x) \:,
\eeq
where~$\B_\as$ are again classical potentials and~$L_\as$ are certain symmetric kernels.
In this description, there is a conserved current and a corresponding conserved inner product
on the Dirac solutions which has a similar structure as the commutator inner product~\eqref{cli}.
We refer the interested reader for detailed explanations to~\cite{fockdynamics}.
We finally remark that the nonlocal operator~$\B$ of the form~\eqref{Bnonloc}
composed of many potentials~$\B_\as$ was also derived in~\cite{nonlocal} by a thorough
analysis of the linearized field equations for causal fermion systems describing Minkowski space.
The existence theory for the Dirac equation involving integral operators is studied in~\cite{cauchynonloc}.

\section{A Distinguished Quantum State}
\sindex{quantum state|textbf}%
The constructions outlined in the previous sections make it possible to construct
general measures~$\tilde{\rho}$ which go beyond measures describing a classical spacetime
with classical bosonic fields. The EL equations for these measures
can be understood as equations describing the dynamics in these generalized spacetimes.
With this in mind, the remaining question is how to interpret the resulting measure~$\tilde{\rho}$.
Can it be understood in terms of an interaction via quantum fields?
Or, in more physical terms, what does the measure~$\tilde{\rho}$ tell us about measurements
performed in the corresponding spacetime?
In order to address these questions in a systematic way, in~\cite{fockfermionic}
a {\em{distinguished quantum state}} is constructed. It describes how the interacting measure~$\tilde{\rho}$
looks like if measurements are performed at a given time using the objects of a causal fermion
describing the vacuum. This ``measurement'' can also be understood more generally as a
``comparison'' of the measures~$\tilde{\rho}$ and~$\rho$ at time~$t$.
In technical terms, the quantum state, denoted by~$\omega^t$, is a positive linear functional
on the algebra of fields~$\A$ of the non-interacting spacetime,
\beq \label{statedef}
\omega : \A \rightarrow \C \qquad \text{with} \qquad
\omega(A^* A) \geq 0 \quad \text{for all~$A \in \A$}\:.
\eeq
Here, we use the language of algebraic quantum field theory
(as introduced, for example, in the textbooks~\cite{baer+fredenhagen, brunettibook, rejzner}),
which seems most suitable for describing quantum fields in the needed generality.
This notion of a quantum state is illustrated in Exercise~\ref{exstate}.

We now outline the construction of the quantum state as given in~\cite{fockfermionic}.
We are given two causal fermion systems~$(\tilde{\H}, \tilde{\F}, \tilde{\rho})$
and~$(\H, \F, \rho)$ describing the interacting system and the vacuum, respectively.
Our goal is to ``compare'' these causal fermion systems at a given time.
In order to specify the time, we choose sets~$\tilde{\Omega} \subset \tilde{M}:= \supp \tilde{\rho}$
and~$\Omega \subset M:= \supp \rho$ which can be thought of as the past of this time
in the respective spacetime. We want to relate the two causal fermions systems
with the help of the nonlinear surface layer integral~\eqref{osinl} introduced in Section~\ref{secosinonlin}.
However, we need to take into account that the causal fermion systems are defined on two
different Hilbert spaces~$\tilde{\H}$ and~$\H$. Therefore, in order to make sense of the nonlinear surface layer integral, we need to identify the
Hilbert spaces~$\H$ and~$\tilde{\H}$ by a unitary transformation denoted by~$V$,
\beq %\label{Vident}
V : \H \rightarrow \tilde{\H} \qquad \text{unitary} \:. \eeq
Then the operators in~$\tilde{F}$ can be identified with operators in~$\F$ by the unitary transformation,
\beq %\label{Ftrans}
\F = V^{-1}\, \tilde{\F}\, V \:. \eeq
An important point to keep in mind is that this identification is not canonical, but it leaves the freedom
to transform the operator~$V$ according to
\beq \label{Vtrans}
V \rightarrow V \scrU \qquad \text{with} \qquad \scrU \in \Lin(\H) \text{ unitary} \:.
\eeq
The freedom in choosing~$\scrU$ must be taken into account in the nonlinear surface layer
integral, which now takes the form
\begin{align} %\label{osinlU}
&\gamma^{\tilde{\Omega}, \Omega}(\tilde{\rho}, \scrU \rho) \notag \\
&= \int_{\tilde{\Omega}} \dd\tilde{\rho}(x) \int_{M \setminus \Omega} \Diff\rho(y)\: \L(x, \scrU y \scrU^{-1})
- \int_{\Omega} \Diff\rho(x) \int_{\tilde{M} \setminus \tilde{\Omega}}  \dd\tilde{\rho}(y)\: \L(\scrU y \scrU^{-1},y) \:.
\end{align}

The method for dealing with the freedom in choosing~$\scrU$ is to integrate over the unitary
group. Moreover, it is preferable to consider the exponential of the nonlinear surface layer integral.
This leads us to introduce the {\em{partition function}}~$Z^{\tilde{\Omega}, \Omega}$ by
\sindex{partition function}%
\nindex{aez@$Z^{\tilde{\Omega}, \Omega}(\tilde{\rho}, \rho)$ -- partition function}%
\beq \label{Zdef}
Z^{\tilde{\Omega}, \Omega}\big( \beta, \tilde{\rho} \big) = \fint_\G \exp \Big( \beta \,\gamma^{\tilde{\Omega}, \Omega} \big(\tilde{\rho}, \scrU \rho \big) \Big) \: \Diff\mu_\G(\scrU) \:,
\eeq
where~$\mu_\G$ is the normalized Haar measure on the unitary group
(in order for this Haar measure to be well-defined, one needs to assume that the Hilbert space~$\H$
is finite-dimensional, or else one must exhaust~$\H$ by finite-dimensional subspaces).

In analogy to the path integral formulation of quantum field theory, the quantum state is obtained by
introducing insertions into the integrand of the partition function, that is, symbolically,
\beq \label{statesymbol}
\omega(\cdots) = \frac{1}{Z^{\tilde{\Omega}, \Omega}\big( \beta, \tilde{\rho} \big)}
\fint_\G (\cdots) \exp \Big( \beta \,\gamma^{\tilde{\Omega}, \Omega} \big(\tilde{\rho}, \scrU \rho \big) \Big) \: \Diff\mu_\G(\scrU) \:.
\eeq
These insertions have the structure of surface layer integrals involving linearized solutions
in the vacuum spacetime. Likewise, the argument of the state on the left-hand side is formed of
operators that are parametrized by the same linearized solutions which enter the insertions on the right side.
More precisely, they are operators of the field algebra~$\A$, being defined as the $*$-algebra
generated by the linearized solutions, subject to the canonical commutation and anti-commutation
relations. The commutation relations involve the causal fundamental solution
of the linearized solutions which can be constructed with energy methods as outlined in Chapter~\ref{seclinhyp}
(for details see~\cite{linhyp}). Likewise, for the anti-commutation relations, we use the
causal fundamental solutions of the dynamical wave equation mentioned at the end of Section~\ref{seccip}
in~\eqref{dwe} (for more details see~\cite{dirac}).
The positivity property of the state is ensured by the specific form of the insertions.
We refer the interested reader to~\cite{fockfermionic}.
We remark that, as is worked out in~\cite{fockentangle}, the above quantum state allows for the
description of general entanglement. Moreover, the dynamics of the quantum state is studied in~\cite{fockdynamics}.
%For more details on the connection to quantum field theory we refer to~\cite{qftlimit}.

We finally note that the definition of the partition function~\eqref{Zdef} and of the insertions
in the definition of the state~\eqref{statesymbol} bears a similarity with the path integral
formulation of quantum theory (see, for example, \cite{kleinert, glimm+jaffe}). However, this similarity
does not seem to go beyond formal analogies. In particular, one should keep in mind that,
in contrast to the integral over field configurations in the path integral formulation,
in~\eqref{Zdef} one integrates over the unitary transformations arising from the freedom in
identifying the Hilbert spaces~$\H$ and~$\tilde{\H}$ (see~\eqref{Vtrans}).
This is a major conceptual difference that, at least at present, prevents us from getting a tighter
connection to path integrals and the functional integral approach.

\section{Exercises}

\begin{Exercise} \label{exstate}
\sindex{quantum state}%
{\em{ The purpose of this exercise is to get familiar with
the notion of a {\em{quantum state}} as defined by~\eqref{statedef}.
 In quantum mechanics, the system is usually described by a unit vector~$\psi$
in a Hilbert space~$(\H, \la .|. \ra)$. An observable corresponds to a symmetric operator~$A \in \Lin(\H)$
on this Hilbert space (for simplicity, we here restrict attention to bounded operators). The expectation value of a measurement is given by the expectation~$\la \psi | A | \psi \ra$.
\bitem
\item[(a)] Show that the linear operator~$W \in \Lin(\H)$ defined by
\beq \label{Wpure}
W \phi = \la \phi | \psi \ra\: \psi \qquad \text{or, in bra/ket notation,} \qquad
W = |\psi \ra \la \psi |
\eeq
is a projection operator (i.e.\ it is symmetric and idempotent). Show that the expectation value of a measurement
can be written as
\beq \la \psi | A | \psi \ra = \tr_\H \big( W A \big) \:. \eeq
\item[(b)] Show that the mapping
\beq \label{omegaex}
\omega \::\: A \mapsto \tr_\H \big(W A \big)
\eeq
is a quantum state in the sense~\eqref{statedef} (here for the
algebra~$\A$, we take the {\em{algebra of observables}}; that is, the set of all operators obtained from
all observables by taking products and linear combinations).
\item[(c)] Let~$\psi_1$ and~$\psi_2$ be two distinct, non-zero vectors of~$\H$. Show that, choosing
\beq \label{Wmixed}
W := |\psi_1 \ra \la \psi_1| + |\psi_2 \ra \la \psi_2| \:,
\eeq
the mapping~\eqref{omegaex} again defines a quantum state in the sense~\eqref{statedef}.
Show that this quantum state cannot be written in the form~\eqref{Wpure}.
\sindex{quantum state!pure}%
\sindex{quantum state!mixed}%
One refers to~\eqref{Wpure} as a {\em{pure state}}, whereas~\eqref{Wmixed} is a {\em{mixed state}}.
\item[(d)] Is the quantum state in~(c) properly normalized in the sense that~$\omega(\1)=1$?
If not, how can this normalization be arranged?
\eitem
}} \end{Exercise}

%%% Local Variables:
%%% mode: latex
%%% TeX-master: t
%%% End:

%% file: appendix.tex
%!TEX root = intro.tex
\appendix
\chapter{The Spin Coefficients}
\sindex{spin coefficient}%
\label{Anhang_eichfix}
In this appendix, we verify by explicit computation that the
matrices~$E_j$ containing the spin coefficients as given by~\eqref{Ejform},
\beq \label{AZ}
    E_j = \frac{\cI}{2}\: \pseudo \, \partial_j\pseudo - \frac{\cI}{16} \Tr
    \left( G^m \, \nabla_j G^n \right) G_m G_n + \frac{\cI}{8} \Tr \left( \pseudo
    G_j \,\nabla_m G^m \right) \pseudo \:,
\eeq
have the following behavior under gauge transformations:
\begin{align}
E_j &\rightarrow U E_j U^{-1} && {\mbox{for~$\U(1)$ gauge transformations}} \label{Aa} \\
E_j &\rightarrow U E_j U^{-1} + \cI U (\partial_j U^{-1}) && {\mbox{for~$\SU(2,2)$ gauge transformations}}
\label{Ab}
\end{align}
Under $U(1)$ gauge transformations, all the terms in~\eqref{AZ} remain unchanged
because~$U$ and its partial derivatives commute with~$\pseudo$ as well as with the~$G^j$.
Therefore, the relation~\eqref{Aa} is obvious.
Thus it remains to consider $\SU(2,2)$ gauge transformations.
Our goal is to verify~\eqref{Ab} for a fixed spacetime point~$p$.

We decompose the gauge transformation~$U$ as~$U = U_2 \: U_1$ with
\begin{align}
   U_1(x) &= U(p)   \\
   U_2(x) &= U(x) \: U^{-1}(p) \:.
\end{align}
Being constant, the first transformation clearly satisfies the
transformation law~\eqref{Ab}. Therefore, it suffices to consider
a gauge transformation~$U$ with~$U(p)=\1$.
Then~\eqref{Ab} can be written as
\beq \label{Af}
    \tilde{E}_j = E_j - \cI \partial_j U \:.
\eeq

We now compute the transformation law of each of the summands in~\eqref{AZ}
after each other:
\begin{itemize}[leftmargin=2em]
 \item[(1)] $\displaystyle \frac{\cI}{2} \:\pseudo\, \partial_j \pseudo$
    \begin{itemize}[leftmargin=2.5em]
      \item[(i)] odd transformations:
\begin{align}
   \frac{\cI}{2} \:\tilde{\pseudo} \, \partial_j \tilde{\pseudo}
        &= \frac{\cI}{2}\: \pseudo \left( \partial_j \pseudo + [\partial_j U, \pseudo] \right) \notag \\
        &= \frac{\cI}{2}\: \pseudo \, \partial_j \pseudo
            + \frac{\cI}{2}\: \pseudo \,\Big( (\partial_j U)\, \pseudo - \pseudo \,(\partial_j U) \Big) \notag \\
        &= \frac{\cI}{2}\: \pseudo \,\partial_j \pseudo - i \pseudo \pseudo\, \partial_j U   \notag \\
        &= \frac{\cI}{2}\: \pseudo \,\partial_j \pseudo - \cI \partial_j U
\end{align}
      \item[(ii)] even transformations:
\beq  \frac{\cI}{2} \:\tilde{\pseudo}\, \partial_j \tilde{\pseudo}
      = \frac{\cI}{2} \:\pseudo \Big( \partial_j \pseudo + \big[ \partial_j U, \pseudo \big] \Big)
      = \frac{\cI}{2} \: \pseudo \, \partial_j \pseudo \eeq
    \end{itemize}
Thus, for odd transformations, we get the correct transformation law,
whereas for even transformations, the desired term~$\cI \partial_j U$ is still missing.
 \item[(2)] $\displaystyle -\frac{\cI}{16}\: \Tr( G^m \, \nabla_j G^n)\, G_m G_n$ \\[-0.5em]
    \begin{itemize}[leftmargin=2.5em]
\item[(i)] odd transformations:
\begin{align}
-\frac{\cI}{16} & \Tr \left( \tilde{G}^m \, \nabla_j \tilde{G}^n \right) \tilde{G}_m \tilde{G}_n \notag \\
     & =  -\frac{\cI}{16} \Tr \left( G^m \, \nabla_j G^n \right) G_m G_n \\
      &\quad\: -\frac{\cI}{16} \Tr \Big( G^m \big[\partial_j U, G^n \big] \Big) G_m G_n \label{tracezero} \\
     & =  -\frac{\cI}{16} \Tr \left( G^m \, \nabla_j G^n \right) G_m G_n \:,
\end{align}
where we used that~$G^m$, $G^n$ and $\partial_j U$ are odd,
implying that the trace in the last summand in~\eqref{tracezero} vanishes.
 \item[(ii)] $\partial_j U = \cI \sigma_{kl}$ for arbitrary indices~$k$,$l$:
\begin{align}
-\frac{\cI}{16} & \Tr \left( \tilde{G}^m \, \nabla_j \tilde{G}^n \right) \tilde{G}_m \tilde{G}_n \notag \\
     &= -\frac{\cI}{16} \Tr \left( G^m \,\nabla_j G^n \right) G_m G_n
        -\frac{\cI}{16} \Tr \left( G^m \big[i \sigma_{kl} , G^n \big] \right) G_m G_n \notag \\
     &= -\frac{\cI}{16} \Tr \left( G^m \,\nabla_j G^n \right) G_m G_n
        -\frac{\cI}{16} \Tr \left( [G^n,G^m] \:i \sigma_{kl} \right) G_m G_n \notag \\
     &= -\frac{\cI}{16} \Tr \left( G^m \, \nabla_j G^n \right) G_m G_n
        -\frac{\cI}{8} \Tr \left( \sigma^{mn} \:i \sigma_{kl} \right) \sigma_{mn} \notag \\
     &= -\frac{\cI}{16} \Tr \left( G^m \:\nabla_j G^n \right) G_m G_n - \cI \partial_j U
\end{align}
      \item[(iii)] $\partial_j U = \pseudo$:
\begin{align}
-\frac{\cI}{16} &\Tr \left( \tilde{G}^m \: \nabla_j \tilde{G}^n \right) \tilde{G}_m \tilde{G}_n \notag \\
    &= -\frac{\cI}{16} \Tr \left( G^m \:\nabla_j G^n \right) G_m G_n
        -\frac{\cI}{16} \Tr \left( G^m \,\big[\pseudo, G^n \big] \right) G_m G_n  \notag \\
    &= -\frac{\cI}{16} \Tr \left( G^m \:\nabla_j G^n \right) G_m G_n
        -\frac{\cI}{16} \Tr \left( \pseudo \,[G^n,G^m] \right) G_m G_n   \notag \\
    &= -\frac{\cI}{16} \Tr \left( G^m \, \nabla_j G^n \right) G_m G_n \:,
\end{align}
because~$\Tr \left( \pseudo \sigma^{mn} \right)=0$ for all~$m$,$n$.
    \end{itemize}
Thus we get the correct transformation law for bilinear
transformations~$\partial_j U = \cI \sigma_{kl}$.
 \item[(3)] $\displaystyle \frac{\cI}{8} \Tr \left( \pseudo \,G_j \:\nabla_m G^m \right) \pseudo$
\begin{align}
\frac{\cI}{8} \Tr \left( \tilde{\pseudo} \,\tilde{G}_j \:\nabla_m \tilde{G}^m \right) \tilde{\pseudo}
    &= \frac{\cI}{8} \Tr \left( \pseudo \,G_j \:\nabla_m G^m \right) \pseudo
        + \frac{\cI}{8} \Tr \left( \pseudo \,G_j \,[\partial_m U,G^m] \right) \pseudo    \notag \\
    &= \frac{\cI}{8} \Tr \left( \pseudo \,G_j \:\nabla_m G^m \right) \pseudo
        + \frac{\cI}{8} \Tr \left( \partial_m U \, [G^m, \pseudo G_j] \right) \pseudo   \notag \\
    &= \frac{\cI}{8} \Tr \left( \pseudo \,G_j\: \nabla_m G^m \right) \pseudo
        - \frac{\cI}{4} \Tr \left( \partial_m U \,\pseudo \,\delta^m_j \right) \pseudo    \notag \\
    &= \frac{\cI}{8} \Tr \left( \pseudo \,G_j \:\nabla_m G^m \right) \pseudo
        - \frac{\cI}{4} \Tr \big( (\partial_j U)\: \pseudo \big) \pseudo \:,
\end{align}
where we used the relations
\beq \big[ G^j, \pseudo G^k \big] =-\pseudo\: \big\{ G^j, G^k \big\}  \eeq
as well as the anti-commutation relations for Dirac matrices.
We again distinguish different cases:
\begin{itemize}
\item[(i)]  $\partial_j U$ is odd or~$\partial_j U = i \sigma_{kl}$:
\beq \frac{\cI}{8} \Tr \left( \tilde{\pseudo}\, \tilde{G}_j\: \nabla_m \tilde{G}^m \right) \tilde{\pseudo}
= \frac{\cI}{8} \Tr \left( \pseudo \,G_j \: \nabla_m G^m \right) \pseudo \eeq
      \item[(ii)]  $\partial_j U = \pseudo$:
\beq \frac{\cI}{8} \Tr \left( \tilde{\pseudo}\, \tilde{G}_j\: \nabla_m \tilde{G}^m \right) \tilde{\pseudo}
= \frac{\cI}{8} \Tr \left( \pseudo \,G_j \: \nabla_m G^m \right) \pseudo - \cI \partial_j U   \eeq
\end{itemize}
Hence, we get the correct transformation law if~$\partial_j U = \pseudo$.
\end{itemize}
Adding all the terms gives the desired transformation law~\eqref{Af}.

%% file: backmatter.tex
%!TEX root = intro.tex
\chapter*{Back Cover}

\centerline{\large{\bf{About this book}}}

\vspace*{.5em}
This textbook introduces the basic concepts of the theory of causal fermion systems, a recent approach to the description of fundamental physics. The theory yields quantum mechanics, general relativity and quantum field theory as limiting cases and is therefore a candidate for a unified physical theory. From the mathematical perspective, causal fermion systems provide a general framework for describing and analyzing non-smooth geometries and ``quantum geometries''. The dynamics is described by a novel variational principle, the causal action principle.

The book includes a detailed summary of the mathematical and physical preliminaries. It explains the physical concepts behind the causal fermion system approach from the basics. Moreover, all the mathematical objects and structures are introduced step by step. The mathematical methods used for the analysis of causal fermion systems and the causal action principle are introduced in depth. Many examples and applications are worked out.

The textbook is addressed to master and graduate students in mathematics or physics. Furthermore, it serves as a reference work for researchers working in the field.

\vspace*{3em}
\centerline{\bf{\large{About the authors}}}
\vspace*{.5em}

Felix Finster studied physics and mathematics at the University of Heidelberg, where he graduated in 1992 with Claus Gerhardt and Franz Wegner. In 1992-1995 he wrote his PhD thesis at ETH Z\"urich with Konrad Osterwalder. In 1996-1998 he was a post-doc with Shing-Tung Yau at Harvard University. From 1998-2002 he was member of the Max Planck Institute for Mathematics in the Sciences in Leipzig in the group of Eberhard Zeidler. Since 2002 he has been a full professor of mathematics at the University of Regensburg. He works on problems in general relativity and quantum field theory. 

Sebastian Kindermann studied physics and mathematics at the University of Regensburg,
where he graduated in 2020. He is teacher at the Comenius Gymnasium in Deggendorf.

Jan-Hendrik Treude studied physics and mathematics at the University of Freiburg and the University of Vienna. He 
wrote his PhD thesis at the University of Regensburg, graduating in~2015. 
Currently, he belongs to the permanent academic staff and is department manager (Fach\-be\-reichs\-re\-fe\-rent)
of the Department of Mathematics and Statistics at the University of Konstanz.